%% file: thesis.tex
\documentclass[a4paper,oneside,11pt]{book}
\usepackage{epsfig}
\usepackage{pslatex}
\usepackage{graphics,amsmath,amssymb,graphicx, upgreek,natbib,bm,multirow}
\usepackage[bookmarks=false]{hyperref}
\newcommand{\Rff}[1]{Figure \ref{#1}}
\newcommand{\mbold}[1]{\mbox{\emph{\textbf{#1}}}}
\newcommand{\rf}[1]{eq.~(\ref{#1})}
\newcommand{\rs}[1]{sec.~\ref{#1}}
\newcommand{\rff}[1]{fig.~\ref{#1}}
\newcommand{\rft}[1]{tab.~\ref{#1}}
\newcommand{\rfch}[1]{chap.~\ref{#1}}

\newcommand{\lb}{\;\;\;\;\;}

\mathchardef\mhyphen="2D

\begin{document}
\frontmatter

\pagenumbering{roman}
\pagestyle{empty}
\begin{center}
\fontsize{18pt}{18pt}\selectfont Charles University in Prague\\
\vspace{1cm}
\fontsize{18pt}{18pt}\selectfont Faculty of Mathematics and Physics\\
\vspace{1cm}
\fontsize{20pt}{20pt}
DOCTORAL THESIS

\vspace{1.3cm} \resizebox{0.35\hsize}{!}{\includegraphics{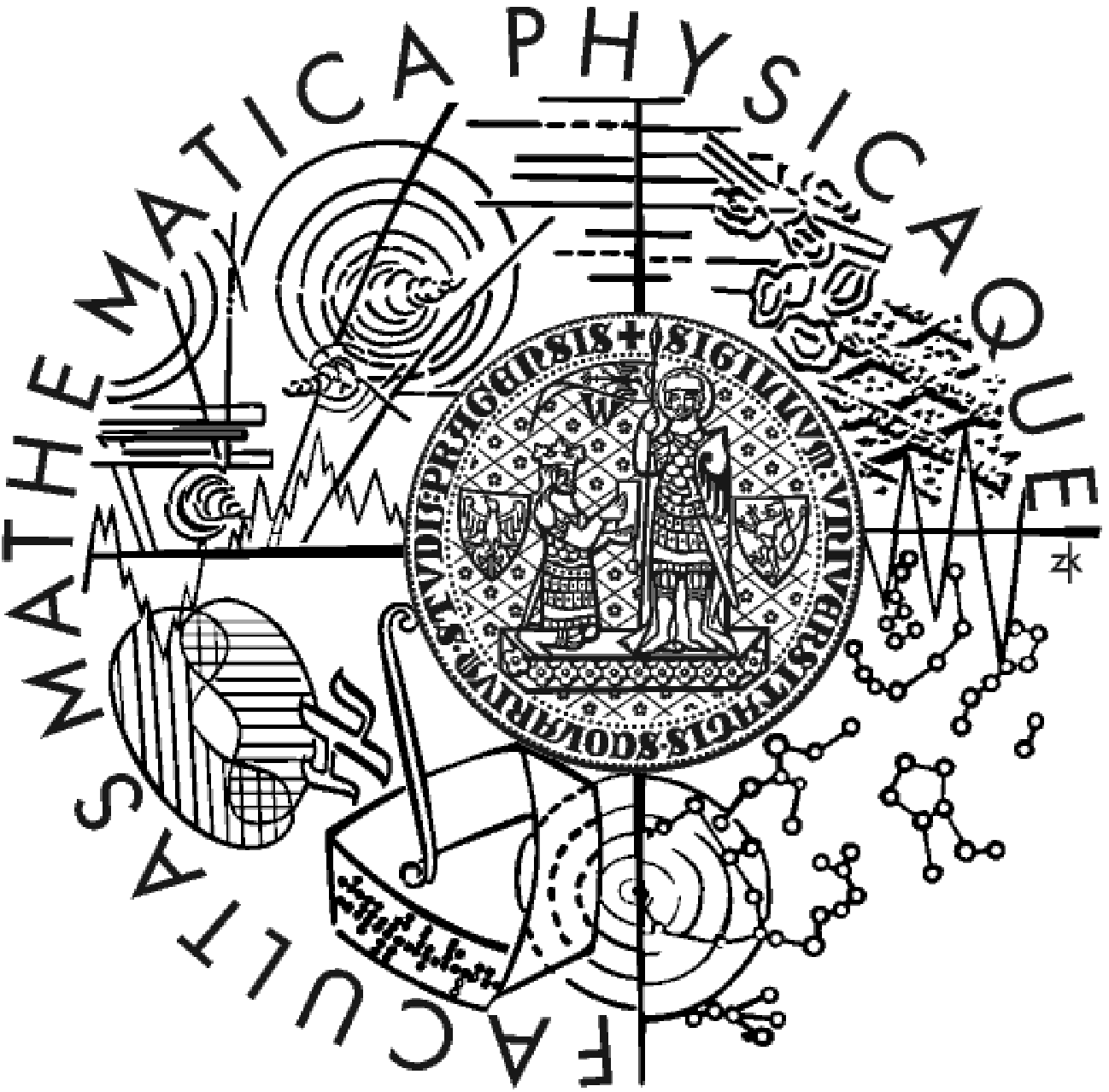}}

\vspace{1.5cm} 
\fontsize{18pt}{18pt}\selectfont{Ond\v{r}ej Kop\'{a}\v{c}ek}\\
\vspace{1cm}
\fontsize{20pt}{20pt}\selectfont{\textbf{Transition from regular to chaotic motion\\ in black hole magnetospheres}}

\vspace{1.1cm}
\fontsize{14pt}{14pt}\selectfont{Astronomical Institute of the Academy of Sciences\\ of the Czech Republic}  \\

\vspace{1cm}
\fontsize{14pt}{14pt}\selectfont{Supervisor: doc.\,RNDr.\,Vladim\'{i}r Karas, DrSc.}\\
\vspace{1.9cm}Prague 2011
\end{center}

\newpage
\pagestyle{plain} 
\noindent  The thesis summarizes the results of the research conducted during the doctoral studies of the branch Theoretical physics, Astronomy and Astrophysics at the Faculty of Mathematics and Physics of Charles University in Prague during the years 2007-2011.

\vspace{3cm}
\noindent\textbf{Student}\\[2mm] 
Mgr. Ond\v{r}ej Kop\'{a}\v{c}ek\\

\vspace{.5cm}
\noindent\textbf{Supervisor}\\[2mm]
doc.\,RNDr.\,Vladim\'{i}r Karas, DrSc.\\
Astronomical Institute of the Academy of Sciences of the Czech Republic\\

\vspace{.5cm}
\noindent\textbf{Affiliation}\\[2mm] 
Astronomical Institute of the Academy of Sciences of the Czech Republic\\Bo\v{c}n\'{i} II 1401/1a, 14131 Prague\\

\vspace{.5cm}
\noindent\textbf{Referees}\\[2mm] 
Prof. Luciano Rezzolla, Ph.D. \\Albert Einstein Institut, Potsdam\\[4mm]
Prof. RNDr. Petr Kulh\'{a}nek, CSc. \\FEL \v{C}VUT, Prague\\
\vspace{3cm}

\noindent
The thesis was successfully defended on 14 September 2011 in Prague before the committee of the study branch 4F1 (Theoretical physics, Astronomy and Astrophysics).
\newpage
\pagestyle{plain} 
\addcontentsline{toc}{section}{Acknowledgements}
\noindent \begin{center}{\large\bf Acknowledgements}\end{center}
I would like to thank my supervisor, Dr. Vladim\'{i}r Karas, for the patient guidance, encouragement
and advice he has provided throughout my doctoral studies.

\newpage \pagestyle{plain} 

\vspace*{14cm} \noindent
I declare that I carried out this doctoral thesis independently, and only with the cited
sources, literature and other professional sources.

I understand that my work relates to the rights and obligations under the Act No.
121/2000 Coll., the Copyright Act, as amended, in particular the fact that the Charles
University in Prague has the right to conclude a license agreement on the use of this
work as a school work pursuant to Section 60 paragraph 1 of the Copyright Act.

\vspace{1.5cm}
\line(1,0){120} \hfill \line(1,0){120} \hspace*{0.85cm}\\
\noindent \hspace*{1.5cm} in Prague \hfill Ond\v{r}ej
Kop\'{a}\v{c}ek \hspace*{1.5cm}
\newpage \pagestyle{plain} 
\addcontentsline{toc}{section}{Abstract in Czech}
\noindent N\'{a}zev pr\'{a}ce: \textit{Regul\'{a}rn\'{i} a chaotick\'{e} pohyby v magnetosf\'{e}\v{r}e \v{c}ern\'{y}ch d\v{e}r}\\
Autor: \textit{Ond\v{r}ej Kop\'{a}\v{c}ek} \\
Katedra\,(\'{u}stav): \textit{Astronomick\'{y} \'{u}stav Univerzity Karlovy}\\
\v{S}kolic\'{i} pracovi\v{s}t\v{e}: \textit{Astronomick\'{y} \'{u}stav Akademie v\v{e}d \v{C}esk\'{e} republiky}\\
Vedouc\'{i} doktorsk\'{e} pr\'{a}ce: \textit{doc.\,RNDr.\,Vladim\'{i}r Karas, DrSc., Astronomick\'{y} \'{u}stav Akademie v\v{e}d \v{C}esk\'{e} republiky, Praha}\\
Abstrakt:  \textit{P\r{u}soben\'{i} siln\'{e} gravitace v okol\'{i} \v{c}ern\'{y}ch d\v{e}r m\r{u}\v{z}e v\'{e}st k urychlen\'{i} hmoty. V t\'{e}to pr\'{a}ci zkoum\'{a}me vlastnosti syst\'{e}mu tvo\v{r}en\'{e}ho rotuj\'{i}c\'{i} \v{c}ernou d\'{i}rou ve velkorozm\v{e}rov\'{e}m uspo\v{r}\'{a}dan\'{e}m magnetick\'{e}m poli. Nabit\'{e} \v{c}\'{a}stice v bl\'{i}zkosti horizontu jsou krom\v{e} siln\'{e} gravitace ovliv\v{n}ov\'{a}ny magnetick\'{y}m polem a indukovan\'{y}m polem elektrick\'{y}m. Oproti ji\v{z} d\v{r}\'{i}ve v literatu\v{r}e diskutovan\'{y}m situac\'{i}m p\v{r}id\'{a}v\'{a}me n\v{e}kter\'{a} v\'{y}znamn\'{a} zobecn\v{e}n\'{i}. Magnetick\'{e} pole v na\v{s}em p\v{r}\'{i}pad\v{e} nemus\'{i} b\'{y}t koaxi\'{a}l\-n\'{i}~s ro\-ta\v{c}n\'{i} osou \v{c}ern\'{e} d\'{i}ry, tak\v{z}e syst\'{e}m ztr\'{a}c\'{i} osovou symetrii. Krom\v{e} toho p\v{r}ed\-po\-kl\'{a}\-d\'{a}me trans\-la\v{c}n\'{i} pohyb \v{c}ern\'{e} d\'{i}ry s obecn\'{y}m sm\v{e}rem i rychlost\'{i}. Uk\'{a}\v{z}eme, \v{z}e d\'{i}ky tomu do\-ch\'{a}z\'{i} k nov\'{y}m efekt\r{u}m. V komplikovan\'{e} struktu\v{r}e v\'{y}sledn\'{e}ho magnetick\'{e}ho pole pozorujeme v ergo\-sf\'{e}\v{r}e jeho rychl\'{e} prostorov\'{e} zm\v{e}ny prov\'{a}zen\'{e} vznikem nulov\'{y}ch bod\r{u}, kter\'{e} dokazuj\'{i}, \v{z}e gravita\v{c}n\'{i} p\r{u}soben\'{i} rotuj\'{i}c\'{i}ho zdroje m\r{u}\v{z}e podn\v{e}covat rekonekci magnetick\'{y}ch silo\v{c}ar. D\'{a}le zkoum\'{a}me dynamick\'{e} vlastnosti nabit\'{y}ch \v{c}\'{a}stic vystaven\'{y}ch p\r{u}soben\'{i} tohoto typu pol\'{i}. P\v{r}edev\v{s}\'{i}m se zaj\'{i}m\'{a}me o p\v{r}echody mezi regul\'{a}rn\'{i}m re\v{z}imem a deterministick\'{y}m chaosem, ke kter\'{y}m do\-ch\'{a}\-z\'{i} v z\'{a}vislosti na volb\v{e} parametr\r{u}. P\v{r}i numerick\'{e}m zkoum\'{a}n\'{i} \v{c}\'{a}sticov\'{e} dynamiky aplikujeme v kontextu obecn\'{e} relativity  zat\'{i}m nepou\v{z}itou metodu rekuren\v{c}n\'{i} anal\'{y}zy.}\\
Kl\'{i}\v{c}ov\'{a} slova: \textit{obecn\'{a} relativita, kompaktn\'{i} t\v{e}lesa,  astrofyzik\'{a}ln\'{i} kor\'{o}na, deterministick\'{y} chaos} \\

\newpage\pagestyle{plain}
\addcontentsline{toc}{section}{Abstract in English}
\noindent Title: \textit{Transition from regular to chaotic motion in black hole magnetospheres}\\
Author: \textit{Ond\v{r}ej Kop\'{a}\v{c}ek}\\
Department: \textit{Astronomical Institute of Charles University}\\
Affiliation: \textit{Astronomical Institute of Academy of Sciences of the Czech Republic}\\
Supervisor: \textit{doc.\,RNDr.\,Vladim\'{i}r Karas, DrSc., Astronomical Institute of the Aca\-demy of Sciences of the Czech Republic, Prague}\\
Abstract: \textit{Cosmic black holes can act as agents of particle acceleration. We study properties of a system consisting of a rotating black hole immersed in a large-scale organized magnetic field. Electrically charged particles in the immediate neighborhood of the horizon are influenced by strong gravity acting together with magnetic and induced electric components. We relax several constraints which were often imposed in previous works: the magnetic field does not have to share a common symmetry axis with the spin of the black hole but they can be inclined with respect to each other, thus violating the axial symmetry. Also, the black hole does not have to remain at rest but it can instead perform fast translational motion together with rotation. We demonstrate that the generalization brings new effects. Starting from uniform electro-vacuum fields in the curved spacetime, we find separatrices and identify magnetic neutral points forming in certain circumstances. We suggest that these structures can represent signatures of magnetic reconnection triggered by frame-dragging effects in the ergosphere. We further investigate the motion of charged particles in these black hole magnetospheres. We concentrate on the transition from the regular motion to chaos, and in this context we explore the characteristics of chaos in relativity. For the first time, we apply recurrence plots as a suitable technique to quantify the degree of chaoticness near a black hole.}


\newpage \pagestyle{plain} 
\pagestyle{headings}
\tableofcontents
\newpage

\mainmatter
\setcounter{page}{1}
\pagenumbering{arabic}

\input chap1.tex

\input chap2n.tex

\input chap3nn.tex
\input chap4n.tex
\input chap5n.tex
\addcontentsline{toc}{chapter}{Appendices}
\appendix
\input appendix.tex
\input appendix2n.tex
\input appendix3.tex

\end{document}

%% file: chap1.tex
\chapter{\sffamily{Introduction}}
\section{Astrophysical black holes}
Nowadays it is a consensus that black holes are a vivid part of
physical reality. Astrophysicist's attitude toward the possibility of a
real existence of completely gravitationally collapsed bodies, i.e.
black holes (BHs), has undergone profound changes since 1935 when sir
Eddington commented recent theoretical results of \citet{chandra35}, suggesting that black hole could be the endpoint of star evolution, by these words: ``I think there should be a law of Nature
to prevent a star from behaving in this absurd way''. At present, due to discoveries and observations made in recent decades, it seems highly probable that there is actually no such power which would save a heavy star from collapsing into the black hole.


Until the late 1960's there had not been much progress made in this
field. Although Karl Schwarzschild had given an exact solution to
Einstein's equations for a spherically symmetrical source
(Swcharzschild black hole in the case of collapsed body) already in
1916 (just one year after publishing the theory of relativity), it was not
believed that such objects really exist. Almost a half century later, \citet{kerr63} gave a new exact solution to Einstein's field equations
describing the geometry around a rotating compact object, which
naturally appears to be more relevant in astrophysical context than
prior non-rotating Schwarzschild solution. But this progress still
would not become a concern of astrophysicists unless there had not
been made important observational discoveries during 1960's.

Astronomers discovered sources of radio waves with highly
redshifted (with redshift factor $z \approx 0.1 - 6$) spectrum \citep{schmidt63} which
proved them (according to Hubble's law) to be very distant from us
($\approx 0.3 - 3 \:\mbox{Gpc}$ ). Combined with observed visual
magnitude of these objects -- called \textbf{quasars} (quasi-stellar
radio sources) -- there comes a conclusion of their huge power output
of about 100 times that of the total luminosity of average galaxy
($L_{\rm{quasar}} \approx 10^{35}-10^{40}\:\rm{W}$). Quasars also
appeared to emit significantly in X-ray and even in gamma part of
the spectrum.

Quasars were not the only peculiar objects observed during those
decades. According to the type of their spectrum and luminosity time
dependence of those ``new'' objects they were classified as
\textbf{Seyfert galaxies} (being observed and studied since the
1940's actually), \textbf{blazars}, or \textbf{radio galaxies} (RGs)
with all of them consisting of further subclasses. They are all
characterized by extraordinarily high luminosity coming from a small
volume ($\approx10^{-6} \:\mbox{pc}^{3}$). As there is no
consistent way to explain the mechanism of those energy sources
conventionally (i.e. regarding stars and therein running nuclear
synthesis as the most efficient energy source in the universe),
attention was turned to hypothesis employing strong gravitational
fields considering compact objects and subsequently the black holes.
Now it is generally believed that above mentioned phenomenons are of
the same origin which was given the name \textbf{active galactic nuclei} (AGN).
Galaxies with active galactic nuclei and subsequently also those with non-active
nuclei are suspected of hosting a supermassive black hole (MBH) of $M \approx
10^{6}-10^{9}\:M_{\odot}$ in their centres, e.g. M87 -- AGN of
$\approx 10^9\:M_{\odot}$ or Sagittarius A* -- nonactive nucleus of
our Galaxy -- with $M \approx 4.4\times10^6\: M_{\odot}$ \citep{genzel10}.

\begin{figure}[h]
    \centering
        \includegraphics[scale=.12,clip]{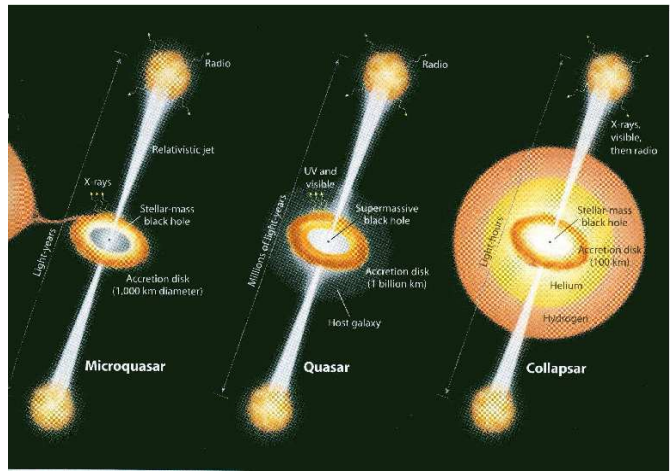}
\caption{Same physical mechanism is supposed to operate in different types of cosmic objects. Central black hole is accreting matter which forms an accretion disk. Collimated jets are launched from the central part of the system. Illustration credit: \citet{mirabel07}.}
    \label{model}
\end{figure}

The main reason for such an assumption is that it provides a clarification of the observed
luminosity of active galaxies. When the Kerr black hole is employed to
model the properties of AGN we conclude that the matter accreted
from its vicinity would form an \textbf{accretion disk} which becomes
heated by accretion process and subsequently emits radiation of
various types (depending on temperature and many other properties of
the disk). Doing so there could be as much as $\approx 40\%$
(\cite{mtw} p. 885) of accreted material rest mass turned into
emitted radiation which is considerably more than in the case of
thermonuclear synthesis of helium ($\approx 7\%$ of the rest mass).

Whether we regard observed AGN as quasar, radio galaxy, blazar or
Seyfert galaxy depends mostly on the spatial orientation
 of the surrounding galaxy which acts as the shield for some parts
 of the spectrum while being a source in some other parts \citep[see][for a review of unified schemes of radio sources]{urry95}.

Tolman-Oppenheimer-Volkoff (TOV) limit sets the upper bound to the mass of neutron star \citep{opp}. Today's estimates of the value of TOV limit range among $\approx 1.44-3
\:M_{\odot}$ \citep{lattimer04} and the uncertainty is due to unknown equation of state. Pressure of degenerate
neutron gas in the neutron star reaching this limit cannot oppose
gravitational pressure and inevitably collapses into the black hole
(while collapse is being accompanied with gamma ray
burst, GRB). Stellar mass black hole ($M \approx 4-15
\:M_{\odot}$) should thus represent the final state of stellar evolution of
every heavier star. To be observed it needs to be a part of the
binary system. Under certain circumstances black hole accretes
matter from its companion star (flowing via Lagrange point when
boundaries of a Roche lobe are exceeded). Similarly like in the
supermassive black hole case there establishes accretion disk emitting in X-ray part of the
spectrum. Due to recent observation from the satellites Chandra and
XMM-Newton equipped with high resolution X-ray detectors we now have
many stellar mass black hole candidates in our Galaxy (e.g.
Cygnus X-1). The key to distinguish accretion disk of a compact object
(typically neutron star) from the one surrounding the black hole is
the nonexistence of the surface in the black hole case. Thus irregular
flares of gamma rays accompanying thermonuclear reaction of accreted
material on the compact object surface are not detected.

Intermediate-mass black holes (IMBHs) are those with $M\approx
10^2-10^4 \:M_{\odot}$. Existence of IMBHs is still uncertain although a number of candidates was identified. IMBH might be possibly formed in globular star clusters \citep[][]{maccarone07}. Ultra-luminous
X-ray sources (ULXs) in close galaxies are suspected to be powered by IMBH. An ULX source HLX1 located on the edge of spiral galaxy ESO243-49 was claimed \citep{farrell09} to host an IMBH of over $500\: M_{\odot}$, though the interpretation was recently disputed by \citet{soria11}. A stellar complex IRS 13E residing close to our galactic center
Sagittarius A* is yet another candidate for the IMBH system \citep{imbh}.

Primordial black holes (PBHs) are hypothetical objects whose origin differs
fundamentally from all above mentioned types. They have not been
established by gravitational collapse of stars or other astronomical
bodies since they might have been born from density fluctuation
during early stages of the evolution of the Universe. On the theoretical grounds it has been argued \citep{carr74} that PBHs of mass from $10^{-5}\:\rm{g}$ upwards might exist in the present Universe. Updated constraints on PBHs were given recently by \citet{carr10}. Primordial black holes could be detected due to the Hawking radiation which black holes emit according to the quantum relativity \citep{page76}.

Hawking radiation has ordinary thermal spectra and energy it radiates away goes on account of the mass -- the black hole evaporates. Predicted evaporation times for stellar mass black holes exceed present age of
the universe by many orders but for primordial black holes with the
mass $M\approx10^{12}\:\mbox{kg}$ evaporation time approaches its
current age. Rate of evaporation escalates as the mass decreases
(power output $P\propto\frac{1}{M^{2}}$). Endpoint of the primordial
black hole existence should thus be explosive \citep{hawking74} and is believed to be
accompanied with loud GRB. Search for these GRB signatures is one of the key scientific objectives of current mission of Fermi Gamma-ray Space Telescope which is operating since 2008.

\begin{figure}[h]
    \centering
        \includegraphics*[width=.6\textwidth,trim=15 0 30 0]{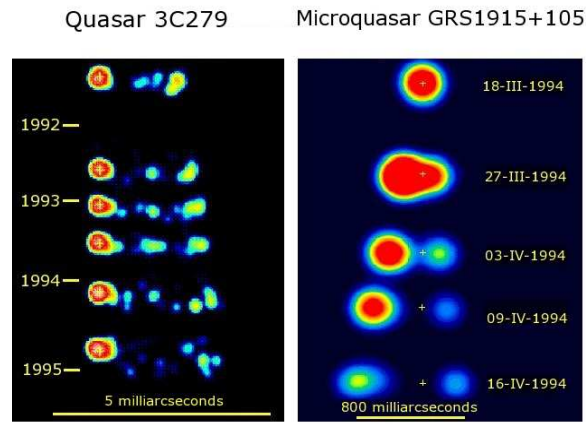}
\caption{Evidence for fast motion of the emerging jets in the
microquasar GRS 1915+105 (observed at radio frequency of 8.6 GHz)
and in the quasar 3C 279 (at 22 GHz). Synchrotron emission has been
reported in infrared wavelengths and, in some cases, even up to
X-rays, implying the presence of electrons in the jets with TeV
energies. Figure credit: \citet{mirabel}.}
    \label{jet}
\end{figure}

\section{Magnetic field: trigger for accretion and outflow}
Recent observations of microquasars, pulsars, gamma-ray bursts
 indicate that the astrophysical jets play an important role
everywhere (not only in the case of AGNs). There is plenty of
observational evidence suggesting that the initial acceleration of
jets takes place very near black holes (or other compact object) and
proceeds via electromagnetic forces. Jets and accretion disks in the vicinity of compact objects probably create symbiotic magnetically driven system \citep[e.g.][]{falcke95}. 

The current promising model of the dynamics (i.e. launching,
accelerating and collimating) of the astrophysical jets is based on the
magnetohydrodynamics (MHD). The results of the simulations
employing general relativistic MHD equations \citep[][]{krolik10} correlate with
observations of M87 \citep{junor99} where the formation and the
collimation of the jet were analyzed.

Moreover, the 3D relativistic MHD simulations carried out by \citet[][]{hawley06} demonstrate clearly the essential role which the accretion disk's coronae play in the collimation and acceleration of the jet. Indeed the dominant force accelerating the matter outward in a given numerical model originates from the coronal pressure. Regions above and below the equatorial plane become dominated by the magnetic pressure and large-scale magnetic fields may also develop by the dynamo action.

Recent numerical relativistic study by \citet{rezzolla11} reveals the formation of ordered jet-like structure of ultrastrong magnetic field in the merger of binary neutron stars. Such system thus might serve as an astrophysical engine for observed short gamma ray bursts.

In accretion models the magnetic field was also employed -- the so called
magnetorotational instability (MRI) must operate in accretion disc,
generating the effective viscosity necessary for the accretion
process \citep{balbus}. Magnetic reconnection is likely to be
responsible for rapid flares, which are observed in X-rays. Finally,
Faraday rotation measurements suggest that tangled magnetic fields
are present in jets \citep{begelman}. 

Observations of the Galactic Center (GC) reveal the presence of another remarkable large-scale magnetic structure -- nonthermal filaments (NTFs). NTFs cross the Galactic plane and their length reaches tens of parsecs while they are only tenths of parsec wide. The strength of the magnetic field within the NTF may approach $\approx1\:\rm{mG}$ while the typical interstellar value is $\approx10\:\mu \rm{G}$ \citep{larosa04}. Initially, it was thought that NTFs trace the pervasive poloidal magnetic field present throughout the GC \citep{morris90}. Later, however, it became apparent that the structure of the magnetic field in the central region of the Galaxy is more complex \citep{ferr10}. See \rff{ntf} for the snapshot of the GC at $90\: \rm{cm}\;\;(330\:\rm{MHz})$ which shows the NTFs clearly.

\begin{figure}[htb]
\centering
\includegraphics[scale=.44 ,clip]{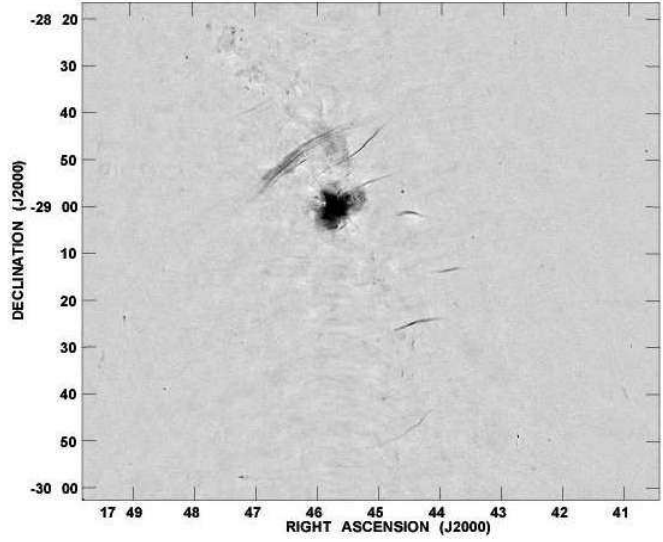}
\caption{Milky way's Galactic Center penetrated by the narrow nonthermal filaments (NTFs). The strength of ordered magnetic field  may approach $\approx1\:\rm{mG}$ within NTFs. Length of NTFs reaches tens of parsecs.  Inner region of $0.8^{\circ}\, \times\, 1.0^{\circ}$ is shown at wavelength $90\: \rm{cm} \:(330\,\rm{MHz})$. Snapshot was taken by the Very Large Array (VLA). Credit: \citet{nord04, larosa04}.}
\label{ntf}
\end{figure}

Overall it is quite likely that electromagnetic mechanisms play a
major role and operate both near supermassive black holes in quasars
as well as stellar-mass black holes and neutron stars in accreting binary systems (see
figures \ref{model} and \ref{jet}). Besides that a faint magnetic field is present throughout the interstellar medium, being locally intensified in NTFs.

\subsubsection{Electro-vacuum fields}
The survey of the vacuum electromagnetic (EM) fields may be regarded as the fundamental starting point in studying  the dynamics of astrophysical plasma. If the examination of the fields itselves represents the initial step in a given direction, at succeeding stage we would consider the motion of a non-interacting test particles exposed to these fields. In this work we will be dealing with both issues. Structure of a particular astrophysically motivated EM field emerging in the vicinity of rotating black hole will be studied in detail.  Subsequently we shall discuss the motion of charged particles exposed to the field representing a special case of a general solution explored before. Primarily we concern ourselves with the stable orbits occupying off-equatorial potential lobes. Particles on these orbits are relevant for the description of astrophysical corona comprising of diluted plasma residing outside the equatorial plane in the inner parts of accreting black hole systems. 

Gaseous corona is supposed to play a key role in the formation of observed X-ray spectra of both active galactic nuclei (AGNs) and microquasars \citep{done}. Power law component of the spectra is believed to result from the inverse Compton scattering of the thermal photons emitted in the inner parts of the disk. Relativistic electrons residing in the corona serve as a scatterers in this process. Their dynamic properties (e.g. resonances) thus shall have imprint on the observed spectra.

The role of magnetic fields near strongly gravitating objects has been
subject of many investigations \citep[e.g.][]{punsly08}. They are relevant for 
accretion disks that may be embedded in large-scale magnetic fields, for
example when the accretion flow penetrates close to a neutron star
\citep{lipunov92,halo2_11}. Outside the main body of the accretion
disk, i.e.\ above and below the equatorial plane, the accreted material
forms a highly diluted environment, a corona, where the density of
matter is low and the mean free path of particles is large in comparison
with the characteristic length-scale, i.e.\ the gravitational radius of
the central body, $R_{\rm g}\equiv GM/c^2\approx 1.5(M/M_\odot)\;$km,
where $M$ is the central mass. The origin of the coronal flows and the
relevant processes governing their structure are still unclear. In this
context we discuss motion of electrically charged particles outside the
equatorial plane.

\subsubsection{Regular and chaotic dynamics}
Accretion onto black holes and compact stars brings material in a zone
of strong gravitational and electromagnetic fields. We study dynamical
properties of motion of electrically charged particles forming a 
highly diluted medium (a corona) in the regime of strong gravity and
large-scale (ordered) magnetic field.

We start our discussion from a system that allows regular motion, then we
focus on the onset of chaos. To this end,
we investigate the case of a rotating black hole immersed in a
weak, asymptotically uniform magnetic field. We also consider
a magnetic star, approximated by the Schwarzschild metric and a test
magnetic field of a rotating dipole. These are two model examples of
systems permitting energetically bound, off-equatorial motion of matter
confined to the halo lobes that encircle the central body. Our approach
allows us to address the question of whether the spin parameter of the
black hole plays any major role in determining the degree of the
chaoticness.

The both dynamic systems may be regarded as different instances of the originally integrable systems which were perturbed by the
electromagnetic test field. Complete integrability of geodesic
motion of a free particle in Schwarzschild spacetime is easy to
verify \citep{mtw}. To some surprise it was later found that also
free particle motion in Kerr spacetime and even the charged particle
motion in Kerr-Newman is completely integrable \citep{carter68} since
separation of the equations of motion is possible as there exists
additional integral of motion -- Carter's constant $\mathcal{L}$. Trajectories found in such a systems are merely regular.

In the non-integrable system, however, both regular and chaotic trajectories may coexist in the phase space. Standard method of a qualitative survey of the non-linear dynamics is based on the construction of Poincar\'{e} surfaces of section which allow to visually discriminate between the chaotic and regular regime of motion. 

On the other hand quantifying the chaos by Lyapunov characteristic exponents (LCEs), as its standard and commonly used indicator, becomes problematic in the general relativity (GR) since LCEs are not invariant under the coordinate transformations. Besides that the usual method of computing LCEs involves evaluation of the distances between the neighboring trajectories which becomes intricate in GR. Although there are operational workabouts to partially  overcome these difficulties \citep[e.g.][]{wu2003} the need for a consistent treatment is apparent. Perhaps the geometrical approach suggested recently by \citet{stach10} could eventually provide a covariant method of the evaluation of the Lyapunov spectra in GR.

In this context we adopt a different tool to investigate the dynamic system -- Recurrence Analysis \citep{marwan}.
To characterize the motion, we construct the Recurrence Plots (RPs) and
we compare them with Poincar\'e surfaces
of section. We describe the Recurrence Plots in terms of
the Recurrence Quantification Analysis (RQA), which allows us to
identify the transition between different dynamical regimes. We demonstrate
that this new technique is able to detect the chaos onset very efficiently,
and to provide its quantitative measure. The chaos
typically occurs when the conserved energy is raised to a sufficiently
high level that allows the particles to traverse the equatorial plane.
We find that the role of the black-hole spin in setting the chaos is
more complicated than initially thought.

\section{Structure of Thesis}
The thesis is organized as follows. Most of its contents are contained in \rfch{kapEM}. We begin with technical preliminaries, namely in \rs{lines} we introduce several alternative definitions of electric and magnetic vector fields. In \rs{emfield} we give explicitly the components of the electromagnetic tensor $F_{\mu\nu}$ describing the field around the Kerr black hole drifting in the arbitrary direction through the asymptotically homogeneous magnetic field which is generally inclined with respect to the rotation axis of the BH. Choice of the observer's frame is discussed in \rs{tetrads}. Structure of the both electric and magnetic fields is explored in detail in \rs{structure}. First in \rs{stationary} we revisit the issue of the magnetic expulsion (Meissner effect) which is present in the case of aligned field. Subsequently we study field structures in the case of inclined field. Finally, in \rs{driftingEM} we introduce a drift of the black hole and explore the both electric and magnetic fields emerging in this general setup. 

In the remaining part of the \rfch{kapEM} we shall deal with the dynamics of the charged test particles exposed to the test fields analyzed in the previous sections. In \rs{pohrce} we review the equations of particle motion, which we then integrate to obtain trajectories. In \rs{ra} we introduce the basic properties of Recurrence Plots. Sec.\ \ref{sectionwald} analyses the motion around a Kerr black hole endowed with a uniform magnetic test field. We employ Poincar\'e surfaces of section and Recurrence Plots. The two approaches allow us to show the onset of chaos in different, complementary ways. We examine  the motion in off-equatorial lobes, pay special attention to the spin dependence of the stability of motion, and we notice the emergence of `potential valleys' that allow the particles to escape from the equatorial plane along a narrow collimated corridor. Analysis of the off-equatorial motion around a magnetic star is presented in \rs{sectionmagnetized}. We consider a dipole-type magnetic  field, which sets different limits on the off-equatorial range of allowed motion of charged particles. It also defines different regimes of
chaoticness in the comparison with the uniform magnetic field. 

Finally, results of the analysis are summarized in \rfch{conclusion}. Chapter \ref{future} represents a brief outlook to the future as it specifies several topics suggesting the direction in which our research could continue.

In Appendix \ref{gu} we list conversion factors between SI and geometrized units. Scaling of the quantities by the central mass $M$ is also discussed. Appendix \ref{integ} gives detailed comparison of numerical integrators applicable to our system. Benefits of symplectic routines are discussed therein. In Appendix \ref{vfexplorer2} we introduce our program tool {\em vfexplorer2} which provides a simple graphic user interface (GUI) for the effective exploration of a complex vector fields.

Most of the scientific contents presented in the Thesis were published in the papers \citet{kopacek10} and \citet{karas09}. Investigation of the particle motion in the astrophysical corona was preceded by the study of the topology of off-equatorial potential lobes performed by \citet{halo2}. Discussion of the charged particle motion in such lobes was also a subject of the contribution \citet{kopacek10b}. Initial steps of the investigation of electromagnetic fields around drifting Kerr black hole were described by \citet{kopacek08}.

At different phases of my doctoral studies I took part in several research projects. Namely I acknowledge support from the doctoral student program of the Czech Science Foundation (project No. 205/09/H033), Plan for European Cooperating States of the European Space Agency (ESA PECS 98040) and two projects of the Grant Agency of Charles University (GAUK 119210/2010 and SVV-263301).

%% file: chap2n.tex
\chapter{\sffamily{Regular and chaotic motion in~black hole magnetospheres}}\chaptermark{Motion in black hole magnetospheres}
\label{kapEM}
\pagestyle{headings}

\section{Electromagnetic field around drifting Kerr black hole}
In this section we construct a test field solution describing the electromagnetic field around a Kerr black hole which is drifting in an arbitrary direction with respect to the asymptotically uniform magnetic field with the general orientation with respect to the rotation axis. Definition of the electric and magnetic field intensities is discussed as well as the choice of the physical observer. Structure of the field is explored in detail.

Electromagnetic (EM) test field solutions of Maxwell equations in curved spacetime play an important role in astrophysics since we can usually suppose that astrophysically relevant EM fields are weak enough, so that their influence upon background geometry may be neglected. 

We are interested in the solutions describing an originally uniform magnetic field under the influence of the Kerr black hole. Since the Kerr metric is asymptotically flat, this EM field reduces to the original homogeneous magnetic field in the asymptotic region. First such a test field solution was given by \citet{wald74} for the special case of perfect alignment of the asymptotically uniform magnetic field with the symmetry axis. Using a different approach of Newman Penrose formalism a more general solution for an arbitrary orientation of the asymptotic field was inferred by \citet{bicak80}. We use their solution to construct the EM field around the Kerr black hole which is drifting through the asymptotically uniform magnetic field. 

Such generalized setup shall represent an astrophysically relevant model. In the actual accreting BH system the misaligned ordered field may arise if the accretion disk is inclined with respect to the rotation axis of the BH and the Bardeen-Petterson effect does not operate to align the axes. Such misaligned accretion was observed in some numerical simulations \citep[e.g.][]{rockefeller05}. A possible scenario how the BH could receive substantial velocity with respect to its accretion disk is suggested by the simulations of a merger process \citep[e.g.][]{rezzolla09, gonzales07}.

In the context of black hole mergers \citet{lyutikov11} recently studied electrodynamic properties of the simplified model consisting of the Schwarzschild black hole in the uniform transversal motion with respect to the homogeneous magnetic field. The author considers the interaction of the field with the plasma generated from the pair production due to the vacuum breakdown. Analyzing the resulting situation in the force-free approximation, Lyutikov concludes that observed electrodynamic properties resemble in many aspects the pulsar magnetosphere.


\section{Lines of force}
\label{lines}
Kerr metric in Boyer-Lindquist coordinates $x^{\mu}= (t,\:r, \:\theta,\:\varphi)$ may be expressed as follows \citep{mtw}:
\begin{equation}
\label{kn}
ds^2=-\frac{\Delta}{\Sigma}\:[dt-a\sin{\theta}\,d\varphi]^2+\frac{\sin^2{\theta}}{\Sigma}\:[(r^2+a^2)d\varphi-a\,dt]^2+\frac{\Sigma}{\Delta}\;dr^2+\Sigma d\theta^2,
\end{equation}
where
\begin{equation}
\label{knleg} {\Delta}\equiv{}r^2-2Mr+a^2,\;\;\;
\Sigma\equiv{}r^2+a^2\cos^2\theta.
\end{equation}
We stress that geometrized units $G=c=k=k_C=1$ are used throughout the text. See Appendix \ref{gu} for details.

For the sake of reference it might be useful to list both covariant and contravariant components of the Kerr metric explicitly
\begin{align}
\label{kncov}
g_{tt}=\frac{a^2\,\sin^2{\theta}-\Delta}{\Sigma},\lb{}g_{rr}=\frac{\Sigma}{\Delta},\lb{}g_{\theta\theta}=\Sigma,\lb g_{\varphi\varphi}=\frac{A\:\sin^2{\theta}}{\Sigma},\\ \nonumber g_{t\varphi}=g_{\varphi{}t}=\frac{a\,\sin^2{\theta}}{\Sigma}\;\left({}\Delta-r^2-a^2\right){},\\
g^{tt}=\frac{-A}{\Delta\:\Sigma},\lb{}g^{rr}=\frac{\Delta}{\Sigma},\lb{}g^{\theta\theta}=\frac{1}{\Sigma},\lb g^{\varphi\varphi}=\frac{1}{\Sigma}\;\left(\frac{1}{\sin^2{\theta}}-\frac{a^2}{\Delta} \right){},\\
\nonumber{}g^{t\varphi}=g^{\varphi{}t}=\frac{a}{\Delta\:\Sigma}\;\left({}\Delta-r^2-a^2\right){},
\end{align}
denoting $A\equiv(r^2+a^2)^2-a^2\sin\theta^2\Delta$. Following relation proved useful in the calculations: $\Delta\:\Sigma=A-2Mr(r^2+a^2)$.

Radial metric component $g_{rr}$ and contravariant components $g^{tt}$, $g^{\varphi\varphi}$ and $g^{t\varphi}$ are singular at $\Delta=0$ which defines the outer (+) and the inner (-) horizon $r_\pm=M\pm\sqrt{M^2-a^2}$ of the black hole. By querying the curvature scalars one finds that the singularities present at $r_\pm$ are purely coordinate singularities rather than physical ones. Both horizons merge at $r=M$ in the case of extreme Kerr black hole $a=M$. For $a>M$ the horizon disappears and the central singularity becomes naked. In the following, however, we will be concerned with the cases $a\leq M$ only. 

Dragging of the inertial frames caused by the rotation of the source may be characterized by the quantity $\Omega=-g_{t\varphi}/g_{\varphi\varphi}=2Mra/A$. Coordinate angular velocity $\mathrm{d}\varphi/\mathrm{d}t=u^{\varphi}/u^{t}$ (``angular velocity as measured at infinity``) of the observer freely falling from the rest at infinity reads $\Omega$ (see eq. \ref{FFOFIspeed}). 

In the classical electrodynamics we spontaneously define electric and magnetic lines of force as a field lines of three-vectors $\vec{E}=(E_x,E_y,E_z)$ and $\vec{B}=(B_x,B_y,B_z)$ which form a solution of Maxwell equations for a given problem. Definition condition of the field line $\vec{r}=(x(s),y(s),z(s))$ parametrized by the parameter $s$ is such that its tangent vector $\frac{\mathrm{d}\vec{r}}{\mathrm{d}s}$ is at each point parallel to the vector field itself. Thus for (e.g. electric) lines of force we obtain equation
\begin{equation}
 \label{linesofforce}
\frac{\mathrm{d}\vec{r}}{\mathrm{d}s}\;\times\;\vec{E}=0\\
\Longrightarrow\;\frac{\mathrm{d}x}{E_x}=\frac{\mathrm{d}y}{E_y}=\frac{\mathrm{d}z}{E_z},
\end{equation}
where $\times$ stands for the ordinary cross product of two vectors.

If we recall Lorentz relation specifying the force felt by a particle with electric test charge $q_e$ and hypothetical magnetic monopole test charge $q_m$ moving with the velocity $\vec{v}$ in the external fields $\vec{E}$, $\vec{B}$
\begin{equation}
 \label{clalorentz}
\vec{F}=q_{e}(\vec{E}+\vec{v}\times\vec{B})+q_m(\vec{B}-\vec{v}\times\vec{E}),
\end{equation}
we conclude that in a given reference frame we can identify the electric intensity $\vec{E}$ with the force felt by the unit electric charge and magnetic induction $\vec{B}$ with the force felt by the unit magnetic monopole charge provided that these charges are static in a given reference frame.


In the covariant language of general relativity the Lorentz force felt by the test particle of mass $m$ and the electric charge $q_e$ or the magnetic monopole charge $q_m$ is expressed as follows:
\begin{align}
 \label{lorentzgr}
a^{\mu}&=\frac{\mathrm{D}u^{\mu}}{\mathrm{d}\tau}=\frac{\mathrm{d}u^{\mu}}{\mathrm{d}\tau}+\Gamma^{\mu}_{\alpha\beta}u^{\alpha}u^{\beta}=\frac{q_e}{m}F^{\mu}_{\;\;\:\nu}u^{\nu}\\
a^{\mu}&=\frac{\mathrm{D}u^{\mu}}{\mathrm{d}\tau}=\frac{\mathrm{d}u^{\mu}}{\mathrm{d}\tau}+\Gamma^{\mu}_{\alpha\beta}u^{\alpha}u^{\beta}=-\frac{q_m}{m}^{*}\!\! F^{\mu}_{\;\;\:\nu}u^{\nu},
\end{align}
where $\tau$ stands for the proper time, \mbold{a} is particle's four-acceleration, \mbold{u} its four-velocity and $\Gamma^{\mu}_{\alpha\beta}$ are Christoffel symbols. $F_{\mu\nu}$ is electromagnetic tensor and $^{*}F_{\mu\nu}$ its dual.
 
In the close analogy with the classical case it appears natural to define {\bf coordinate components of magnetic and electric fields} as follows \citep{hanni73} 
\begin{equation}
\label{lorentzf} 
B^{\mu}=- ^{*}\!\!F^{\mu}_{\;\:\nu}u^{\nu},\;\;\;E^{\mu}=F^{\mu}_{\;\:\nu}u^{\nu}.
\end{equation}
However, the demand on the test charge to be static in a given reference frame, i.e. to have 4-velocity of a form $u^{\mu}=(u^t,0,0,0)$, becomes highly problematic when dealing with the non-static geometry of the Kerr spacetime. It is well known that in this case no physical observer may remain static inside the ergosphere whose boundary is defined by
\begin{equation}
 \label{staticl}
r_{\rm{s}}=M+\sqrt{M^2-a^2\cos{\theta}^2}.
\end{equation}
Since the region near the horizon is typically of a big interest when investigating the effects of a strong gravity we do not content us with being limited to the region outside the ergosphere. To this end we generalize the above definition \rf{lorentzf} in such a way that we allow general four-velocity of the test charge $u^{\mu}=(u^t,u^r,u^{\theta},u^{\varphi})$. In the classical analogy this would mean not insisting on $\vec{v}=0$ when defining the vector fields $\vec{E}$ and $\vec{B}$ using the classical form of the Lorentz force equation (\ref{clalorentz}). Nevertheless all the cases we shall discuss actually have $u^{\theta}=0$ therefore we may write explicitely
\begin{align}
\label{explicitvf1}
B^r&=-g^{rr}\left( ^{*}\!\!F_{rt}u^t + ^{*}\!\!F_{r\varphi}u^{\varphi} \right),\\ 
\label{explicitvf2}
B^{\theta}&=-g^{\theta\theta}\left( ^{*}\!\!F_{\theta{}t}u^t+ ^{*}\!\!F_{\theta{}r}u^r +^{*}\!\!F_{\theta\varphi}u^{\varphi}\right),\\
\label{explicitvf3}
B^{\varphi}&=-\left[ g^{\varphi\varphi} \left( ^{*}\!\!F_{\varphi t}u^{t}+^{*}\!\!F_{\varphi r}u^r \right)+ g^{\varphi t} \left( ^{*}\!\!F_{tr}u^r+^{*}\!\!F_{t \varphi}u^{\varphi} \right) \right].
\end{align}
Equation for the lines of force \rf{linesofforce} takes following form in the poloidal plane: $\mathrm{d}r/\mathrm{d}\theta=B^r/B^{\theta}$ and similarly in the equatorial plane: $\mathrm{d}r/\mathrm{d}\varphi=B^r/B^{\varphi}$. To obtain analogical expressions for the components of electric field $E^r$, $E^{\theta}$, $E^{\varphi}$ we just need to omit the minus sign at the beginning of each term and replace $^{*}\!\!F_{\mu\nu}$ components with $F_{\mu\nu}$ in eqs. (\ref{explicitvf1})--(\ref{explicitvf3}).

Another issue connected with the definition (\ref{lorentzf}) is a question of the normalization of the basis vectors. We notice that components of magnetic and electric fields $B^{\mu}$ and $E^{\mu}$ are provided in the canonical basis of Boyer-Lindquist coordinates, namely in the basis of vectors $\frac{\partial}{\partial t}$, $\frac{\partial}{\partial r}$, $\frac{\partial}{\partial \theta}$, $\frac{\partial}{\partial \varphi}$ which are not normalized. This problem may be seemingly easily overcome by expressing {\bf "physical components'' of the fields}$^1$\footnotetext[1]{The adjective {\em physical} is sometimes used to refer to the quantities measured in the frame attached to a given physical observer. We stress that our usage of the term differs as our physical components are expressed in the coordinate frame.} \citep{hanni73} as follows
\begin{align}
\label{physicalc1}B^{r}_{\rm{physical}}&=\mathrm{sign}(B^r)\sqrt{B^{r}B_{r}}=\sqrt{g_{rr}}\;B^r,\\
\label{physicalc2}B^{\theta}_{\rm{physical}}&=\mathrm{sign}(B^{\theta})\sqrt{B^{\theta}B_{\theta}}=\sqrt{g_{\theta\theta}}\;B^{\theta},\\
\label{physicalc3}B^{\varphi}_{\rm{physical}}&=\mathrm{sign}(B^{\varphi})\sqrt{B^{\varphi}B_{\varphi}}=\mathrm{sign}(B^{\varphi})\sqrt{B^{\varphi}\left( ^{*}\!\!F_{t \varphi}u^{t}-^{*}\!\!F_{\varphi r}u^r \right)},
\end{align}
and analogically for physical components of electric field $E^{r,\theta,\varphi}_{\rm{physical}}$.

However, in the case of Boyer-Lindquist coordinate system which is singular at the horizon (namely the radial metric component $g_{rr}$ diverges here) using above defined physical components turns out to be a controversial choice since it introduces this divergence directly into the radial component of electric and magnetic fields. 

To fix this problem we suggest to define {\bf ``renormalized components`` of the fields} as follows
\begin{align}
\label{semiphysicalc1}B^{r}_{\rm{renormalized}}&=B^{r}\:h_r,\;\;\;\;\;E^{r}_{\rm{renormalized}}=E^{r}\:h_r,\\
\label{semiphysicalc2}B^{\theta}_{\rm{renormalized}}&=B^{\theta}\:h_{\theta},\;\;\;\;\;E^{\theta}_{\rm{renormalized}}=E^{\theta}\:h_{\theta},\\
\label{semiphysicalc3}B^{\varphi}_{\rm{renormalized}}&=B^{\varphi}\:h_{\varphi}, \;\;\;\;\;E^{\varphi}_{\rm{renormalized}}=E^{\varphi}\:h_{\varphi},
\end{align}
where $h_r=1$, $h_{\theta}=r$, $h_{\varphi}=r\sin\theta$ are ordinary Lam\'{e} coefficients of spherical coordinates in flat space. Correspondence between physical and renormalized components lies in the asymptotic region where they become identical since the Kerr spacetime is asymptotically flat and Boyer-Lindquist coordinates $r,\theta,\varphi$ asymptotically turn into spherical coordinates describing the spatial part of Minkowskian spacetime. Explicitly given the asymptotic behaviour of the metric coefficients is $g_{rr}\rightarrow 1=h_r^2$, $g_{\theta\theta}\rightarrow r^2=h^2_{\theta}$ and $g_{\varphi\varphi}\rightarrow r^2\sin^2{\theta}=h^2_{\varphi}$ and $g_{t\varphi}\rightarrow 0$.   

Going even further in ''operational flattening`` of the curved background one could eventually identify the components of magnetic and electric fields directly with the components of the electromagnetic tensor $F_{\mu\nu}$. Such an identification is fully justified in any local Lorentz frame where $F_{\mu\nu}$ takes the form given by \rf{tetradF}. Nevertheless although the asymptotics of Kerr background is flat, its Boyer-Lindquist coordinate basis $1$-forms are not normalized and normalization factors $h_r$, $h_{\theta}$ and $h_{\varphi}$ must be used accordingly. We define {\bf asymptotically motivated (AMO) components of the fields}
\begin{align}
 \label{amo1}
B^r_{\rm{AMO}}&=\frac{F_{\theta\varphi}}{h_{\theta}h_{\varphi}}, \;\;\;\;\; E^r_{\rm{AMO}}=\frac{F_{rt}}{h_{r}},\\ 
 \label{amo2}
B^{\theta}_{\rm{AMO}}&=\frac{F_{\varphi r}}{h_{\varphi}h_{r}}, \;\;\;\;\;E^{\theta}_{\rm{AMO}}=\frac{F_{\theta{}t}}{h_{\theta}},\\
 \label{amo3}
B^{\varphi}_{\rm{AMO}}&=\frac{F_{r\theta}}{h_{r}h_{\theta}} \;\;\;\;\;E^{\varphi}_{\rm{AMO}}=\frac{F_{\varphi{}t}}{h_{\varphi}}.
\end{align}
AMO components asymptotically coincide with the physical and renormalized components provided that the test charge is static. Although AMO components do not allow for a direct physical interpretation when applied outside the asymptotic region, it may still be useful to explore them since they do not employ any particular observer (four-velocity of a test charge) in their definition and therefore it may be easier to acquire intuitive insight into the nature of a given EM field. 

Nevertheless a consistent way to define the electric and magnetic fields should provide obvious physical interpretation of the observables measured by a certain physical observer at any distance from the center. We let such an observer with four-velocity $u^{\mu}$ equipped with the orthonormal tetrad $e_{(\alpha)}^{\mu}$ measure the Lorentz force of \rf{lorentzf} using his tetrad basis. {\bf Tetrad components of the vector fields} determining desired lines of force are given as the spatial part of the projection
\begin{align}
\label{tetradB}
B^{(i)}&=B_{(i)}=-e^{(i)\;*}_{\;\,\mu} \!F^{\mu}_{\;\;\nu}u^{\nu}=-e_{(i)}^{\;\,\mu\;*}\!F_{\mu\nu}e_{(t)}^{\nu}=-^{*}\!F_{(i)(t)},\\
\label{tetradE}
E^{(i)}&=E_{(i)}=e^{(i)}_{\;\,\mu}F^{\mu}_{\;\;\nu}u^{\nu}=e_{(i)}^{\mu}F_{\mu\nu}e_{(t)}^{\nu}=F_{(i)(t)},
\end{align}
where $e^{(\alpha)}_{\;\,\mu}$ are 1-forms dual to the tetrad vectors $e_{(\alpha)}^{\mu}$. Lowering/raising of the spatial tetrad indices does not matter since the tetrad is supposed to be orthonormal: $g_{(\mu)(\nu)}=\eta_{(\mu)(\nu)}$.

For the sake of the future reference we review both tetrad components of EM tensor $F_{(\alpha)(\beta)}$ and its dual $^{*}\!\!F_{(\alpha)(\beta)}$. We remind that such an interpretation of these tensors is possible only in any local Lorentz frame which may be attached to any physical observer, not necessarily inertial. Magnetic and electric test charges which are used to measure the fields are at rest in this frame. 

\begin{equation}
\label{tetradF}
F_{(\alpha)(\beta)}=
\begin{pmatrix}
0&-E_{(r)}&-E_{(\theta)}&-E_{(\varphi)}\\
E_{(r)}&0&B_{(\varphi)}&-B_{(\theta)}\\
E_{(\theta)}&-B_{(\varphi)}&0&B_{(r)}\\
E_{(\varphi)}&B_{(\theta)}&-B_{(r)}&0
\end{pmatrix},
\end{equation}

\begin{equation}
\label{tetraddF}
^{*}F_{(\alpha)(\beta)}=
\begin{pmatrix}
0&B_{(r)}&B_{(\theta)}&B_{(\varphi)}\\
-B_{(r)}&0&E_{(\varphi)}&-E_{(\theta)}\\
-B_{(\theta)}&-E_{(\varphi)}&0&E_{(r)}\\
-B_{(\varphi)}&E_{(\theta)}&-E_{(r)}&0
\end{pmatrix}. 
\end{equation}

Several ways to define electric and magnetic field have been described in this section. Coordinate components of the Lorentz force felt by the unit ele\-ctric/mag\-netic test charge (eq. \ref{lorentzf}) provide a natural generalization of the classical definition. In the case of the Kerr background, however, we cannot insist upon the usage of the static test charges in the definition as no static observers may penetrate inside the ergosphere. Since the structure of the fields in the vicinity of the horizon is usually of the utmost interest we have assumed more general four-velocity of the test charge in the form $u^{\mu}=(u^t,u^r,0,u^{\varphi})$. The Lorentz force may be projected onto the tetrad basis attached to the test charge. 

Nevertheless if we remain in the Boyer-Lindquist coordinate basis we  note that the fields  do not come with the proper dimension because the coordinate basis vectors are not normalized. To correct this in a rigorous manner we define {\it physical components} of the fields by eqs. (\ref{physicalc1})--(\ref{physicalc3}). It appears, however, that this definition amplifies the effect of the coordinate singularity at the horizon. Thus we suggested to define {\it renormalized components} (eqs. \ref{semiphysicalc1}--\ref{semiphysicalc3}) as an useful approximation which is less problematic near the horizon. Far more approximative (in the sense of treating the curved background as flat) are {\it asymptotically motivated AMO components} (eqs. \ref{amo1}--\ref{amo3}) which in fact map $F_{\mu\nu}$ onto the flat surface directly. Brief discussion regarding the definition of the lines of force of electric and magnetic fields was also held by \citet{bicak80}.

\section{Electromagnetic field}
\label{emfield}
Stationary and axisymmetric test field solutions to the Maxwell equations on the Kerr background traditionally attract attention by both relativistic theoreticians and astrophysicists. The latter are usually concerned with the solutions describing some potentially realistic electrodynamic scenario. In particular we mention test fields of axisymmetric current loops \citep{petterson, moss11} and uniform magnetic field \citep{wald74,bicak76}. 

We start out from the $F_{\mu\nu}$ describing the test field with asymptotic form of a general (i.e. not necessarily parallel) uniform magnetic field given by \citet{bicak85}. Due to the axial symmetry of Kerr space-time only two components of asymptotic field were considered in that paper without any loss of generality (asymptotic components $B_0$ (parallel) and $B_1$ (equatorial) to be specific). We rewrite components of EM tensor \citep[eq.~(A3) of][]{bicak85} denoting $B_{x}\equiv{}B_{1}$, $B_{z}\equiv{}B_{0}$ and splitting the result into two parts according to the asymptotic component. We obtain the asymptotically perpendicular part of the field:

\begin{align}
\label{BFtr(Bx)}
\nonumber F_{tr}^{B_x}=&B_{x}aMr\Sigma^{-2}\Delta^{-1}\sin\theta\cos\theta[(r^3-2Mr^2+ra^2(1+\sin^2{\theta})+2Ma^2\cos^2\theta)\cos\psi\\\notag&-a(r^2-4Mr+a^2(1+\sin^2\theta))\sin\psi],\\
\nonumber F_{t\theta}^{B_x}=&B_{x}aM\Sigma^{-2}(r^2\cos2\theta+a^2\cos^2\theta)(a\sin\psi-r\cos\psi),\\
\nonumber F_{t\varphi}^{B_x}=&B_{x}aM\Sigma^{-1}\sin\theta\cos\theta(a\cos\psi+r\sin\psi),\\
\nonumber F_{r\theta}^{B_x}=&-B_{x}(a\cos\psi+r\sin\psi)\\ &-B_{x}a\Delta^{-1}\left[(Mr-a^2\sin^2\theta)\cos\psi-a(r\sin^2\theta+M\cos^2\theta)\sin\psi\right],\\ \nonumber
F_{r\varphi}^{B_x}=&-B_{x}\sin\theta\cos\theta\left[(r-Ma^2\Delta^{-1})\cos\psi-a(1+rM\Delta^{-1})\sin\psi\right]\\\nonumber&+a\sin^2\theta F_{tr}^{B_x},\\
\nonumber F_{\theta\varphi}^{B_x}=&B_{x}\left[(r^2\sin^2\theta+Mr\cos2\theta)\cos\psi-a(r\sin^2\theta+M\cos^2\theta)\sin\psi\right]\\ \nonumber &+(r^2+a^2)B_{x}M\Sigma^{-2}(r^2\cos2\theta+a^2\cos^2\theta)(a\sin\psi-r\cos\psi)
\end{align}
and the part which approaches uniform field aligned along the axis: 
\begin{align}
\label{BFtr(Bz)}
\nonumber F_{tr}^{B_z}=&B_{z}aM\Sigma^{-2}(r^2-a^2\cos^2\theta)(1+\cos^2\theta),\\
F_{t\theta}^{B_z}=&2B_{z}aMr\Sigma^{-2}\sin\theta\cos\theta(r^2-a^2),\\
\nonumber F_{r\varphi}^{B_z}=&B_{z}r\sin^2\theta+B_{z}a^2\sin^2\theta{}M\Sigma^{-2}(r^2-a^2\cos^2\theta)(1+\cos^2\theta),\\
\nonumber F_{\theta\varphi}^{B_z}=&B_{z}\Delta\sin\theta\cos\theta+2(r^4-a^4)B_{z}Mr\Sigma^{-2}\sin\theta\cos\theta,
\end{align}
where we use the azimuthal coordinate $\psi$ of Kerr ingoing coordinates, which is related to Boyer--Lindquist coordinates as follows:
\begin{equation}
\label{kicpsi}
\psi=\varphi+\frac{a}{r_{+}-r_{-}}\ln{\frac{r-r_{+}}{r-r_{-}}},
\end{equation}
with $r_{\pm}\equiv M \pm\sqrt{M^2-a^2}$ denoting the outer and the inner horizon. We notice that $\lim_{r\to \infty}\psi=\varphi$.

As we shall introduce a drift of the black hole in the general direction we lose axial symmetry and need to consider all spatial components of the asymptotic magnetic field. We obtain $F_{\mu\nu}^{B_y}$ (which may only appear due to nonzero drift) by rotating $F_{\mu\nu}^{B_x}$ along the $z$-axis by angle $\frac{\pi}{2}$ - i.e. $F_{\mu\nu}^{B_y}=F_{\mu\nu}^{B_x}\left(\varphi\rightarrow\varphi-\frac{\pi}{2}, B_{x}\rightarrow B_{y} \right)$ which only causes $\sin\psi\rightarrow-\cos\psi$ and $\cos\psi\rightarrow\sin\psi$.

Since the drift shall induce uniform electric field in the asymptotic region we need to have appropriate $F_{\mu\nu}^{E_{x,y,z}}$ handy. We get them easily by performing dual transformation of $F_{\mu\nu}^{B_{x,y,z}}$. Dual transformation is carried out as follows:
\begin{equation}
 \label{dualobecne}
^{*}F_{\alpha\beta}=\frac{1}{2}F^{\mu\nu}\varepsilon_{\mu\nu\alpha\beta},
\end{equation}
where $\varepsilon_{\mu\nu\alpha\beta}$ is the Levi-Civita tensor whose components are given as:
\begin{equation}
 \label{levicivita}
\varepsilon_{\mu\nu\alpha\beta}=\sqrt{-det||g_{\sigma\omega}||}[\mu\nu\alpha\beta]\equiv\sqrt{-g}[\mu\nu\alpha\beta],
\end{equation}
with $[\mu\nu\alpha\beta]$ denoting completely antisymmetric symbol. Determinant of the Kerr metric is $g=g_{tt}g_{rr}g_{\theta\theta}g_{\varphi\varphi}-g_{\varphi{}t}^2g_{rr}g_{\theta\theta}=-\sin^2\theta\;\Sigma^2$.

Performing the dual transformation we immediately obtain EM tensors with desired asymptotics of uniform electric field:
\begin{equation}
 \label{dualkonkretne}
F_{\mu\nu}^{E_{x,y,z}}=^{*}\negthickspace F_{\mu\nu}^{B_{x,y,z}}(B_{x,y,z}\rightarrow -E_{x,y,z}).
\end{equation}
In the explicit form we get:
\begin{align}
\nonumber F_{tr}^{E_{x,y,z}}&=\sin\theta\;\Sigma\left(F_{\theta{}t}^{B_{x,y,z}}(B_{x,y,z}\rightarrow -E_{x,y,z})g^{\varphi{}t}+F_{\theta\varphi}^{B_{x,y,z}}(B_{x,y,z}\rightarrow -E_{x,y,z})g^{\varphi\varphi}\right)g^{\theta\theta},\\
\nonumber F_{t\theta}^{E_{x,y,z}}&=\sin\theta\;\Sigma\left(F_{tr}^{B_{x,y,z}}(B_{x,y,z}\rightarrow -E_{x,y,z})g^{\varphi{}t}+F_{\varphi{}r}^{B_{x,y,z}}(B_{x,y,z}\rightarrow -E_{x,y,z})g^{\varphi\varphi}\right)g^{rr},\\
\label{BF(E)} F_{t\varphi}^{E_{x,y,z}}&=\sin\theta\;\Sigma\; F_{r\theta}^{B_{x,y,z}}(B_{x,y,z}\rightarrow -E_{x,y,z})g^{rr}g^{\theta\theta},\\
\nonumber F_{r\theta}^{E_{x,y,z}}&=\sin\theta\;\Sigma\; F_{\varphi t}^{B_{x,y,z}}(B_{x,y,z}\rightarrow -E_{x,y,z})\left((g^{\varphi t})^2- g^{\varphi\varphi}g^{tt}\right),\\
\nonumber F_{r\varphi}^{E_{x,y,z}}&=\sin\theta\;\Sigma\left(F_{\theta\varphi}^{B_{x,y,z}}(B_{x,y,z}\rightarrow -E_{x,y,z})g^{\varphi{}t}+F_{\theta t}^{B_{x,y,z}}(B_{x,y,z}\rightarrow -E_{x,y,z})g^{tt}\right)g^{\theta\theta},\\
\nonumber F_{\theta\varphi}^{E_{x,y,z}}&=\sin\theta\;\Sigma\left(F_{\varphi r}^{B_{x,y,z}}(B_{x,y,z}\rightarrow -E_{x,y,z})g^{\varphi{}t}+F_{tr}^{B_{x,y,z}}(B_{x,y,z}\rightarrow -E_{x,y,z})g^{tt}\right)g^{rr}.
\end{align}  

Now we are fully equipped to construct any asymptotically uniform test field on the Kerr background just by linear superposing of the above EM tensors. As we are concerned in constructing $F_{\mu\nu}$ which describes the test field around the black hole drifting through asymptotically uniform magnetic field in general direction, we shall employ Lorentz transformation to find the correct asymptotic components of such a field. Once obtained we just use them to replace the original ``non-drifting'' quantities $E_x , E_y, E_z, B_x, B_y$ and $B_z.$
Matrix of general Lorentz transformation is \citep{jackson}:
\begin{equation}
\label{lorentz}
||\Lambda^{\nu '}_{\;\mu}||=
\begin{pmatrix}
\gamma&-\gamma v_{x}&\gamma v_{y}&-\gamma v_{z}\\
-\gamma v_{x}&1+\frac{(\gamma-1)v_{x}^2}{v^2}&\frac{(\gamma-1)v_{x}v_{y}}{v^2}&\frac{(\gamma-1)v_{x}v_{z}}{v^2}\\
-\gamma v_{y}&\frac{(\gamma-1)v_{x}v_{y}}{v^2}&1+\frac{(\gamma-1)v_{y}^2}{v^2}&\frac{(\gamma-1)v_{y}v_{z}}{v^2}\\
-\gamma v_{z}&\frac{(\gamma-1)v_{x}v_{z}}{v^2}&\frac{(\gamma-1)v_{y}v_{z}}{v^2}&1+\frac{(\gamma-1)v_{z}^2}{v^2}
\end{pmatrix},
\end{equation}
where $v=(v_{x}^2+v_{y}^2+v_{z}^2)^{\frac{1}{2}}$ and $\gamma=(1-v^2)^{-\frac{1}{2}}$.

Our original field, $F_{\mu\nu}=F^{B_{x}}_{\mu\nu}+F^{B_{z}}_{\mu\nu}$, has a simple asymptotic form in Min\-kows\-kian coordinates,
\begin{equation}
\label{Fmink}
||F^{\rm{asymptotic}}_{\mu\nu}||=
\begin{pmatrix}
0&0&0&0\\
0&0&B_{z}&0\\
0&-B_{z}&0&B_{x}\\
0&0&-B_{x}&0
\end{pmatrix}.
\end{equation}
 To transform the covariant tensor $F_{\mu\nu}$, the inverse Lorentz transformation $\Lambda^{\mu}_{\;\nu '}=(\Lambda^{\nu '}_{\;\mu})^{-1}$ would be used. But we realize that the Boyer--Lindquist coordinate system which we use to perform all the calculations (and also to express the EM tensor of the final field) is centered around black hole and the rest frame of the black hole is thus our ``laboratory'' reference frame. As we consider a drift of the black hole against the field \rf{Fmink}, we need to perform inverse Lorentz transformation. Quantities $B_{x}$ and $B_{z}$ appearing therein would be primed in standard notation. For inverse transformation of covariant tensors we use original $\Lambda^{\nu '}_{\;\mu}$. 
Thus for ``drifting'' $F_{\mu\nu}$ we have (denoting $F_{\mu ' \nu '}$ that of \rf{Fmink}):
\begin{equation}
 \label{transform}
F^{\rm{asymptotic}}_{\mu\nu}=F^{\rm{asymptotic}}_{\mu ' \nu '}\Lambda^{\mu '}_{\;\mu}\Lambda^{\nu '}_{\;\nu},
\end{equation}
which may be written in the matrix formalism as follows:
\begin{equation}
 \label{transformmatrix}
||F^{\rm{asymptotic}}_{\mu\nu}||=||\Lambda^{\mu '}_{\;\mu}||^{\rm{t}}\;||F^{\rm{asymptotic}}_{\mu ' \nu '}||\;||\Lambda^{\nu '}_{\;\nu}||=||\Lambda^{\mu '}_{\;\mu}||\;||F^{\rm{asymptotic}}_{\mu ' \nu '}||\;||\Lambda^{\nu '}_{\;\nu}||,
\end{equation}
and results in:
\begin{equation}
\label{transvysl}
||F^{\rm{asymptotic}}_{\mu\nu}||=
\begin{pmatrix}
0&v_{y}\gamma B_{z}&-\gamma (v_{x}B_{z}-v_{z}B_{x})&-v_{y}\gamma B_{x}\\
-v_{y}\gamma B_{z}&0&\gamma B_{z}-v_{z}N&v_{y}N \\
\gamma (v_{x}B_{z}-v_{z}B_{x})&-\gamma B_{z}+v_{z}N&0& \gamma B_{x}-v_{x}N\\
v_{y}\gamma B_{x}&-v_{y}N&-\gamma B_{x}+v_{x}N&0
\end{pmatrix},
\end{equation}
with $N\equiv \frac{\gamma^2}{\gamma+1}(v_{z}B_{z}+v_{x}B_{x}).$

Final step of the derivation is thus substitution of ``non-drifting'' quantities $E_{x,y,z}$ and $B_{x,y,z}$ in the tensors $F_{\mu\nu}^{E_{x,y,z}}$ and $F_{\mu\nu}^{B_{x,y,z}}$ by Lorentz transformed values from \rf{transvysl} and superposing the components to acquire general EM tensor describing the field around the Kerr black hole drifting through asymptotically uniform magnetic field of general orientation:
\begin{align}
\nonumber F_{\mu\nu}=&F^{E_{x}}_{\mu\nu}(E_{x}\rightarrow-v_{y}\gamma B_{z})+F^{E_{y}}_{\mu\nu}(E_{y}\rightarrow\gamma(v_{x}B_{z}-v_{z}B_{x}))+\\
\label{finalF} &+F^{E_{z}}_{\mu\nu}(E_{z}\rightarrow v_{y}\gamma B_{x})+F^{B_{x}}_{\mu\nu}(B_{x}\rightarrow \gamma B_{x}-v_{x}N)+\\
\nonumber &+F^{B_{y}}_{\mu\nu}(B_{y}\rightarrow -v_{y}N)+F^{B_{z}}_{\mu\nu}(B_{z}\rightarrow \gamma B_{z}-v_{z}N).
\end{align}  

\section{Choice of the tetrad}
\label{tetrads}
To settle our notation of the tetrad formalism we review basic relations between the tetrad basis vectors $e^{\mu}_{(\alpha)}$ and dual basis 1-forms $e_{\mu}^{(\alpha)}$
\begin{align}
 e^{\mu}_{(\alpha)}e_{\nu}^{(\beta)}&=\delta^{(\beta)}_{(\alpha)} \delta^{\mu}_{\nu},\\
\eta_{(\mu)(\nu)}&=g_{\alpha\beta}e^{\alpha}_{(\mu)}e^{\beta}_{(\nu)},\\
g_{\mu\nu}&=\eta_{(\alpha)(\beta)}e^{(\alpha)}_{\mu}e^{(\beta)}_{\nu},\\
e_{(\alpha)\mu}e_{(\beta)\nu}&=g_{\mu\nu}\eta_{(\alpha)(\beta)},
\end{align}
i.e. tetrad indices are raised/lowered using the Minkowskian metric tensor $\eta_{(\alpha)(\beta)}$.

Tetrad basis vector $e^{\mu}_{(t)}$ is defined by the four-velocity of the observer carrying the tetrad $e^{\mu}_{(t)}=u^{\mu}$. Covariant components $u_{\mu}$ are thus related to the tetrad basis as follows $u_{\mu}=e_{(t)\:\mu}=-e^{(t)}_{\mu}$.

\subsection{Locally non-rotating frame}
One of the standard tetrads which are often used when dealing with Kerr geometry is the one carried by the zero angular momentum observer (ZAMO). ZAMO's rest frame is usually called {\bf locally non-rotating frame (LNRF)} since it seemingly suppresses intrinsic rotation of the geometry. LNRF basis vectors and dual basis 1-forms may be expressed as follows \citep{bardeen72}
\begin{align}
\label{zamotetrad1}
 e_{(t)}^{\mu}&= u^{\mu}=\frac{A^{1/2}}{\Delta^{1/2}\Sigma^{1/2}}\left[1,0,0,\Omega\right],\\
e_{(r)}^{\mu}&=\left[0,\frac{\Delta^{1/2}}{\Sigma^{1/2}},0,0\right],\\
e_{(\theta)}^{\mu}&=\left[0,0,\frac{1}{\Sigma^{1/2}},0\right],\\
\label{zamotetrad4}
e_{(\varphi)}^{\mu}&=\left[0,0,0, \frac{\Sigma^{1/2}}{A^{1/2}\sin\theta}\right],\\
e^{(t)}_{\mu}&= u_{\mu}=\left[\frac{\Sigma^{1/2} \Delta^{1/2}}{A^{1/2}},0,0,0\right],\\
e^{(r)}_{\mu}&=\left[0,\frac{\Sigma^{1/2}}{\Delta^{1/2}},0,0\right],\\
e^{(\theta)}_{\mu}&=\left[0,0,\Sigma^{1/2},0\right],\\
e^{(\varphi)}_{\mu}&=\frac{A^{1/2}\sin\theta}{\Sigma^{1/2}}\left[-\Omega,0,0,1\right],
\end{align}
where $\Omega=-\frac{g_{t\varphi}}{g_{\varphi\varphi}}$. The way in which ZAMO ``fits the geometry'' may be expressed by relation $g^{t\varphi}u^t_{\rm{ZAMO}} = g^{tt}u^{\varphi}_{\rm{ZAMO}}$ which appears useful in the calculations. For a detailed discussion of the properties of LNRF and other stationary frames see \citet{semerak93}.

Elegant way to calculate tetrad basis vectors of any other frame is Lorentz transformation of LNRF basis vectors. In order to do this one needs to express linear velocity $v^{(i)}$ of the new frame relative to ZAMO:
\begin{equation}
\label{linspeed}
 v^{(i)}=\frac{u^{(i)}}{u^{(t)}}=\frac{e^{(i)}_{\mu}u^{\mu}}{e^{(t)}_{\mu}u^{\mu}}.
\end{equation}
This speed defines the Lorentz boost with factor $\gamma=(1-v^2)^{-1/2}$. Change of the basis matrix for this transformation is the inverse Lorentz matrix $(\Lambda^{\nu '}_{\;\mu})^{-1}$. For the transformation of basis 1-forms the direct $\Lambda^{\nu '}_{\;\mu}$ is used.

\subsection{Frame of the free-falling observer}
We shall calculate the basis vectors of the frame attached to the {\bf inertial observer who is free-falling from the rest at infinity (FFOFI)}. Substituting $L=0$ and $E=m$ (particle's rest energy in geometrized units) into Carter's equations of motion \citep{carter68} yields
\begin{align}
 \label{FFOFIspeed}
u^t_{\rm{FFOFI}}=\frac{A}{\Delta\Sigma},\;\;\; u^r_{\rm{FFOFI}}=-\left( \frac{2Mr(r^2+a^2)}{\Sigma^2} \right)^{1/2},\;\;\;u^{\theta}_{\rm{FFOFI}}=0,\;\;\;u^{\varphi}_{\rm{FFOFI}}=\frac{2Mar}{\Delta\Sigma},
\end{align}
 which we substitute into \rf{linspeed} obtaining single nonzero component $v^{(r)}=-\left( \frac{2r(r^2+a^2)}{A} \right)^{1/2}$. The boost is indeed purely radial: $v^{(\varphi)}=0$ although $u^{\varphi}_{\rm{FFOFI}}\neq u^{\varphi}_{\rm{LNRF}}$. Lorentz factor reads $\gamma=\frac{A^{1/2}}{\Delta^{1/2}\Sigma^{1/2}}$. New (primed) basis is then obtained straightforwardly from the LNRF (unprimed) basis by the matrix multiplication

\begin{equation}
 \begin{pmatrix}
e'^{\:t}_{(t)}&e'^{\:t}_{(r)}&e'^{\:t}_{(\theta)}&e'^{\:t}_{(\varphi)}\\
e'^{\:r}_{(t)}&e'^{\:r}_{(r)}&e'^{\:r}_{(\theta)}&e'^{\:r}_{(\varphi)}\\
e'^{\:\theta}_{(t)}&e'^{\:\theta}_{(r)}&e'^{\:\theta}_{(\theta)}&e'^{\:\theta}_{(\varphi)}\\
e'^{\:\varphi}_{(t)}&e'^{\:\varphi}_{(r)}&e'^{\:\varphi}_{(\theta)}&e'^{\:\varphi}_{(\varphi)}
\end{pmatrix}
=
 \begin{pmatrix}
e^{\:t}_{(t)}&e^{\:t}_{(r)}&e^{\:t}_{(\theta)}&e^{\:t}_{(\varphi)}\\
e^{\:r}_{(t)}&e^{\:r}_{(r)}&e^{\:r}_{(\theta)}&e^{\:r}_{(\varphi)}\\
e^{\:\theta}_{(t)}&e^{\:\theta}_{(r)}&e^{\:\theta}_{(\theta)}&e^{\:\theta}_{(\varphi)}\\
e^{\:\varphi}_{(t)}&e^{\:\varphi}_{(r)}&e^{\:\varphi}_{(\theta)}&e^{\:\varphi}_{(\varphi)}
\end{pmatrix}
\begin{pmatrix}
\gamma&v^{(r)}\gamma&0&0\\
v^{(r)}\gamma&\gamma&0&0\\
0&0&1&0\\
0&0&0&1
\end{pmatrix}.
\end{equation}

We read FFOFI tetrad basis vectors from the above-given matrix equation. Corresponding 1-forms are obtained from the analogous equation by arranging their components in a very same manner as those of basis vectors and switching to direct Lorentz transformation ($v^{(r)}\rightarrow -v^{(r)}$ in the matrix of transformation). 
\begin{align}
 e_{(t)}^{\mu}&= u^{\mu}=\left[\frac{A}{\Delta\Sigma},-\left( \frac{2Mr(r^2+a^2)}{\Sigma^2} \right)^{1/2},0,\frac{2Mar}{\Delta\Sigma}\right],\\
e_{(r)}^{\mu}&=\left[-\frac{(2Mr(r^2+a^2))^{1/2}A^{1/2}}{\Delta\Sigma},\frac{A^{1/2}}{\Sigma},0,-\frac{2Mar(2Mr(r^2+a^2))^{1/2}}{\Delta\Sigma A^{1/2}}\right],\\
e_{(\theta)}^{\mu}&=\left[0,0,\frac{1}{\Sigma^{1/2}},0\right],\\
e_{(\varphi)}^{\mu}&=\left[0,0,0, \frac{\Sigma^{1/2}}{A^{1/2}\sin\theta}\right],\\
e^{(t)}_{\mu}&= u_{\mu}=\left[1,\frac{(2Mr(r^2+a^2))^{1/2}}{\Delta},0,0\right],\\
e^{(r)}_{\mu}&=\left[\frac{(2Mr(r^2+a^2))^{1/2}}{A^{1/2}},\frac{A^{1/2}}{\Delta},0,0\right],\\
e^{(\theta)}_{\mu}&=\left[0,0,\Sigma^{1/2},0\right],\\
e^{(\varphi)}_{\mu}&=\frac{A^{1/2}\sin\theta}{\Sigma^{1/2}}\left[-\Omega,0,0,1\right],
\end{align}
we notice that $e_{(\theta)}^{\mu}$ and $e_{(\varphi)}^{\mu}$ are common to both LNRF and FFOFI frames. 

Both ZAMO and FFOFI observers  are physical everywhere above the horizon. In the following, however, we shall set up an observer which is restricted to the equatorial plane $\theta=\pi/2$. 

\subsection{Frame of Keplerian observer}
As we are primarily interested in astrophysically relevant situations we will employ the orthonormal tetrad carried by the inertial observer on the circular Keplerian orbit around the black hole (KEP tetrad). Such an orbit is specified by the values of constants of motion -- by specific angular momentum $\tilde{L}\equiv u_{\varphi}$ and specific energy $\tilde{E}\equiv -u_{t}$ which are expressed as follows \citep{bardeen72}:
\begin{equation}
 \label{kepconst}
\tilde{E}(r)=\frac{r^2-2Mr\pm a \sqrt{Mr}}{r\sqrt{r^2-3Mr\pm 2a\sqrt{Mr}}},\;\;\;\;\tilde{L}(r)=\frac{\pm\sqrt{M} (r^2+a^2\mp 2a\sqrt{Mr})}{\sqrt{r(r^2-3Mr\pm 2a\sqrt{Mr})}},
\end{equation}
where the upper signs are valid for the prograde (direct) orbits and the lower ones for the retrograde (counter--revolving) orbits. 
Such a tetrad is physical only above marginally stable orbit $r_{\rm{ms}}$ which represents a radial boundary for the stationary geodesic motion in the equatorial plane:
\begin{equation}
 \label{rms}
r_{\rm{ms}}=M\left(3+Z_2\mp\sqrt{(3-Z_1)(3+Z_1+2Z_2)}\right),
\end{equation}
where $Z_1\equiv1+\left(1-\frac{a^2}{M^2}\right)^{1/3}\left[\left(1+\frac{a}{M}\right)^{1/3}+\left(1-\frac{a}{M}\right)^{1/3}\right]$ and $Z_{2} \equiv \sqrt{\frac{3a^2}{M^2}+Z_1^2}$. Position of the marginally stable orbit is plotted as a function of spin $a$ in \rff{rms_fig}.

\begin{figure}[htb!]
\centering
\includegraphics[scale=0.4, clip]{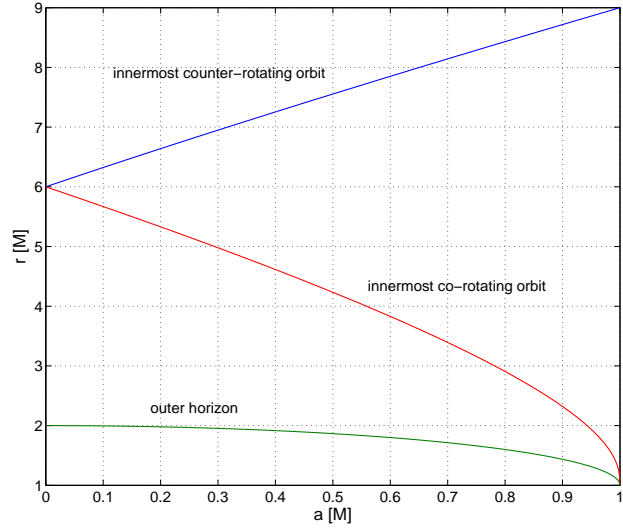}
\caption{Position of the marginally stable orbit for both co-rotating and counter-rotating equatorial circular geodesics as well as the position of the outer horizon is plotted as a function of BH spin parameter $a$.}
\label{rms_fig}
\end{figure}

Angular velocity of a circular orbit is $\Omega_{\rm{Kep}}=\frac{\pm 1}{M^{-1/2}r^{3/2}\pm a}$ where the upper signs are for prograde orbits and the lower ones for the retrograde orbits. Linear velocity \rf{linspeed} of the orbiting tetrad as measured by ZAMO observer in LNRF reads $v^{(\varphi)}=\frac{A\sin\theta}{\Sigma\Delta^{1/2}}\left(\Omega_{\rm{Kep}}-\Omega\right)$. Lorentz factor of this azimuthal boost is $\gamma=(1-[v^{(\varphi)}]^2)^{-1/2}$. By means of Lorentz transformation of LNRF basis we obtain basis vectors of the orbiting frame \citep{yoko2005}:
\begin{align}
\nonumber &e_{(t)}^{\mu}=u^{\mu}=\gamma\left( \frac{A}{\Delta\Sigma} \right)^{1/2} [1,0,0,\Omega_{\rm{Kep}}],\\
\label{orbitingtetrad} &e_{(r)}^{\mu}=\left(\frac{\Delta}{\Sigma}\right)^{1/2}[0,1,0,0],\\
\nonumber &e_{(\theta)}^{\mu}=\frac{1}{\sqrt{\Sigma}}[0,0,1,0],\\
\nonumber &e_{(\varphi)}^{\mu}=\gamma\left[v^{(\varphi)}\left( \frac{A}{\Delta\Sigma} \right)^{1/2},0,0,\frac{\sqrt{\Sigma}}{\sin\theta\sqrt{A}}+v^{(\varphi)}\Omega\left( \frac{A}{\Delta\Sigma} \right)^{1/2}\right].
\end{align}  
 where we define $A\equiv(r^2+a^2)^2-a^2\Delta\sin^2\theta$ to express the coordinate angular velocity of LNRF (i.e. angular velocity of the frame-dragging) $\Omega=\frac{2a}{A}Mr$. 

For $r<r_{\rm{ms}}$ there are no more circular orbits. Thus we suppose that the orbiting observer who reaches this limit performs a free fall to the black hole keeping the values of the constants of motion corresponding to the marginally stable orbit at $r_{\rm{ms}}$ (FFO tetrad). He is falling with $\tilde{E}_{\rm{ms}}\equiv\tilde{E}(r_{\rm{ms}})$ and $\tilde{L}_{\rm{ms}}\equiv\tilde{L}(r_{\rm{ms}})$ given by eqs. (\ref{kepconst}) -- (\ref{rms}).
Having fixed $u_{t}(r<r_{\rm{ms}})=-\tilde{E}_{\rm{ms}}$, $u_{\varphi}(r<r_{\rm{ms}})=\tilde{L}_{\rm{ms}}$ and $u_{\theta}=0$ we get radial component $u_{r}$ easily from the normalisation condition $u^{\mu}u_{\mu}=-1$. Contravariant components of the 4-velocity are then:
\begin{align}
\nonumber u^{t}&=\frac{1}{r\Delta}[(r[r^2+a^2]+2Ma^2)\tilde{E}_{\rm{ms}}-2Ma\tilde{L}_{\rm{ms}}],\\
\label{falling4velocity} u^{r}&=-\frac{1}{r^{3/2}}\sqrt{[r(r^2+a^2)+2Ma^2]\tilde{E}^2_{\rm{ms}}-4 Ma \tilde{E}_{\rm{ms}}\tilde{L}_{\rm{ms}}-(r-2M)\tilde{L}^2_{\rm{ms}}-r\Delta},\\
\nonumber u^{\theta}&=0,\\
\nonumber u^{\varphi}&=\frac{1}{r\Delta}[2M^2a\tilde{E}_{\rm{ms}}+M(r-2M)\tilde{L}_{\rm{ms}}].
\end{align}  
 Spatial 1-forms of the tetrad of this falling observer may be expressed as follows \citep{dovciak_thesis}:
\begin{align}
\label{falling1forms}
\nonumber e^{(r)}_{\mu}&=\frac{r}{\sqrt{\Delta(1+u^{r}u_{r})}}(u^{r}[u_{t},u_{r},0,u_{\varphi}]+[0,1,0,0]),\\
e^{(\theta)}_{\mu}&=[0,0,r,0],\\
\nonumber e^{(\varphi)}_{\mu}&=\sqrt{\frac{\Delta}{1+u^{r}u_{r}}}[-u^{\varphi},0,0,u^{t}].
\end{align}

We conclude the we are now equipped with three distinct tetrads to be applied in the equatorial plain, namely non-inertial LNRF and two inertial tetrads FFOFI and KEP+FFO. Outside the equatorial plain only LNRF and FFOFI may be applied. 

\section{Structure of the electromagnetic field}
\label{structure}
\subsection{Stationary electromagnetic field}
\label{stationary}
First we shall review several aspects of the stationary, i.e. non-drifting, asymptotically homogeneous test fields given by eqs. (\ref{BFtr(Bx)}) and (\ref{BFtr(Bz)}). Since the stationary fields of this type has already been discussed thoroughly in the literature \citep[see e.g. ][and references therein]{bicak89} we will focus mainly on the comparison of the field line structures resulting from the alternative definitions of the field components which were given in \rs{lines}. We also discuss the choice of the test charge four-velocity profile (ZAMO versus FFOFI basically). Besides that we shall introduce various techniques to visualize the vector fields in both,  two-dimensional plane sections as well as the stereometric projections of three-dimensional space.
 
\subsubsection{Expulsion of the aligned magnetic field (Meissner effect)}
We briefly revisit the issue of the expulsion of the axisymmetric stationary magnetic field out of the horizon of the extreme Kerr black hole ($a=M$) which is known as Meissner or Meissner-type effect in the analogy with similar effect which superconducting bodies exhibit upon the external magnetic fields \citep[e.g.][and references therein]{dovciak00,bicak00}. In particular we shall discuss the role of the definition of the lines of force and the observer dependence of this effect. 


\begin{figure}[h!]
\centering
\includegraphics[scale=0.4, clip]{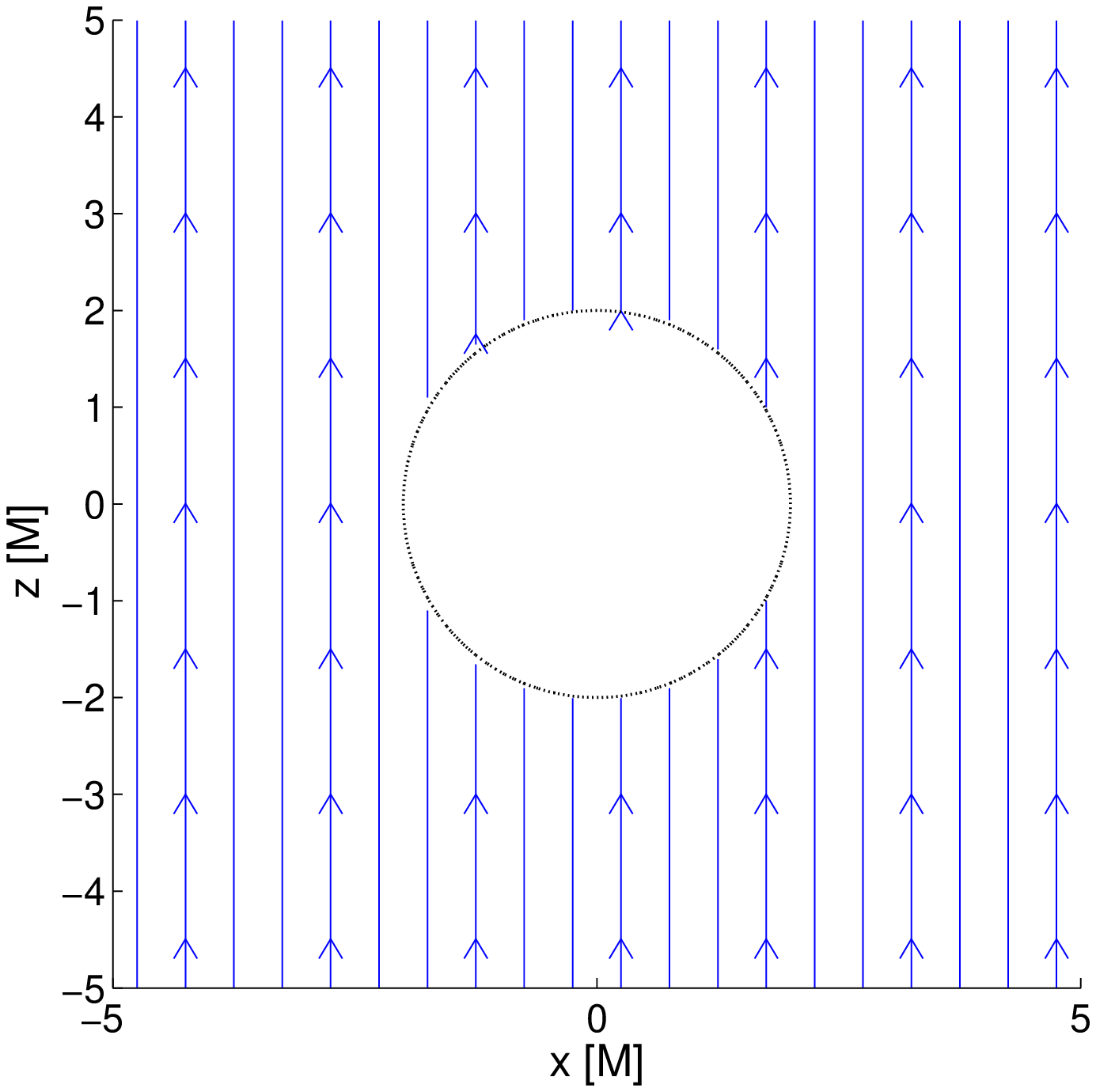}\includegraphics[scale=0.4, clip]{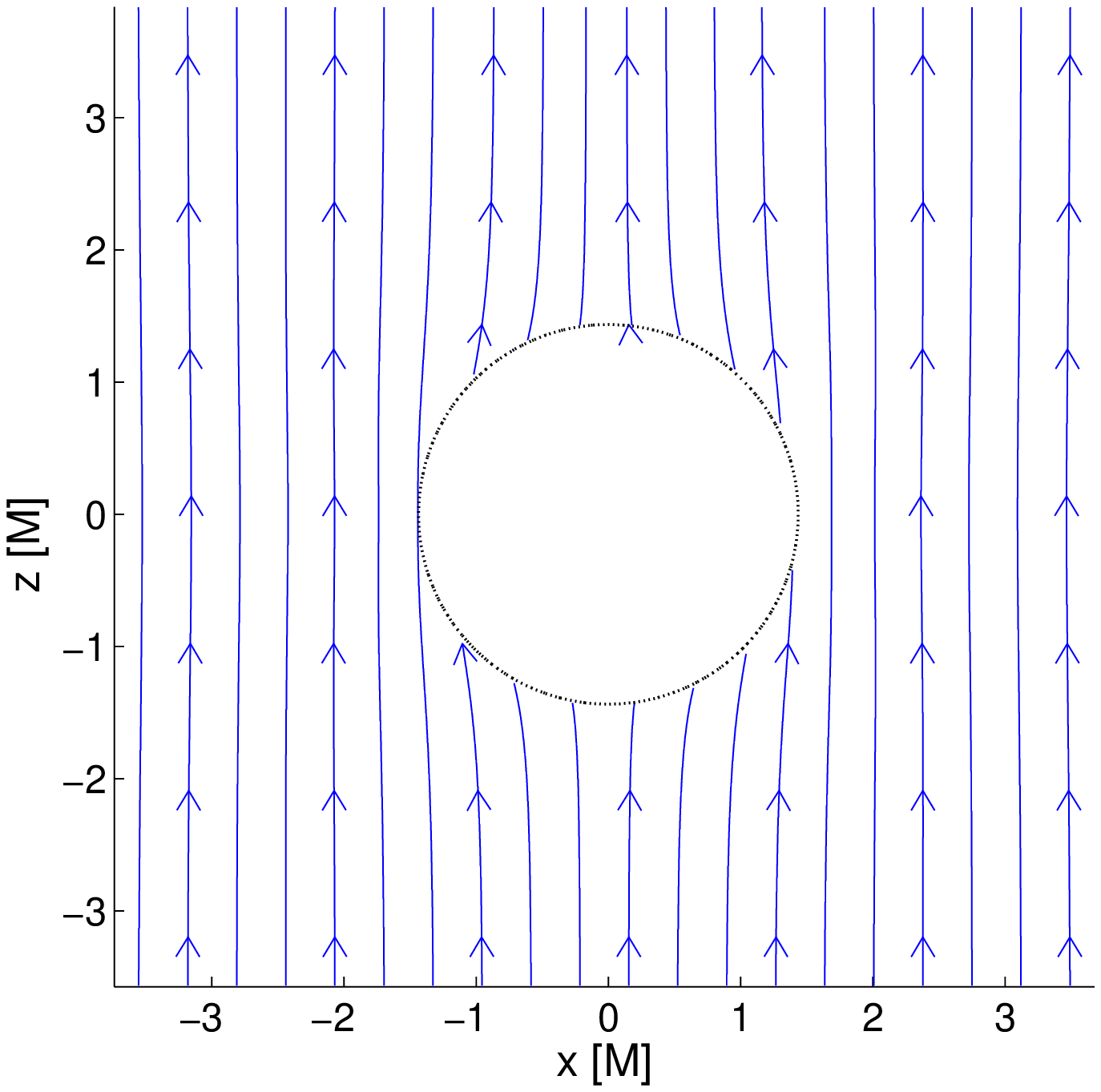}\\
\includegraphics[scale=0.4, clip]{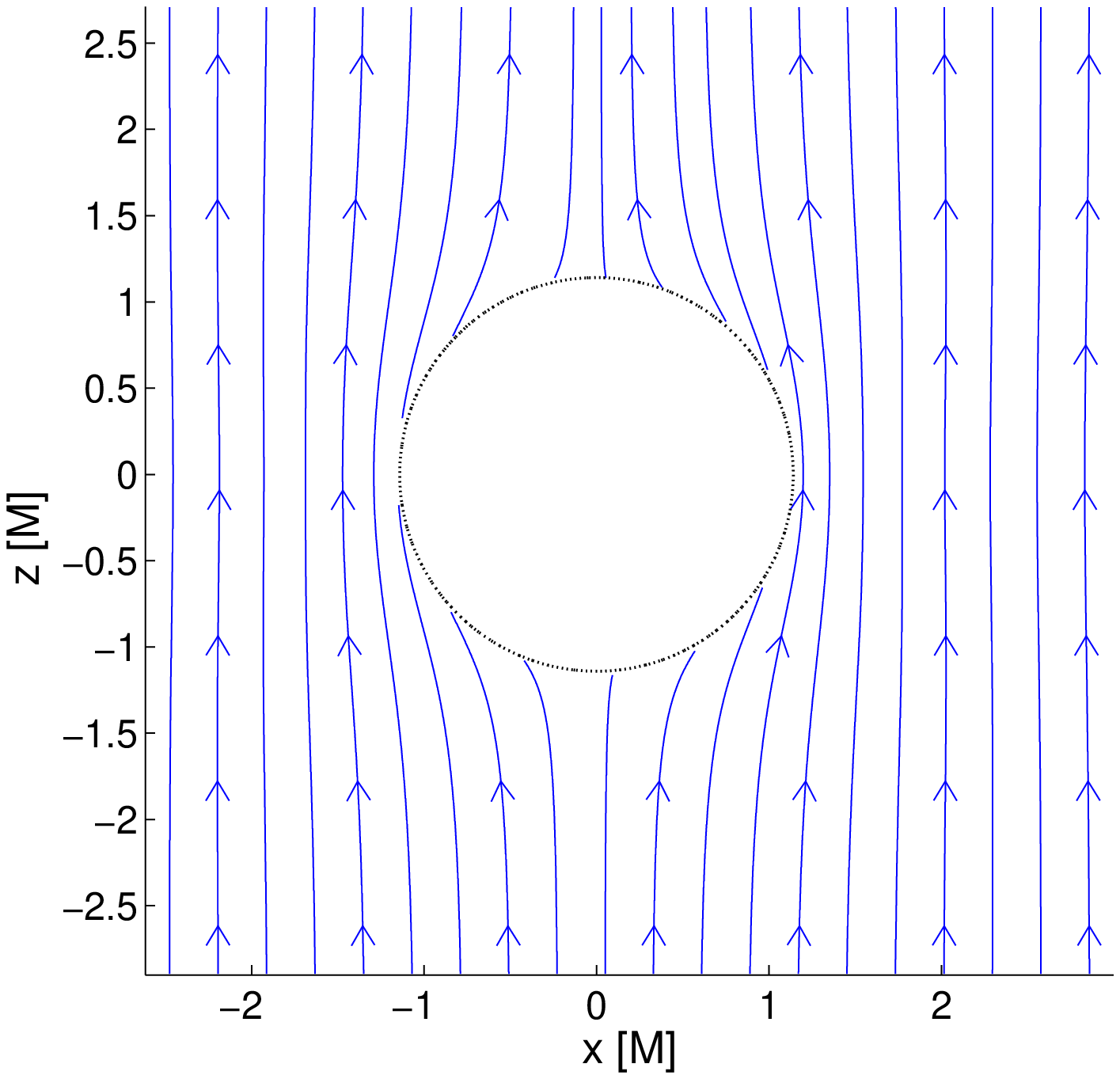}\includegraphics[scale=0.4, clip]{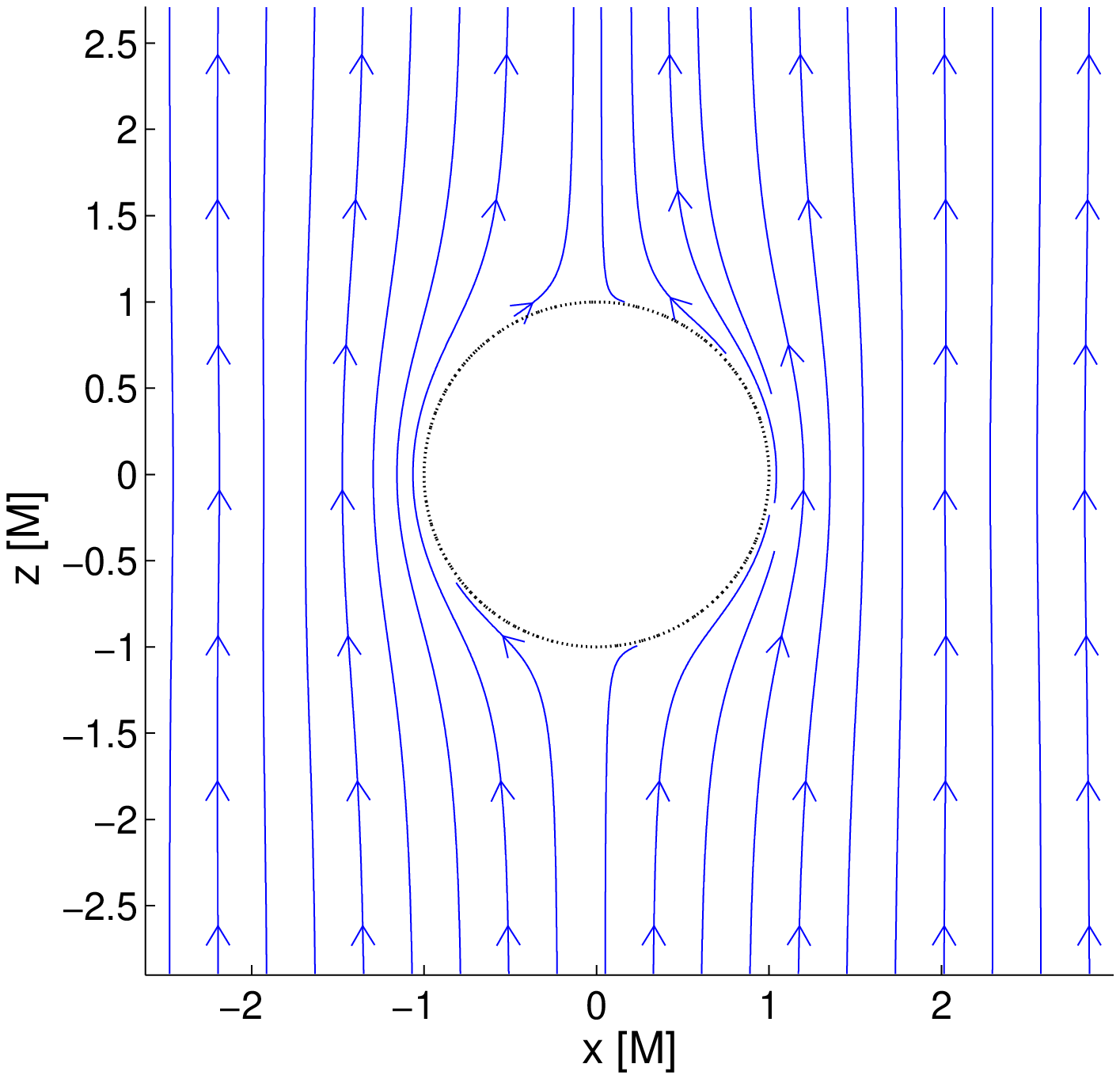}
\caption{Meissner effect captured by AMO components which translate the vanishing of the $F_{\theta\varphi}$ at the horizon directly into the radial component of the magnetic field $B^r_{\rm{AMO}}$. Upper left panel shows the case $a=0$ (Schwarzschild limit) where the field lines penetrate the horizon completely indifferently. Increasing the spin the lines of force start to bend slightly which becomes obvious for higher spin values only. Upper right panel captures the situation for $a=0.9\;M$. For $a=0.99\;M$ the effect becomes more apparent (bottom left panel) and in the extreme case $a=M$ the magnetic field is expelled from the horizon completely (bottom right panel).}
\label{vytl_amo}
\end{figure}

\begin{figure}[h!]
\centering
\includegraphics[scale=0.32, clip]{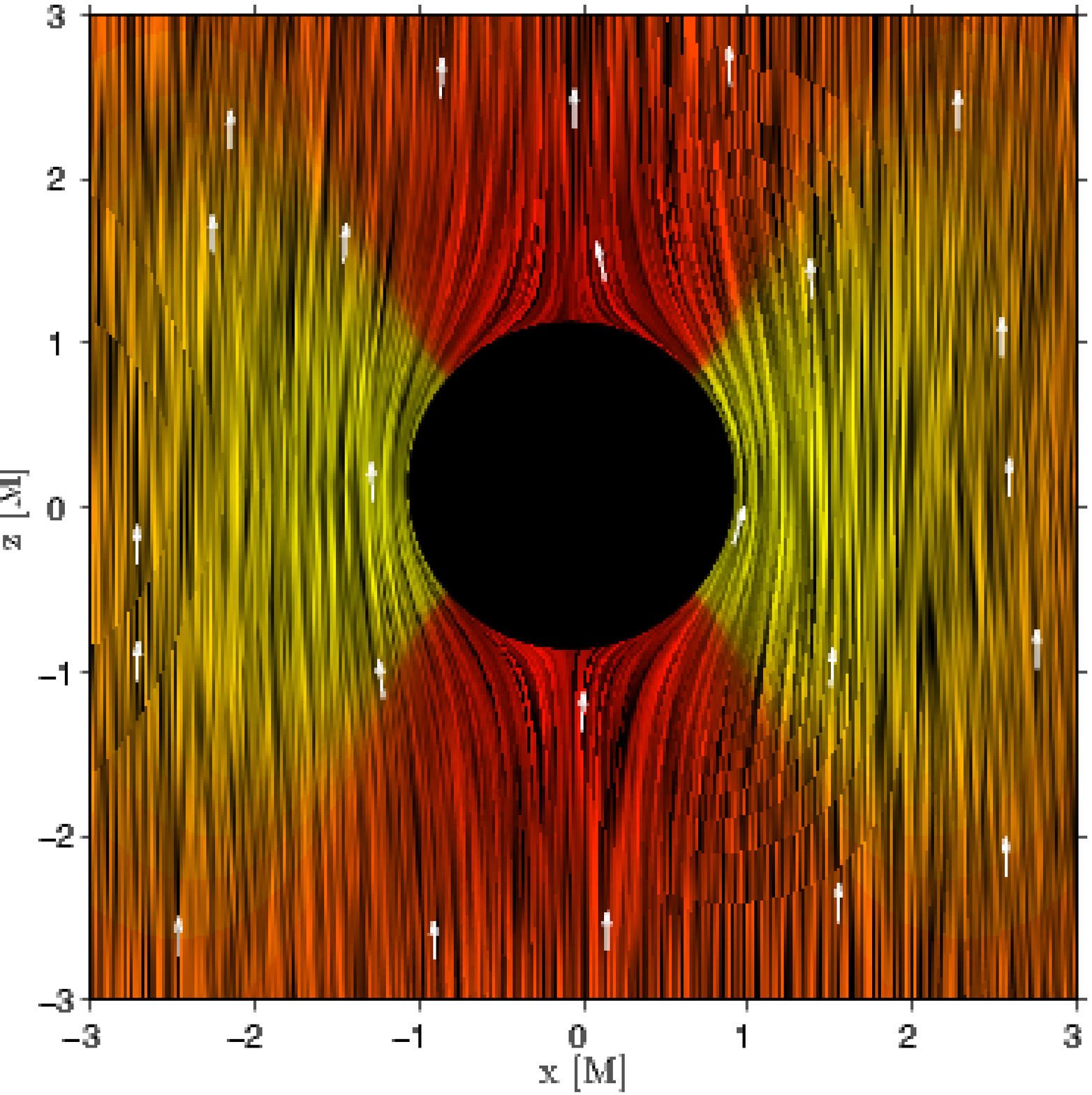}
\includegraphics[scale=0.32, clip]{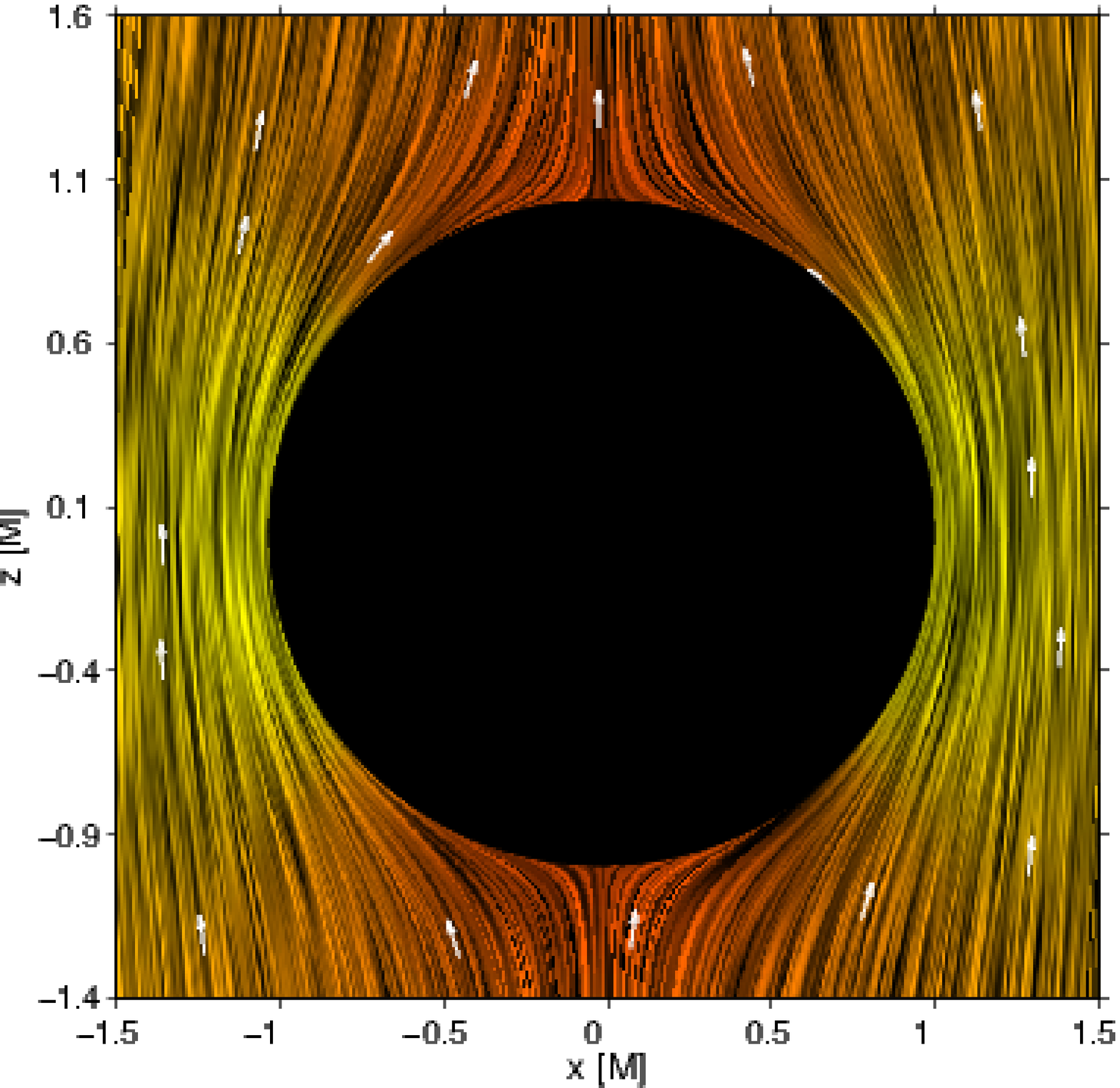}
\caption{Alternative visualization of the field lines employing linear integral convolution  (LIC) method \citep{shambo05}. Strength of the field is encoded by a colormap: brighter color means stronger field. Amplification of the magnetic field in the equatorial zone in the vicinity of the horizon is observed.}
\label{vytl_amo_LIC}
\end{figure}

Scalar magnetic/electric flux $\phi_{\rm{e/m}}$ through the given surface $S$ may be calculated as another quantity revealing the structure of the field. In particular the general expression for the magnetic flux
\begin{equation}
\label{magneticflux}
 \phi_{\rm{m}}=\int_S \mbold{F}\wedge \mbold{d} S,
\end{equation}
reduces to the simple form $\phi_{\rm{m}}=\int F_{\theta\varphi} \;\mathrm{d}\theta\,\mathrm{d}\varphi$ if we choose the surface to be a part of the sphere of constant $r$. If we consider the case of aligned field of the strength $B_{\rm{z}}$ given by \rf{BFtr(Bz)} and further specify the surface to be an axisymmetric polar cap determined by the polar angle $\theta$ we may integrate straightforwardly to obtain $\phi_{\rm{m}}=\pi B_{z}\left[\Delta+\frac{2Mr}{\Sigma}(r^2-a^2)\right]\:\sin^2\theta$. The flux through the horizon vanishes in the case of extreme black hole and magnetic field is fully expelled from the horizon. Poloidal section of the surfaces of the constant flux discussed by \citet{dovciak00} coincide with the AMO field lines since they both directly reflect the behaviour of the $F_{\theta\varphi}^{B_z}$ which is zero at the horizon of the extreme BH. Expulsion of the magnetic field in AMO definition is visualized in figs. \ref{vytl_amo} and \ref{vytl_amo_LIC}. The latter employs the linear integral convolution (LIC) method \citep{shambo05} which encodes the field structure into the texture resembling iron filings. The strength of the field is expressed by the color scale. Visualisation of the vector fields using LIC method proves especially useful when dealing with complex field structures for which the usual field lines do not keep up with the sudden spatial changes (e.g. dense zigzag structures, crack propagation etc).

\begin{figure}[hp!]
 \centering
\includegraphics[scale=0.34, clip]{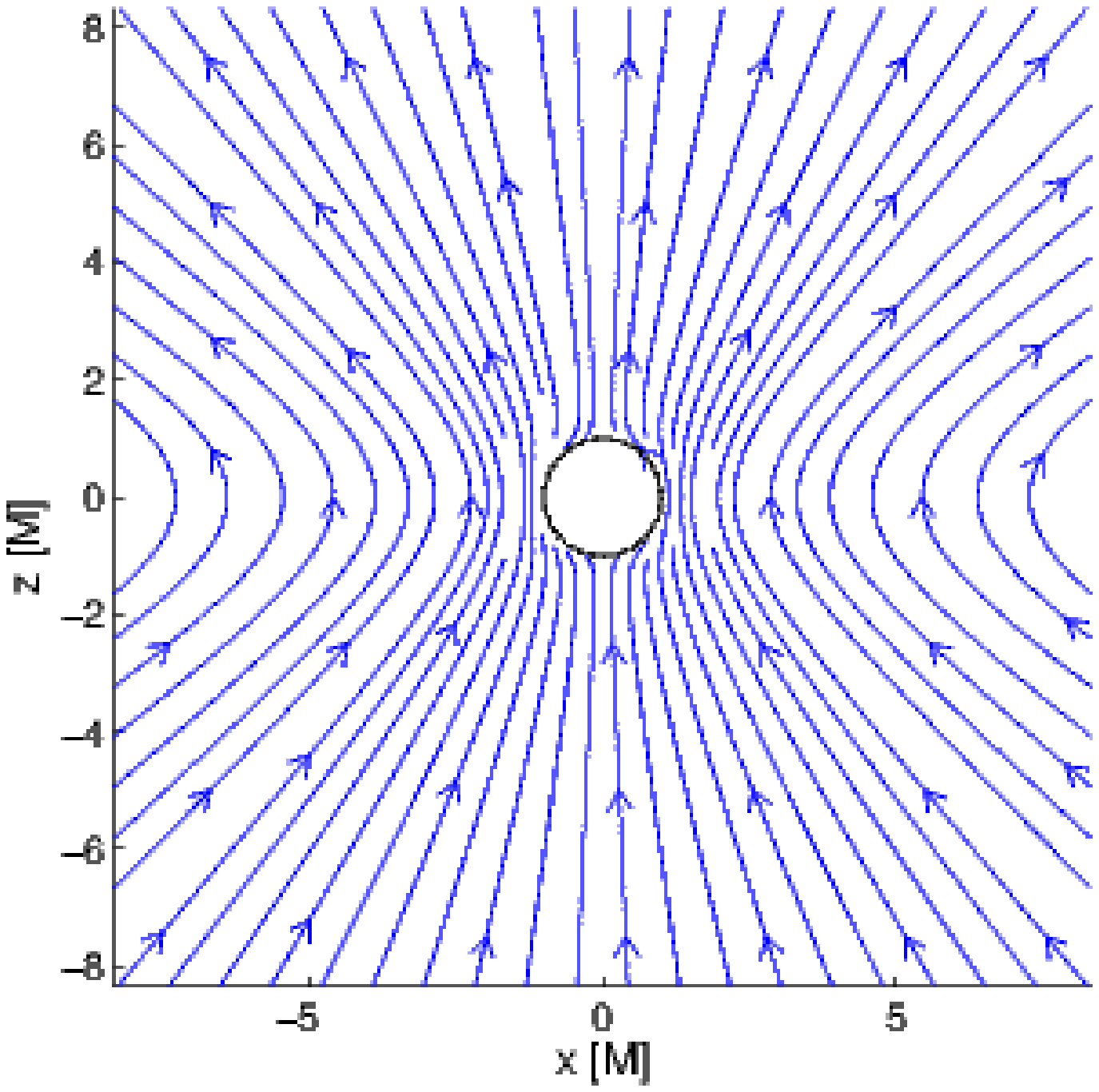}~~~~
\includegraphics[scale=0.34, clip]{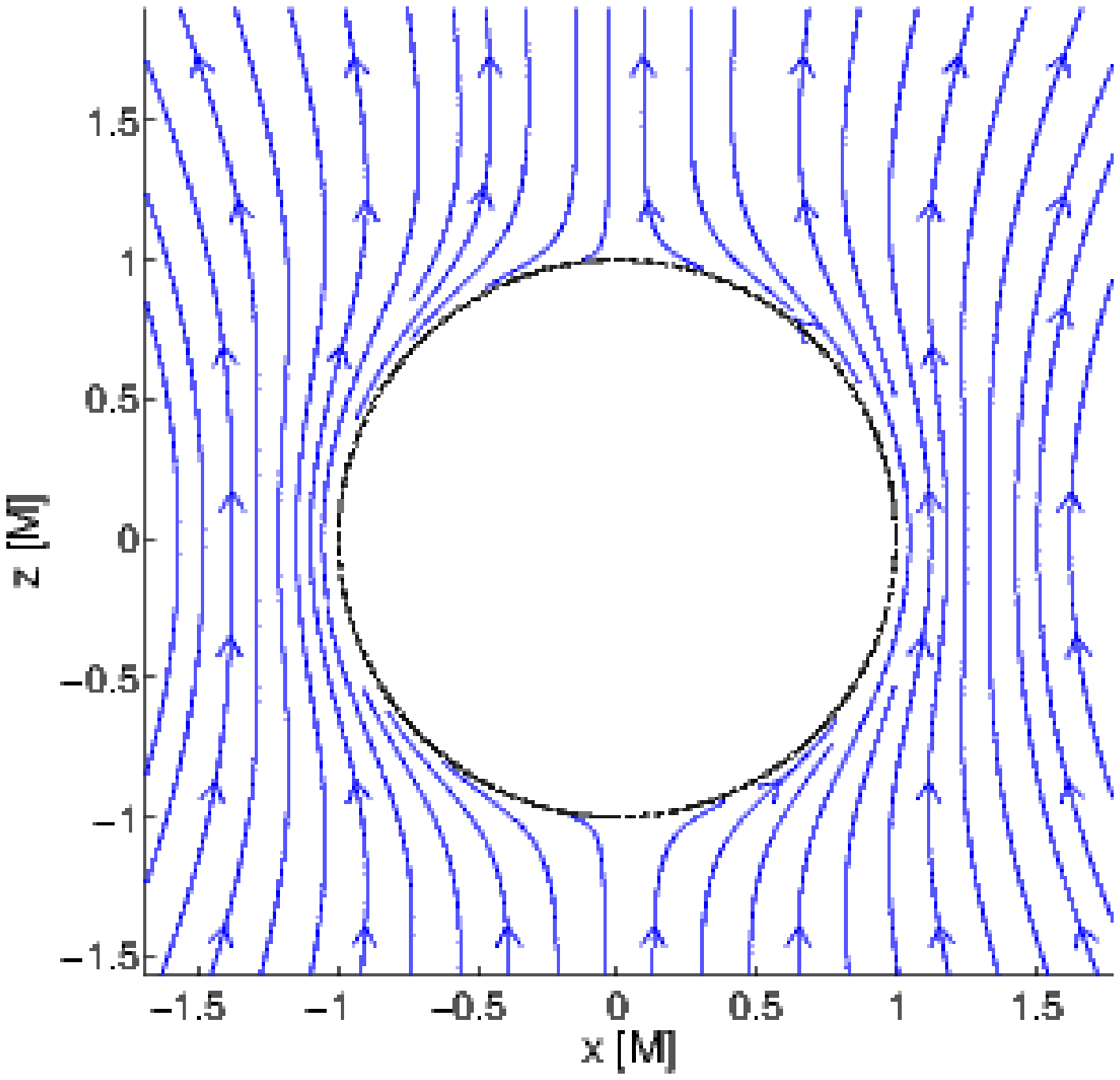}\\
\includegraphics[scale=0.34, clip]{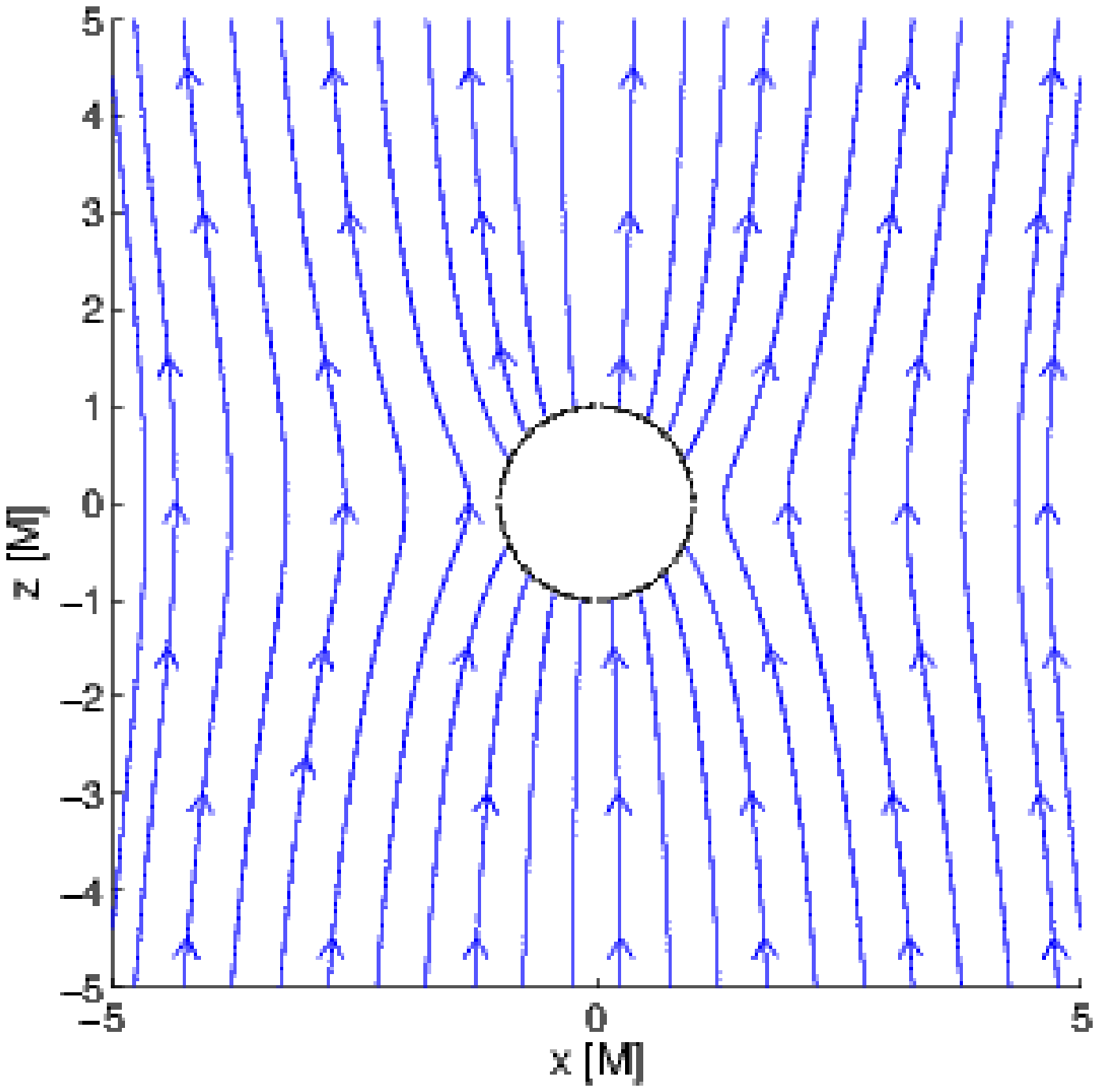}~~~~
\includegraphics[scale=0.34, clip]{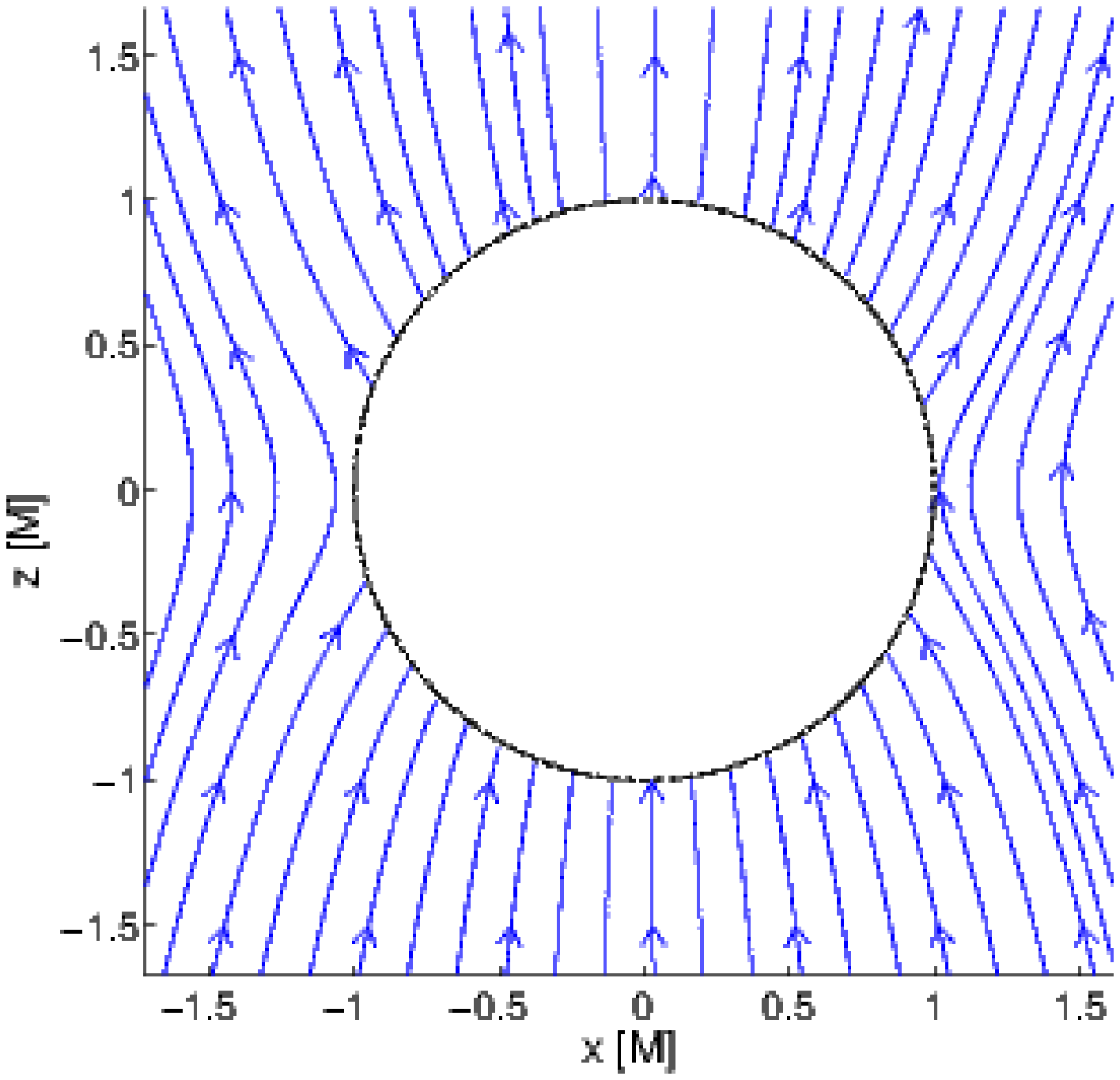}\\
\includegraphics[scale=0.34, clip]{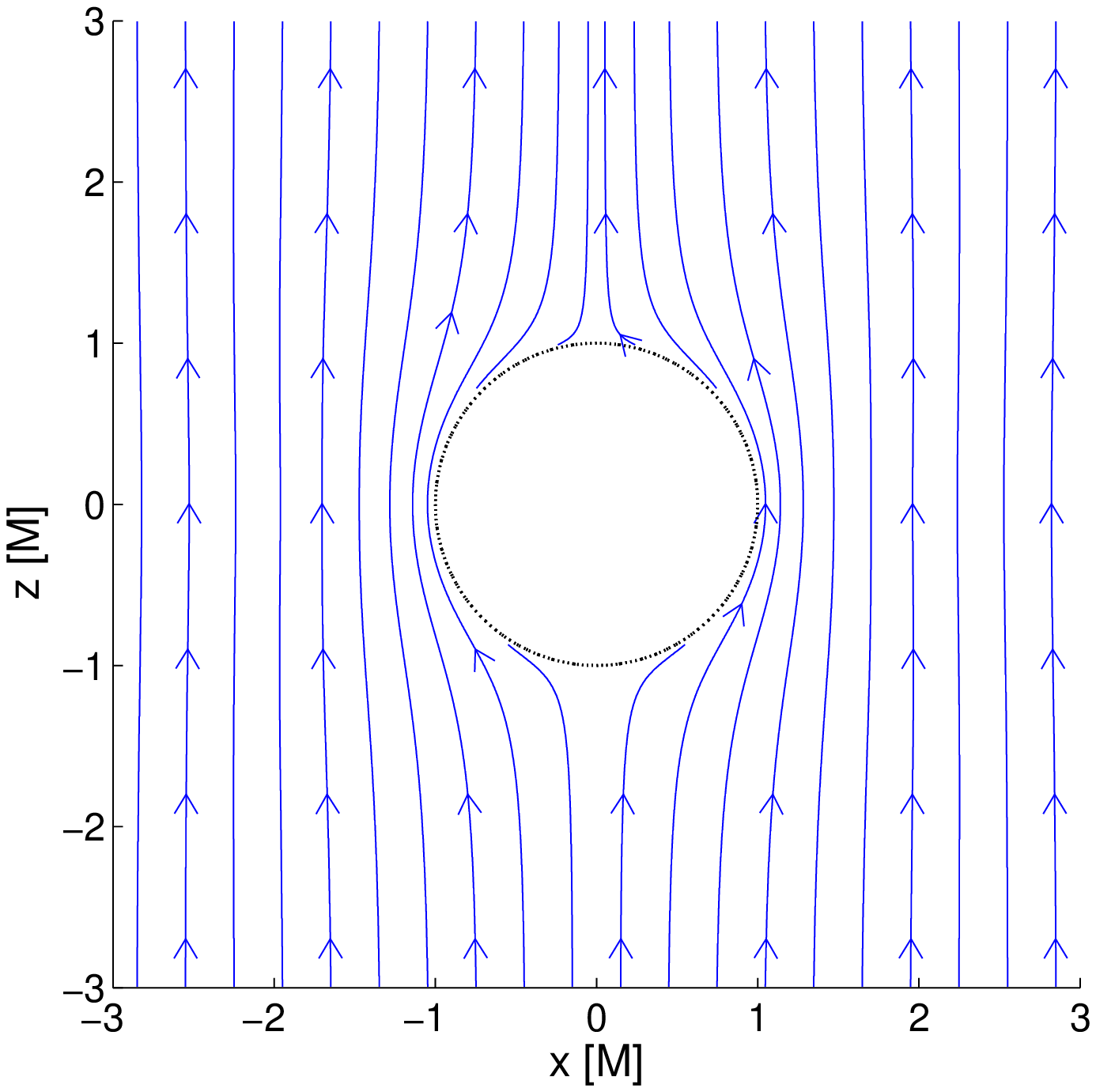}~~~~
\includegraphics[scale=0.34, clip]{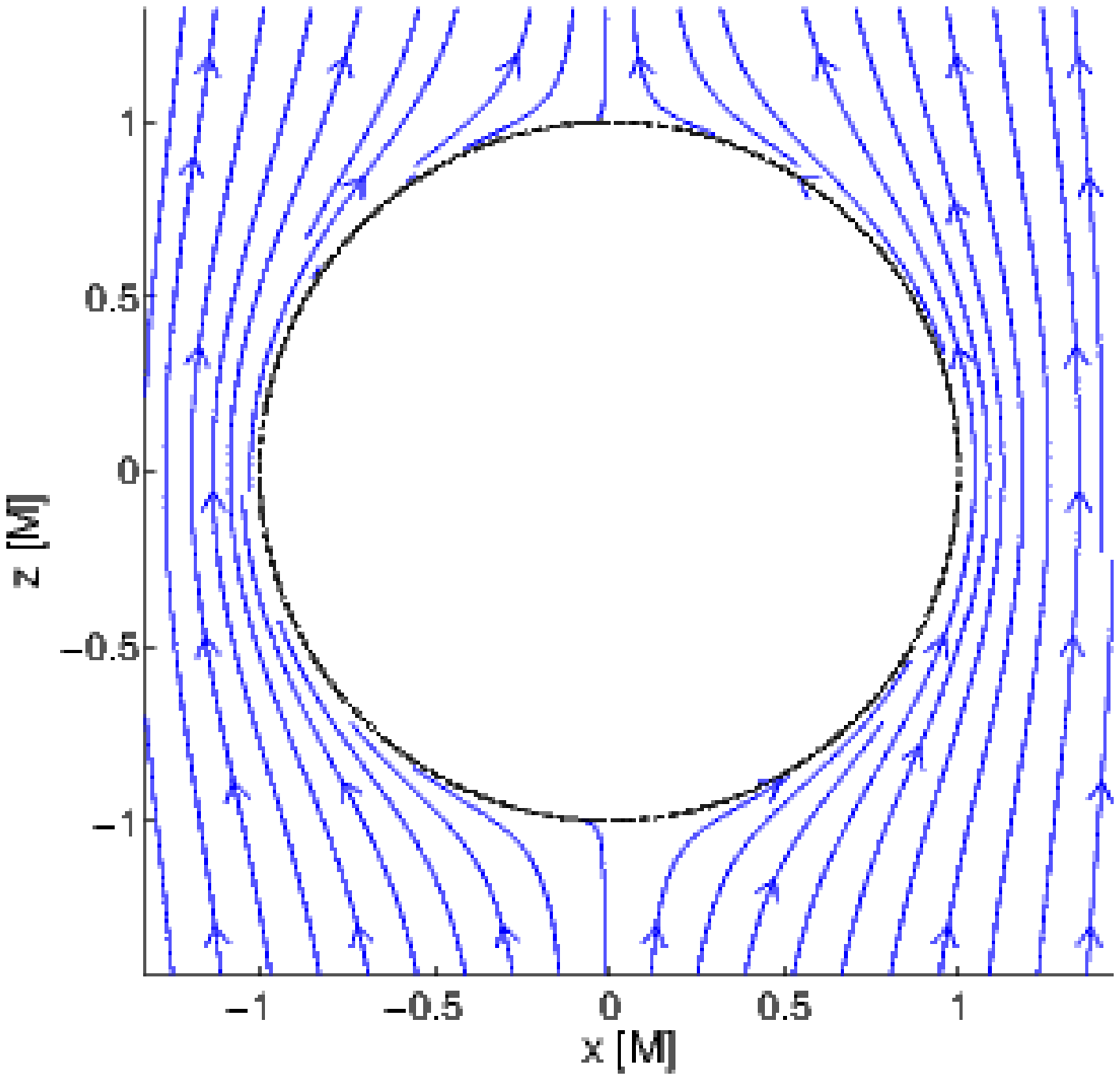}\\
\includegraphics[scale=0.34, clip]{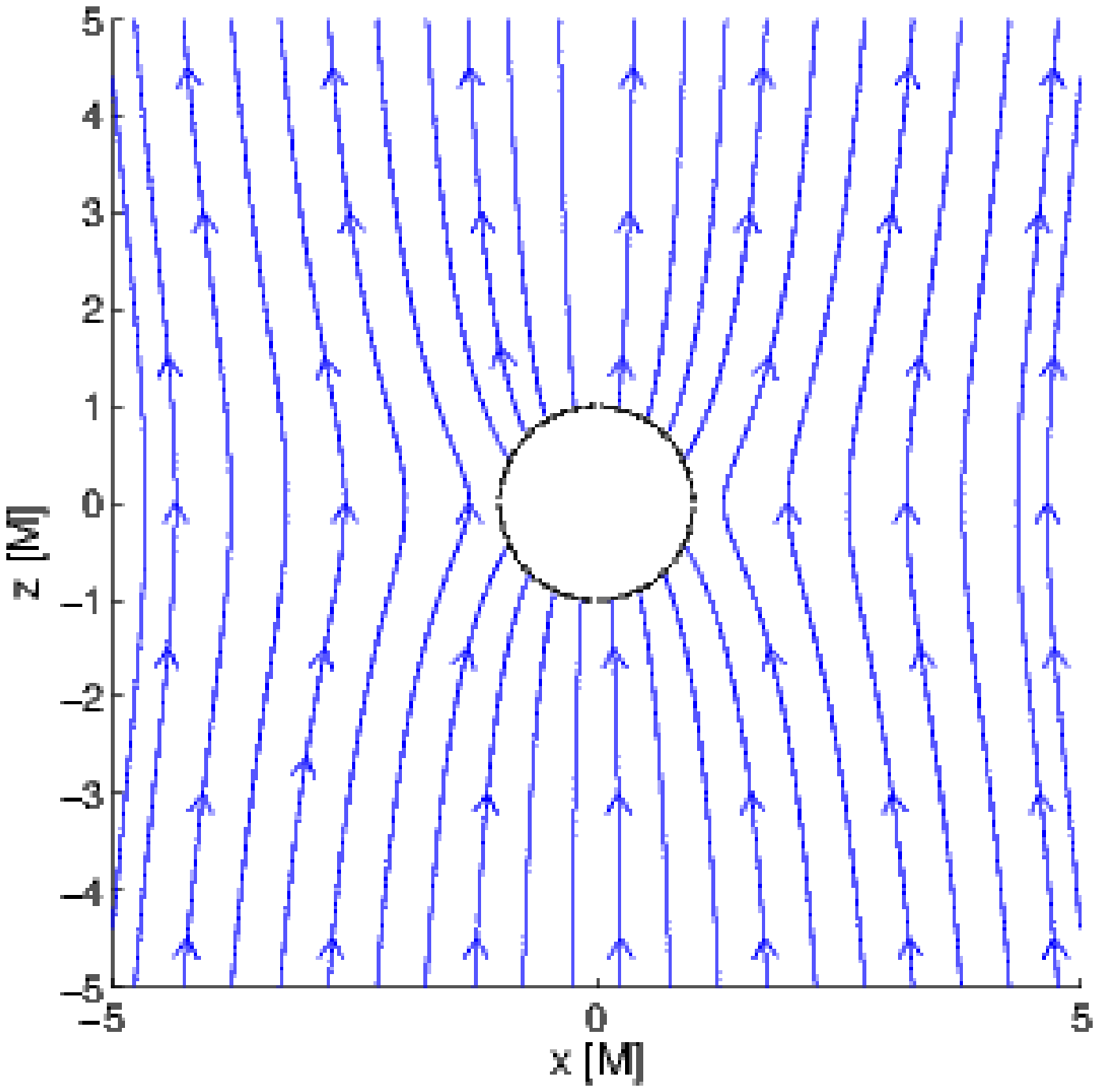}~~~~
\includegraphics[scale=0.34, clip]{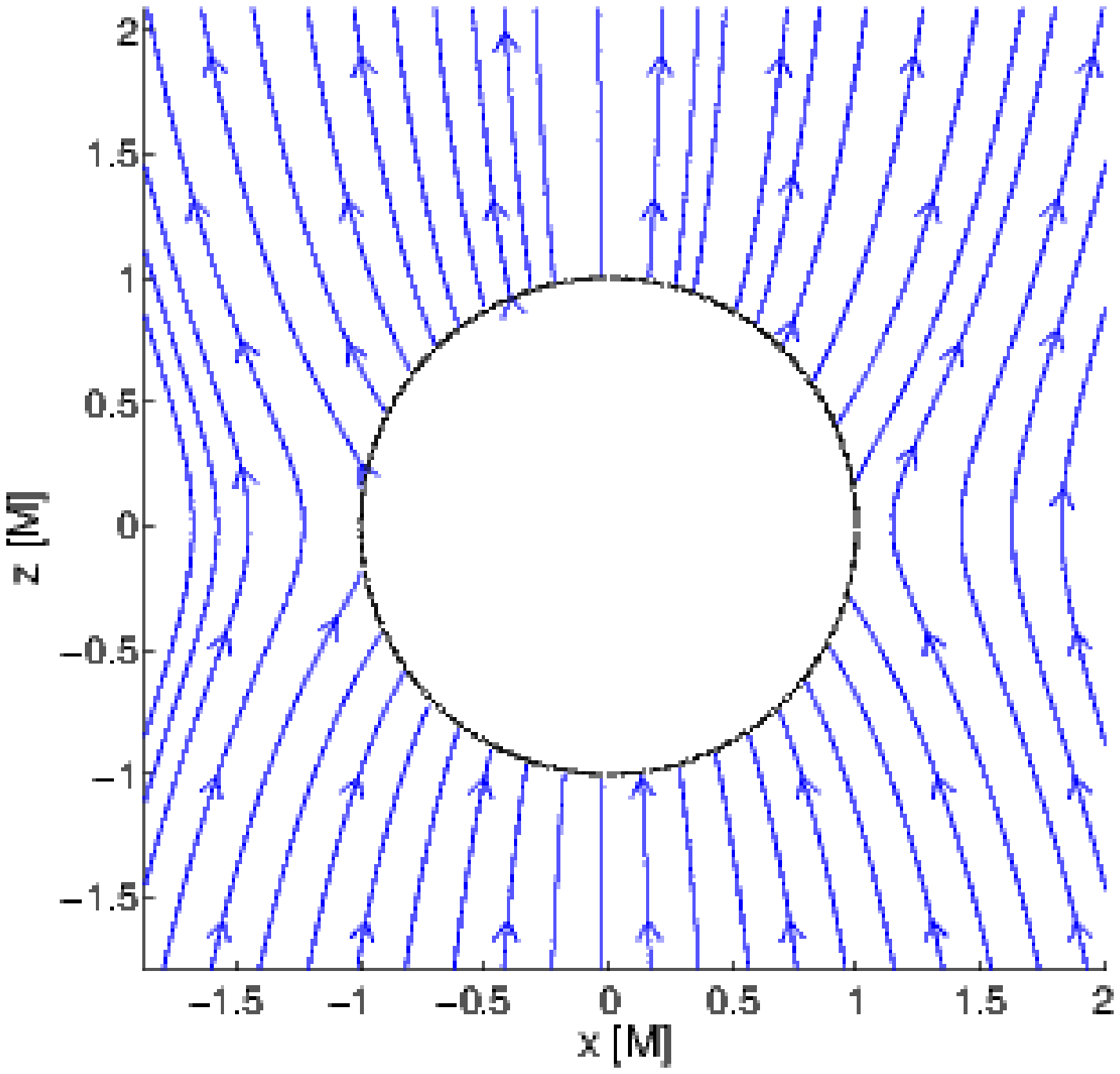}
\caption{Magnetic field felt by ZAMO test charge orbiting above the horizon of the extremal Kerr black hole immersed into the aligned field. We compare (top to bottom) coordinate, physical, renormalized and tetrad components of the field showing both, the overall shape of the field lines as well as the detailed bahaviour in the vicinity of the horizon.}
\label{vytl_c_ZAMO}
\end{figure}

\begin{figure}[hp!]
 \centering
\includegraphics[scale=0.34, clip]{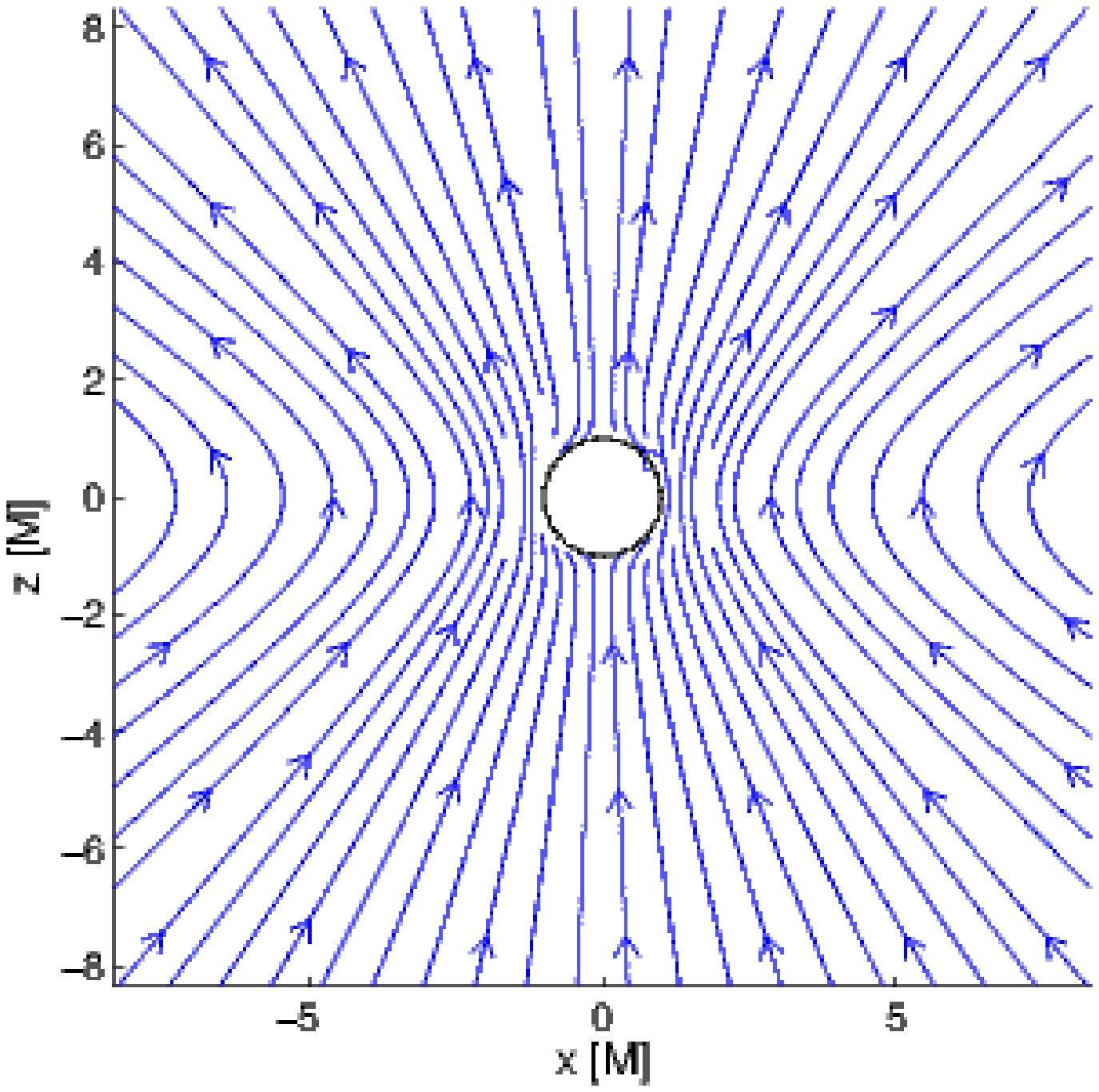}~~~~
\includegraphics[scale=0.34, clip]{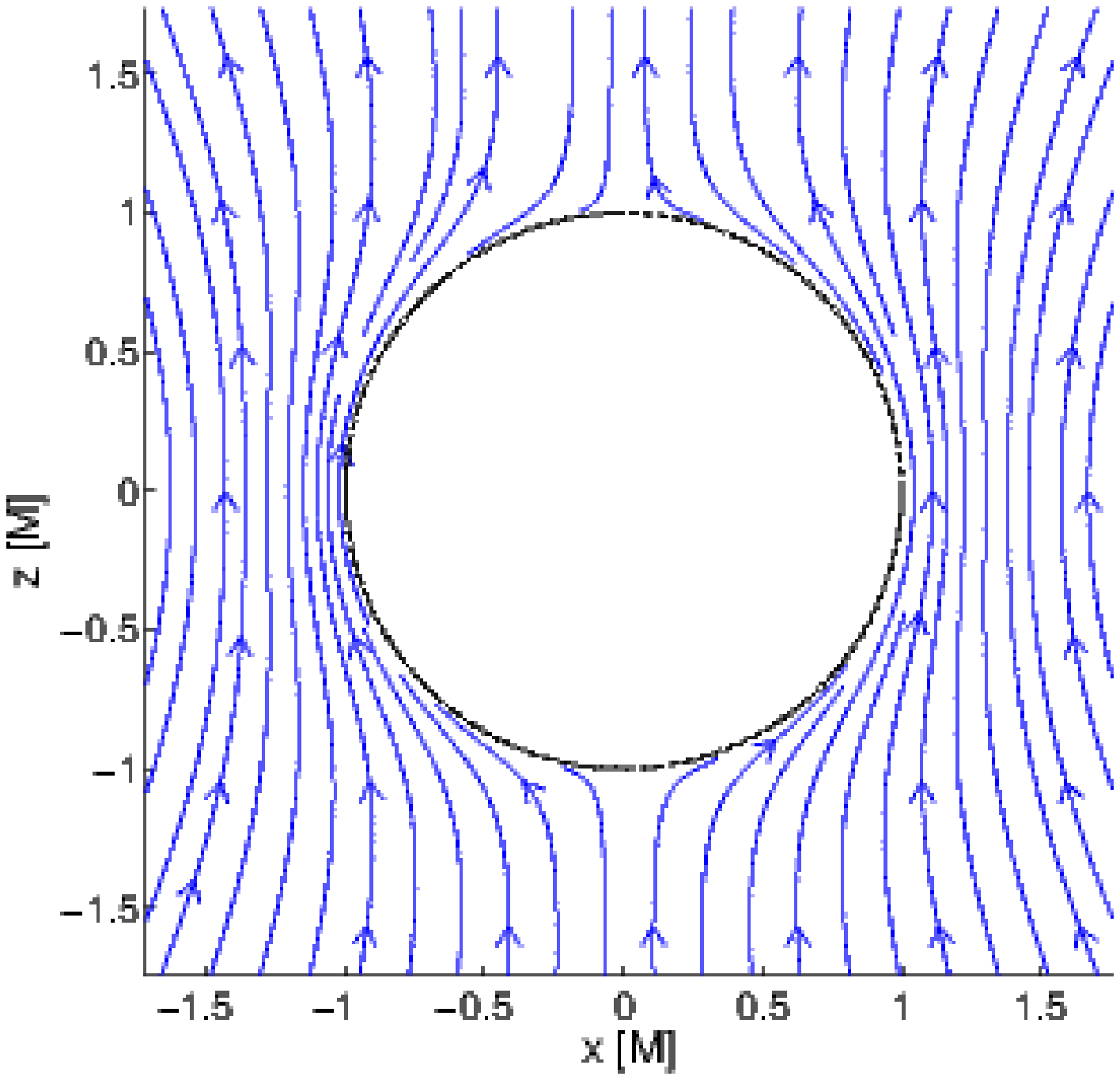}\\
\includegraphics[scale=0.34, clip]{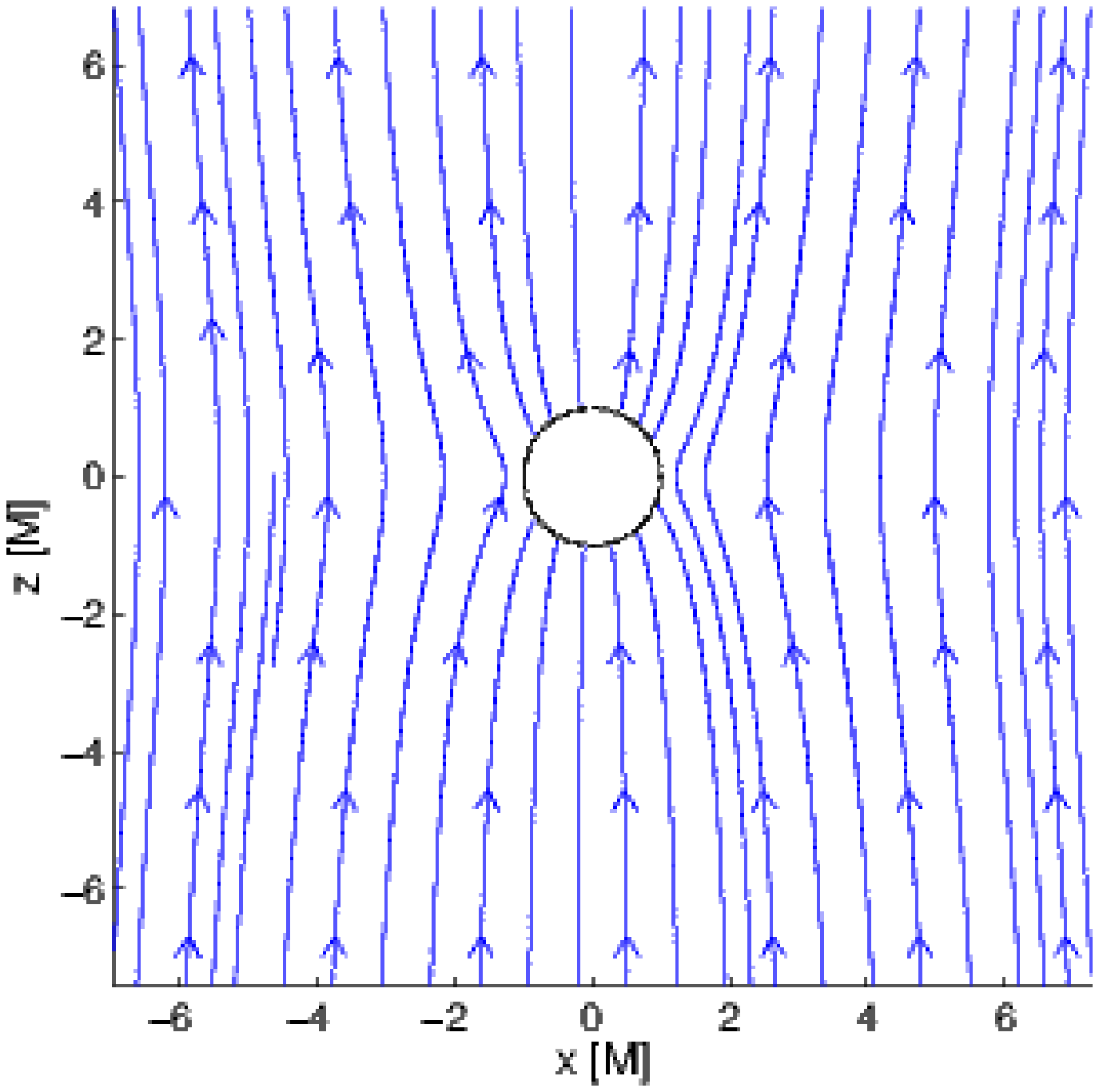}~~~~
\includegraphics[scale=0.34, clip]{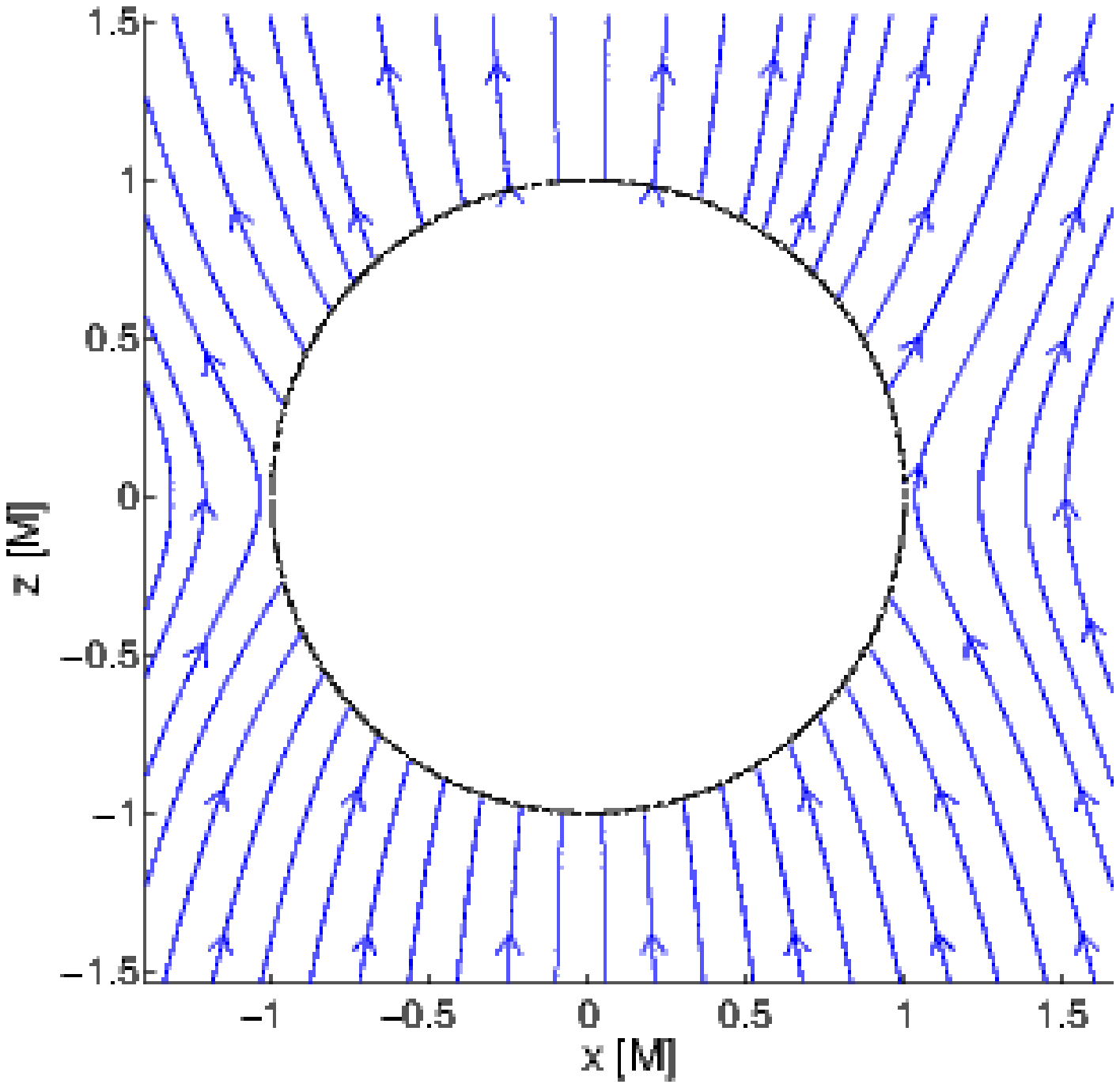}\\
\includegraphics[scale=0.34, clip]{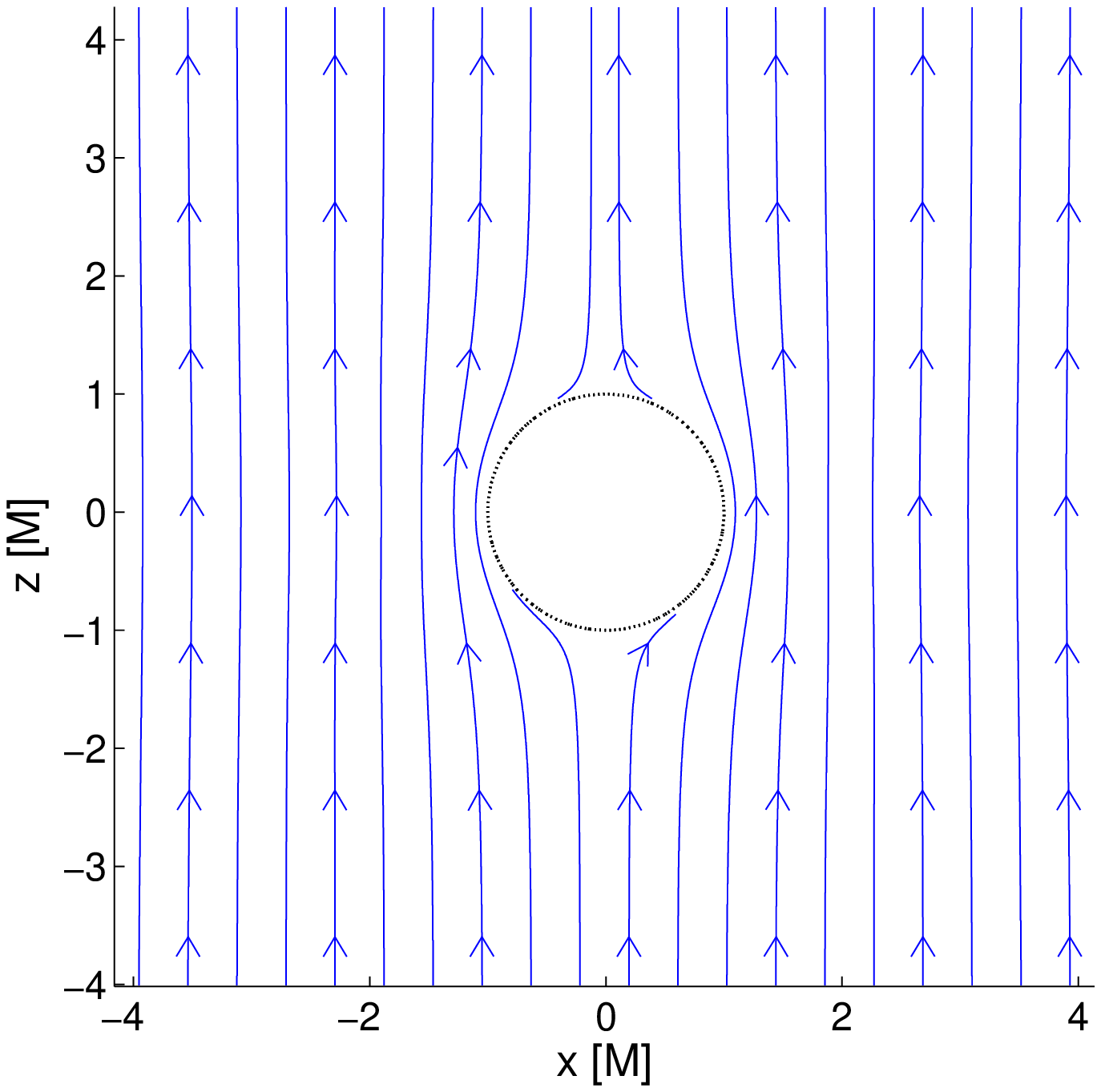}~~~~
\includegraphics[scale=0.34, clip]{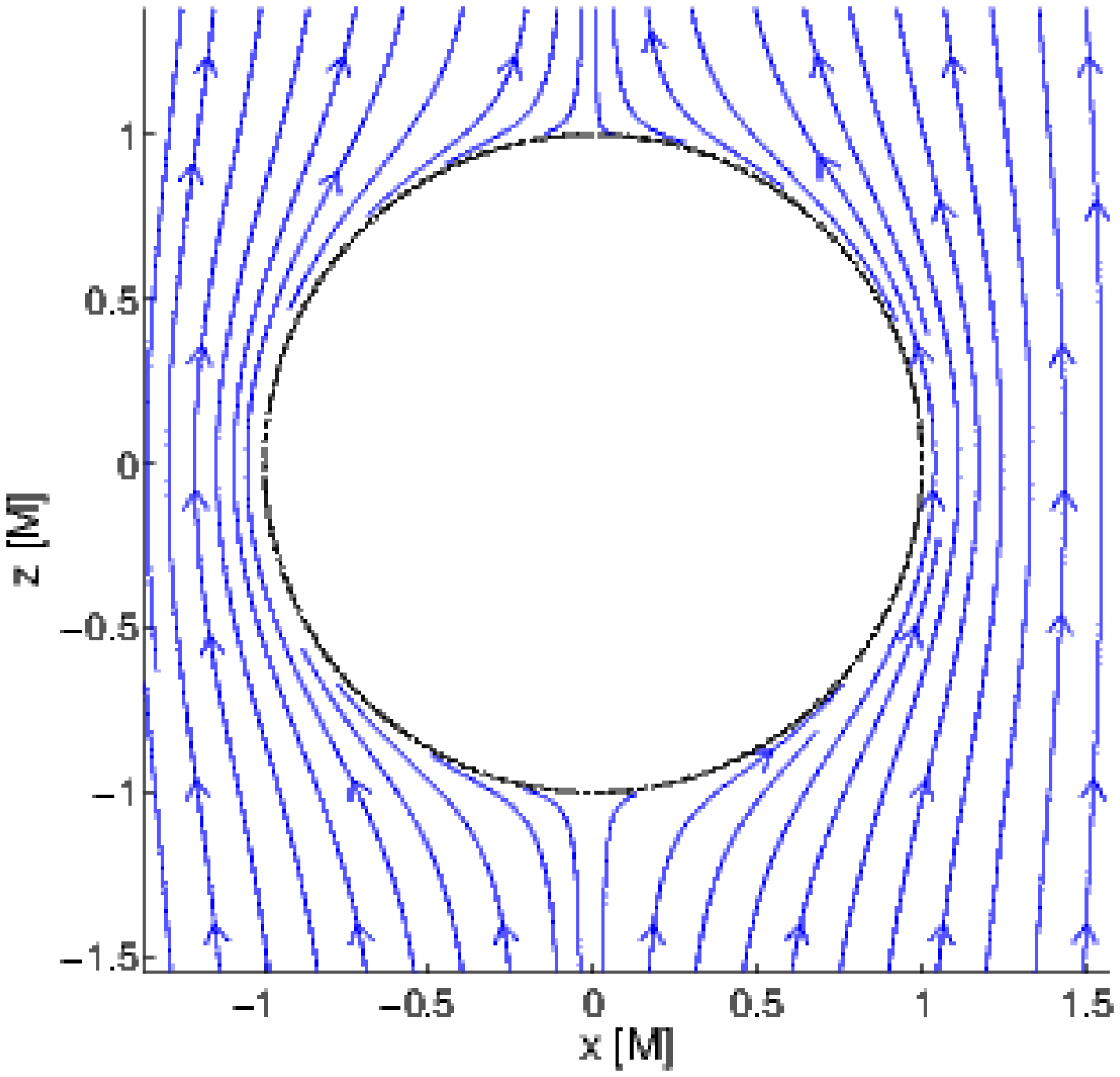}\\
\includegraphics[scale=0.34, clip]{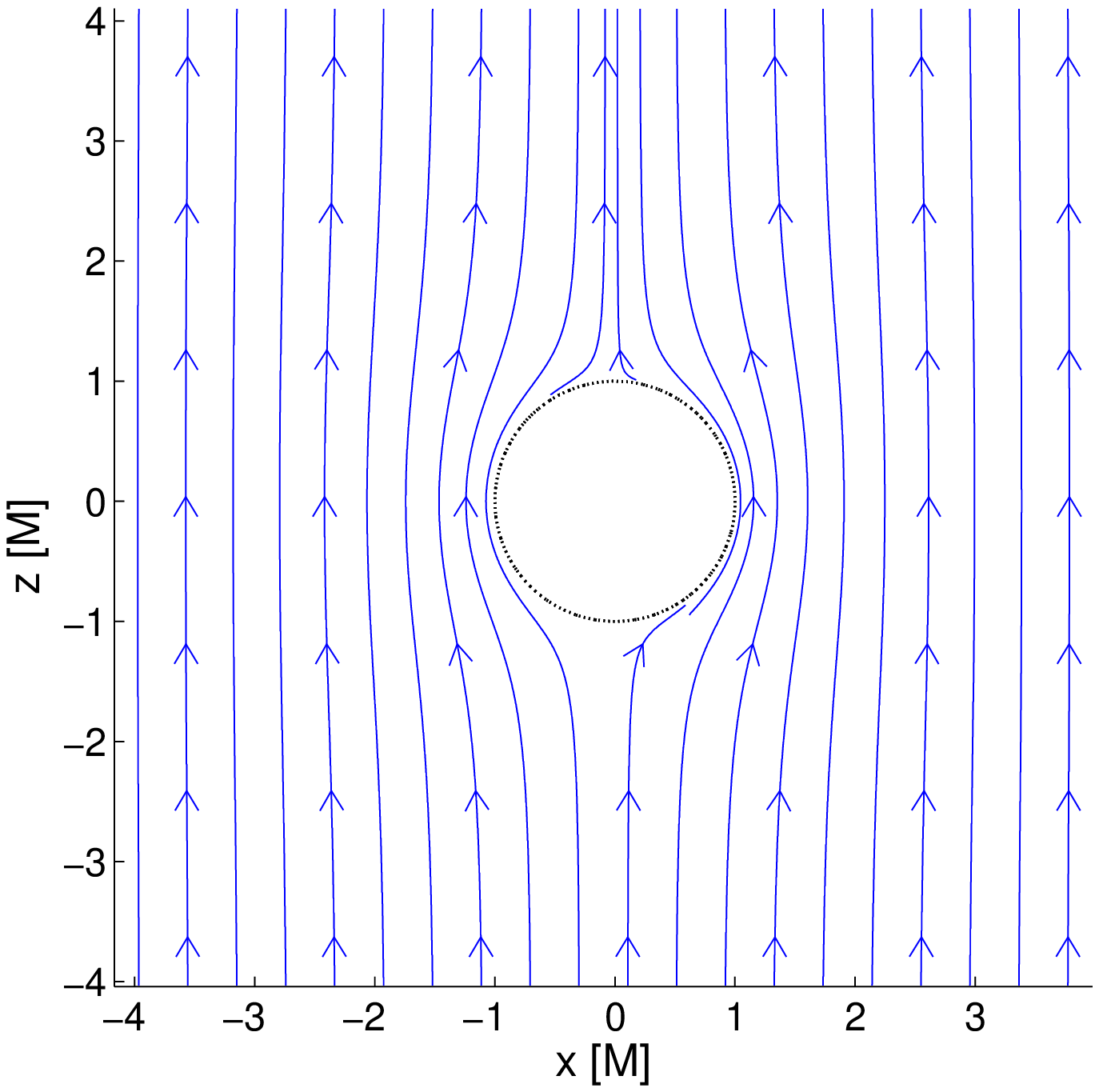}~~~~
\includegraphics[scale=0.34, clip]{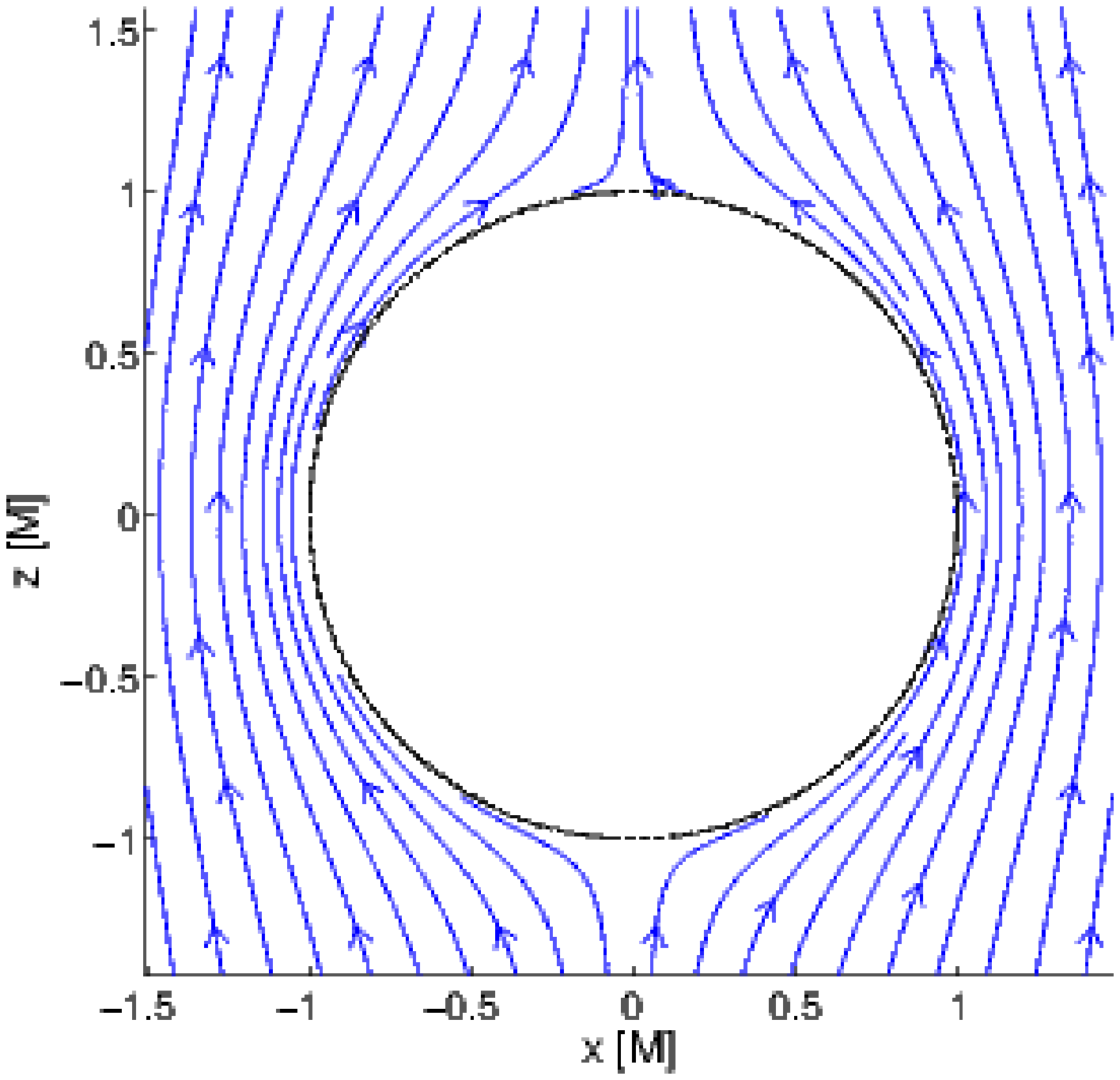}
\caption{Magnetic field felt by the FFOFI test charge freely falling onto the extremal Kerr black hole immersed into the aligned field. We compare (top to bottom) coordinate, physical, renormalized and tetrad components of the field showing both, the overall shape of the field lines as well as the detailed bahaviour in the vicinity of the horizon.}
\label{vytl_c_FFOFI}
\end{figure}

In figs. \ref{vytl_c_ZAMO} and \ref{vytl_c_FFOFI} we review behaviour of the magnetic field lines of various definitions for both ZAMO and FFOFI velocity profiles of the test charge in the case of extreme BH $a=M$. We observe that coordinate components $B^{i}$ for both ZAMO and FFOFI four-velocities exhibit the Meissner effect and magnetic field is expelled from the horizon (top panels of figs. \ref{vytl_c_ZAMO} and \ref{vytl_c_FFOFI}). 

Large-scale bending of the field lines is necessarily artificial since both ZAMO and FFOFI are asymptotically static and the distant test charge thus shall perceive asymptotic shape of the field which is aligned uniform magnetic field in this case. Bending is caused by the lack of normalization of the coordinate basis vectors which led us to the definition of the physical components given by eqs. (\ref{physicalc1})--(\ref{physicalc3}). In the second rows of figs. \ref{vytl_c_ZAMO} and \ref{vytl_c_FFOFI} we observe, however, that physical components do not exhibit the expulsion of the magnetic field; neither for ZAMO nor for FFOFI. This is caused by the fact that the definition of physical components  incorporates the coordinate singularity at the horizon directly into the field component $B^r_{\rm{physical}}$ as commented above. By exploring the behaviour of the underlying expressions we learn that in ZAMO case the radial component has finite limit at the horizon $\lim_{r\to r_{+}} B^r_{\rm{physical}}=-\frac{2\,\cos^3\!\theta\: B}{(1+\cos^2\theta)^2}$ while in FFOFI case it diverges as $B^r_{\rm{physical}}\propto \frac{1}{r-r_{+}}$. We thus confirm that {\bf physical components $\mathbf{B^{i}_{\rm{physical}}}$, $\mathbf{E^{i}_{\rm{physical}}}$ are inappropriate when working in Boyer-Lindquist coordinates}. 



Renormalized components which we suggested to use to overcome this problem actually eliminate the divergence of the radial component at the horizon (in the limit $r\to r_{+}$ renormalized components behave in the same manner as the coordinate ones)  and also provide the proper asymptotic behaviour (see the third rows in figs. \ref{vytl_c_ZAMO} and \ref{vytl_c_FFOFI}). Finally we notice that FFOFI observer in his frame observes the expulsion (bottom panels of \rff{vytl_c_FFOFI}) while the ZAMO observer sees the magnetic field lines penetrating the horizon (\rff{vytl_c_ZAMO}). Expression for $B^{(r)}$ involves factor terms $e^t_{(r)}u^r-e^{r}_{(r)}u^t$ and $e^r_{(r)}u^{\varphi}-e^{\varphi}_{(r)}u^r$ which are both zero in the limit $r\to r_{+}$ in the FFOFI case and $B^{(r)}$ thus cancels out. For ZAMO which has $u^r=0$, however, these terms do not cancel and thus allow $B^{(r)}$ to acquire nonzero value at the horizon of the extreme Kerr BH. 

On the other hand $u^r\ne 0$ causes that FFOFI measures nonzero azimuthal component of the magnetic field given as $B_{\rm{FFOFI}}^{(\varphi)}=-e_{(\varphi)}^{\varphi\;*}\!F_{\varphi r}u^r$ which has real valued limit $\lim_{r\to r_{+}} B_{\rm{FFOFI}}^{(\varphi)}=-\frac{B\:\sin2\,\theta}{(1+\cos^2\theta)^{3/2}}$ at the horizon of the extreme Kerr BH. We further notice that renormalized components of the field felt by the FFOFI test charge encompass the azimuthal component $B_{\rm{renormalized}}^{\varphi}=-{u^r r\sin\theta}\left( g^{\varphi\varphi\;*} \!F_{\varphi r}+ g^{\varphi t\;*}\!F_{tr} \right)$ which actually diverges  as $B_{\rm{renormalized}}^{\varphi}\propto\frac{1}{(r-r_{+})^2}$ for $r\to r_{+}$ in the extreme Kerr geometry $a=M$ where $r_{+}=M$. 

Due to the presence of the azimuthal component the FFOFI field lines are not confined to the given poloidal plane $\varphi=\{const,\: const+\pi\}$ and the ordinary two-dimensional plane plots of the lines of force thus represent a projection of the vector field rather then the true field lines. Under these circumstances one should occasionally check the stereometric projection of the field lines in order to reveal its azimuthal component since it may change the picture profoundly. We emphasize, however, that the axial symmetry is maintained and the azimuthal component is independent of $\varphi$.

 We introduce stereometric plots by showing the simple AMO case (which has no azimuthal component since $F^{\rm{B_z}}_{r\theta}=0$) for the two values of spin $a=0.9M$ and $a=M$ in the upper panels of \rff{meissner_3d_AMO}. These may be directly compared with the corresponding plane figures of \rff{vytl_amo}. Then we move to the FFOFI case where the azimuthal component is present. In bottom panels of \rff{meissner_3d_AMO} we compare FFOFI tetrad components with renormalized FFOFI components. The latter exhibit divergence in the azimuthal component at the horizon which causes strong winding of nearby field lines while the tetrad field lines are twisted only partially, shifted by the finite angle in the $r\to r_{+}$ asymptotics. 

In the following sections we shall preferently work with the tetrad components of the fields as they provide a consistent interpretation of quantities measurable by a physical observer. When occasionally switching to the coordinate basis we use renormalized components for their properties. Sometimes it appears useful to visualize the components of the EM tensor itselves without notion of the specific four-velocity of the test charge -- then the AMO components are employed.

\begin{figure}[hp]
\centering
\includegraphics[scale=0.56, clip, trim=0mm 10mm 0mm 10mm]{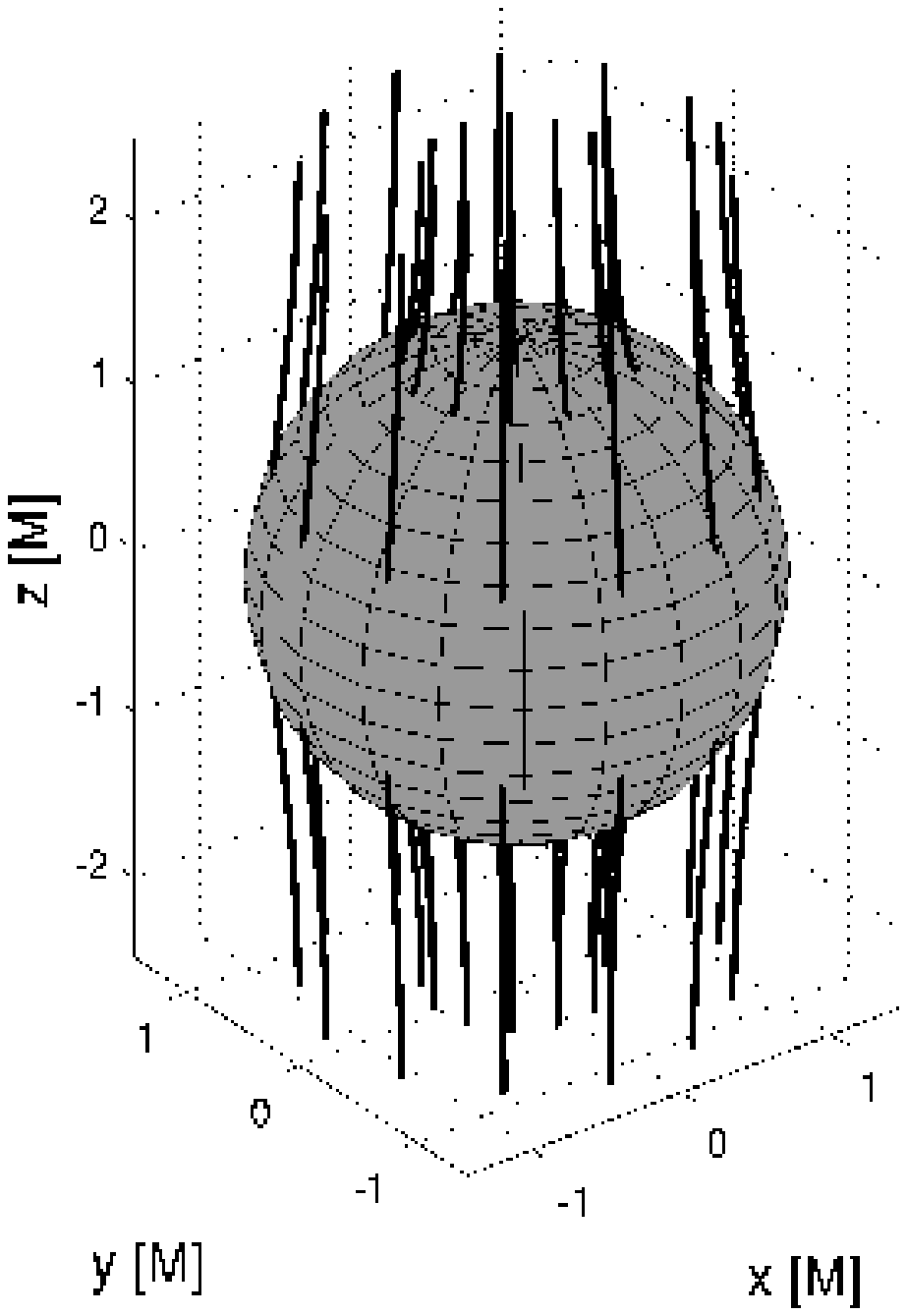}
\includegraphics[scale=0.56, clip, trim=0mm 8mm 0mm 10mm]{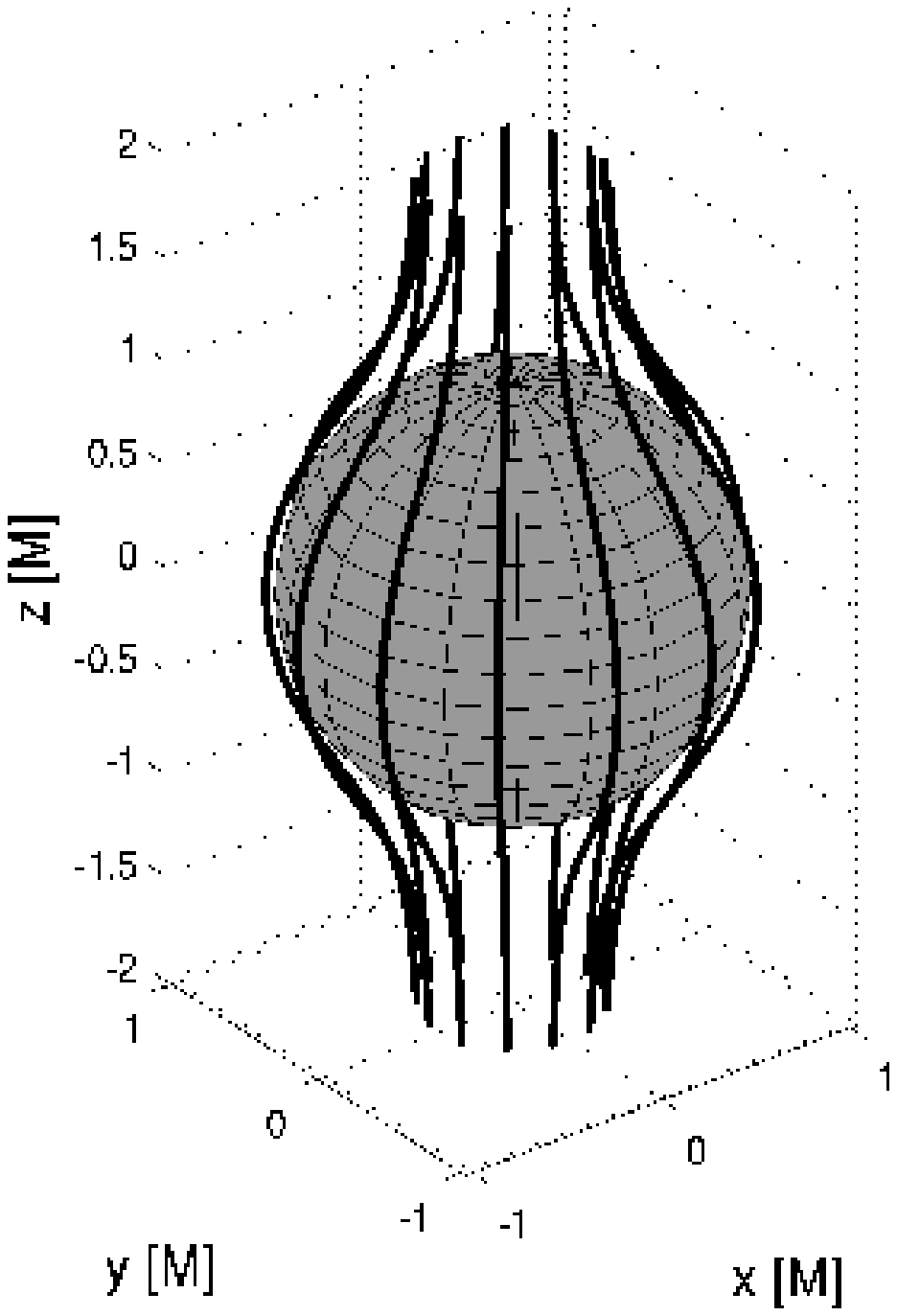}
\includegraphics[scale=0.51, clip, trim= 0mm 0mm 0mm 10mm]{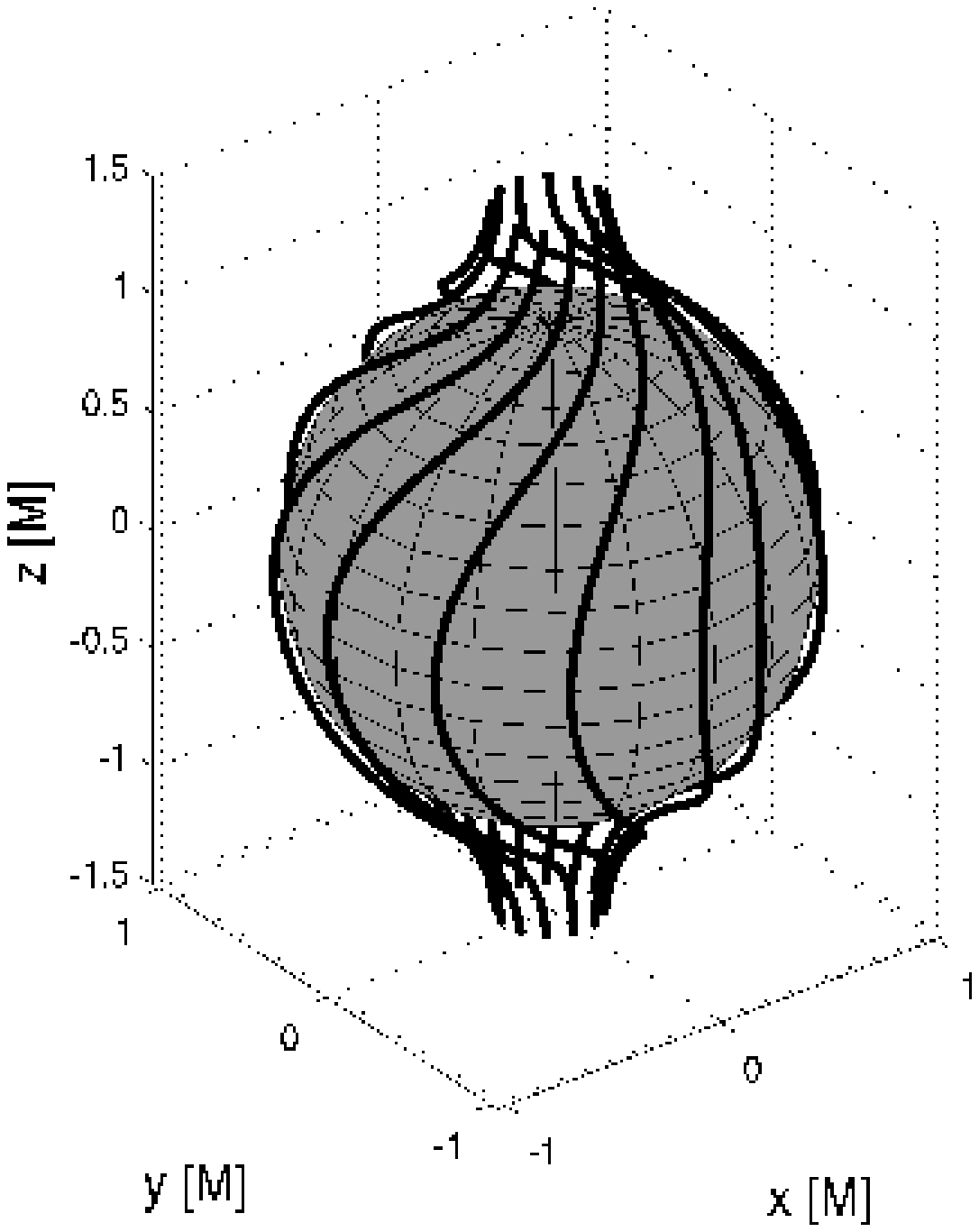}
\includegraphics[scale=0.54, clip, trim= 0mm 0mm 0mm 10mm]{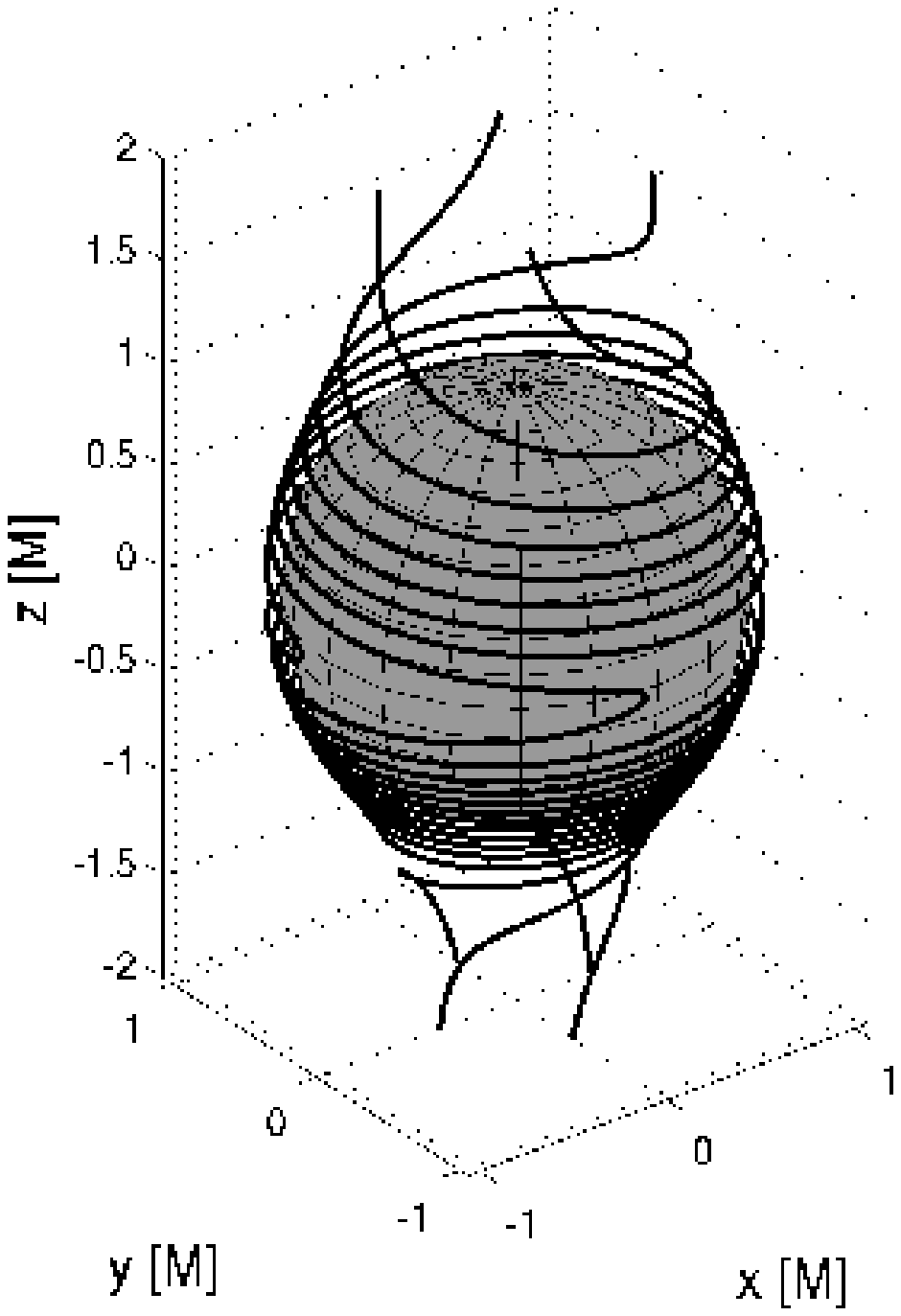}
\caption{Meissner-type expulsion of the magnetic field from the horizon of the extreme Kerr black hole. In the upper left panel for $a=0.9 M$ we observe slightly bent AMO field lines piercing the horizon. Upper right panel shows the extreme case $a=M$ for which the AMO magnetic field is fully expelled. Bottom panels show the field as felt by FFOFI. In the bottom left panel we observe tetrad components of the magnetic field with nonzero azimuthal component which approaches finite nonzero value on the horizon. Renormalized FFOFI components, however, exhibit divergence in the azimuthal component as we approach the horizon. In the close neighborhood of the horizon we observe strong azimuthal winding of the field lines (bottom right panel). Width of the ``tubes of force`` is chosen arbitrarily to optimize legibility of the plots.}
\label{meissner_3d_AMO}
\end{figure}

\newpage

\subsubsection{Non-aligned uniform magnetic field}
In the following we shall in brief consider also a field component $B_{x}$ perpendicular to the rotation axis. We therefore lose the axial symmetry. Lines of force are not anymore confined to the poloidal sections as they were in the case of stationary observers/test charges in the axisymmetric situation. We therefore plot the sections in couples of $(x,z)$ and $(y,z)$ plane projections (by setting $\varphi=\{0,\pi\}$ and $\varphi=\{\pi/2,3\pi/2\}$, respectively) of the field lines to capture at least fractionally their azimuthal dependence. Stereometric projections are often employed to provide more insightful view on the complex field structure. 

\begin{figure}[h!]
\centering
\includegraphics[scale=0.36, clip]{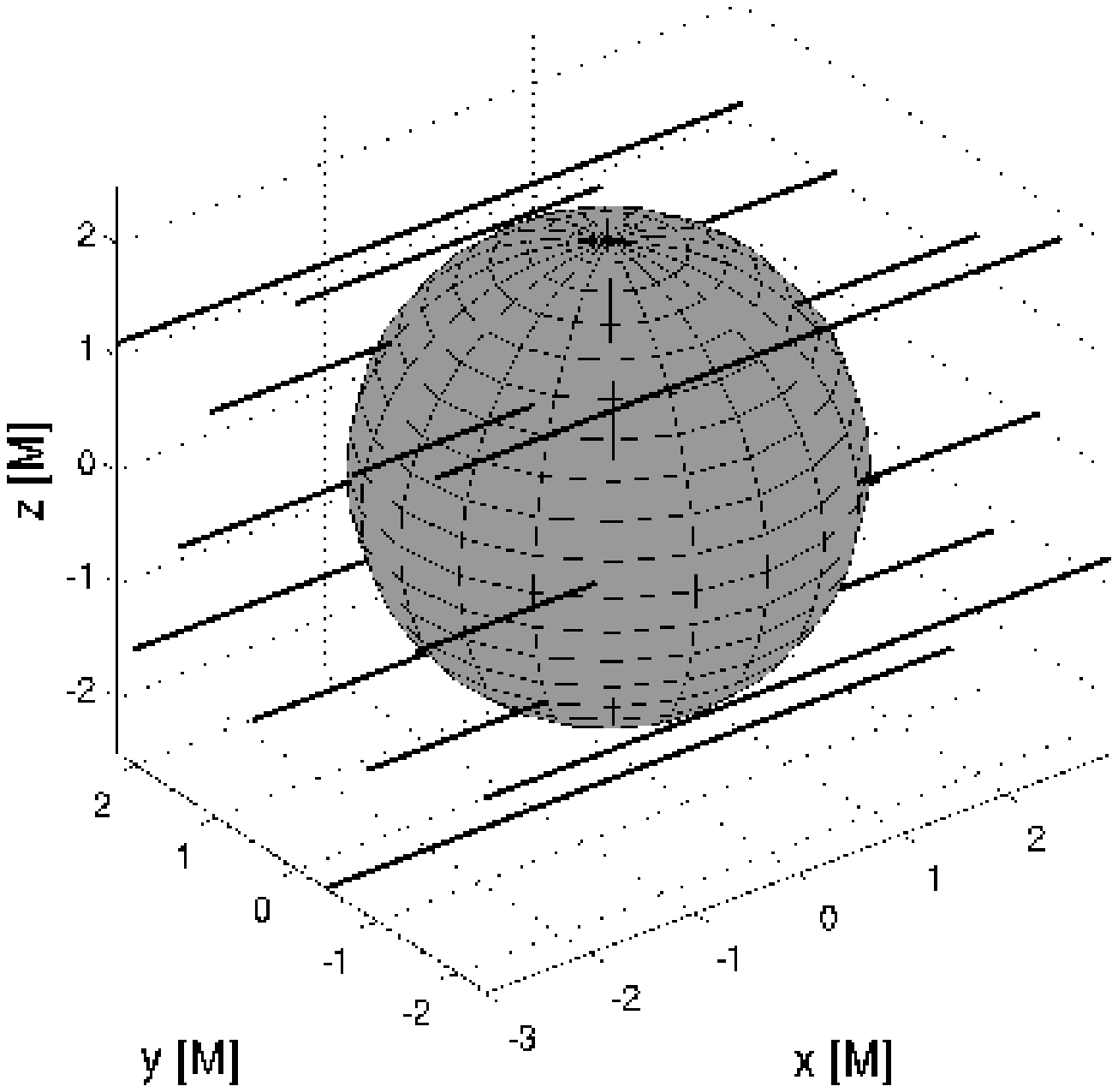}
\includegraphics[scale=0.43, clip, trim= 9mm 0mm 0mm 0mm]{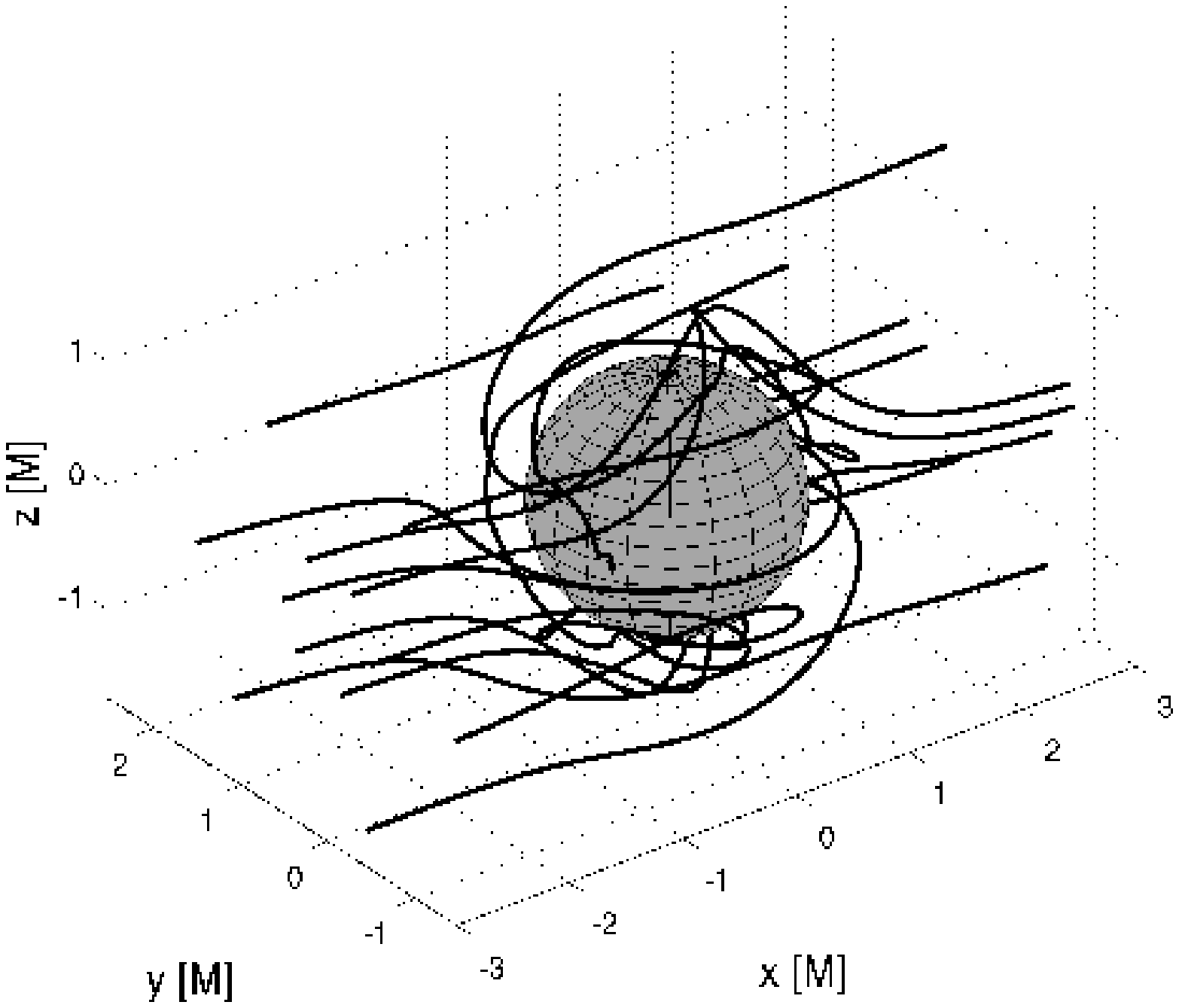}
\caption{AMO components of the asymptotically homogeneous magnetic field perpendicular to the symmetry axis ($B_z=0$). Left panel depicts the Schwarzschild limit $a=0$ for which the black hole does not affect the field at all as we have already observed in the analogical aligned case in \rff{vytl_amo}. The behaviour of the field lines, however, differs profoundly once the rotation is switched on. In the right panel we observe the field lines twisted by the extreme spin $a=M$. Unlike the aligned case, here we observe no expulsion of the field.}
\label{mag_amo_Bx}
\end{figure}

First we mention that by \rf{BFtr(Bx)} the $F^{B_x}_{\theta\varphi}$ component does not diminish at the horizon of the extreme Kerr black hole as $F^{B_z}_{\theta\varphi}$ does. Magnetic field lines thus penetrate the horizon; effect of the magnetic expulsion is not present in the AMO field components which we illustrate in \rff{mag_amo_Bx} where we compare AMO magnetic field lines for two extremal values of spin, namely $a=0$ and $a=M$, setting purely perpendicular magnetic field on the background: $B_x=M^{-1}$, $B_z=0$. In the Schwarzschild limit the field lines are not affected by the black hole since the $F^{B_x}_{\mu\nu}$ components reduce to its asymptotic form immediately by setting $a=0$. Setting nonzero spin the symmetry is lost and the field lines are twisted in a rather complicated way as we observe in the right panel of the \rff{mag_amo_Bx}. In general the field lines may penetrate the horizon even in the extreme case $a=M$ which we confirm also for other definitions of the field lines and for both ZAMO and FFOFI four-velocity profiles by exploring the underlying formulae.

\begin{figure}[htbp!]
\centering
\includegraphics[scale=0.28, clip]{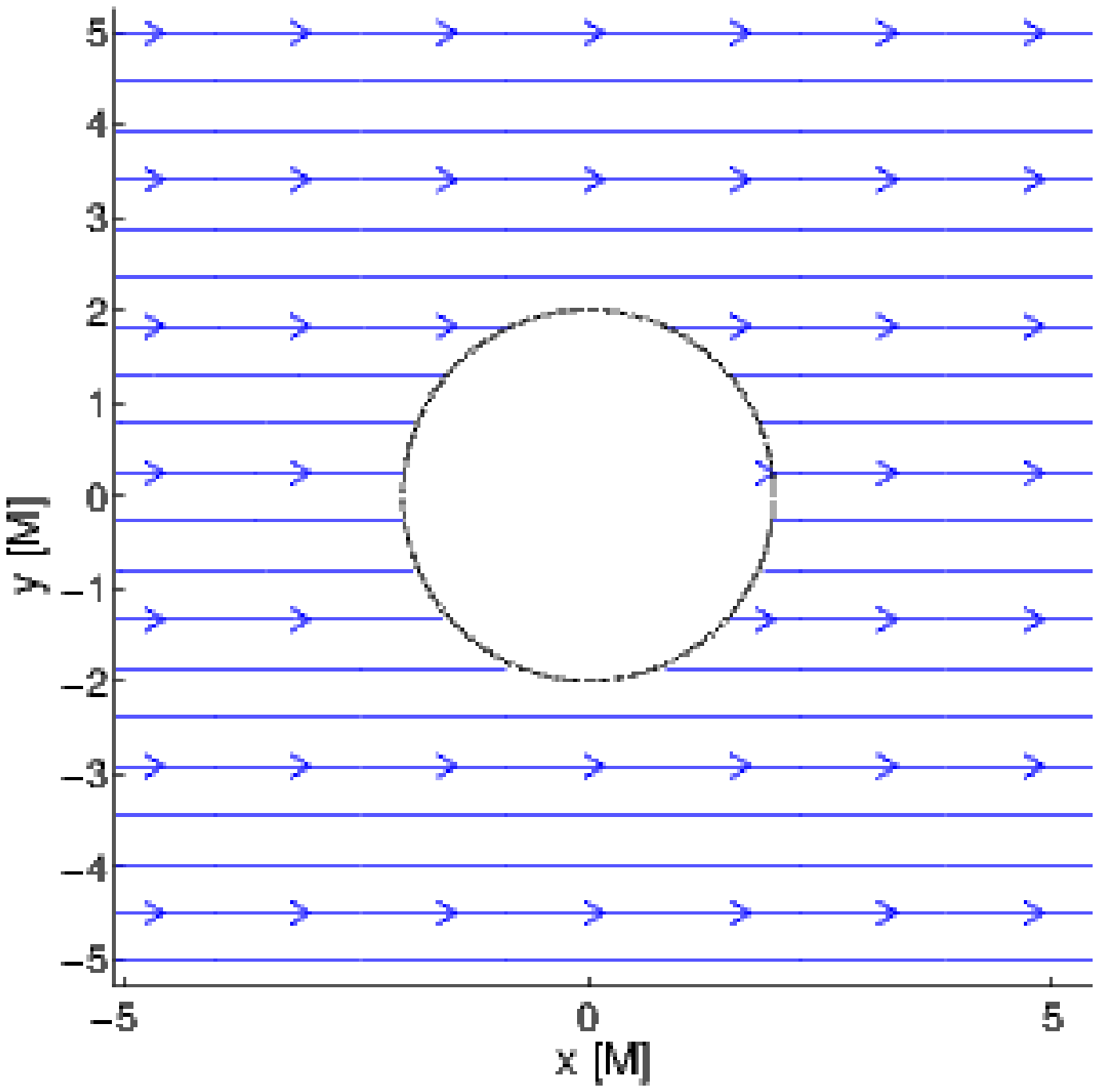}
\includegraphics[scale=0.28, clip]{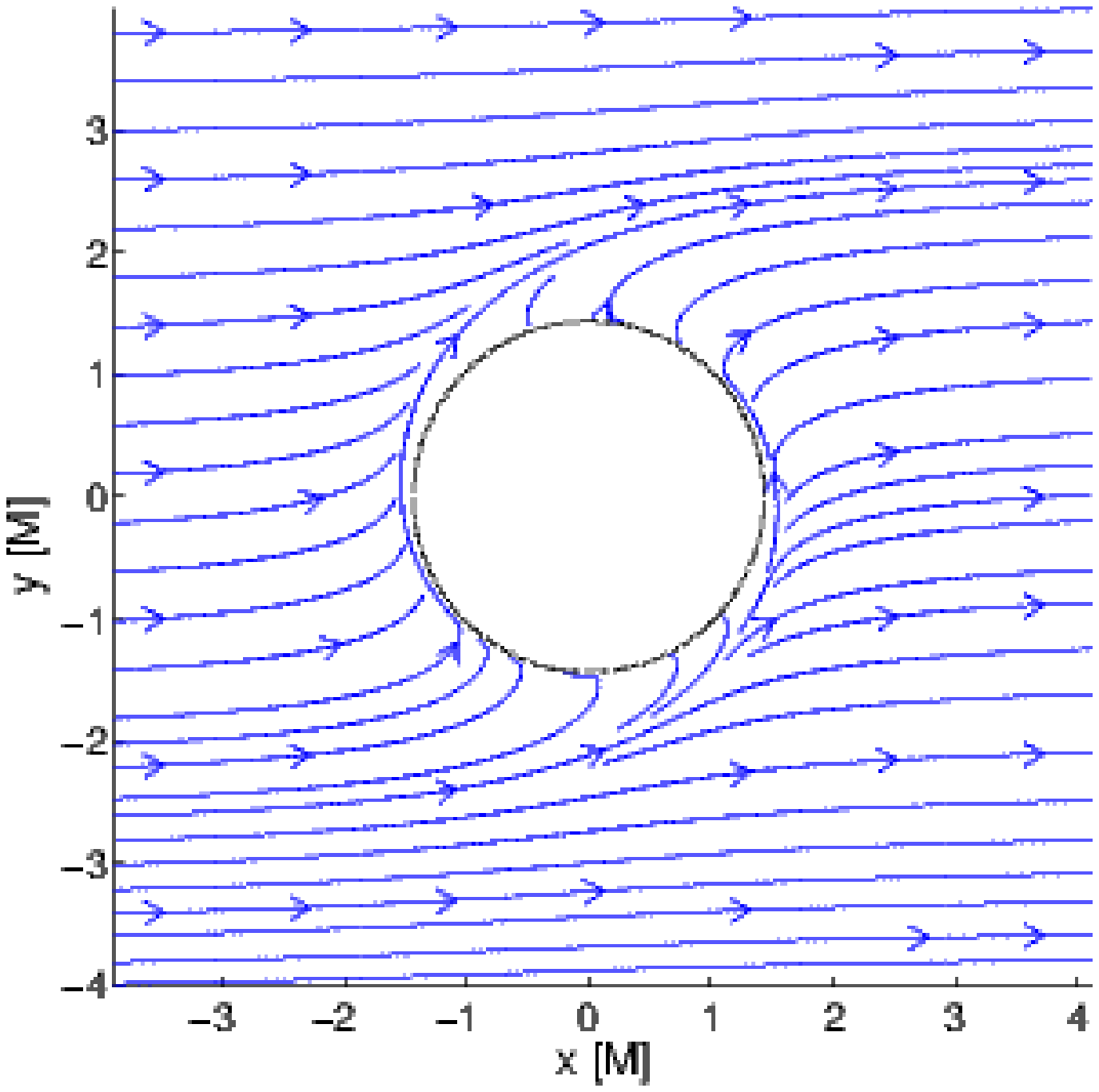}
\includegraphics[scale=0.28, clip]{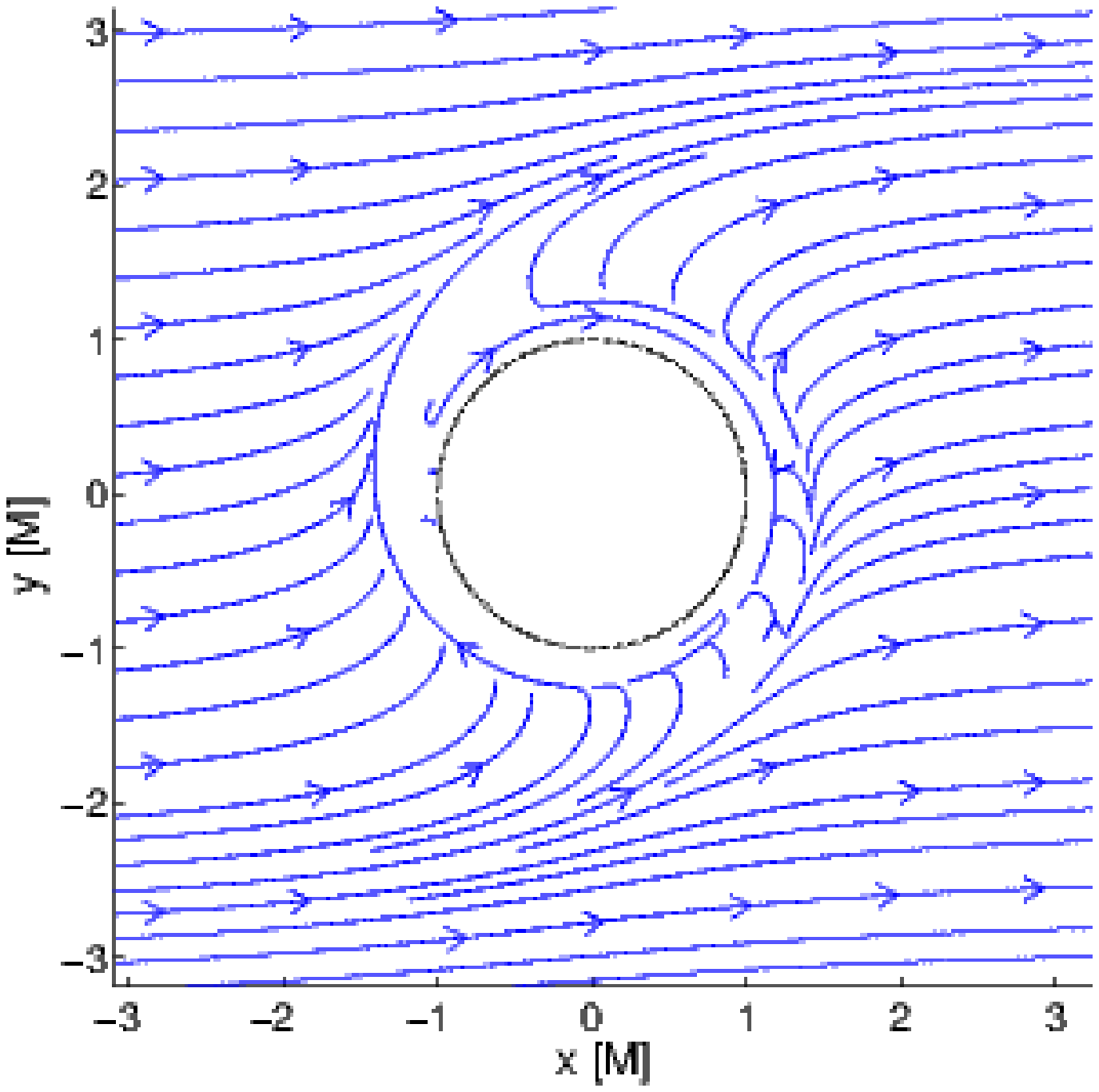}
\includegraphics[scale=0.28, clip]{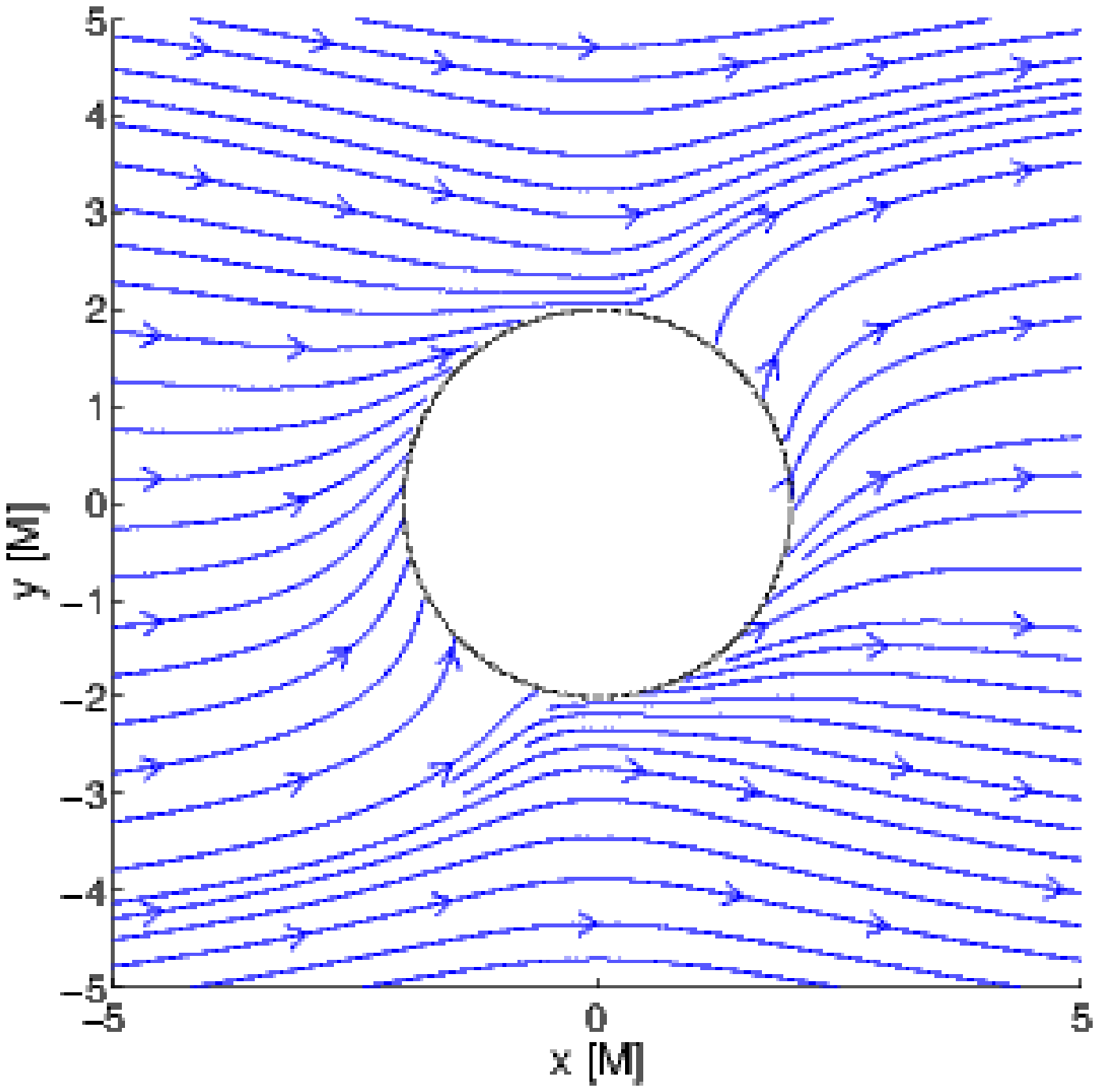}
\includegraphics[scale=0.28, clip]{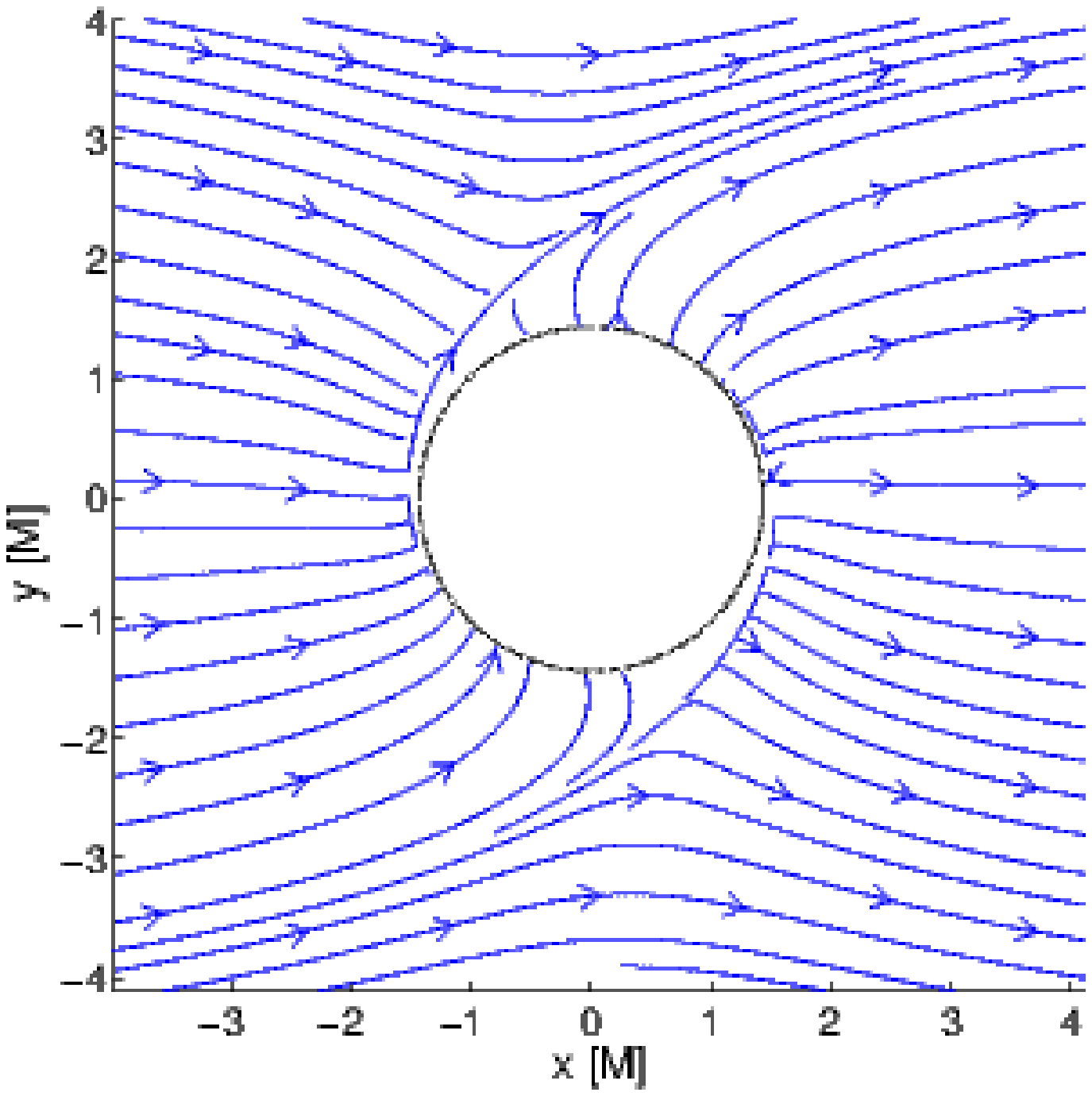}
\includegraphics[scale=0.28, clip]{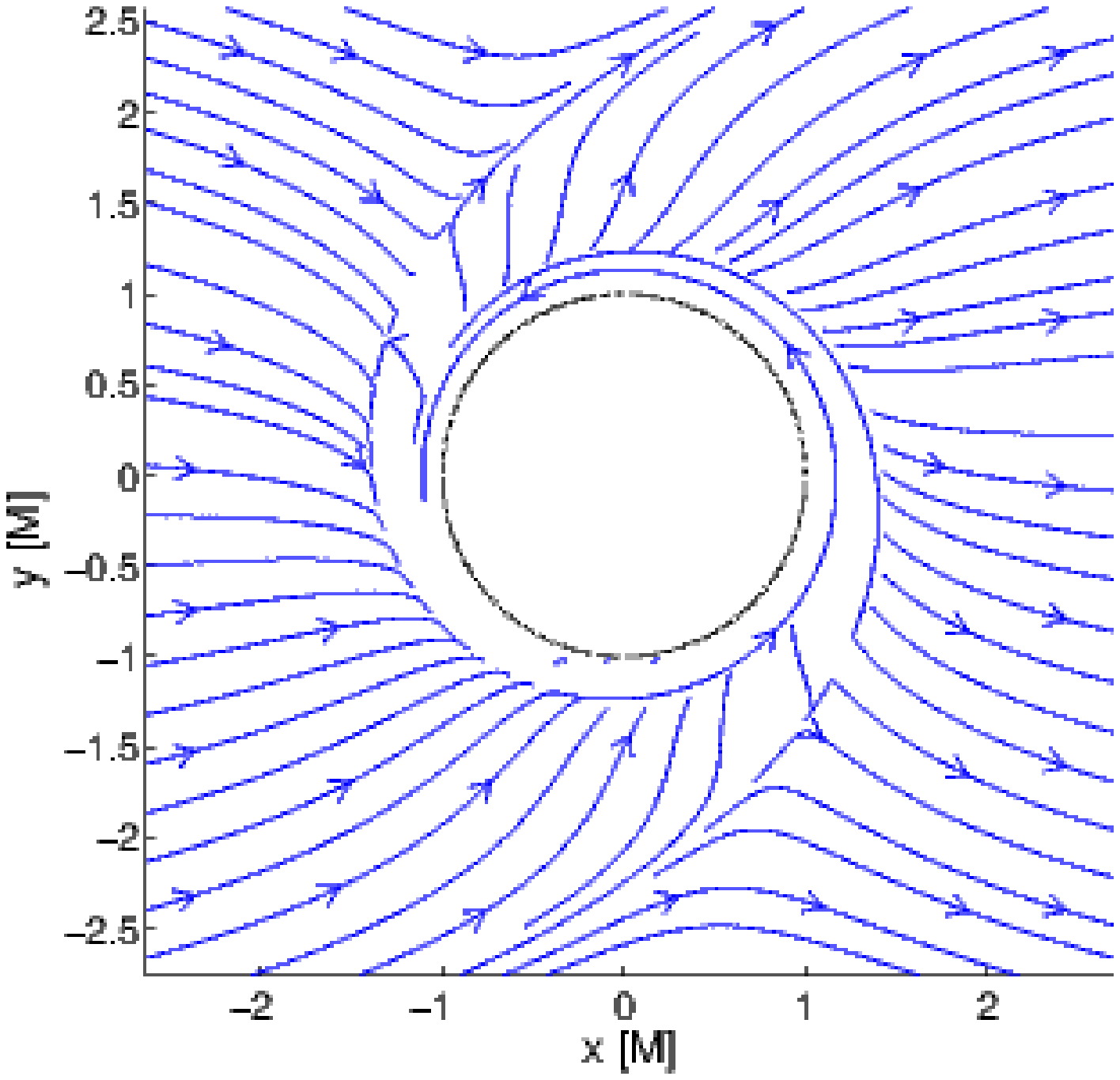}
\includegraphics[scale=0.28, clip]{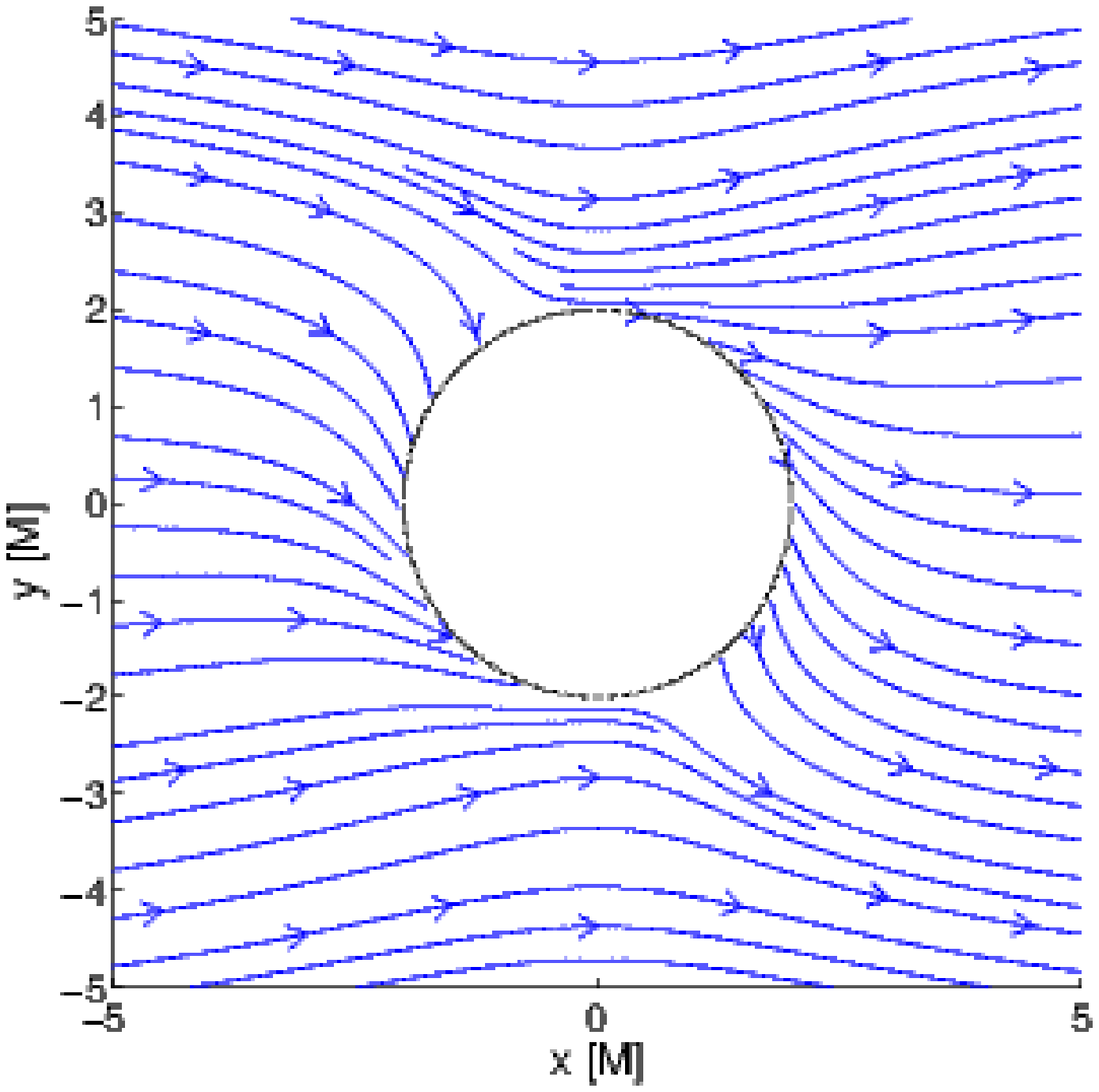}
\includegraphics[scale=0.28, clip]{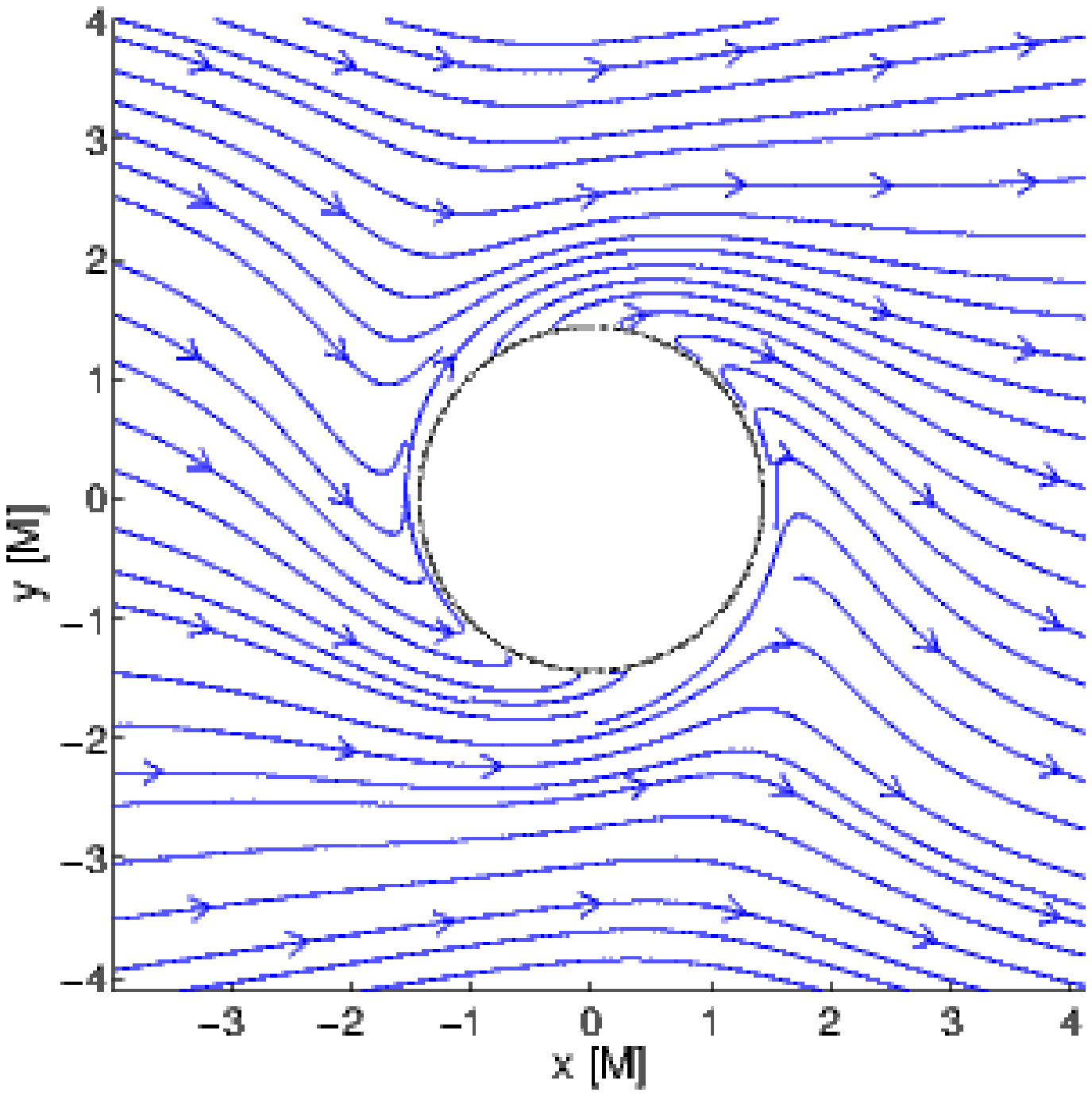}
\includegraphics[scale=0.28, clip]{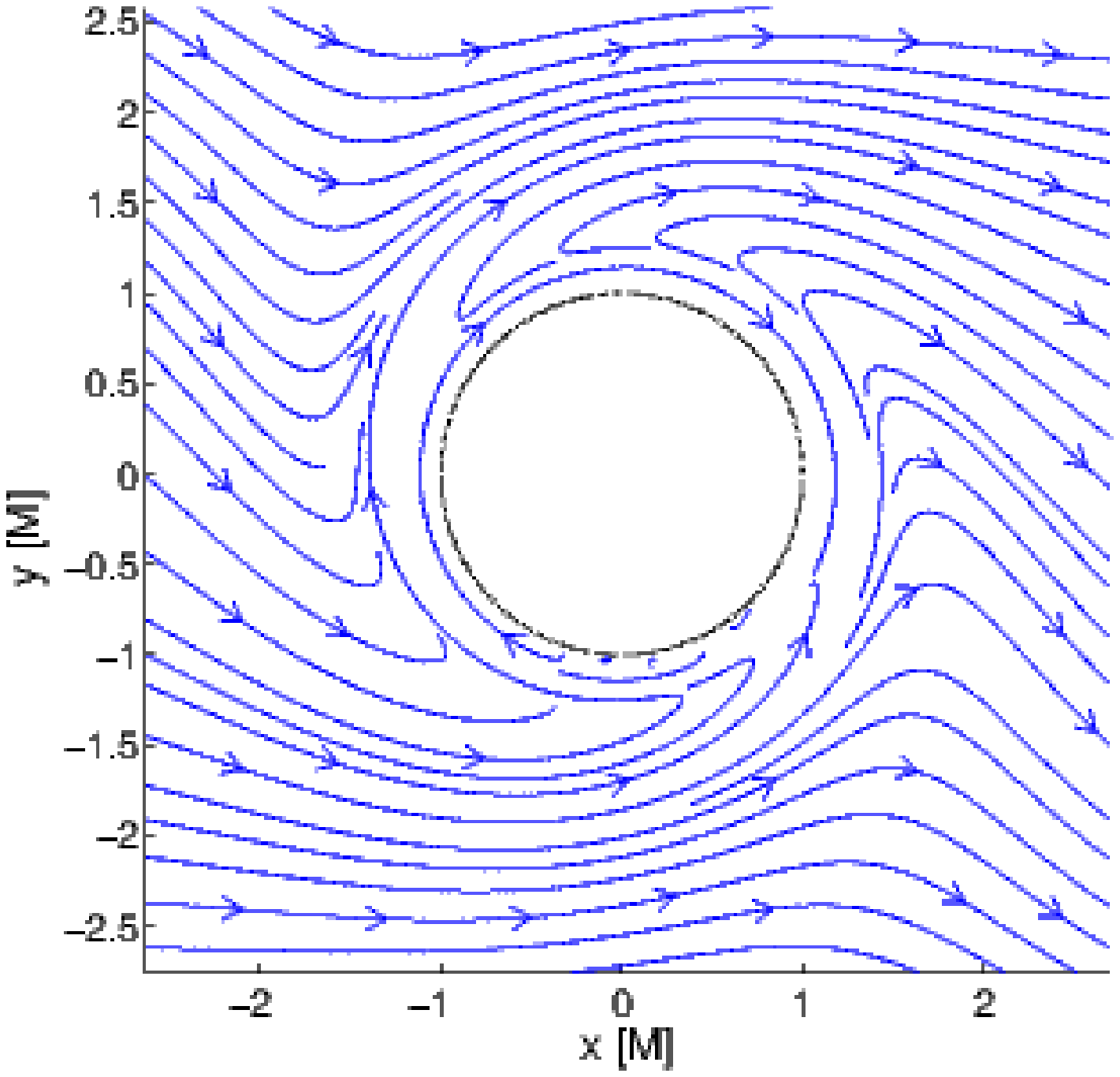}
\includegraphics[scale=0.28, clip]{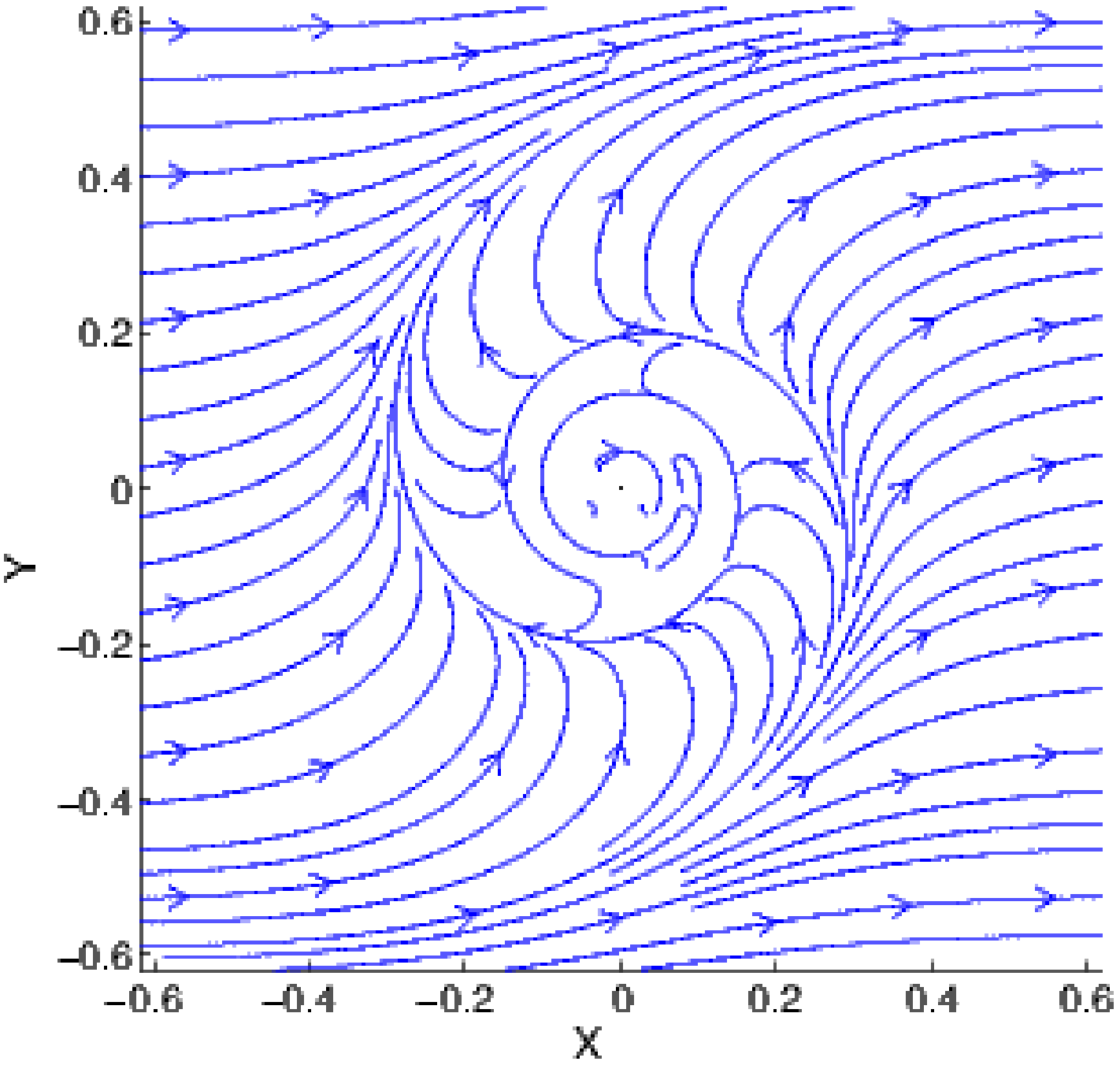}
\includegraphics[scale=0.28, clip]{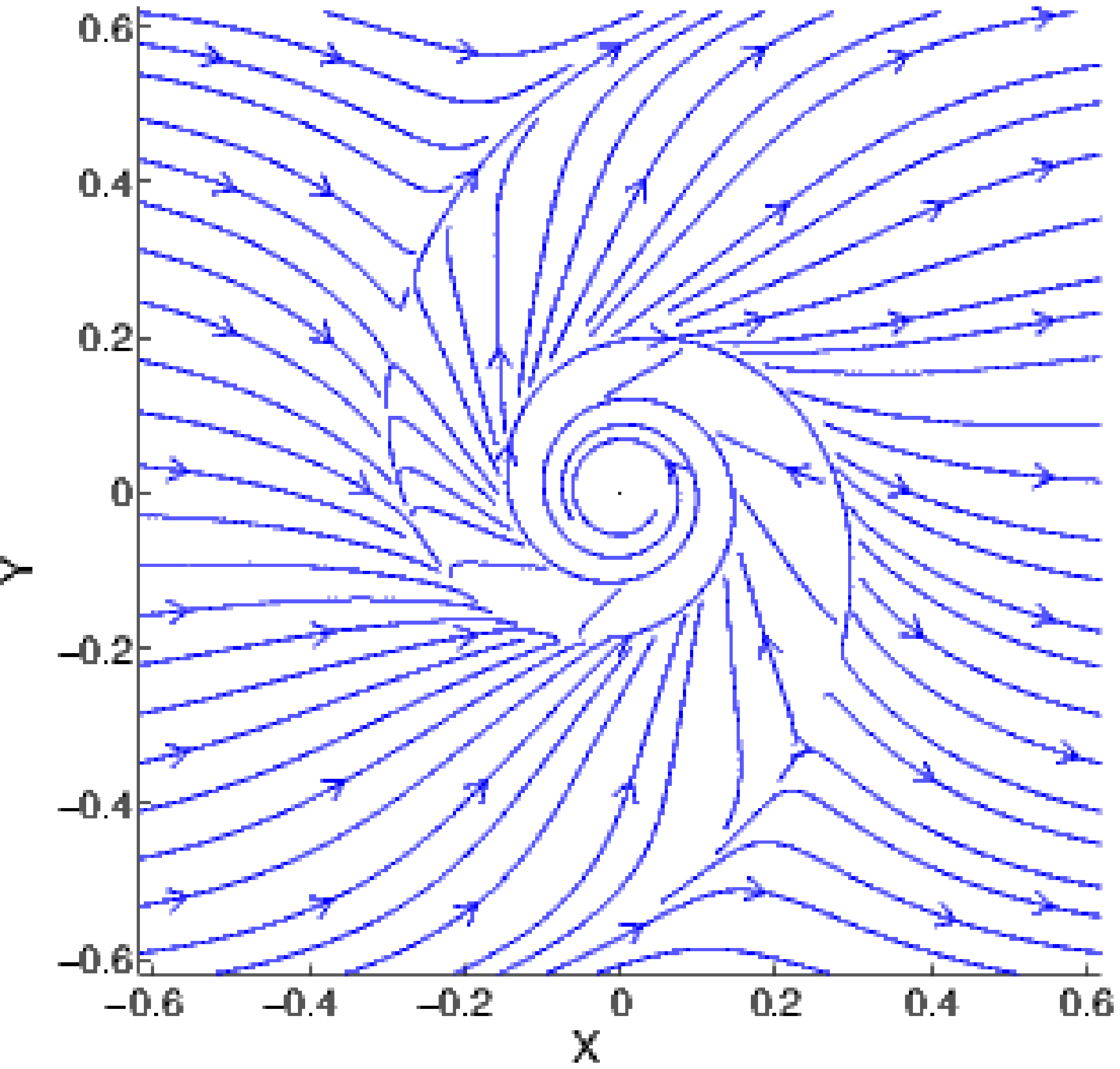}
\includegraphics[scale=0.28, clip]{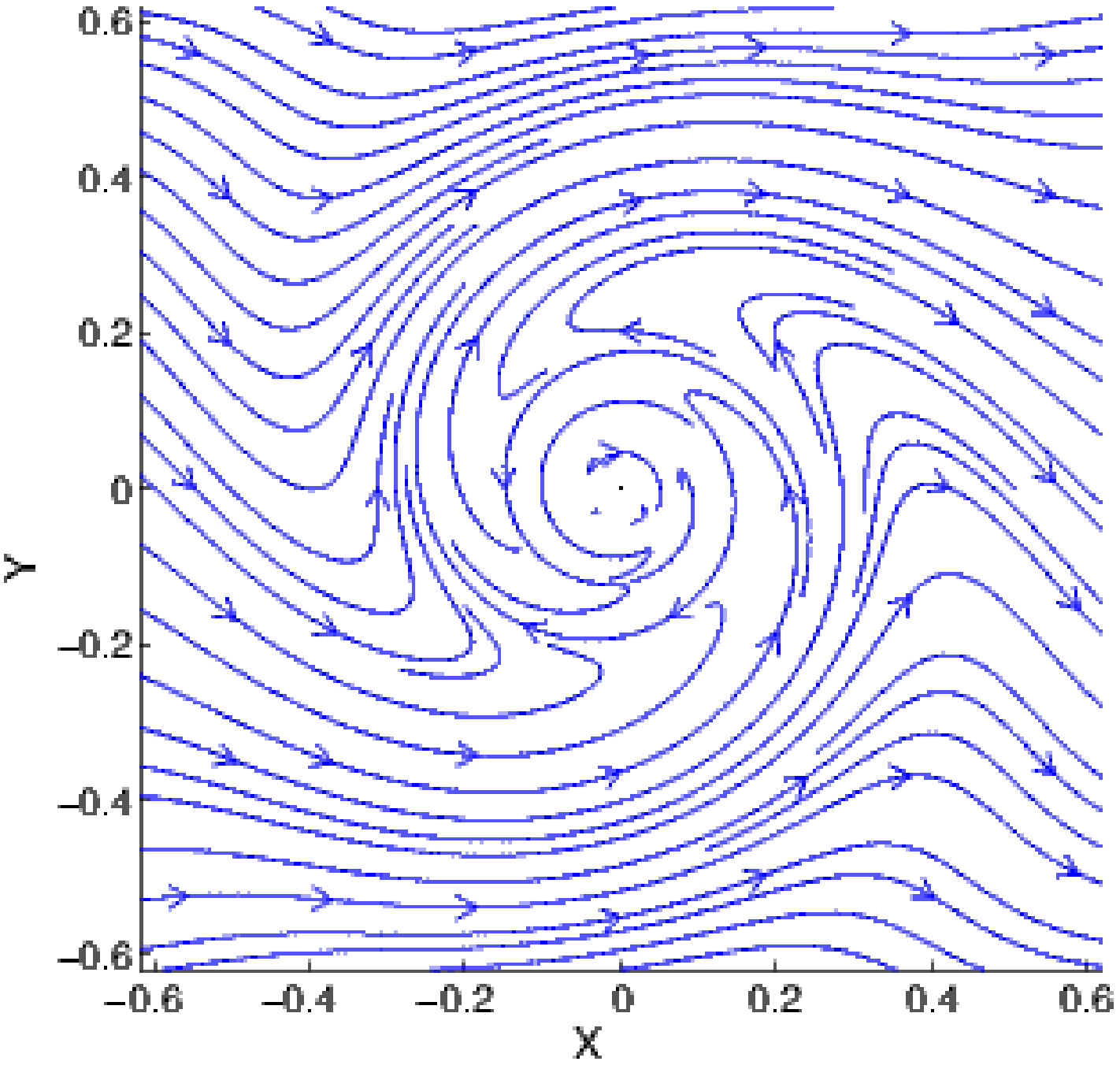}
\caption{Equatorial behaviour of the magnetic field with purely perpendicular asymptotics ($B_x\ne 0$, $B_z=0$) for three slightly distinct tetrads. We remind that FFOFI stands for the observer which is freely falling from the rest at infinity (angular momentum $L=0$) while KEP+FFO is the tetrad attached to the Keplerian observer being stable above the marginally stable orbit $r_{\rm{ms}}$ (eq. \ref{rms}) and falling freely below this orbit keeping $L(r_{\rm{ms}})$. The top row presents the FFOFI tetrad, the second the prograde KEP+FFO and the third is for the retrograde KEP+FFO. Schwarzschild limit $a=0$ is shown in the first column, middle column corresponds with $a=0.9M$ and the last one represents the extreme case $a=M$. In the bottom row the rescaled dimensionless radial coordinate $R\equiv\frac{r-r_{+}}{r}$ is used to stretch the region close to the horizon which is of the utmost interest. Panels of the bottom row show the fields measured by FFOFI and the both, co-rotating and counter-rotating KEP+FFO for extremal BH case.}
\label{mag_ekv}
\end{figure}

\begin{figure}[hp!]
\centering
\includegraphics[scale=0.53, clip]{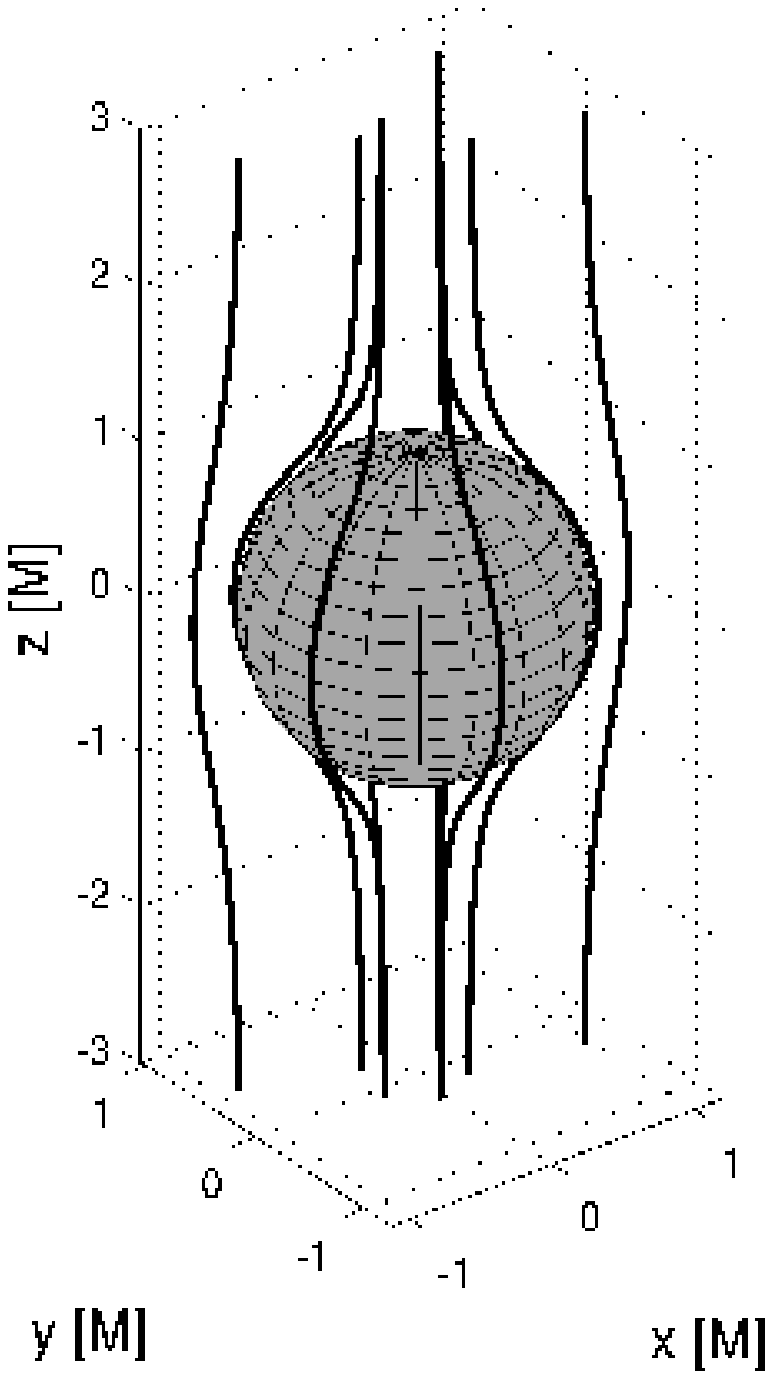}
\includegraphics[scale=0.56, clip]{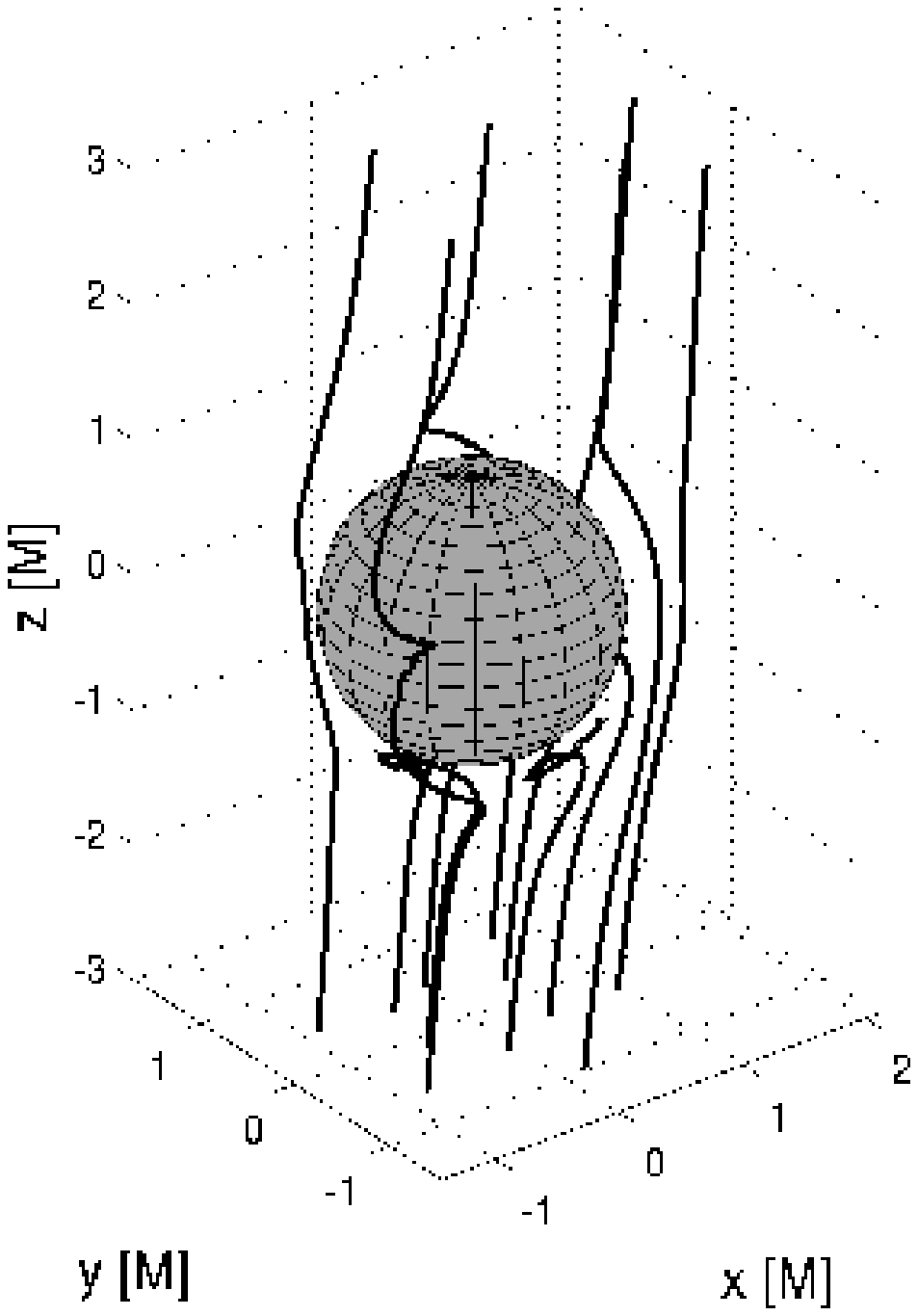}
\includegraphics[scale=0.47, clip]{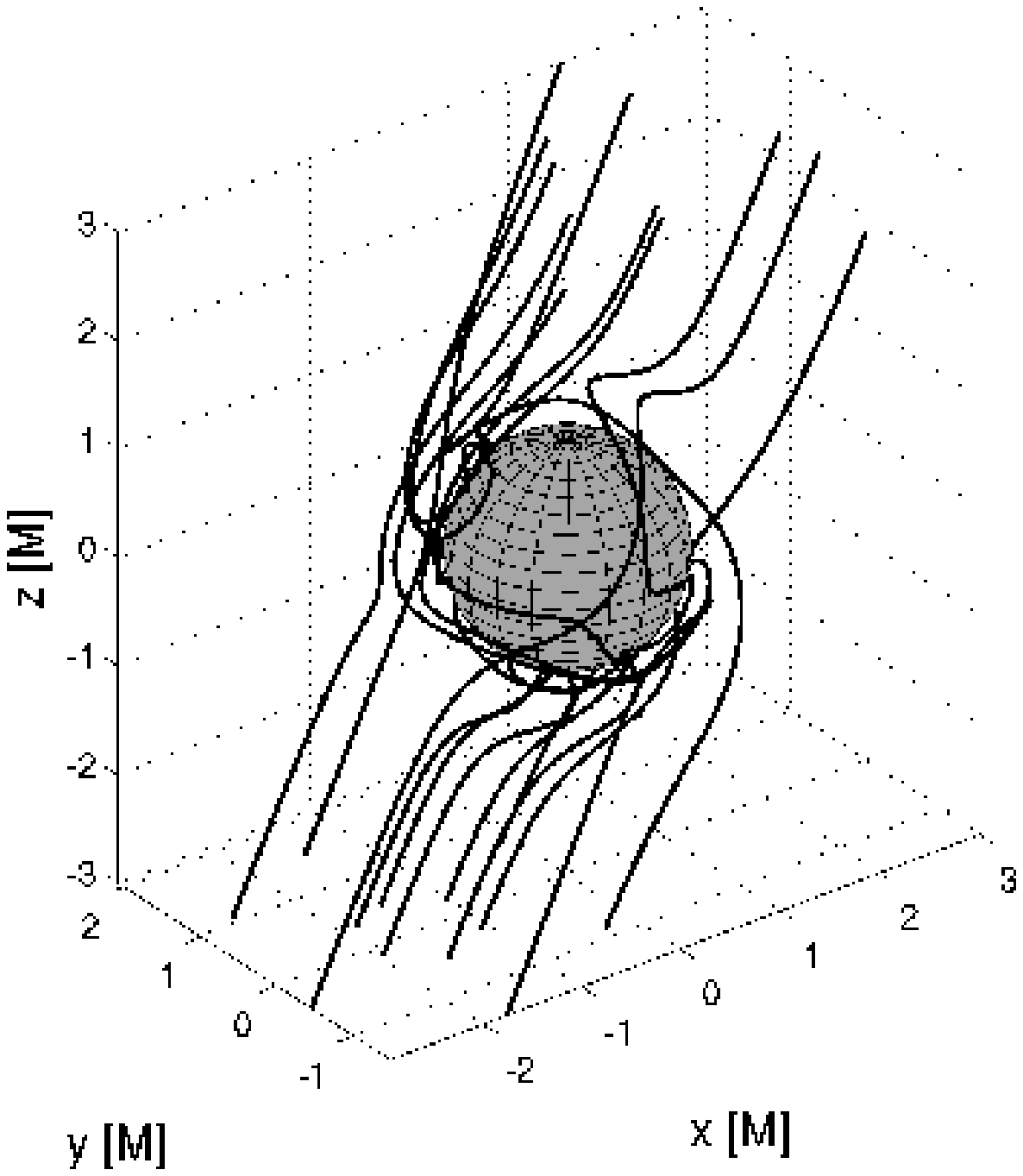}
\includegraphics[scale=0.52, clip]{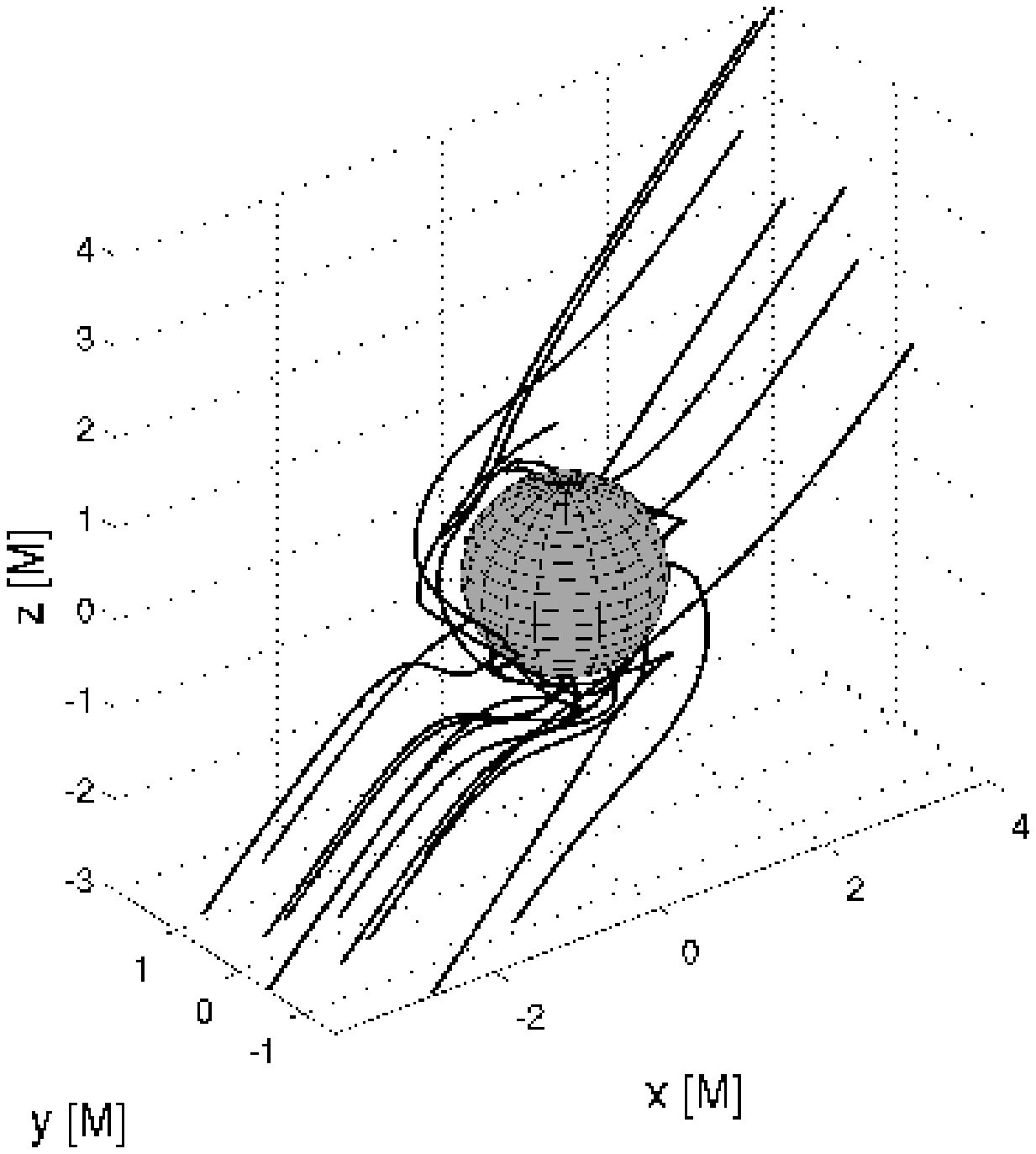}
\caption{Series of stereometric projections illustrates the impact of the perpendicular component $B_x$ upon the originally axisymmetric magnetic AMO field lines of $B_z$ origin in the case of extreme Kerr BH. Upper left panel shows aligned case ($B_x=0$) where the field is expelled, upper right $\frac{B_x}{B_z}=0.1$, bottom left $\frac{B_x}{B_z}=0.5$ and bottom right with $B_x=B_z$. Perpendicular component $B_x$ generally causes complex twisting of the field lines and allows the field to penetrate the horizon even in the extreme case.}
\label{magBx_3d}
\end{figure}

\begin{figure}[htb]
\centering
\includegraphics[scale=0.285, clip]{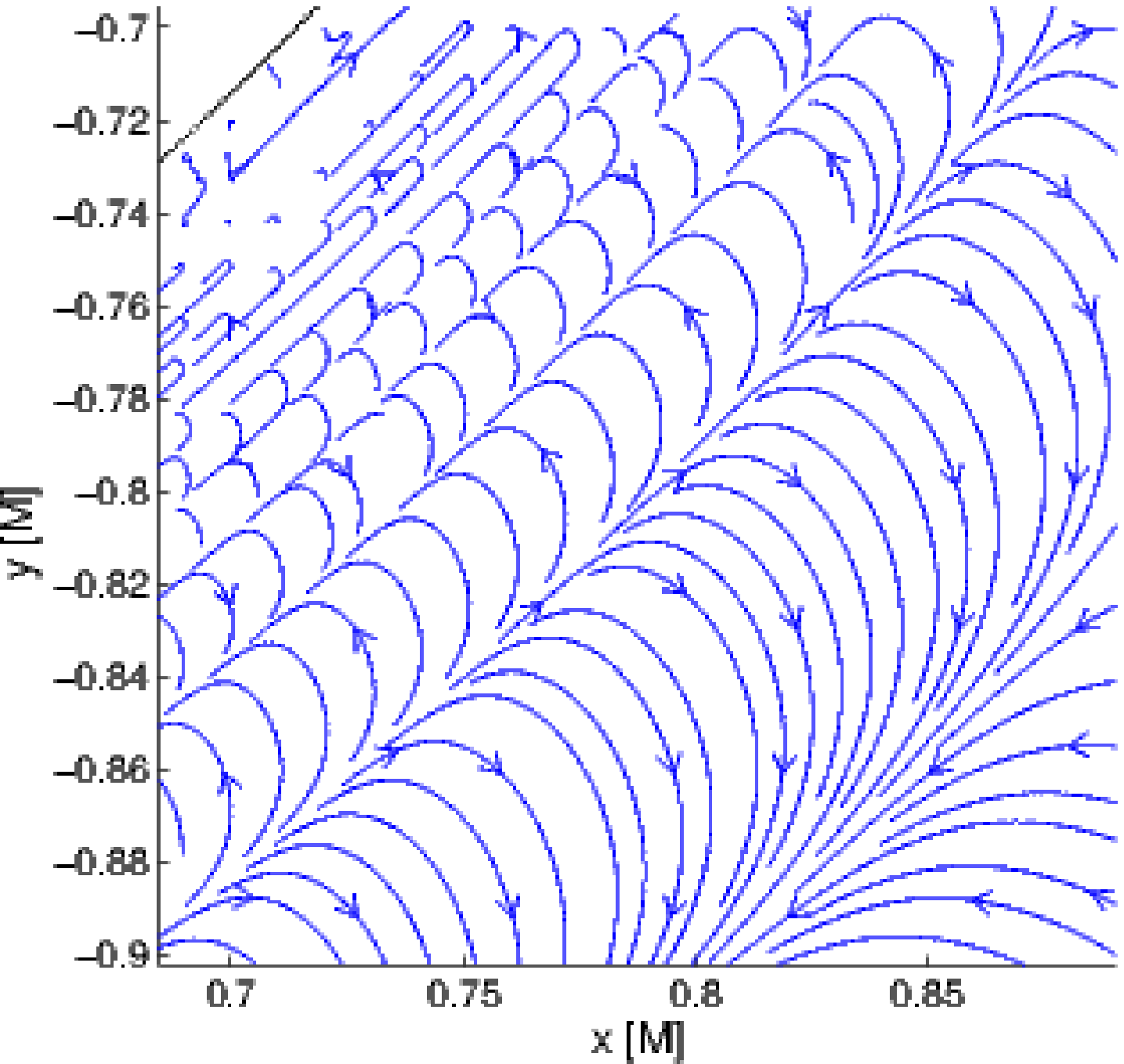}
\includegraphics[scale=0.285, clip]{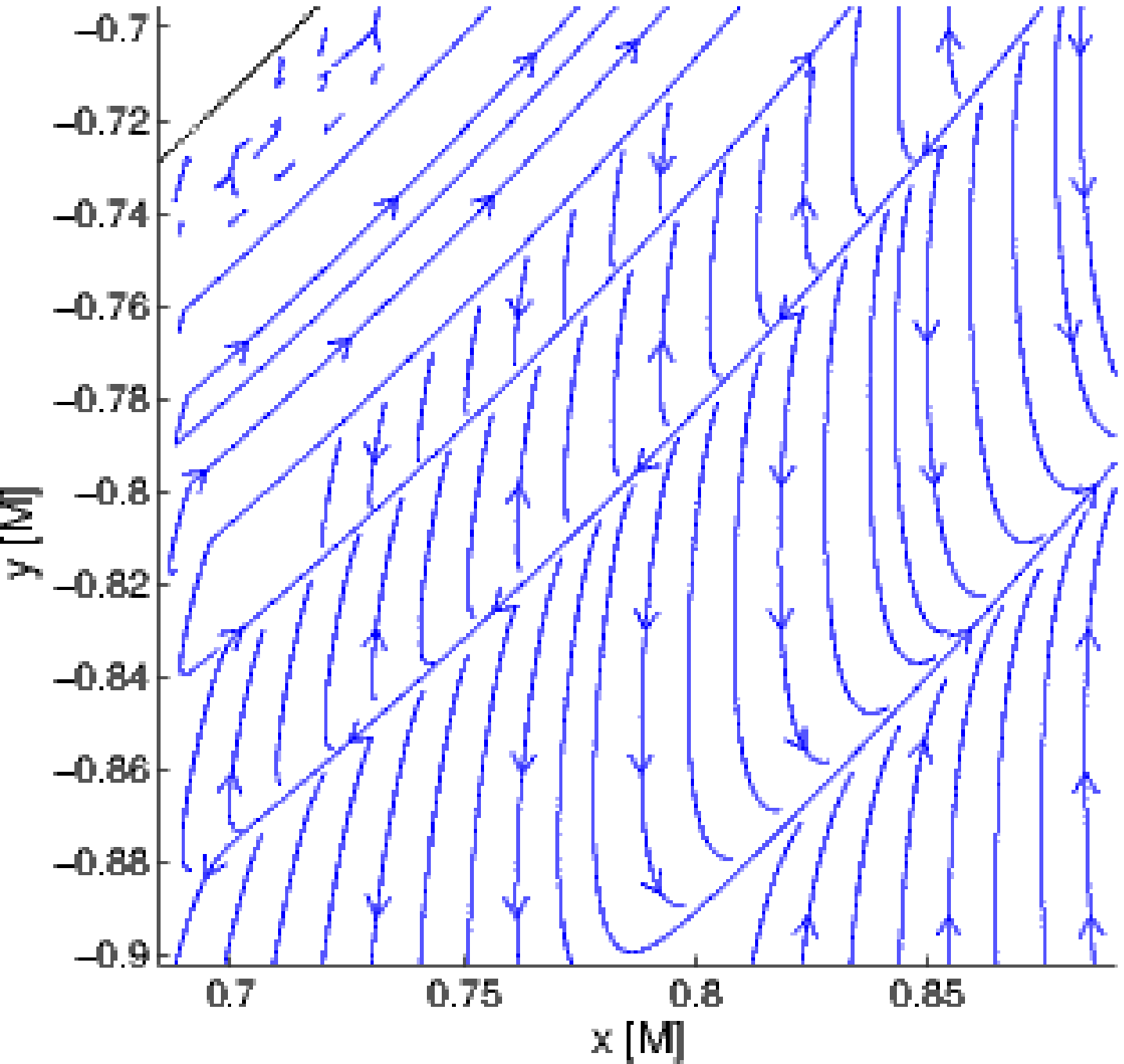}
\includegraphics[scale=0.285, clip]{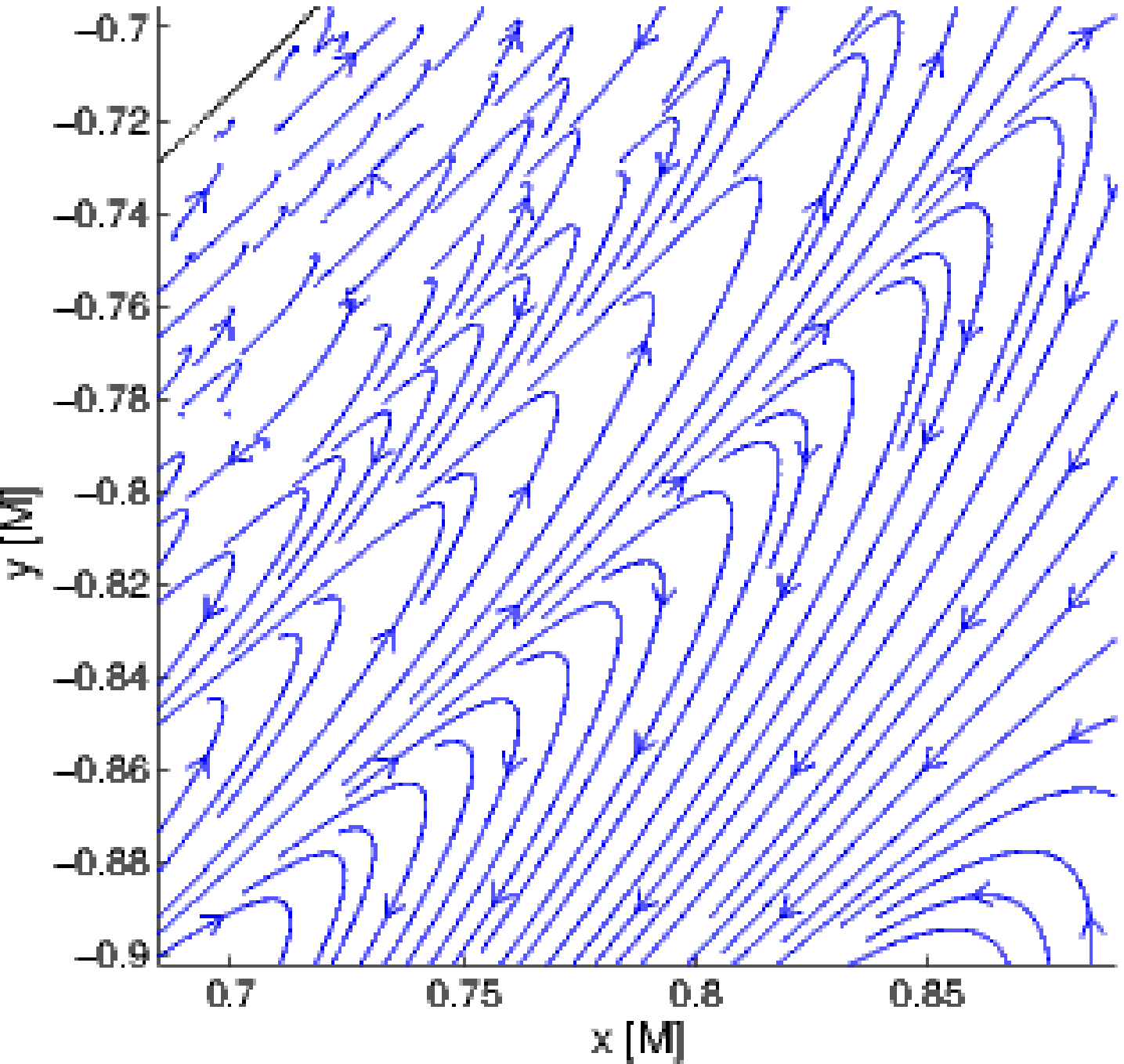}
\includegraphics[scale=0.285, clip]{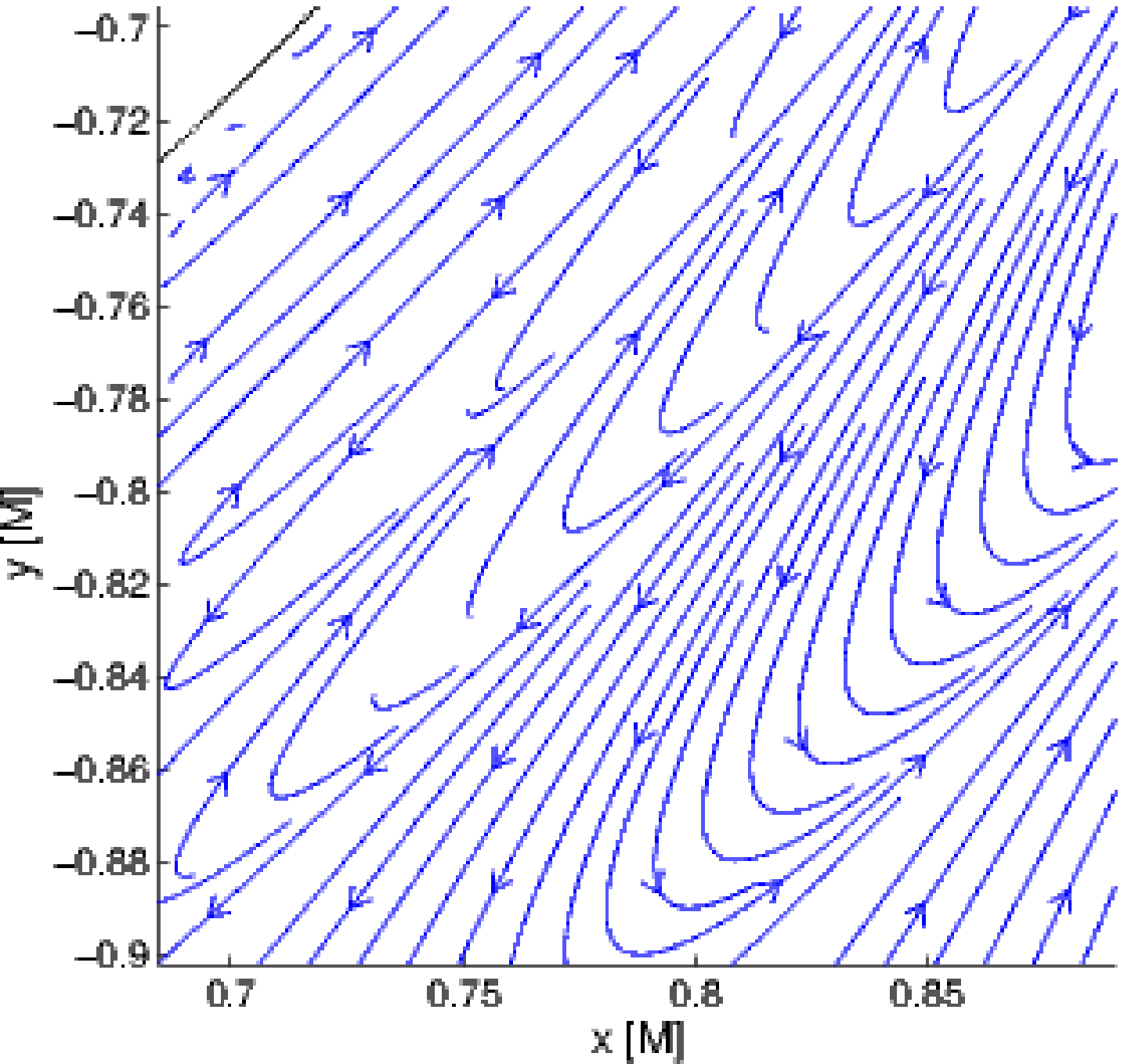}
\includegraphics[scale=0.285, clip]{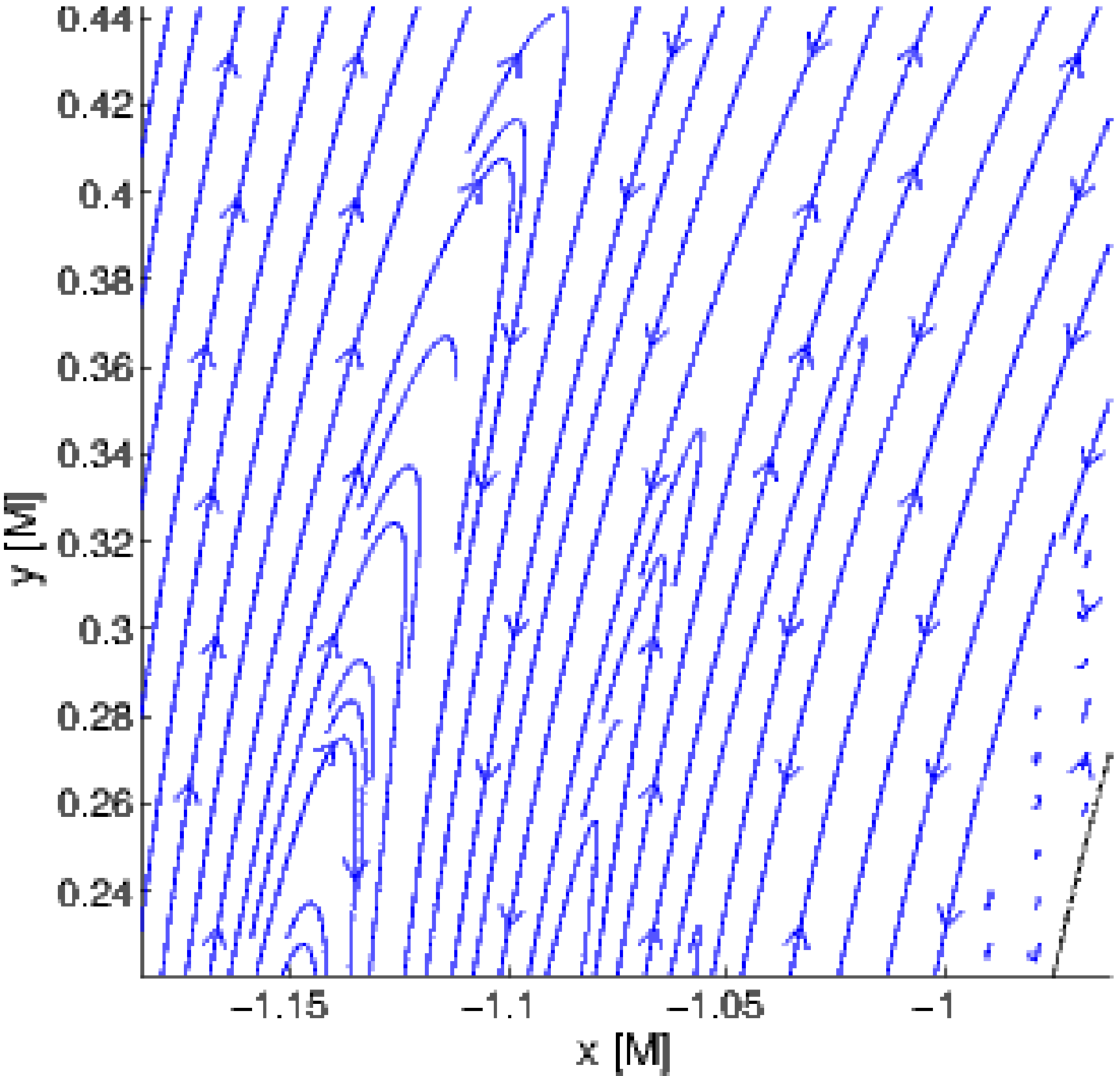}
\includegraphics[scale=0.285, clip]{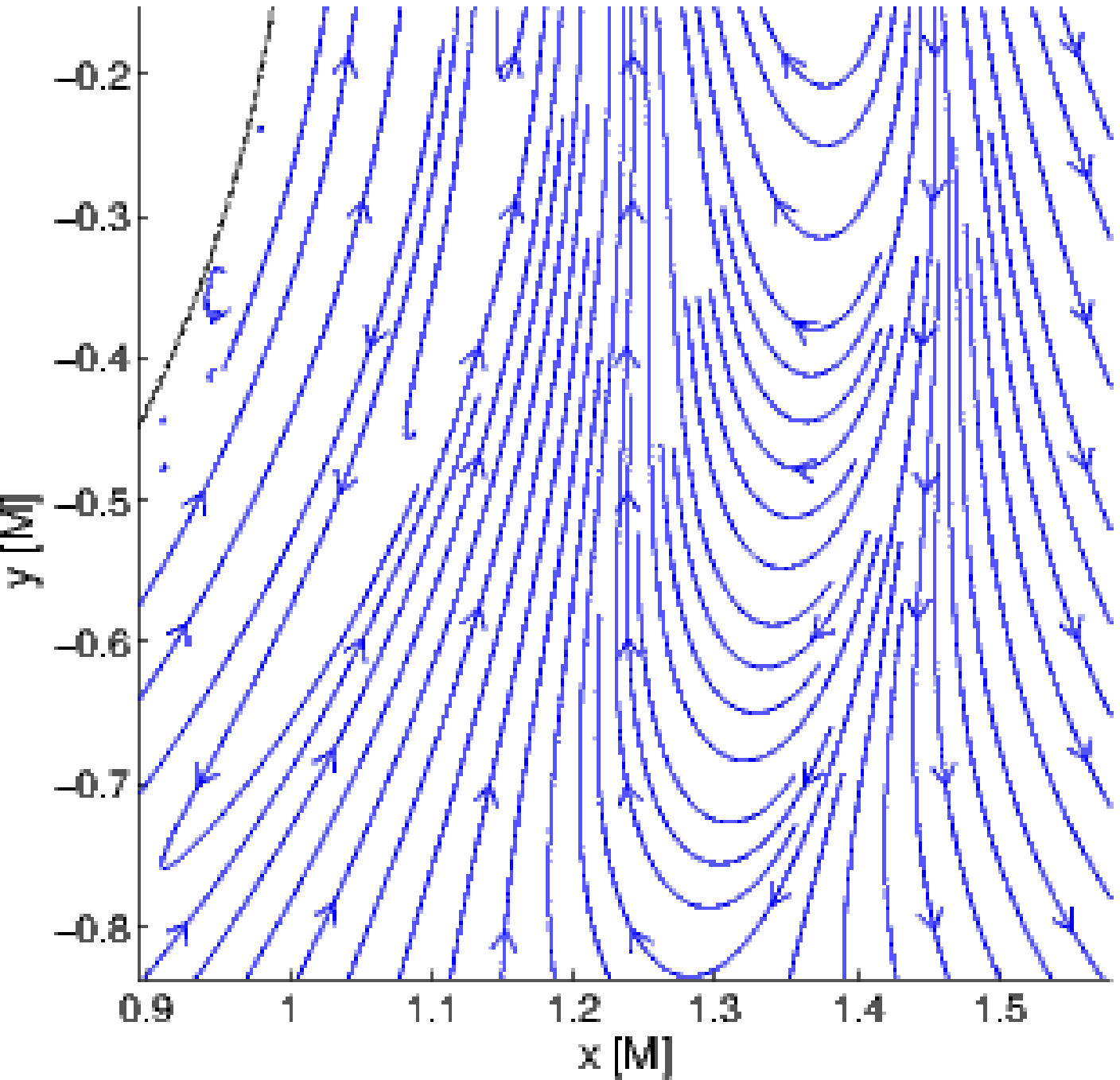}
\caption{Magnetic layers above the horizon of extreme Kerr BH that develop as a consequence of an interplay between the spin induced frame-dragging effect and perpendicular component of the prescribed magnetic field. Fraction of the layered zone is shown. We remind that in this case the field lines reside in the equatorial plane. Panels show the field measured by (from the top left to the bottom right): FFOFI observer, co-rotating Keplerian observer, observer freely falling from the counter-rotating marginally stable orbit and ZAMO. Last two panels show observer independent AMO components of the field in which the layering is also apparent proving it to be an intrinsic feature of the system rather than a pure observer effect.}
\label{ml_nodrift}
\end{figure}

We further notice that in the equatorial plane the asymptotically perpendicular magnetic field ($B_x\ne 0$, $B_z=0$) will have zero longitudinal component $B^{\theta}=B^{(\theta)}=0$ regardless the field definition also in the vicinity of the black hole (only equatorially  vanishing terms from (\ref{BFtr(Bx)}) appear in the definition of $B^{\theta}$ in eq. \ref{explicitvf1}). In other words the field lines reside in the equatorial plane in this special case. This might not be that visually apparent from the right panel of \rff{mag_amo_Bx}. Therefore we add a series of equatorial sections in \rff{mag_ekv} to reveal the field in the equatorial plane. Three equatorial observers, namely FFOFI, co-rotating KEP+FFO and counter-rotating KEP+FFO are involved in the comparison. We observe that free fall with $L=0$ (FFOFI) results in the aligned magnetic field lines in the Schwarzschild limit which we observed also in AMO components in \rff{mag_amo_Bx} as a result of $F_{\mu\nu}^{B_x}$ taking its asymptotic form for $a=0$. On the other hand, nonzero angular momentum of KEP+FFO observers which are freely falling with $L=L(r_{\rm{ms}})$ under $r_{\rm{ms}}$ causes twisting of the field lines even in the Schwarzschild case since the four-velocity and other tetrad vectors components do not take their asymptotic form with $a=0$ although the $F_{\mu\nu}$ does. When the spin is employed the lines of force twist in a rather complex way regardless the choice of the observer. Complex layering of the field lines emerges in the narrow zone above the horizon. In order to stretch this tiny yet very interesting area and make it more convenient for visualisation we rescale the radial component introducing the dimension-less coordinate $R\equiv\frac{r-r_{+}}{r}$ which squeezes the horizon into a single point $R=0$ and radial infinity into the circle $R=1$. Rescaled coordinate $R$ is employed in bottom panels of \rff{mag_ekv}.

In the series of stereometric projections in \rff{magBx_3d} we observe the interplay between the aligned and perpendicular magnetic field components using AMO definition of the field on the extreme Kerr background. We start with the aligned situation where the expulsion of the field is observed and gradually increase the perpendicular $B_{x}$ component to see that it generally causes intricate deformations of the field lines and allows them to penetrate the horizon even in the extreme Kerr case.

In \rff{mag_ekv} we observe that the field twisting caused by the interplay between the black hole rotation and the horizontal component of the prescribed magnetic field is accompanied by intensive layering of the magnetic field in the narrow zone above the horizon. In \rff{ml_nodrift} we compare the detailed structure of the magnetic layer zone as measured by several distinct observers. We observe that magnetic layers are present regardless the choice of the frame although their actual shape changes as we switch between them. It is thus apparent that magnetic layers do not emerge as a observer effect. They rather originate directly from the interplay between the frame-dragging effect and prescribed magnetic field which is confirmed by observing layered magnetic structures also in observer independent AMO components.

Simulations of non-vacuum magnetospheres carried out in the framework of force-free electrodynamics (FFE) often reveal similar magnetic layers in which the direction of the field changes sharply. FFE assumes that plasma is streaming solely along the magnetic field lines and such layers thus become current sheets. See \citet{spitkovsky06} for the simulation of pulsar magnetosphere in which the current sheets develop.


\subsubsection{Electric field}
Until now we have paid hardly any attention to the electric fields felt by the test charges in a given setup of Kerr black hole embedded in the asymptotically uniform magnetic field with general orientation with respect to the rotation axis. To correct this we present a summary of nonzero electric field components for various field definitions, observers and values of the spin parameter $a$ in \rft{el_pole}. We distinguish whether $F^{B_z}_{\mu\nu}$ or $F^{B_x}_{\mu\nu}$ components (or actually both) act as a source terms of a given electric field component. We treat separately the case of equatorial plane $\theta=\frac{\pi}{2}$ where part of the components diminish. Exploring \rft{el_pole} we notice particularly that nonzero electric field is measured even in the Schwarzschild limit $a=0$ in the case of FFOFI and KEP observers. 

\begin{figure}[hp]
\centering
\includegraphics[scale=0.48, clip]{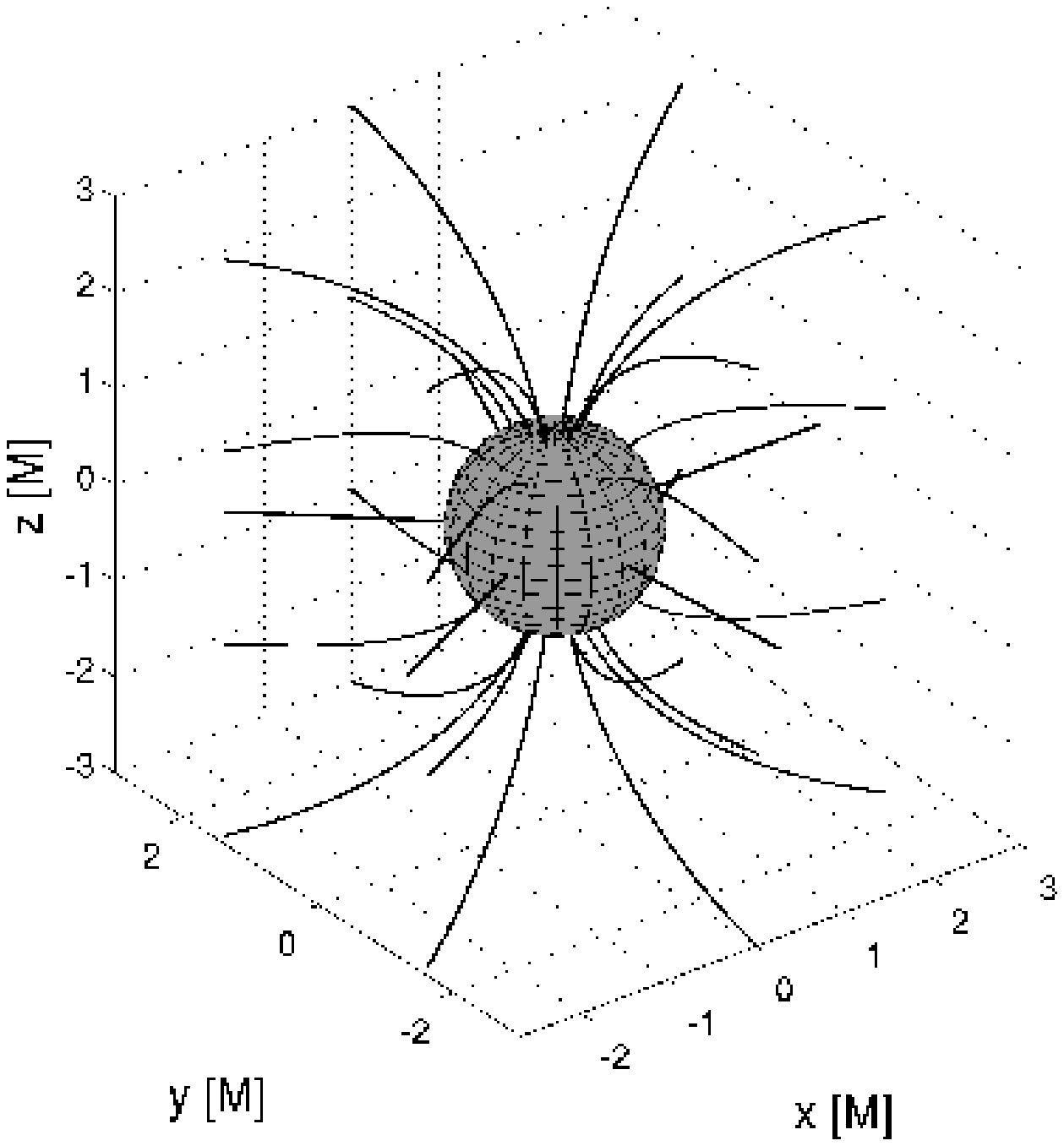}
\includegraphics[scale=0.36, clip, trim= 35mm 5mm 50mm 15mm]{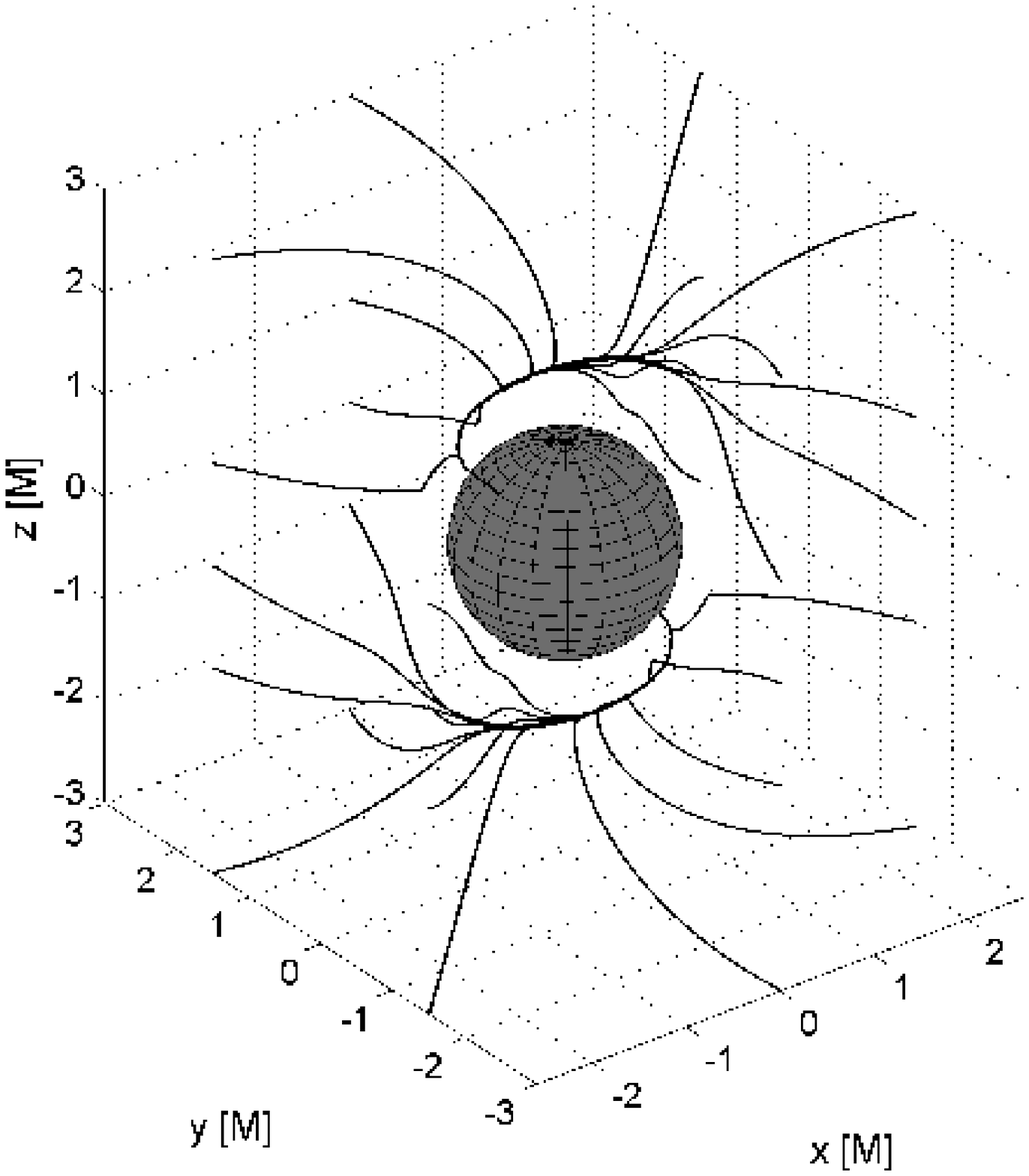}
\caption{AMO components of the electric field above the extreme Kerr BH. In the left panel we depict the aligned case $B_z=M^{-1}$, $B_x=0$ for which no azimuthal field components is measured. In the equatorial plane this field becomes purely radial. On the other hand the non-aligned situation $B_z=B_x=M^{-1}$ presented in the right panel brings azimuthal electric component into the play.}
\label{el_amo_3d}
\end{figure}


Stereometric projections of AMO electric fields above extreme Kerr BH are presented in \rff{el_amo_3d} to illustrate the profound changes in the field line structure which accompany the onset of perpendicular magnetic field component $B_x$ which disrupts the symmetry of aligned case.

\begin{table}[hp!]
\label{el_pole}
\begin{center}
\begin{tabular}{|c|c|c|c|c|}
\hline
  field definition& $a[M]$ & $\theta$ & $B_z$ &  $B_x$\\
\hline\hline
\multirow{4}{*}{AMO}&\multirow{2}{*}{0}&$\frac{\pi}{2}$, $0$, $\pi$&$\emptyset$&$\emptyset$\\
\cline{3-5}
&&$\ne \frac{\pi}{2}$, $0$, $\pi$&$\emptyset$&$\emptyset$\\
\cline{2-5}
&\multirow{2}{*}{$\ne 0$}&$\frac{\pi}{2}$, $0$, $\pi$&$E^r$&$E^{\theta}$\\
\cline{3-5}
&&$\ne \frac{\pi}{2}$, $0$, $\pi$&$E^r$, $E^{\theta}$&$E^r$, $E^{\theta}$, $E^{\varphi}$\\
\hline
\multirow{4}{*}{ZAMO coord.}&\multirow{2}{*}{0}&$\frac{\pi}{2}$, $0$, $\pi$&$\emptyset$&$\emptyset$\\
\cline{3-5}
&&$\ne \frac{\pi}{2}$, $0$, $\pi$&$\emptyset$&$\emptyset$\\
\cline{2-5}
&\multirow{2}{*}{$\ne 0$}&$\frac{\pi}{2}$, $0$, $\pi$&$E^r$&$E^{\theta}$\\
\cline{3-5}
&&$\ne \frac{\pi}{2}$, $0$, $\pi$&$E^r$, $E^{\theta}$&$E^r$, $E^{\theta}$, $E^{\varphi}$\\
\hline
\multirow{4}{*}{ZAMO tetrad}&\multirow{2}{*}{0}&$\frac{\pi}{2}$, $0$, $\pi$&$\emptyset$&$\emptyset$\\
\cline{3-5}
&&$\ne \frac{\pi}{2}$, $0$, $\pi$&$\emptyset$&$\emptyset$\\
\cline{2-5}
&\multirow{2}{*}{$\ne 0$}&$\frac{\pi}{2}$, $0$, $\pi$&$E^{(r)}$&$E^{(\theta)}$\\
\cline{3-5}
&&$\ne \frac{\pi}{2}$, $0$, $\pi$&$E^{(r)}$, $E^{(\theta)}$&$E^{(r)}$, $E^{(\theta)}$, $E^{(\varphi)}$\\
\hline
\multirow{4}{*}{FFOFI coord.}&\multirow{2}{*}{0}&$\frac{\pi}{2}$, $0$, $\pi$&$E^{\varphi}$ &$E^{\theta}$\\
\cline{3-5}
&&$\ne \frac{\pi}{2}$, $0$, $\pi$&$E^{\varphi}$&$E^{\theta}$, $E^{\varphi}$\\
\cline{2-5}
&\multirow{2}{*}{$\ne 0$}&$\frac{\pi}{2}$, $0$, $\pi$&$E^r$, $E^{\varphi}$ &$E^{\theta}$\\
\cline{3-5}
&&$\ne \frac{\pi}{2}$, $0$, $\pi$&$E^r$, $E^{\theta}$, $E^{\varphi}$&$E^r$, $E^{\theta}$, $E^{\varphi}$\\
\hline
\multirow{4}{*}{FFOFI tetrad}&\multirow{2}{*}{0}&$\frac{\pi}{2}$, $0$, $\pi$&$E^{(\varphi)}$ &$E^{(\theta)}$\\
\cline{3-5}
&&$\ne \frac{\pi}{2}$, $0$, $\pi$&$E^{(\varphi)}$&$E^{(\theta)}$, $E^{(\varphi)}$\\
\cline{2-5}
&\multirow{2}{*}{$\ne 0$}&$\frac{\pi}{2}$, $0$, $\pi$&$E^{(r)}$, $E^{(\varphi)}$&$E^{(\theta)}$\\
\cline{3-5}
&&$\ne \frac{\pi}{2}$, $0$, $\pi$&$E^{(r)}$, $E^{(\theta)}$, $E^{(\varphi)}$&$E^{(r)}$, $E^{(\theta)}$, $E^{(\varphi)}$\\
\hline
\multirow{2}{*}{KEP coord.}&0&$\frac{\pi}{2}$&$E^{r}$ &$E^{\theta}$\\
\cline{2-5}
&$\ne 0$&$\frac{\pi}{2}$&$E^{r}$&$E^{\theta}$\\
\hline
\multirow{2}{*}{KEP tetrad}&0&$\frac{\pi}{2}$&$E^{(r)}$ &$E^{(\theta)}$\\
\cline{2-5}
&$\ne 0$&$\frac{\pi}{2}$&$E^{(r)}$&$E^{(\theta)}$\\
\hline
\end{tabular}
\caption{Summary of the electric field components depending on the definition of the field, spin of the black hole and employed preset magnetic field (which may be aligned or not). We treat separately the situation in the equatorial plane $\theta=\frac{\pi}{2}$ since the field tends to simplify here profoundly. Separate columns labeled $B_{z}$ and $B_{x}$ suggest for which components of measured electric field is given component of the asymptotic magnetic field being "responsible''. In other words electric field components found in e.g. $B_x$ column are ``generated'' by $B_x$. We notice particularly that in the case of free-falling test particle (for both coordinate and tetrad components) as well as for Keplerian four-velocity we measure nonzero  components of electric field even if there is zero spin (Schwarzschild limit). We emphasize that the fact that we arrive at the very same components being nonzero (though not of the same values!) both in the coordinate frame and the tetrad frame for a given four-velocity of the test charge is not in any sense automatic or generally guaranteed as we could easily think of such $F_{\mu\nu}$ for which it would not be the case.}
\end{center}
\end{table}

\begin{figure}[hp]
\centering
\includegraphics[scale=0.29, clip]{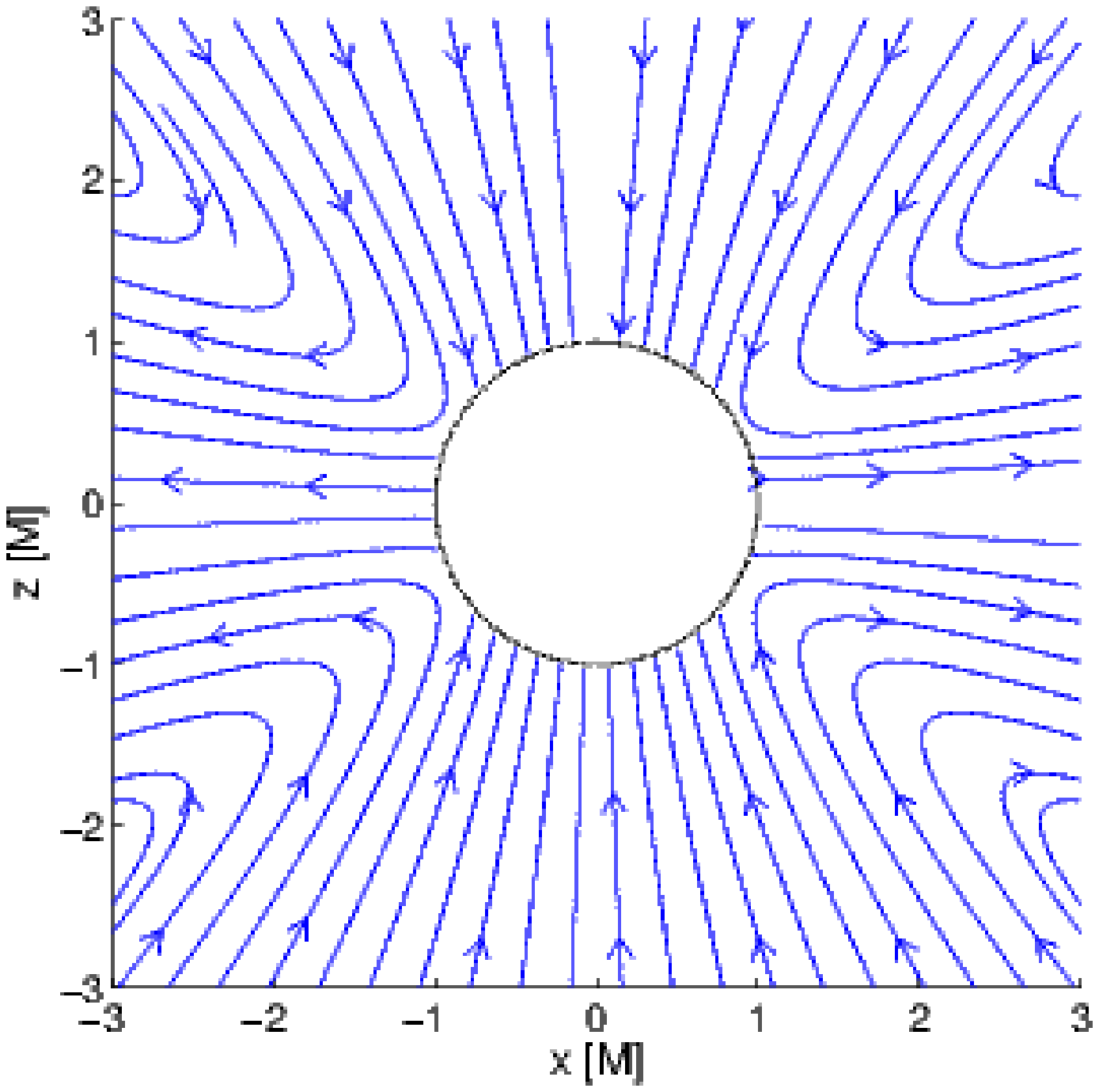}
\includegraphics[scale=0.29, clip]{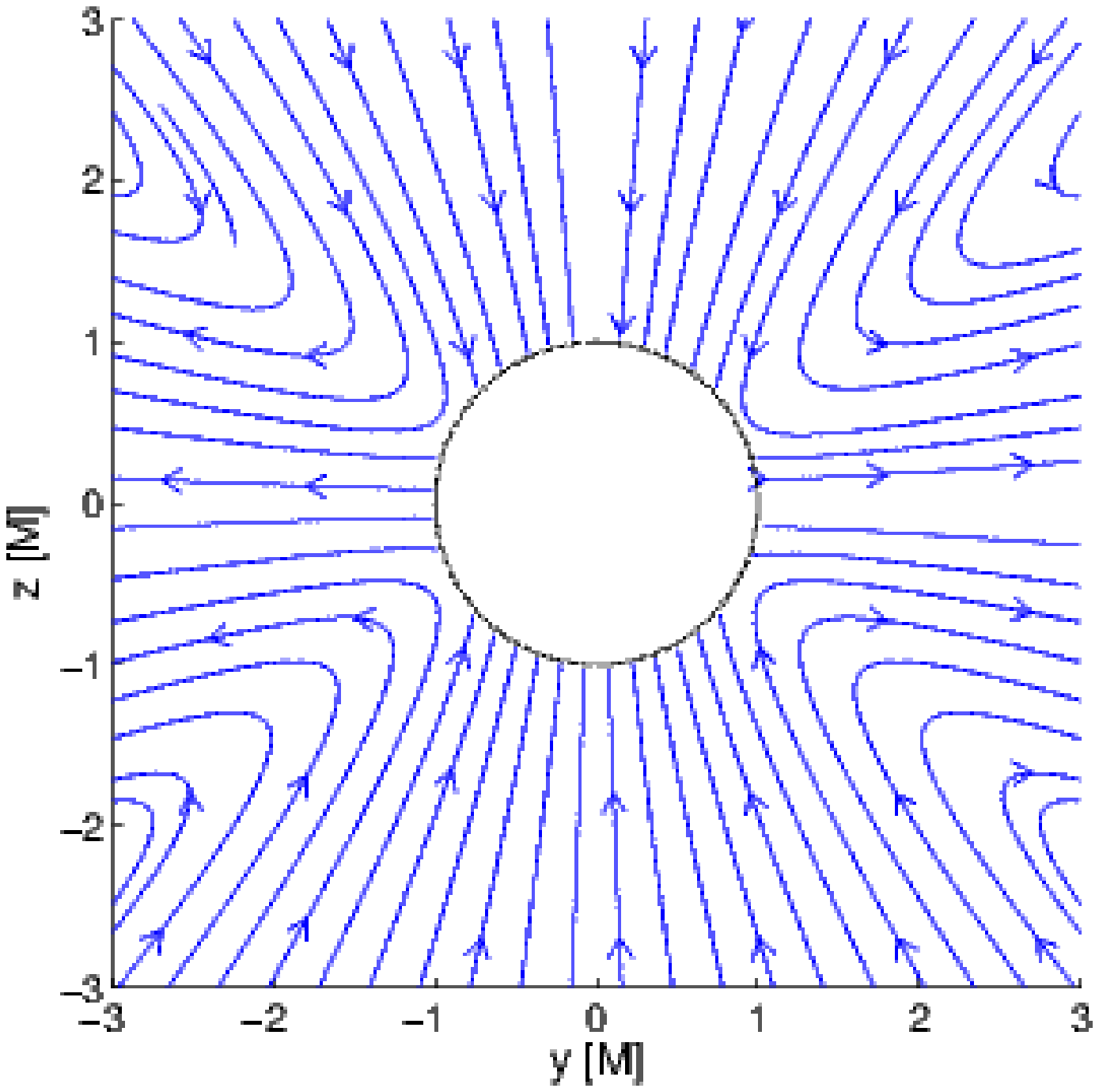}
\includegraphics[scale=0.29, clip]{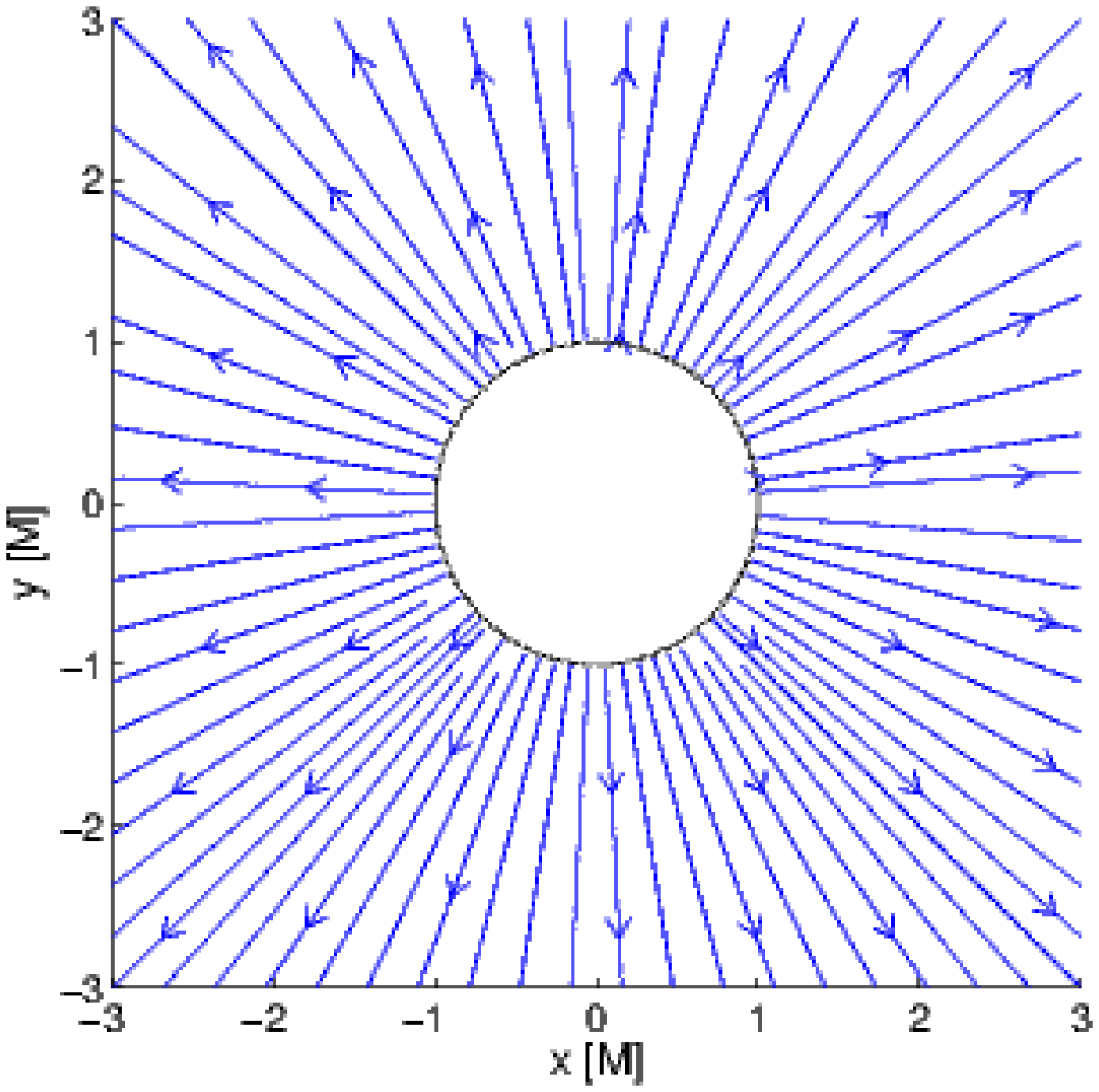}
\includegraphics[scale=0.29, clip]{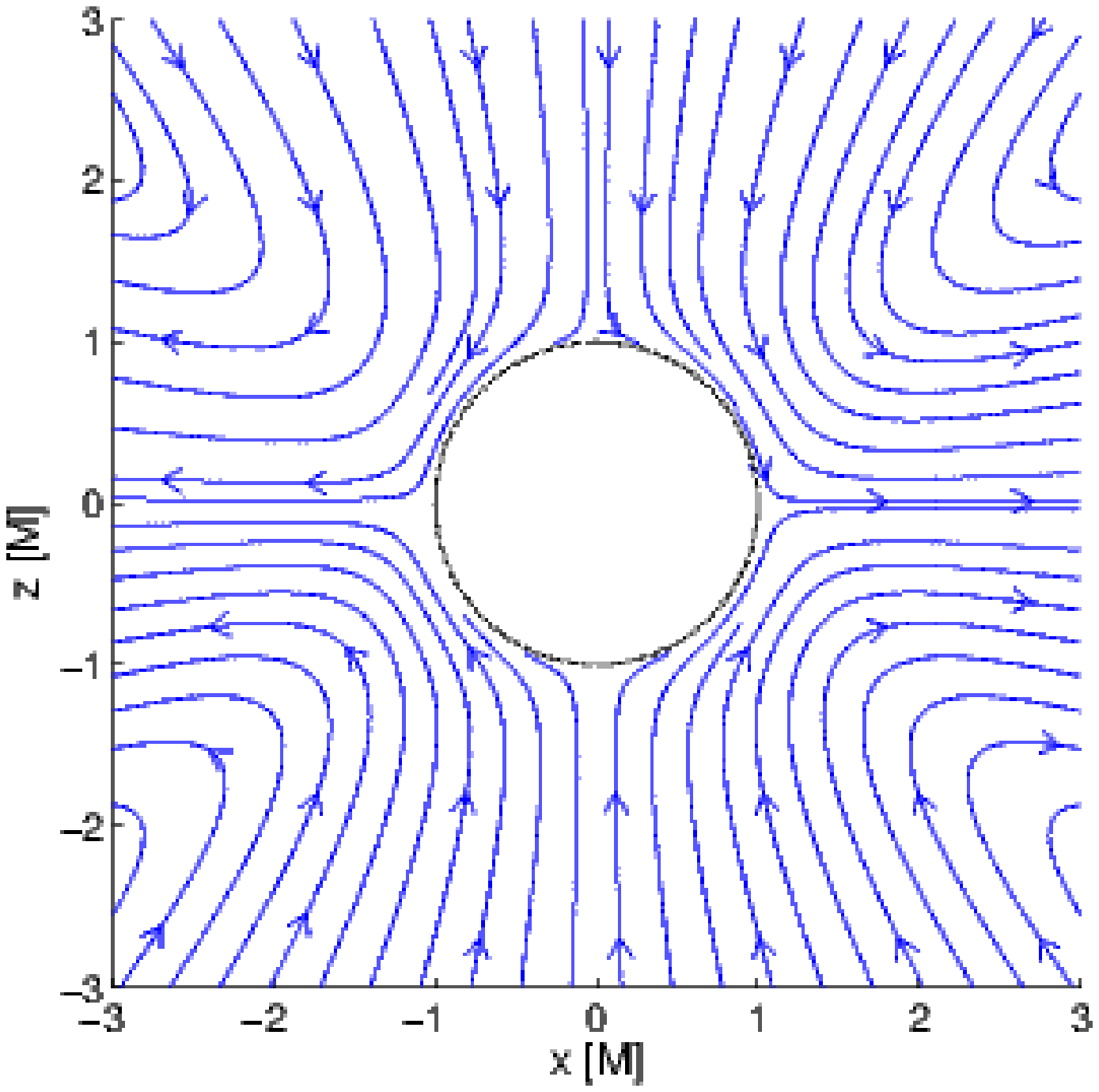}
\includegraphics[scale=0.29, clip]{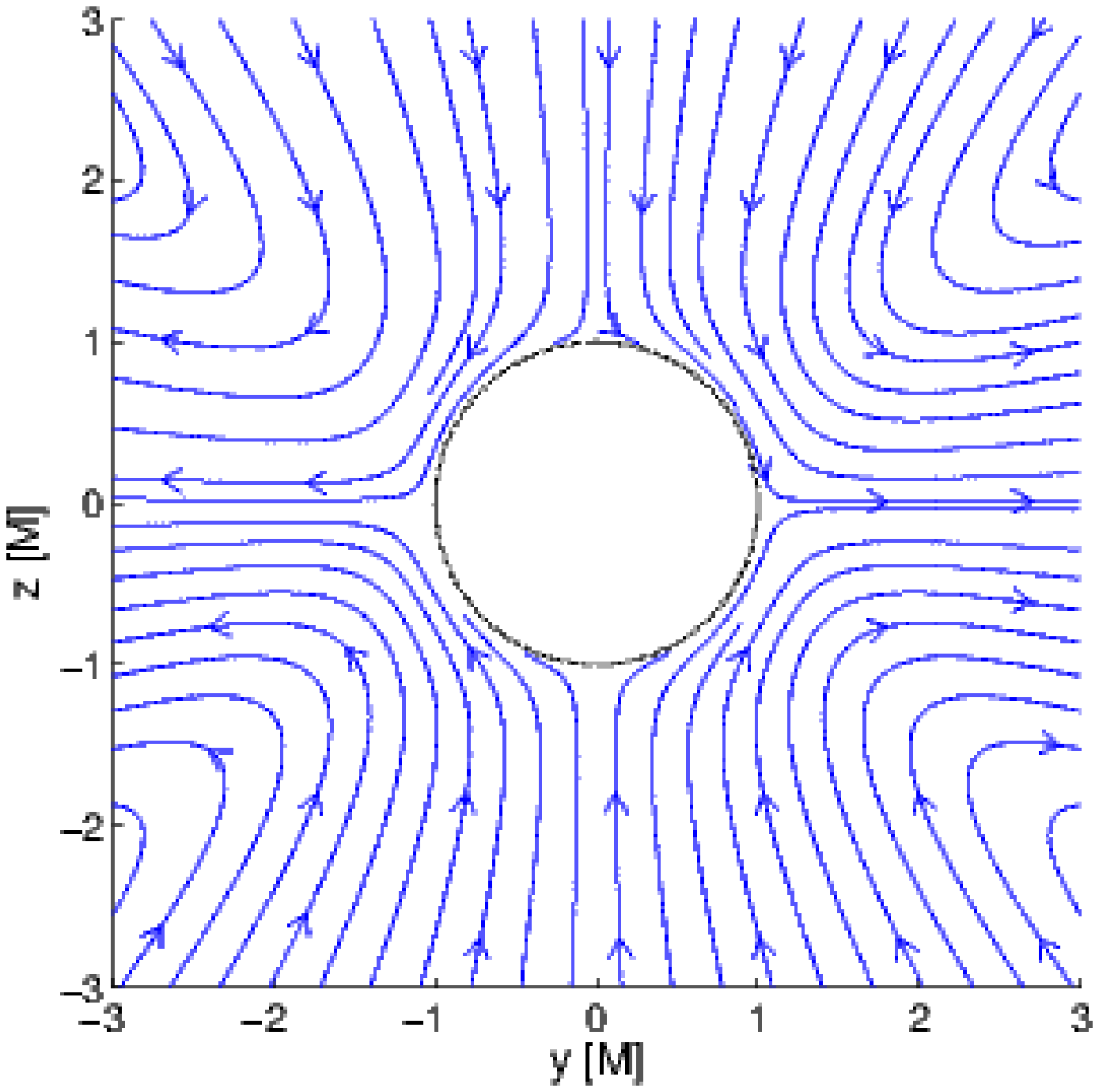}
\includegraphics[scale=0.29, clip]{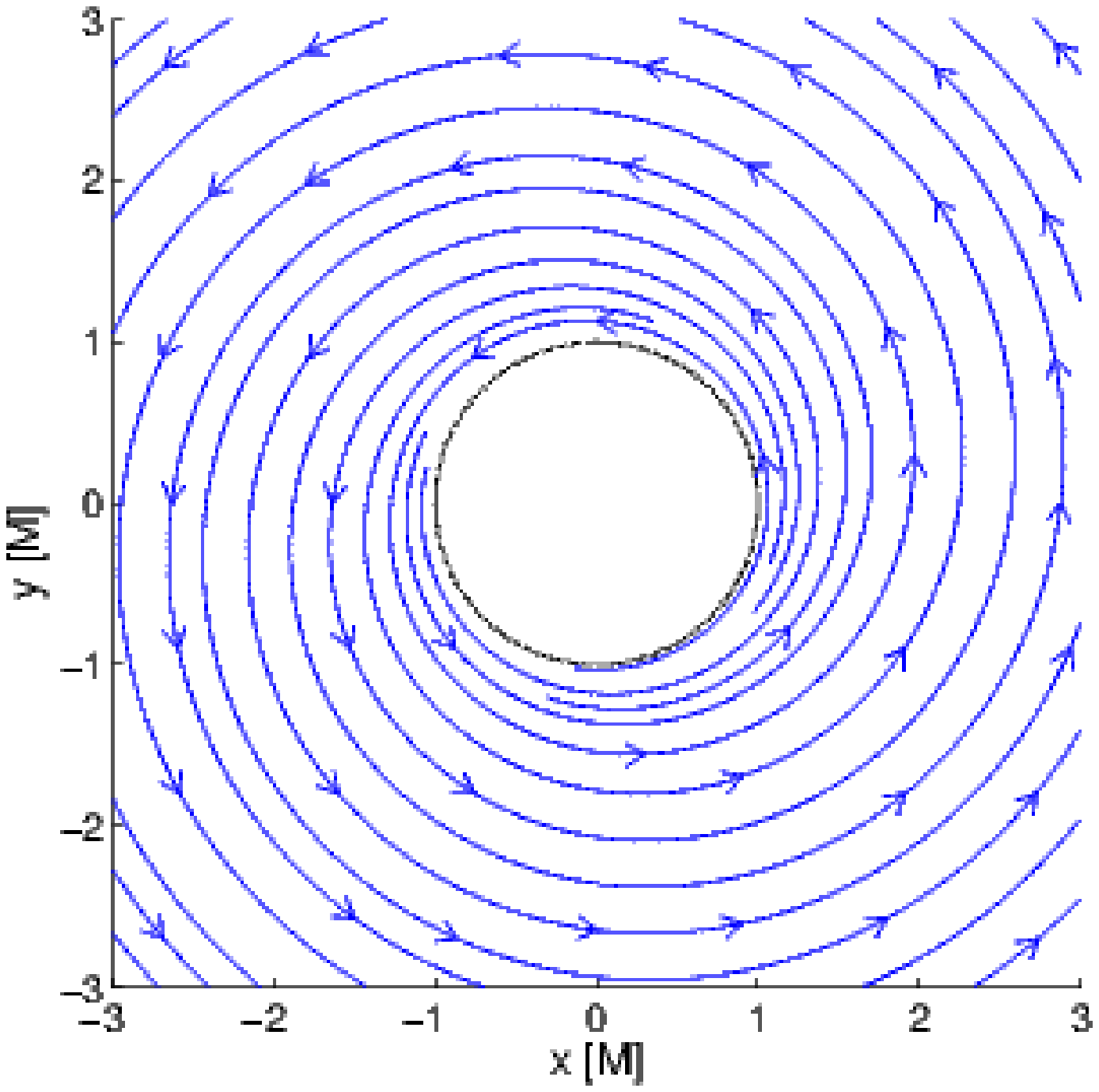}
\includegraphics[scale=0.29, clip]{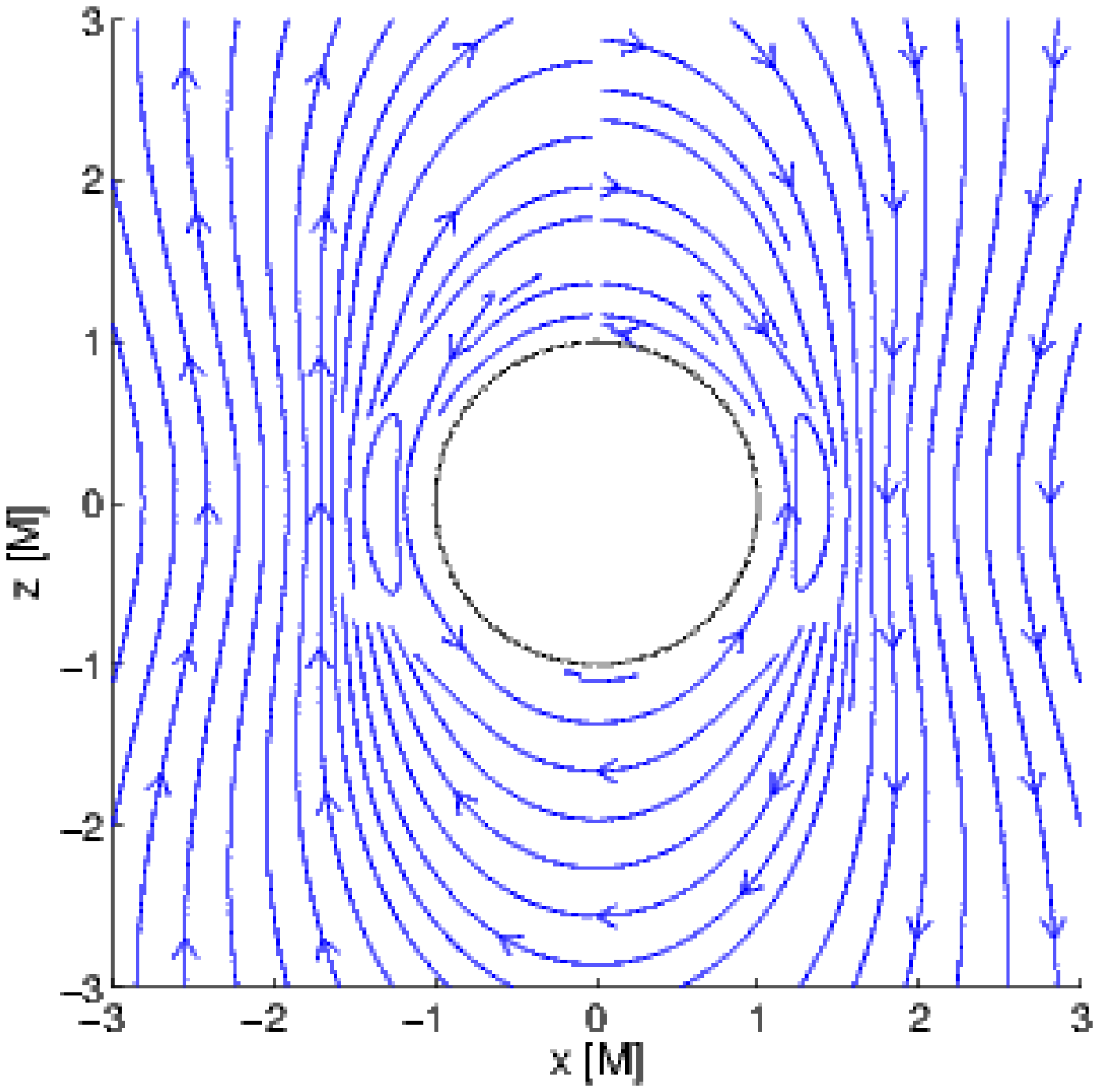}
\includegraphics[scale=0.29, clip]{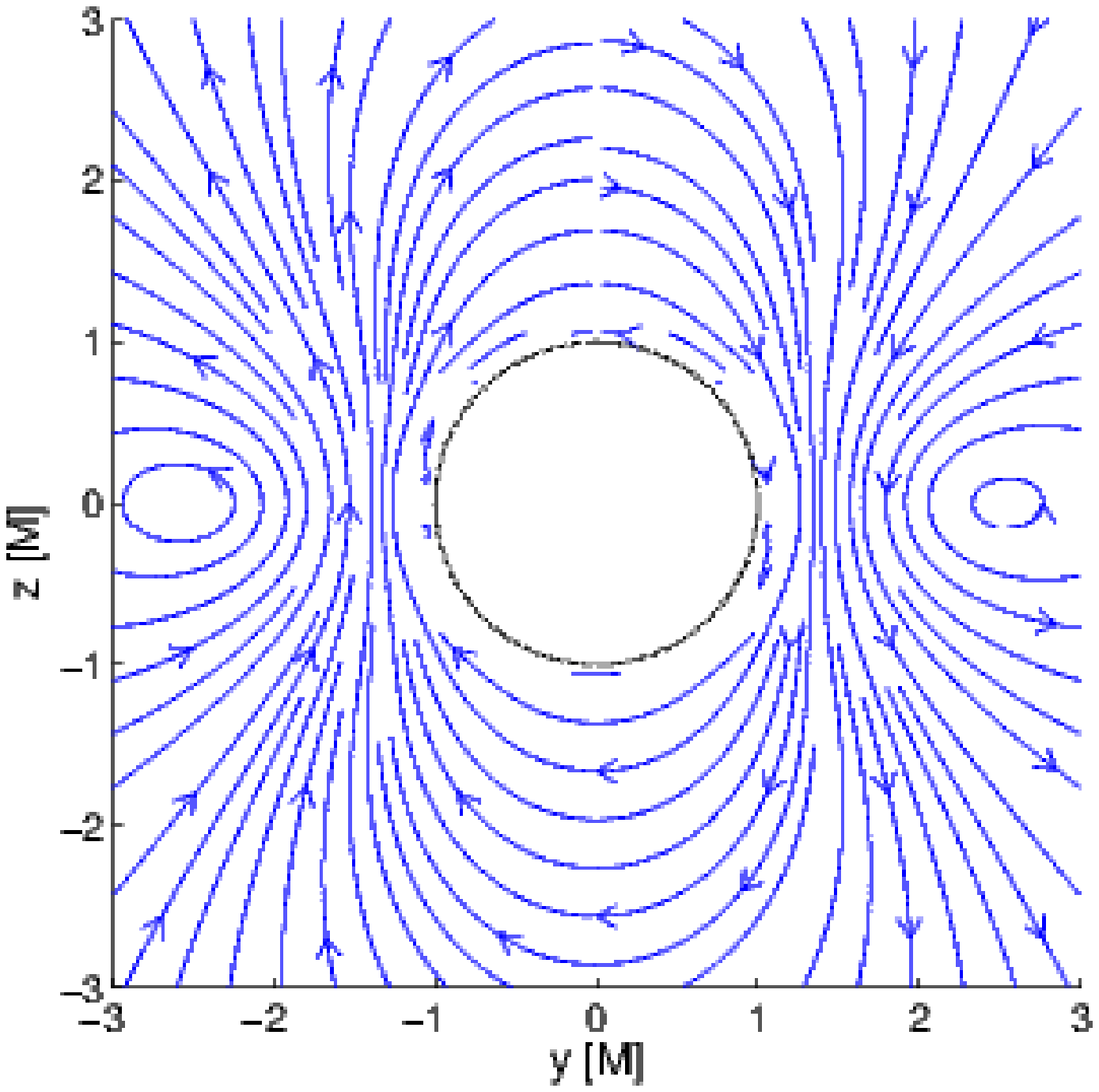}
\includegraphics[scale=0.29, clip]{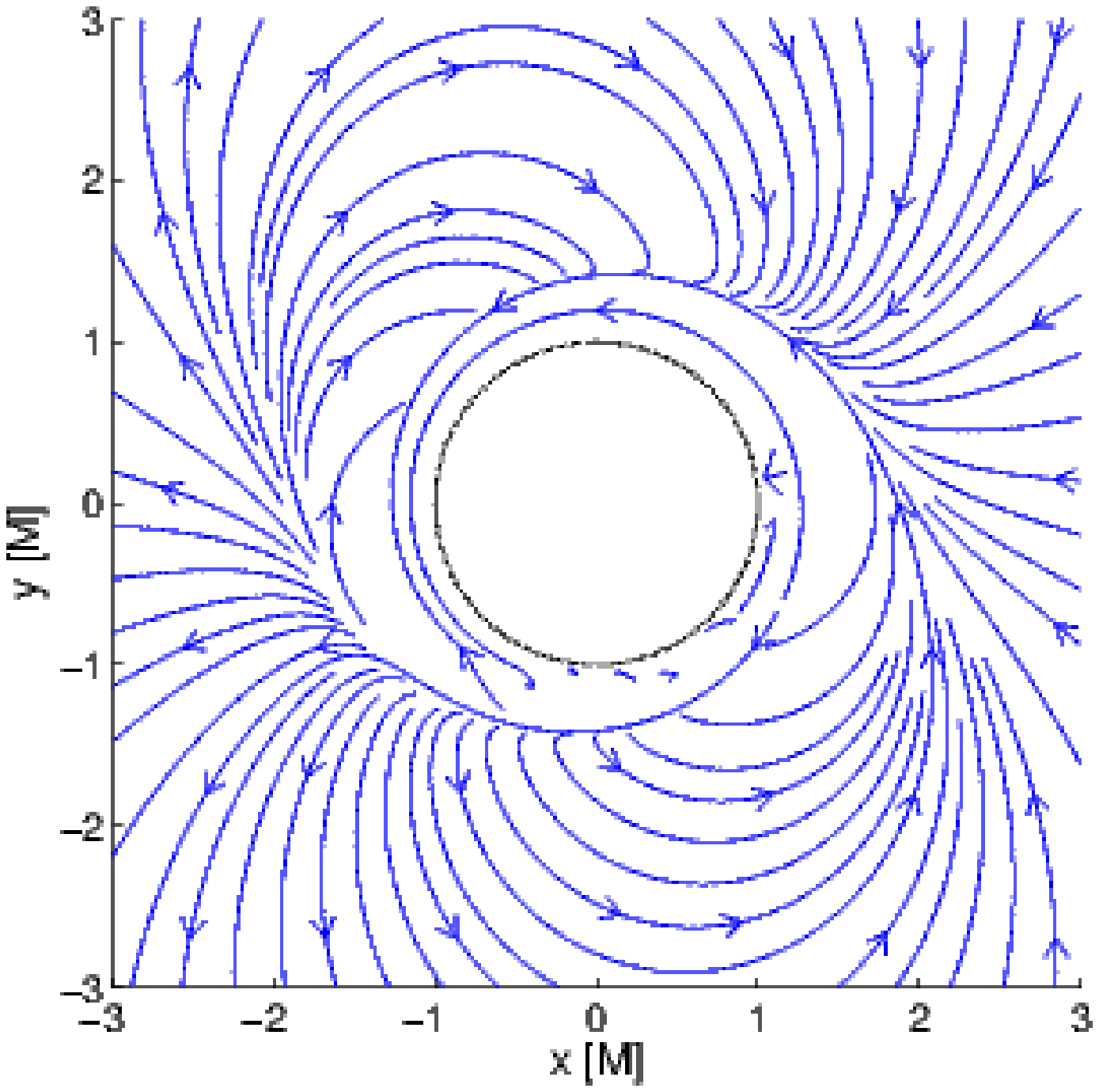}
\includegraphics[scale=0.29, clip]{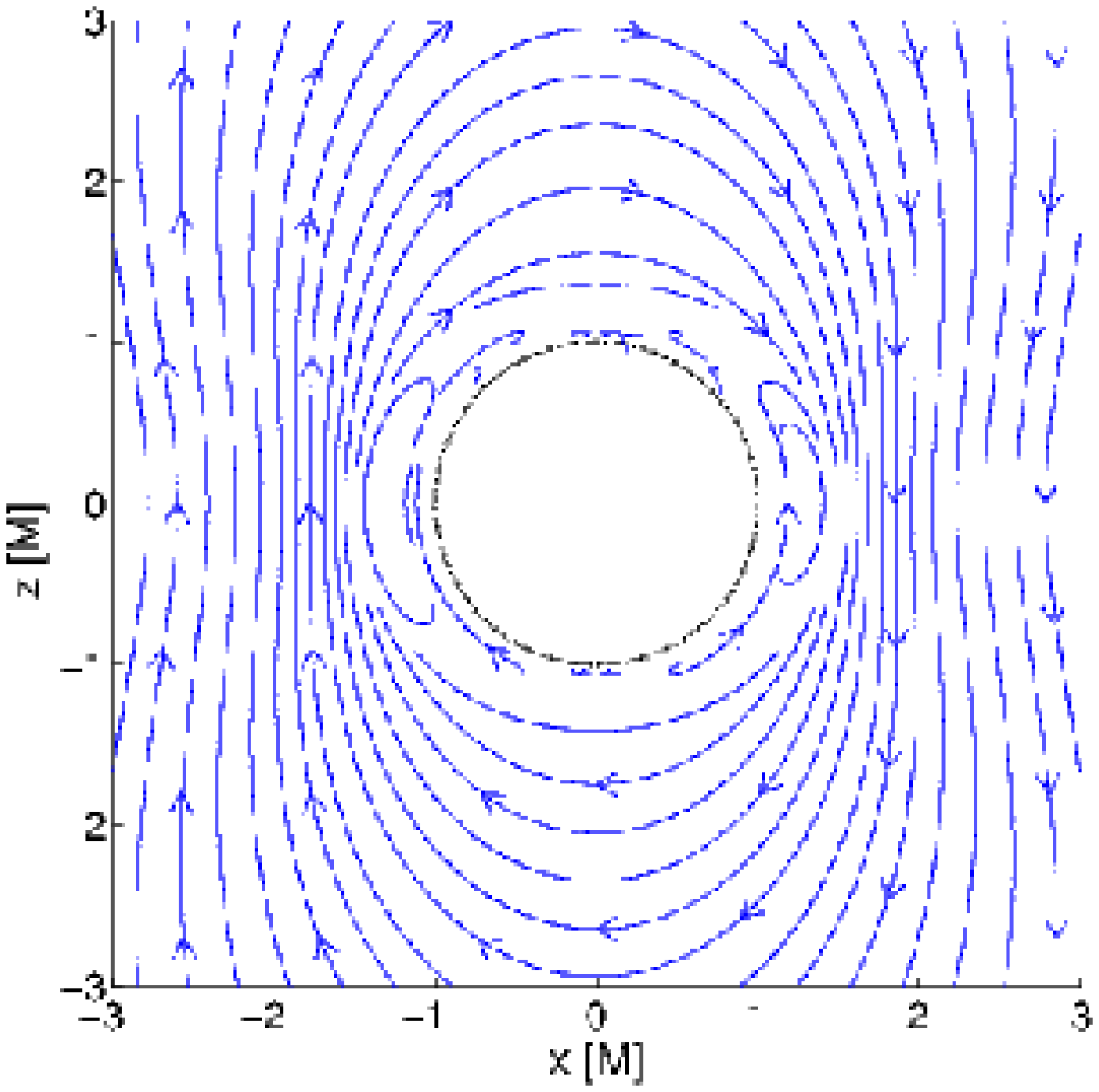}
\includegraphics[scale=0.29, clip]{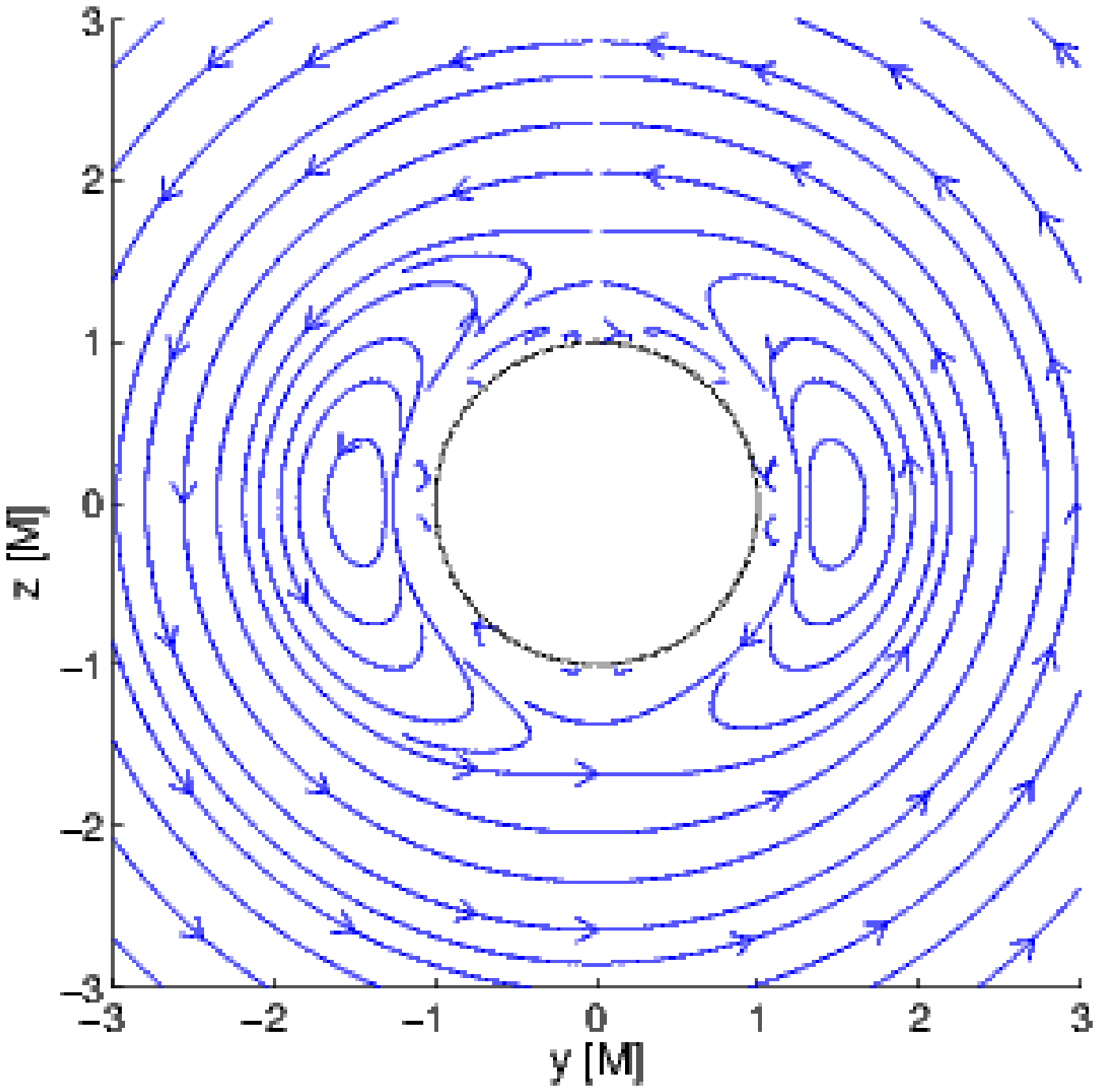}
\includegraphics[scale=0.29, clip]{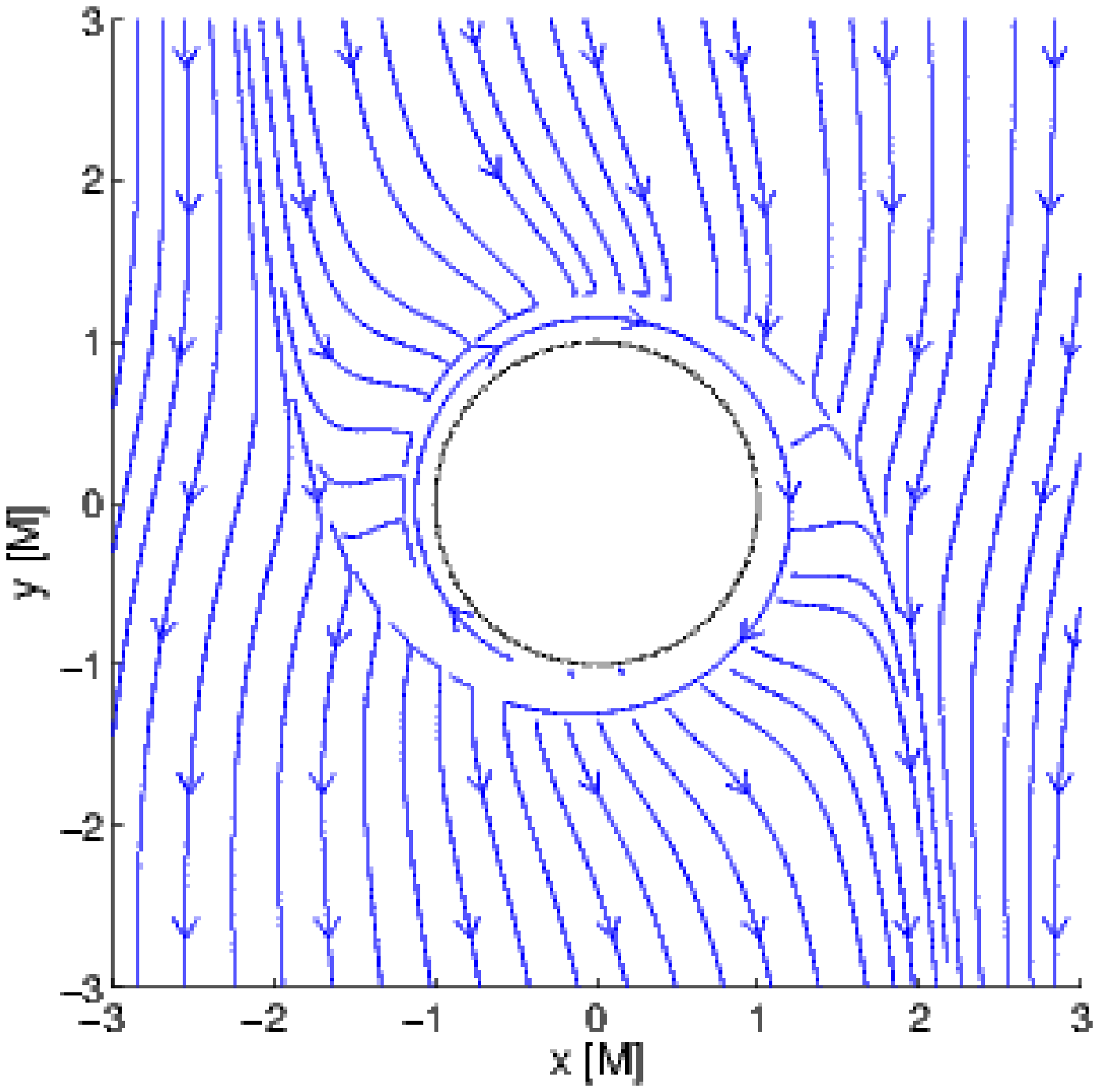}
\caption{Projections of the electric field onto the $(x,z)$, $(y,z)$ and $(x,y)$ planes in the case of the extreme Kerr BH with the asymptotic magnetic field. Top six panels capture the situation in the aligned setup, $B_x=0$ and $B_z=M^{-1}$. Due to the axial symmetry the poloidal projections $(x,z)$ and $(y,z)$ coincide in this case. We compare ZAMO tetrad components in the first row with the FFOFI tetrad components in the second one. FFOFI differs from ZAMO  most strikingly in measuring nonzero azimuthal components $E^{(\varphi)}$. Second series of six panels represent the situation of transversal orientation of the magnetic field, $B_x=M^{-1}$ and $B_z=0$. Axial symmetry is lost, projections onto different poloidal planes generally differ. We compare ZAMO tetrad components in the third row with the FFOFI tetrad components in the fourth one.}
\label{el_a1_Bx0_Bz1}
\end{figure}

\begin{figure}[hp]
\centering
\includegraphics[scale=0.29, clip]{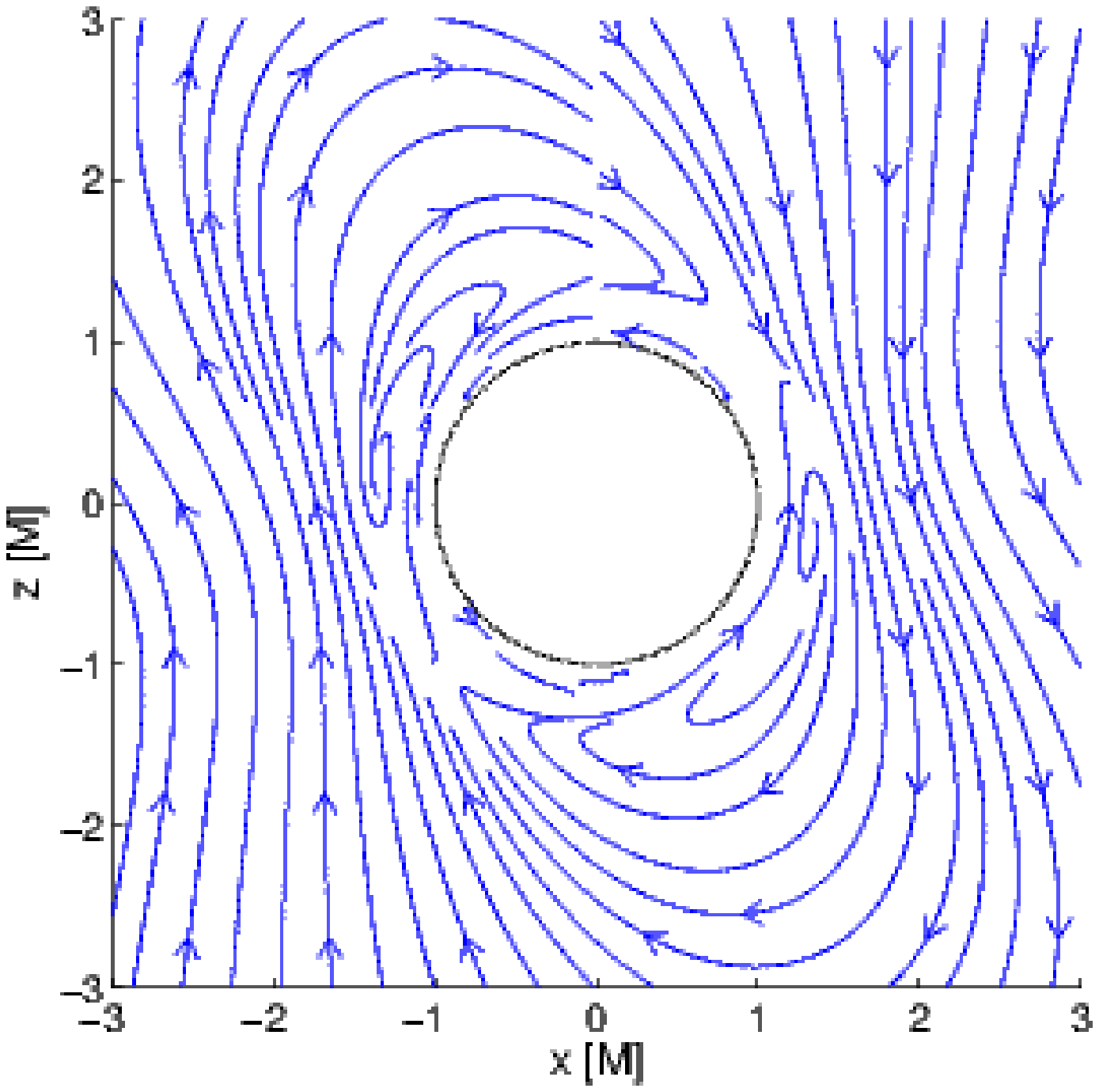}
\includegraphics[scale=0.29, clip]{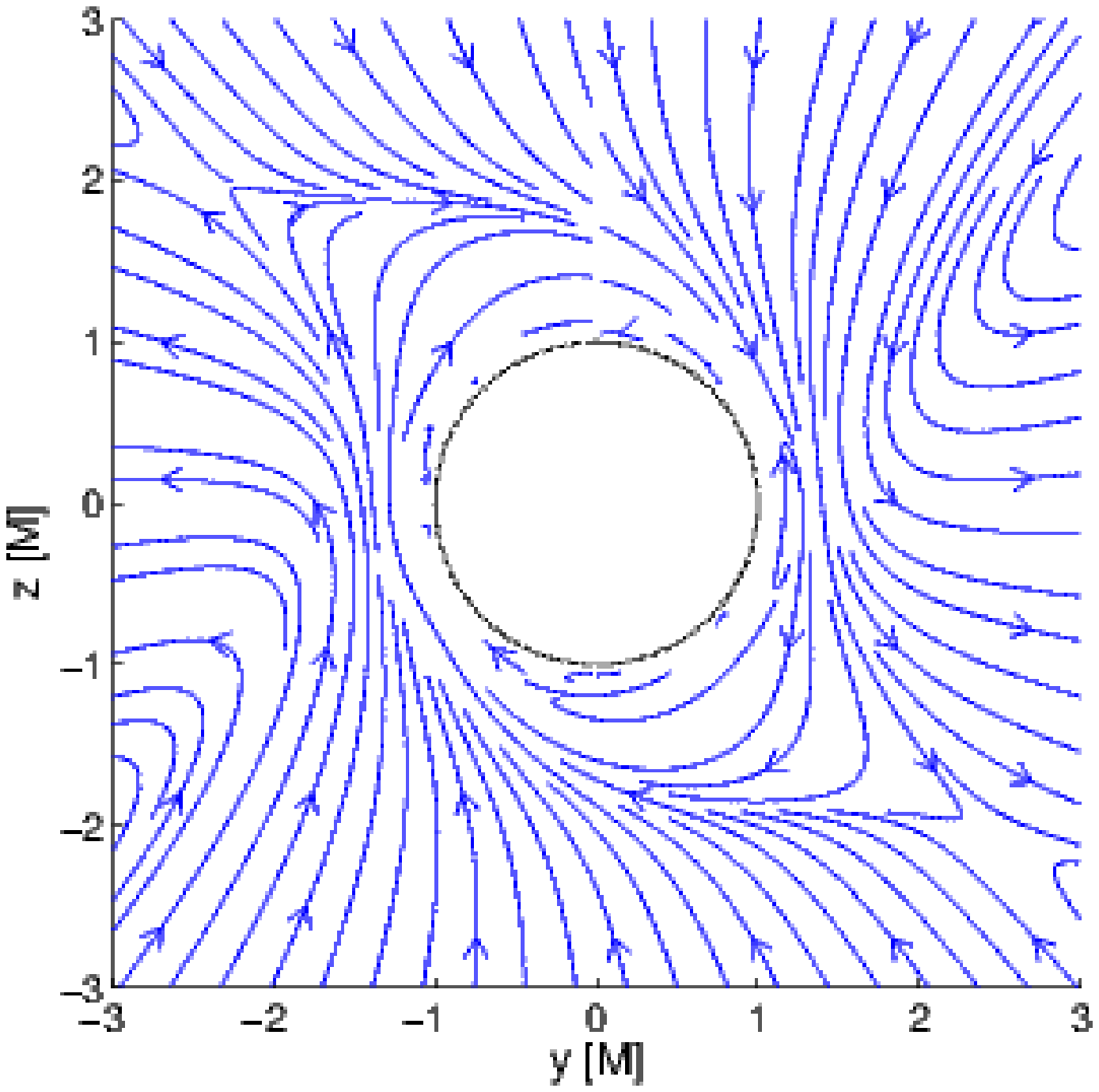}
\includegraphics[scale=0.29, clip]{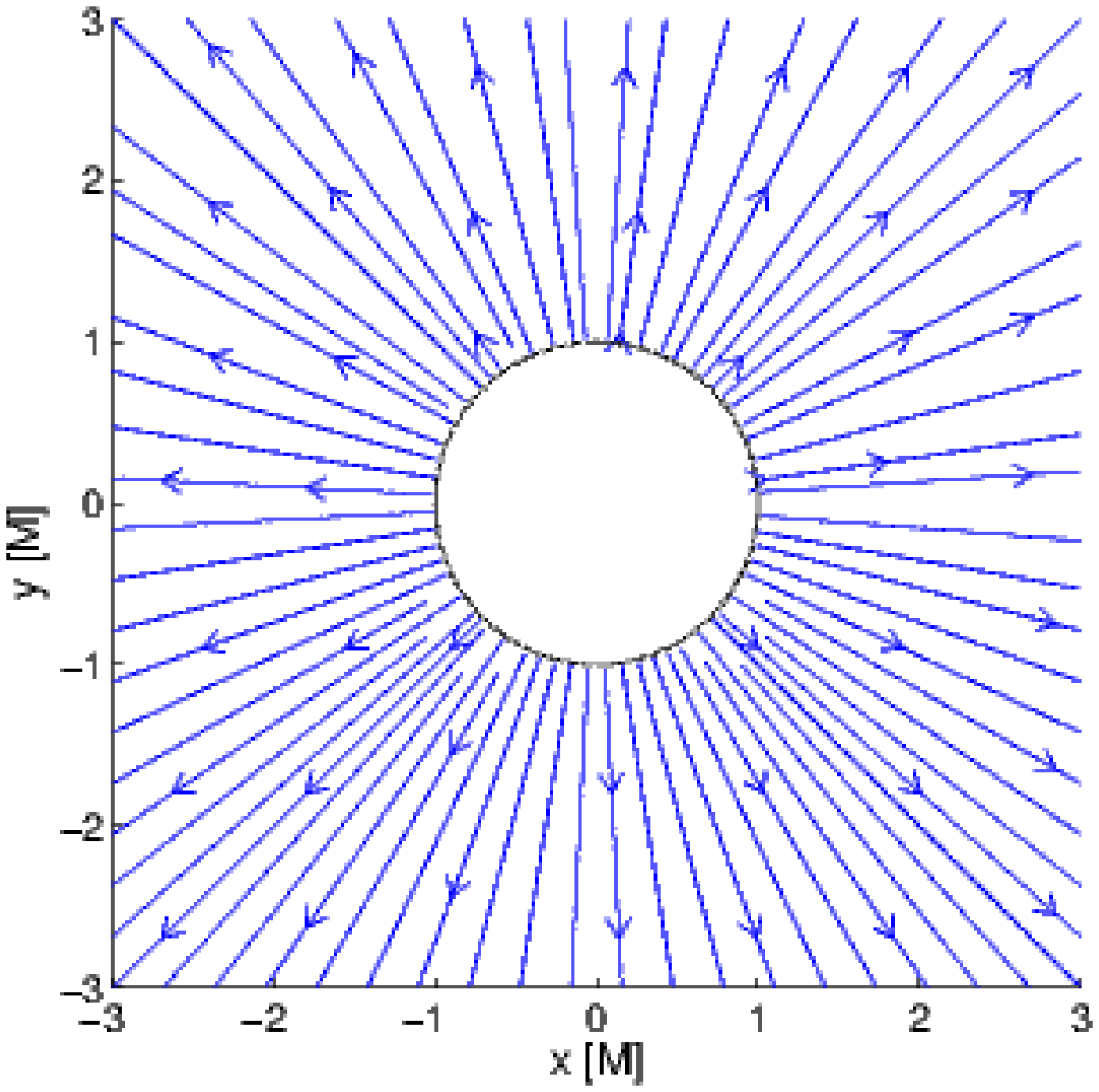}
\includegraphics[scale=0.29, clip]{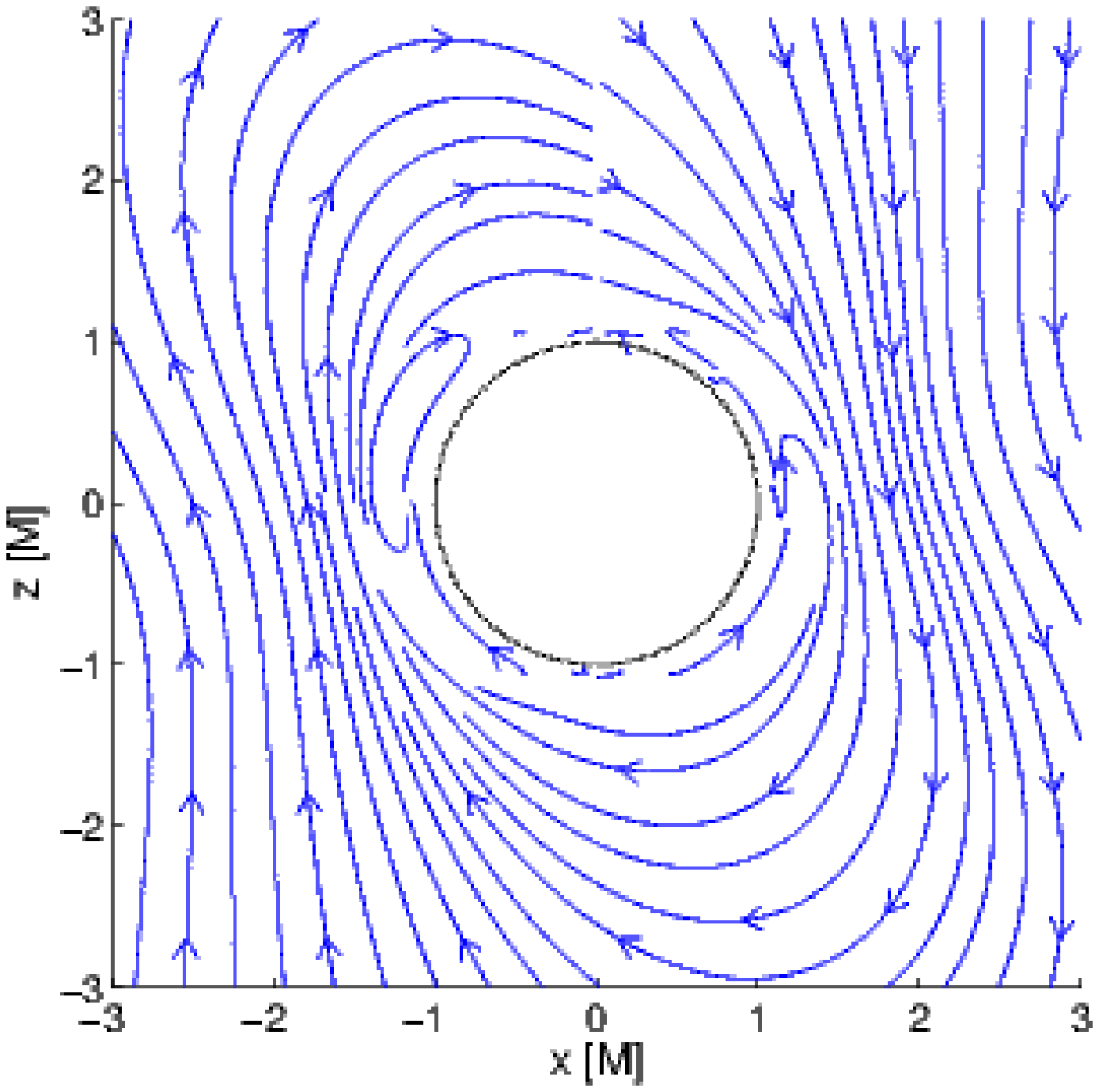}
\includegraphics[scale=0.29, clip]{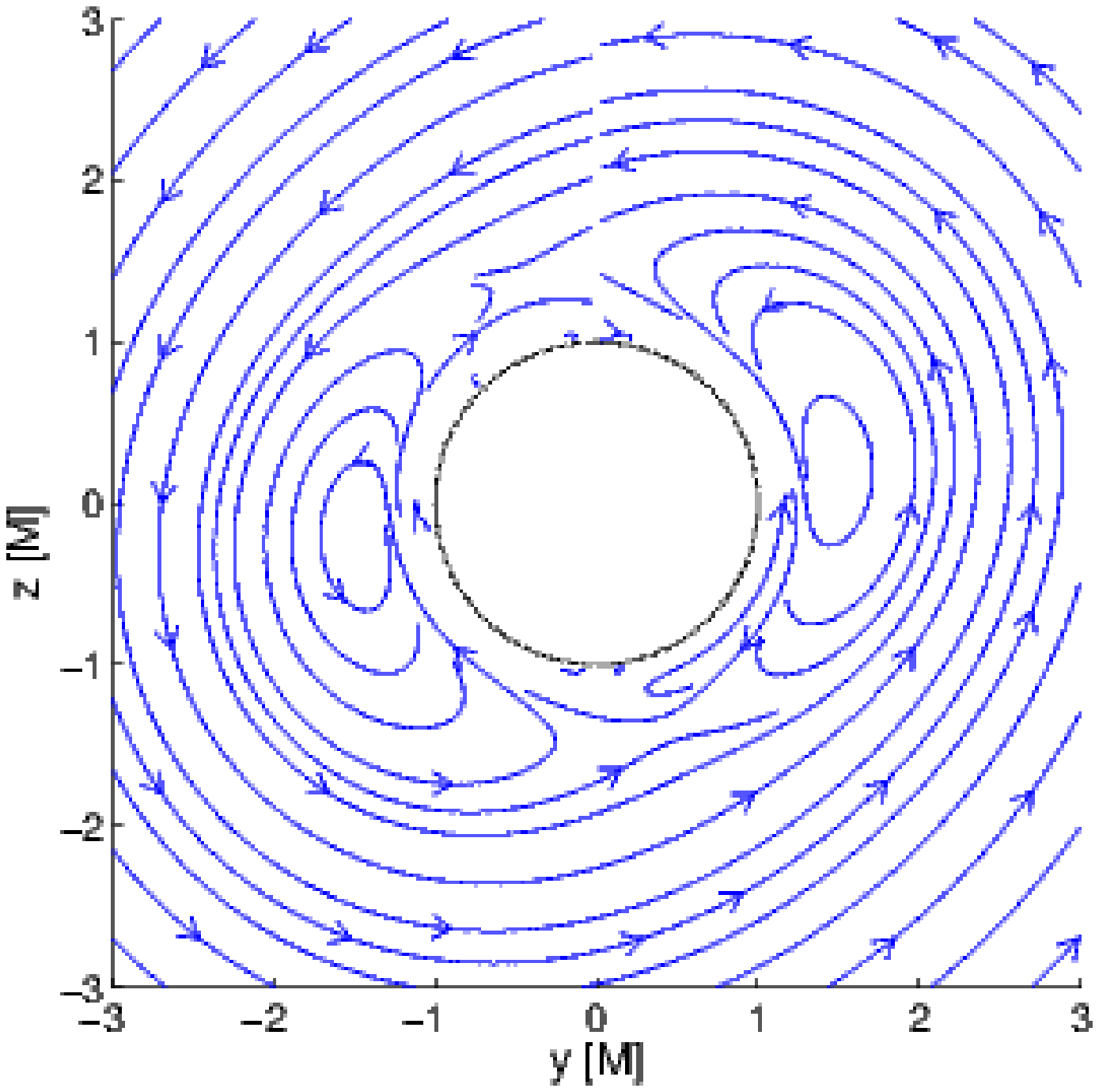}
\includegraphics[scale=0.29, clip]{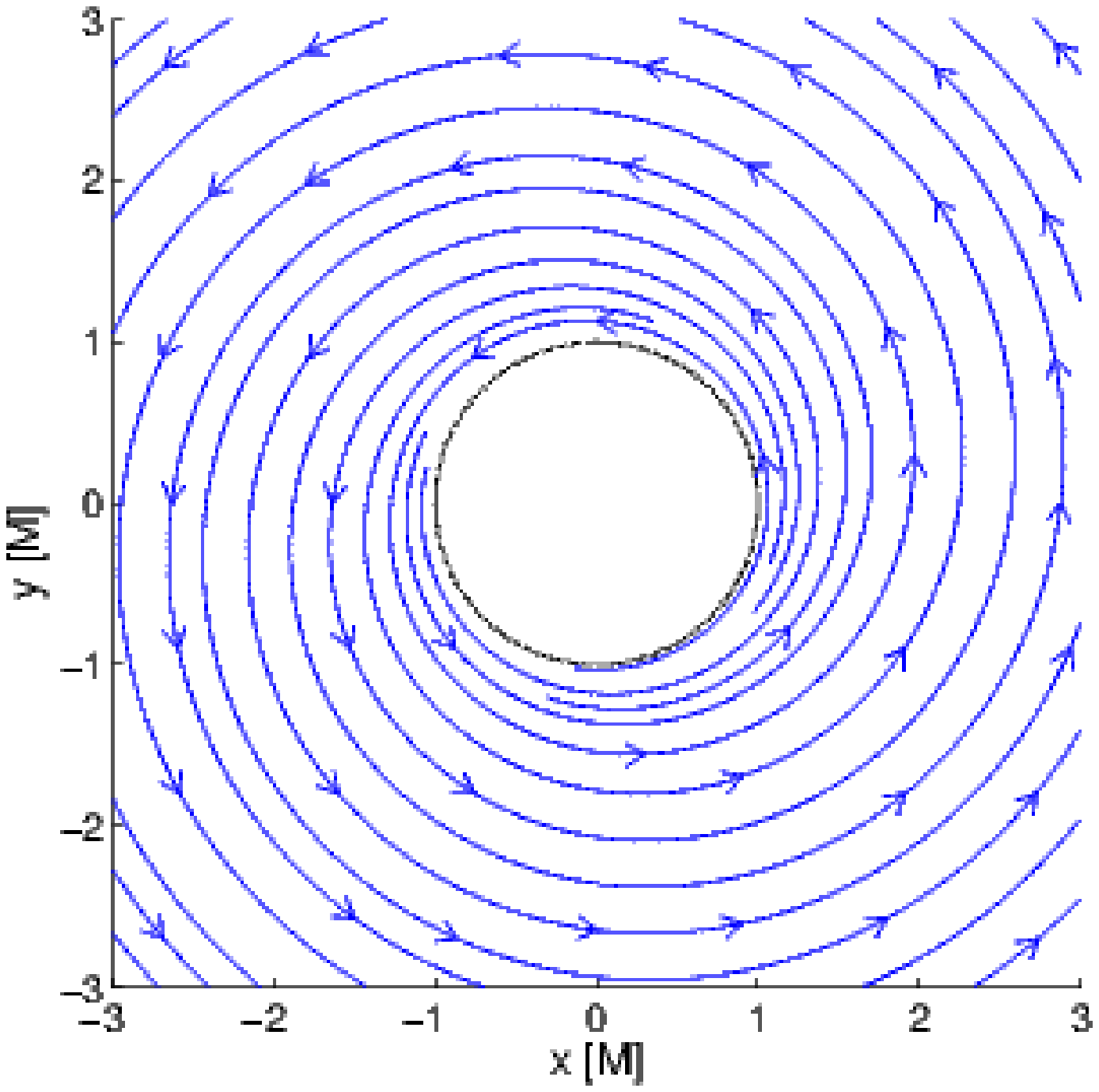}
\caption{Projections of the electric field onto the $(x,z)$, $(y,z)$ and $(x,y)$ planes in the case of the extreme Kerr BH with the asymptotic magnetic field components $B_x=M^{-1}$ and $B_z=M^{-1}$. We compare ZAMO tetrad components in the upper panels with the FFOFI tetrad components in the bottom ones. ZAMO measures azimuthal field $E^{(\varphi)}$ outside the equatorial plane only, see \rft{el_pole} for the summary.}
\label{el_a1_Bx1_Bz1}
\end{figure}

In the series of plane sections in figs. \ref{el_a1_Bx0_Bz1} and \ref{el_a1_Bx1_Bz1} we compare the structures of the electric fields as measured by ZAMO and FFOFI observers. First we present the situation of aligned field ($B_x=0$) in which all poloidal sections of the fields coincide due to the symmetry (top panels of \rff{el_a1_Bx0_Bz1}). We notice that in the FFOFI tetrad (unlike the ZAMO tetrad) the electric field is expelled from the horizon of the extreme BH. Once the non-aligned field is considered (by setting $B_x\ne 0$) the symmetry is lost and poloidal sections for different values of $\varphi$ generally differ (see the bottom panels of  \rff{el_a1_Bx0_Bz1} and \rff{el_a1_Bx1_Bz1}). The effect of electric expulsion is gone once the non-aligned field is considered.

Stereometric projections in \rff{el_meissner_3d} reveal that in the renormalized components the expulsion of the electric field out of the horizon of the extreme Kerr is present not only for the FFOFI test charge but also for the test charge with ZAMO four-velocity (while in the ZAMO tetrad the field lines are not expelled). Azimuthal component $E^{\varphi}$ which is present in the FFOFI case causes winding of the field lines while the ZAMO field lines remain confined to a given poloidal sections. Bottom panels of \rff{el_meissner_3d} illustrate that setting $B_x\ne 0$ generally disrupts the symmetry and allows the field lines to penetrate the horizon regardless the spin value. 

The expulsion of the electric field as observed in the FFOFI tetrad is further studied in \rff{vytl_el} where we provide series of poloidal plane sections for the aligned background field which differ in the spin value. Such discussion allows us to visually reveal the mechanism of the expulsion which takes place in the case of the extreme spin value. LIC patterns are also employed to help to reveal this effect.


Electric field is generally very sensitive to the value of spin parameter which is not that surprising since the field itself has gravitomagnetic origin by which we mean that it arises from the interplay between the background magnetic field and spin induced frame-dragging effect. As a consequence we measure no AMO electric field in the Schwarzschild limit since all $F_{ti}$ vanish with $a=0$. Naturally this does not necessarily mean that all observers will measure zero electric field. We have seen, for instance, that even in the case of zero spin the non-stationary FFOFI observer may observe nonzero electric field induced due to his radial velocity component $u^r$. Nevertheless these electric components asymptotically diminish since the observer is asymptotically static.

The question of the expulsion of the electric field has been addressed in the case of aligned background field. Unlike the magnetic expulsion where the radial AMO component $B^r_{\rm{AMO}}=\frac{F^{Bz}_{\theta\varphi}}{r^2\sin\theta}$ vanishes at the horizon for $a=M$ here the analogical component $E^r_{\rm{AMO}}=F^{Bz}_{rt}$ does not vanish. Thus the electric expulsion is merely an observer effect. We have seen that in the ZAMO tetrad the electric field was not expelled while in the FFOFI one it was. We recall that also the magnetic field is fully expelled in the FFOFI frame.

\begin{figure}[hp]
\centering
\includegraphics[scale=0.42, clip, trim= 40mm 5mm 49mm 15mm]{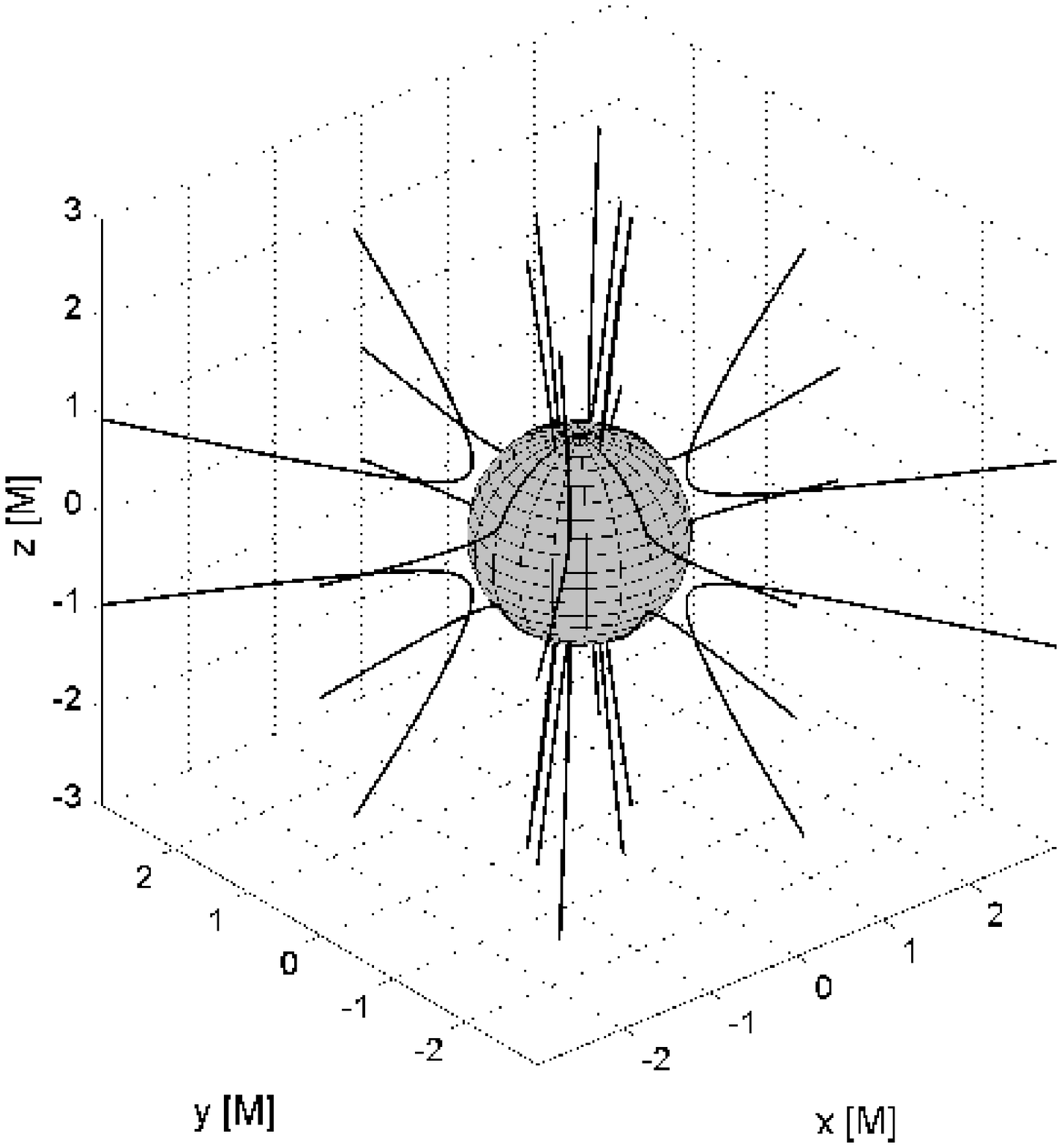}\includegraphics[scale=0.42, clip, trim= 50mm 5mm 50mm 15mm]{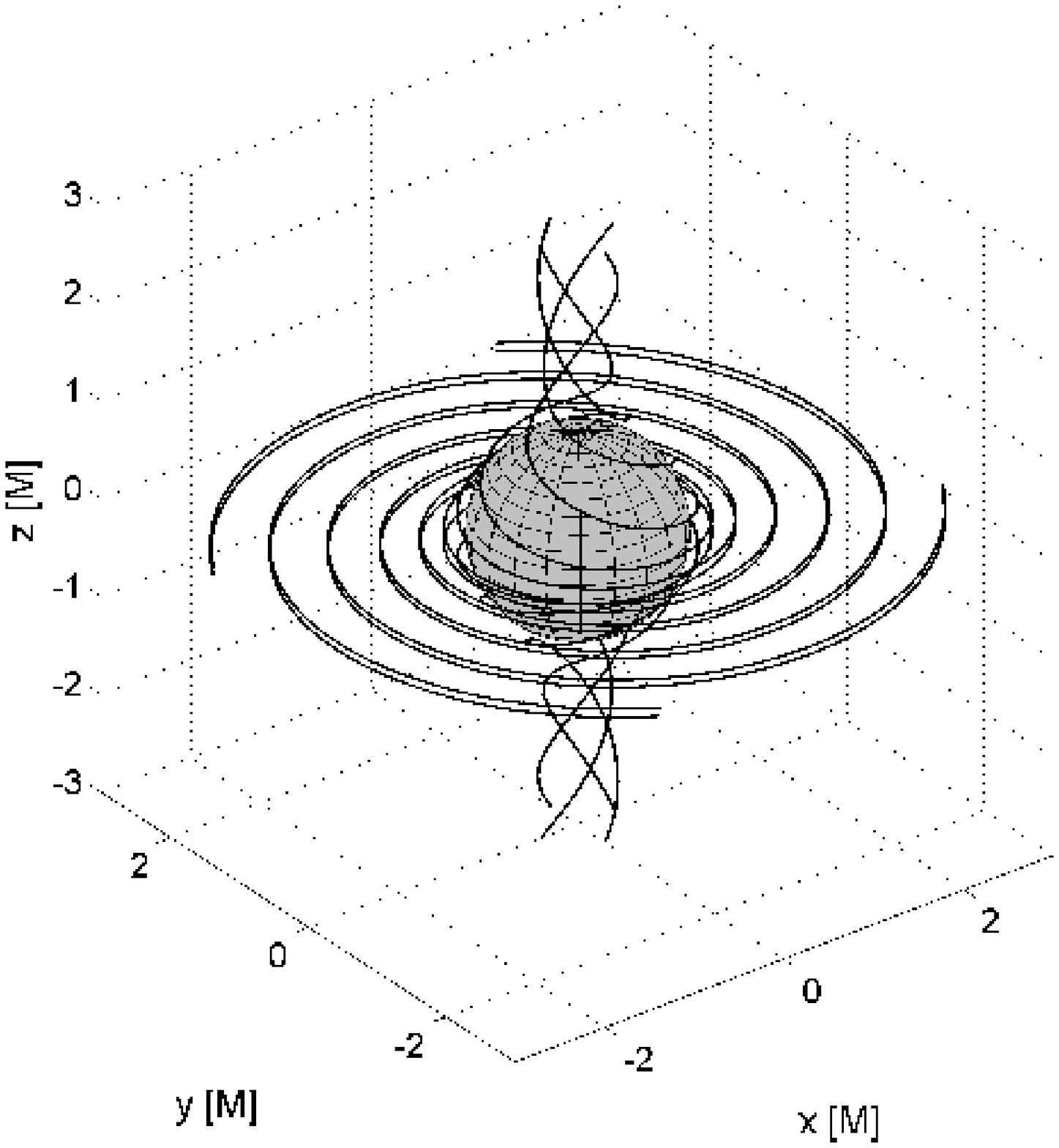}\\
\includegraphics[scale=0.45, clip, trim= 32mm 5mm 49mm 15mm]{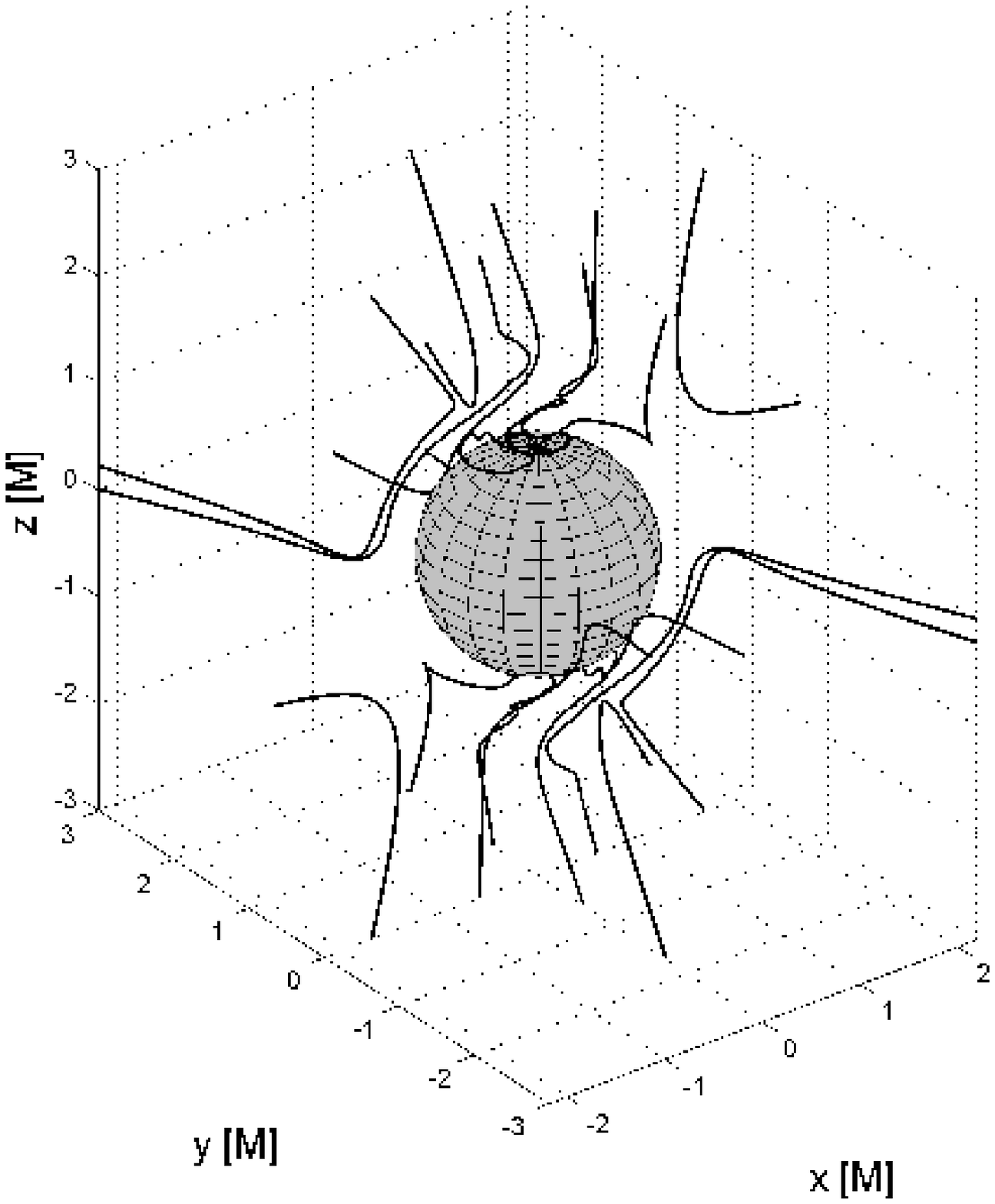}\includegraphics[scale=0.45, clip, trim= 37mm 5mm 50mm 15mm]{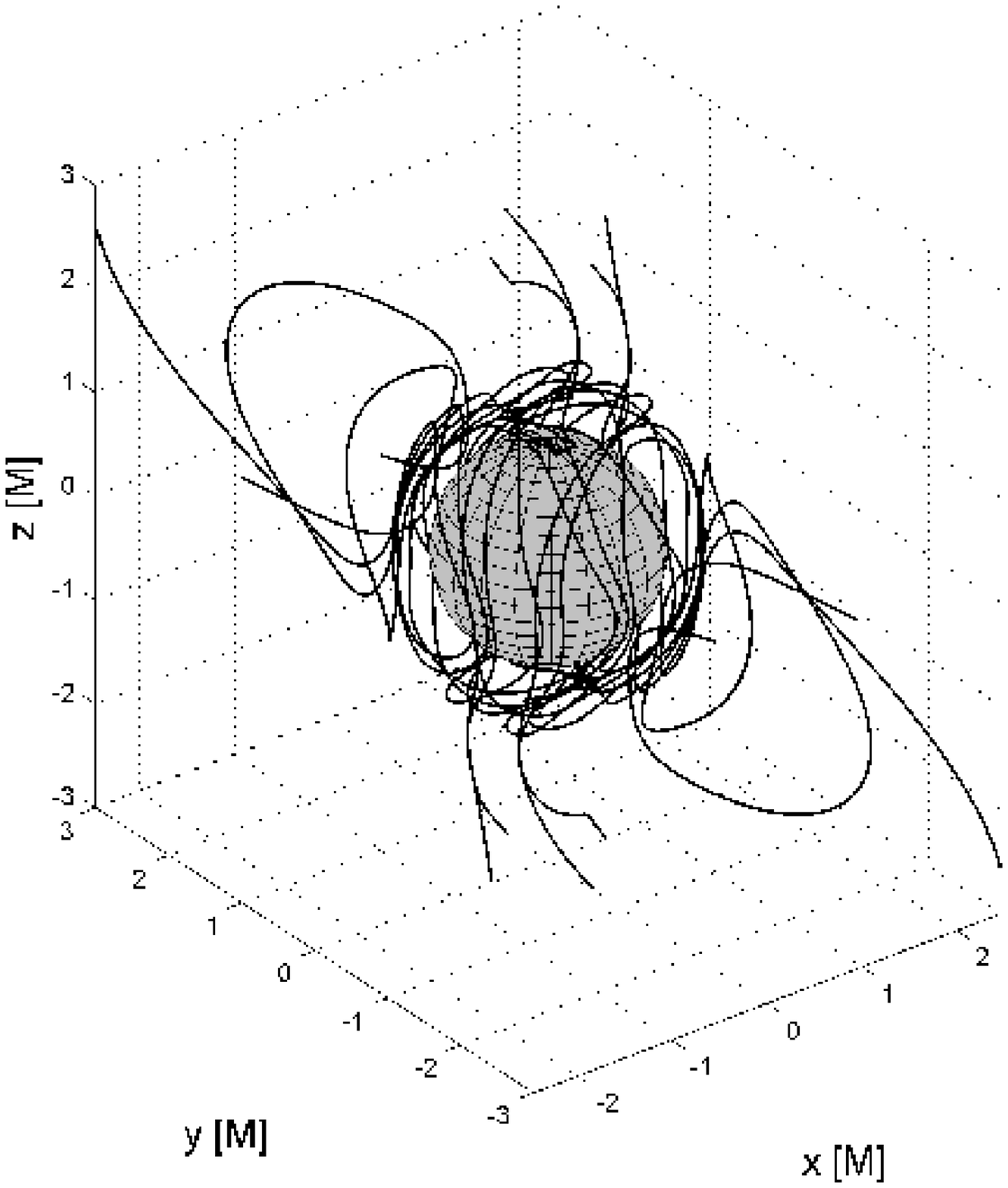}
\caption{Electric field is expelled from the horizon of the extreme Kerr black hole in both ZAMO and FFOFI renormalized components in the case of aligned magnetic field on the background ($B_x=0$). In the case of ZAMO (upper left panel) there is no azimuthal component while the FFOFI observer measures also azimuthal electric field causing winding of the field lines (upper right panel). However, considering $B_x\ne 0$ brings azimuthal electric component also for ZAMO as we observe in the bottom panels. The bottom left shows  the situation for $\frac{B_x}{B_z}=0.1$ while the right one for $\frac{B_x}{B_z}=0.3$. Electric field is not expelled in general, some field lines penetrate the horizon.}
\label{el_meissner_3d}
\end{figure}

\begin{figure}[hp]
\centering
\includegraphics[scale=0.38, clip]{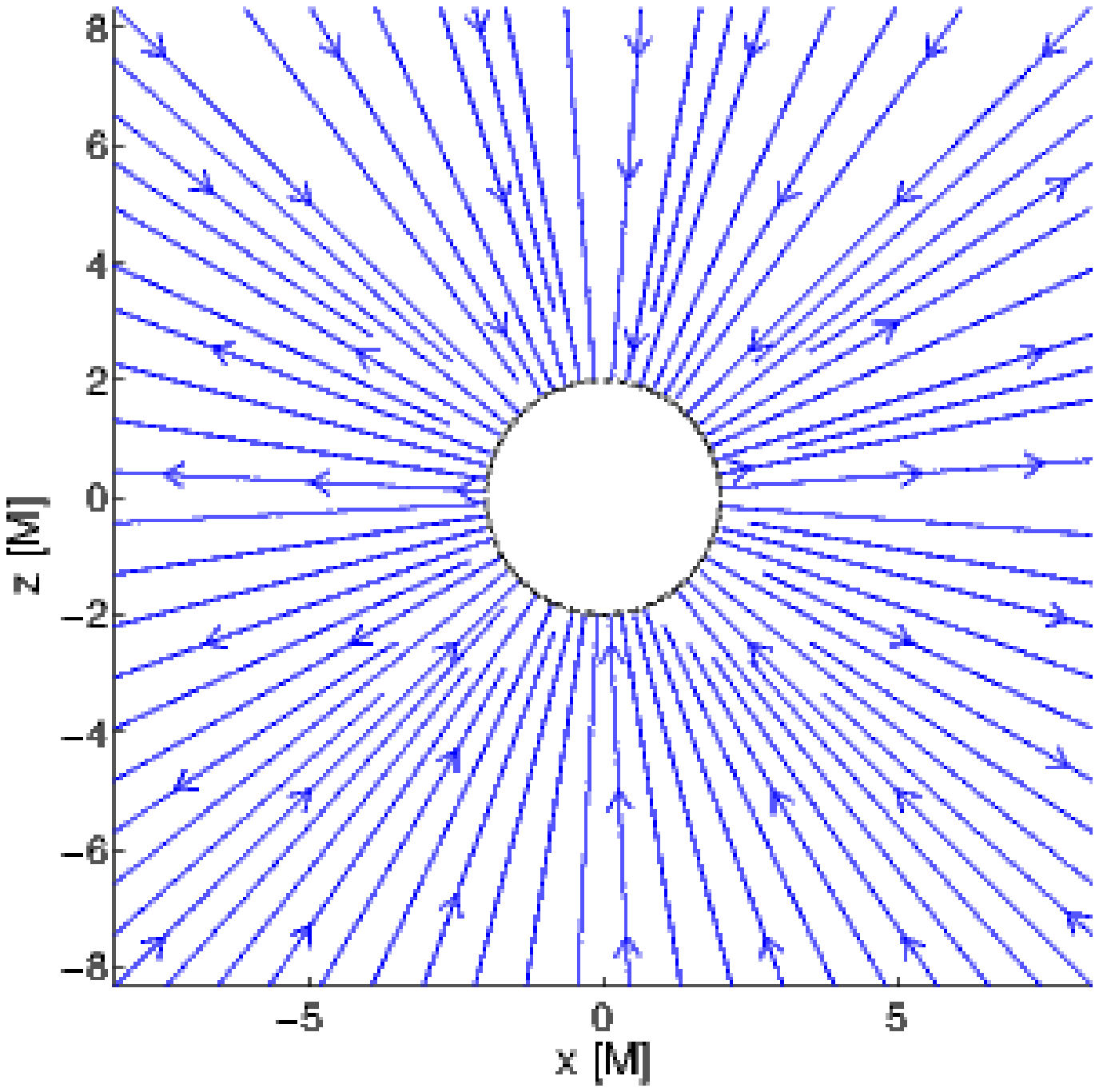}
\includegraphics[scale=0.38, clip]{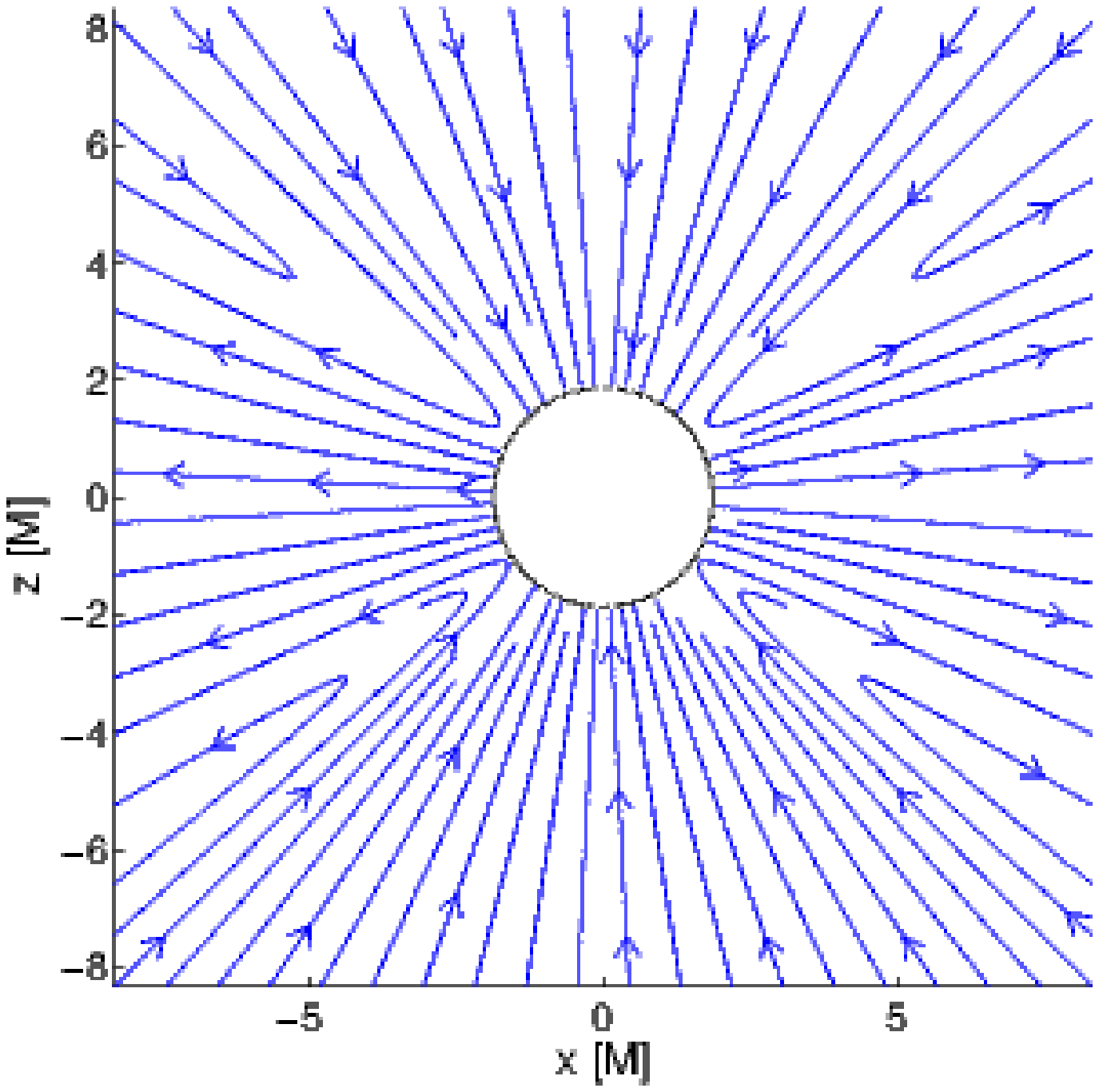}
\includegraphics[scale=0.38, clip]{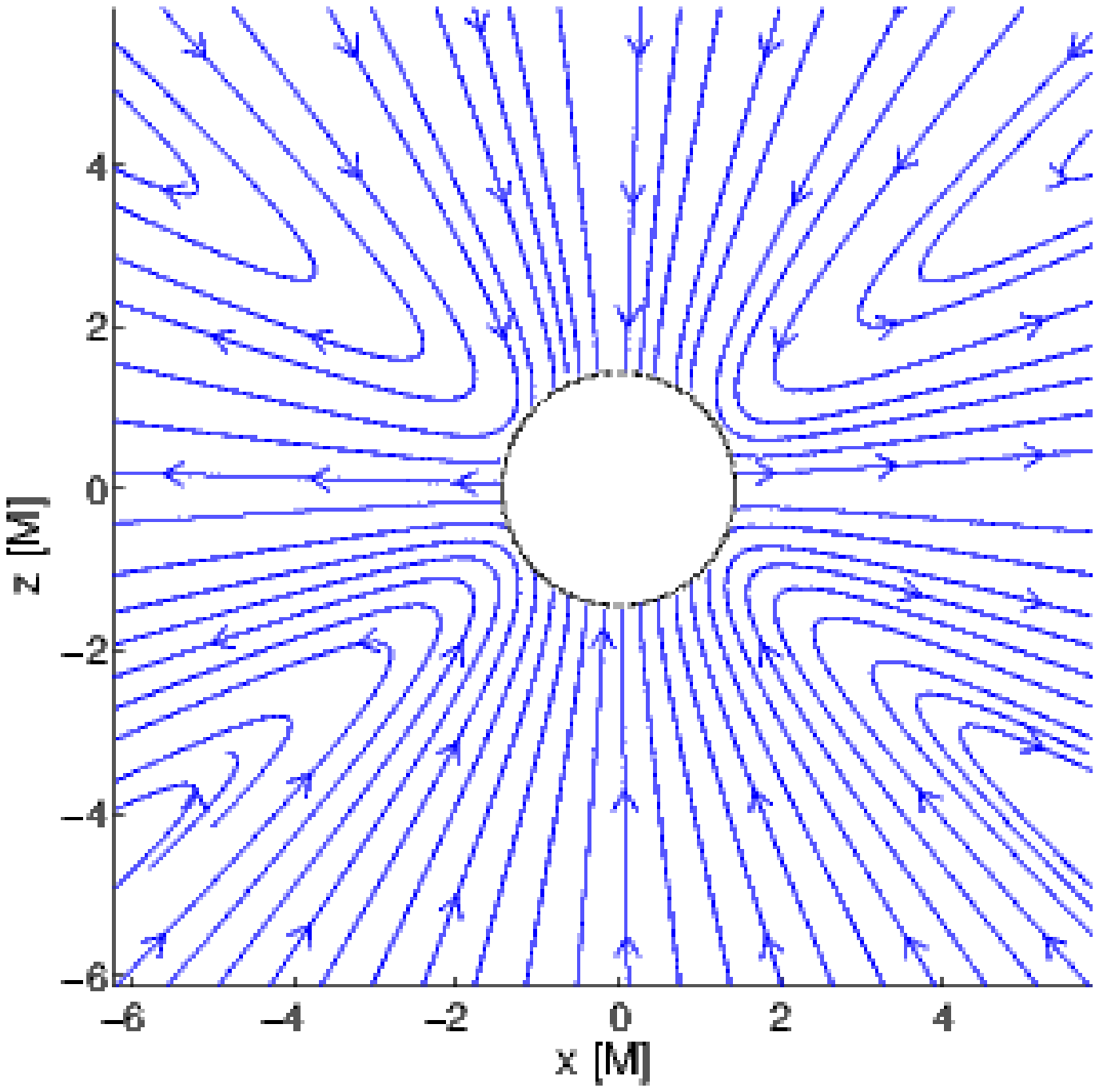}
\includegraphics[scale=0.38, clip]{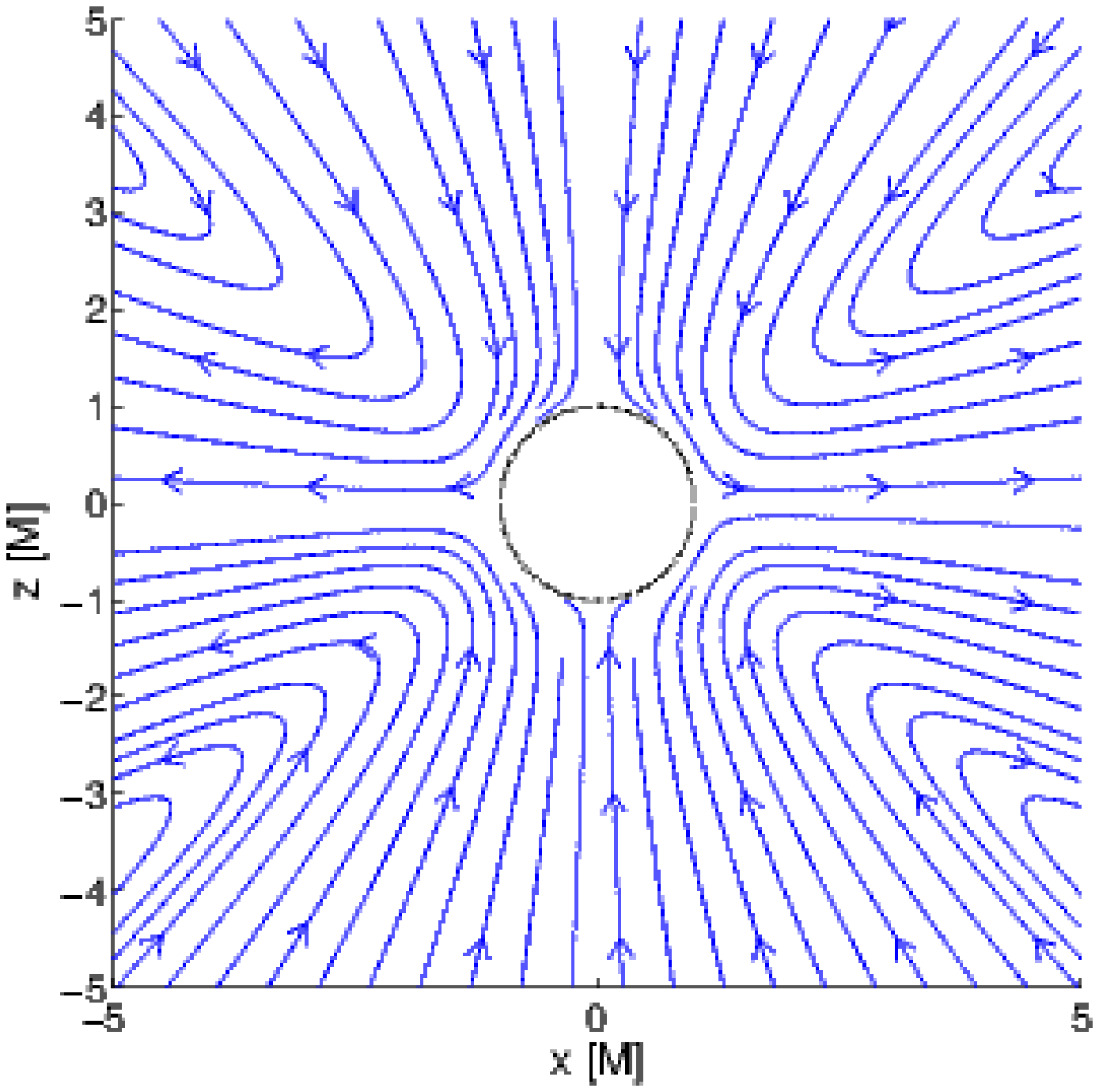}
\includegraphics[scale=0.3, clip]{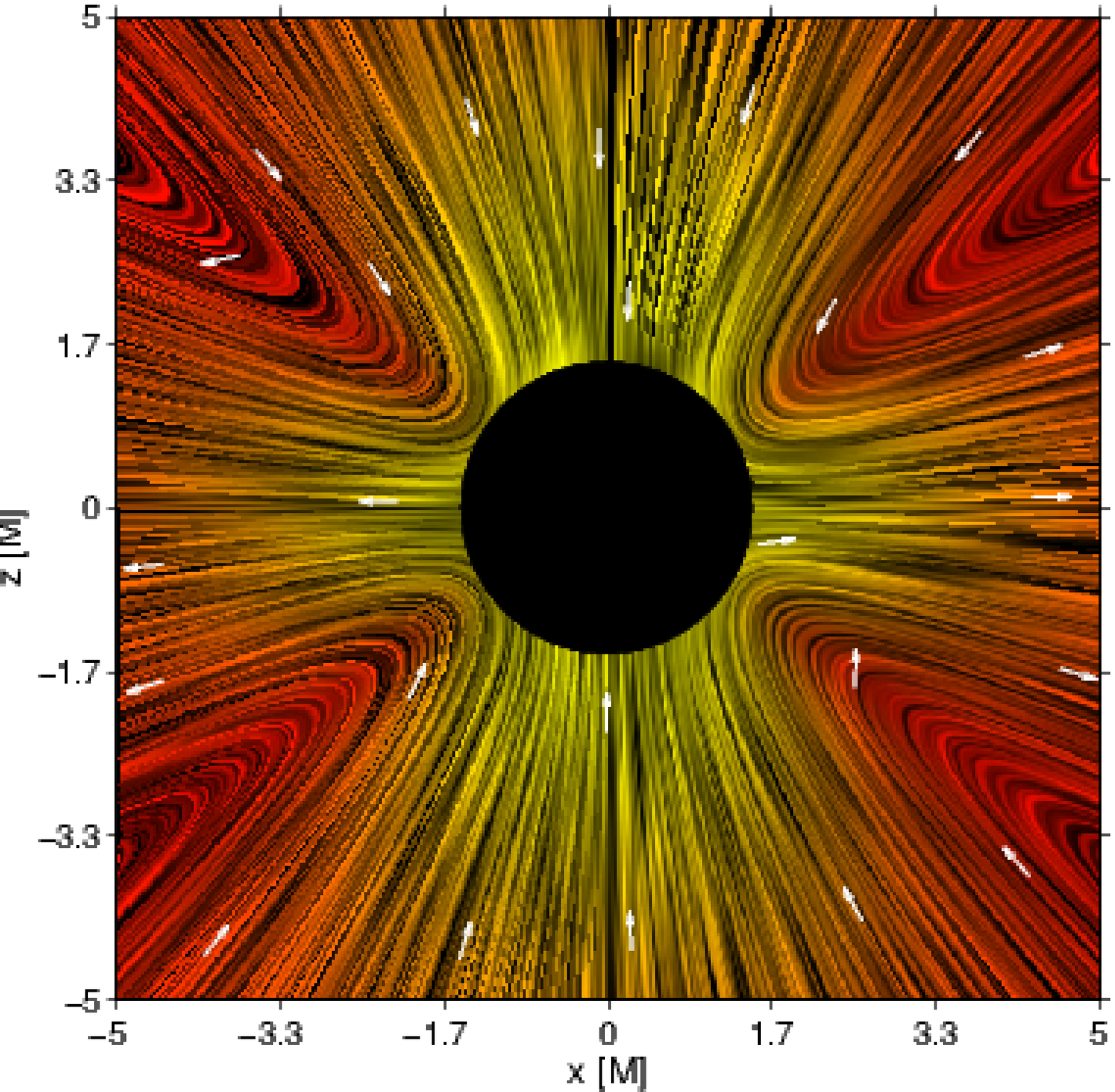}
\includegraphics[scale=0.3, clip,trim= 8mm 0mm 0mm 0mm]{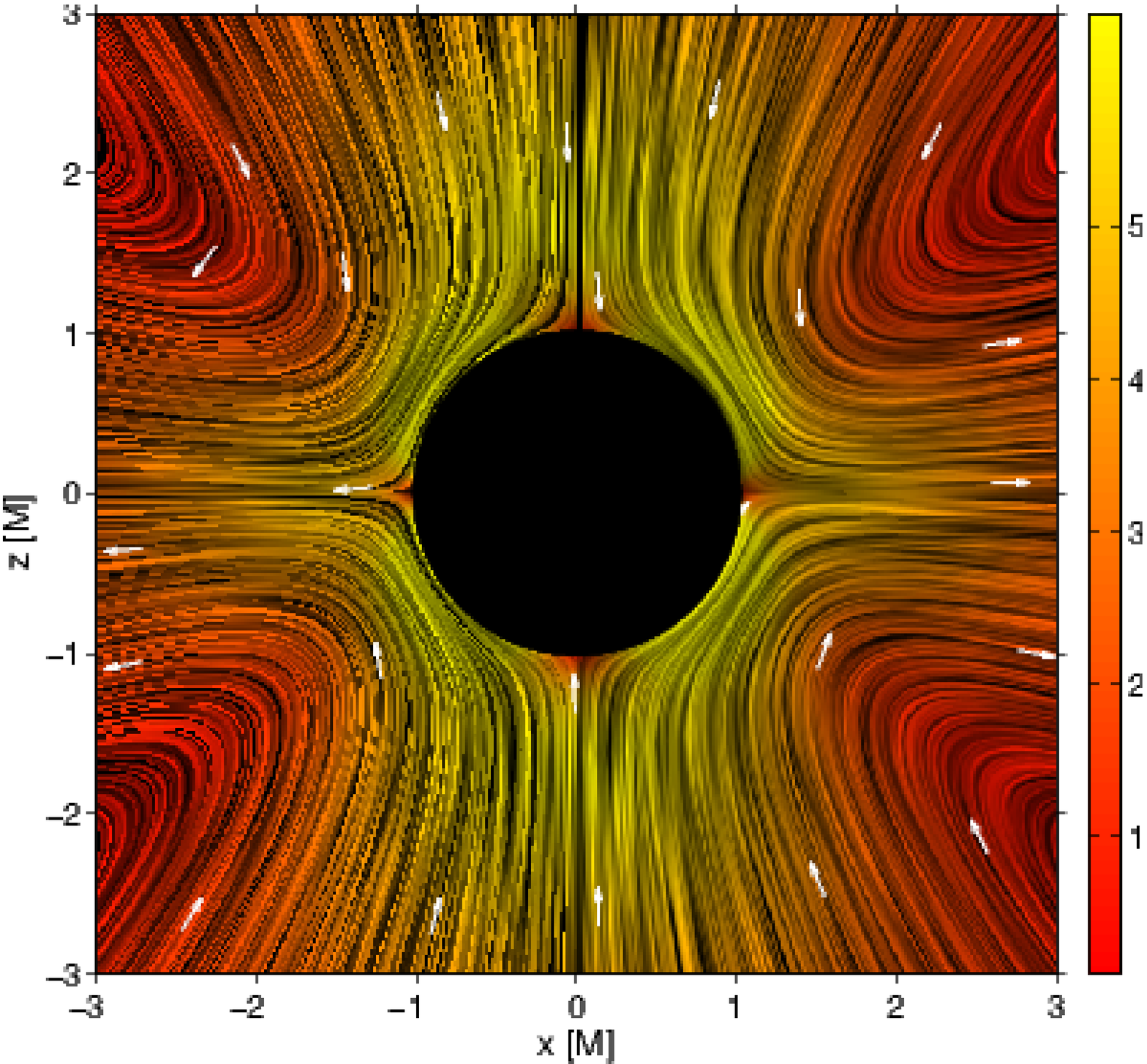}
\caption{Evolution of the FFOFI measured electric field in the case of aligned background magnetic field ($B_x=0$) when the spin is gradually increased is shown in the series of poloidal plane sections. We begin with the low value $a=0.1M$ (upper left panel) which is increased to the medium value $a=0.5M$ (upper right panel), high value $a=0.9M$ (middle left) and eventually to the extreme value $a=M$ in the middle right panel. We observe the effect of the expulsion of the electric field in the extreme case. Bottom panels show LIC patterns reflecting the expulsion. Bottom left panel shows $a=0.9M$ case with only partial expulsion, the right one captures the complete expulsion occurring for $a=M$. Color bar suggests the strength scale of the field (units are arbitrary).}
\label{vytl_el}
\end{figure}

\newpage
\subsection{Electromagnetic field around drifting black hole}
\label{driftingEM}
In this section we shall discuss the electromagnetic fields around Kerr black hole which is drifting in a general direction through the oblique uniform magnetic field. Components of EM tensor $F_{\mu\nu}$ describing such field were derived in \rs{emfield} in terms of Lorentz transformation of former non-drifting solution given by eqs. \ref{BFtr(Bx)}, \ref{BFtr(Bz)}.

Resulting system provides highly simplified though not unrealistic model of conditions occurring in a diluted gaseous medium in the vicinity of the inner edge of the accretion disk. This region may be threaded by the external large-scale magnetic field which we model by oblique uniform field in our setup. Gravitomagnetic effect acts on the field enriching profoundly its structure as we have seen in previous section. Observer dependence of the EM field components has also been discussed previously. In addition to these effects we introduce the translational uniform motion of the black hole itself bringing extra parameters to the system. See \citet{karas09} for a discussion of astrophysical relevance of our model.

Lorentz transformation of the field changes its asymptotical form -- in general it changes the direction of the former magnetic field and induces electric field which was not present in the asymptotic region of non-drifting BH (for asymptotically static observers at least). In the previous section we observed that interaction between BH's rotation and perpendicular component of the magnetic field causes complex twisting of the field lines and that a narrow zone of magnetic layers emerges above the horizon regardless the choice of the observer. 

In \rff{expul_drift} we introduce the drift in the case of extreme BH embedded in the aligned field in order to detect the drift impact upon the field most clearly. We observe that the Meissner effect is suppressed by the drift. We note that with the drift velocity increasing the shock front develops in which the field has complex layered structure which gradually enhances as the velocity rises. Besides the magnetic layers we newly observe the pair of magnetic null points which emerge for the sufficiently rapid drift. Sites of zero magnetic intensity leave the charged particles prone to the acceleration by the electric field which is generally not vanishing here. The layered field structures which surround the neutral point present the sites of possible magnetic reconnection \citep[][and references therein]{karas09}. Magnetic layers and neutral points are further explored in \rff{mag_drift_LIC} using LIC patterns with the strength of the field being encoded by the color scale.

From \rff{ml_nodrift} we recall that magnetic layers emerge also in the non-drifting case as a consequence of a perpendicular $B_x$ component of the field. Layering accompanied with the drift through the aligned field thus may be attributed to the tilt arising from the Lorentz transformation. The neutral points, however, only appear in the drifting case.

In stereometric projections \rff{mag_drift_3d} we compare magnetic field measured in the FFOFI frame to that captured by the ZAMO renormalized components. In both the cases the aligned non-drifting  field above the extremal BH is fully expelled, the only difference being that the FFOFI field is slightly azimuthally twisted (see \rff{meissner_3d_AMO}). When the drift is introduced, however, the behaviour of the field lines differs strikingly as they are dragged in the opposite direction. FFOFI lines stretch in the drift direction in an intuitive way while in the ZAMO renormalized components they are dragged backwards manifesting profound differences between the two field definitions. 

Electric field also reacts markedly when the drift is introduced. In \rff{el_drift} we explore its impact upon the FFOFI electric components above the horizon of the extremal BH in the poloidal $(x,z)$ plane section. Gradually increasing the drift velocity we observe how the topology of the field evolves. We note that without the drift the electric field asymptotically decays whereas the drift induces non-vanishing electric field which overruns the original one. Thus the global impact of the drift is more dramatic compared to what we observed in the case of magnetic field. Besides overall changes of the field we also notice progressive layering in the narrow region above the horizon. Similarly to their magnetic analogue, the electric layers are also present regardless of the choice of the observer and their origin may be attributed to the tilting of the originally aligned magnetic field due to the drift.

We also explore the electric field in the equatorial plane in \rff{el_drift_turtle}.  Since we start out from the aligned magnetic field and the drift is restricted to be perpendicular to the axis ($v_z=0$) the field lines will reside in the plane (see \rft{el_pole}).  The rescaled radial coordinate $R\equiv\frac{r-r_{+}}{r}$ is employed. We compare the tetrad components of ZAMO, FFOFI and both co-rotating and counter-rotating KEP+FFO. The drift changes the structure of the field completely including the asymptotic region. Although the fields differ from each other noticeably for different frames, in all the cases the neutral points are formed (at different location, though) once the sufficiently rapid drift is introduced.

We conclude that the translational motion of the black hole through the uniform magnetic field enriches profoundly the structure of the resulting electric and magnetic fields. We observed that in the case of axisymmetric background the drift causes a formation of a narrow zone in which the fields (both electric and magnetic) are heavily layered and the layering progressively enhances with the rising drift velocity. However, we observe such effect also for nondrifting oblique fields (see \rff{ml_nodrift}) and therefore we attribute the formation of the layers to the perpendicular component which arises from the field's tilt in the case of drift through the aligned magnetic field. Nevertheless only in the drifting scenario we can observe such topology of layers which leads to the formation of neutral points. These were observed for both, magnetic as well as electric fields.

\begin{figure}[hp!]
\centering
\includegraphics[scale=0.29, clip]{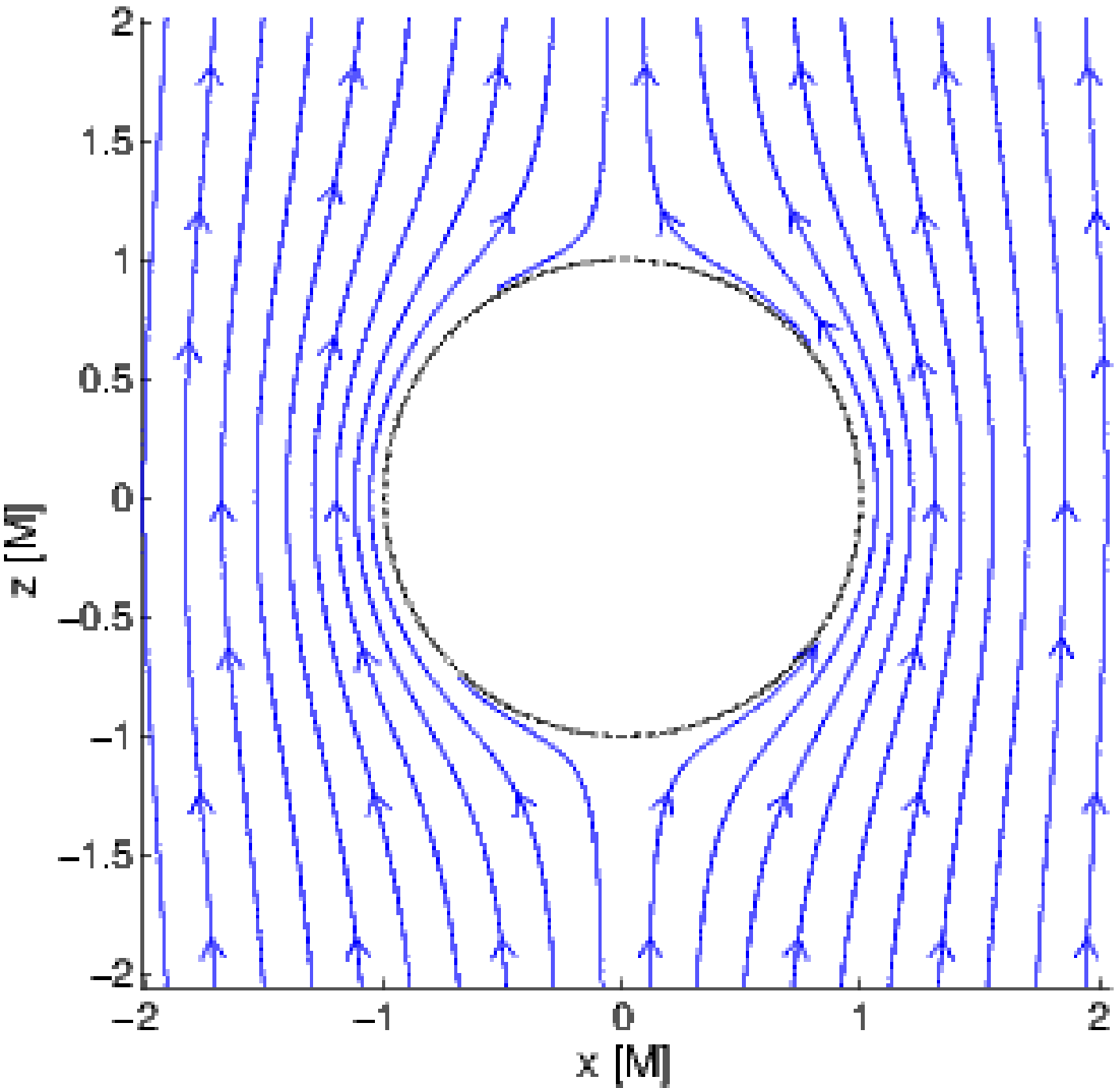}
\includegraphics[scale=0.29, clip]{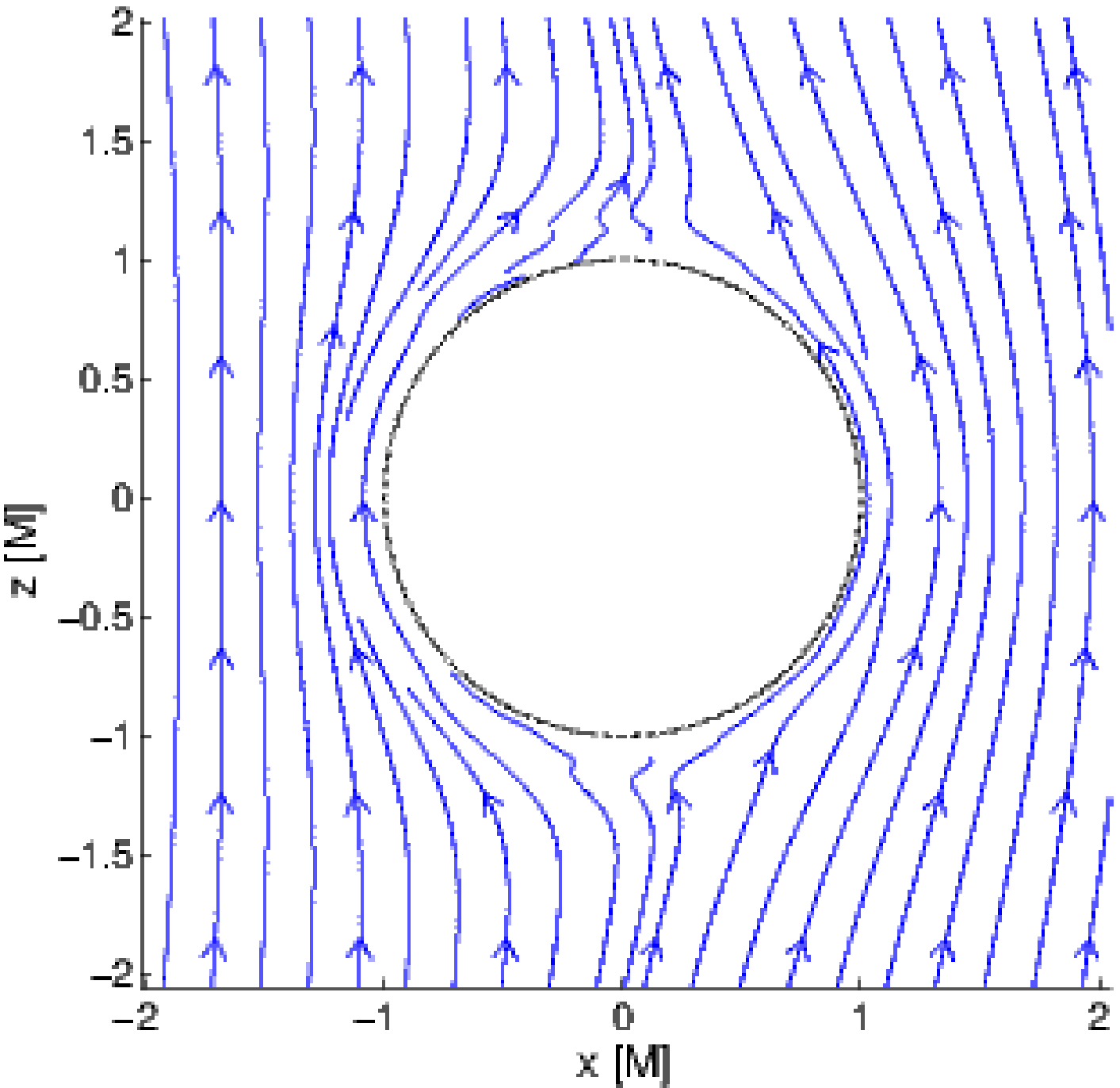}
\includegraphics[scale=0.29, clip]{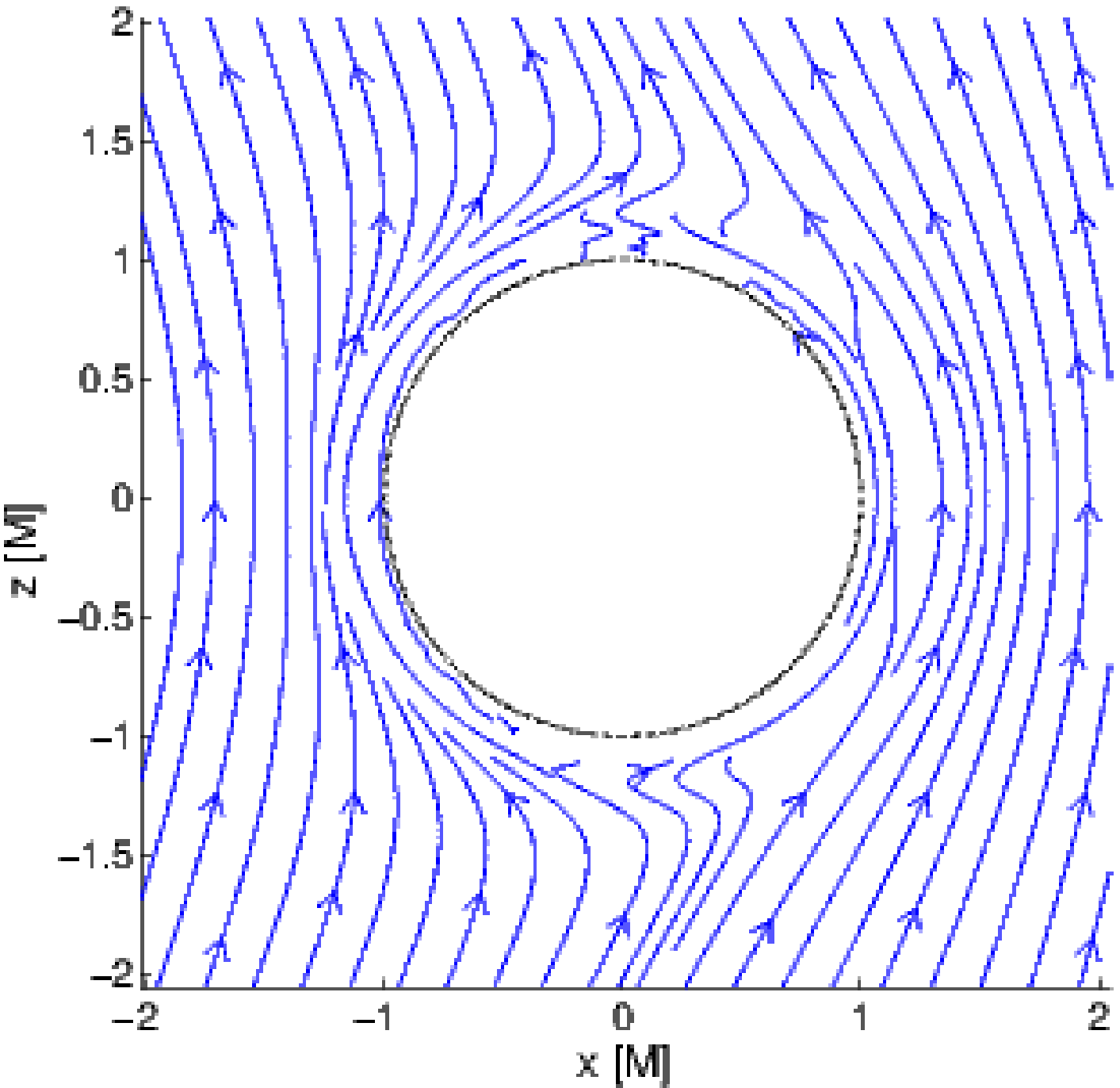}
\includegraphics[scale=0.29, clip]{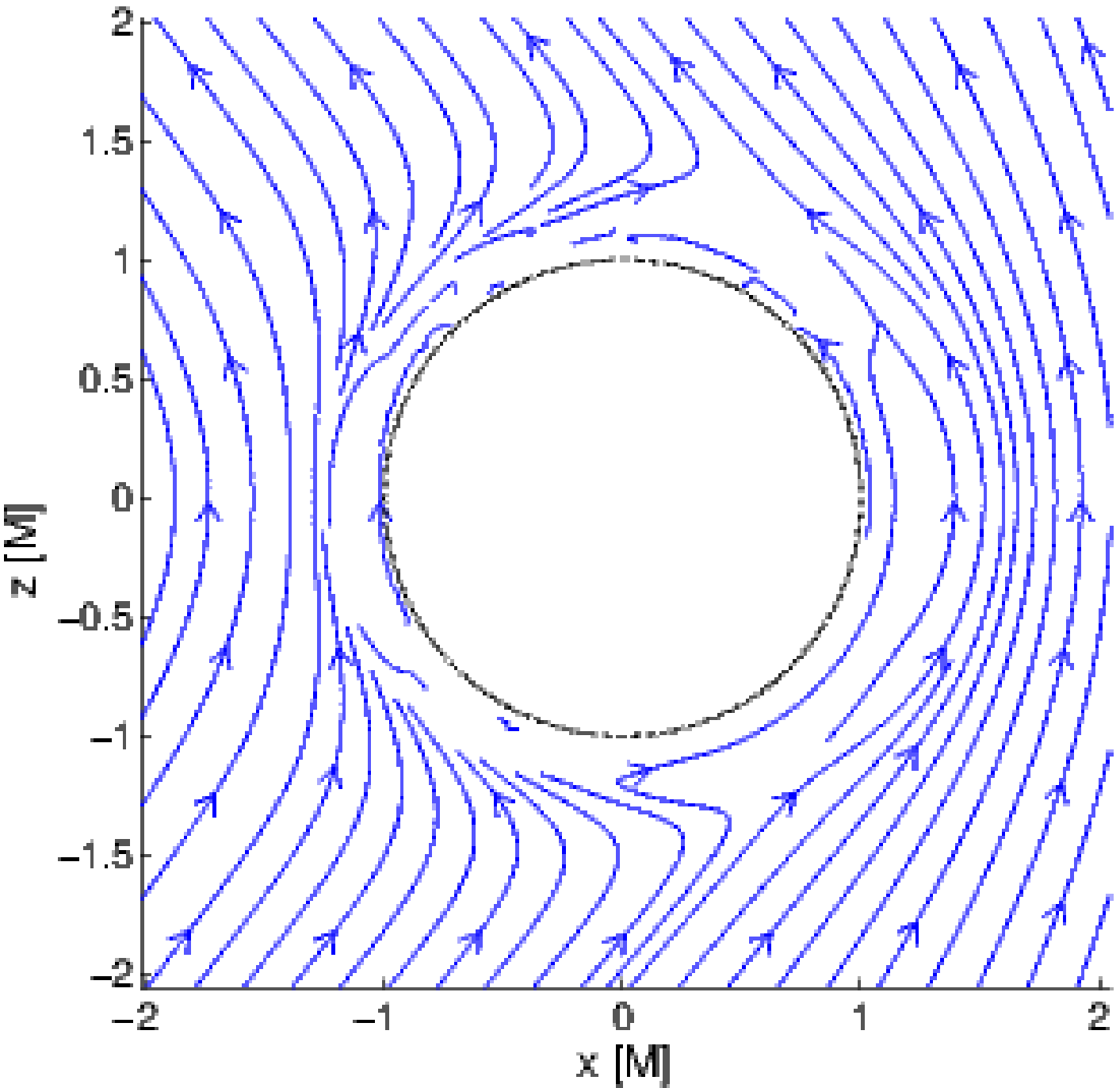}
\includegraphics[scale=0.29, clip]{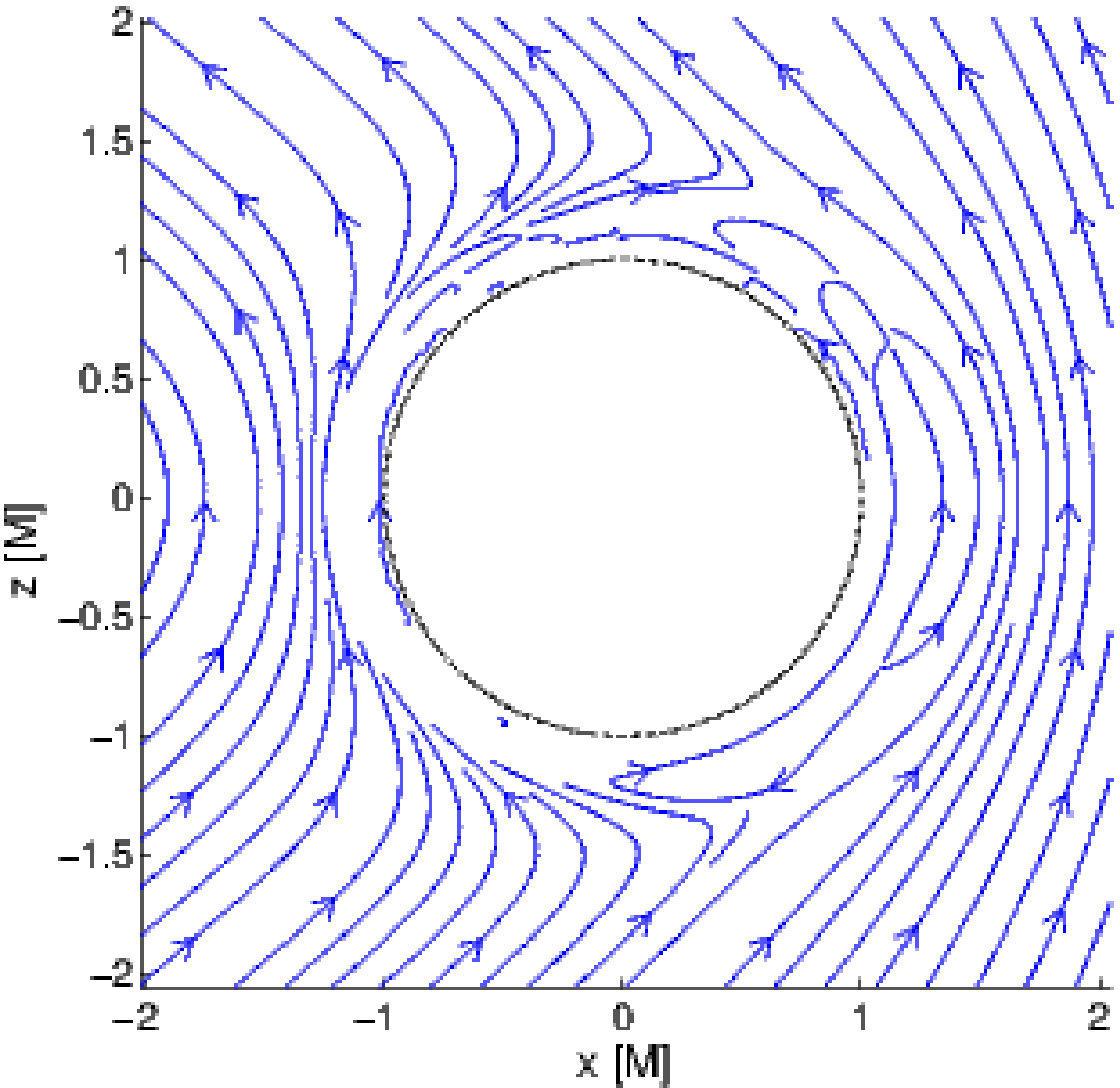}
\includegraphics[scale=0.29, clip]{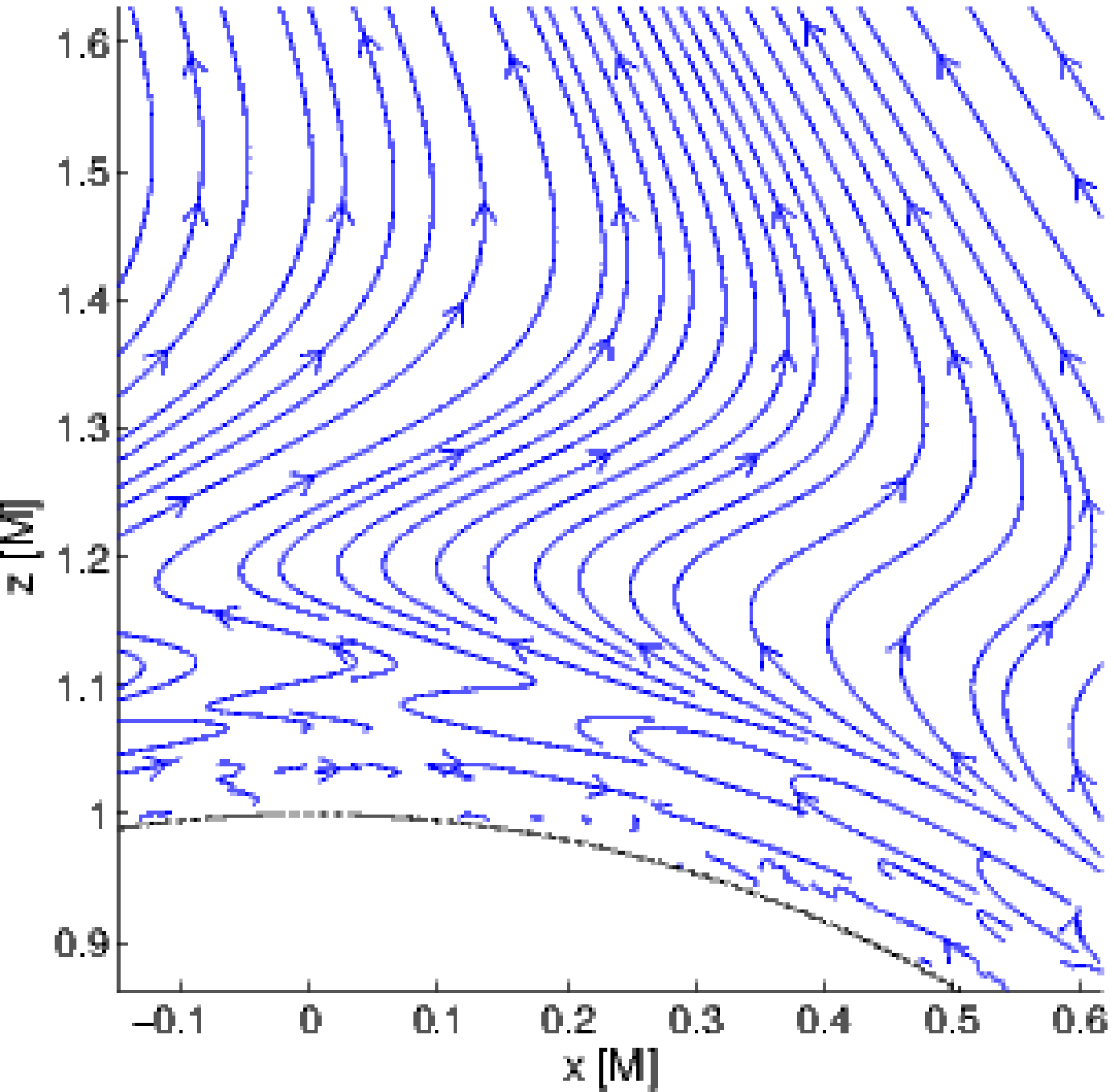}
\includegraphics[scale=0.29, clip]{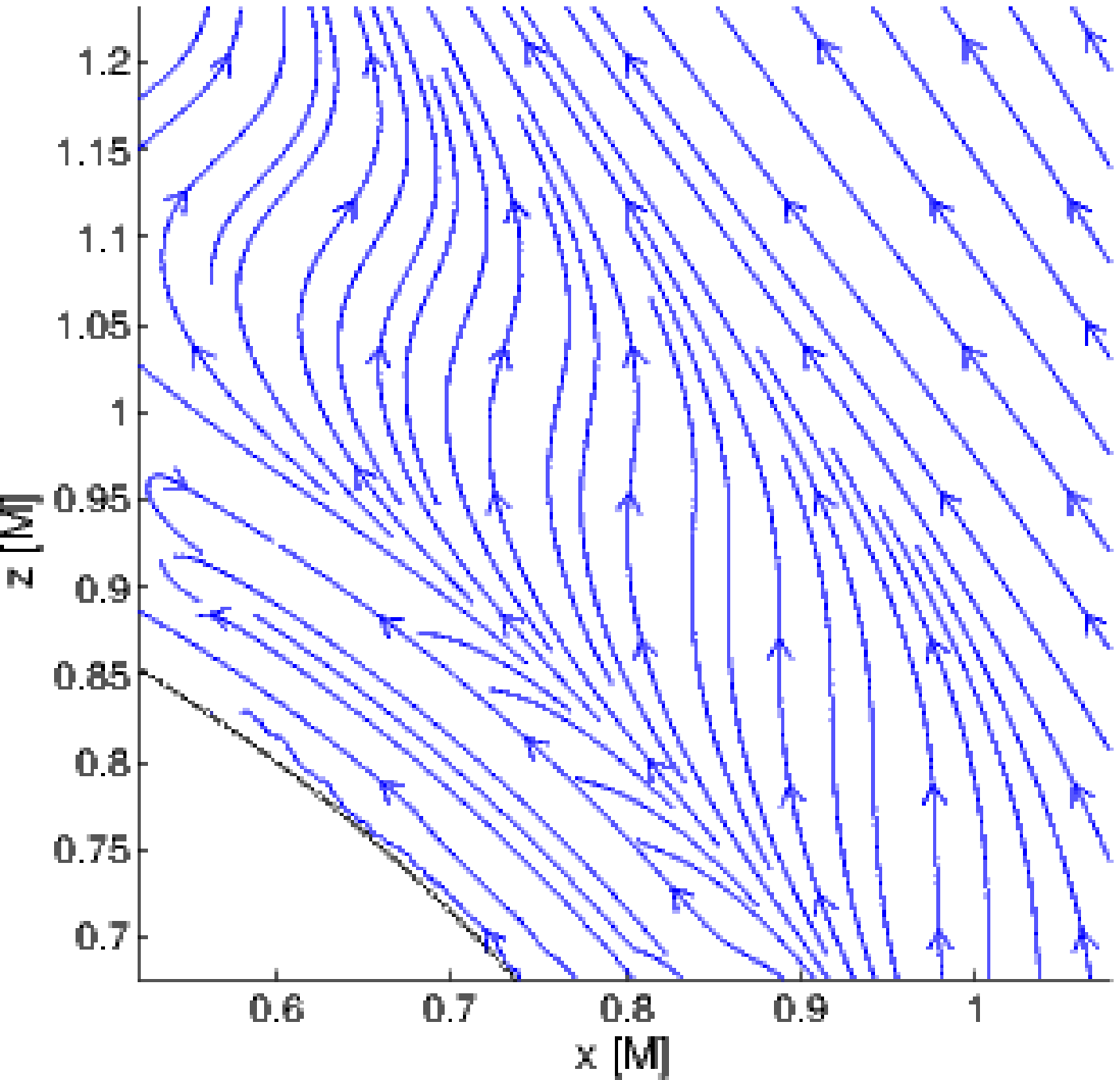}
\includegraphics[scale=0.29, clip]{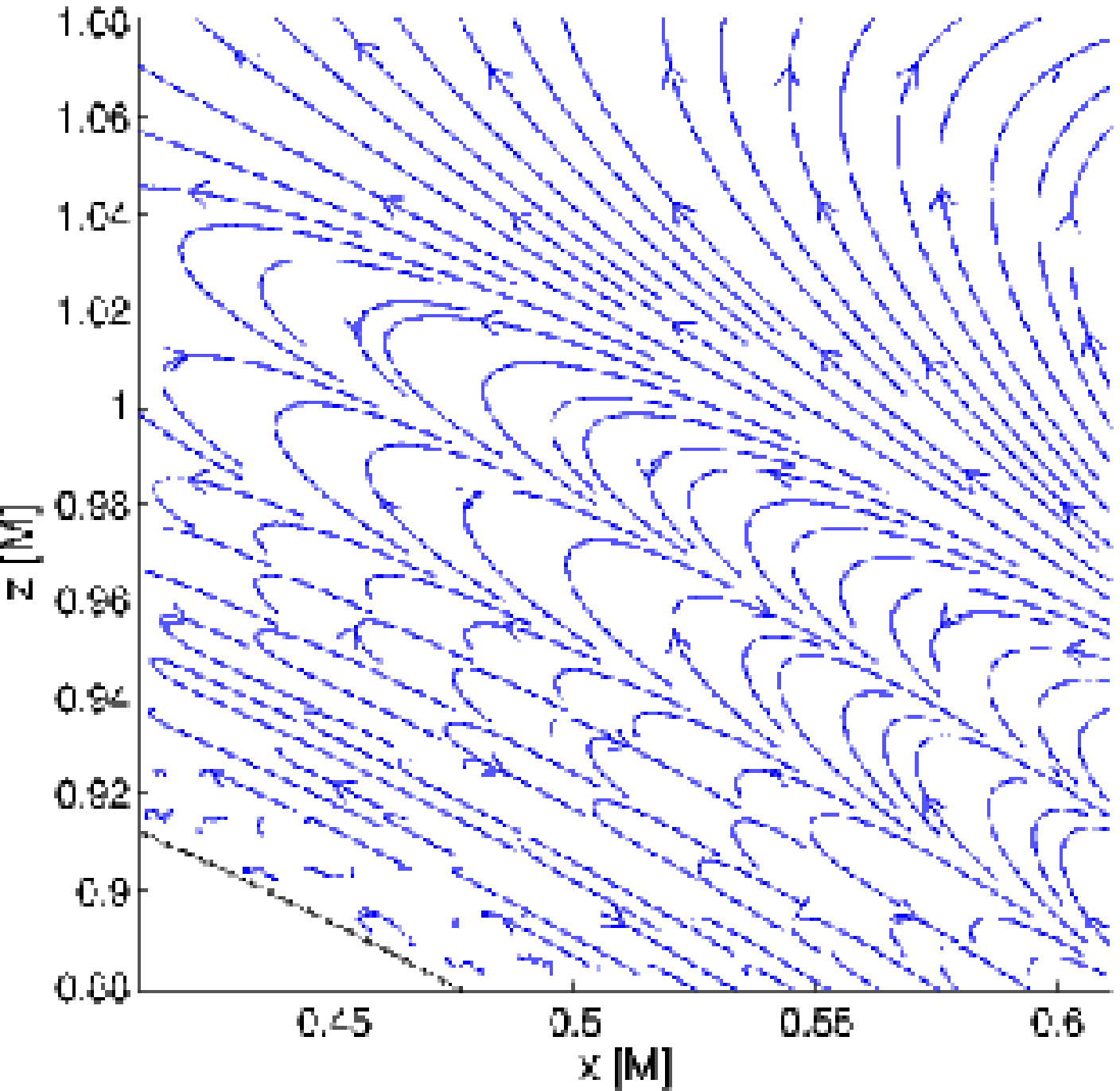}
\includegraphics[scale=0.29, clip]{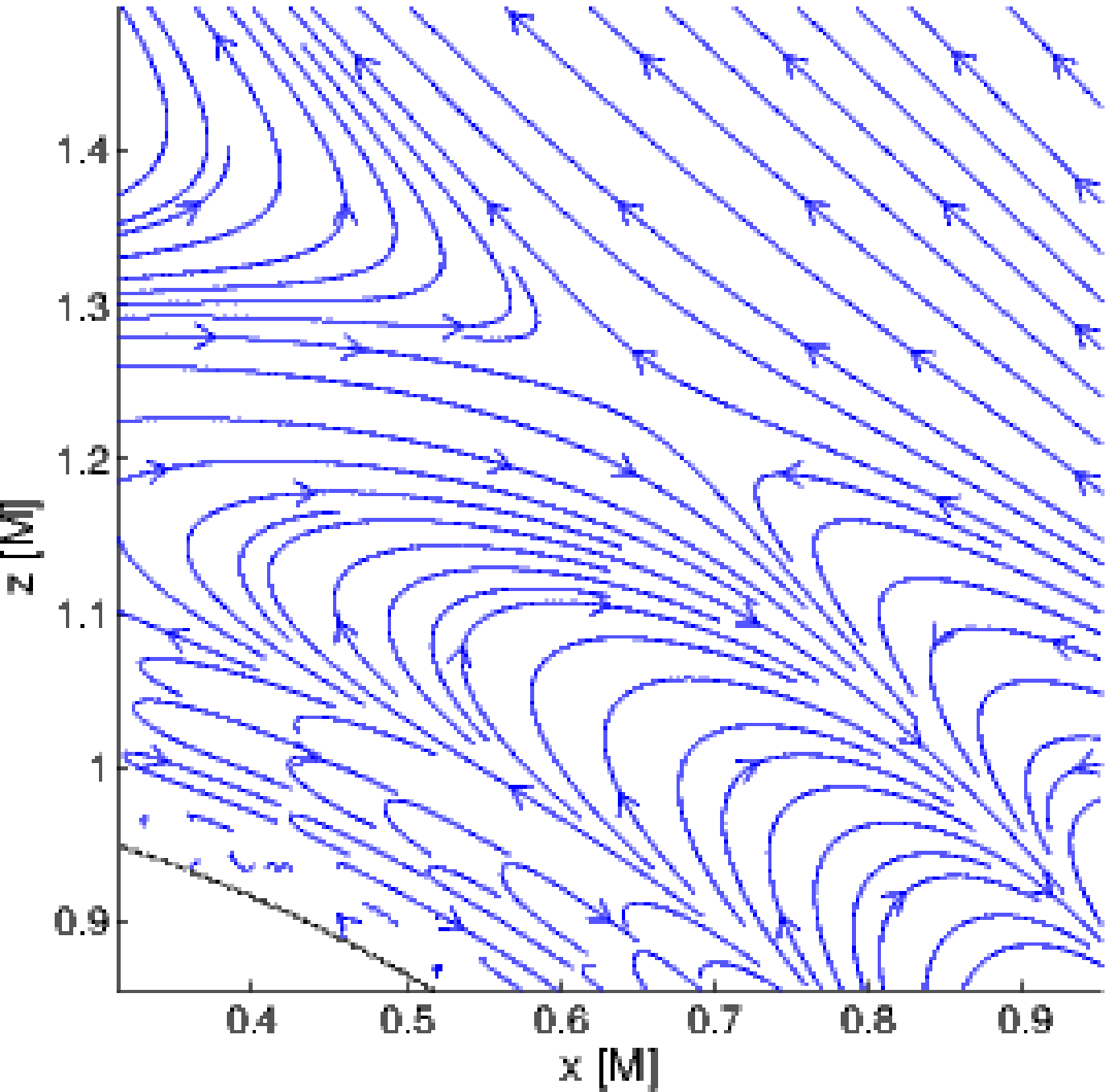}
\caption{In the series of first five panels we observe the impact which the gradually increased velocity $v_x=0,\:0.2,\:0.5,\:0.8$ and $0.99$ of the drift along the positive direction of the horizontal axis has upon the FFOFI measured magnetic field. Spin is set to the extreme value $a=M$. Prescribed aligned magnetic field is originally expelled from the horizon. When the drift is introduced, however,  the field lines are allowed to penetrate the horizon. Increasing the drift speed we observe that the layered bumper zone develops  being further enhanced with rising drift velocity. Right panel of the middle row and the left one on the bottom row show the magnetic layer structures which arise for $v_x=0.5$ in detail. Layered patterns further enhance when the speed increases -- next panel (middle bottom) shows the rapidly moving case $v_x=0.99$. Last panel reveals the null point of the magnetic field which develops with high drift velocity. }
\label{expul_drift}
\end{figure}

\begin{figure}[hp!]
\centering
~~~~\includegraphics[scale=0.313,trim=0mm 0mm 12mm 0mm ,clip]{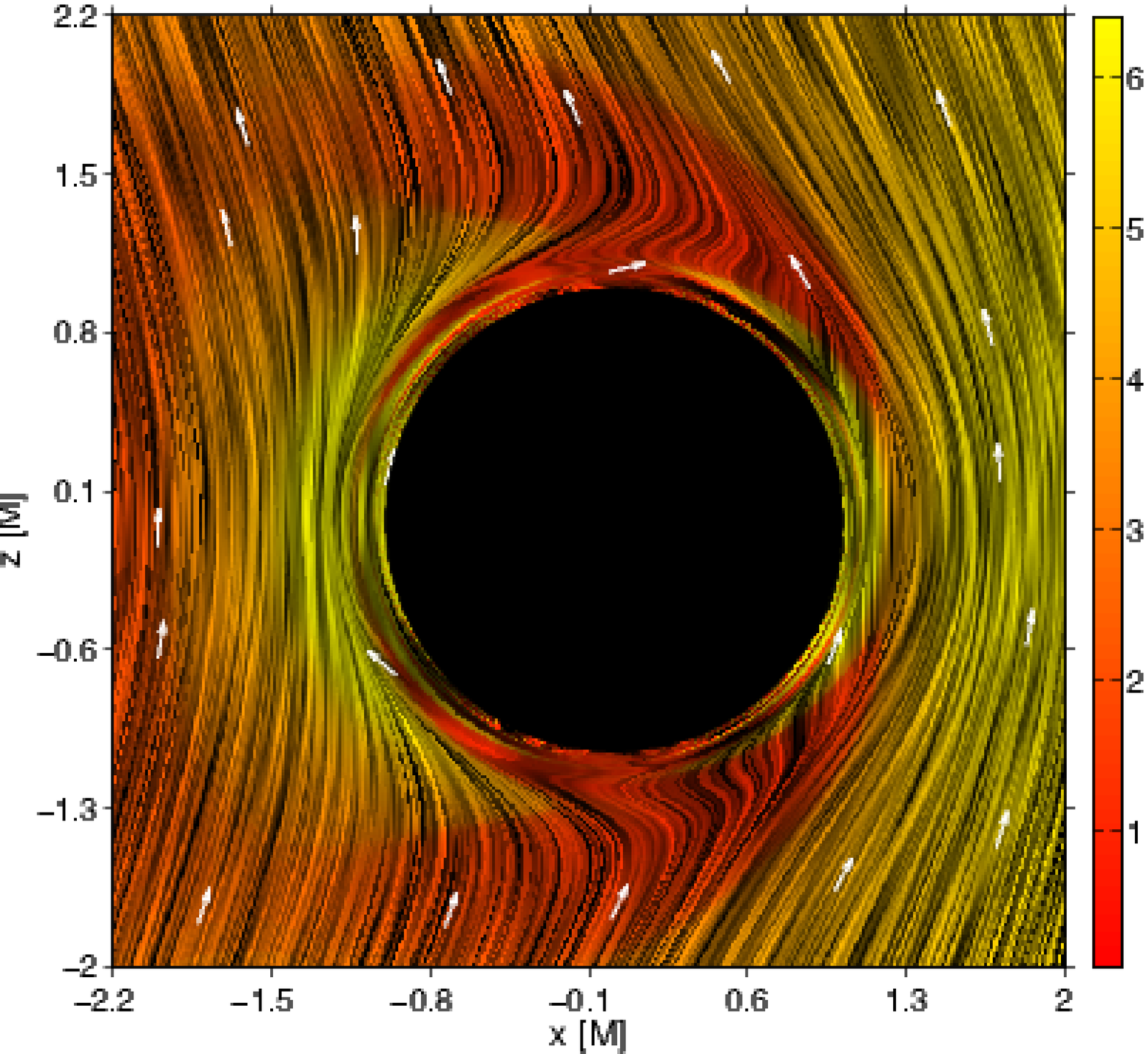}~
\includegraphics[scale=0.313, trim=25mm 0mm 20mm 0mm, clip]{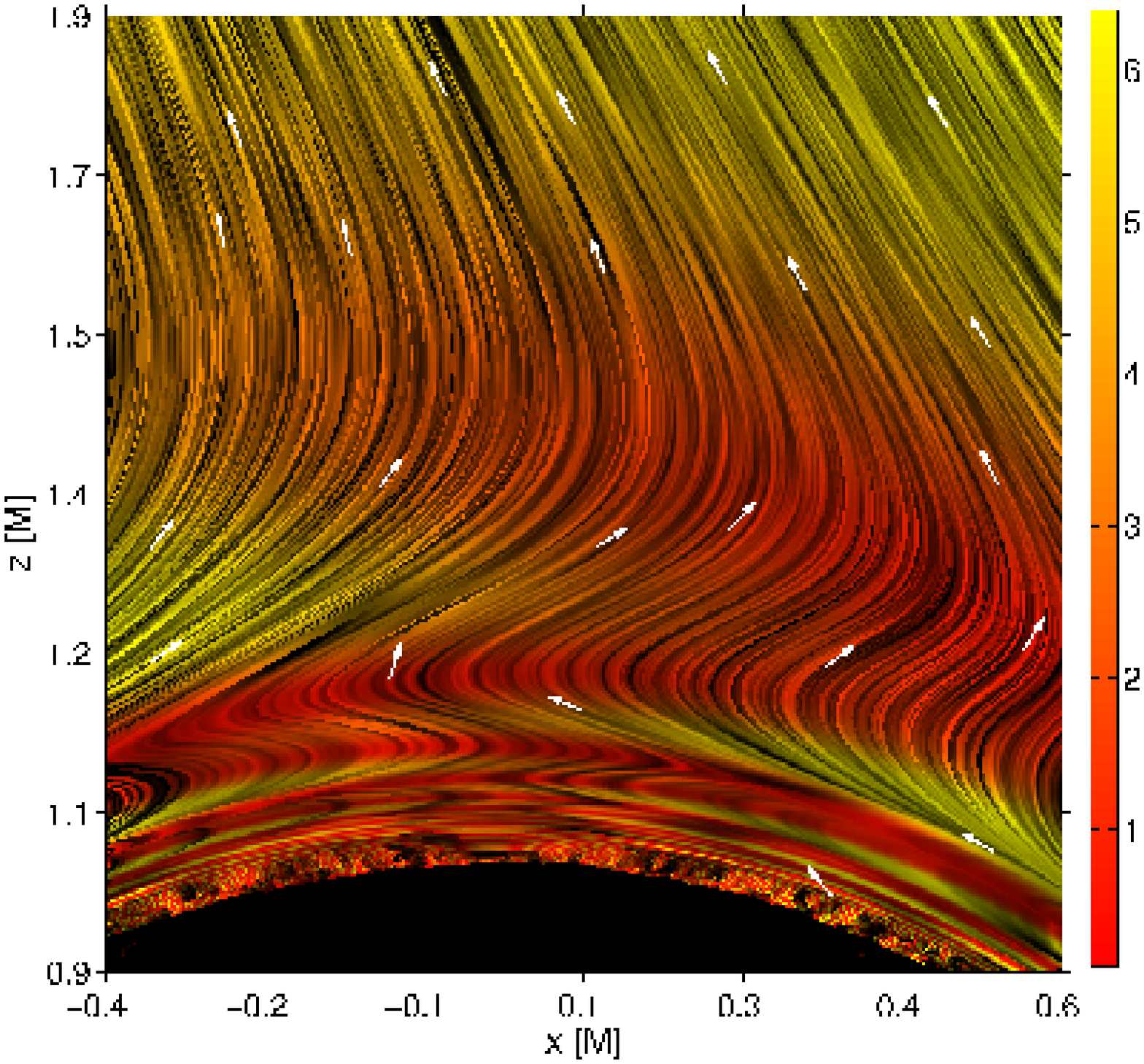}
\includegraphics[scale=0.313, trim=19.5mm 0mm 35mm 0mm, clip]{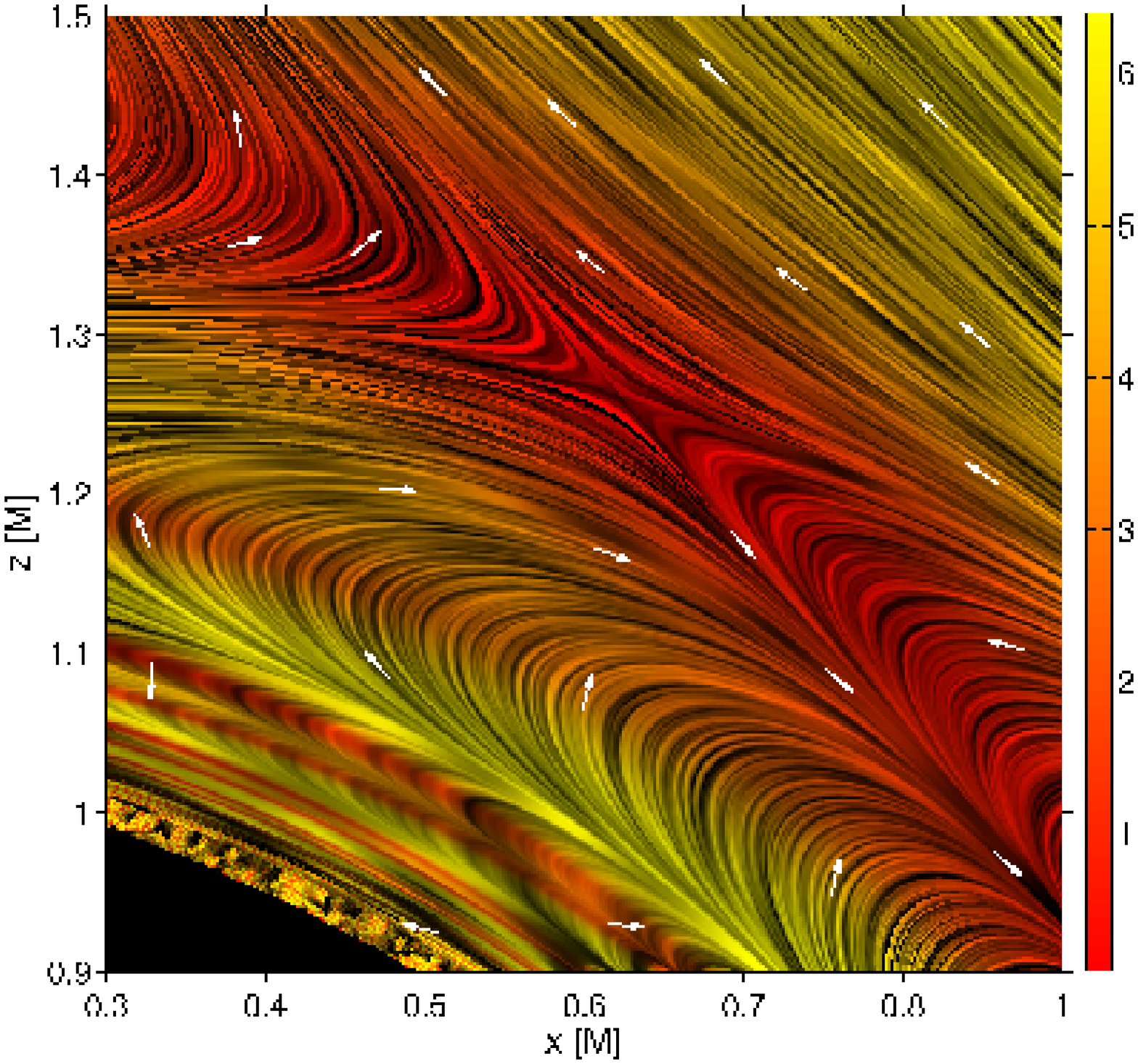}
\includegraphics[scale=0.313, trim=23mm 0mm 35mm 0mm, clip]{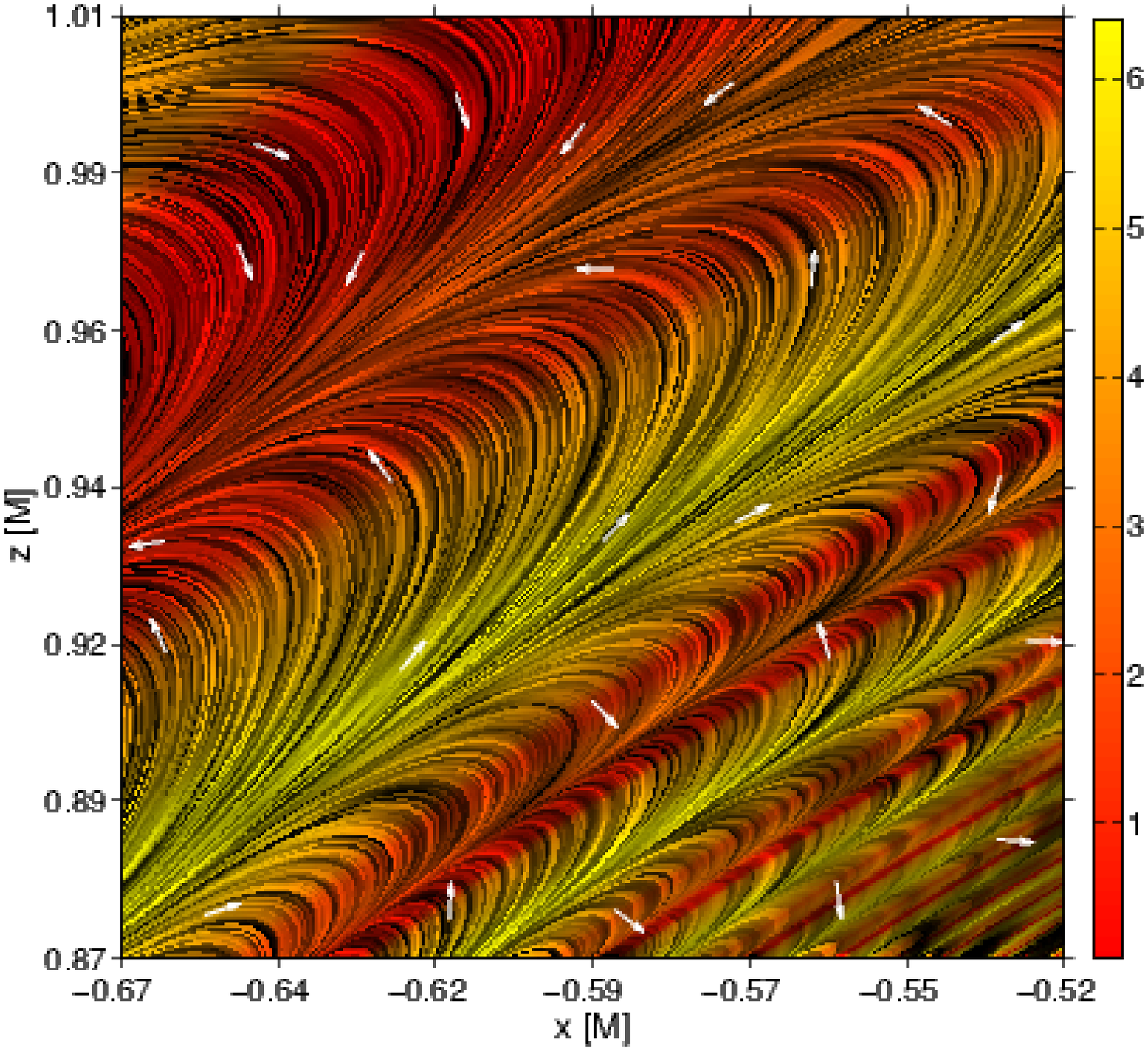}
\caption{Drift induced effects upon the structure of the FFOFI tetrad components of the aligned magnetic field in the case of extreme spin value $a=M$. Translational motion is restricted to be parallel with the horizontal axis, $v_x\ne 0$, $v_{y}=v_{z}=0$. Upper figures reveal the situation for $v_{x}=0.5$. We observe that a narrow buffer zone develops above the horizon in which the field lines have complex layered structure. Zones of antiparallely oriented magnetic field are brought to the close contact here. Bottom panels show the case of extremely rapid motion $v_x=0.99$. Null point of the magnetic field appears in this case (bottom left panel). In the right panel we observe self-similar tightly folded layered structures which are considerably enhanced when compared to the case of slower motion in the upper right panel.}
\label{mag_drift_LIC}
\end{figure}

\begin{figure}[hp!]
\centering
\includegraphics[scale=0.33, clip,trim=30mm 5mm 65mm 5mm]{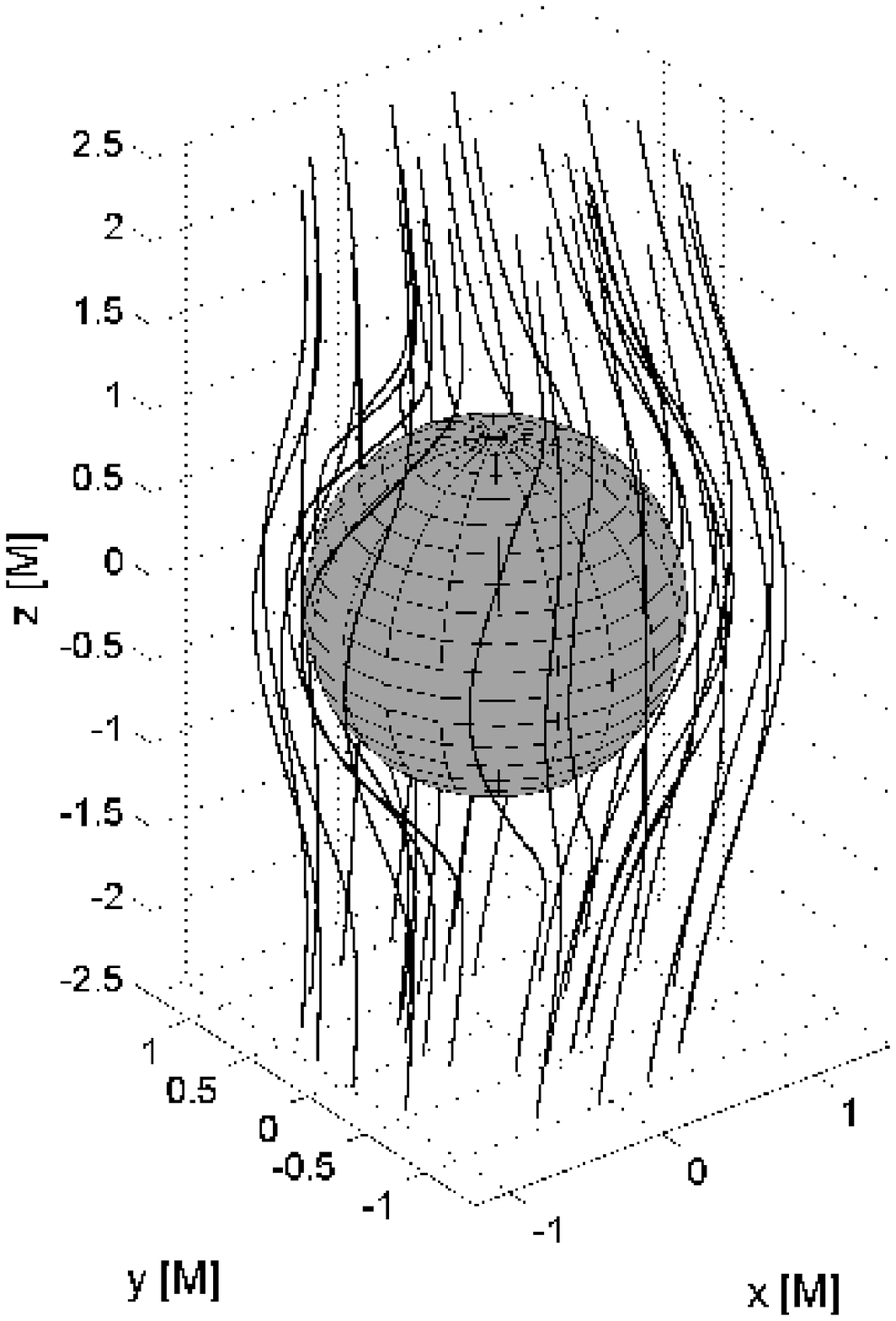}
\includegraphics[scale=0.33, clip,trim=57mm 5mm 65mm 5mm]{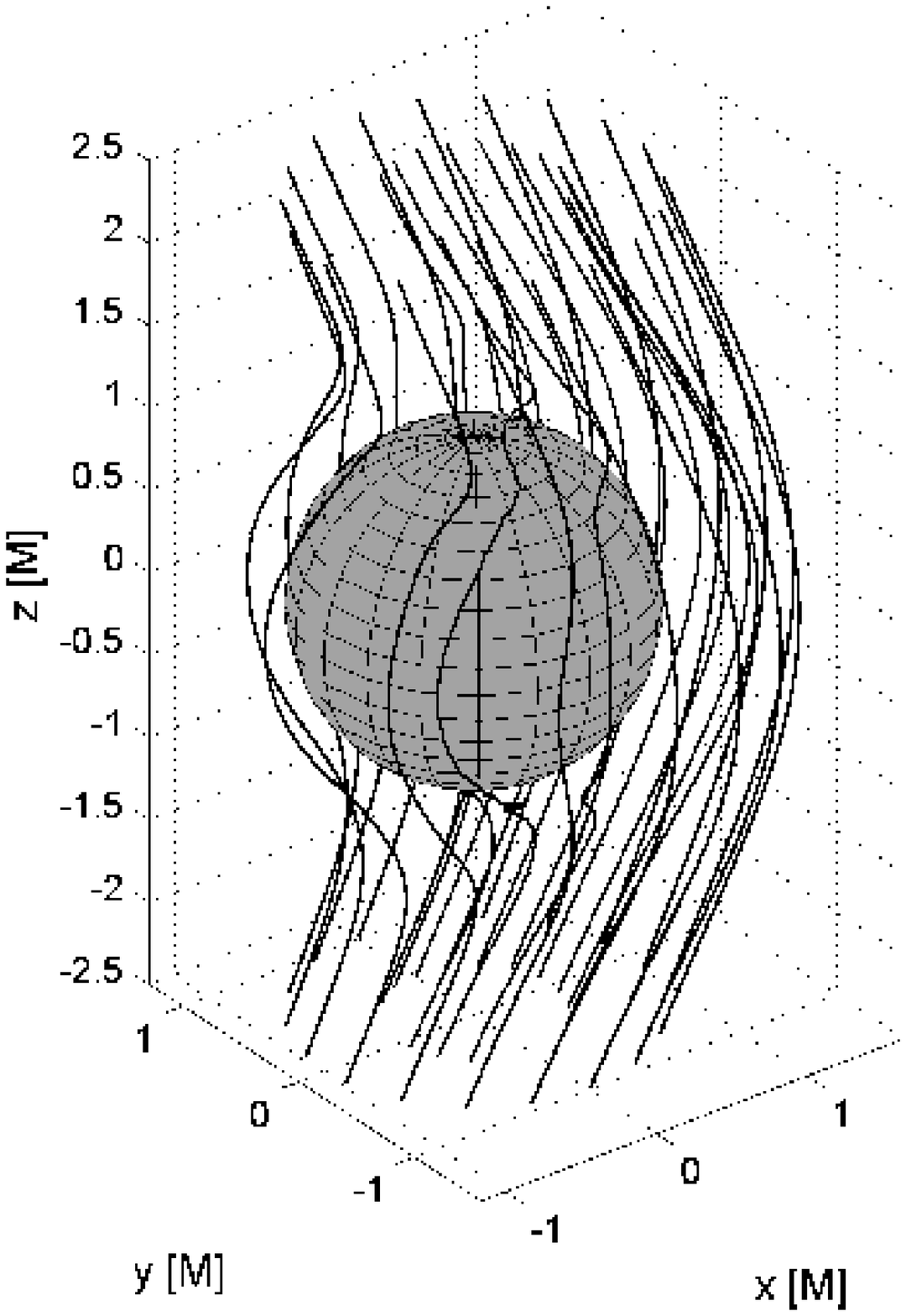}
\includegraphics[scale=0.33, clip,trim=45mm 5mm 65mm 5mm]{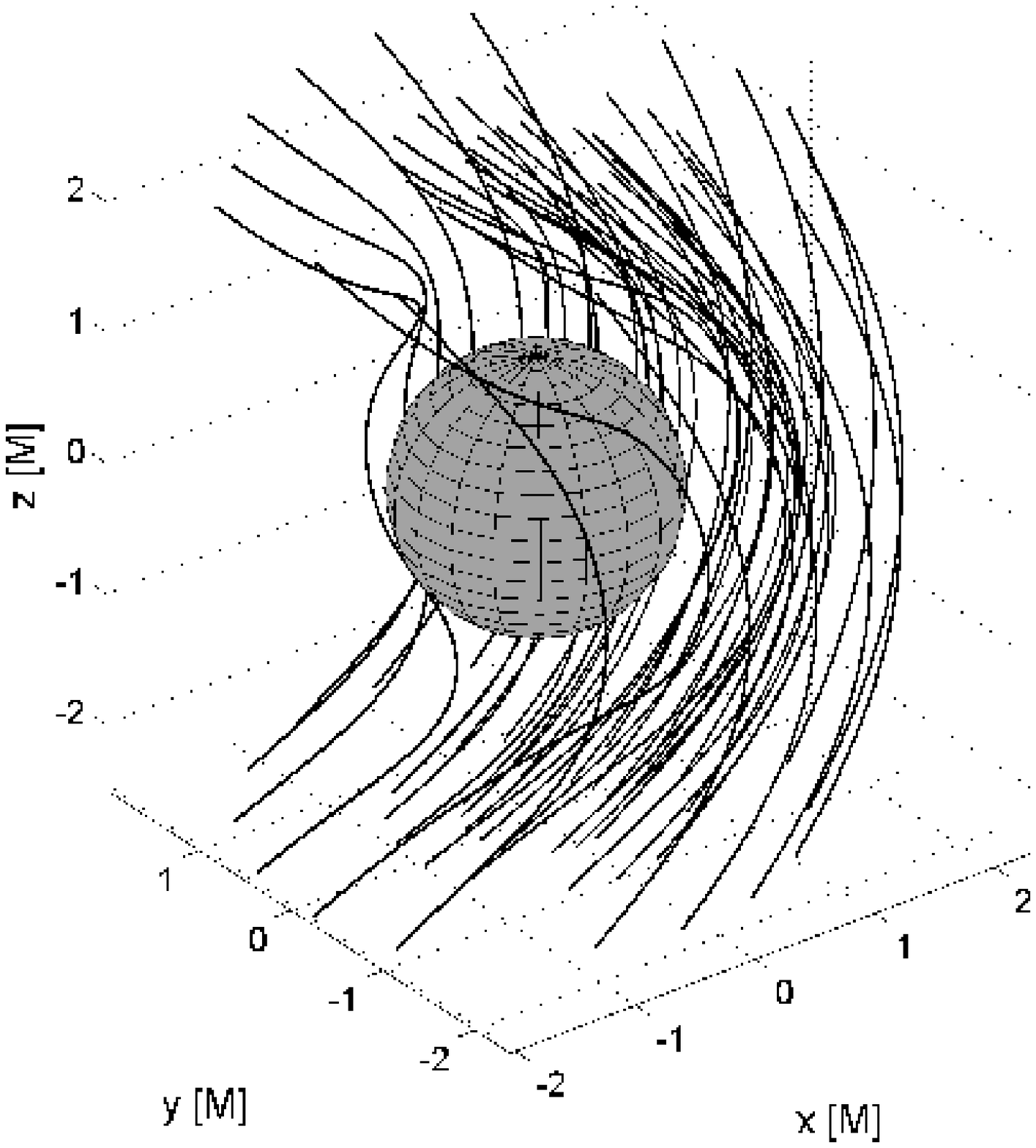}
\includegraphics[scale=0.33, clip,trim=30mm 5mm 67mm 5mm]{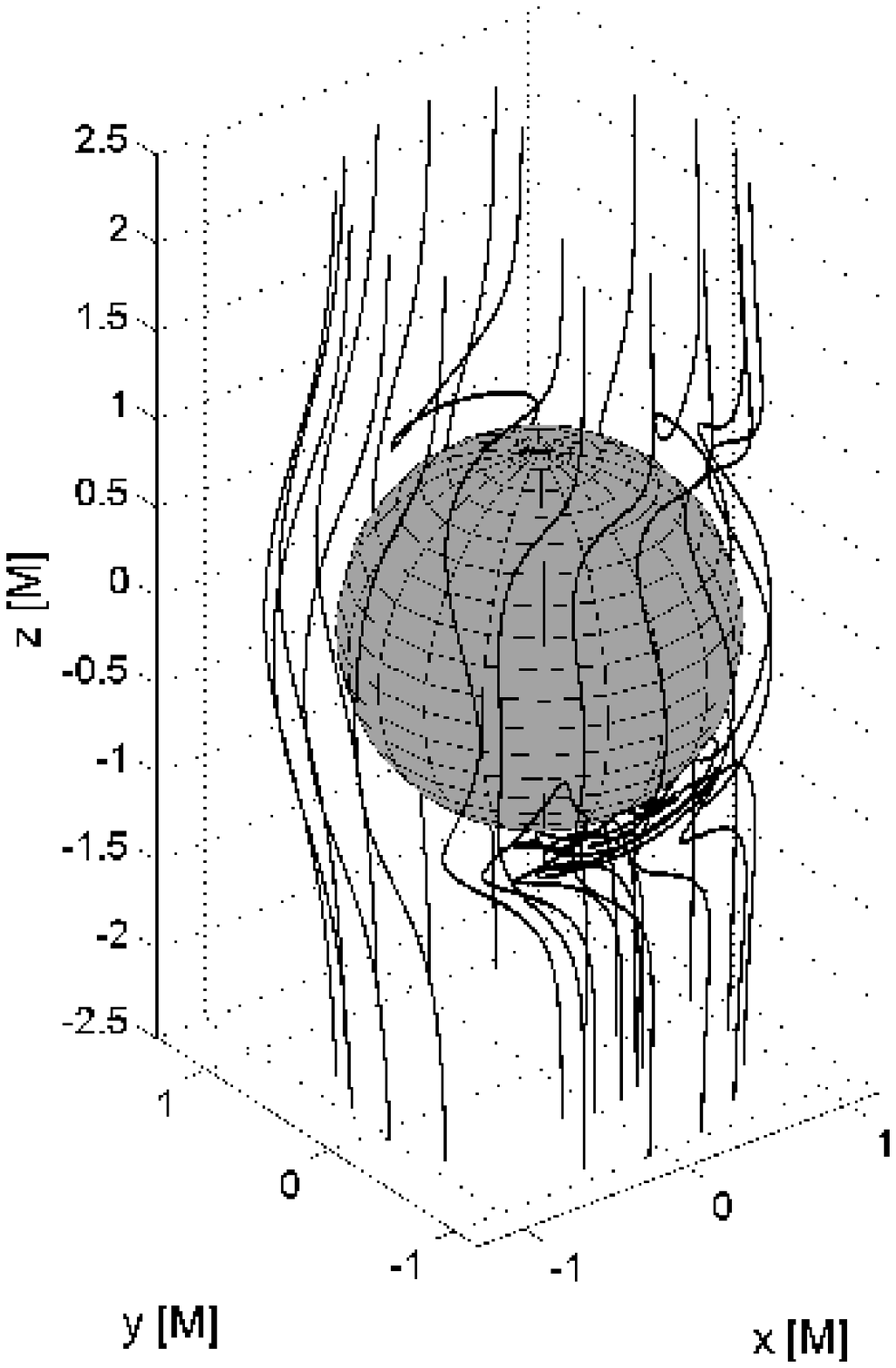}
\includegraphics[scale=0.33, clip,trim=56mm 5mm 65mm 5mm]{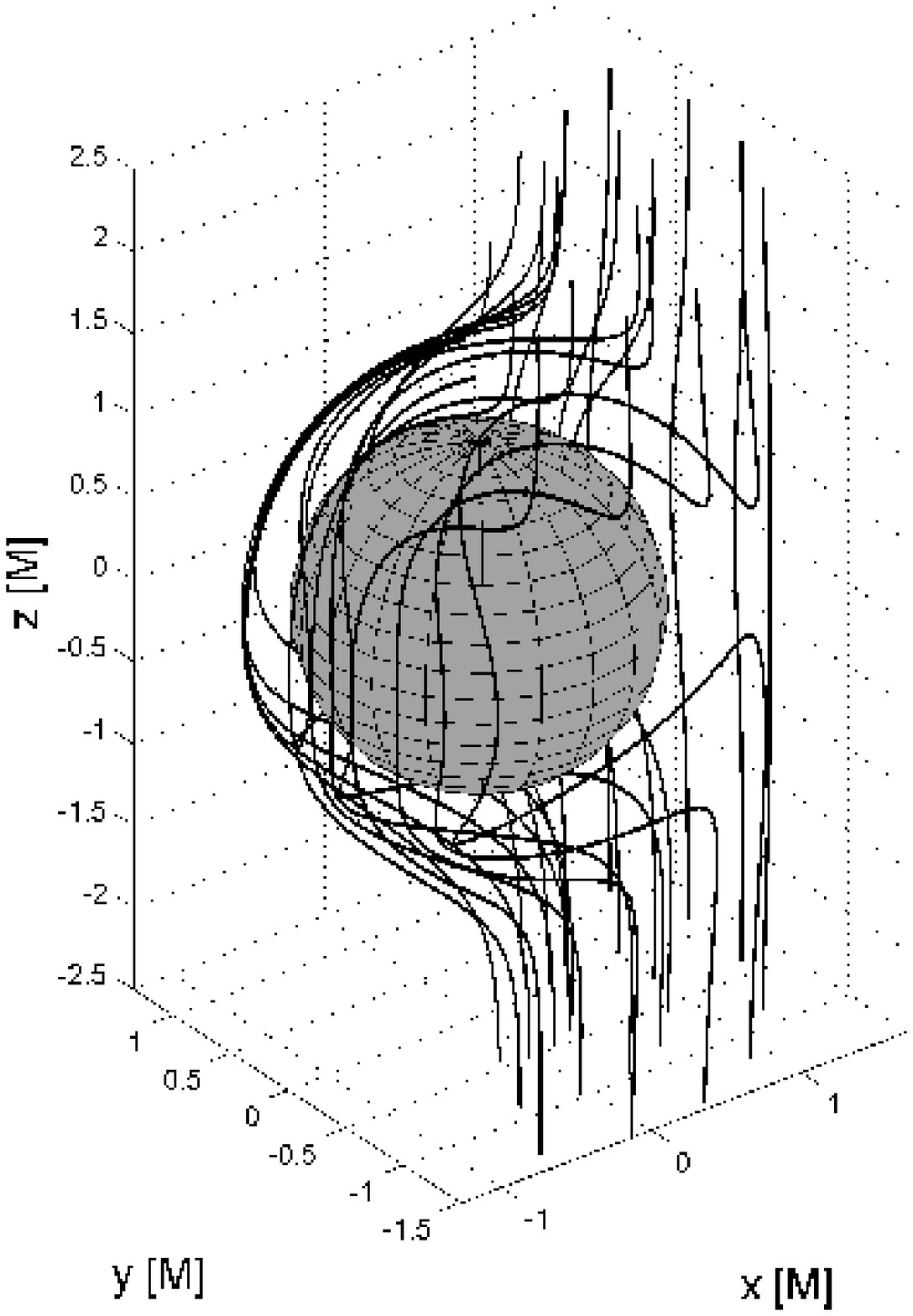}
\includegraphics[scale=0.33, clip,trim=55mm 5mm 65mm 5mm]{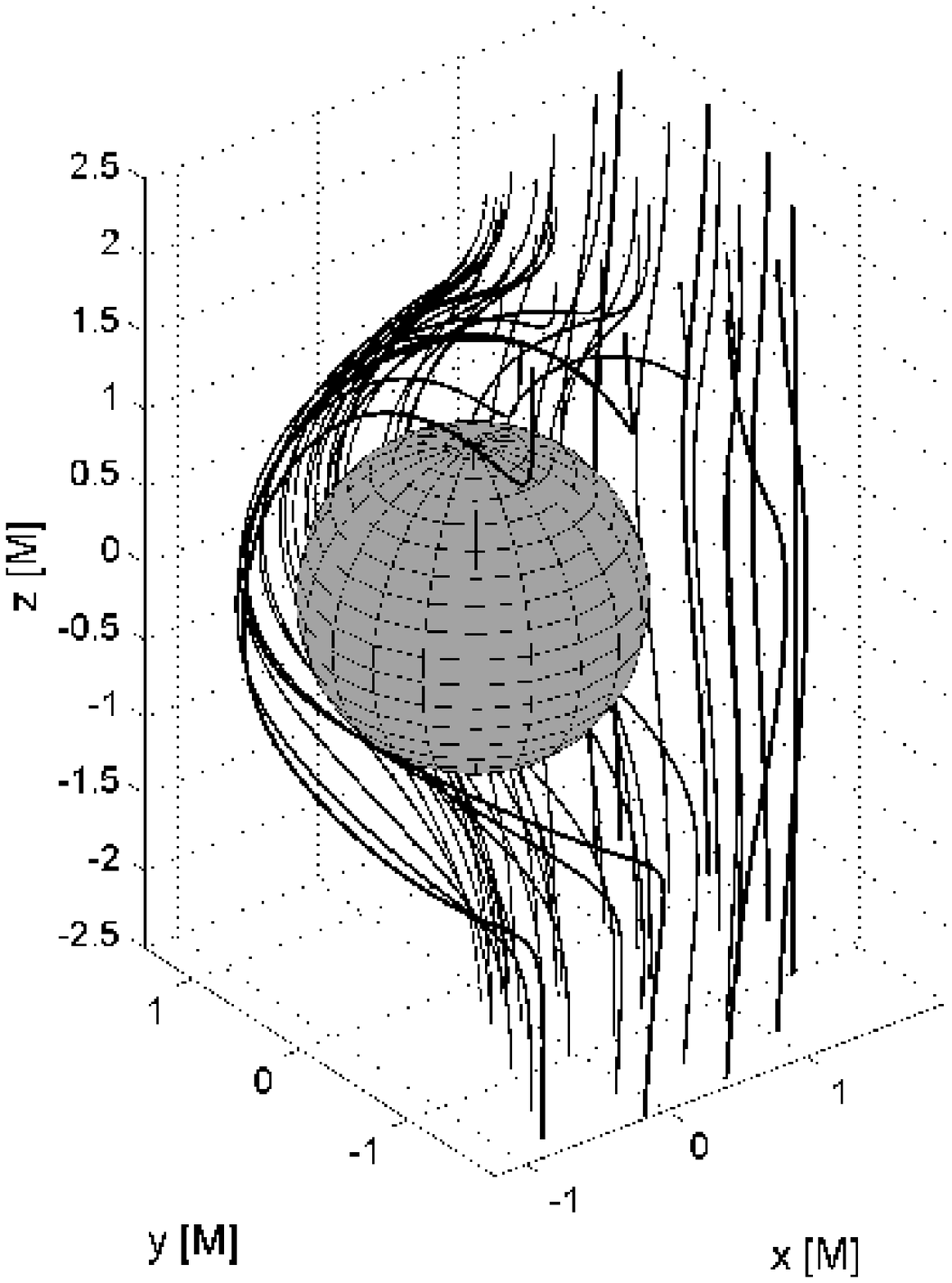}
\caption{Drift induced deformations of the originally aligned magnetic field lines above the horizon of the extreme Kerr black hole. FFOFI tetrad components in the upper row are compared to the renormalized ZAMO components in the bottom row. In both the cases the field is expelled for the zero drift velocity (see \rff{meissner_3d_AMO}) the only difference being that FFOFI field is slightly twisted in the azimuthal direction. Drift direction is set to coincide with the horizontal axis of the stereometric projection, $v_x>0,\;v_y=-v_x$ and $v_z=0$. Velocity is increased in the following sequence: $v_x=0.1,\;0.3,\;0.7$. FFOFI tetrad and ZAMO renormalized components exhibit completely different behaviour when the drift is introduced as ZAMO field is dragged in the opposite direction compared to FFOFI.}
\label{mag_drift_3d}
\end{figure}

\begin{figure}[hp!]
\centering
\includegraphics[scale=0.29, clip]{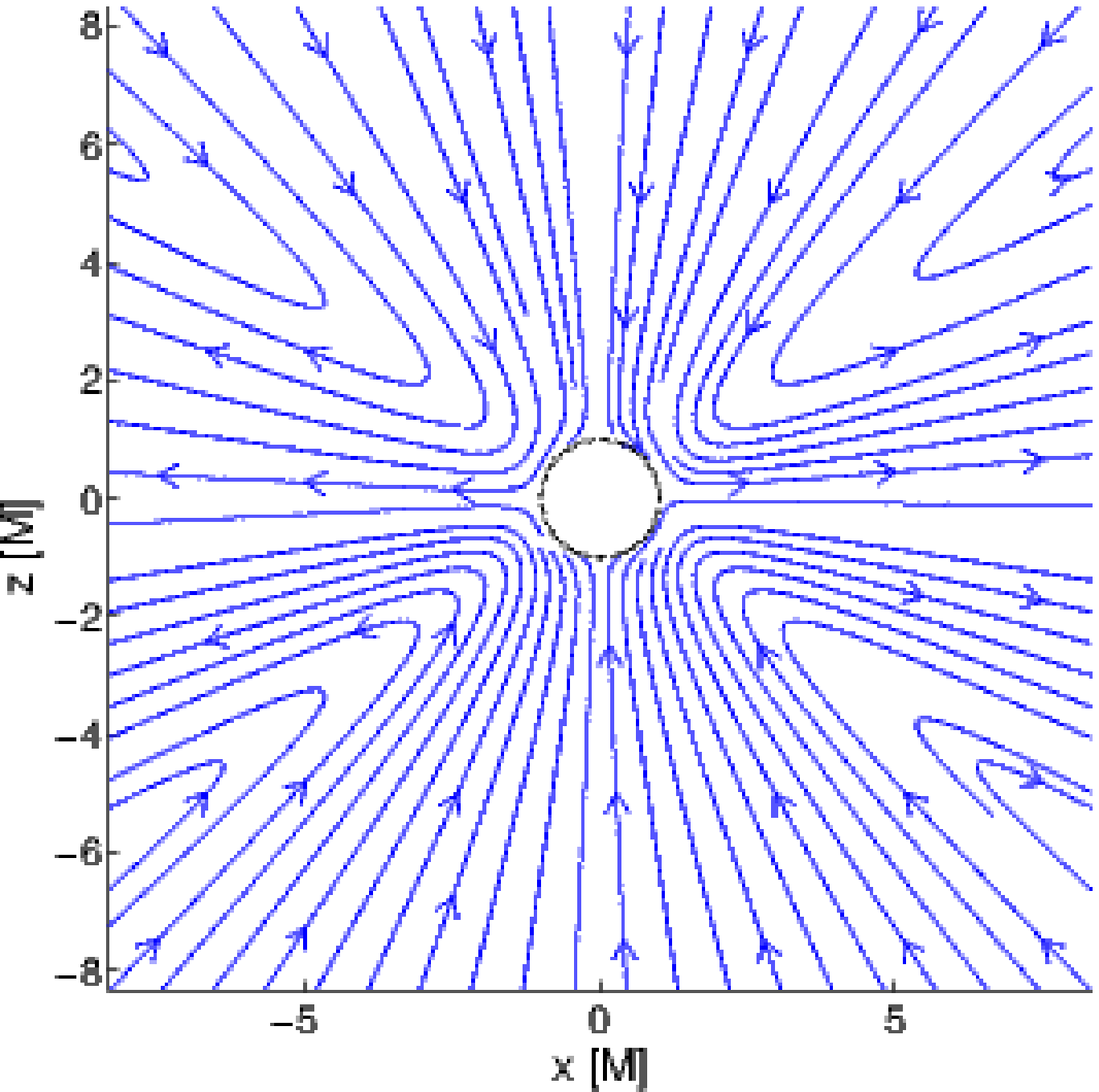}
\includegraphics[scale=0.29, clip]{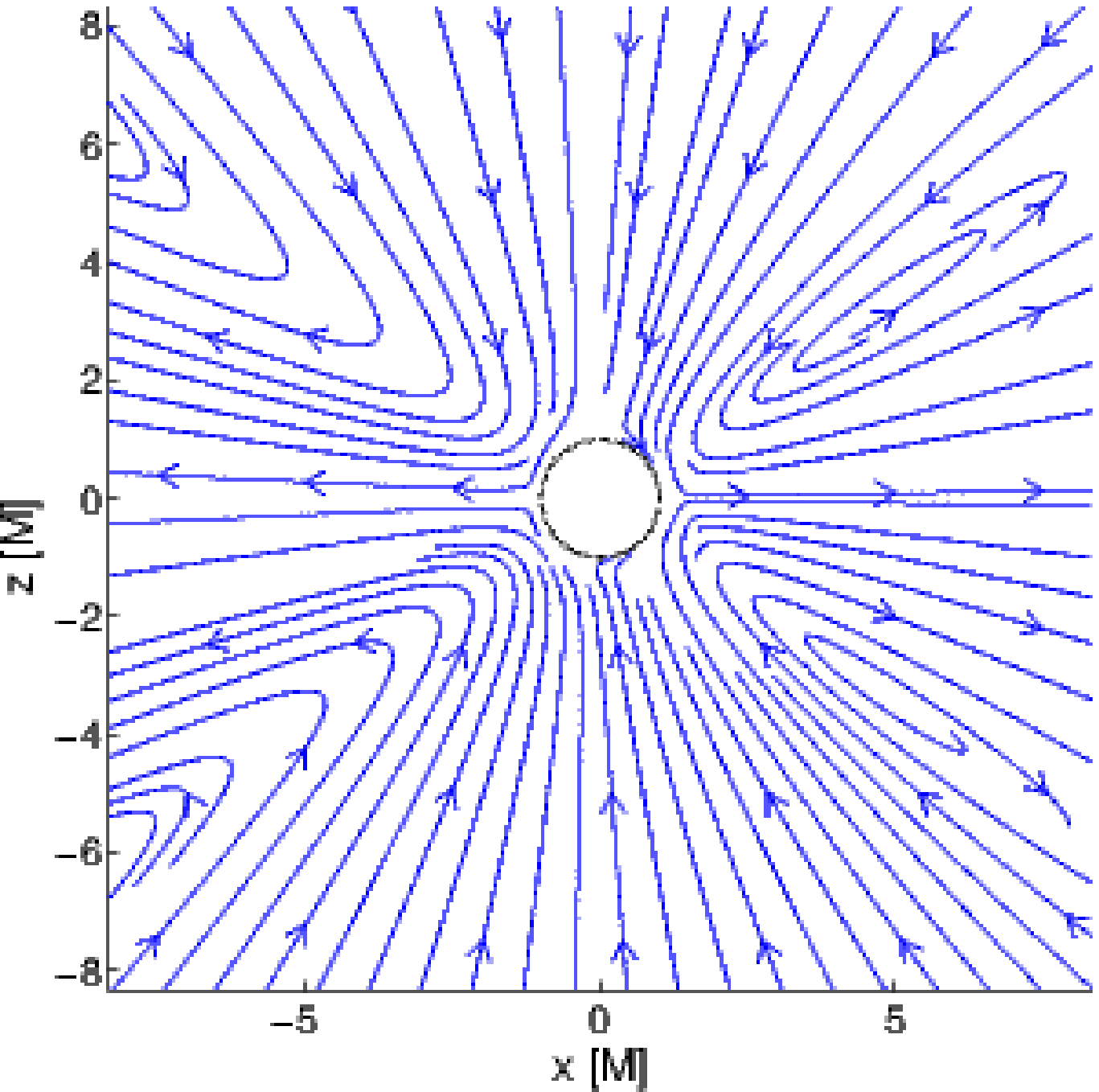}
\includegraphics[scale=0.29, clip]{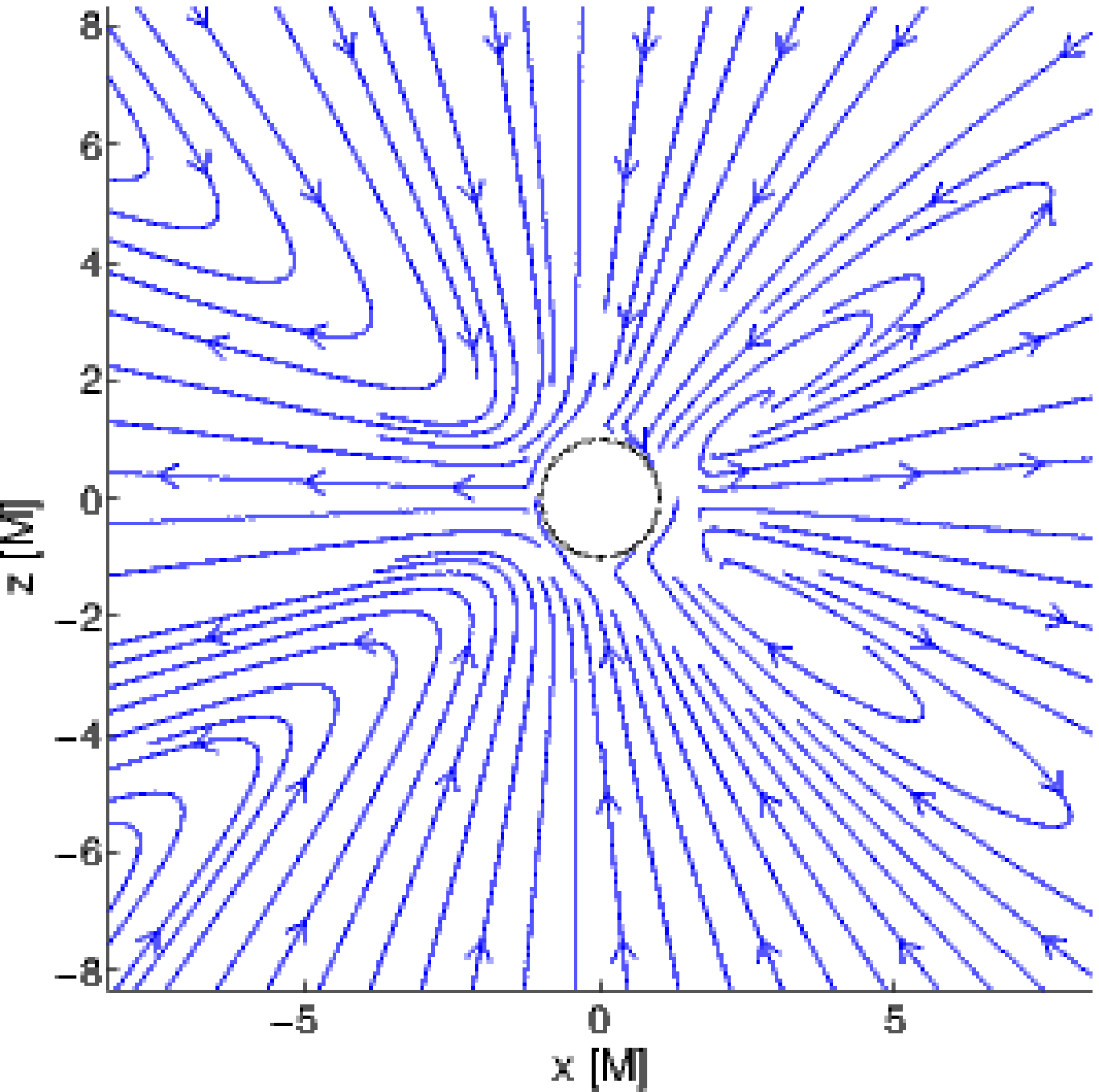}
\includegraphics[scale=0.29, clip]{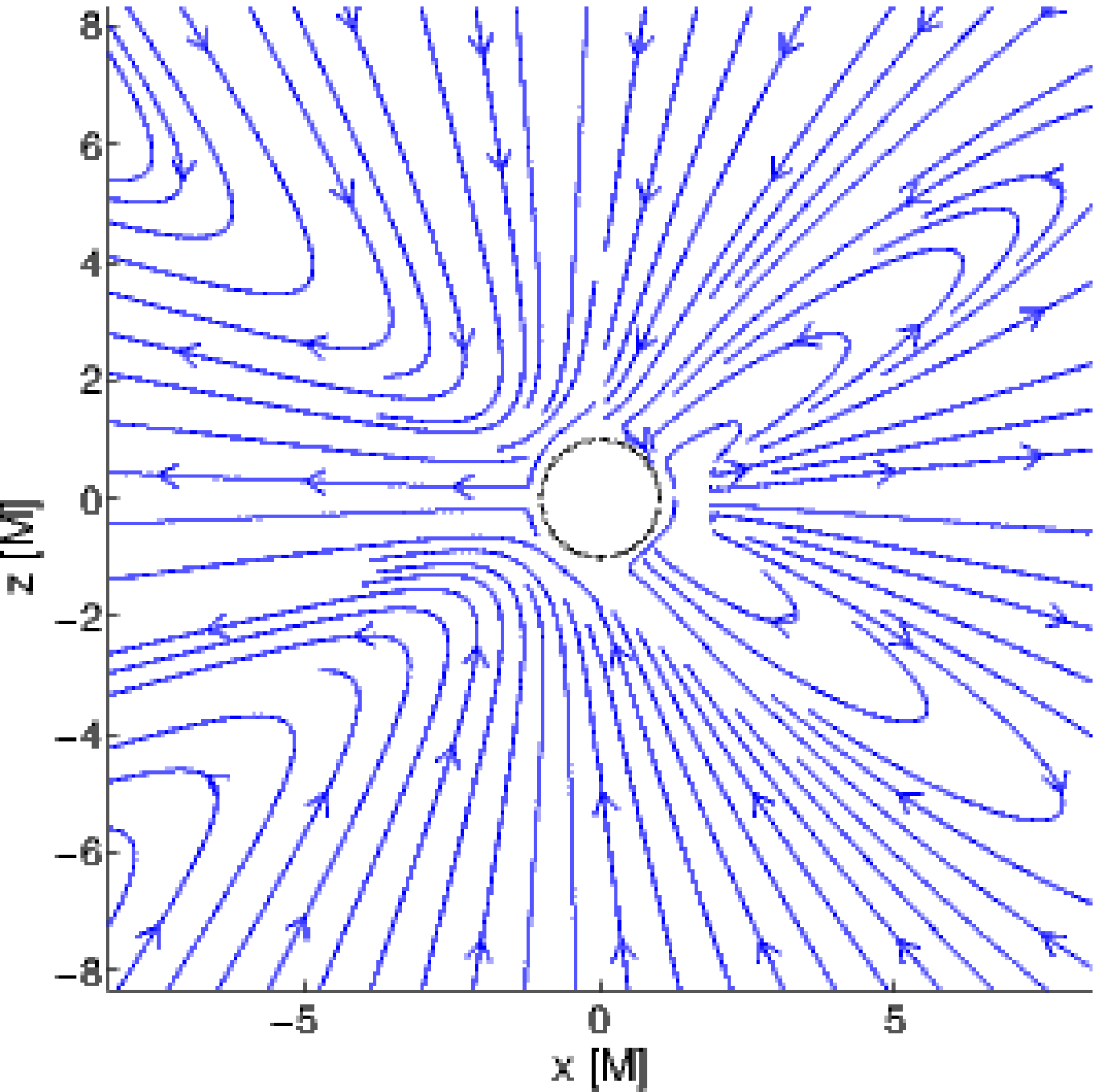}
\includegraphics[scale=0.29, clip]{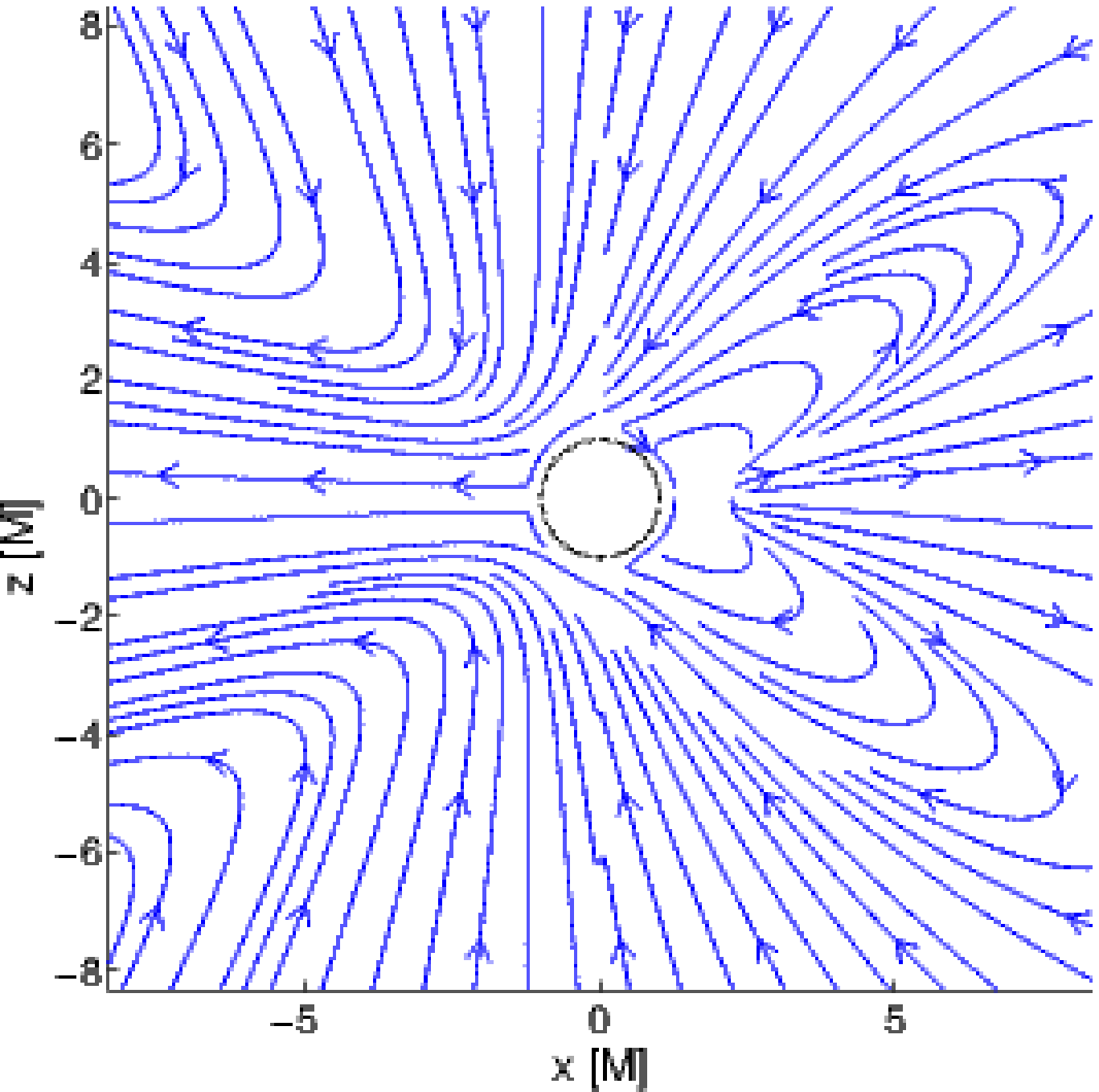}
\includegraphics[scale=0.29, clip]{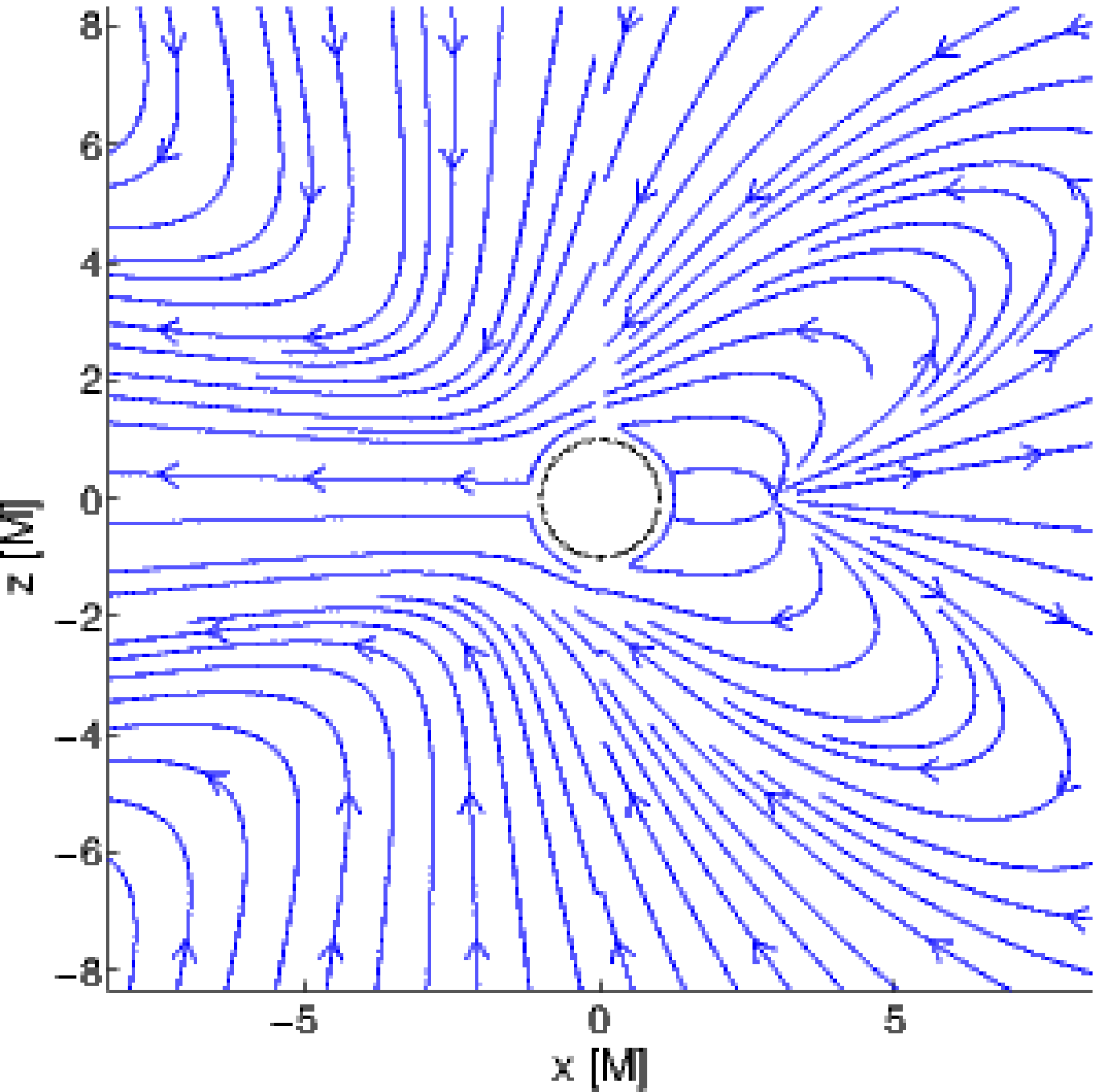}
\includegraphics[scale=0.29, clip]{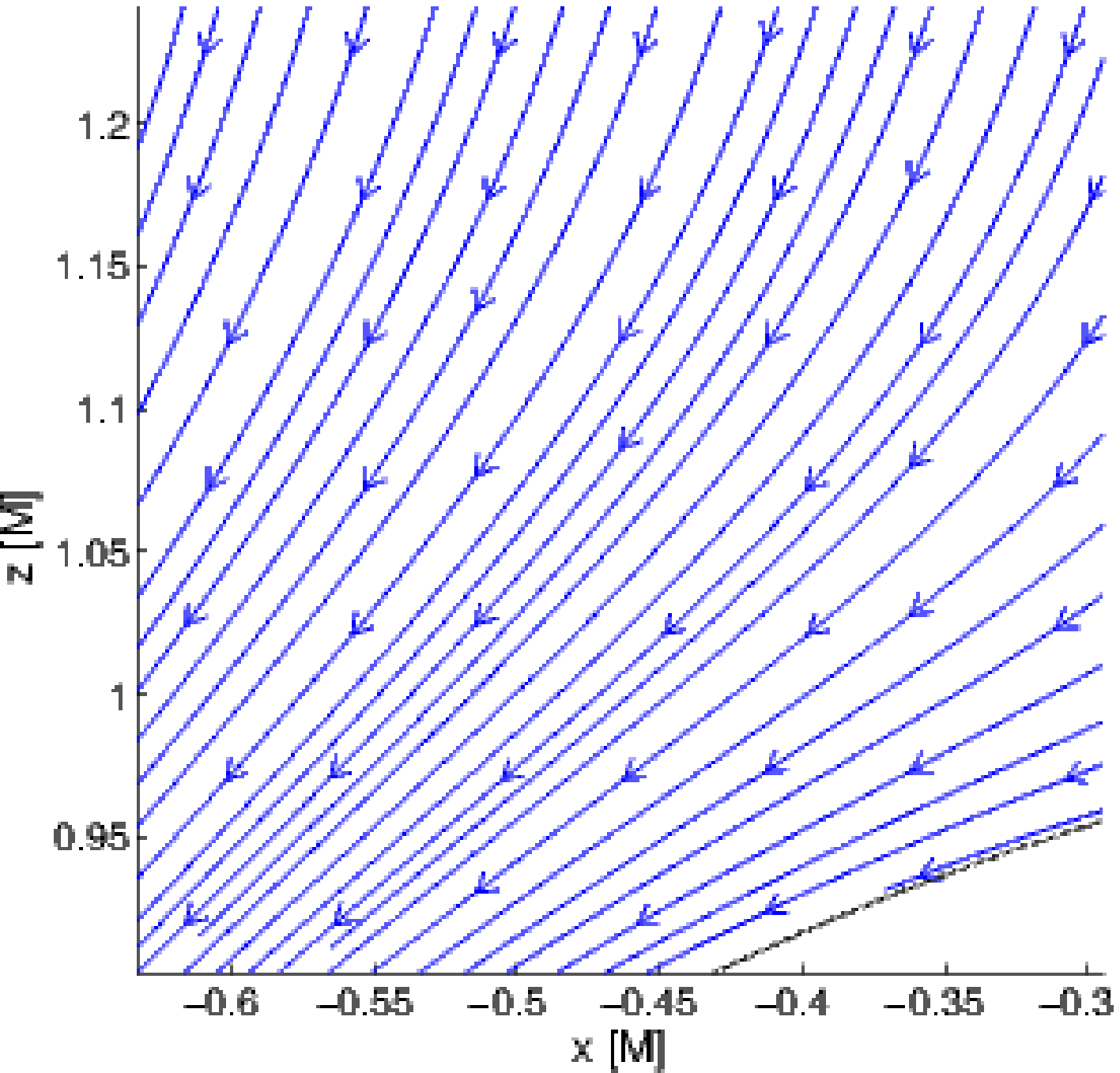}
\includegraphics[scale=0.29, clip]{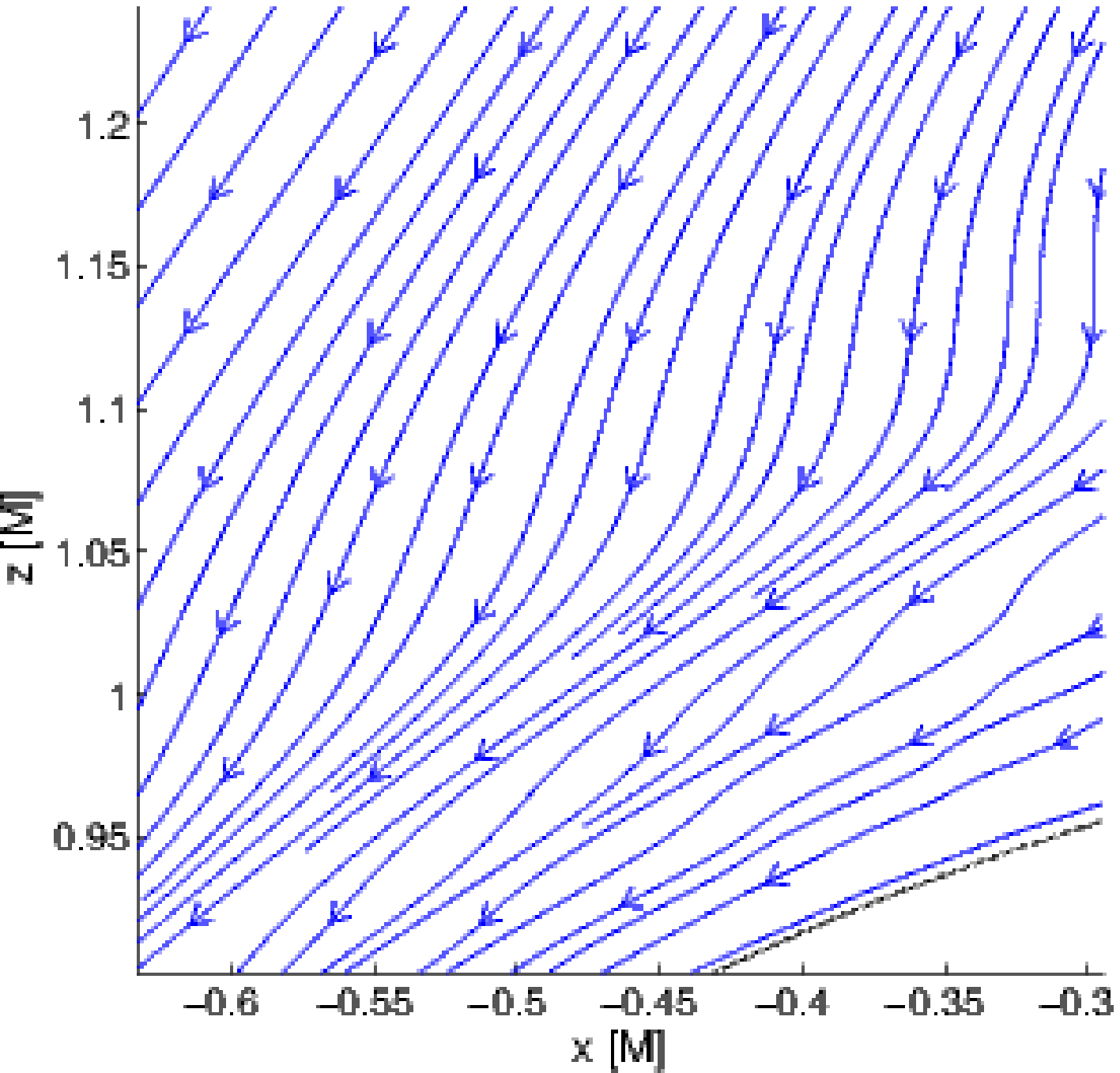}
\includegraphics[scale=0.29, clip]{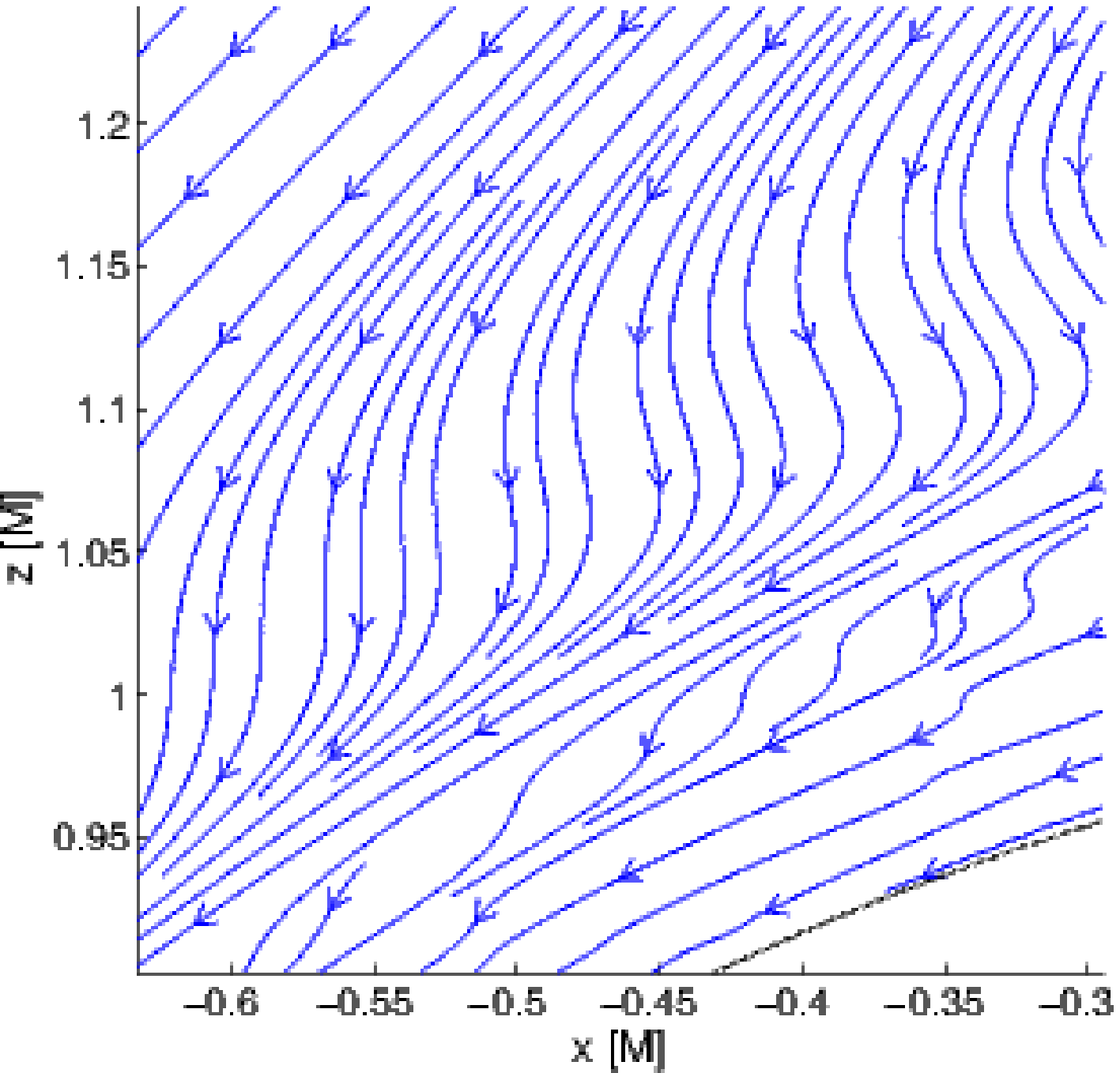}
\includegraphics[scale=0.29, clip]{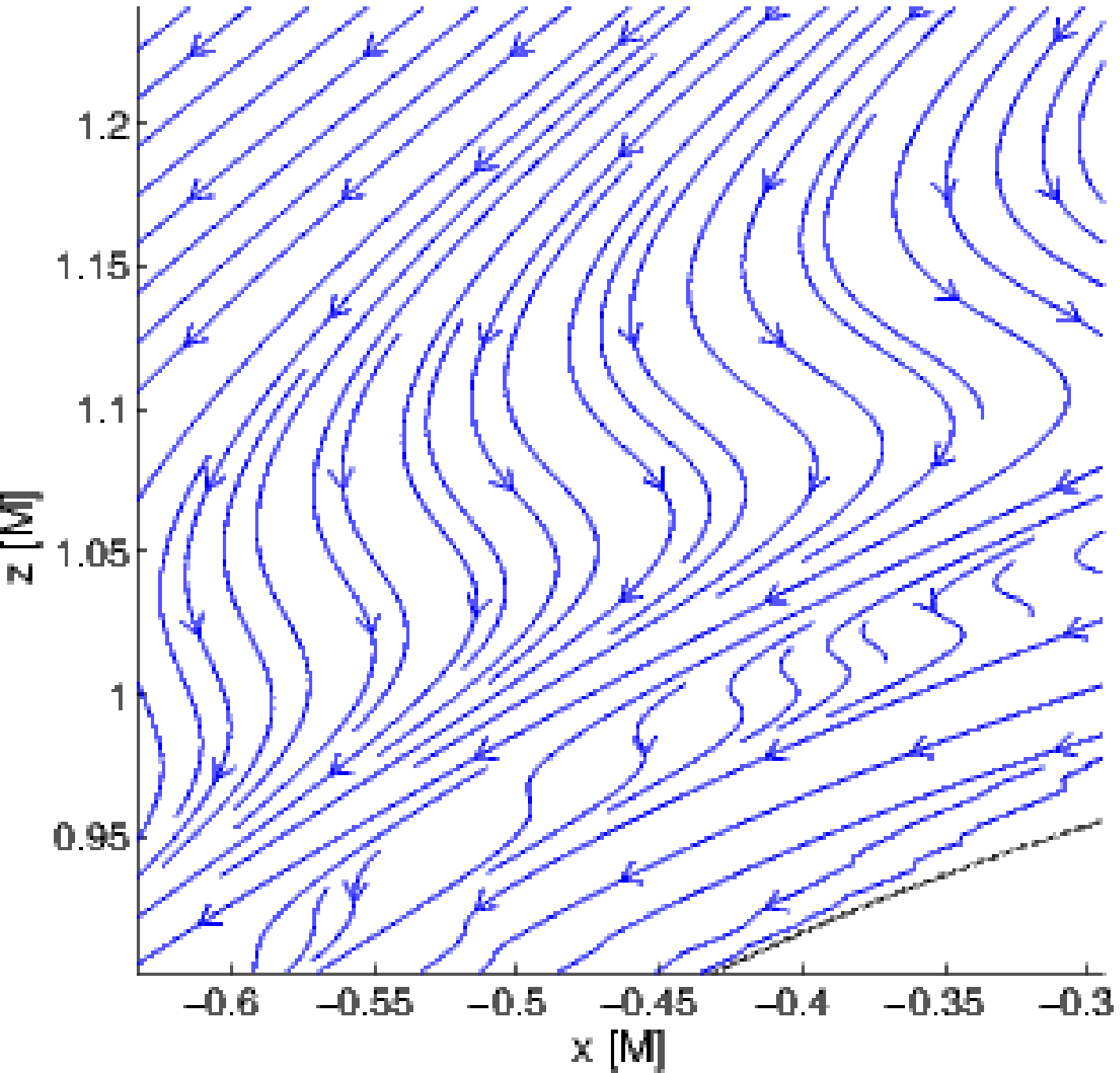}
\includegraphics[scale=0.29, clip]{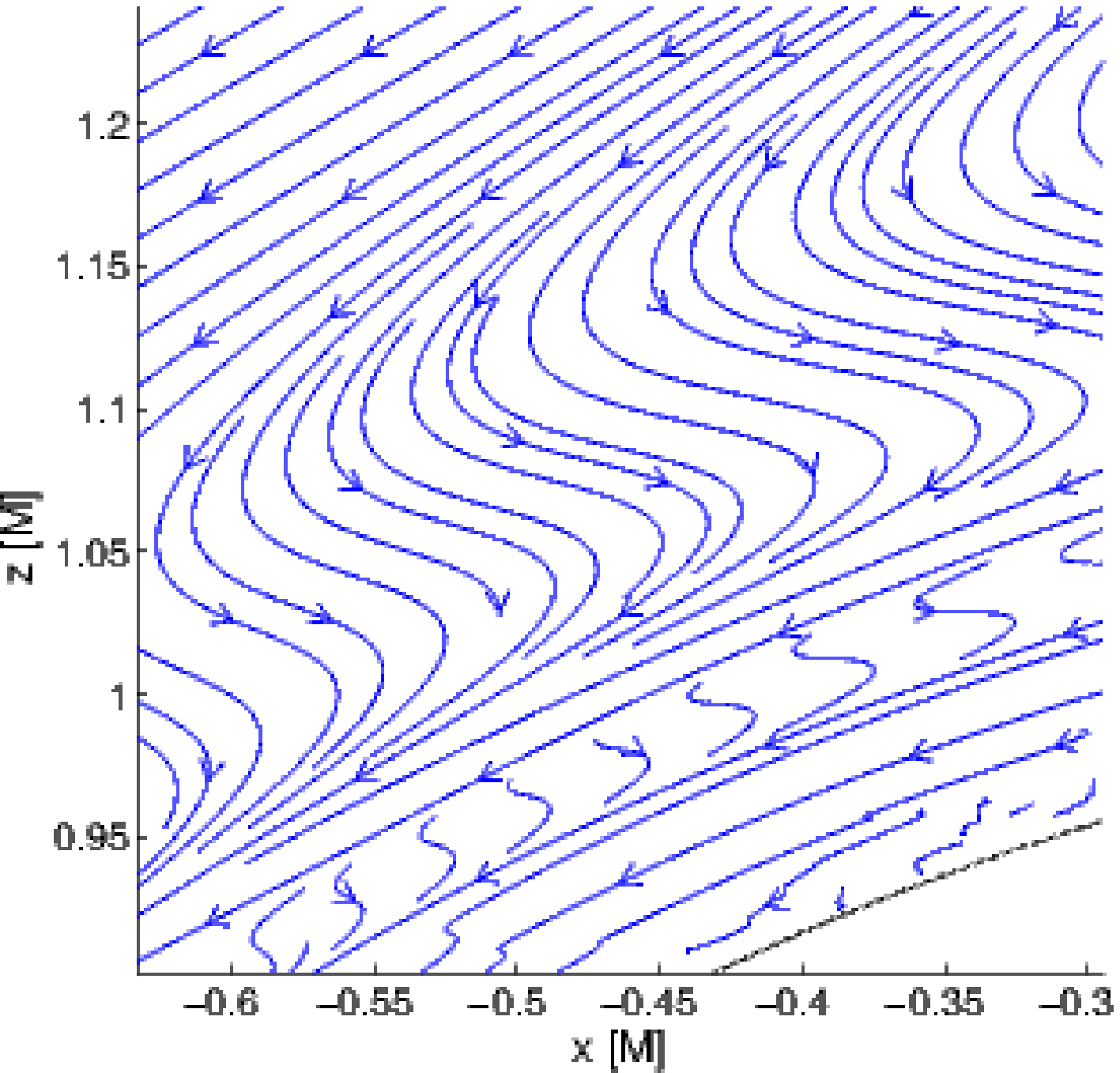}
\includegraphics[scale=0.29, clip]{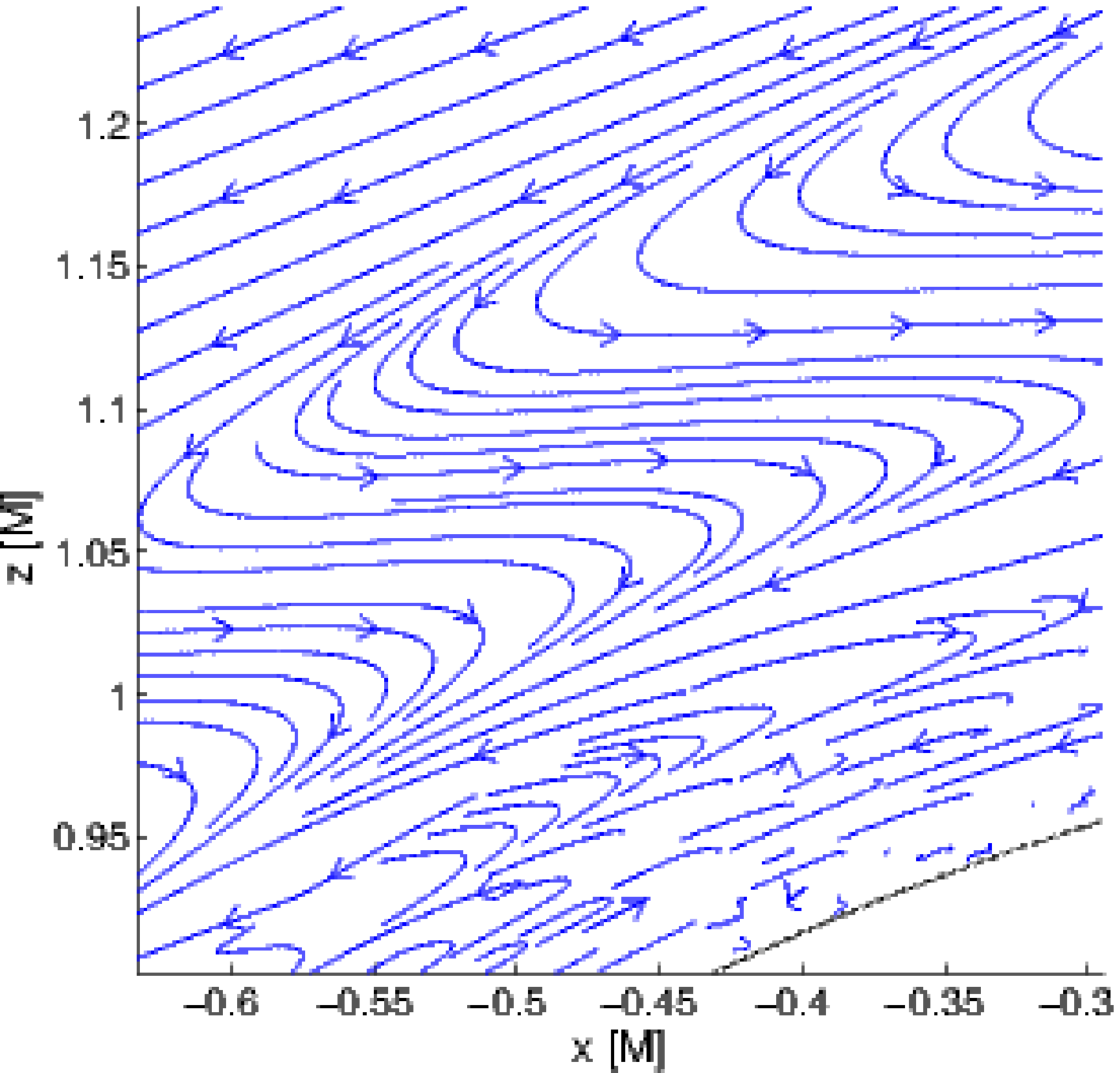}
\caption{Poloidal $(x,z)$ plane sections of FFOFI measured electric field around drifting extreme Kerr BH. First six panels show the overall changes of the field topology as the drift velocity increases in a following sequence $v_x=0,\:0.1,\:0.2,\:0.3,\:0.5$ and $0.99$. Bottom series of six panels presents detailed view on the progressive layering which takes place in a narrow region above the horizon. We observe that as the drift velocity rises the layering enhances profoundly.}
\label{el_drift}
\end{figure}

\begin{figure}[htbp!]
\centering
\includegraphics[scale=0.295, clip]{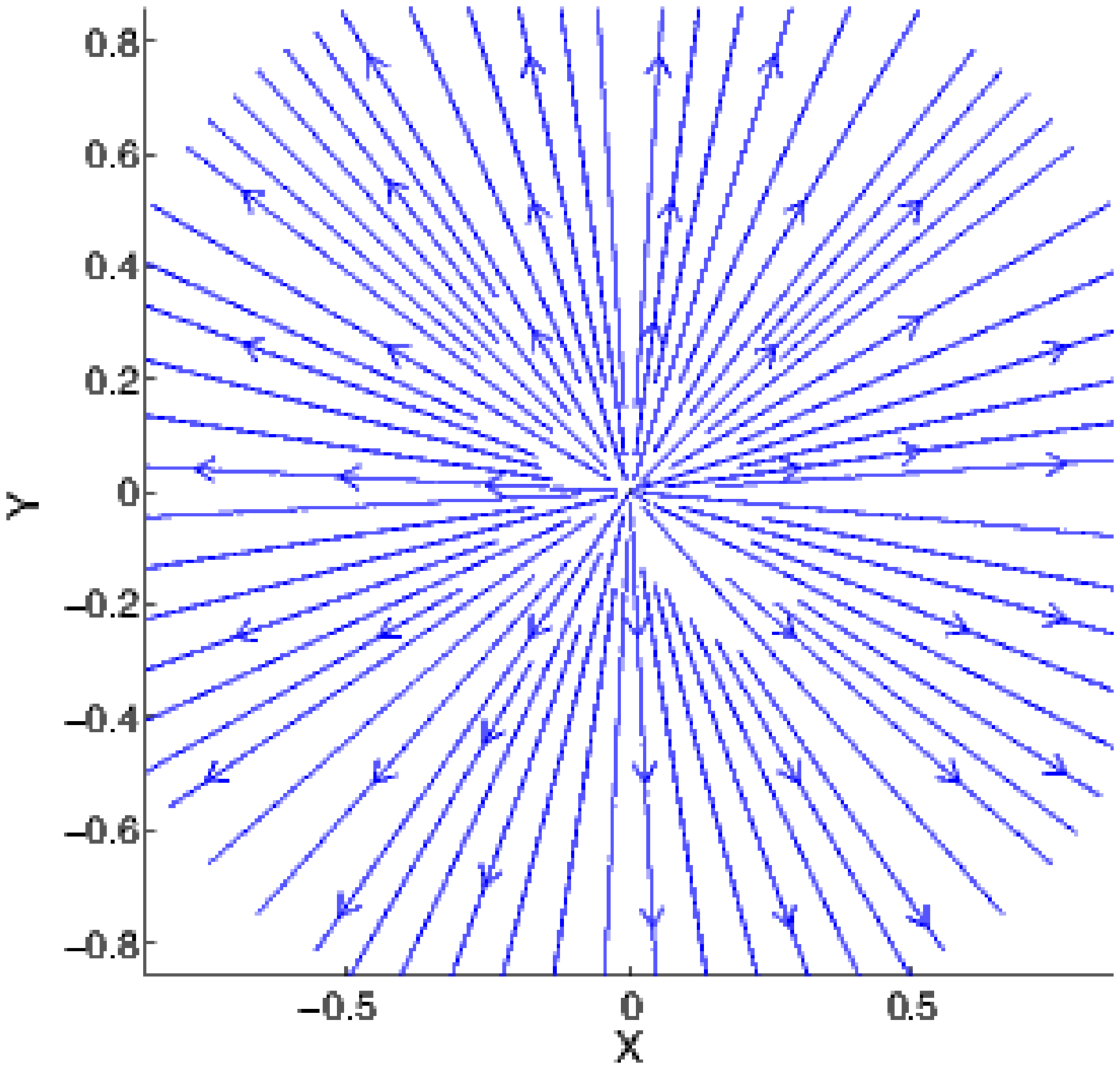}
\includegraphics[scale=0.295, trim= 8mm 0mm 0mm 0mm, clip]{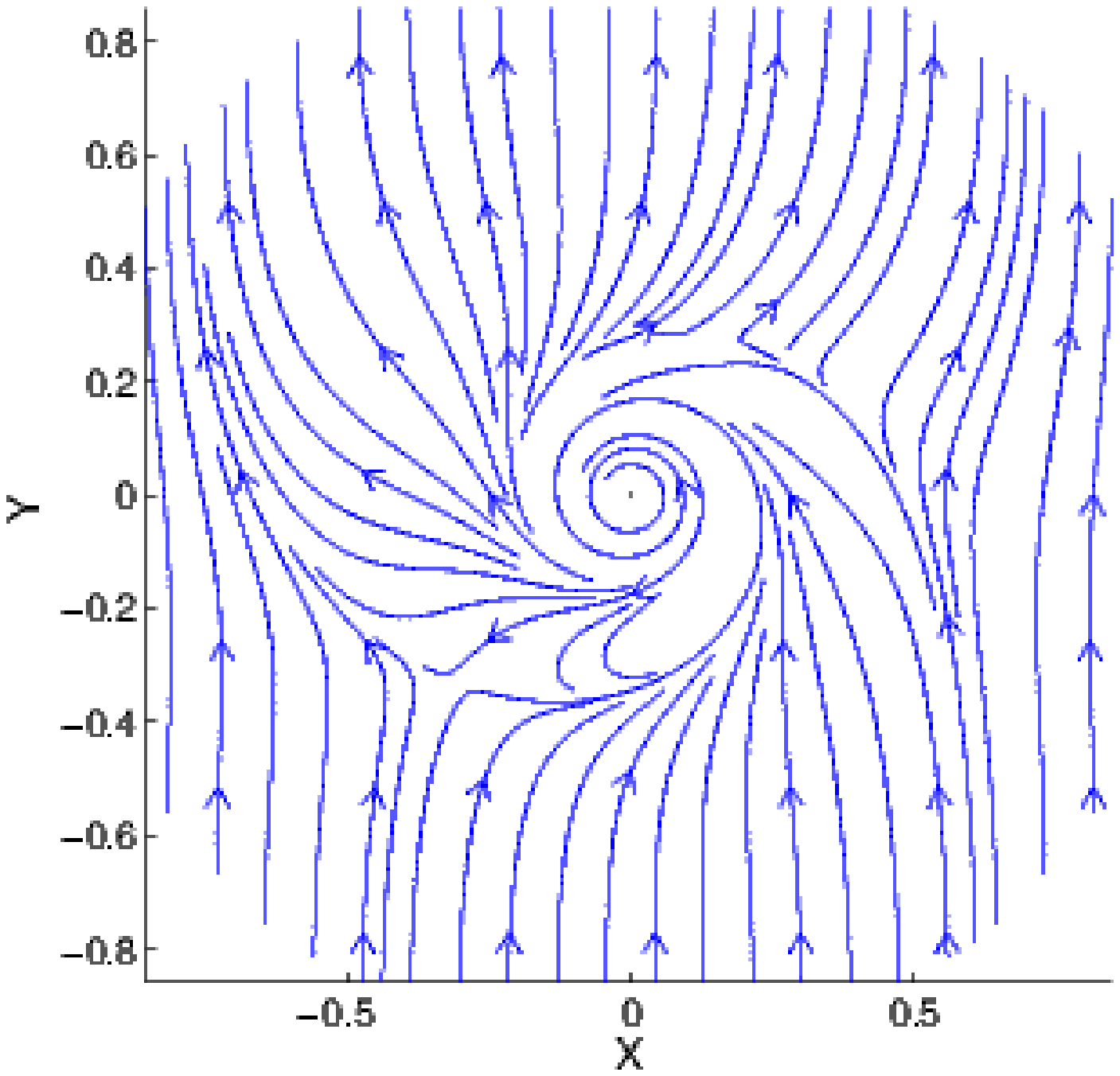}
\includegraphics[scale=0.295, trim= 8mm 0mm 0mm 0mm, clip]{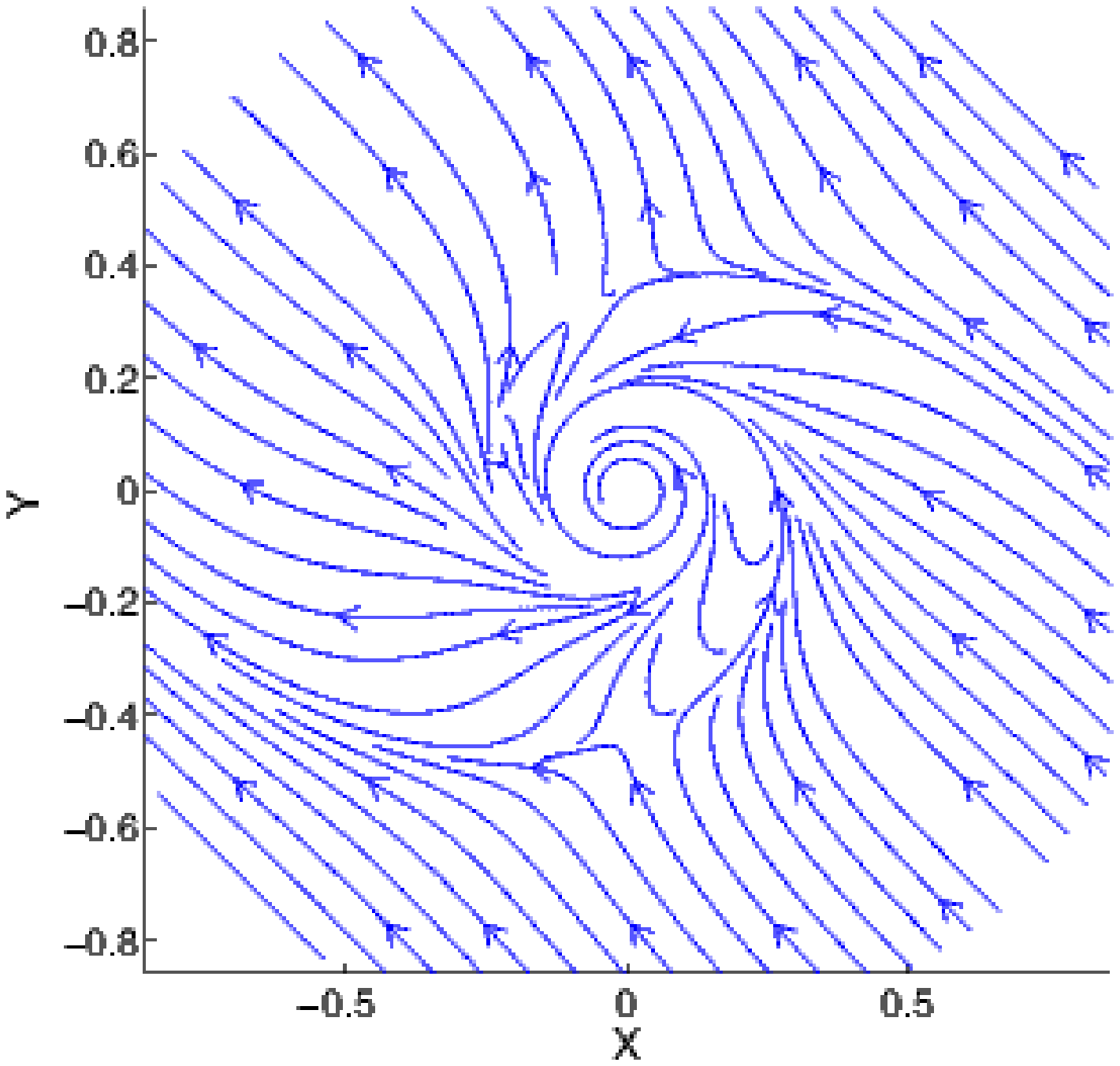}
\includegraphics[scale=0.295, clip]{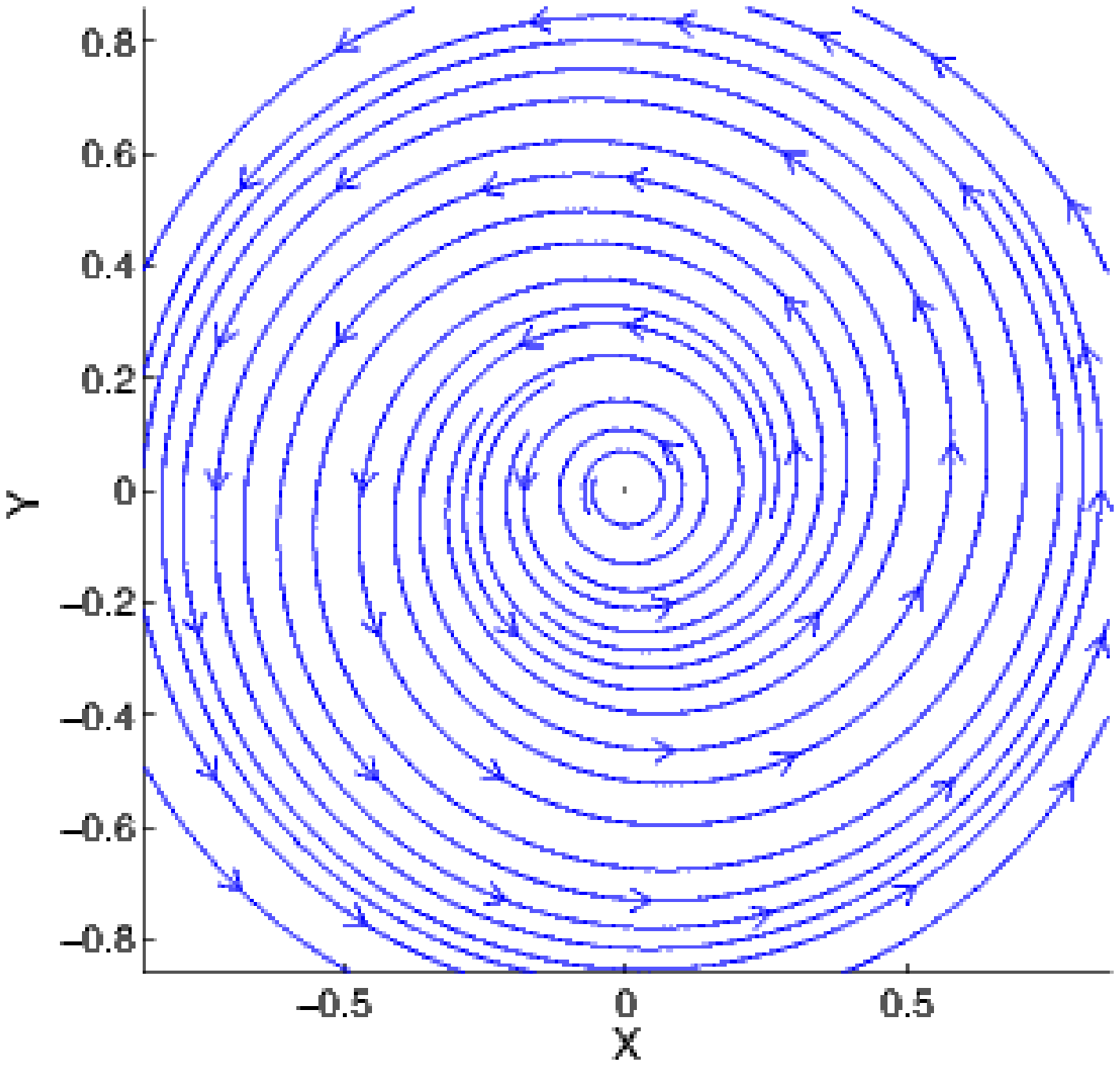}
\includegraphics[scale=0.295, trim= 8mm 0mm 0mm 0mm,clip]{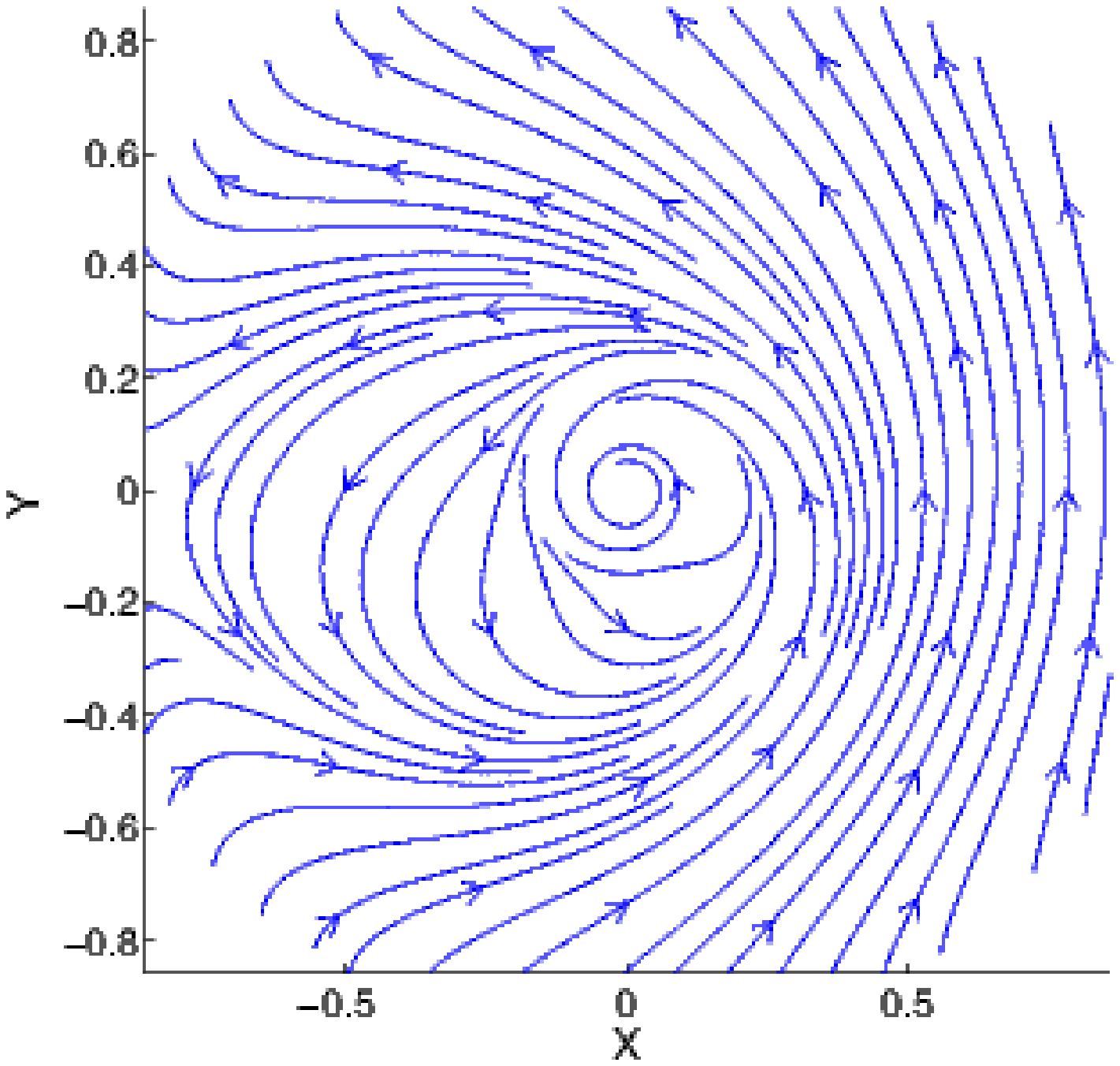}
\includegraphics[scale=0.295, trim= 8mm 0mm 0mm 0mm,clip]{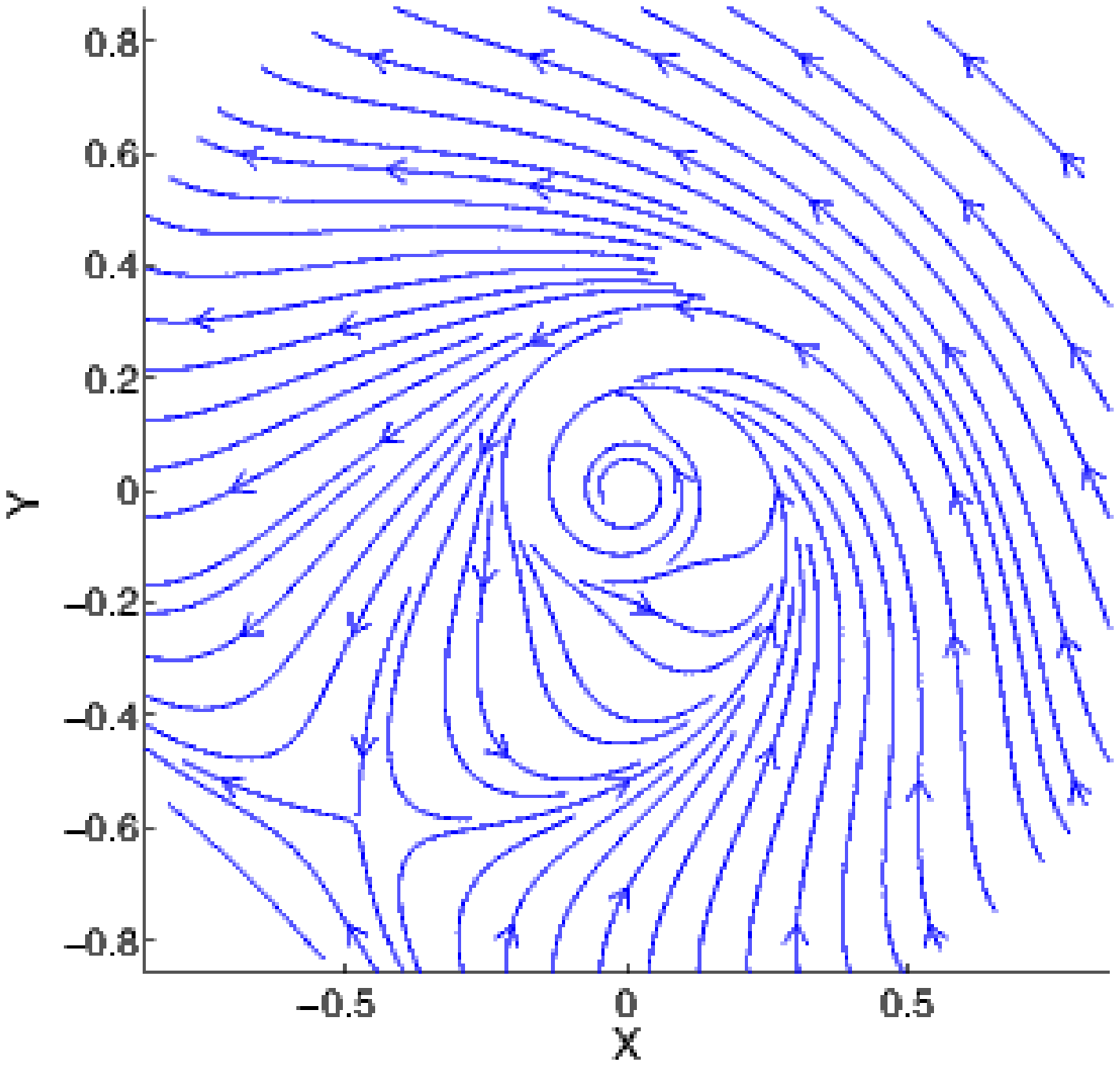}
\includegraphics[scale=0.295, clip]{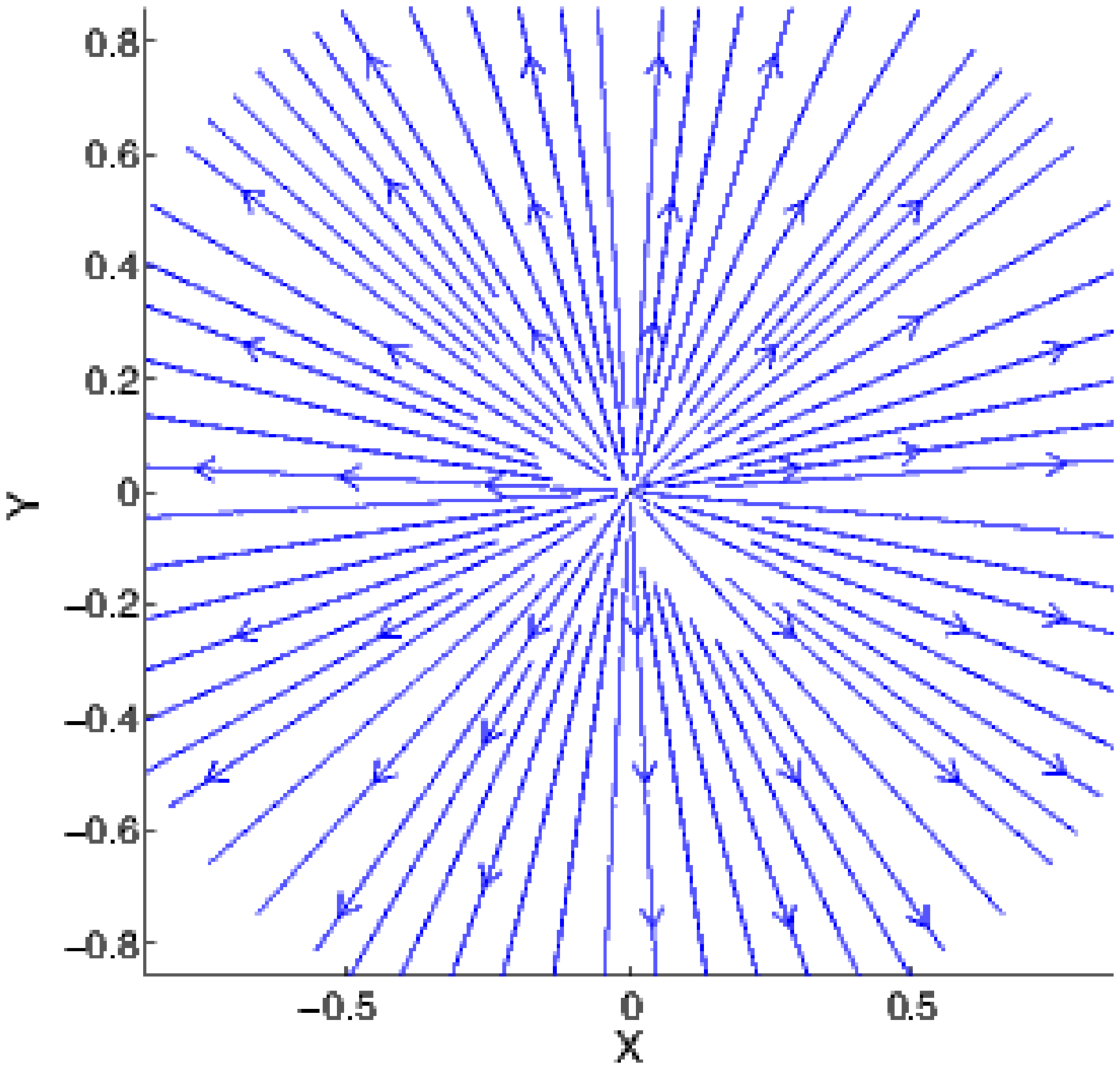}
\includegraphics[scale=0.295, trim= 8mm 0mm 0mm 0mm,clip]{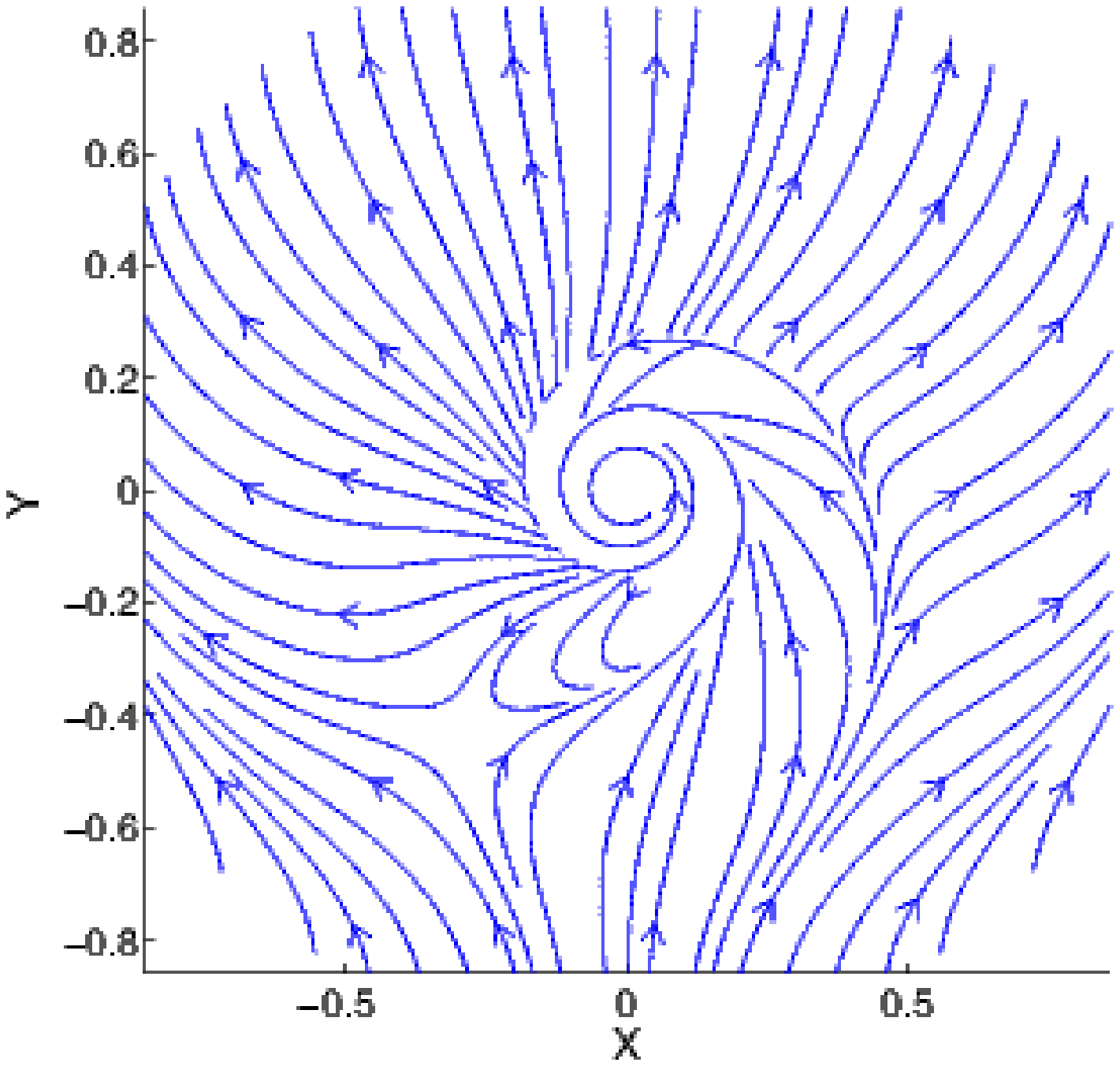}
\includegraphics[scale=0.295, trim= 8mm 0mm 0mm 0mm,clip]{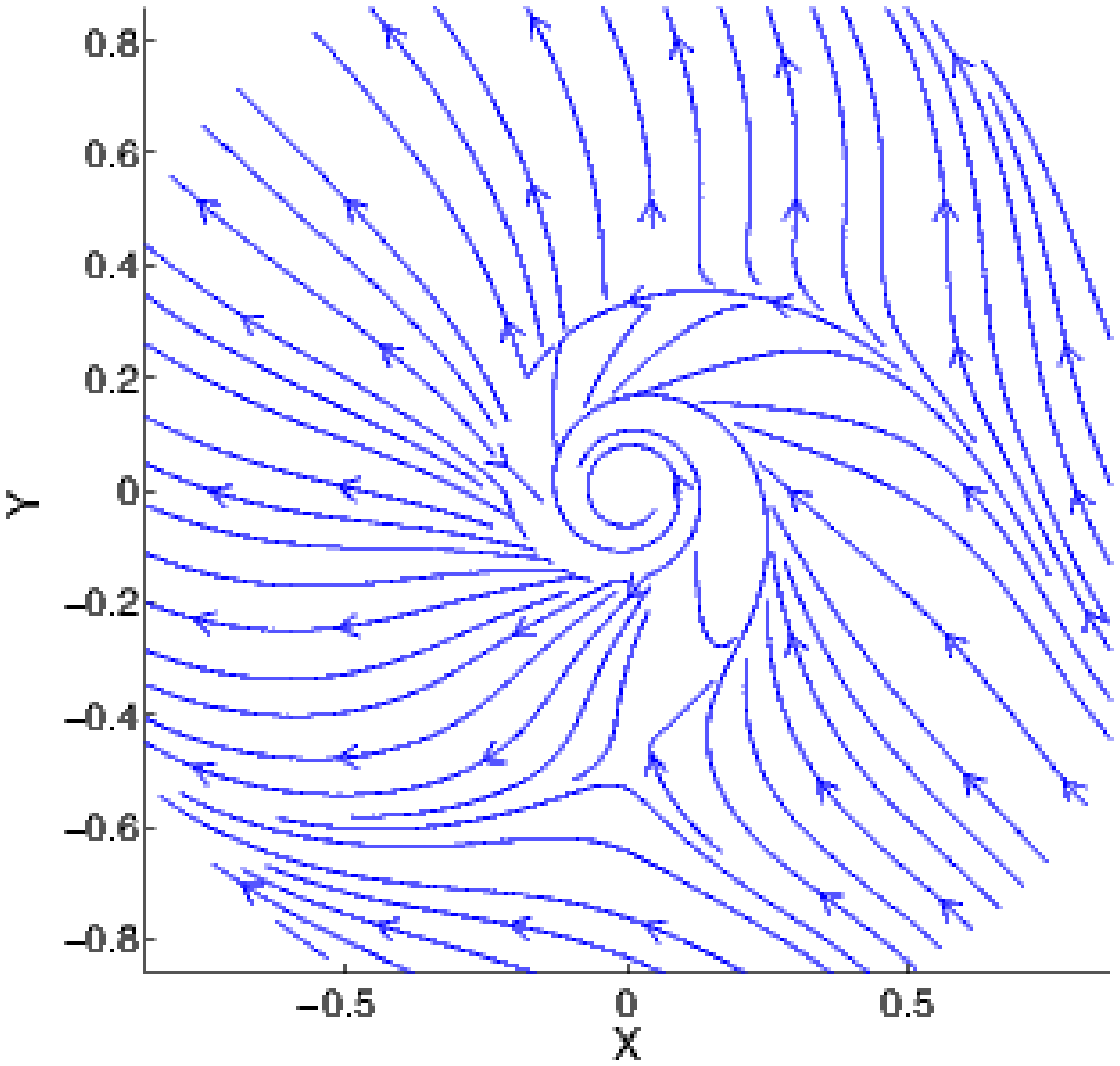}
\includegraphics[scale=0.295, clip]{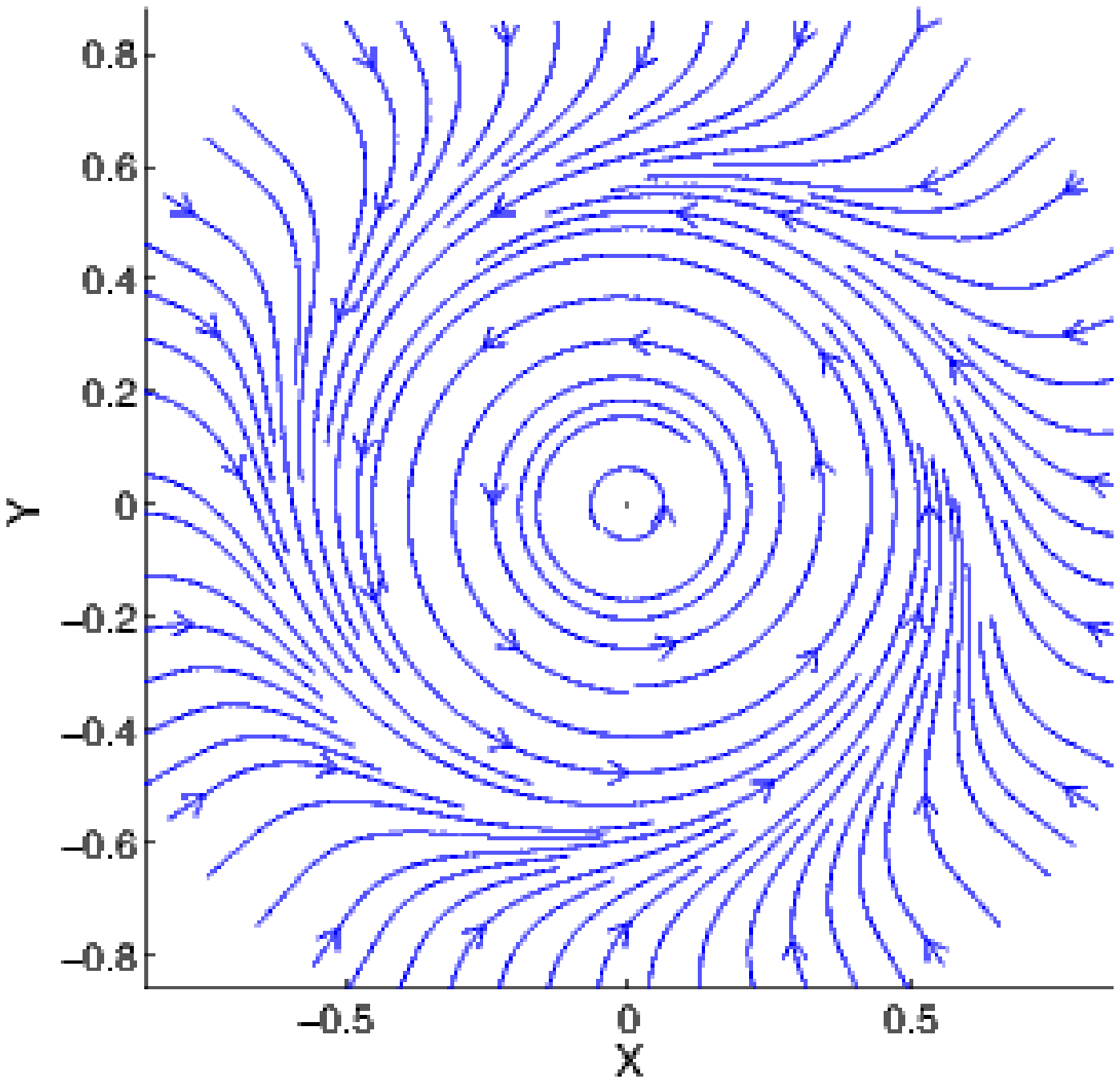}
\includegraphics[scale=0.295, trim= 8mm 0mm 0mm 0mm,clip]{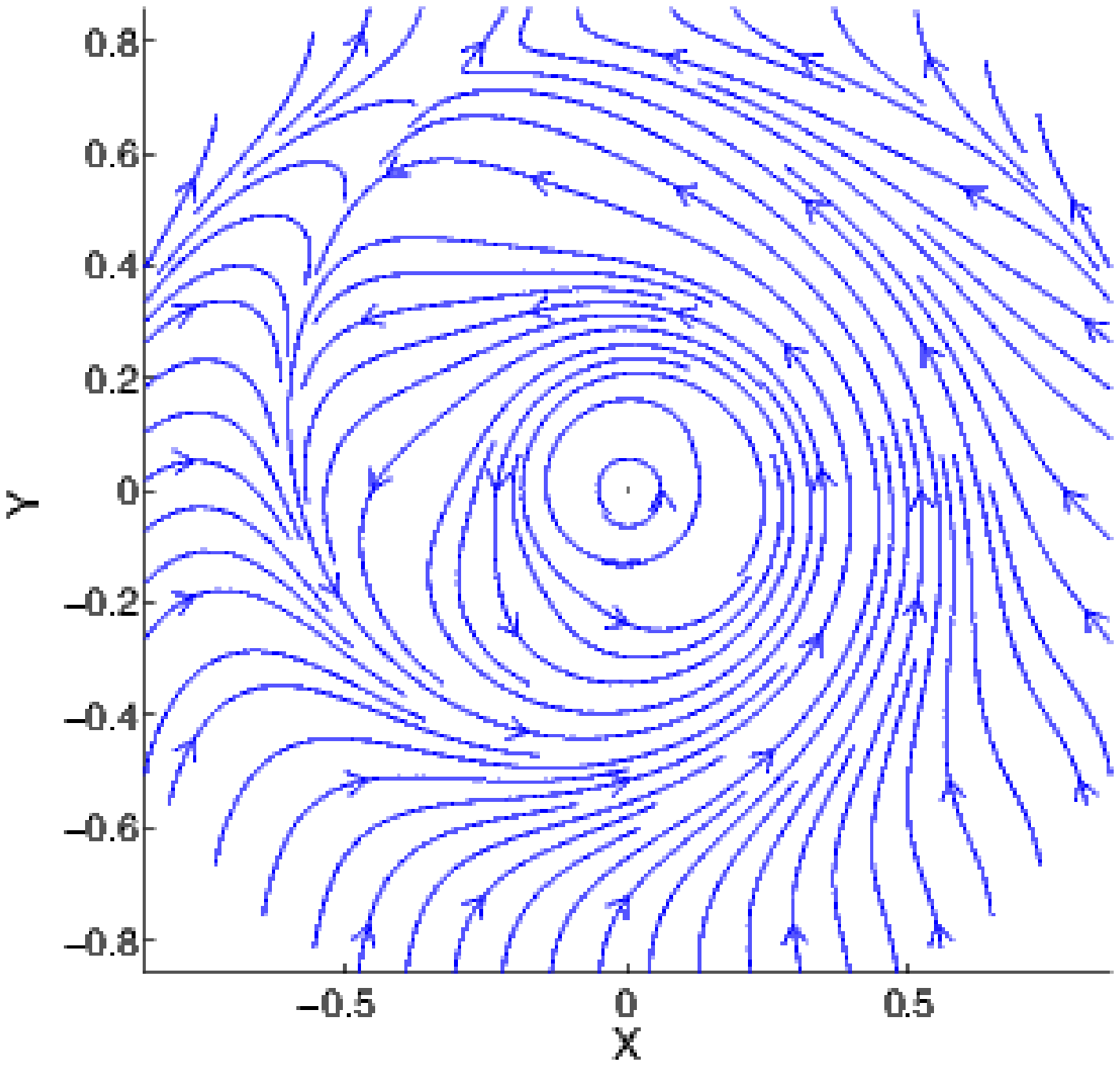}
\includegraphics[scale=0.295, trim= 8mm 0mm 0mm 0mm,clip]{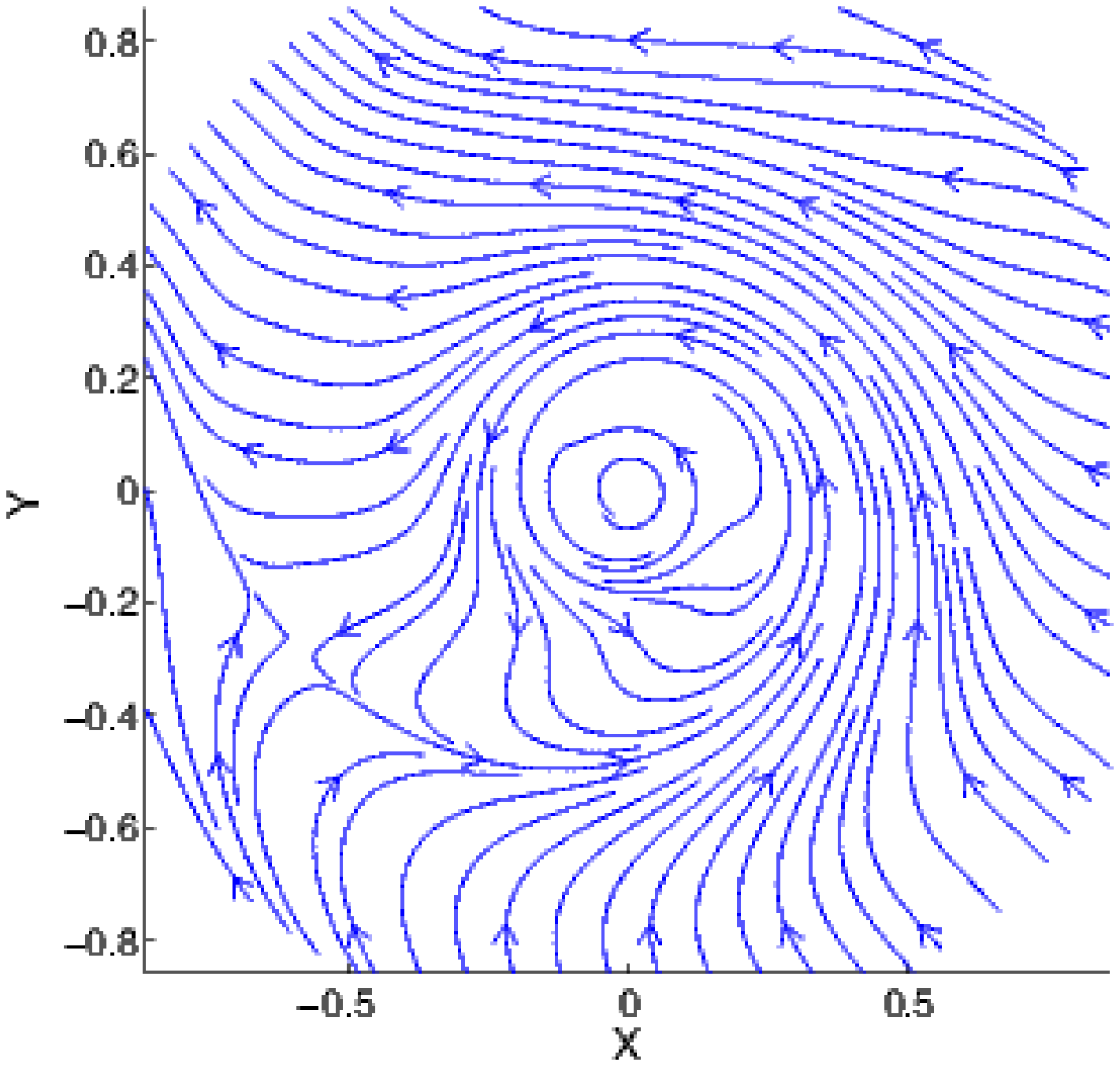}
\caption{Equatorial behaviour of the electric field around extremal Kerr BH drifting in the aligned magnetic field. Employing the rescaled radial coordinate $R\equiv\frac{r-r_{+}}{r}$ makes the horizon shrink into the single point residing at the origin. First column presents non-drifting case, in the second we set $v_x=0.5$, $v_y=0$ and in the third $v_x=0.5$, $v_y=0.5$. Four distinct frames are compared in the rows (top to bottom): ZAMO, FFOFI, co-rotating KEP+FFO and counter-rotating KEP+FFO. We observe that in all the cases the neutral electric points develop as the drift is introduced. We stress that as the original field is aligned ($B_x=0$) we measure $E^{(\theta)}=0$ in the equatorial plane provided that $v_z=0$. In other words above plots present true field lines rather than mere section.}
\label{el_drift_turtle}
\end{figure}
\clearpage

%% file: chap3nn.tex

\section{Motion of charged test particles}
\label{kaptraj}
Off-equatorial, energetically bound motion of charged particles in
strong gravitational and electromagnetic fields is pertinent to the
description of accretion disk coronae around black holes and compact
stars. In recent papers \citep{halo1, halo2}  we
discussed the existence of energetically-bound stable orbits of charged
particles occurring outside the equatorial plane, extending thus a large
variety of complementary studies
\citep[e.g.,][and further references cited therein]{bicak89,halo2_5,halo2_3,halo2_6,aliev02}. Particles on
off-equatorial stable trajectories form a coronal flow that is possible
at certain radii and for certain combinations of the model parameters,
namely, the specific charge of the particles, the conserved energy and
the angular momentum of the particle motion, the strength and
orientation of the magnetic field, and the spin of the central body.

We assume that the magnetic field permeating the corona has a
large-scale (ordered) component \citep{bisnovatyi07}. In this case,
charged particles can be trapped in toroidal regions, extending symmetrically
above and below the equatorial plane and forming two halo lobes. However,
this trapping happens only for certain combinations of model parameters
\citep{halo2}.

We consider two types of the model setup: a rotating (Kerr) black hole
in an asymptotically uniform magnetic field parallel to the symmetry
axis \citep{wald74,tomimatsu01,koide04,koide06}, and a non-rotating star (described
by the Schwarzschild metric) endowed with a rotating magnetic dipole
field \citep{petterson,sengupta}. Both cases can be regarded as integrable 
systems with the electromagnetic field acting as a perturbation.

\begin{figure}[htb]
\centering
\includegraphics[scale=0.54, clip]{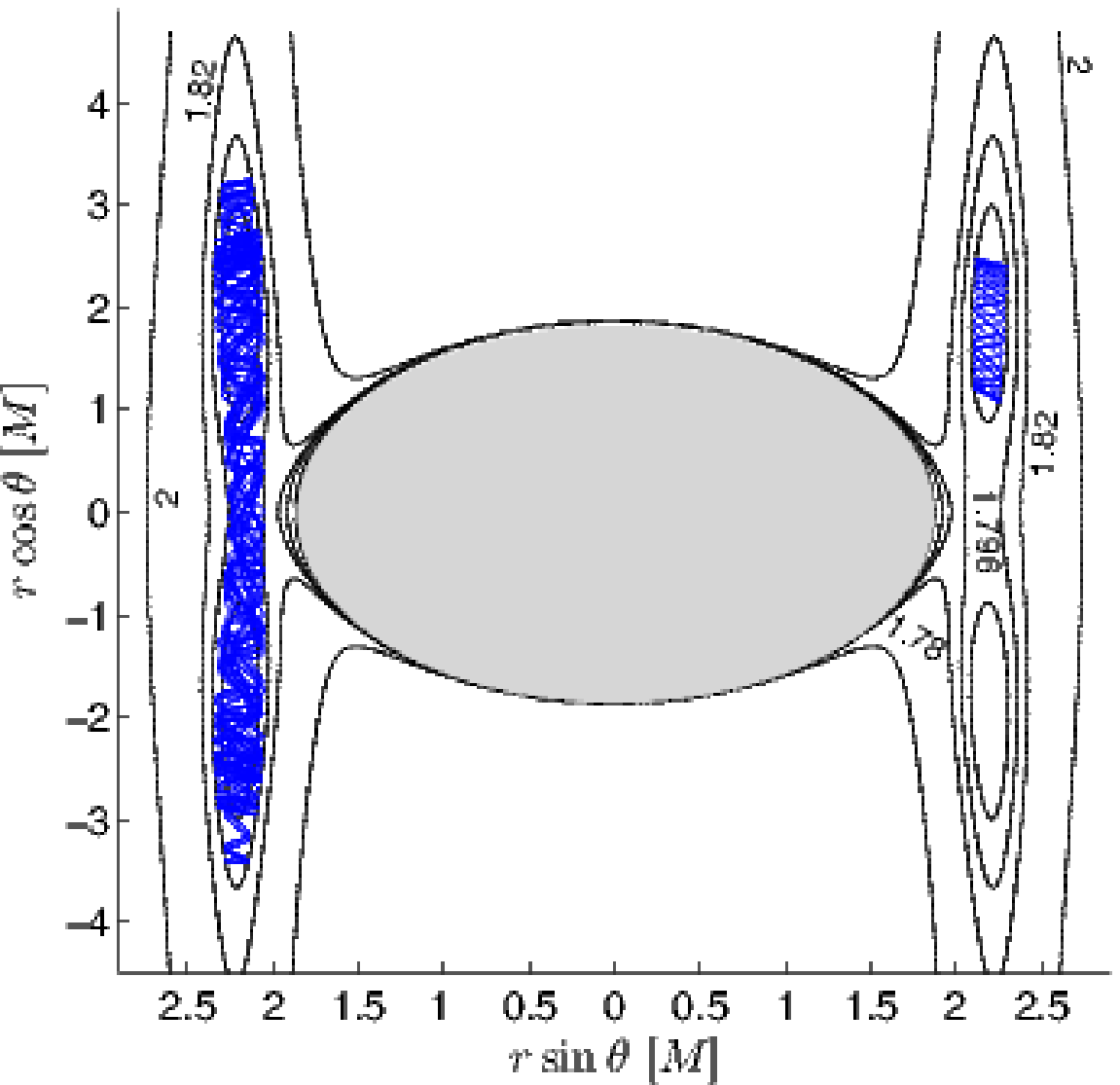}\includegraphics[scale=0.55, clip]{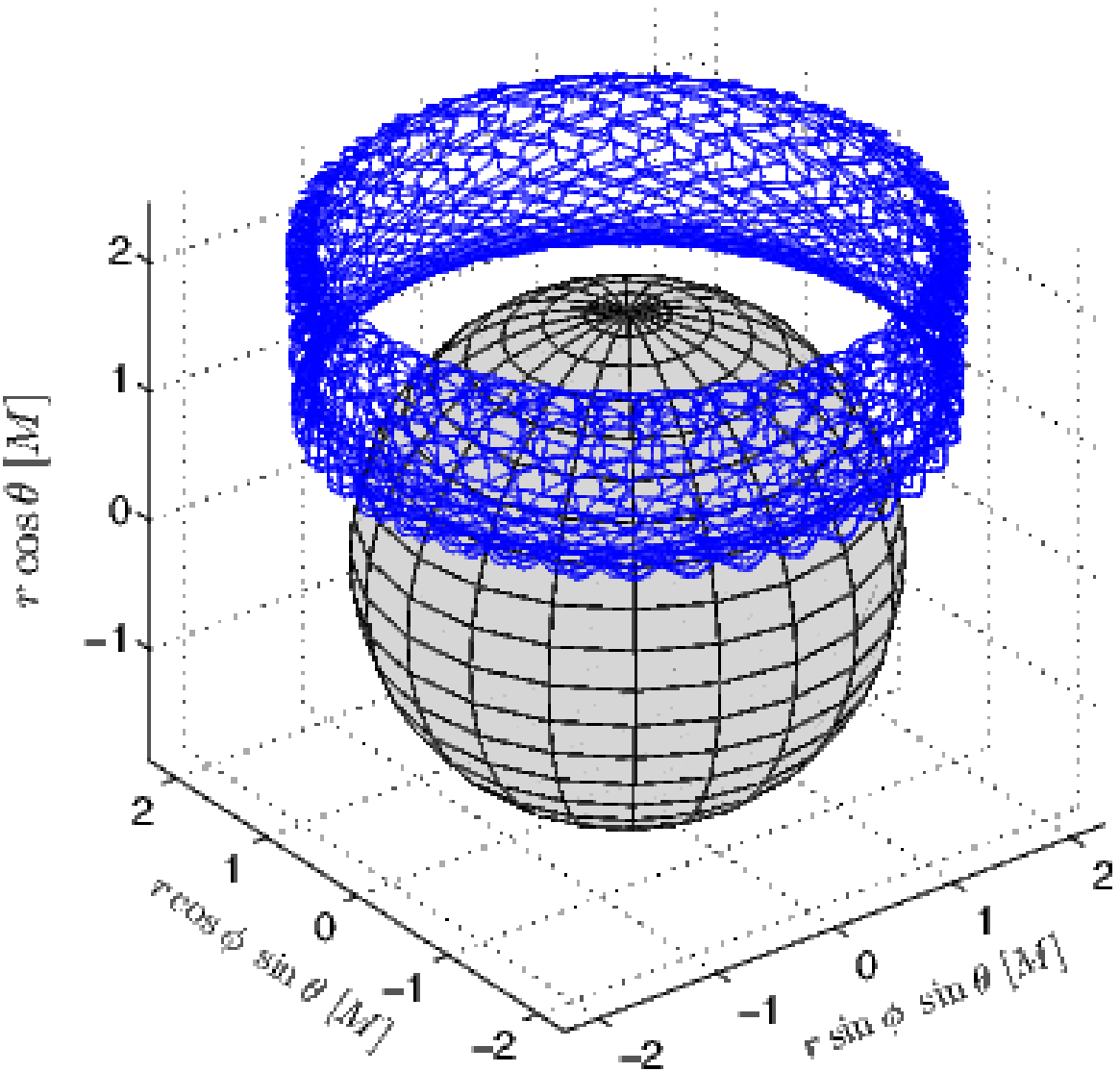}
\caption{In the left panel we present a poloidal section of the
selected isocontours of the effective potential
$V_{\rm{}eff}(r,\theta)$, \rf{effpot},  for a charged
particle ($\tilde{q}\tilde{Q}=2$, $\tilde{L}=5\;M$) on the Kerr background
($a=0.5\;M$). We assume  the presence of Wald uniform magnetic field
($\tilde{q}B_{0}=2M^{-1}$). The off-equatorial potential lobes
are present, allowing stable motion. Two exemplary trajectories of
test particles are shown -- in the left lobe a chaotic orbit of energy
$\tilde{E}=1.796$, while in the right lobe the regular, purely
off-equatorial trajectory of $\tilde{E}=1.78$. Both particles were
launched at $r(0)=3.11$, $\theta(0)=\pi/4$ with $u^r(0)=0$ and their
trajectories interweave with each other.  We plot the poloidal
$(r,\theta)$ projection of the trajectory; what appears as a lobe in the
poloidal plane is an axially symmetric 3-dimensional rotational
structure. The latter is illustrated in the right panel where the case
of the off-equatorial regular trajectory is shown.}
\label{fig1}
\end{figure}

The above-mentioned lobes are defined by the figures of the effective
potential in the poloidal plane. These were previously studied in the
context of charge separation that is expected to occur in pulsar
magnetospheres \citep[e.g.][]{neukirch93}. Here, we address whether the
trajectories within these lobes are regular (i.e., whether the system is
integrable), or if they instead exhibit a chaotic behavior. A related
problem was studied recently by \citet{japonci} in an attempt to find a
connection between chaoticness of the motion and the spin of a rotating
black hole residing in the center. These authors suggest that chaotic
behavior occurs for certain values of the black hole spin, while for
others the system is indeed regular.

The idea of investigating the connection between the spin of a black
hole and chaoticness of motion of matter near its horizon is very
interesting for the following reason. Because of high degree of symmetry of the background
spacetime, the unperturbed motion is regular \citep{carter68}; no chaos is
present. The electromagnetic perturbation may trigger the chaos,
however, its effect can be expected to diminish very near the horizon,
where strong gravity of the black hole should prevail. This is also the
region where the spin effects are most prominent. Further out various
other influences become important due to distant matter and the
turbulence in accreted material. Therefore the connection between the spin
and the motion chaoticness is best applicable in the immediate vicinity
of the black hole, i.e.\ within the inner parts of corona.

The recurrence analysis \citep{marwan} provides us with a powerful tool
for the investigation of complex dynamical systems. The method examines
the recurrences of the system to the vicinity of previously reached
phase space points. It has been typically adopted to study the
experimental data, where often only some (if not just one) of the phase
space variables are known from the measurements. Takens' embedding
theorems \citep{takens} are then used to reconstruct the phase space
portrait of such a system. In our study we are equipped with the full
phase space trajectory from the numerical integration of the equations
of motion, so that we can use the recurrence analysis directly.

It appears that the method of Recurrence Plots has not been employed in
the context of relativistic astrophysical systems yet. To this end, one
needs a consistent definition of the neighborhood of a point in the
phase space in a curved spacetime. Below, we discuss the phase space distance
and suggest a form of the distance norm suitable in such circumstances.

\begin{figure}[htb]
\centering
\includegraphics[scale=0.72, clip]{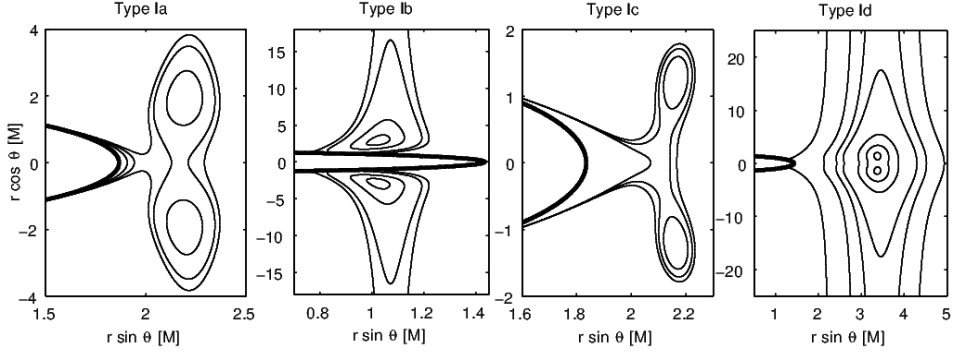}
\caption{The overview of possible topologies of the off-equatorial
potential structure above the event horizon (thick line in plots) of
Kerr black hole endowed with the Wald test field.}
\label{wald_abc}
\end{figure}

\section{Equations of motion and the effective potential}
\label{pohrce}
The phase space trajectories of integrable systems are regular, meaning
that they are bound to the surface of an $n$-dimensional torus, where
$n$ is the number of degrees of freedom. The torus is determined
uniquely by $n$ constants of motion that are present in such a system.
Its behavior can be explored by Poincar\'e surfaces of section, which
are defined by  intersections of the phase space trajectory with a
$2$-dimensional plane \citep{lieberman}. On the other hand,
non-integrable chaotic systems generally have fewer integrals
than the number of degrees of freedom. In general, both the regular and
the chaotic orbits may coexist in the phase space of a single system.

Chaotic orbits are ergodic on the given hypersurface. Its dimension is
now larger than $n$, and the section points thus fill areas in the plot
of the Poincar\'e surface. However, depending on the initial conditions,
regular orbits can also appear in non-integrable systems. Such orbits
maintain the value of some additional constant of motion, although it is
not generally possible to write this constant in an explicit form. In
the context of motion around black holes perturbed by (weak) external
sources, various aspects of chaos were studied e.g.\ by
\citet{karas92,nakamura93,podolsky98}, and very recently by
\citet{semerak10}.

A standard approach to an integrable system with a non-integrable
perturbation assumes complete control over the strength of the
perturbation (i.e., the perturbation can be set to be arbitrarily weak). If this were
the case, we could first switch the perturbation completely off, analyze the
orbits, and then observe the impact of gradually increasing the
perturbation strength upon these orbits. However, the class of off-equatorial
bound orbits only exists when the electromagnetic term is strong enough
to balance the gravitational attraction of the central body. Then the
(sufficiently strong) perturbation is by itself the cause of the new
kind of the regular motion that happens outside the equatorial plane. 

\begin{figure}[htb]
\centering
\includegraphics[scale=0.5, clip]{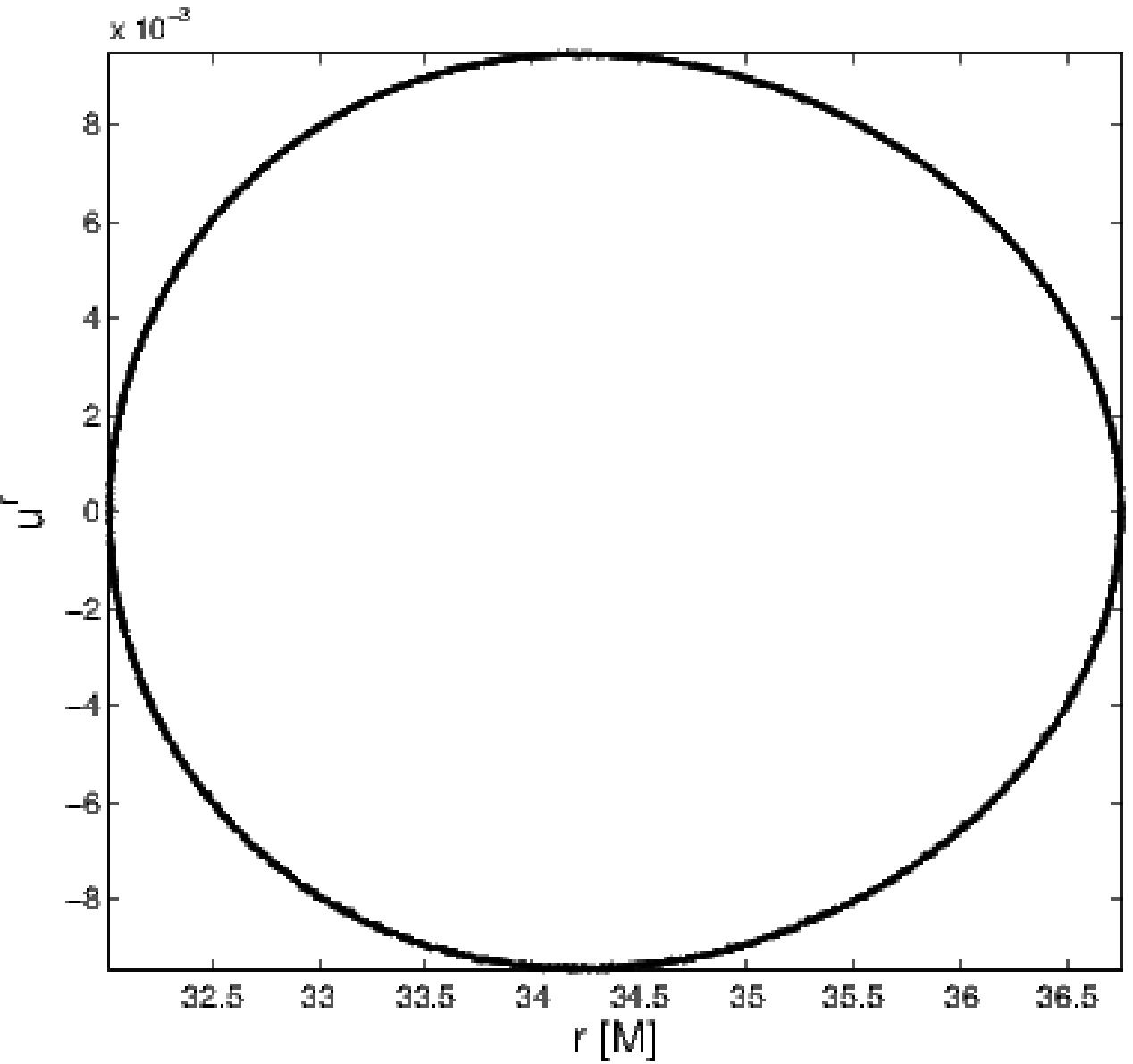}~~
\includegraphics[scale=0.48,  clip]{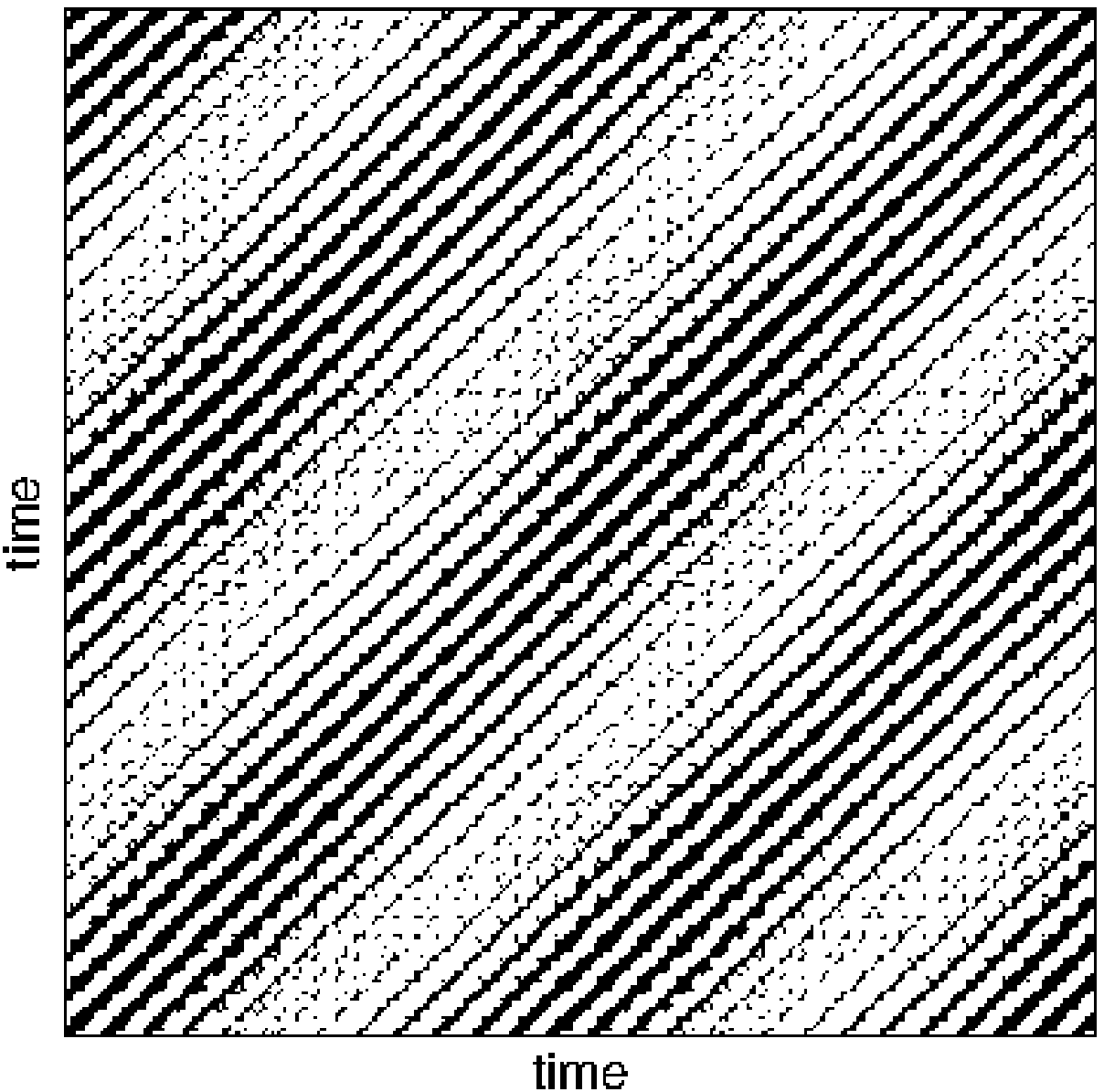}
\caption{Regular 
motion in the equatorial potential lobe in the fully integrable system
of charged test particle ($\tilde{E}=0.99$, $\tilde{L} =5M$,
$\tilde{q}=10^4$, $r(0)=32.02\:M$, $\theta(0)=1.54$) in the pure
Kerr-Newman spacetime ($\tilde{Q}=3\times10^{-5}$, $a=0.5\:M$) endowed
with the fourth Carter constant of motion $\mathcal{L}$. 
Long diagonals parallel to the LOI  are general
hallmark of regularity in the RPs. }
\label{rppoinc1}
\end{figure}

Having this delicacy on mind, we shall use the usual Hamiltonian formalism 
to express equations of motion
governing the trajectories. We first
construct the super-Hamiltonian $\mathcal{H}$ \citep{mtw},
\begin{equation}
\label{SuperHamiltonian}
\mathcal{H}=\textstyle{\frac{1}{2}}g^{\mu\nu}(\pi_{\mu}-qA_{\mu})(\pi_{\nu}-qA_{\nu}),
\end{equation}
where $m$ and $q$ are the rest mass and charge of the test particle,
$\pi_{\mu}$ is the generalized (canonical) momentum, $g^{\mu\nu}$ is the
metric tensor, and $A_{\mu}$ denotes the vector potential of the
electromagnetic field. The latter is related to the electromagnetic
tensor $F_{\mu\nu}$ by $F_{\mu\nu}=A_{\nu,\mu}-A_{\mu,\nu}$. Unless
otherwise stated, we will use geometrical units, $G=c=1$ (see Appendix \ref{gu}).

The Hamiltonian equations are given as
\begin{equation}
\label{HamiltonsEquations}
\frac{{\rm d}x^{\mu}}{{\rm d}\lambda}\equiv p^{\mu}=
\frac{\partial \mathcal{H}}{\partial \pi_{\mu}},
\quad 
\frac{d\pi_{\mu}}{d\lambda}=-\frac{\partial\mathcal{H}}{\partial x^{\mu}},
\end{equation}
where $\lambda=\tau/m$ is the affine parameter, $\tau$ denotes the
proper time, and $p^{\mu}$ is the standard kinematical four-momentum for
which the first equation reads $p^{\mu}=\pi^{\mu}-qA^{\mu}$.

\begin{figure}[htb]
\centering
\includegraphics[scale=0.51, clip]{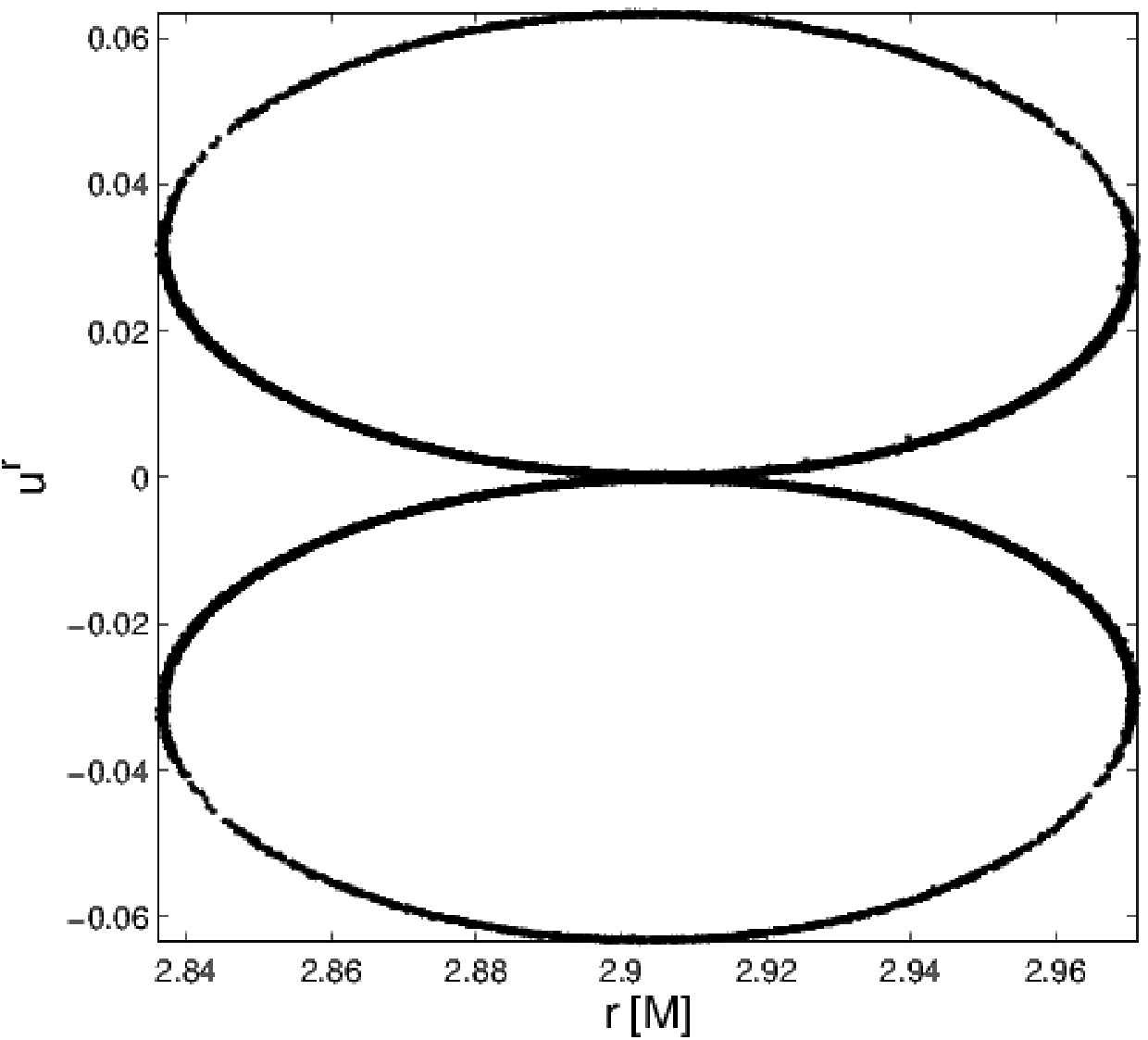}~~
\includegraphics[scale=0.51, clip]{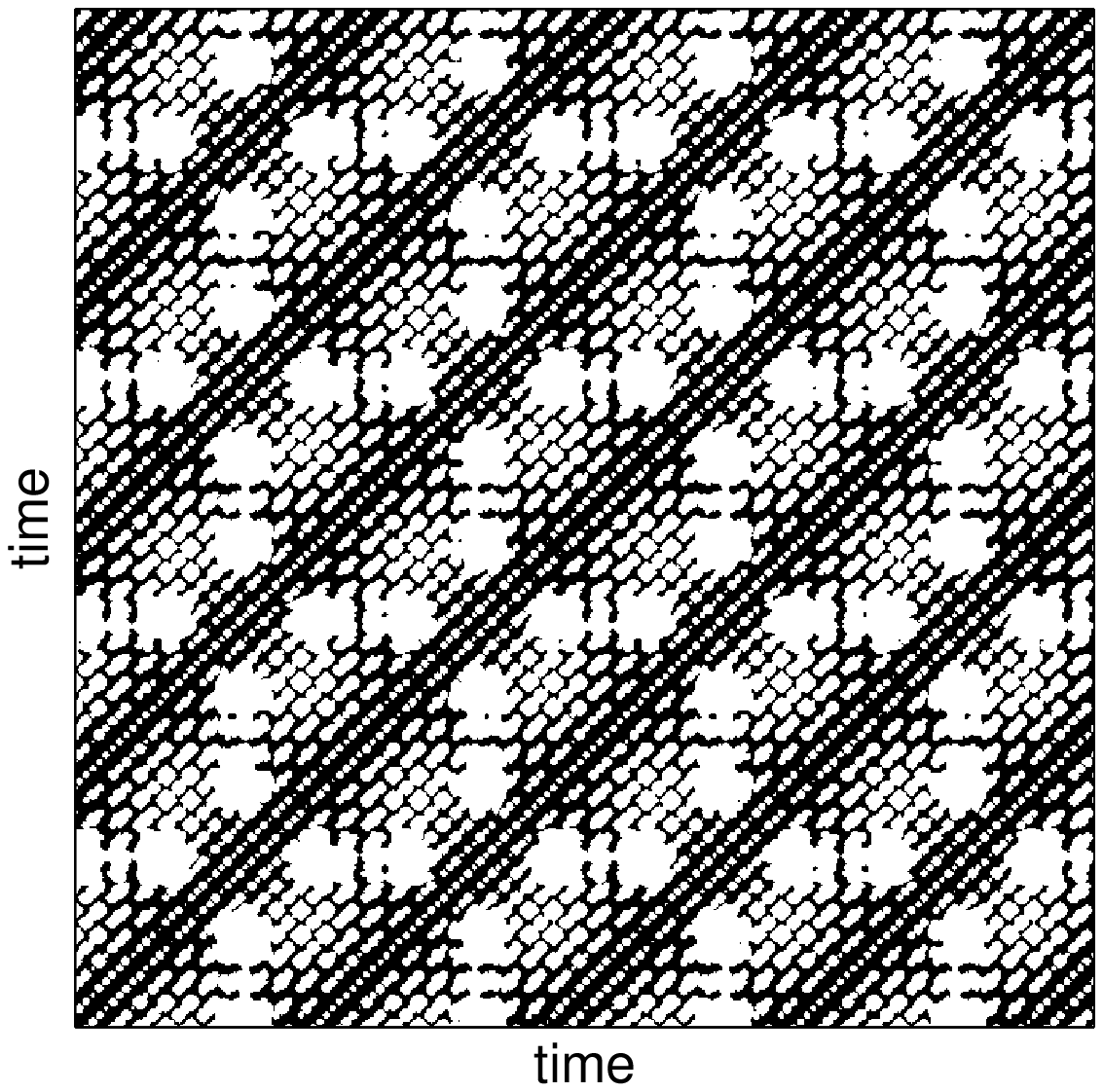}
\caption{Regular off-equatorial motion of a charged test
particle ($\tilde{E}=1.77$, $\tilde{L} =5M$, $r(0)=2.9\:M$, $\theta(0)=0.856$
and $u^r(0)=0$) on the Kerr background ($a=0.5\:M$) enriched with the
Wald test field ($\tilde{q}B_{0}=2M^{-1}$, $\tilde{q}\tilde{Q}=2$). The
diagonal structures typical for trajectories in integrable systems are
preserved, though the pattern is more complicated.}
\label{rppoinc2}
\end{figure}

In the case of stationary and axially-symmetric systems, we
identify two constants of motion, namely, the energy $E$ and angular
momentum $L$. From the second Hamiltonian equation (\ref{HamiltonsEquations})
we obtain
\begin{eqnarray}
\label{Momenta}
\pi_t&=&p_t+qA_t\equiv-E\\
\label{27}
\pi_{\varphi}&=&p_{\varphi}+qA_{\varphi}\equiv L.
\end{eqnarray}

The trajectory is specified by the integrals of motion $E$, $L$, and 
the initial values $r(0)$, $\theta(0)$ and $u^r(0)$. The initial
$u^{\theta}(0)$ can be calculated from the normalization condition,
$g^{\mu\nu}u_{\mu} u_{\nu}=-1$ (we always choose the non-negative
root).

The effective potential can be derived in the following form:
\begin{eqnarray}
\label{effpot}
V_{\rm eff}=\frac{-\beta+\sqrt{\beta^2-4\alpha\gamma}}{2\alpha},
\end{eqnarray}
where
\begin{eqnarray}
\label{EfectivePotential_parts}
\alpha&=&-g^{tt},\\
\beta&=&2[g^{t\varphi}(\tilde{L}-\tilde{q}A_{\varphi})-g^{tt}\tilde{q}A_{t}],\\
\gamma&=&-g^{\varphi\varphi}(\tilde{L}-\tilde{q}A_{\varphi})^2-g^{tt}\tilde{q}^2A_t^2+2g^{t\varphi}\tilde{q}A_t(\tilde{L}-\tilde{q}A_{\varphi})-1,
\end{eqnarray}
and where we introduce specific quantities
$\tilde{L}\equiv{}\frac{L}{m}$, $\tilde{E}\equiv{}\frac{E}{m}$ and the
specific charge $\tilde{q}\equiv{}\frac{q}{m}$. Local minima of 
$V_{\rm eff}(r,\theta)$ reflect the
location of stable orbits of test particles. Off-equatorial potential
minima were identified and various types of potential lobes were
discussed elsewhere \citep[see][]{halo2}\footnote{We have employed the method of effective potential to
study the stability of the motion, however, we note that the 
force formalism \citep{halo2_26,halo2_28} can
serve as a very efficient alternative tool. In particular, 
the off-equatorial motion of charged particles can be examined
via the procedure described in \citet{halo2}. This allows us to
localize minima of the effective potential around which the stable
orbits occur.}.
We can express the effective potential (\ref{effpot}) also as a function of $r$
and $u^r$, and use it to determine the boundaries of allowed regions in
Poincar\'e surfaces of section for a given value of 
$\theta$.

\begin{figure}[htb]
\centering
\includegraphics[scale=0.49, clip]{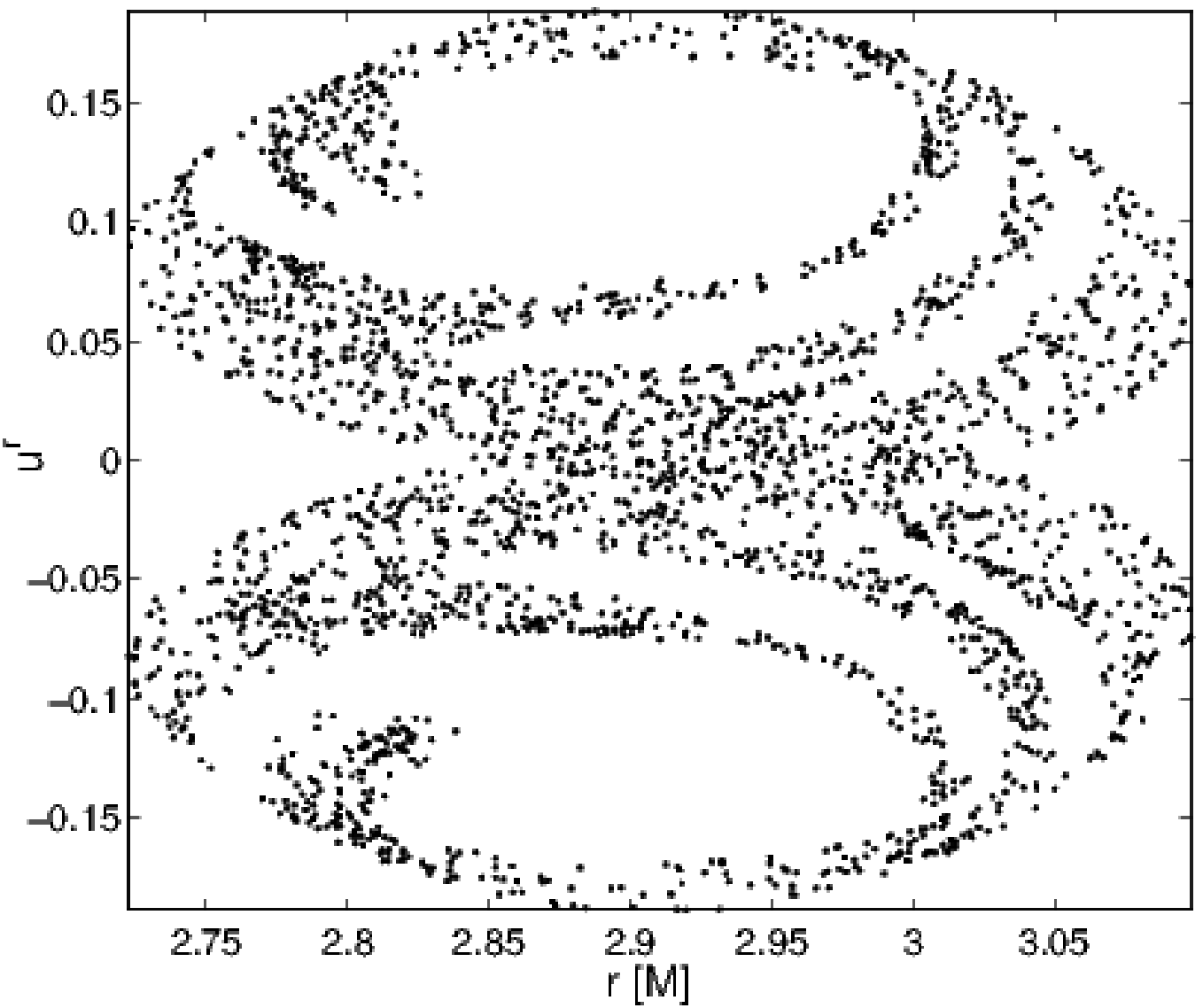}~~\includegraphics[scale=0.49, clip]{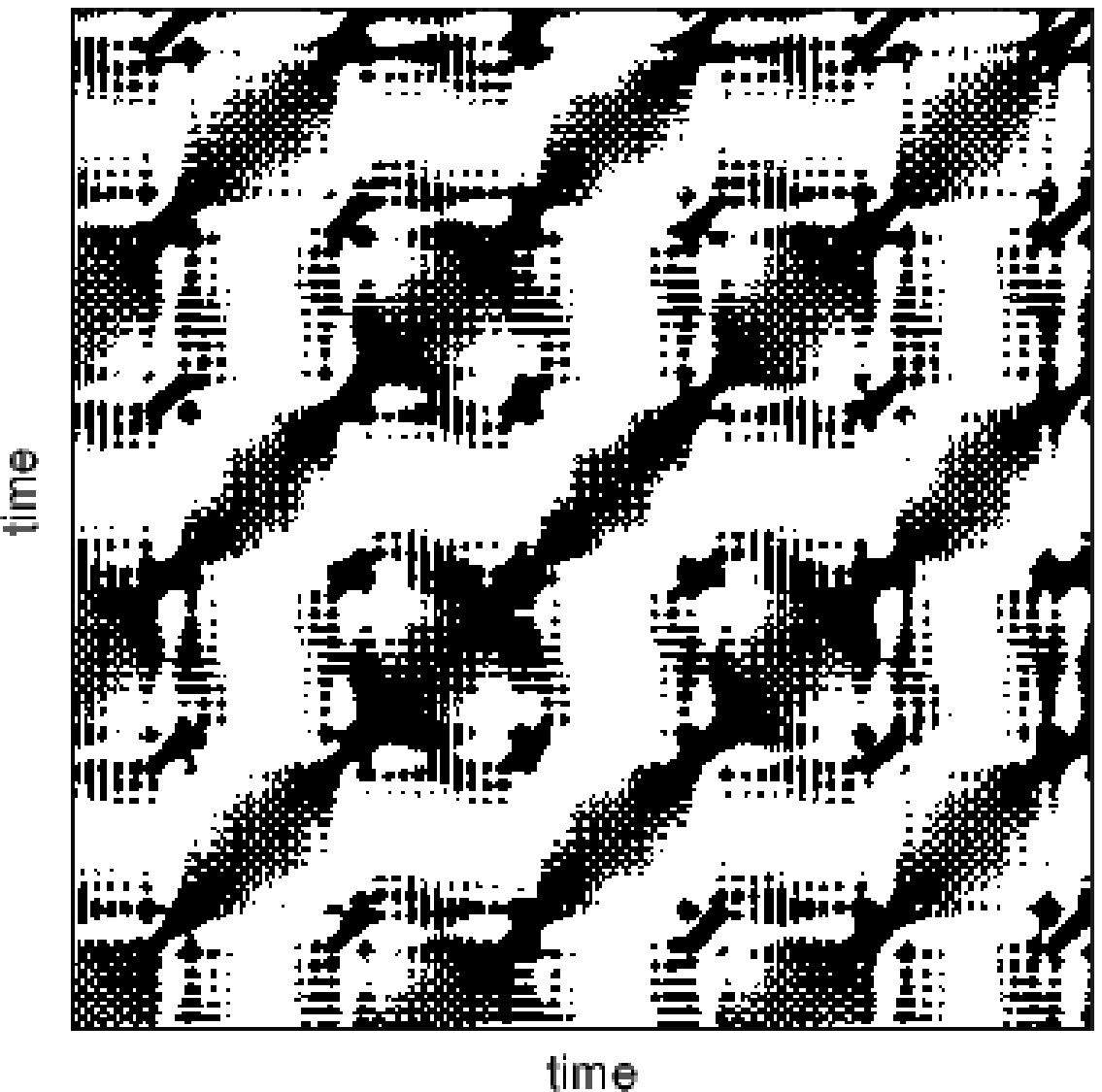}
\caption{A transitional state between the regular and chaotic regimes of
motion of a highly charged test particle which only differs from the
previous case by increasing the energy to $\tilde{E}=1.796$. The diagonal
lines in the RP are partially disrupted, indicating the onset of
chaos.}
\label{rppoinc4}
\end{figure}

We employ the Kerr metric in standard Boyer-Lindquist coordinates $t$, $r$, $\theta$,
$\varphi$ given by \rf{kn}. It is sufficient to consider
positive values of the spin parameter $a$ without loss of generality (the cases of
prograde and retrograde motion are distinguished by the sign of the
particle charge and the orientation of the magnetic field). By setting
$a=0$ the metric (\ref{kn}) reduces to the static one describing the Schwarzschild spacetime.


\begin{figure}[htb]
\centering
\includegraphics[scale=0.51, clip]{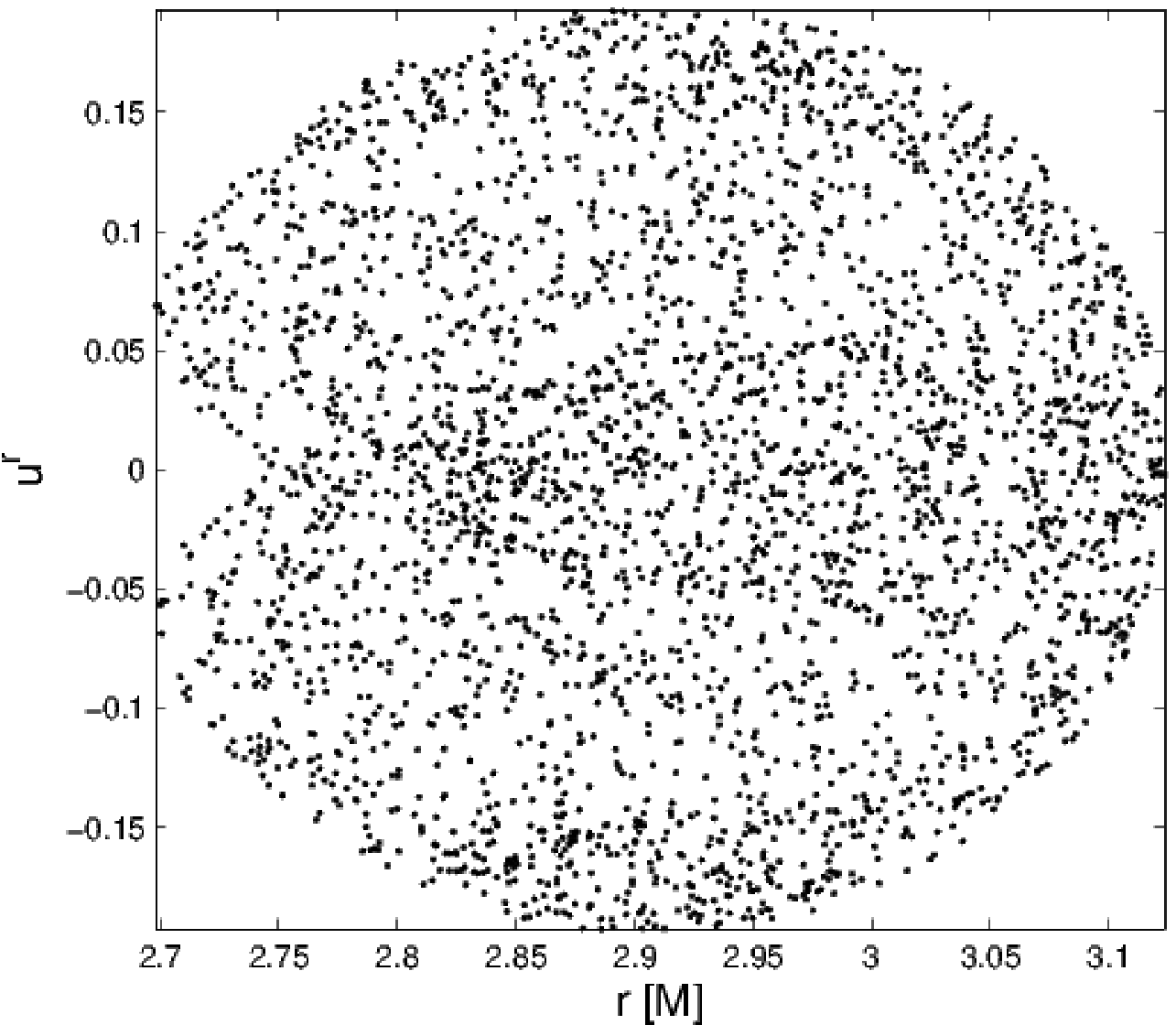}~~\includegraphics[scale=0.5, clip]{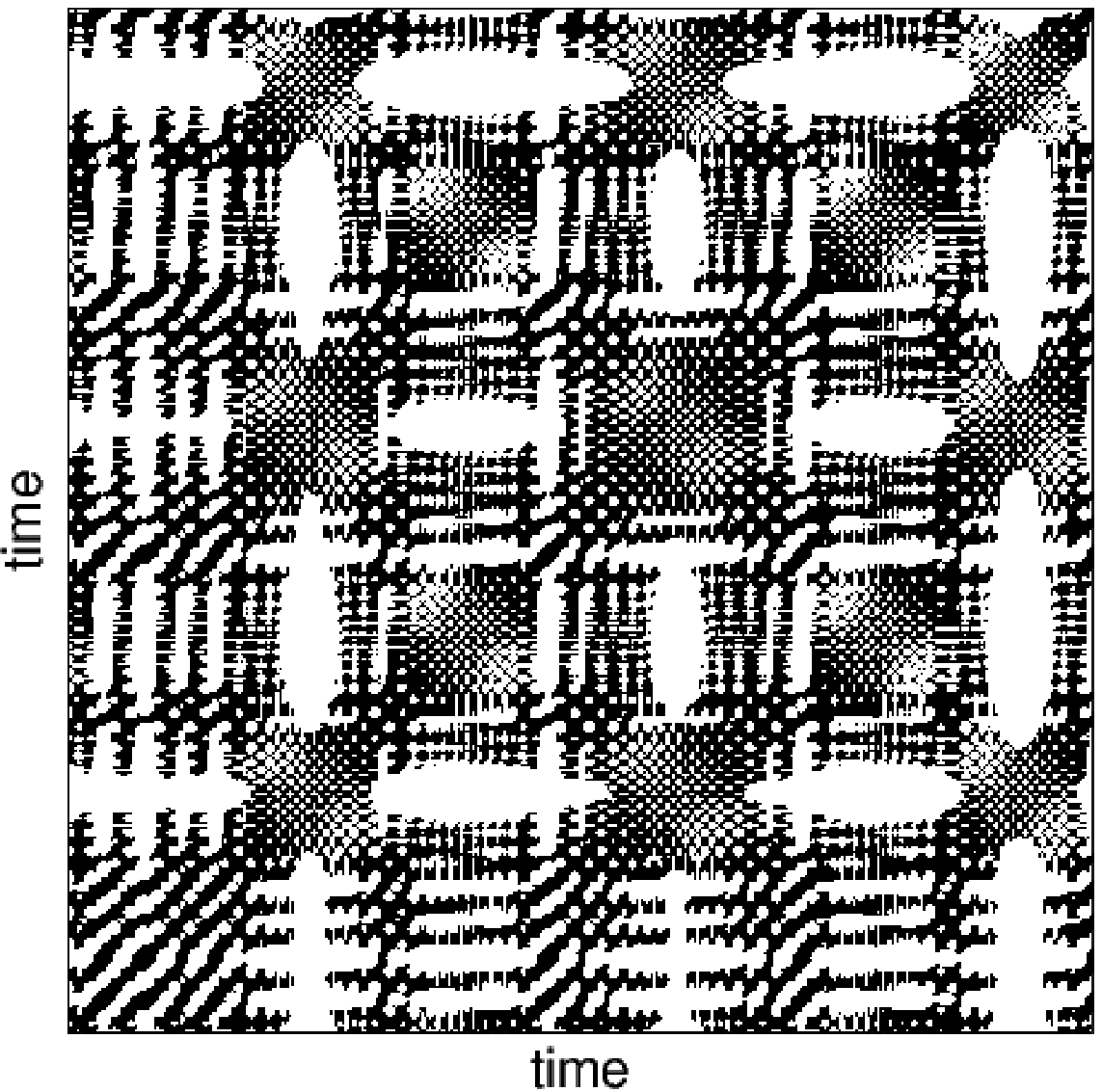}
\caption{The chaotic motion of a highly charged test particle which only
differs from the previous case by increasing the energy to
$\tilde{E}=1.7975$. The diagonal lines in the RP are now disrupted
and complex large-scale structures appear which are a characteristic
indication of deterministic chaos.}
\label{rppoinc3}
\end{figure}

\begin{figure}[htb]
\centering
\includegraphics[scale=.65,clip]{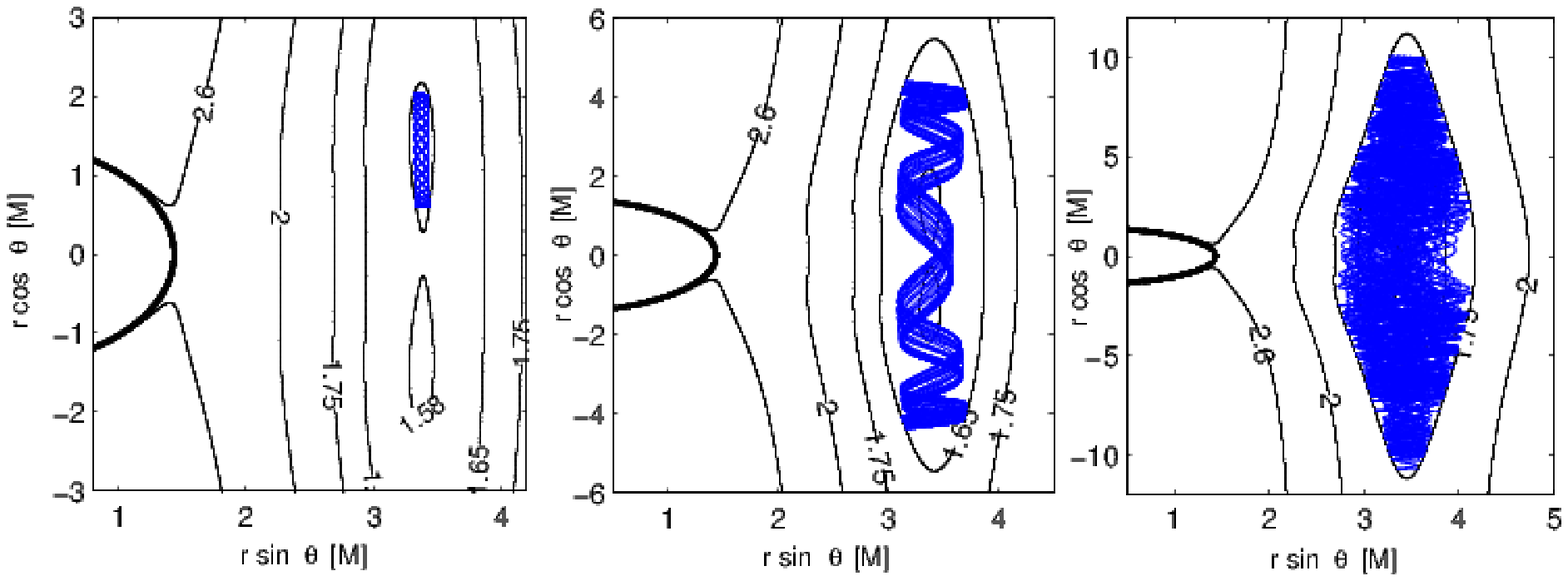}
\caption{The test particle ($\tilde{L} =6M$, $\tilde{q}B_{0}=M^{-1}$ and $\tilde{q}\tilde{Q}=1$) is
launched from the locus of the off-equatorial potential minima $r(0)=3.68\;M$, $\theta(0)=1.18$ with $u^r(0)=0$ and various values of the energy $\tilde{E}$. In the left panel we set $\tilde{E}=1.58$ and we observe ordered off-equatorial motion. For the energy of $\tilde{E}=1.65$ cross-equatorial regular motion is observed (middle panel). The trajectory occupies only a part of allowed potential lobe, regardless the length of the
integration period. Finally in the right panel with $\tilde{E}=1.75$ we observe irregular motion whose trajectory  would ergodically fill whole allowed region after the sufficiently long integration time. We show that the motion
is chaotic in this case. Spin of the black hole is $a=0.9\:M$ and its event horizon is depicted by the bold line. Topology of the potential lobes corresponds to the type Id of our classification (see \rff{wald_abc}).}
\label{wald_d_traj}
\end{figure}

At this point, a note is worth on the adopted computational scheme which
we have employed to study the trajectories and to detect the chaotical
behavior. In order to reach reliable results, we coded several approaches
and we checked their stability and precision. We employ the multi-step
Adams-Bashforth-Moulton solver to determine the phase-space trajectory
by numerical integration of eqs.\ (\ref{HamiltonsEquations}). In some
cases, when a higher precision is demanded, we use the $7$-$8$th order
Dormand-Prince method that belongs to the family of explicit Runge-Kutta
solvers with adaptive stepsize. This method improves the accuracy
significantly, as can be verified by checking the conservation of the
integrals of motion along the trajectory. However, the improved accuracy
comes at the expense of computational time, as the adopted
Dormand-Prince scheme is more computationally demanding than the
Adams-Bashforth-Moulton solver.

\begin{figure}[hp]
\centering
\includegraphics[scale=.4,clip]{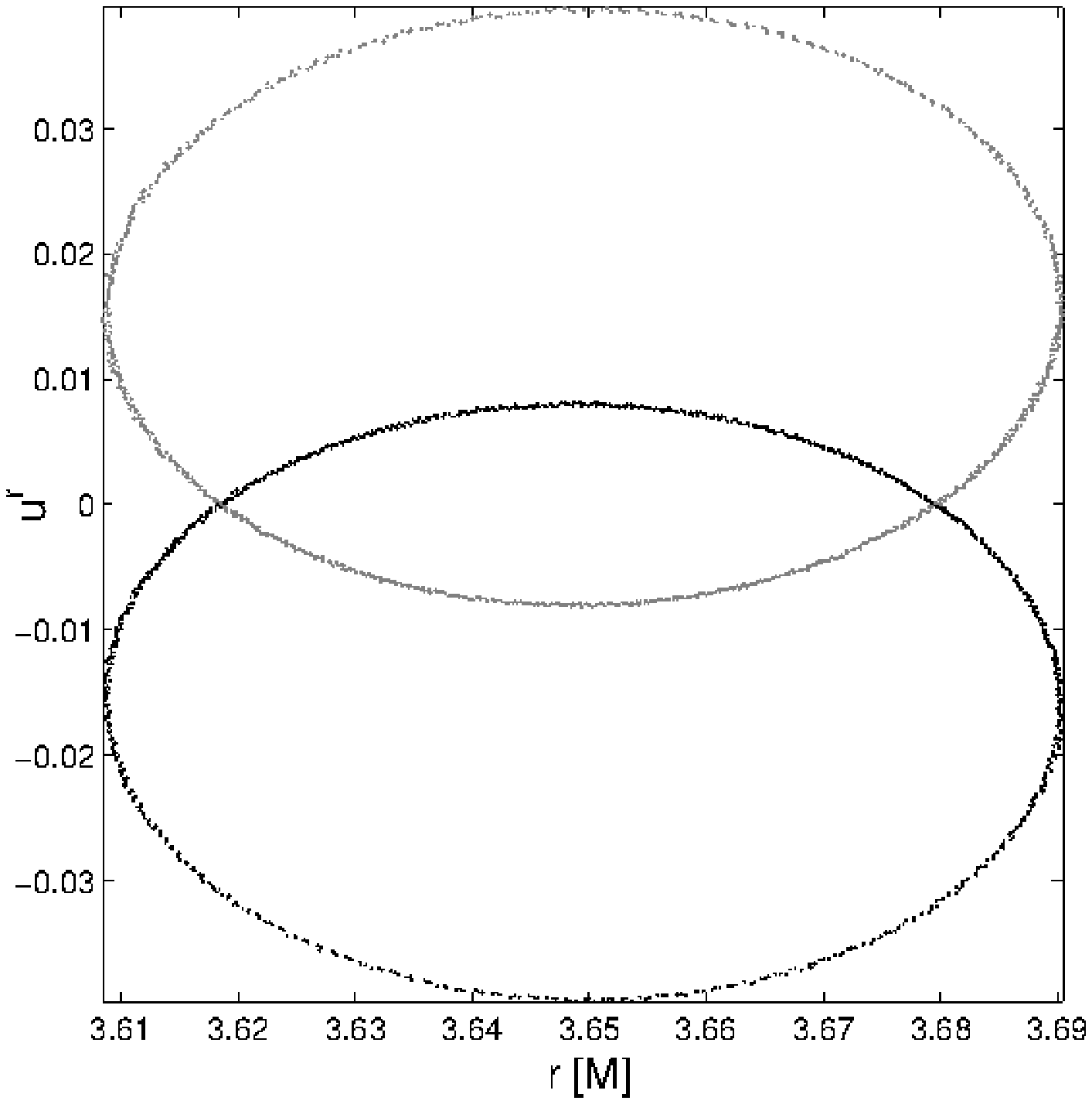}~~~~
\includegraphics[scale=.38,clip]{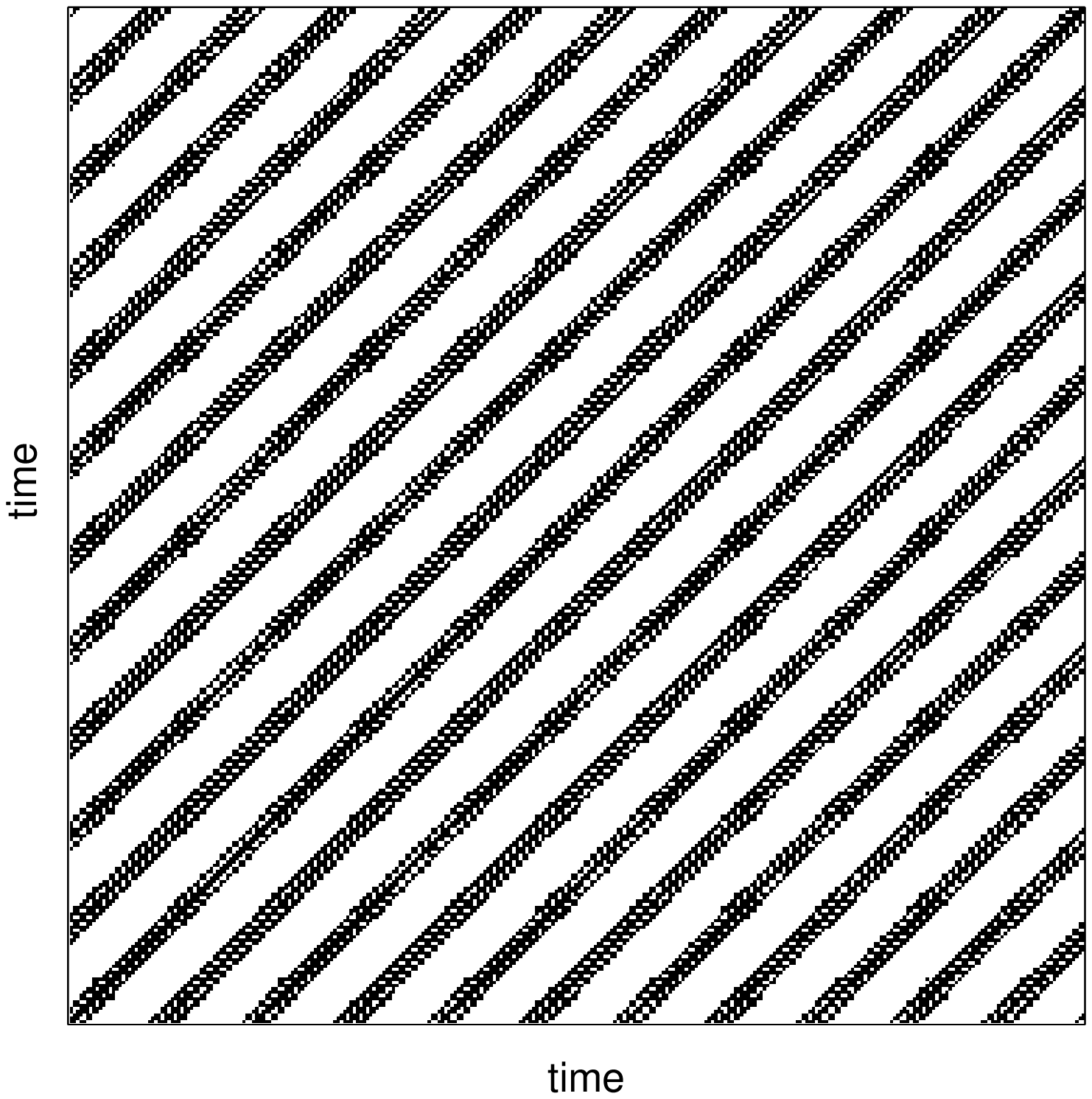}
\includegraphics[scale=0.46,clip]{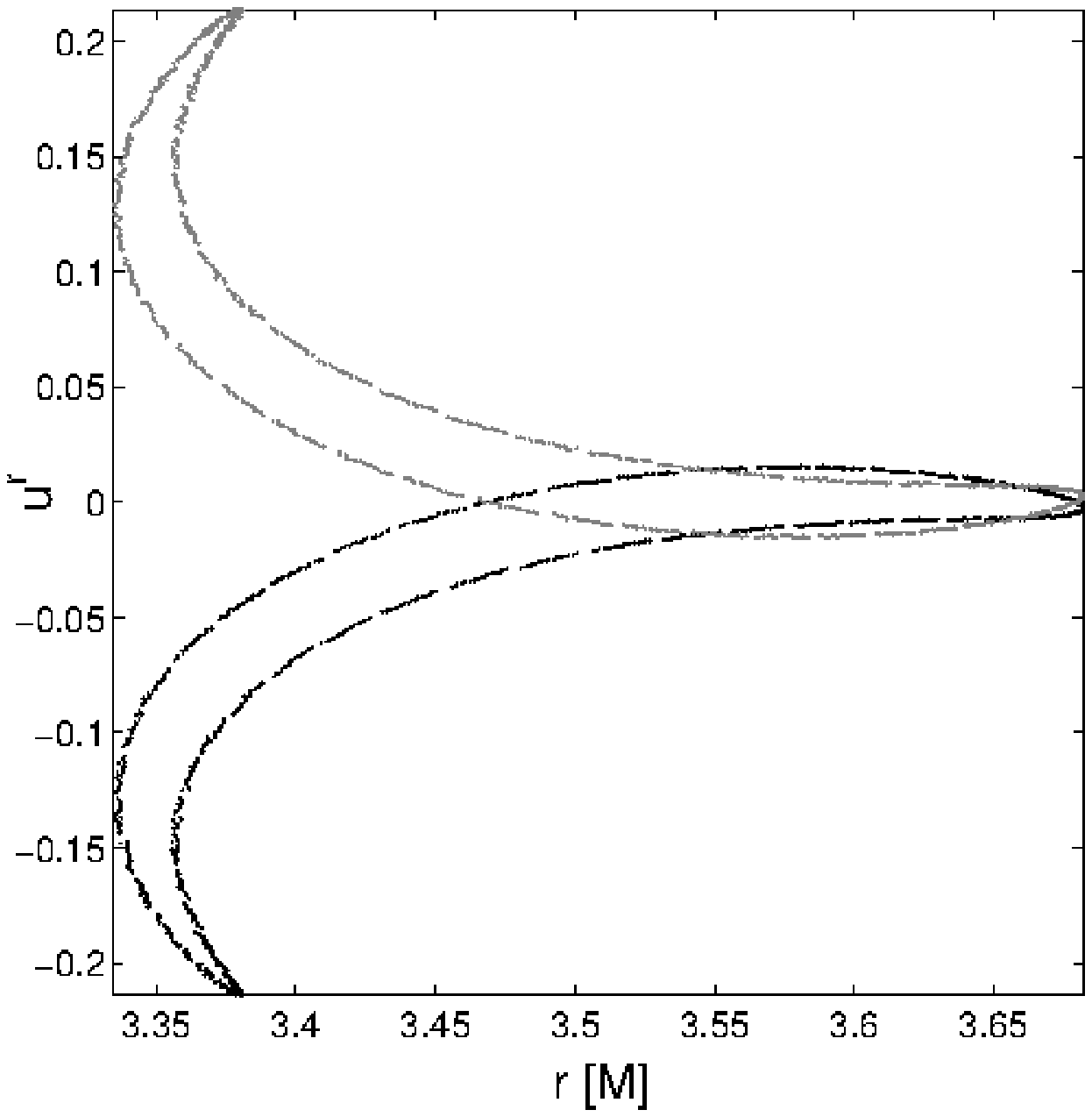}~~~~
\includegraphics[scale=0.375,clip]{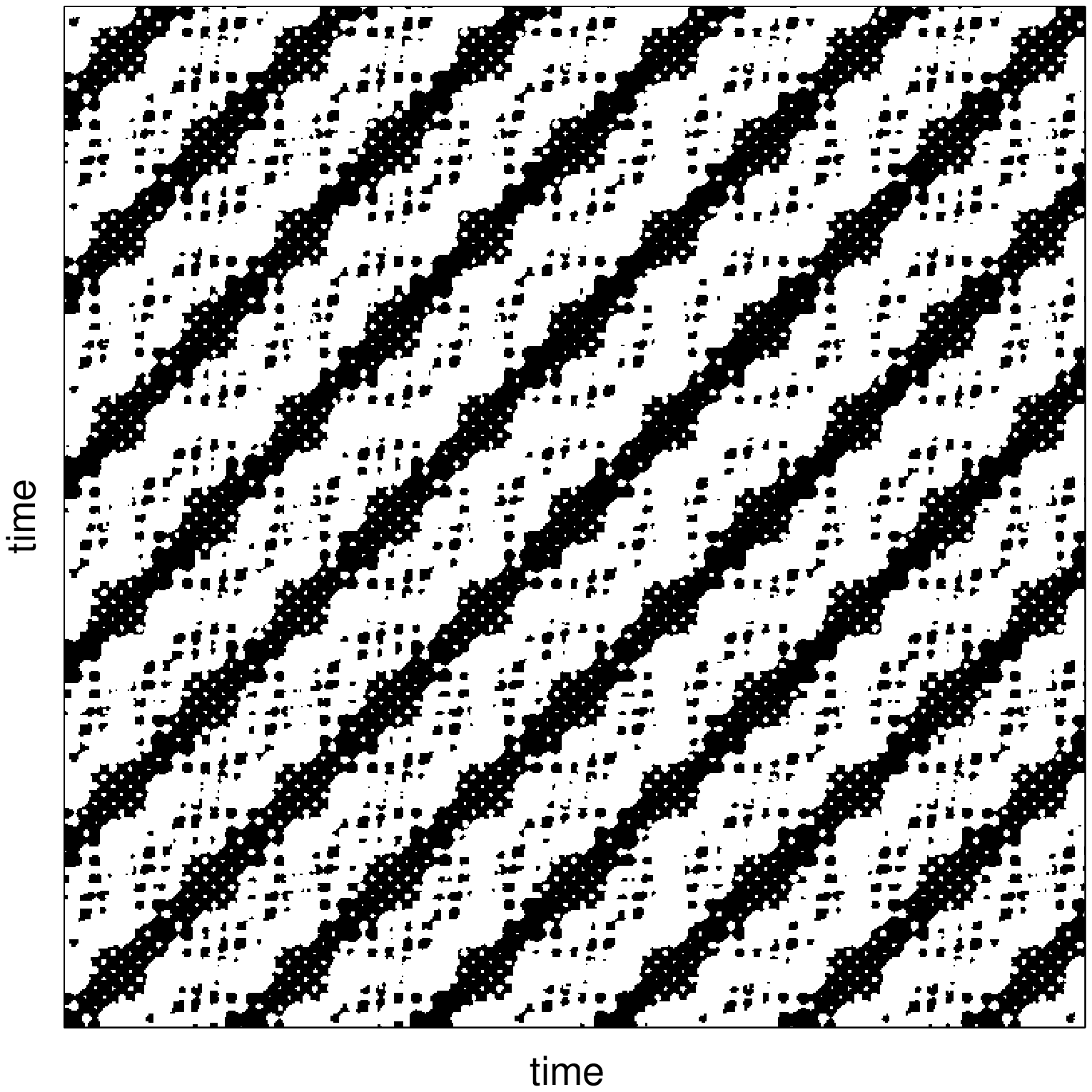}
\includegraphics[scale=0.405,clip]{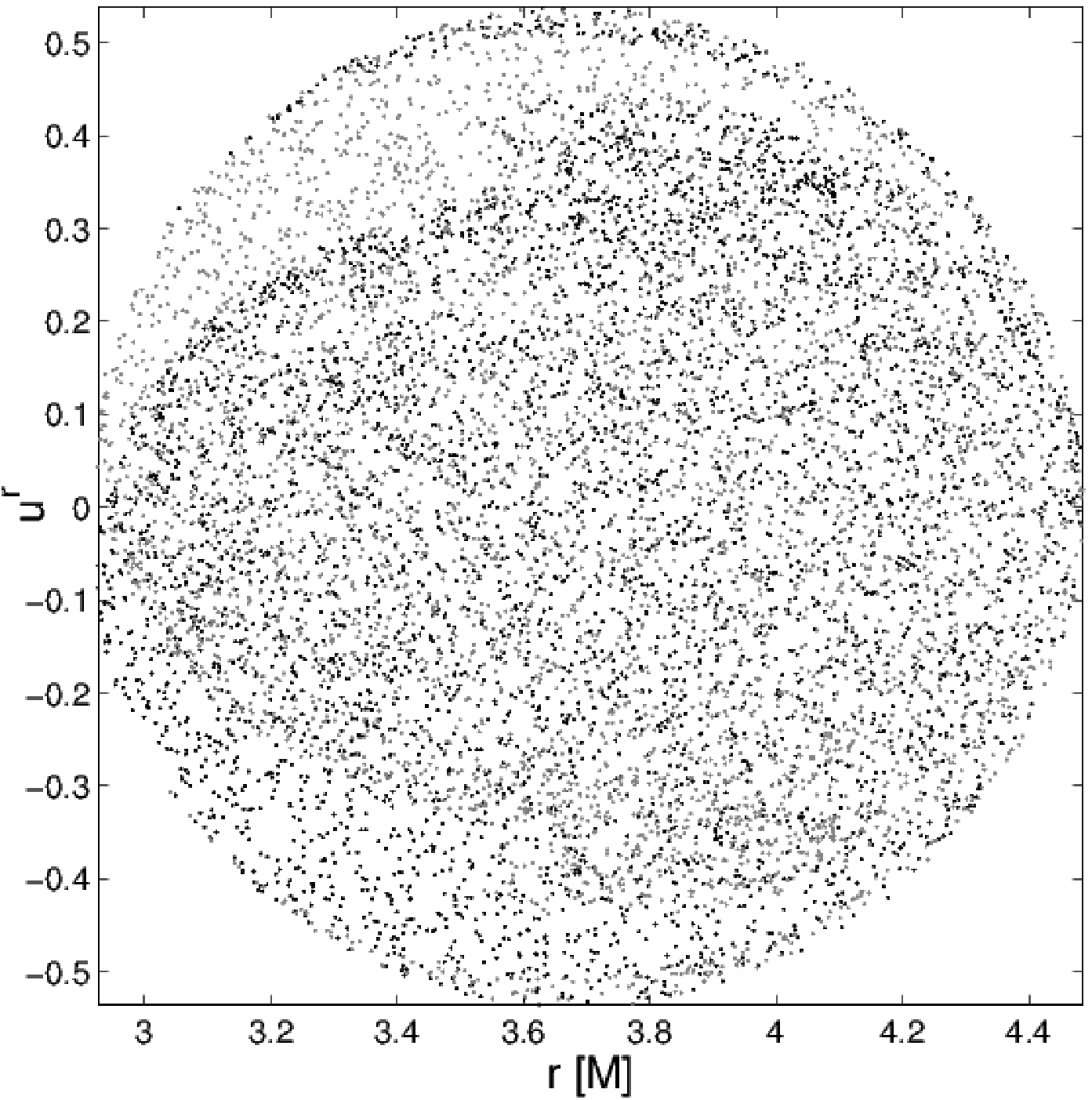}~~~~
\includegraphics[scale=0.405,clip]{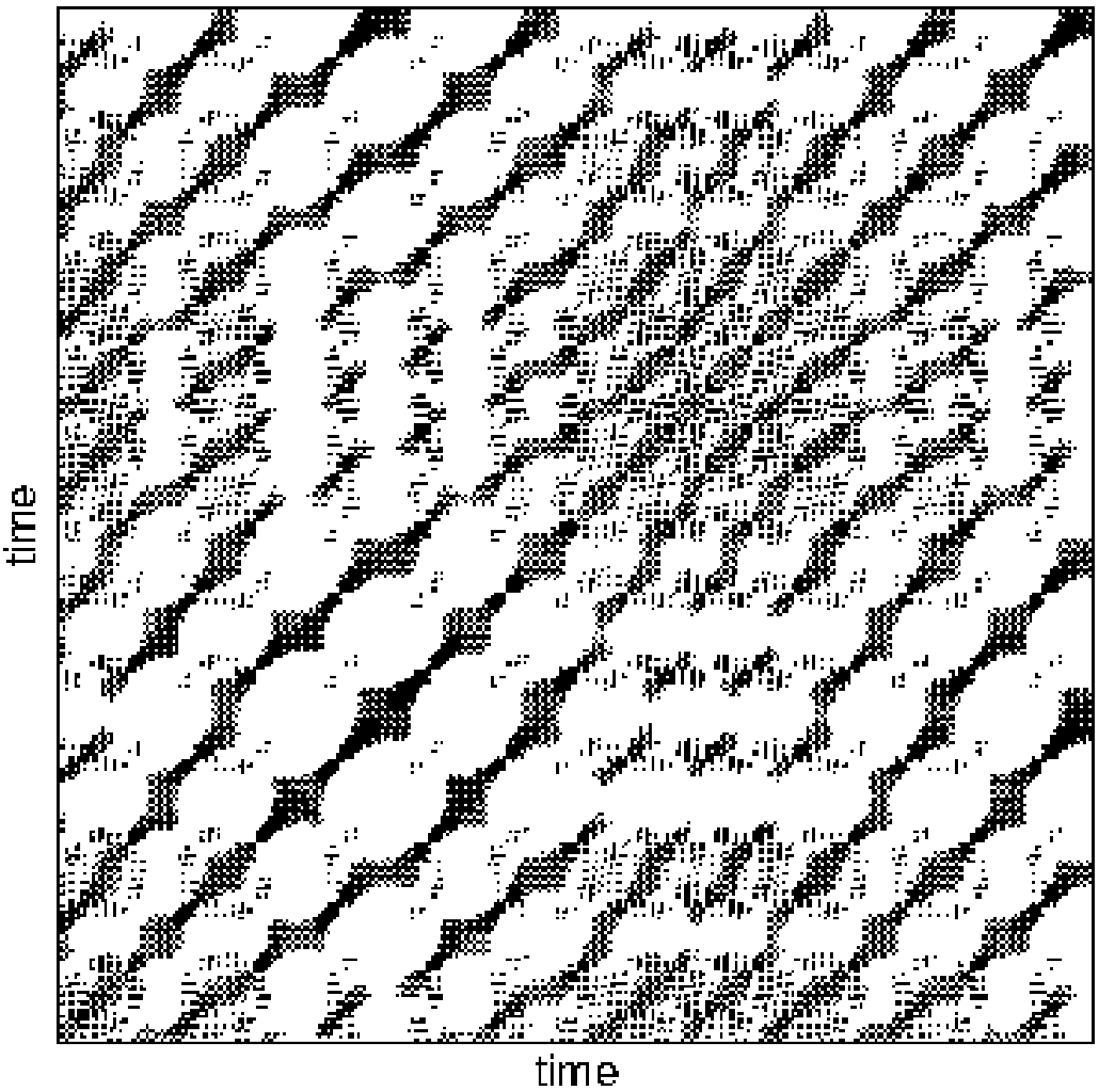}
\caption{Motion of the charged test
particle ($L =6M$,
$\tilde{q}\tilde{Q}=1$, $\tilde{q}B_{0}=M^{-1}$, $r(0)=3.68M$, $u^r(0)=0$ and $\theta(0)=\theta_{\rm section}=1.18$) on the Kerr background
($a=0.9\:M$) with the Wald's test field is discussed. 
For $\tilde{E}=1.578$ we observe regular off-equatorial motion (upper panels). Increasing the energy level to $\tilde{E}=1.65$ we obtain cross-equatorial regular trajectory (middle panels). Rising the energy to $\tilde{E}=1.75$ we get a chaotic cross-equatorial orbit. In the Poincar\'e surface of section we distinguish $u^{\theta}\geq0$ (black point) from $u^{\theta}<0$ (grey point). Analyzed trajectories are of the type Id (\rff{wald_abc}) and correspond to those of \rff{wald_d_traj}.}
\label{wald_d_diskuze}
\end{figure}

Furthermore, it is well-know that, when dealing with Hamiltonian systems, the 
most appropriate solvers are those which respect the symplectic
nature of Hamiltonian dynamics
\citep[e.g.][]{yoshida93}. We therefore employed also 
the implicit Gauss-Legendre Runge-Kutta (GLRK) method, which is a
symplectic scheme. Indeed, we confirm that this code provides the most
reliable results, especially in the case of long-term integration. The
difference in the accuracy between GLRK and non-symplectic solvers reaches
several orders of magnitude and it is
generally more apparent in the case of chaotic trajectories, as
expected. However, the cost in terms of the computational time is also
non-negligible, and so we only use the GLRK method to achieve very accurate 
long-time determination of the trajectory in several exemplary runs. Detailed comparison of the integrators is presented in Appendix \ref{integ}.

\section{Recurrence analysis}
\label{ra}
The Kerr metric is well-known and the analysis of test particle
motion in this spacetime was carried out in many papers \citep{mtw}.
Among important features of the Kerr metric is the fact that the
particle trajectories are integrable, and so the chaos can set in
only when perturbations of the background gravitational field are
introduced or additional electromagnetic interaction with fields of
external sources are allowed. This is also where recurrence analysis
can be helpful.

\begin{figure}[htb]
\centering
\includegraphics[scale=0.585, clip]{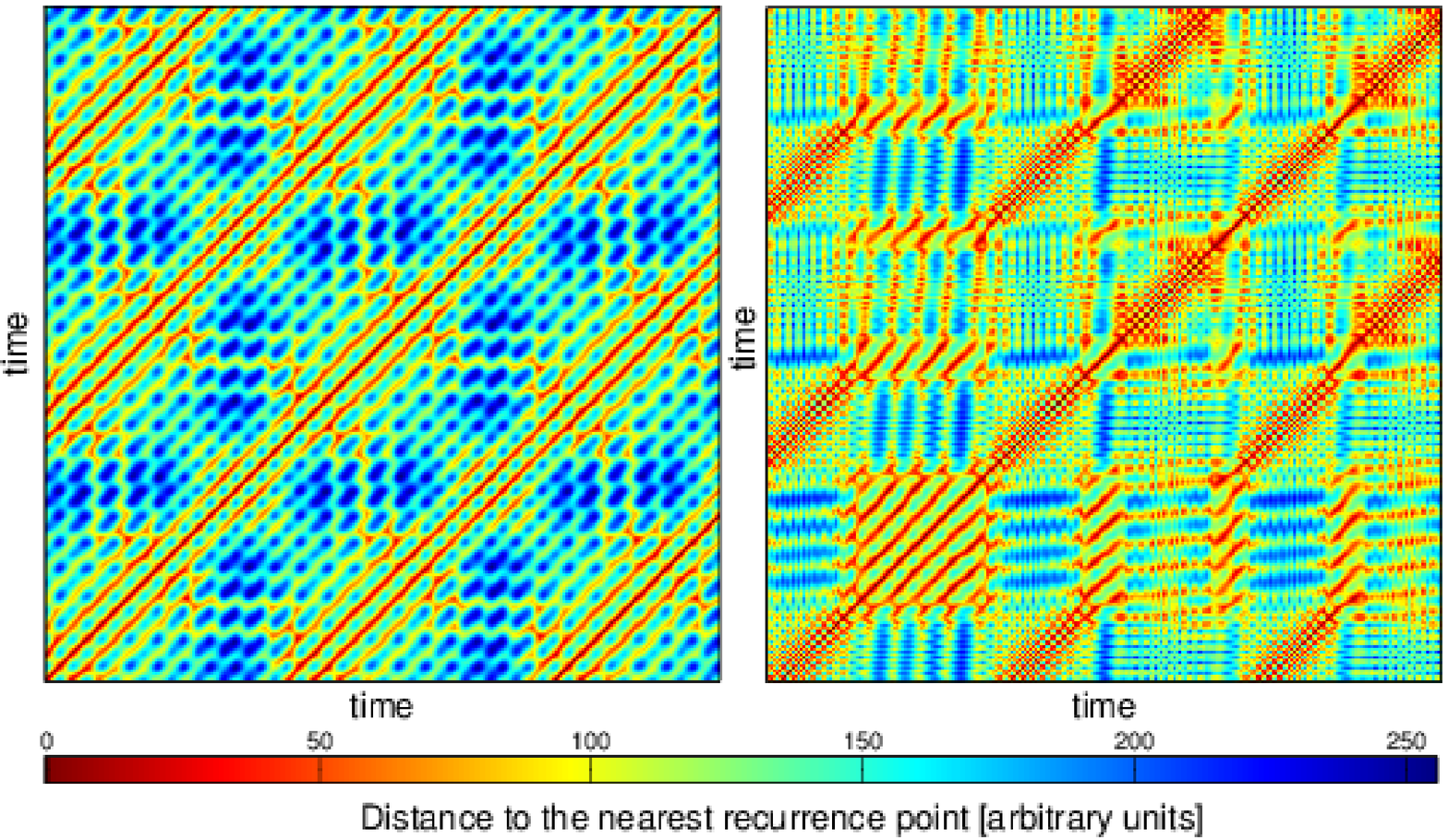}
\caption{Color map of the mutual phase space distance seperating the given pair of points on the trajectory. This example concerns a charged particle  trajectory near the Kerr black hole in the asymptotically uniform magnetic field. Left: the case of regular motion with the energy of $\tilde{E}=1.77$. Right: the case of chaotic motion with $\tilde{E}=1.7975$. In the latter case more complex 
structure appear in the recurrence plot. The common parameters of both panels are $\tilde{L} =5M$, $a=0.5\:M$, $\tilde{q}B_{0}=2M^{-1}$ and $\tilde{q}\tilde{Q}=2$ with the initial condition $r(0)=2.9\:M$, $\theta(0)=0.856$
and $u^r(0)=0$.}
\label{unthresholded}
\end{figure}

Methods of phase space recurrences have been successfully applied to
a wide range of various empirical data, not only in physics but also
related to physiology, geology, finances and other fields.
Recurrence plots are especially suitable for the investigation of
rather short and nonstationary data. On the other hand, the method
of recurrence analysis has not yet been widely applied to study the
dynamical properties of motion in relativistic systems. We thus
briefly summarize this approach for our context.

Besides more traditional methods of the numerical analysis of
dynamical systems, such as a visual survey of Poincar\'{e} surfaces of
section or the evaluation of the Lyapunov spectra \citep{lce}, the
recurrence analysis is a rather novel technique, based on the
analysis of recurrences of the system into the vicinity of its
previous states.

Recurrence Plots (RP)
are introduced as a tool of visualizing the recurrences of a trajectory
in the phase space \citep{eckmann}. The method is based on examination 
of the binary values that are constructed from the trajectory $\vec{x}(t)$. 
Results of the orbit analysis can be quantified statistically in
terms of the Recurrence Quantification Analysis (RQA).\footnote{We use
the CRP ToolBox \citep[p. 321]{marwan} in Matlab (R2009b) to construct RPs
and to evaluate RQA measures.}

The RP construction is straightforward regardless of the dimension of
the phase space. We only need to evaluate the binary values of the
recurrence matrix $\mathbf{R}_{ij}$, which can be formally expressed as
follows:
\begin{equation}
\label{rpdef}
\mathbf{R}_{ij}(\varepsilon)=\Theta(\varepsilon-||\vec{x}(i)-\vec{x}(j)||),\;\;\;
i,j=1,...,N,
\end{equation}
where $\varepsilon$ is a pre-defined threshold parameter, $\Theta$ the
Heaviside step function, and $N$ specifies the sampling frequency. The
sampling frequency is applied to the time segment of the trajectory
$\vec{x}(t)$ under examination. There is, however, 
no unique prescription for the appropriate definition of the
phase space norm $||\;.\;||$ in \rf{rpdef}. We can consider a
purely abstract vector space and apply one of the elementary
norms $L^1$, $L^2$ (Euclidean norm) or $L^{\infty}$ (maximum norm).
Some aspects of the appropriate choice of the norm are deferred to
\rs{appa}.

\begin{figure}[!ht]
\centering
\includegraphics[scale=0.56,clip]{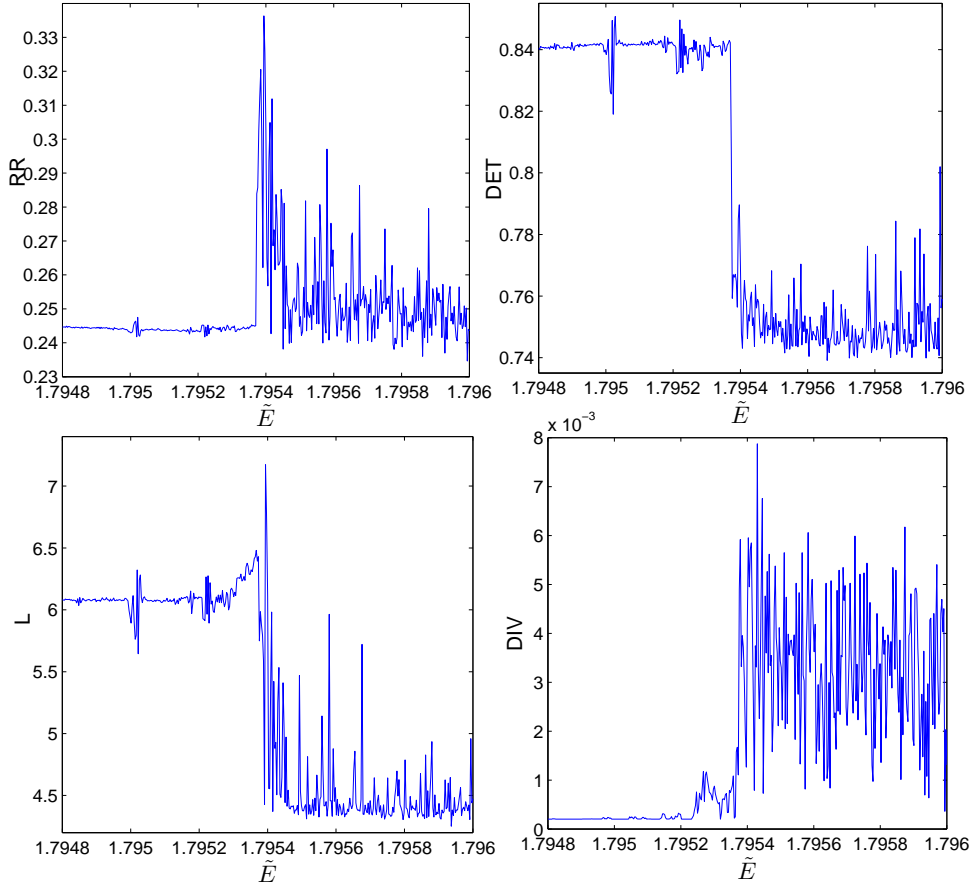}
\caption{Graphs of different RQA
measures based on the diagonal lines in the RP as a function of specific
energy $\tilde{E}$. In each panel, 400 trajectories were analyzed in a
given energetic range. All measures exhibit the evident change of their
behavior at $\tilde{E}\approx{}1.7954$ which we interpret as an onset
of chaos. Other parameters remain fixed at following values: $\tilde{L}
=5M$, $\tilde{q}B_{0}=2$, $r(0)=2.9\:M$, $\theta(0)=0.856$, $u^r(0)=0$,
$a=0.5\:M$ and $\tilde{q}\tilde{Q}=2$.}
\label{rqa1}
\end{figure}

Finally, we need to specify the value of the threshold parameter
$\varepsilon$. To this end we follow the suggestion of 
\citet[sec. 3.2.2]{marwan} and relate $\varepsilon$ to the standard mean deviation,
$\sigma$, of the given data set. Setting $\varepsilon=k\sigma$ is
advantageous because the proportionality constant $k$,
once adjusted to obtain a properly filled Recurrence Plot, remains valid
(with only minor adjustments) for all data sets of other trajectories.
We therefore normalize the time series of each coordinate separately to
zero mean and $\sigma=1$.

The binary valued matrix $\mathbf{R}_{ij}$ represents the RP which we
get by assigning a black dot where $\mathbf{R}_{ij}=1$ and leaving a
white dot where $\mathbf{R}_{ij}=0$. Both axes represent a time segments
over which the data set (the phase space vector) is being examined. RP
is thus symmetric; the main diagonal is always occupied by the line
of identity (LOI).

Recurrence Plots contain wealth of information about the dynamics of the
system \citep{thiel}. Different pieces of knowledge are encoded in
large-scale and short-scale patterns \citep[sec. 3.2.3]{marwan}. To decide if a
particular trajectory is a regular or a chaotic one, the determining
factor is the presence of diagonal structures in the RP. Diagonal lines
in the RP reflect the time segments of phase space trajectory during
which the system evolution proceeds in a regular way. It captures the
epoch when the trajectory proceeds almost parallel to its previous
segment, i.e.\ within the $\varepsilon$-tube around that segment. Hence,
integrable systems exhibit themselves by diagonally oriented structures
in their RP. On the other hand, if the motion is chaotic the diagonal
lines disappear and the diagonal features become shorter, as the
trajectories tend to diverge quickly. As a result, more complicated
structures appear in the RP. 

\begin{figure}[!ht]
\centering
\includegraphics[scale=0.56,clip]{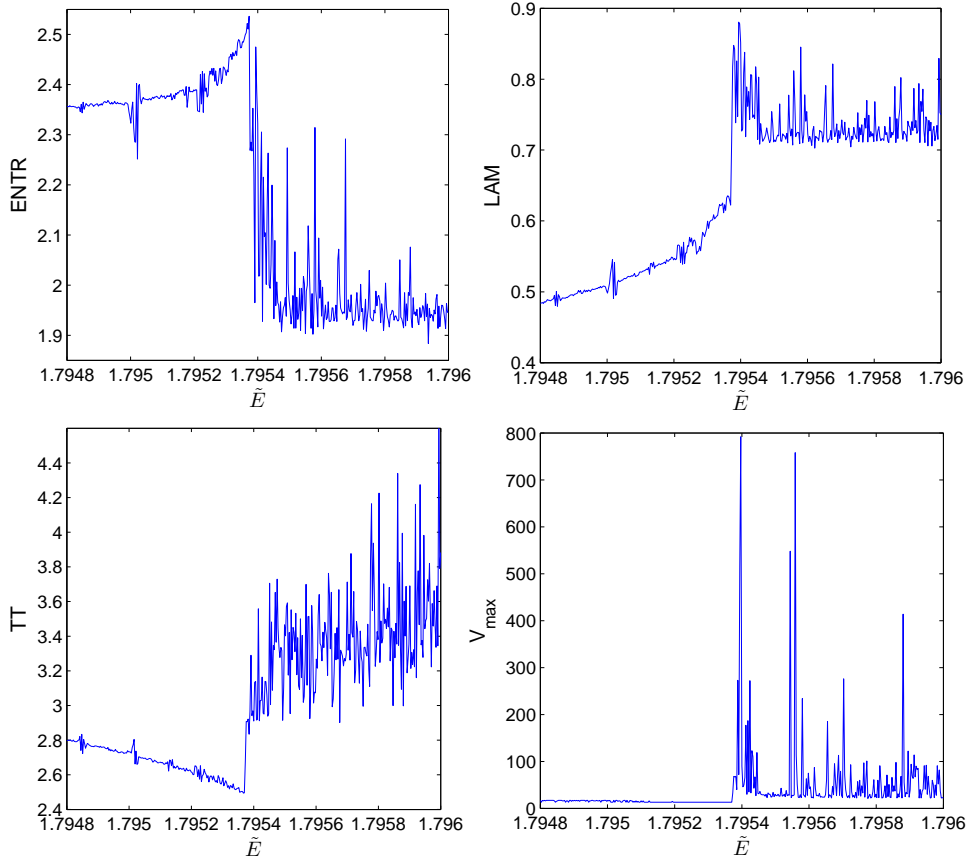}
\caption{Shannon entropy of probability distribution of diagonal
lines lengths $\mathrm{ENTR}$ and three RQA measures based on the vertical
lines of RPs as a function of specific energy $\tilde{E}$ (details in the 
text). To some
surprise, the vertical measures also react dramatically to the onset of
chaos at $\tilde{E}=1.7954.$} \label{rqa2}
\end{figure}

We stress that the interpretation of the RP is primarily intuitive. We refer to the review paper \citet[sec.\ 3.2.3]{marwan} where the patterns appearing in the RP and their relation to the current dynamic regime are analyzed in detail. Although basic conclusions may be inferred in general (e.g. distinction between regular versus chaotic regime) the fine structure of the RP depends heavily on the properties of a given dynamic system. In order to gain more insight into the way in which various dynamical regimes  manifest themselves in our system we typically present RPs accompanied by corresponding Poincar\'{e} surface of section throughout this text.

\begin{figure}[!ht]
\centering
\includegraphics[scale=0.66, clip]{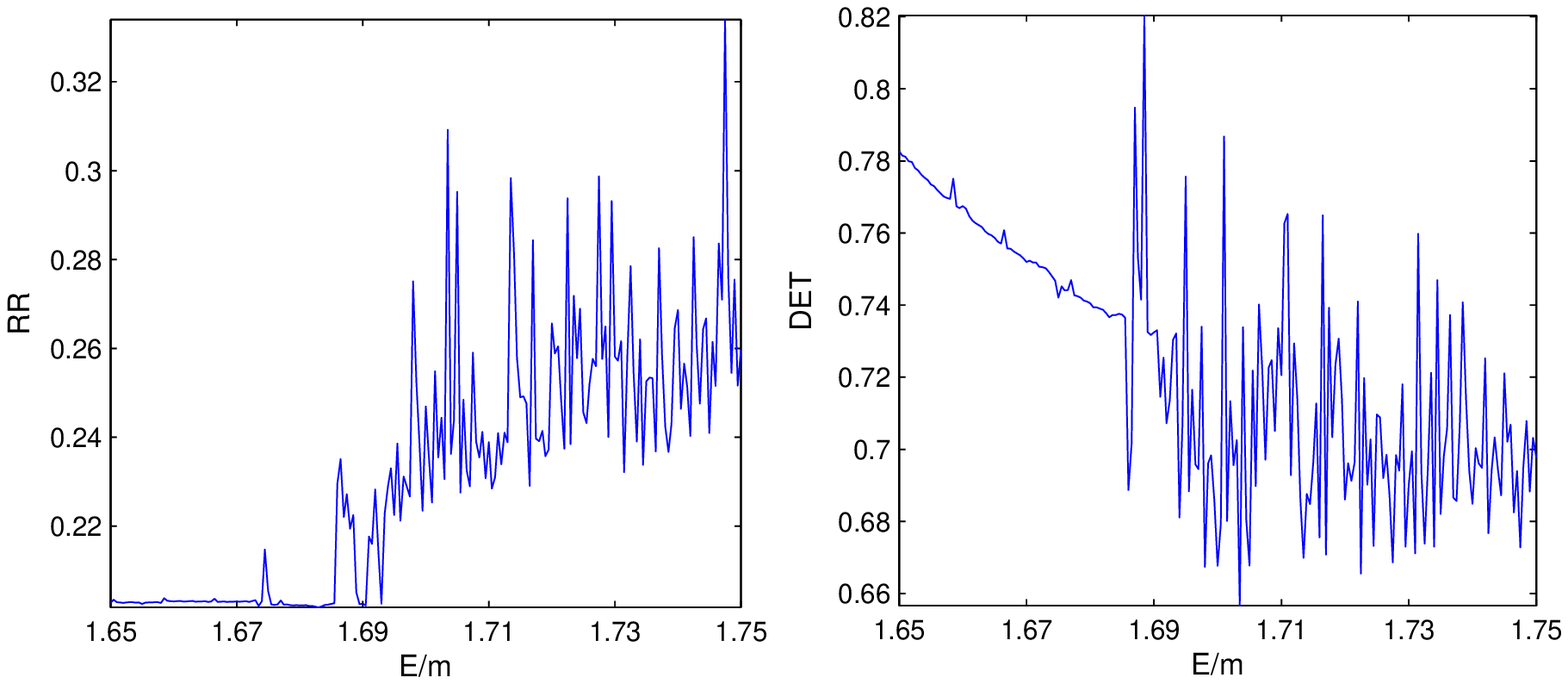}
\includegraphics[scale=0.66, clip]{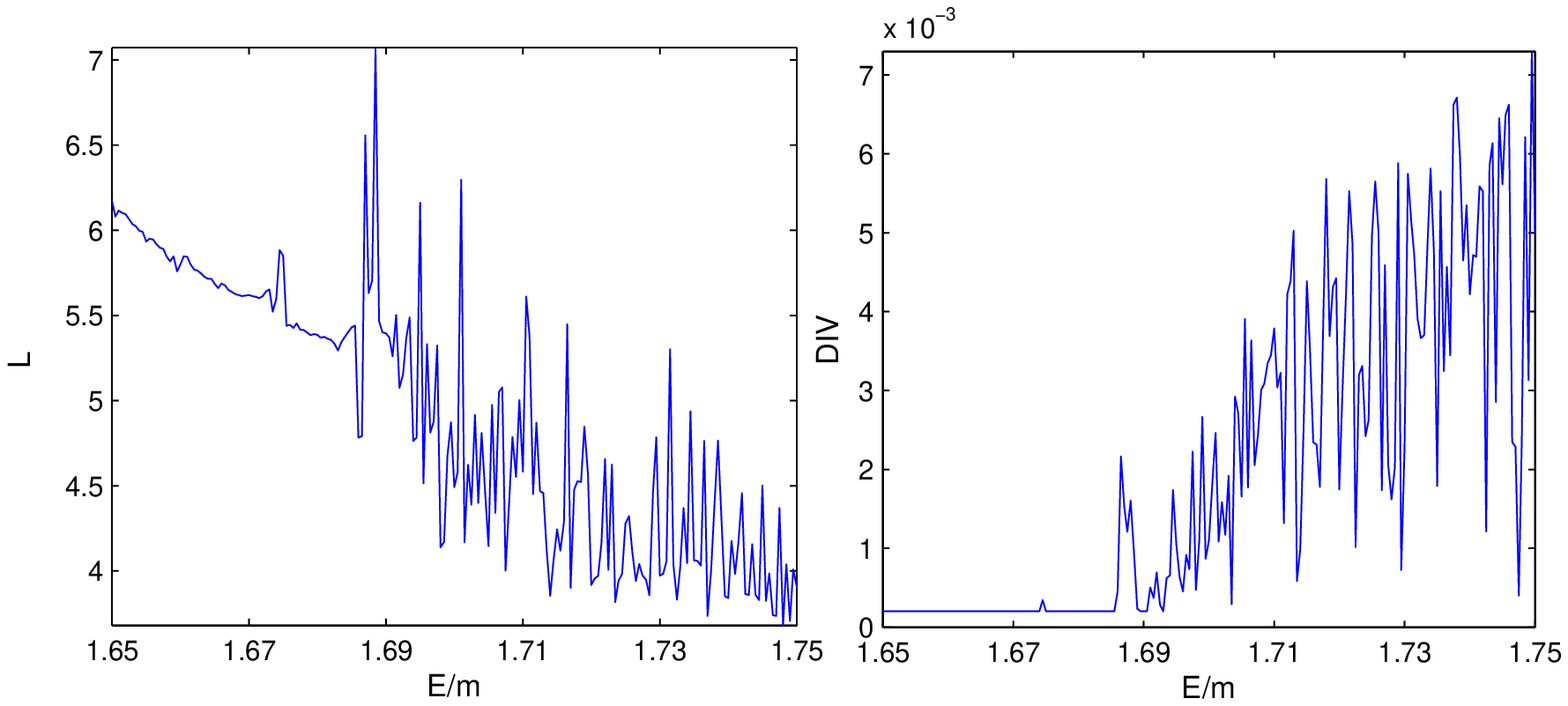}
\caption{Diagonal RQA measures  $RR$, $DET$, $L$ and $DIV$ as a function of specific
energy $\tilde{E}$. Dramatic change of the behaviour at $\tilde{E}\approx1.685$ is apparent for
all the quantities. This is where the chaos
sets on.}
\label{rqa_d1}
\end{figure}

\begin{figure}[!ht]
\centering
\includegraphics[scale=0.325, trim= 0mm 0mm 4mm 0mm, clip]{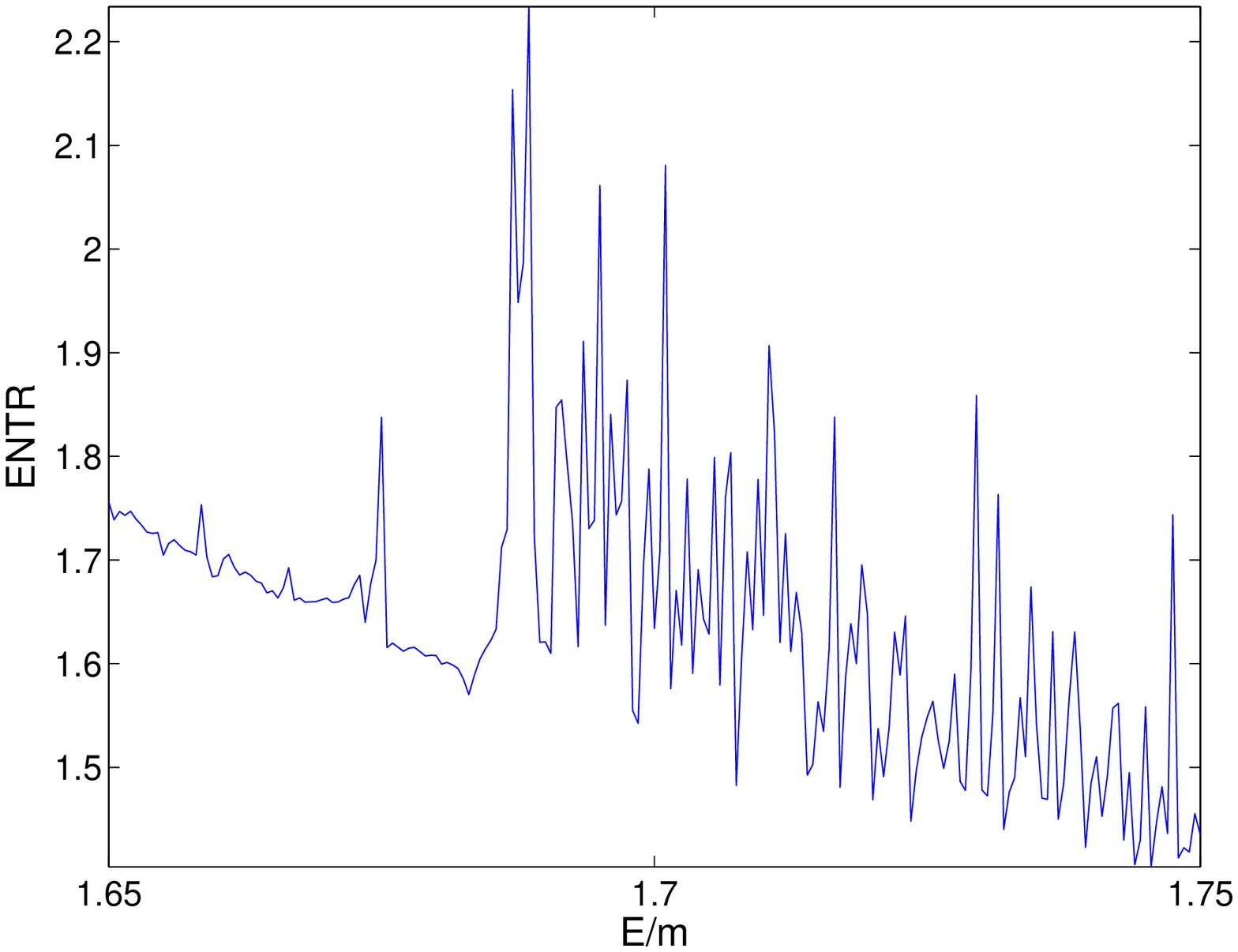}
\includegraphics[scale=0.63, trim= 0mm 0mm 0mm 0mm ,clip]{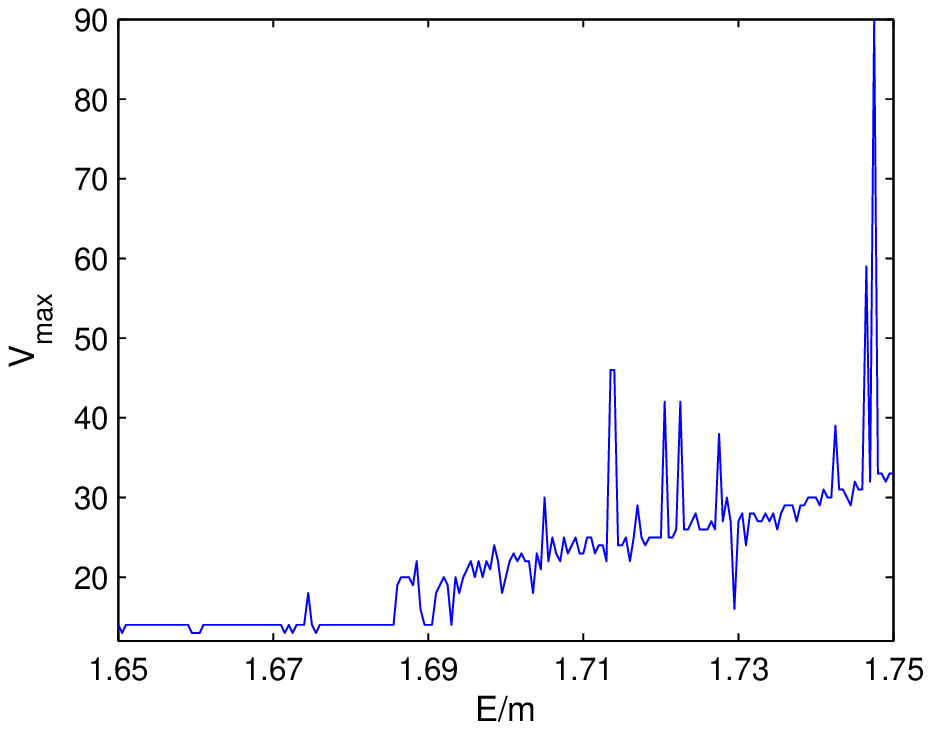}
\includegraphics[scale=0.63, clip]{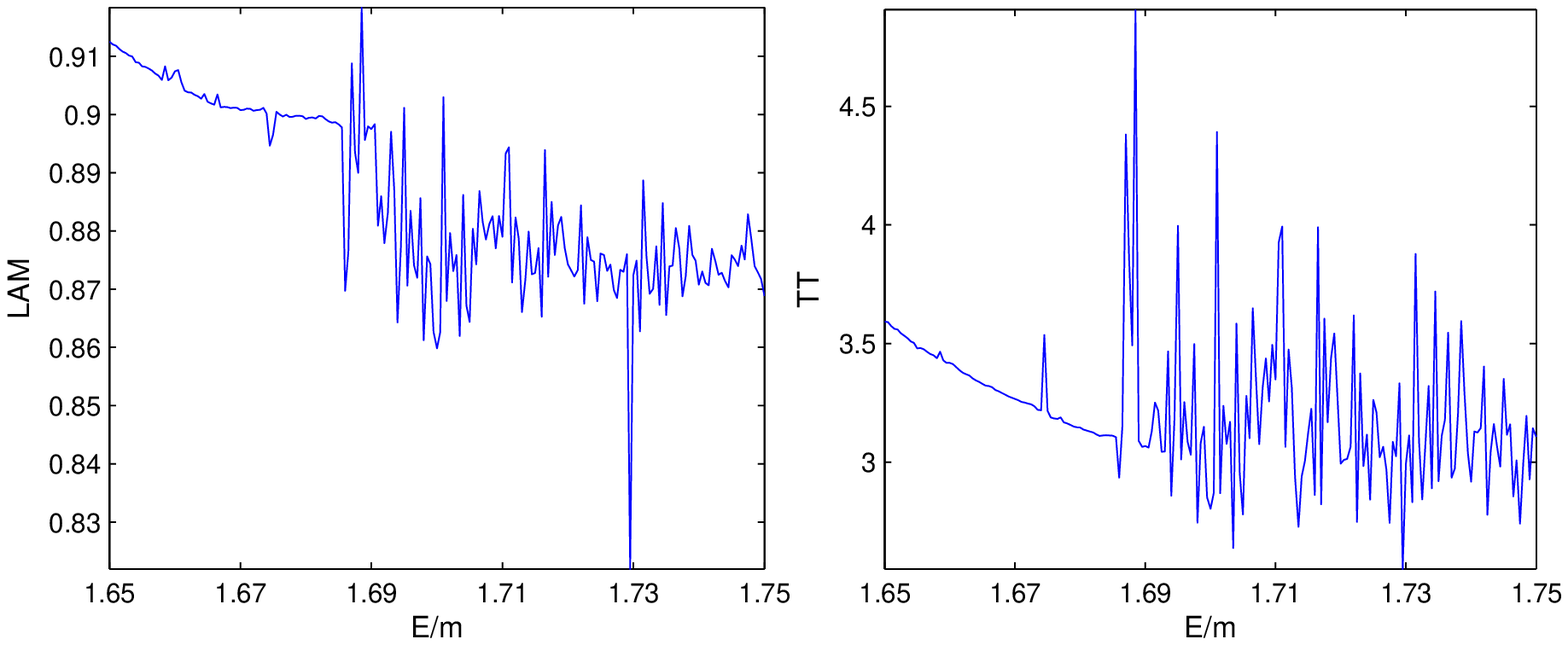}
\caption{Shannon entropy of probability distribution of diagonal
lines lengths $\mathrm{ENTR}$ and three RQA measures based on the vertical
lines of RPs as a function of specific energy. Sudden change of the behaviour at $\tilde{E}\approx1.685$ is apparent for
all the quantities.}
\label{rqa_d2}
\end{figure}

Visual behavior of RP and its complexity is
quantitatively reflected in RQA. The RQA evaluates statistical characteristics of the recurrence matrix
$\mathbf{R}_{ij}$. First of all, we define the recurrence rate
($\mathrm{RR}$) as a density of points in RP,
\begin{equation}
 \mathrm{RR}(\varepsilon)\equiv\frac{1}{N^2}\sum_{i,j=1}^N\mathbf{R}_{i,j}(\varepsilon).
\end{equation}
Now we can turn our attention to diagonal segments in RP. Their length
draws distinction between regularity and chaos. The histogram
$P(\varepsilon,l)$ records the number of diagonal lines of length $l$.
It is formally given as follows:
\begin{equation}
\label{diaghist}
P(\varepsilon,l)=\sum^N_{i,j=1}(1-\mathbf{R}_{i-1,j-1}(\varepsilon))(1-\mathbf{R}_{i+l,j+l}(\varepsilon))\prod_{k=0}^{l-1}\mathbf{R}_{i+k,j+k}(\varepsilon).
\end{equation}
This histogram defines the determinism factor ($\mathrm{DET}$), defined
as a fraction of recurrence points, which form the diagonal lines of
length at least $l_{\rm{min}}$ to all recurrence points,
\begin{equation}
\label{det}
\mathrm{DET}\equiv\frac{\sum^{L_{\rm{max}}}_{l=l_{\rm{min}}}lP(\varepsilon,l)}{\sum^{L_{\rm{max}}}_{l=1}lP(\varepsilon,l)}.
\end{equation}

\begin{figure}[!ht]
\centering
\includegraphics[scale=0.61, clip]{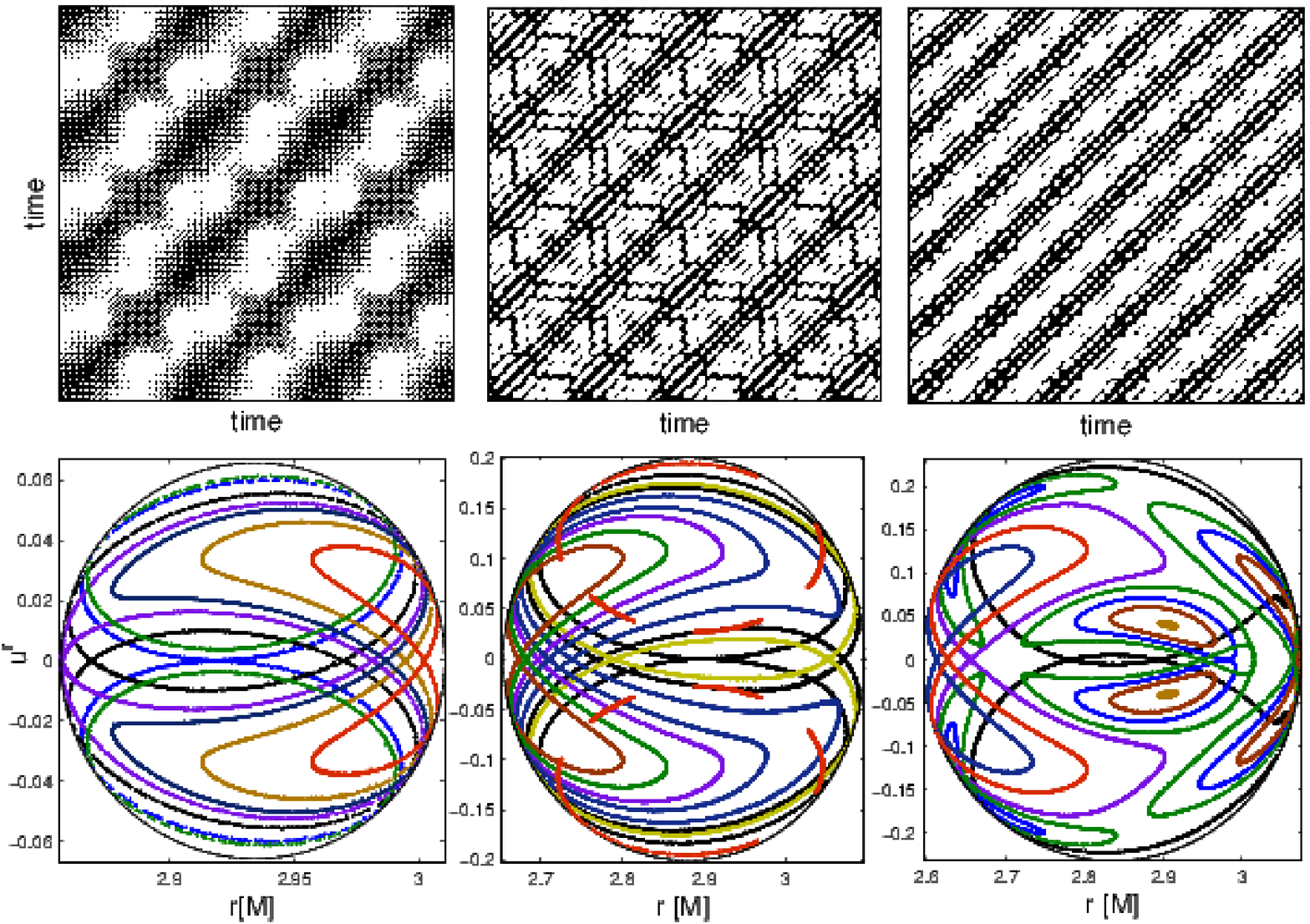}
\caption{Comparison
of purely off-equatorial trajectories in spacetimes differing by the
spin parameter $a$ (left panels: $a=0.3M$; middle: $a=0.6M$; right: $a=M$) 
which is linearly linked to the energy $\tilde{E}$ (left
panels: $\tilde{E}=1.56$; middle: $\tilde{E}=1.9$, and $\tilde{E}=2.35$ in
the right panels). Other parameters remain fixed: $\tilde{L} =5M$, $M^{-1}$,
$\theta(0)=\theta_{\rm{section}}=0.856$, $\tilde{q}B_{0}=2M^{-1}$ and
$\tilde{q}\tilde{Q}=2$. RPs are taken for trajectories with
$r(0)=2.9\:M$ and $u^{r}(0)=0$. Although the structures in the
Recurrence Plots clearly differ from each other,  all of them represent
diagonally oriented patterns that are characteristic of regular motion.
The regularity of the motion is confirmed by  surfaces of section in the
bottom panels, where several trajectories (for each value of spin
 and energy) are presented. Different colours are used to distinguish the orbits originating from
different initial conditions.}
\label{spin_diskuze}
\end{figure}

The average length of diagonal lines $L$ (where only lines of length at
least $l_{\rm{min}}$ count) is
\begin{equation}
\label{Lav}
L\equiv\frac{\sum^{L_{\rm{max}}}_{l=l_{\rm{min}}}lP(\varepsilon,l)}{\sum^{L_{\rm{max}}}_{l=l_{\rm{min}}}P(\varepsilon,l)},
\end{equation}
and the corresponding divergence ($\mathrm{DIV}$) is defined as inverse
of the length of the longest diagonal line $L_{\rm{max}}$,
\begin{equation}
\label{div}
\mathrm{DIV}\equiv\frac{1}{L_{\rm{max}}}.
\end{equation}
$\mathrm{DIV}$ is in its very nature closely related to the divergent
features of the phase space trajectory, and so it was originally
\citep{eckmann} claimed to be directly related to the largest
positive Lyapunov characteristic exponent $\lambda_{\rm{max}}$. On the
other hand, theoretical considerations justify the use of $\mathrm{DIV}$
as an estimator only for the lower limit of the sum of the positive
Lyapunov exponents \citep[sec. 3.6]{marwan}. Nevertheless, a strong correlation
between $\mathrm{DIV}$ and $\lambda_{\rm{max}}$ arises in numerical
experiments \citep{trulla}.

The quantification measure $\mathrm{ENTR}$ is defined as the Shannon
entropy of the probability $p(\varepsilon,l)=P(\varepsilon,l)/N_{l}$ of
finding a diagonal line of length $l$ in the Recurrence Plot,
\begin{equation}
\label{entr}
\mathrm{ENTR}\equiv-\sum_{l=l_{\rm{min}}}^{L_{\rm{max}}}p(\varepsilon,l)\ln{p(\varepsilon,l)},
\end{equation}
where $N_{l}$ is a total number of diagonal lines:
$N_{l}(\varepsilon)=\sum_{l\geq{}l_{\rm{min}}}P(\varepsilon,l)$.

Analogous statistics may be performed for vertical as well as the
horizontal segments (RP is symmetric with respect to the main diagonal).
These segments are generally connected with periods in which the system
evolves during its laminar state. To this end, the histogram
$P(\varepsilon,v)$ records the number of vertical lines of length $v$
and it can be constructed as follows:
\begin{equation}
\label{verthist}
P(\varepsilon,v)=\sum^N_{i,j=1}(1-\mathbf{R}_{i,j}(\varepsilon))(1-\mathbf{R}_{i,j+v}(\varepsilon))\prod_{k=0}^{v-1}\mathbf{R}_{i,j+k}(\varepsilon).
\end{equation}

In analogy with the diagonal statistics histogram, $P(\varepsilon,v)$ is
used to define the vertical RQA measures. Laminarity ($\mathrm{LAM}$) is
defined as a fraction of recurrence points that form vertical lines of
length at least $v_{\rm{min}}$ to all recurrence points,
\begin{equation}
\label{lam}
\mathrm{LAM}\equiv\frac{\sum^{V_{\rm{max}}}_{v=v_{\rm{min}}}vP(\varepsilon,v)}{\sum^{V_{\rm{max}}}_{v=1}vP(\varepsilon,v)}.
\end{equation}
The trapping time ($\mathrm{TT}$) is an average length of vertical lines,
\begin{equation}
\label{tt}
\mathrm{TT}\equiv\frac{\sum^{V_{\rm{max}}}_{v=v_{\rm{min}}}vP(\varepsilon,v)}{\sum^{V_{\rm{max}}}_{v=v_{\rm{min}}}P(\varepsilon,v)}.
\end{equation}
Finally, the length of the longest vertical line ($V_{\rm{max}}$) can
also be of some interest.

RQA measures the crucial dependence of RP on the value of the
threshold parameter, $\varepsilon$, which must be adjusted appropriately
to a given data set. This lack of invariance is a drawback of both RPs
and RQA. Nevertheless, it was shown \citep{thiel2} that stable estimates
of various dynamical invariants, such as the second order R\'{e}nyi
entropy and the correlation dimension, can be inferred if $\varepsilon$
is kept within a reasonable range. Since we shall use the standard RQA
measures to compare the dynamics between test particles with different
initial conditions, we have to eliminate the numerical effect of
variances in the range of coordinate values spanned by these
trajectories. We achieve this by fixing the value of $\varepsilon$.

\begin{figure}[htb]
\centering
\includegraphics[scale=0.53,clip]{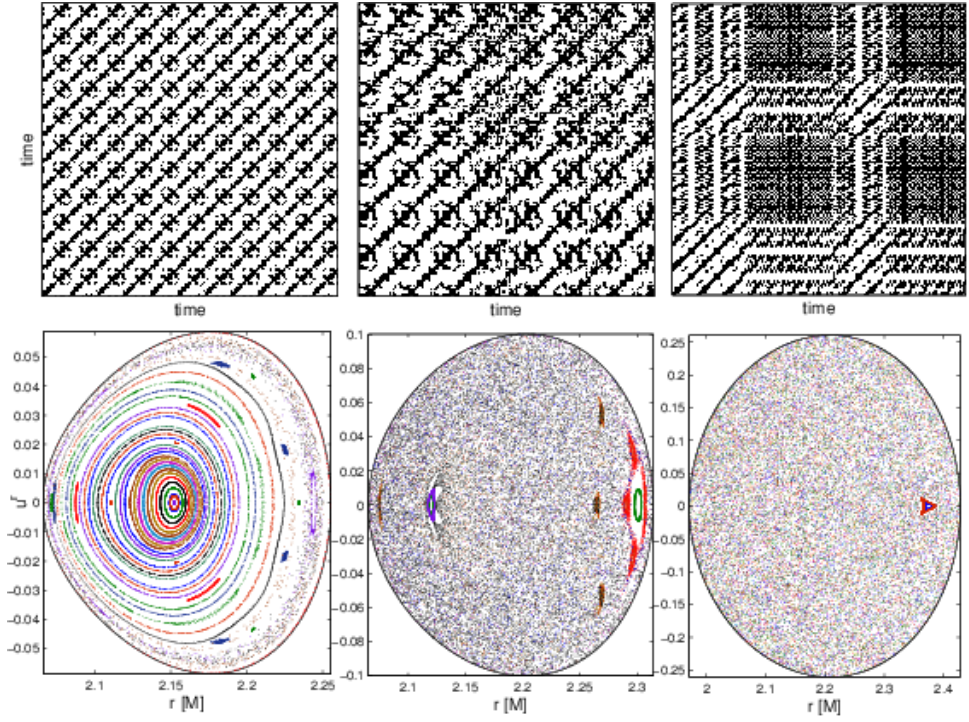}
\caption{Comparison
of trajectories of particles launched from the equatorial plane with
different spin values. Also in this case we have to link linearly the
value of spin $a$ with $\tilde{E}$ in order to maintain the existence of
the potential lobe. In left panels we set $a=0.5M$, $\tilde{E}=1.795$,
in middle panels $a=0.6M$, $\tilde{E}=1.92$ and in right panels $a=M$,
$\tilde{E}=2.42$. For all three cases we show surfaces of section of
several trajectories differing in initial values $r(0)$ and $u^r(0)$.
Recurrence Plots are taken for trajectories with $r(0)=2.15M$,
$u^r(0)=0$. Other parameters remain fixed: $\tilde{L} =5M$,
$\tilde{q}B_{0}=2M^{-1}$, $\theta(0)=\theta_{\rm{section}}=\frac{\pi}{2}$, 
$\tilde{q}\tilde{Q}=2$.}
\label{spin_diskuze_eq}
\end{figure}

After the brief set of preliminaries we are now prepared to proceed to the
intended application of the recurrence analysis in the next section.

\subsection{Phase space recurrence in the curved spacetimes}
\label{appa}
When analyzing the dynamics in the general relativistic context the fundamental question arises whether the distinction between chaotic and regular motion is coordinate dependent or not. To this end \citet{motter10} infers the transformation law for Lyapunov exponents. He concludes that although the Lyapunov exponents themselves are not invariant they transform in such a way that positive Lyapunov exponents remain positive and vice versa. In other words the distinction between regular and chaotic dynamics may be drawn invariantly.

In order to give the notion of recurrence a rigorous and, at the same
time, an intuitive sense, we can employ the $3+1$ formalism
\citep{thorne} that is based on an appropriate selection of a family of
spacetime-filling three-dimensional spacelike hypersurfaces (foliation)
of constant time $t$. The timelike curves orthogonal (in a spacetime
sense) to the hypersurfaces may be regarded as the world-lines of a
family of fiducial observers (FIDO) who naturally parameterize their
world line by proper time $\tau$ (whose rate of change generally differs
from that of $t$). FIDO identify each spatial hypersurface along his
world line as a slice of simultaneity. The geometry of this spacetime
slice is given by 3-metric $\gamma_{ij}$:
\begin{equation}
\label{3Dmetric}
\gamma_{ij}=g_{ij}+u_i{}u_j,
\end{equation}
where $u_i$ stands for the spatial part of FIDO's four-velocity and
$g_{ij}$ for the spatial part of the spacetime metric. Considering also
the time coordinate, $\gamma_{\mu\nu}$ can be regarded as a projector
onto the three-dimensional spatial hypersurface.

Our selection of FIDOs will be restricted here to those moving
along the Killing direction (thus seeing an unchanging spacetime
geometry in their neigbourhood). We shall also demand that all the
hypersurfaces of simultaneity which they constitute have an identical
spatial geometry. Finally, we require that FIDO's proper time $\tau$
becomes asymptotically identical to the coordinate time $t$. In the case
of Kerr spacetime (as an example of stationary and axisymmetric
spacetime) these restrictions lead to the identification of the FIDOs
with the zero angular momentum observers (ZAMOs), see \citet{thorne}. Orthonormal tetrad of ZAMO is given by eqs.~(\ref{zamotetrad1})--(\ref{zamotetrad4}).

\begin{figure}[htb]
\centering
\includegraphics[scale=0.45, clip]{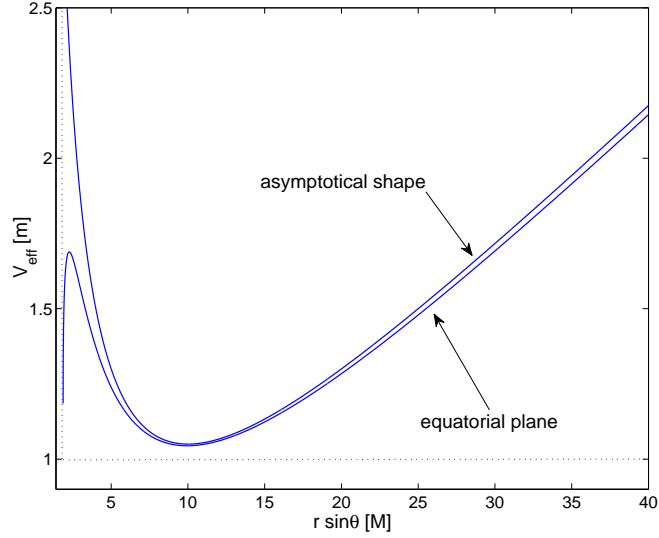}
\caption{Profiles of the effective potential $V_{\rm{eff}}$ for $\tilde{L} =5M$,
$a=0.5\:M$, $\tilde{q}\tilde{Q}=1.03$, $\tilde{q}B_{0}=0.1M^{-1}$, taken
in the equatorial plane and in the asymptotic region
$z=r\cos{\theta}\rightarrow\infty$ (as indicated with the corresponding curves). 
The valley crosses the equatorial
plane, almost unaffected in its bottom parts, although the behavior of
the potential near the symmetry axis is quite different in the
equatorial plane where it approaches the horizon of the black hole
(vertical dotted line at $r=r_+(a)$). The horizontal dotted line at unity measures 
the rest energy of the particle $m$.}
\label{valley}
\end{figure}

Projecting an arbitrary four-vector $C^{\mu}$ onto ZAMO's hypersurface of
simultaneity directly results in the following 3-dimensional quantities,
\begin{eqnarray}
\label{3Dslozky}
^{\rm{3D}}C^i&=&\gamma^{ij}C_j=\left({}C^r{},C^\theta{},\gamma^{\varphi\varphi}C_{\varphi}\right),\\
^{\rm{3D}}C_i&=&\gamma_{ij}^{\rm{3D}}C^j=\left(C_r,C_\theta,g_{\varphi\varphi}C^{\varphi}\right);
\end{eqnarray}
here, $\gamma_{\varphi\varphi}=g_{\varphi\varphi}$ ($u_\varphi=0$ for ZAMO) and,
consequently, $\gamma^{\varphi\varphi}g_{\varphi\varphi}=1$.

ZAMO tetrad components $^{\rm{3D}}C^{(i)}$ are given as
\begin{eqnarray}
\label{projekce}
^{\rm{3D}}C^{(i)}=^{\rm{3D}}C_{(i)}=e_{(i)j}^{\rm{3D}}C_{j}\equiv
&=\left(\sqrt{g^{rr}}C_{r},\sqrt{g^{\theta\theta}}C_{\theta},\frac{1}{\sqrt{g_{\varphi\varphi}}}C_{\varphi}\right)\\
\left(\sqrt{g_{rr}}C^{r},\sqrt{g_{\theta\theta}}C^{\theta},\sqrt{g_{\varphi\varphi}}C^{\varphi}\right).
\end{eqnarray}
The hypersurface components of the phase space constituents $x^{i}$ and $\pi_{i}$,
as measured by ZAMO, are then:
\begin{eqnarray}
\label{FIDOcomp}
^{\rm{3D}}x^{(i)}&=&(\sqrt{g_{rr}}r,\sqrt{g_{\theta\theta}}\theta,\sqrt{g_{\varphi\varphi}}\varphi),\\
^{\rm{3D}}\pi_{(i)}&=&\left(\sqrt{g^{rr}}\pi_{r},\sqrt{g^{\theta\theta}}\pi_{\theta},\frac{1}{\sqrt{g_{\varphi\varphi}}}L\right).
\end{eqnarray}
The spatial 3-metric is
\begin{equation}
\label{ZAMOmetric}
ds^2=\delta_{(i)(j)}dx^{(i)}dx^{(j)}+O(|x^{(k)}|^2)dx^{(i)}dx^{(j)},
\end{equation}
where $x^{(k)}$ represents the spatial distance from the origin of the
tetrad, i.e.\ ZAMO's current location. ZAMO is not an inertial observer,
which generally causes the first order corrections $O(|x^{(k)}|)$ to the
Minkowskian metric $g_{(i)(j)}=\eta_{(i)(j)}$. But these  do not enter
the spatial part of the metric. Thus the 3-metric within the spatial
hypersurface is a Euclidean one, with the deviations of second order in
the distance from the spatial origin on ZAMO's world-line.

\begin{figure}[!ht]
\centering
\includegraphics[scale=0.325,clip]{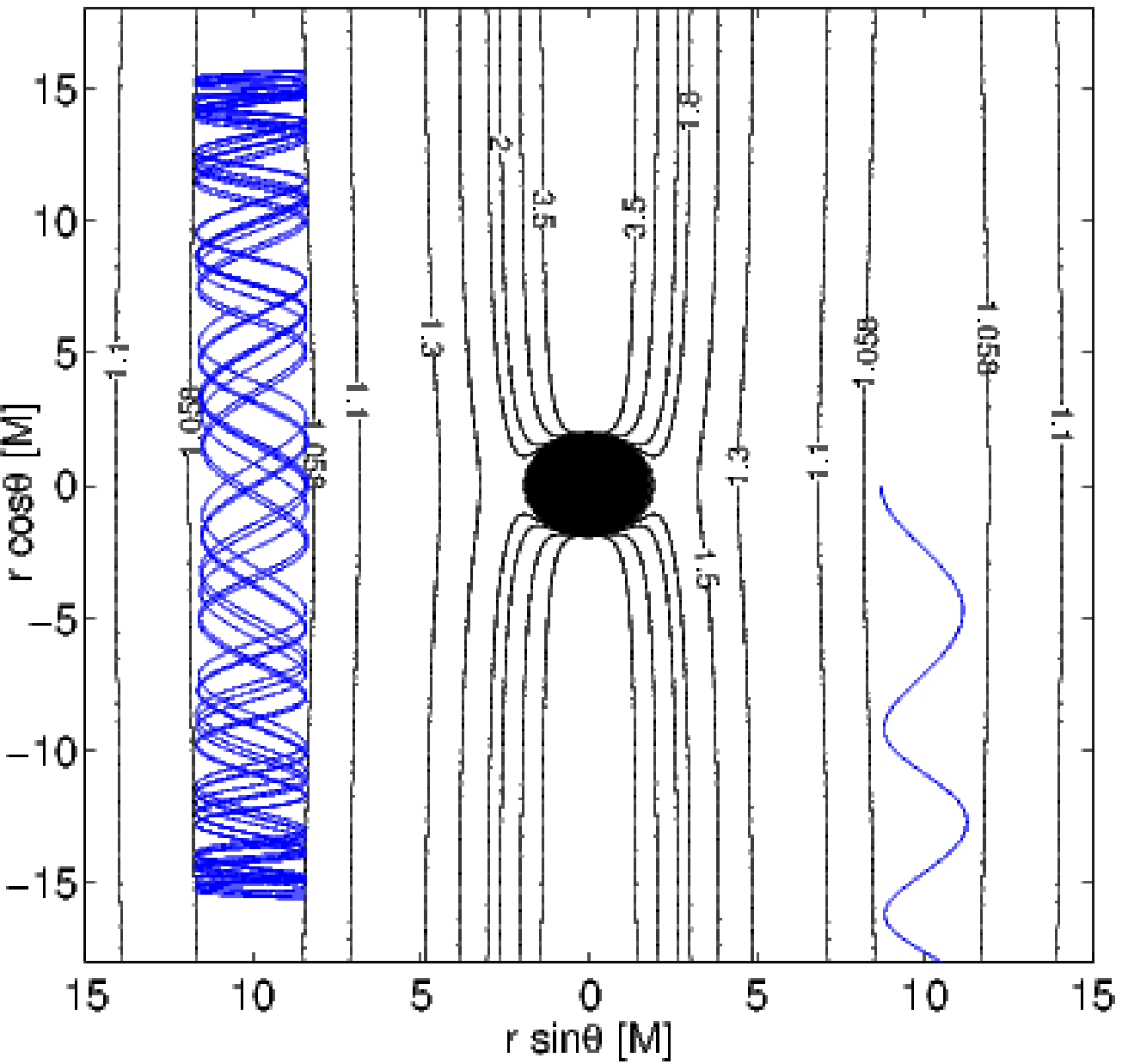}\includegraphics[scale=0.30,clip]{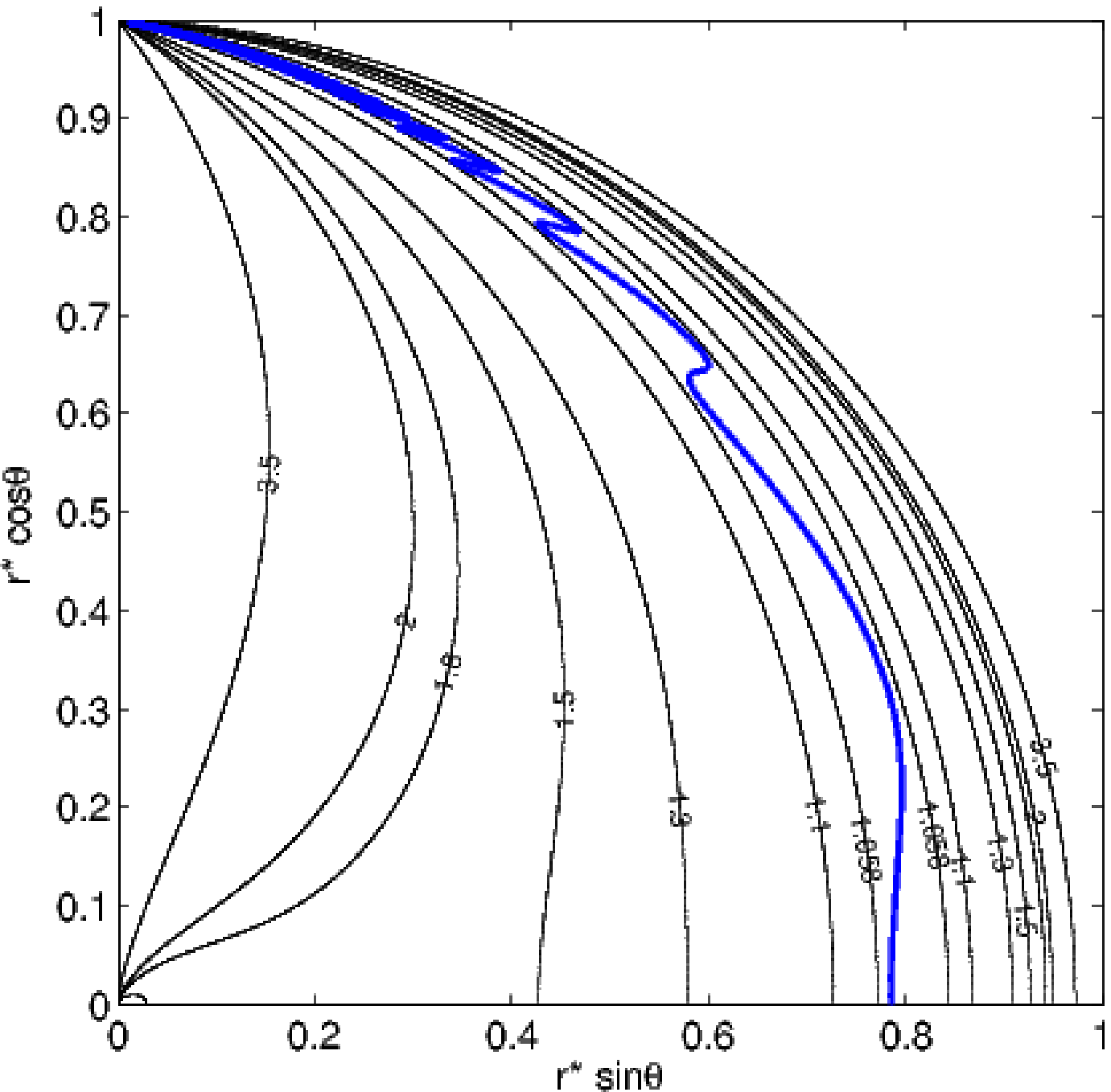}~\includegraphics[scale=0.32,clip]{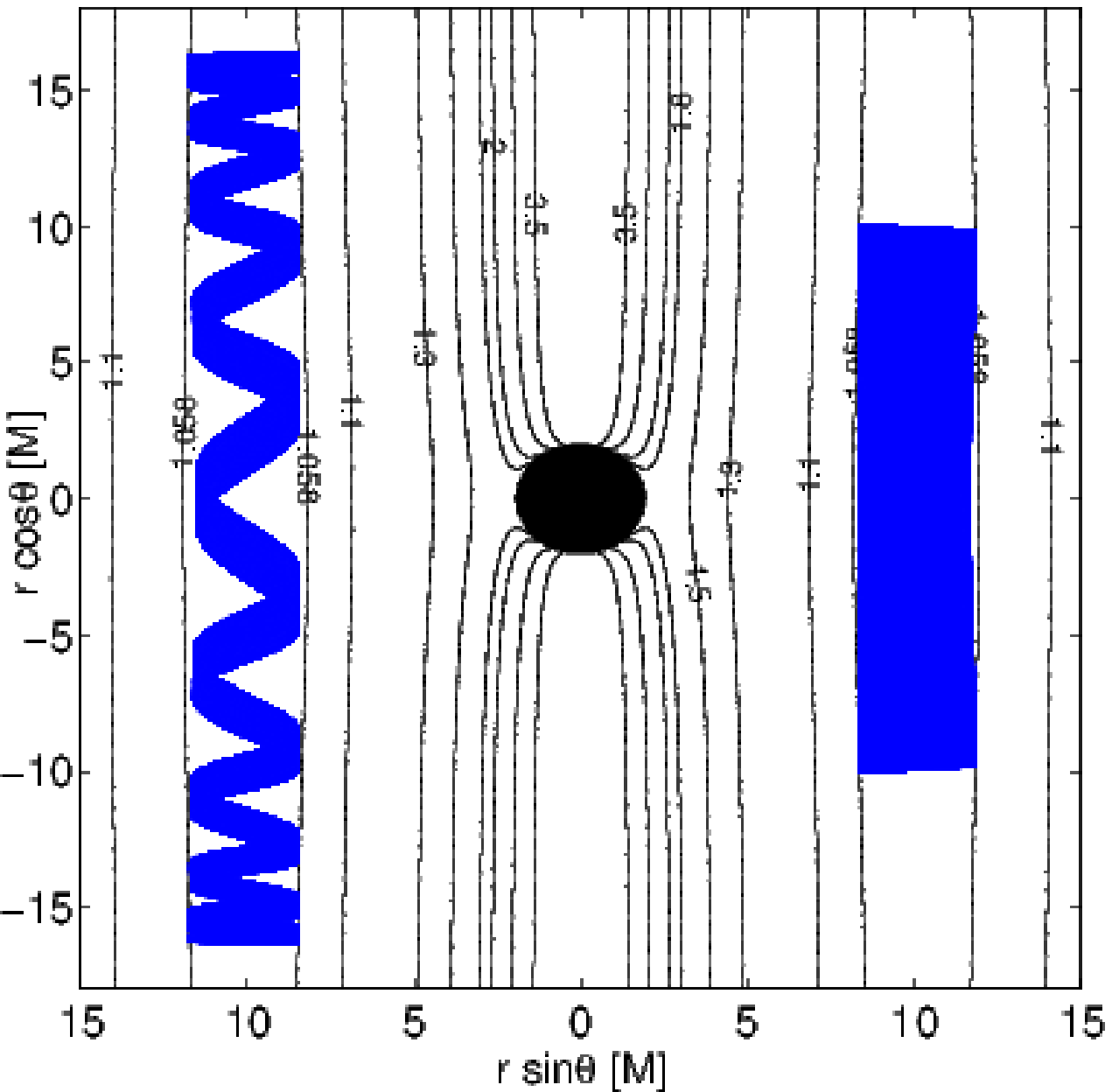}\\[10pt]
\includegraphics[scale=0.375,clip]{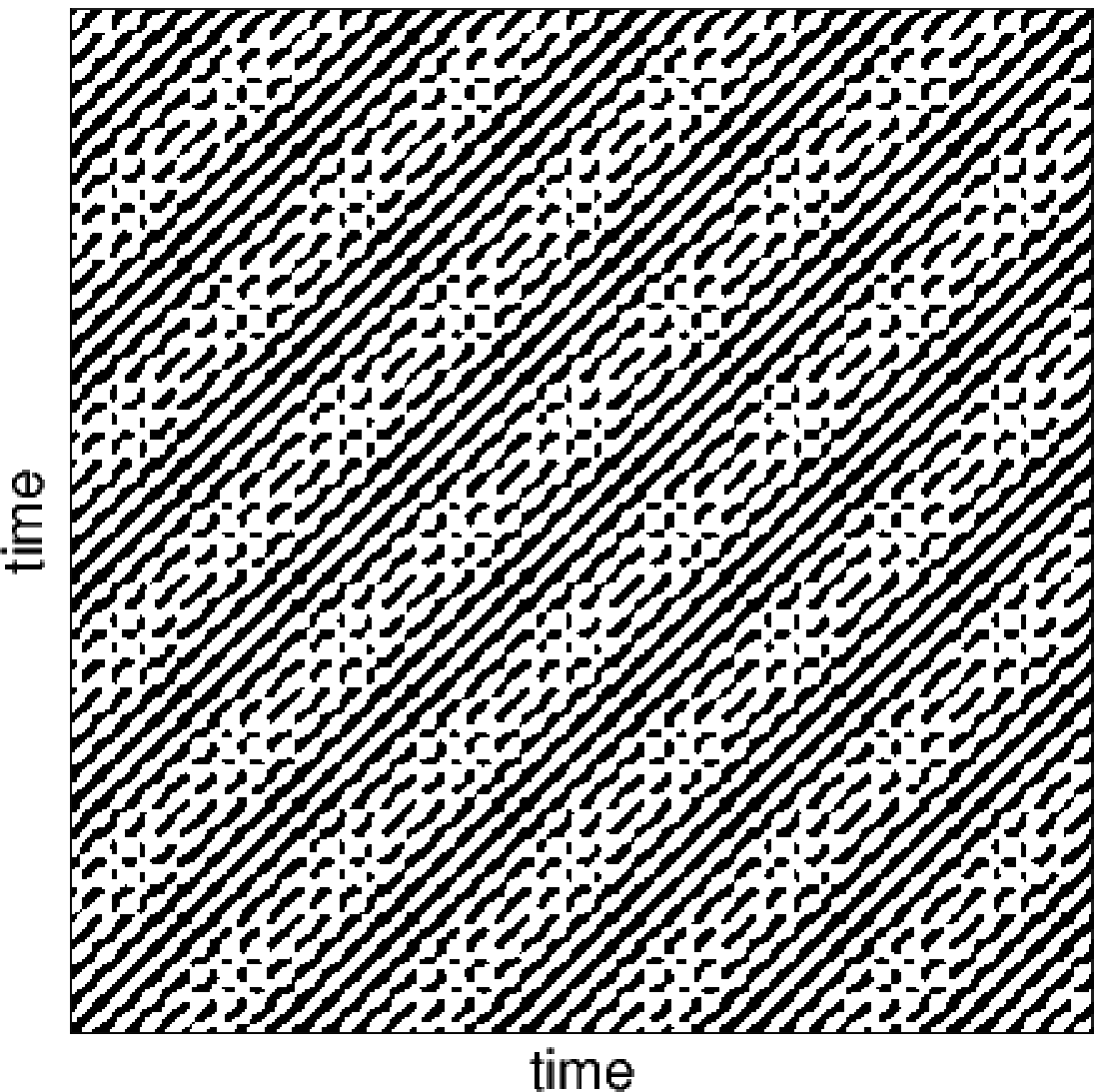}~\includegraphics[scale=0.405, clip]{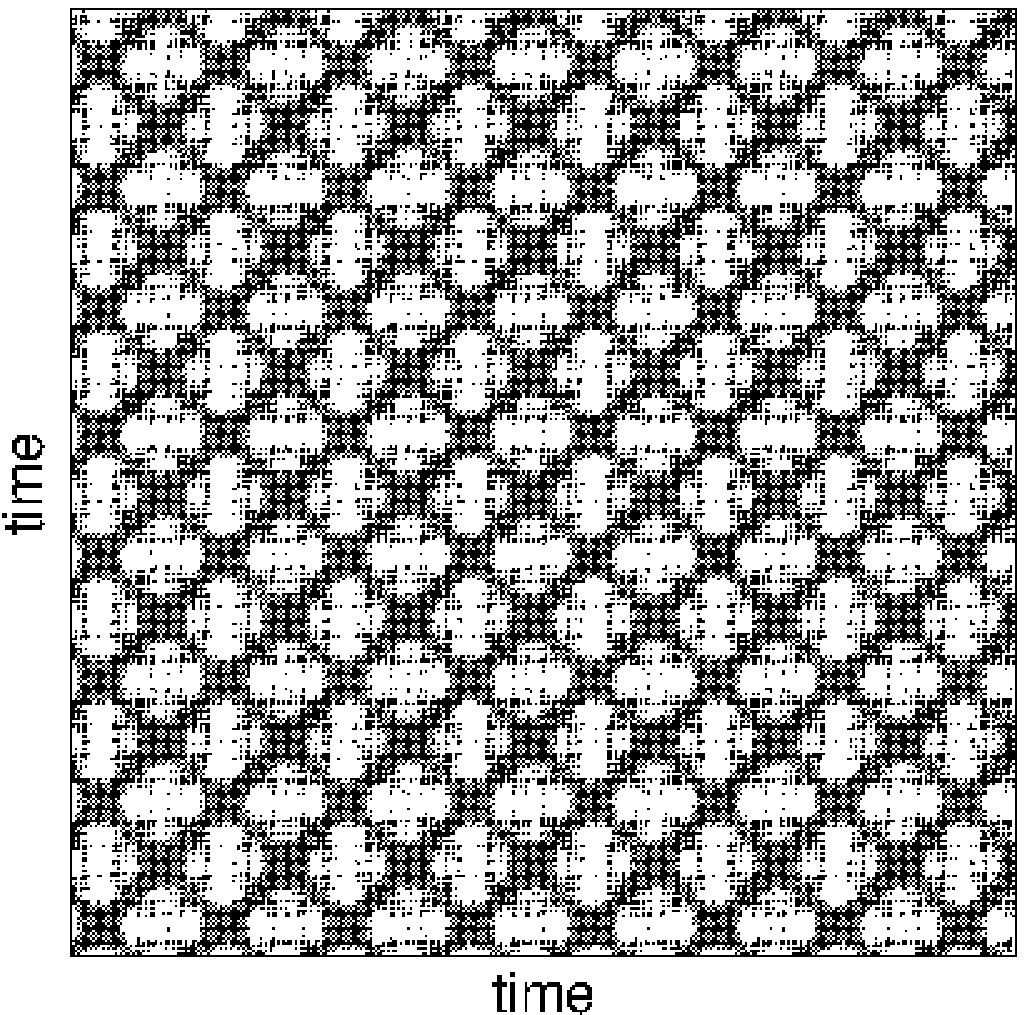}\includegraphics[scale=0.38,clip]{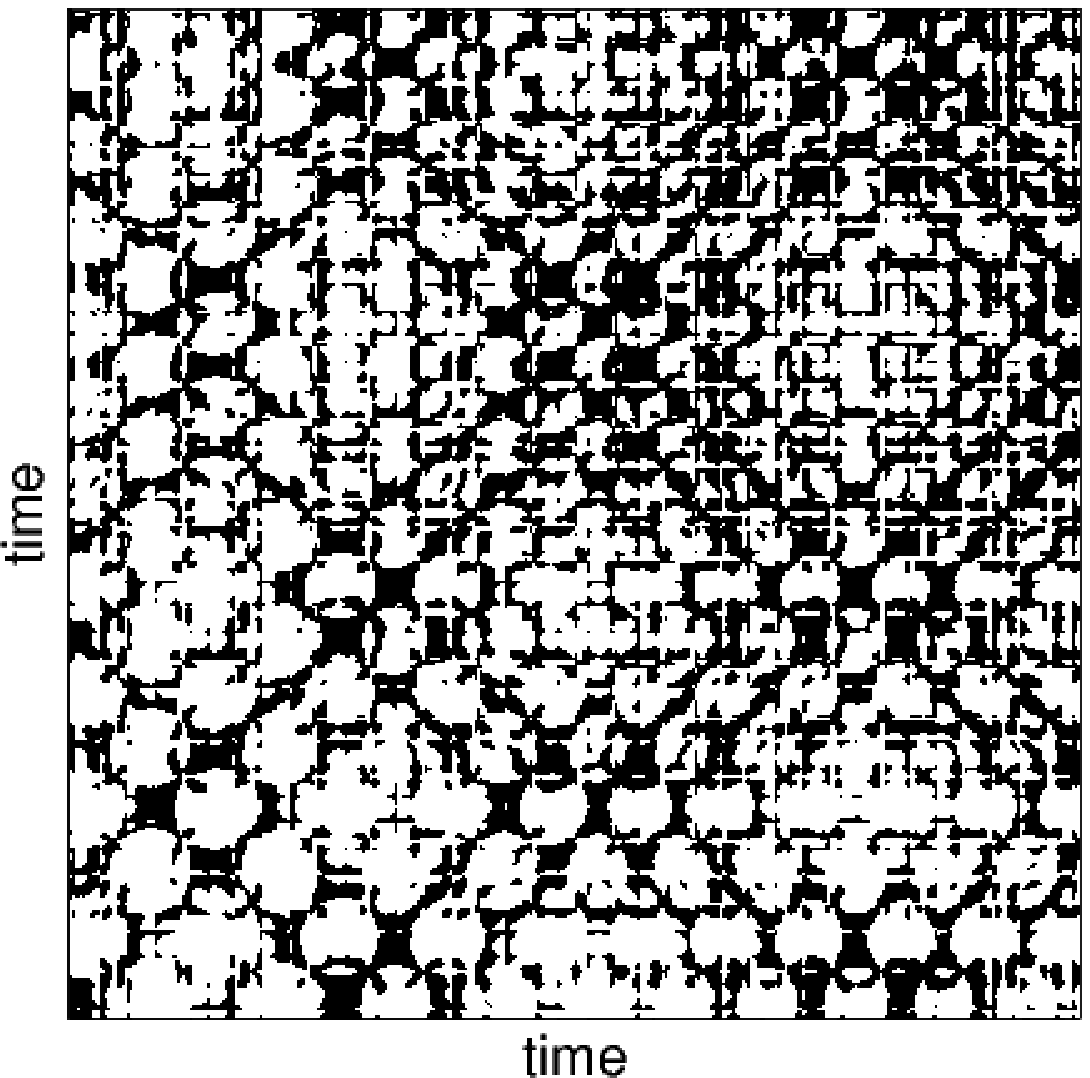}
\caption{An exemplary trajectory ($\tilde{E}=1.058$, $\tilde{L} =5M$)
is launched from the equatorial plane
$\theta(0)=\frac{\pi}{2}$ with $u^r(0)=0$. Parameters of the
background are $a=0.5M$, $\tilde{q}B_{0}=0.1M^{-1}$, $\tilde{q}\tilde{Q}=1.03$. In the upper left panel we observe
that setting $r(0)=8.4M$ results in oscillations around the
equatorial plane while launching it at $r(0)=8.7M$ makes it escape.
In the upper middle panel we examine the trajectory of the escaping
particle in terms of the rescaled radial coordinate
$r^*\equiv\frac{r-r_{+}}{r}$. In the case of oscillating trajectories
two distinct modes of motion are observed (upper right panel). The
first particle ($r(0)=11.5M$) shows a complex ``ribbon-like'' trajectory;
the other one ($r(0)=8.4M$) fills uniformly the given portion of
the potential valley. The Recurrence Plots are also shown (bottom panels). 
We observe a highly ordered regular pattern for the particle with $r(0)=8.4M$
(left panel), a more complicated diagonal pattern of the ribbon--like
trajectory (launched at $r(0)=11.5M$, middle panel), and a
disrupted diagonal pattern of the transitional trajectory ($r(0)=11.4M$,
right panel).}
\label{osc_unik1}
\end{figure}

We will use ZAMO's metric at distances up to the value of the threshold
parameter $\varepsilon$. The Euclidean metric according to \rf{ZAMOmetric} will be therefore justified if
$\frac{\varepsilon^2}{P^2}\ll{}1$, where $P$ stands for a constant (for
a given ZAMO). $P$ has the dimension of length and characterizes the
curvature of the hypersurface. We suggest setting $P\equiv{}K^{-1/4}$,
where $K=R^{\mu\nu\xi\pi}R_{\mu\nu\xi\pi}$ represents the Kretschmann
scalar evaluated from the Riemann curvature tensor. In the case of the
Kerr black hole the Kretschmann scalar may be expressed in a
surprisingly simple form \citep{henry}. While constructing the Recurrence
Plots, we check whether the condition $\frac{\varepsilon^2}{P^2}\ll{}1$
remains satisfied.

The above-mentioned adoption of preferred observers is needed in
order to maintain an operational criterion of chaos and be able to
formulate an explicit form of the equations for RQA measures in a
curved spacetime. Here the notion of the phase space distance plays
a role. In Kerr metric (or another axially symmetric stationary
spacetime), Fiducial Observers (a.k.a. FIDOs) represent a natural
selection of preferred observers. Obviously, this option is not
unique, and so a detailed appearance of the recurrence plots is also
ambiguous to certain extent. But not so the main conclusions that
we infer regarding the chaoticness versus regularity of the system
behavior, because this distinction can be eventually traced down to
the exponential versus polynomial growth of the separation with the
particle proper time along neighboring trajectories.

We can deduce the kind of transformation between different
families of observers that could affect our conclusions: these are
transformations involving exponential dependencies on
observer's phase-space position. For example transformation to
accelerated frames and spacetime points in the vicinity of
singularities may need a special consideration, as well as the
investigation of highly dynamical spacetimes that are lacking
symmetries. On the other hand, selecting LNRF to define ZAMOs in
(weakly perturbed) Kerr metric outside the black hole horizon
appears to be a well-substantiated choice.

\begin{figure}[htb]
\centering
\includegraphics[scale=0.55, clip]{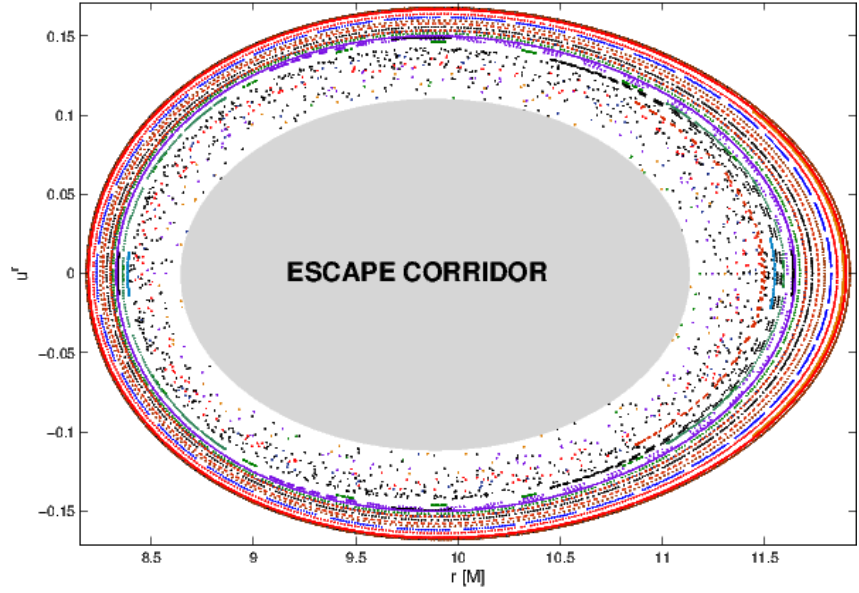}
\caption{The Poincar\'e
surface of section of several trajectories
($\tilde{E}=1.058$, $\tilde{L} =5M$, $\tilde{q}B_{0}=0.1M^{-1}$, 
$\tilde{q}\tilde{Q}=1.03$, $a=0.5M$, $\theta=\frac{\pi}{2}$)
launched from the equatorial plane with
various values of $r(0)$ and  $u^r(0)=0$. Grey colour indicates the
escape corridor which lets the particles escape from the equatorial
plane (see details in the text).} \label{corridor}
\end{figure}

Similar arguments for the adoption of preferred observers on the
basis of spacetime symmetries have been elaborated in greater detail
by \citet{karas92} in the context of Ernst's
magnetized black hole, which is another particularly simple (static)
exact solution of Einstein-Maxwell equations exhibiting the onset of
chaos as the magnetic field strength is increased.

\section{Kerr black hole in uniform magnetic field}
\label{sectionwald}
Large-scale magnetic fields are known to be present in cosmic
conditions. They can exist around black holes, which do not support
their own magnetic field but may be embedded in fields of distant
sources. In the case of neutron stars, dipole-type magnetic fields of
very high strength often arise. We concentrate on black holes in this
section and defer the case of a magnetic star to \rs{sectionmagnetized}.

\begin{figure}[htb]
\centering
\includegraphics[scale=0.54, clip]{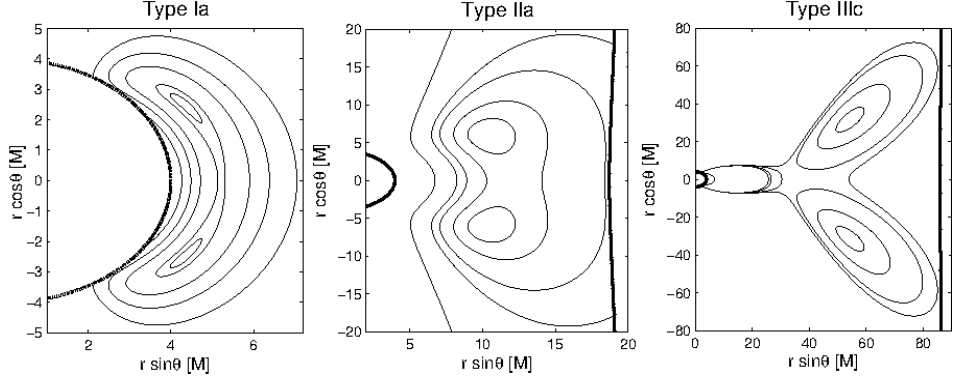}
\caption{Selected 
types of the effective potential behavior in the vicinity of
off-equatorial halo orbits above the surface of magnetic star with
rotating dipole magnetic field. The inner bold line signifies the
surface of the star at $r=4M$. The outer line is the light
surface.}
\label{rotdip_abc}
\end{figure}

By employing the uniform test field solution \citep{wald74} we incorporate
a weak large-scale magnetic field near a rotating black hole. The
vector potential can be expressed in terms of Kerr metric coefficients
(\ref{kn}) as follows,
\begin{eqnarray}
\label{waldpot1}
A_t&=&\textstyle{\frac{1}{2}}B_{0}\left(g_{t\varphi}+2a\,g_{tt}\right)-\textstyle{\frac{1}{2}}\tilde{Q}g_{tt}-\textstyle{\frac{1}{2}}\tilde{Q},\\
\label{waldpot2}
A_{\varphi}&=&\textstyle{\frac{1}{2}}B_{0}\left(g_{\varphi\varphi}+{2a}g_{t\varphi}\right)-\textstyle{\frac{1}{2}}\tilde{Q}g_{t\varphi},
\end{eqnarray}
where $B_0$ is magnetic intensity and  $\tilde{Q}$ stands for the test
charge on the background of Kerr metric. The terms containing
$\tilde{Q}$ can be identified with the components of the vector
potential of Kerr-Newman solution (although the test charge does not
enter the metric itself). An example of an integrated trajectory is
shown in \rff{fig1}. \citet{wald74} has shown that the black hole
selectively accretes charges from its vicinity, until it becomes
itself charged to the equilibrium value
\begin{equation}
\label{waldcharge} \tilde{Q}_{\rm{W}}=2B_{0}a.
\end{equation}

We remark that the particle charge $\tilde{q}$ appears always as a
product with $\tilde{Q}$ or $B_{0}$ in the formula (\ref{effpot}) for
the effective potential, as well as in equations of motion
(\ref{HamiltonsEquations}). Therefore, the simultaneous alteration of
$\tilde{q}$, $\tilde{Q}$ and $B_{0}$ values, preserving the products
$\tilde{q}\tilde{Q}$ and $\tilde{q}B_{0}$, does not affect the particle
dynamics. If we further assume that
$\tilde{Q}=\tilde{Q}_{\rm{W}}=2B_{0}a$ is maintained, we only need to
specify the value of $\tilde{q}B_{0}$ to uniquely determine a particular
trajectory. However, since we do not restrict ourselves to the case
$\tilde{Q}=\tilde{Q}_{\rm{W}}$ we decide to always explicitly specify
the values of $\tilde{q}\tilde{Q}$ and $\tilde{q}B_{0}$.

\subsection{\label{wald_lobes}Motion within the potential lobes}
Our previous analysis \citep{halo2} concludes that the off-equatorial
bound orbits are allowed only for test particles obeying simultaneously
the two conditions, ${\rm sgn}(aL)=1$ and ${\rm sgn}(\tilde{q})={\rm
sgn}(aB_{0})$. Four distinct types were found (see \rff{wald_abc}) and we
examined the dynamics of test particles in all of these types. The results 
for different types are comparable, and so we present here only the analysis of two of them, namely the types Ia and Id. While raising the energy level from the
local minima of the symmetric halo orbits, we observe that the
off-equatorial lobes grow and eventually merge with each other once 
the energy of the saddle point in the equatorial plane is reached. If we further increase the energy the behaviour between the two classes differs profoundly. For Ia we observe that the lobe eventually opens toward the horizon allowing the particle to fall onto the black hole. On the other hand in the Id setup the merged lobe breaks ``toward infinity'' allowing the particle to escape in the axial direction when the energy is raised sufficiently.

The size of the lobes is controlled by the specific energy $\tilde{E}$.
Employing the Poincar\'e surfaces of section and the Recurrence Plots we
investigate how the regime of the particle motion changes with
$\tilde{E}$. This parameter appears as a suitable control parameter
producing a sequence of bound trajectories while all other parameters
(and the initial position) remain fixed. For the sake of comparison we
first present the case of a fully integrable system of a charged test
particle on the pure Kerr-Newman spacetime (\rff{rppoinc1}). The motion
occurs in the potential lobe around the local minimum in the equatorial
plane; there are no halo orbits above the horizon in this case
\citep{halo2, Fel:1979}.

On the Kerr background with Wald test field we first analyze the sequence of trajectories of Ia type. \Rff{rppoinc2} shows regular motion occurring in the off-equatorial lobe. Increasing the energy level (while keeping all other parameters fixed)
above the value in the equatorial saddle point results in a transitional
regime depicted in \rff{rppoinc4}. The onset of chaotic features does
not occur as a direct consequence of the lobe merging. We rather observe
that the orbit bound in merged (cross-equatorial) lobe remains regular
until the particle {\it notices} the possibility of crossing the
equatorial plane which happens when its energy is increased
sufficiently. Once the motion becomes cross-equatorial, chaotic features
appear. By increasing the energy even more we approach the critical
value when the lobe opens and allows the particles to fall onto the
horizon. For energies slightly below this limit we detect a fully
chaotic regime of motion (\rff{rppoinc3}).

Similar sequence of trajectories is analyzed in the case of Id class of the topology of the potential wells. Three typical trajectories found in this case are shown: regular off-equatorial, regular cross-equatorial and chaotic cross-equatorial orbit. First we present their overview in \rff{wald_d_traj} which shows the main difference from the Ia series clearly: the presence of fully regular cross-equtorial orbits. In \rff{wald_d_diskuze} we compare Poincar\'{e} surfaces of section and Recurrence Plots of these trajectories. Unlike previous series we distinguish $u^{\theta}\geq0$ from $u^{\theta}<0$ in the Poincar\'e surface of section.

Figure \ref{unthresholded} shows an alternative representation of the recurrence plots, where the phase space
distance separating the given pair of points of the trajectory is encoded by 
different colors. Again, by comparing the two panels one can clearly recognize how the 
diagonal structures disintegrate into scattered points as the chaos sets in.

From the survey of the first series of Poincar\'e surfaces of section and the
Recurrence Plots (figs. \ref{rppoinc2}--\ref{rppoinc3}) we may only conclude that the transition from the
regular to chaotic regime occurs somewhere close to the
value $\tilde{E}=1.796$. In order to localize this transition more
precisely we evaluate RQA measures for 400 trajectories 
with the energy spread equidistantly over the interval
$\tilde{E}\in(1.7948,1.7968)$. In figs.\ (\ref{rqa1}) and
(\ref{rqa2}), we observe a sudden change of the behavior of
the statistical measures at $\tilde{E}\approx{}1.7954$,
reflecting a dramatic change of the particle dynamics.

Moreover, we know that the divergence $\mathrm{DIV}$ is related to the
Lyapunov exponents, and in \rff{rqa1} we observe that it suddenly rises
at $\tilde{E}\approx{}1.7954$, meaning that the trajectories become more
divergent when this energy is reached. All of these indications combined
lead to the conclusion that this energy level represents a critical value
at which a transition from regular to chaotic regime occurs.

Similarly we perform the RQA for the sequence of orbits of Id type which we analyzed qualitatively in \rff{wald_d_diskuze}. Visual survey suggests that the suspicious interval in which the transition from regular to chaotic dynamics shall occur is  $\tilde{E}\in(1.65,1.75)$. We calculate 200 trajectories with energies equidistantly spread over the given range while other parameters of the system are fixed at values used in \rff{wald_d_diskuze}. In figs. \ref{rqa_d1} and \ref{rqa_d2} we observe that all queried RQA measures exhibit sudden change in its behaviour at $\tilde{E}\approx1.685$. This is where the dynamic transition between the regimes occurs. Transition from the regular motion to the chaos is detected not only by diagonal RQA measures but also by the vertical ones.

We conclude that the energy of the particle $\tilde{E}$ acts as a
governing factor determining the dynamic regime of motion. Our survey
across various initial conditions has shown that motion in potential
wells of the type Ia in a Wald test field is generally regular. Chaos
appears well after the merging point of the lobes and close to the
critical breaking energy. We have verified that all queried RQA
measures, i.e.\ not only those based on diagonal lines in RP, react to
the transition from the regular to chaotic regime of motion, allowing us
to localize precisely the transition. This was fully confirmed also in
the other two types (classes Ib and Ic of our typology in
\rff{wald_abc}).

\subsection{\label{spindep}The effect of spin on the chaoticness of motion}
Much attention has been recently focused towards the problem of
determining the black hole spin from the properties of motion of
surrounding matter \citep{narayan,reynolds03}. The astrophysical
motivation to address these issues arises from the fact that cosmic
black holes are fully described by three parameters -- mass, electric
charge and spin. While the methods of mass determination have been
widely discussed \citep[e.g.][]{casares07,vestergaard10,czerny10}, the
electric charge is considered to be negligible because of rapid
neutralization of black holes via selective accretion. However,
determining the spin is a much more challenging task: the spin is
important, but its influence is apparent only {\em very} near the black hole
horizon \citep{murphy09}.

\begin{figure}[htb]
\centering
\begin{tabular}{rl}
\includegraphics[scale=0.403, clip=true]{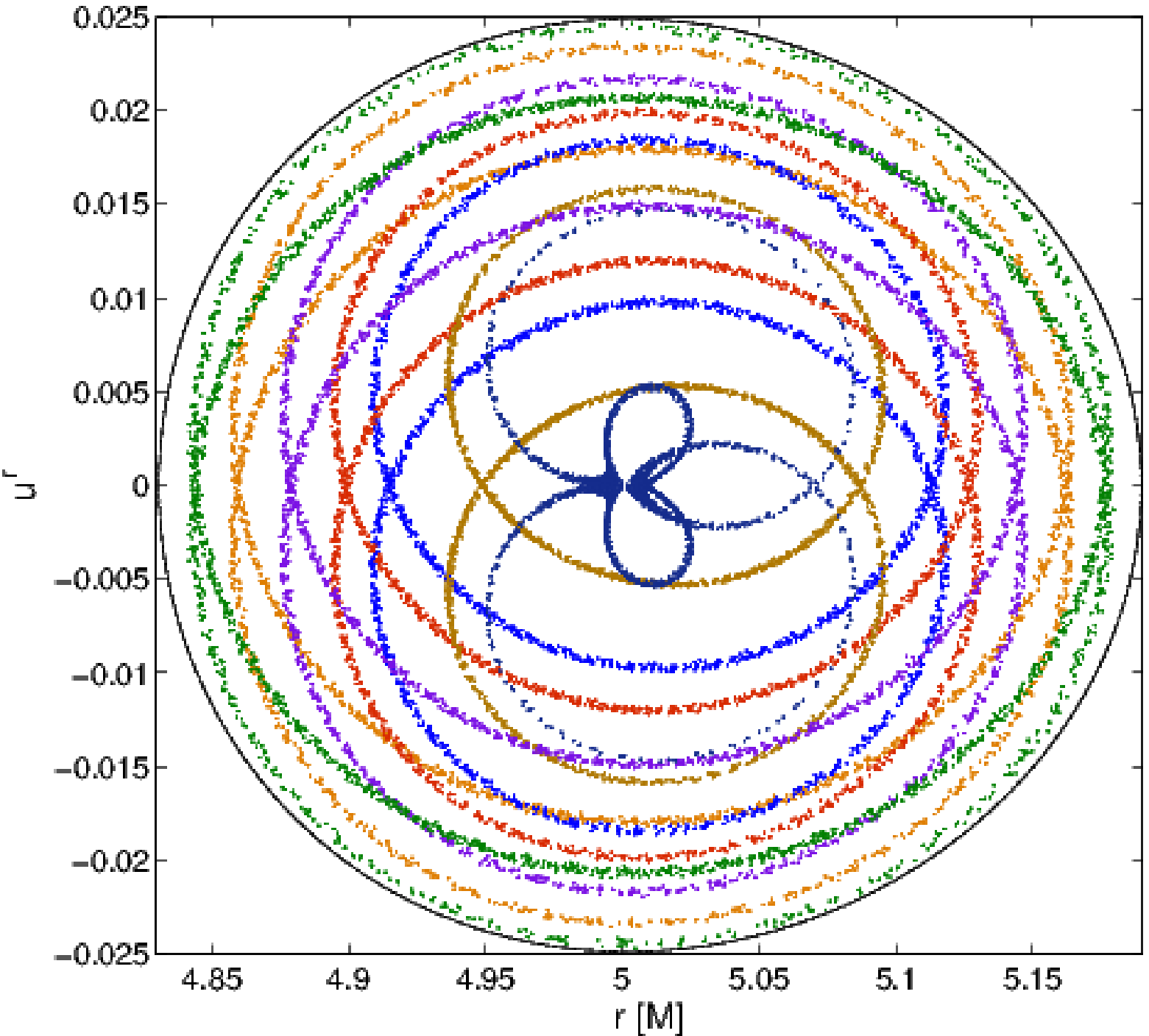} &\includegraphics[scale=0.33,clip=true]{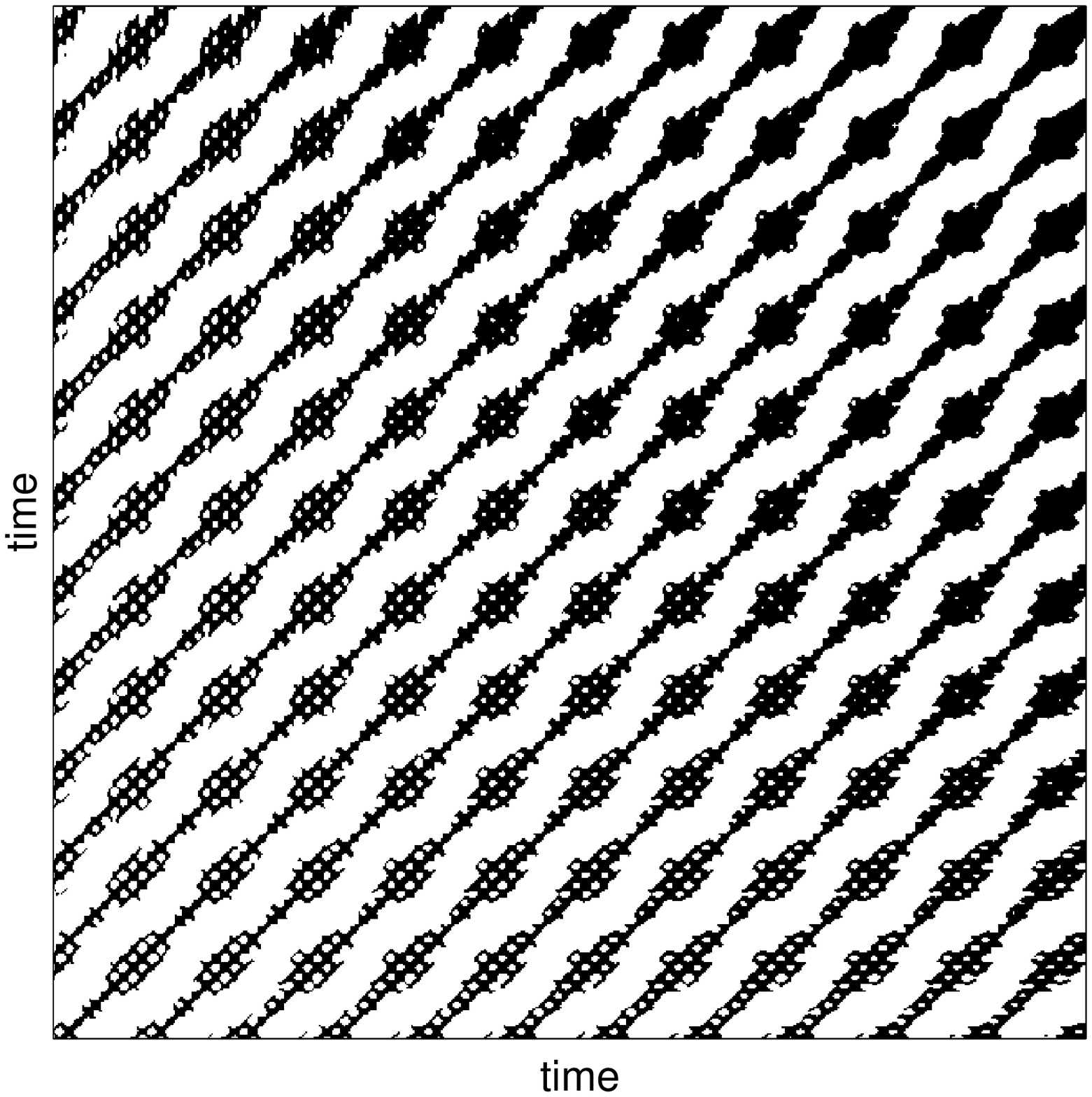}\\
\end{tabular}
\caption{Regular motion in the off-equatorial potential lobe at the
energy level $\tilde{E}=0.8482$. Parameters of the system are
$\tilde{q}\mathcal{M}=-5.71576\; M^2$, $\tilde{L}=0.87643\; M$,
$\Omega=0.011485\; M^{-1}$. The left panel shows sections of several
trajectories launched from
$\theta(0)=\theta_{\rm{section}}=1.0492$. One of them ($r(0)=5.02\;
M$ and $u^r{}(0)=0$) is visualized in the Recurrence Plot in the
right panel. The motion is regular (RP remains diagonal), however,
the density of recurrence points clearly grows
during the analyzed period.} \label{rotdip1_1p}
\end{figure}

One can raise a question of whether the value of spin parameter $a$ of
the Kerr black hole affects the dynamical regime of motion in the
immediate neighborhood of the black hole. In other words, we ask if the
spin parameter $a$ triggers or diminishes the chaoticness of the system.
Answering this question is not straightforward because by altering the
spin across an interval of values ($a^2\leq{}M^2$) we inevitably have
to change some other variables of the system; otherwise the different
cases could not be directly compared. Moreover, the location and the
very existence of the potential lobes is not automatically ensured  over
the whole range of spin because of strong $V_{\rm{eff}}(a)$ dependence.

We found that in order to keep the off-equatorial lobe at the initial
position and the original size we need to increase the energy
$\tilde{E}$, roughly proportionally to the increment of $a$. The effect
of increasing $a$ exhibits itself by lifting the hyperplane of effective
potential. To compensate for this effect we have to elevate the
$\tilde{E}$-plane at which we cut the potential, so that we obtain
(roughly) the original closed contour (the potential lobe) inside of
which the motion is confined. It turns out that this can be achieved by
linking both quantities linearly.

First we compare the dynamics in the off-equatorial lobe in the range
$\frac{a}{M}\in{\langle}0.3,1{\rangle}$ (for $a\lesssim{}0.3M$ the
topology of the effective potential changes) to which we linearly relate
the energy range $\tilde{E}\in{\langle}1.56, 2.35{\rangle}$ (whilst
other parameters are kept fixed as follows: $\tilde{L} =5M$,
$\tilde{q}B_{0}=2M^{-1}$, $r(0)=2.9\:M$, $\theta(0)=0.856$,
$u^{r}(0)=0$, $\tilde{q}\tilde{Q}=2$). By inspecting the Poincar\'e
surfaces of section and performing the recurrence analysis for a large
number of trajectories across the given range of $a$ and $\tilde{E}$
(exemplary cases presented in
\rff{spin_diskuze}) we come to the conclusion that there is no overall
trend that could suggest that $a$ is the unique driving agent affecting
the regime of motion. All trajectories in our survey exhibit a regular
behavior, which is also in agreement with the previous conclusion that
the motion in off-equatorial potential lobes associated with the Wald
test field is generally regular.

\begin{figure}[!ht]
\centering
\includegraphics[scale=0.555, clip=true]{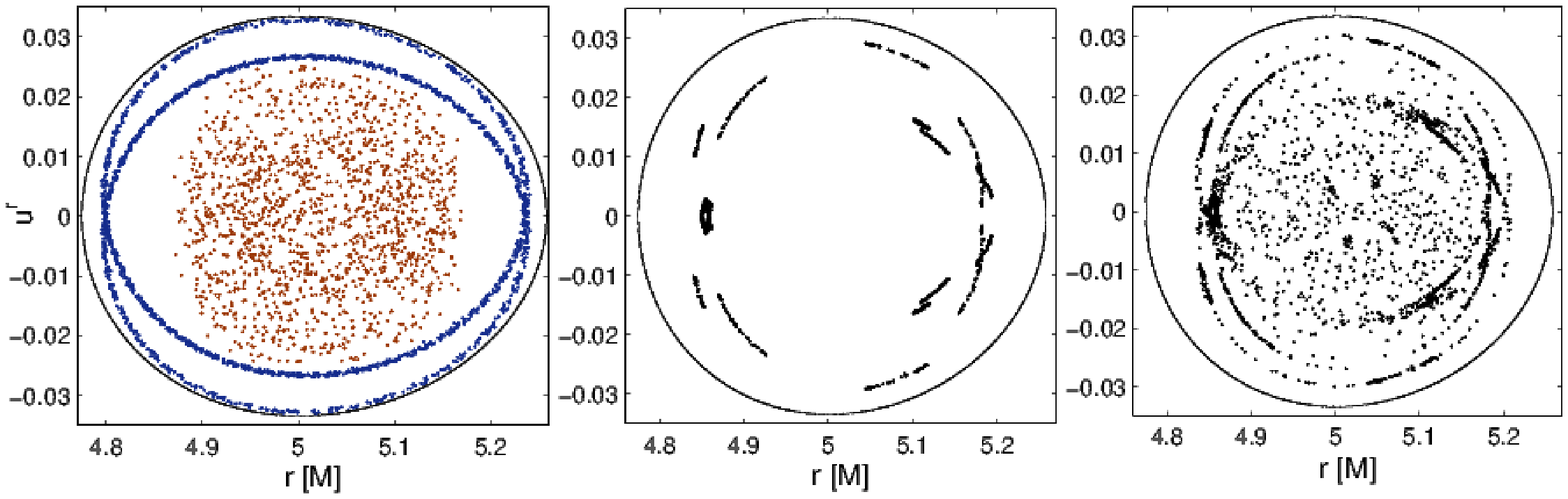}\\[10pt]
\includegraphics[scale=0.325, clip=true]{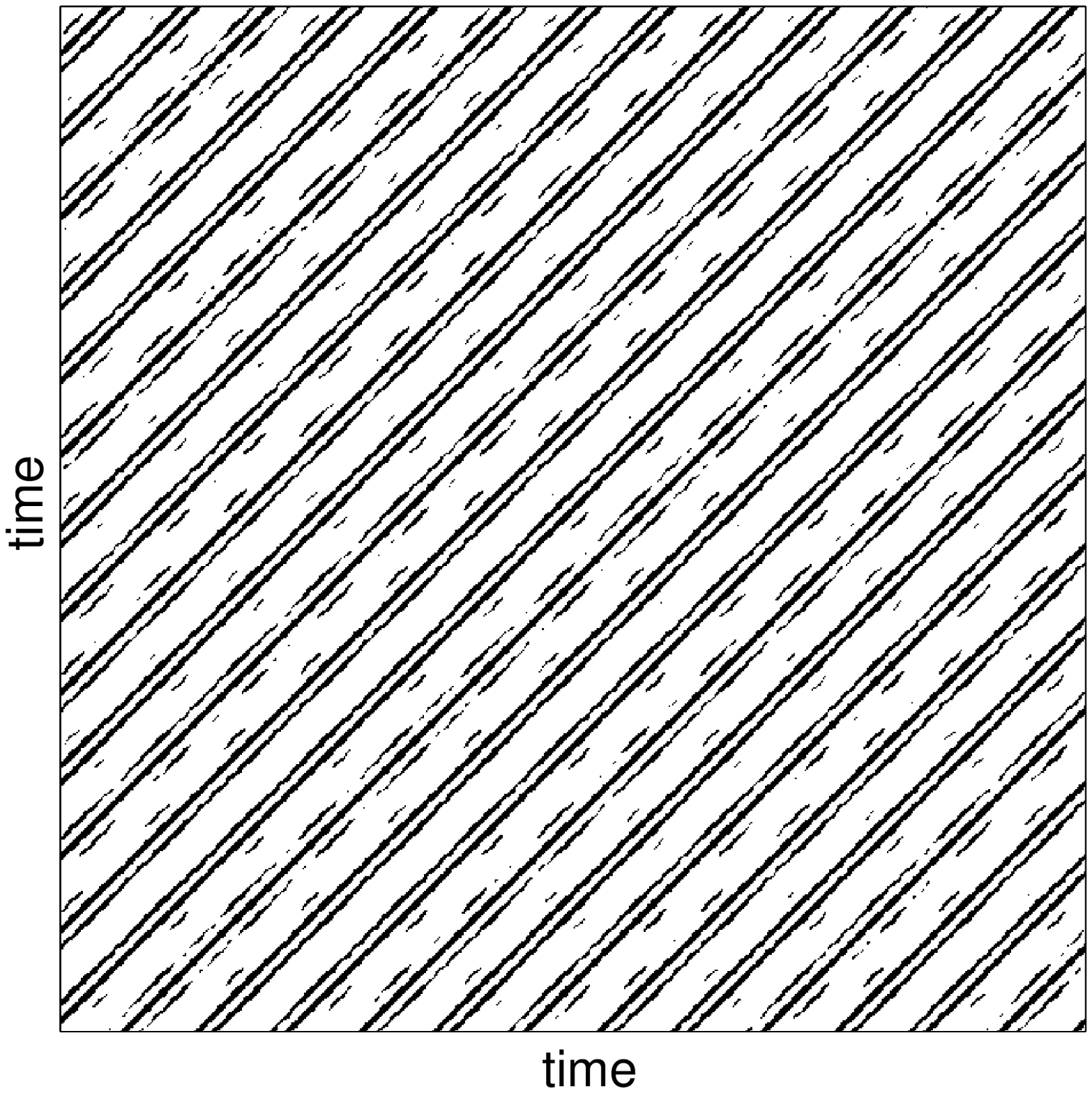}~~
\includegraphics[scale=0.3, clip=true]{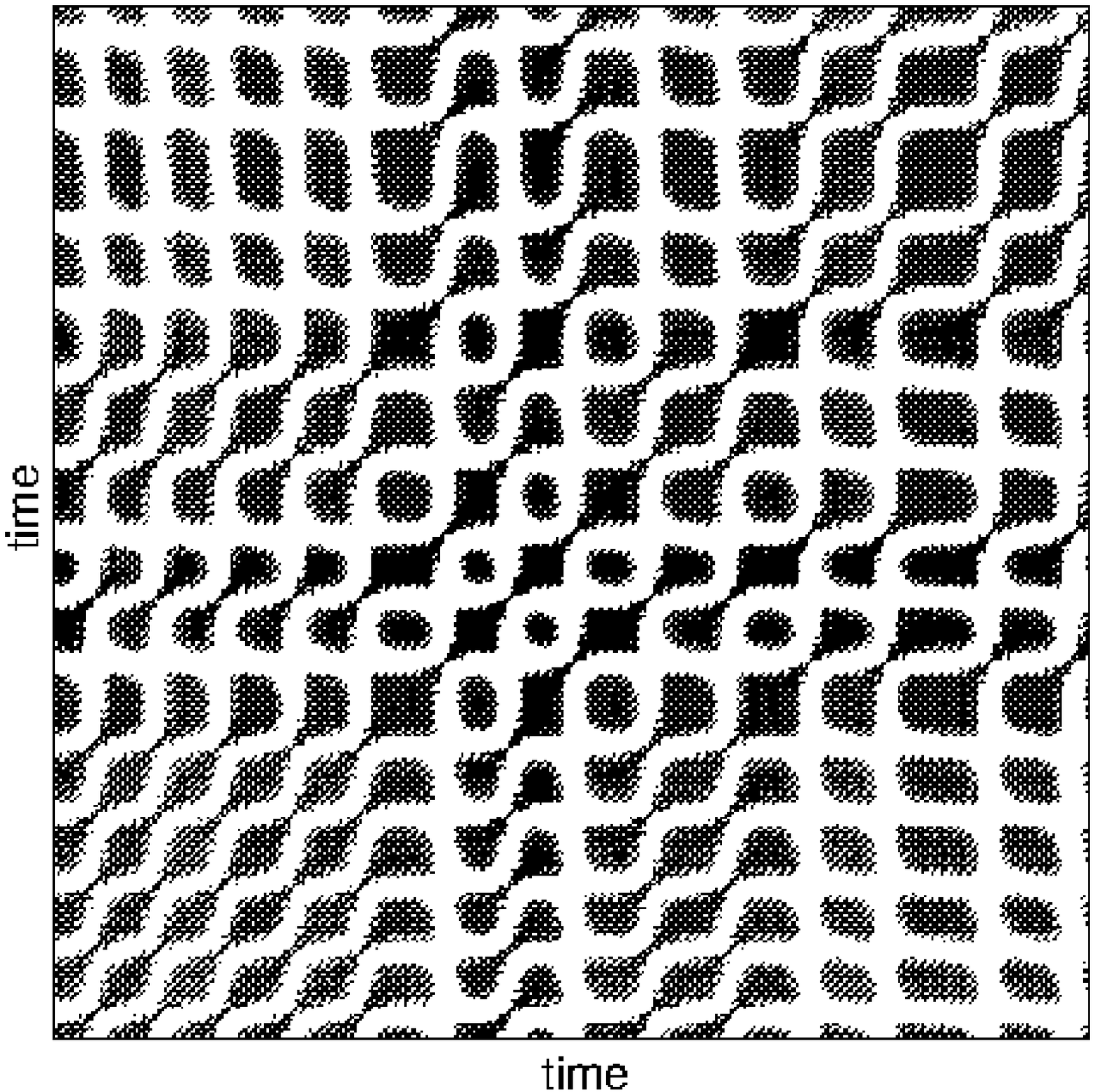} 
\caption{For an energy value of
$\tilde{E}=0.8485$, both off-equatorial lobes merge via the equatorial
plane. The upper-left panel shows Poincar\'e sections of two
trajectories ($\theta(0)=\theta_{\rm{section}}=1.0492$, $u^r(0)=0$). A
particle launched at $r(0)=4.8 \;M$ never crosses the equatorial plane
and moves regularly. Setting $r(0)=5\; M$ we observe a chaotic motion 
crossing the equatorial plane repeatedly. All
particles launched with $r(0)$, $u^r(0)$, corresponding to the inner
parts of the potential curve, move in the same chaotic manner. The
outskirts are occupied by regular trajectories. The upper-middle panel
shows the transient trajectory ($r(0)=4.85\; M$), regular during
the integration period of $\lambda=10^5$. In the upper-right panel, the
integration time is prolonged to
$\lambda=3\times{}10^5$. Here, the onset of chaos is connected with the first
passage through the equatorial plane. The RP of the regular trajectory
with $r(0)=4.8 \;M$ is presented in the bottom-left panel; the RP
on the right belongs to the chaotic trajectory with $r(0)=5 \;M$.}
\label{rotdip1_2}
\end{figure}

We also examined the dynamics of test particles launched from the
equatorial plane whose trajectories occupy the potential lobe extending
symmetrically above and below the equatorial plane. A given lobe
maintains its size for spin values 
$\frac{a}{M}\in{\langle}0.5,1{\rangle}$ and the related interval of
energy $\tilde{E}\in{\langle}1.795, 2.42{\rangle}$. A survey across the
given range of spin (energy) values reveals for this class of
trajectories both chaotic and regular regimes. In
\rff{spin_diskuze_eq} we observe that for the lowest inspected spin,
$a=0.5M$ ($\tilde{E}=1.795$), the regular motion dominates, although
islands of chaotic behavior are also present. Increasing the spin
(energy) we observe that regular trajectories gradually diminish. For
$a=0.6M$ ($\tilde{E}=1.92$) some regular orbits still appear, but they
are already dominated by chaotic trajectories. For higher spins
the traces of regular motion further diminish. We present the extreme case $a=M$
($\tilde{E}=2.42$) in \rff{spin_diskuze_eq} to illustrate this apparent
chaotic takeover. 

We conclude that for the class of orbits originating
in the equatorial plane, spin $a$ could possibly act as a destabilization factor
which triggers the chaotic motion if enhanced sufficiently. However
since the energy $\tilde{E}$ is increased simultaneously it is not
possible to attribute the observed dependence to the spin {\em itself}.
On the other hand we have already seen in the above-given discussion (\rs{wald_lobes}) that energy $\tilde{E}$ may {\em itself} act as a key factor
determining the dynamic regime of motion. Thus we suggest to attribute the observed triggering of the chaos to the increase of energy $\tilde{E}$ rather then spin $a$.

We remark that a similar problem concerning the spin dependence of the
motion chaoticness was addressed recently by \citet{japonci}. Authors of
the quoted paper employ a dipole magnetic test field upon a Kerr
background and perform a study of test particle trajectories, concluding
that increasing the value of the spin parameter stabilizes the motion in
a given setup. Unlike our case, the topology of the effective potential
in their scenario allows the chosen potential lobe to be maintained at a
given location and roughly the same size (but not the same depth) even
if $a$ varies while $\tilde{E}$ is kept constant. However increasing the
spin allows the authors to set gradually lower and lower energies, for
which more regular trajectories are found. This is not at all
surprising in perspective of our results, where the energy $\tilde{E}$
proved to play a key role in determining the stability of motion.
Although a direct comparison of presented surfaces of section differing
only in $a$ value may suggest the spin dependence, there is no
clear and unambiguous correlation. We thus suggest attributing the
observed dependence primarily to the level of energy, which acted as a
motion destabilizer also in our setup. In fact, even if a real trend
with the spin is present, it is hard to disentangle it from a
simultaneous change of $\tilde{E}$.

The question of the dependence of the dynamics upon the other parameters of the system ($\tilde{L}$, $\tilde{q}B_{\rm{0}}$, $\tilde{q}\tilde{Q}$) was also addressed during the analysis. It appeared that above mentioned difficulties accompanying the analysis of the spin dependence became even more serious in this case. Namely, neither it was possible to maintain the given potential lobe for a reasonable range of values of selected parameter nor we were able to fix this problem by binding this parameter in some simple manner to some other parameter (e.g. energy $\tilde{E}$). In other words, none of these parameters itself may be regarded as a trigger for chaos.

\subsection{\label{sec_valley}Motion in potential valleys}
Besides off-equatorial lobes, the potential may form another remarkable
structure -- an endless potential valley of almost constant depth which
runs parallely to the symmetry axis (\rff{valley}). The poloidal
orientation and asymptotical form of the valley are due to the fact that
the Wald test field does not vanish at spatial infinity and it
approaches the uniform magnetic field parallel to the symmetry axis.

The existence of such a potential corridor suggests that test
particles with a particular range of parameter values could escape from
the equatorial plane to large distance. We observe that a test particle
in the potential valley can keep oscillating around the equatorial
plane or it may escape from this plane completely, depending on its
initial position in the phase space (see the upper left panel of
\rff{osc_unik1}). We can examine the motion in the
asymptotic region by rescaling the radial coordinate (upper middle panel
of \rff{osc_unik1}). To achieve this, we use
$r^*\equiv\frac{r-r_{+}}{r}$, where $r_{+}=M+\sqrt{M^2-a^2}$ is the
position of the outer horizon of the black hole.

The normalization condition allows one to express $u^{\theta}$ as a
function of other phase space variables and parameters of the system.
Intuition suggests that, considering particles launched from the
equatorial plane ($\theta(0)=\frac{\pi}{2}$), the initial value
$u^{\theta}(0)$ is a governing parameter which decides whether the
particle remains oscillating around $\theta=\frac{\pi}{2}$ or leaves it
once forever. We can draw the isolines of selected $u^{\theta}$ values
in the ($r$, $u^r$)-plane which we use as a surface of section
($\theta_{\rm{section}}=\frac{\pi}{2}$) for the inspection of the test
particle dynamics. Comparing acquired isolines for various values of
$u^{\theta}$ with the empirically stated escape corridor of
\rff{corridor} leads to the conclusion that they never coincide
perfectly, although the correlation is quite high. In other words there
is no definite threshold value of $u^{\theta}(0)$ which would determine
whether a selected combination of $r(0)$, $u^r(0)$ (while other
parameters are fixed) lets the particle launched at equatorial
plane leave or oscillate around the plane.

In \rff{corridor} we observe several qualitatively different types of
possible particle dynamics. Next to the effective potential contour we
find closed and well-defined curves which represent the regular motion.
Going further inside we notice that fragmented curves are present. Below
them we find progressively more and more blurred patterns, which again
signifies the onset of chaos. The inner parts of the potential lobe are
occupied by the ``escape corridor'' where the particles can stream
freely from the equatorial plane, as seen in \rff{osc_unik1}.

The trajectories represented by the closed curves in the surface of
section differ profoundly from the disconnected curve orbits. The
difference can be seen also in the direct projection onto the poloidal
plane (upper right panel of \rff{osc_unik1}). While the trajectories of
the first type gradually fill each particular compact region of given
section, the latter forms bundles which curl through the projection plane
resembling ribbons that bound regions which are never reached by the
particle.

\begin{figure}[htb]
\centering
\includegraphics[scale=0.5, trim= 0mm 0mm 0mm 0mm, clip]{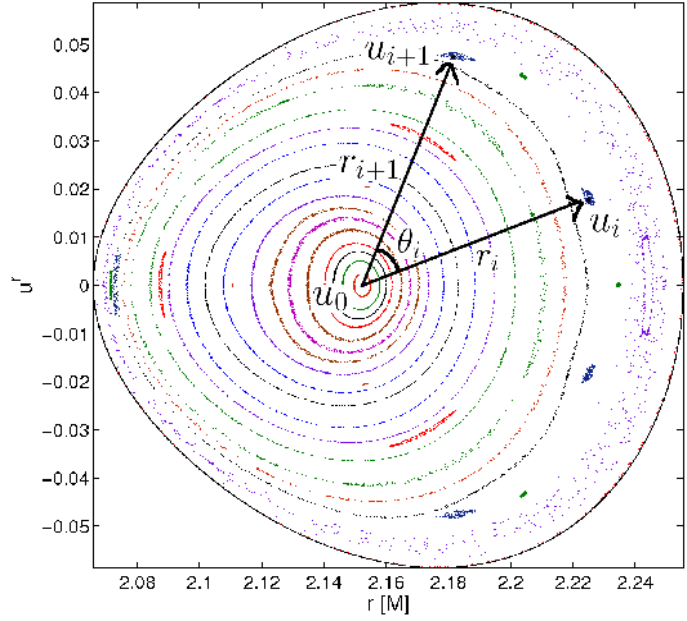}
\caption{Fixed point $u_0$ is localised. Two succeeding section points $u_i$ and $u_{i+1}$ of a given trajectory are marked. The angle $\theta_i$ between the radius vectors $r_i$ and $r_{i+1}$ is evaluated. Rotation number $\nu$ is defined as $\nu=\lim_{N\to\infty}\frac{1}{2\pi N}\sum_{i=1}^N \theta_i$. In practise we compute a mean value of the finite number of angle values $\theta_i$ instead of the limit.}
\label{rotac_intro}
\end{figure}

By employing the Recurrence Plots (bottom panels of \rff{osc_unik1}) we
confirm that the dynamics differs significantly in these distinct modes
of motion. Not surprisingly we obtain a typical regular pattern in the
case of trajectory which forms a closed sharp curve in the surface of
section (bottom left panel of \rff{osc_unik1}). A ``ribbon--like''
trajectory (fragmented curve in \rff{corridor}) results in an ordered
checkerboard pattern (bottom middle panel), which is known to be typical
for periodic and quasi--periodic systems \citep{marwan}. Finally, in the
bottom right panel of \rff{osc_unik1} we observe that a blurred curve
trajectory exhibits slight chaotic behavior in its RP. The diagonal
structures are partially disrupted and we notice that diagonal lines
become bent as they approach the line of identity.


The class of escaping trajectories in the Kerr background was recently
discussed by \citet{preti}. The author suggests that a Wald
electromagnetic field employed in this setup could serve as a charge
separation mechanism for astrophysical black holes since the sign of the
particle charge may determine whether a given particle escapes from the
equatorial plane or becomes trapped in cross-equatorial confinement (or
falls into the horizon).

\subsection{Rotation number and fragmented tori}
\label{rotacnic}
In the integrable system the trajectories in the phase space reside on the surface of the tori characterized by the values of integrals of motion which determine the characteristic frequencies of the orbit. Fundamental frequencies in our axisymmetric system are those of radial and latitudinal motion ($\varphi$ is cyclic coordinate). Once the system is slightly perturbed (by the magnetic field in our model) the tori characterized by the irrational ratio of frequencies $\omega_r/\omega_{\theta}$ survive which is assured by the Kolmogorov-Arnold-Moser (KAM) theorem. On the other hand the Poincar\'e--Birkhoff theorem tells us that the resonant tori with the rational frequency ratio will disintegrate into the chain of islands when we perturb the system \citep{lieberman}. Trajectories belonging to the given chain are characterized by a single frequency ratio. Since the chains have nonzero radial width (whose actual value depends on the 'degree of prominence' of the resonance as well as the values of other parameters of the system) we should be able to detect them in a plot of the ratio $\omega_r/\omega_{\theta}$ as a function of the initial radial coordinate of the orbit $r(0)$. They should appear as the periods of constancy in such a plot.

Rotation number is yet another index which can be used to characterize and detect resonances in this context \citep{contopoulos02,vlny10}. In order to compute it we first need to localise the central fixed point ${u_0}$ in the Poincar\'{e} surface of section, see \rff{rotac_intro}. Equipped with the set of section points of a given trajectory we calculate the angle $\theta_i\equiv {\rm angle} (r_i,\,r_{i+1})$ between the radius vectors of each pair of succeeding section points $u_i$ and $u_{i+1}$. Rotation number $\nu$ is defined as 
\begin{equation}
\label{rotation_no}
\nu=\lim_{N\to\infty}\frac{1}{2\pi N}\sum_{i=1}^N \theta_i.
\end{equation}
 We actually compute the mean of the finite number of values instead of the limit. Since the succeeding section points of a given resonant trajectory always skip between two islands of stability with a constant orientation, we can infer that the overall number of the islands in the chain will be given by the denominator of the rotation number when written in a simplest fractional form. Chaotic orbits are characterized by stochastic behaviour of both indices, $\nu$ as well as $\frac{\omega_r}{\omega_{\theta}}$.

\begin{figure}[hp]
\centering
\includegraphics[scale=0.63, trim= 0mm 0mm 0mm 0mm, clip]{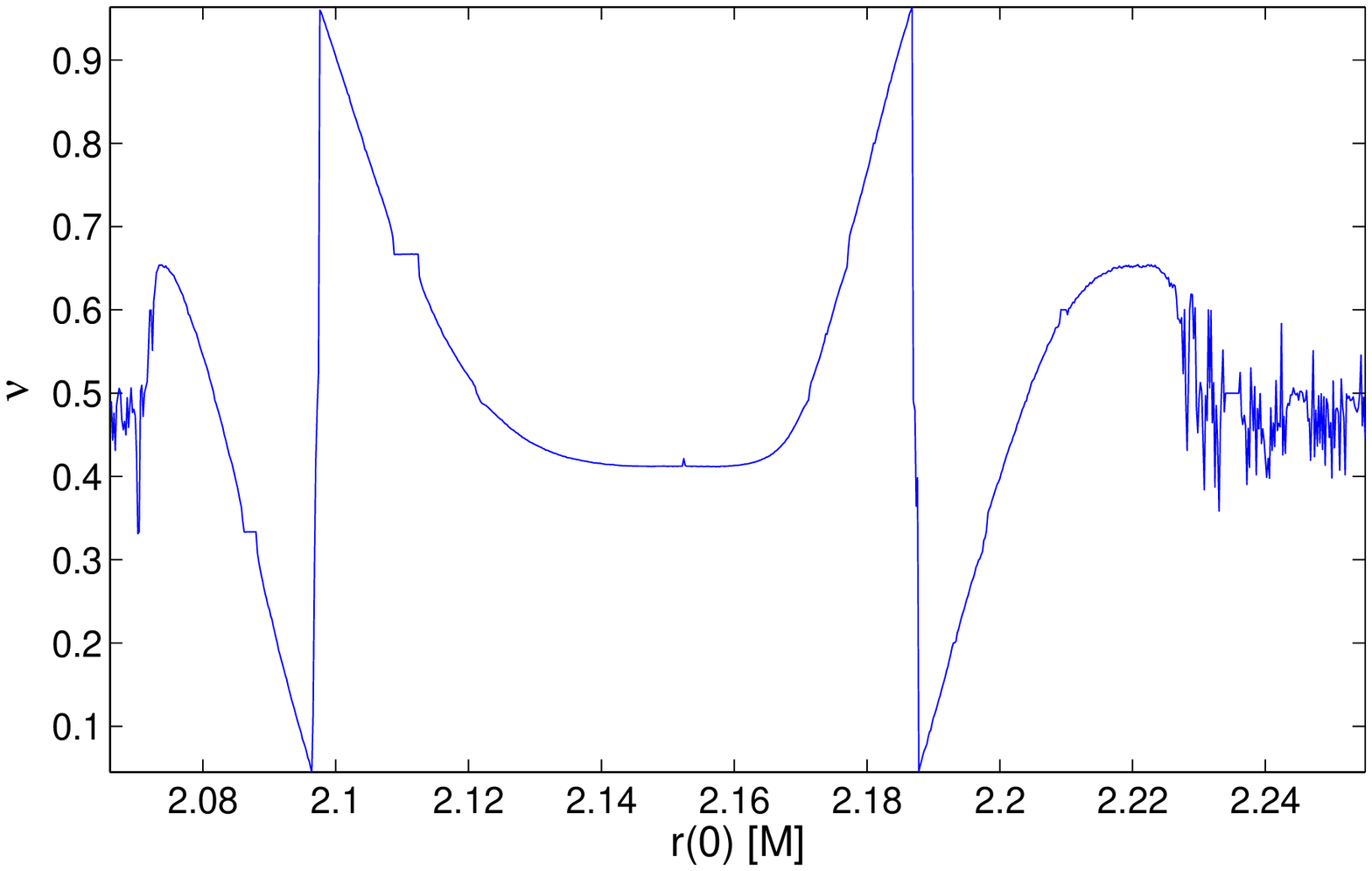}
\includegraphics[scale=0.67, bb=33 248 561 585, clip]{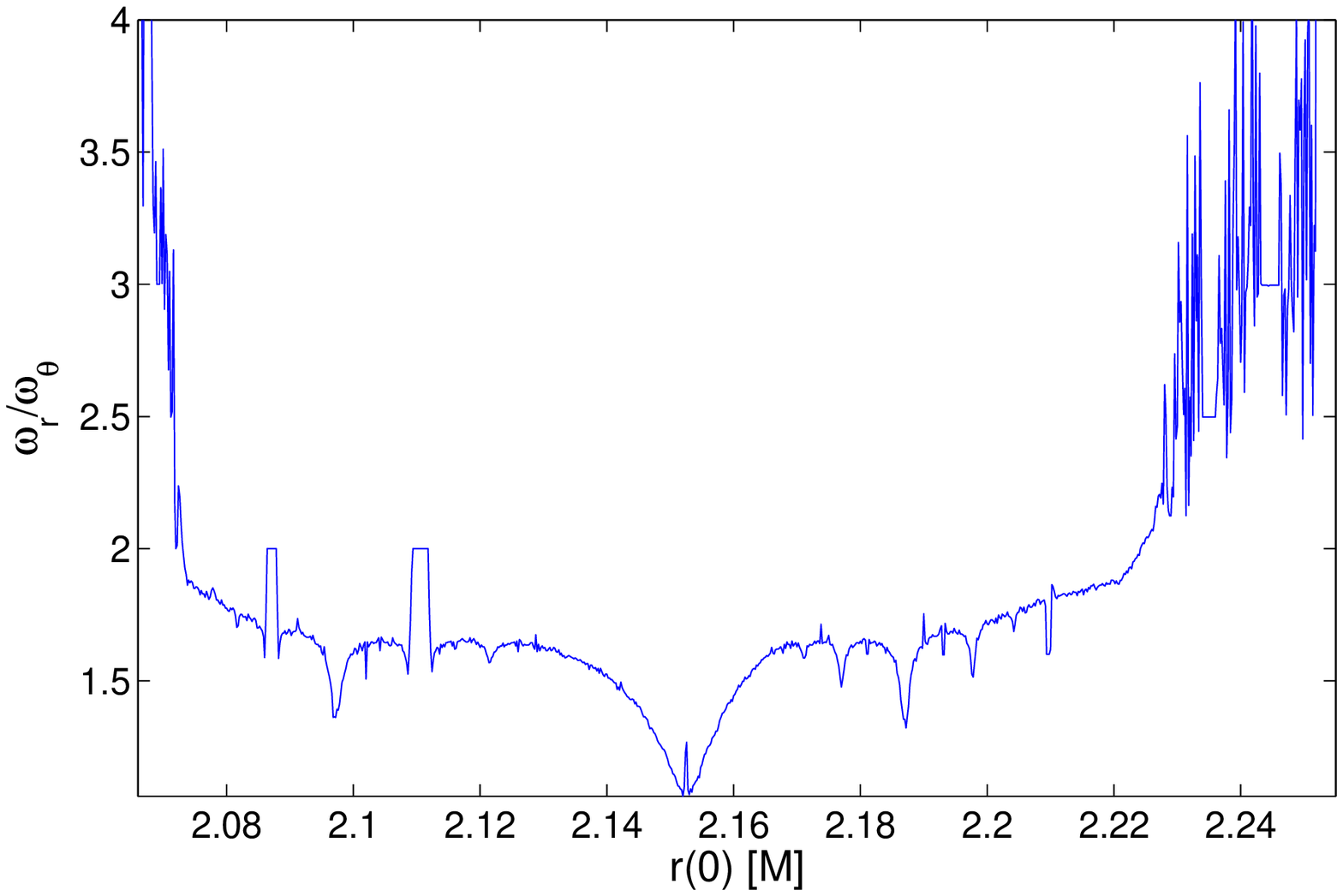}
\caption{Rotation number is compared to the ratio of fundamental frequencies $\frac{\omega_r}{\omega_{\theta}}$. Both are computed for the same set of trajectories originating in the equatorial plane at different $r(0)$ (equidistantly placed with $\Delta r=0.0002\,M$). Other parameters remain fixed: $a=0.5M$, $\tilde{E}=1.795$, $\tilde{L} =5M$,
$\tilde{q}B_{0}=2M^{-1}$, $\theta(0)=\theta_{\rm{section}}=\frac{\pi}{2}$, $\tilde{q}\tilde{Q}=2$. Most prominent resonances correspond with $\nu=1/3\;\left({\omega_r}/{\omega_{\theta}}=2\right)$, $\nu=2/3\;\left({\omega_r}/{\omega_{\theta}}=2\right)$ and $\nu=1/2$ $\left({\omega_r}/{\omega_{\theta}}=5/2\right)$. We note that the resonance at $r(0)\approx 2.245\,M$ having frequency ratio $\frac{\omega_r}{\omega_{\theta}}=3$ is not detected clearly by the rotation number $\nu$ which actually oscillates around $\nu=1/2$ instead of being strictly constant. Number of the islands comprising the resonant chain is given by the denominator of the rotation number. Chaotic regions are characterized by a stochastic behaviour of the both indicators, $\nu$ as well as $\frac{\omega_r}{\omega_{\theta}}$.}
\label{rotac_full}
\end{figure}

In \rff{rotac_full} we compute $\omega_r/\omega_{\theta}$ and $\nu$ for a single set of trajectories bound in the equatorial potential lobe presented in \rff{rotac_intro}. Particles are being launched from the equatorial plane differing only in the initial value of radial coordinate $r(0)$ separated equidistantly by $\Delta r=0.0002\,M$, yielding $945$ trajectories in total. Prominent resonances at $\nu=1/3\;\left({\omega_r}/{\omega_{\theta}}=2\right)$, $\nu=2/3\;\left({\omega_r}/{\omega_{\theta}}=2\right)$ and $\nu=1/2\;\left({\omega_r}/{\omega_{\theta}}=5/2\right)$ are apparent and easily identified with the chains of islands in \rff{rotac_intro}. In the regular regions inhabited by KAM tori the rotation number behaves as non-constant continuous function of $r(0)$ except the intervals of constancy corresponding with the resonant chains and two major discontinuities arising from our convention of taking $\theta_i$ as a positive oriented angle between the radius vectors. In the chaotic border regions both indices fluctuate but resonant chains represented by constant intervals also appear in this zone.

Detailed view in \rff{rotac_det} which zooms the portion of \rff{rotac_full} reveals that a number of faint resonances producing thin Birkhoff chains is present in the chaotic zone. Direct detection of these in the surface of section would be rather difficult since the readability of the surfaces of section decreases rapidly with increasing density of depicted trajectories. In this sense the rotation number appears suitable indicator of resonance.

Resonant chains are in principle detectable in terms of spectral analysis of the observed electromagnetic signal. Presence of the Birkhoff chains allows us to discriminate between perturbed and regular system. Moreover the position and the width of the chains reflects other properties of the system, strength of the perturbing magnetic field for instance. See \citet{vlny10} for a detailed discussion of this approach applied to the different type of system of extreme mass ratio inspiraling sources emitting gravitational radiation.

\begin{figure}[htb]
\centering
\includegraphics[scale=.7, clip]{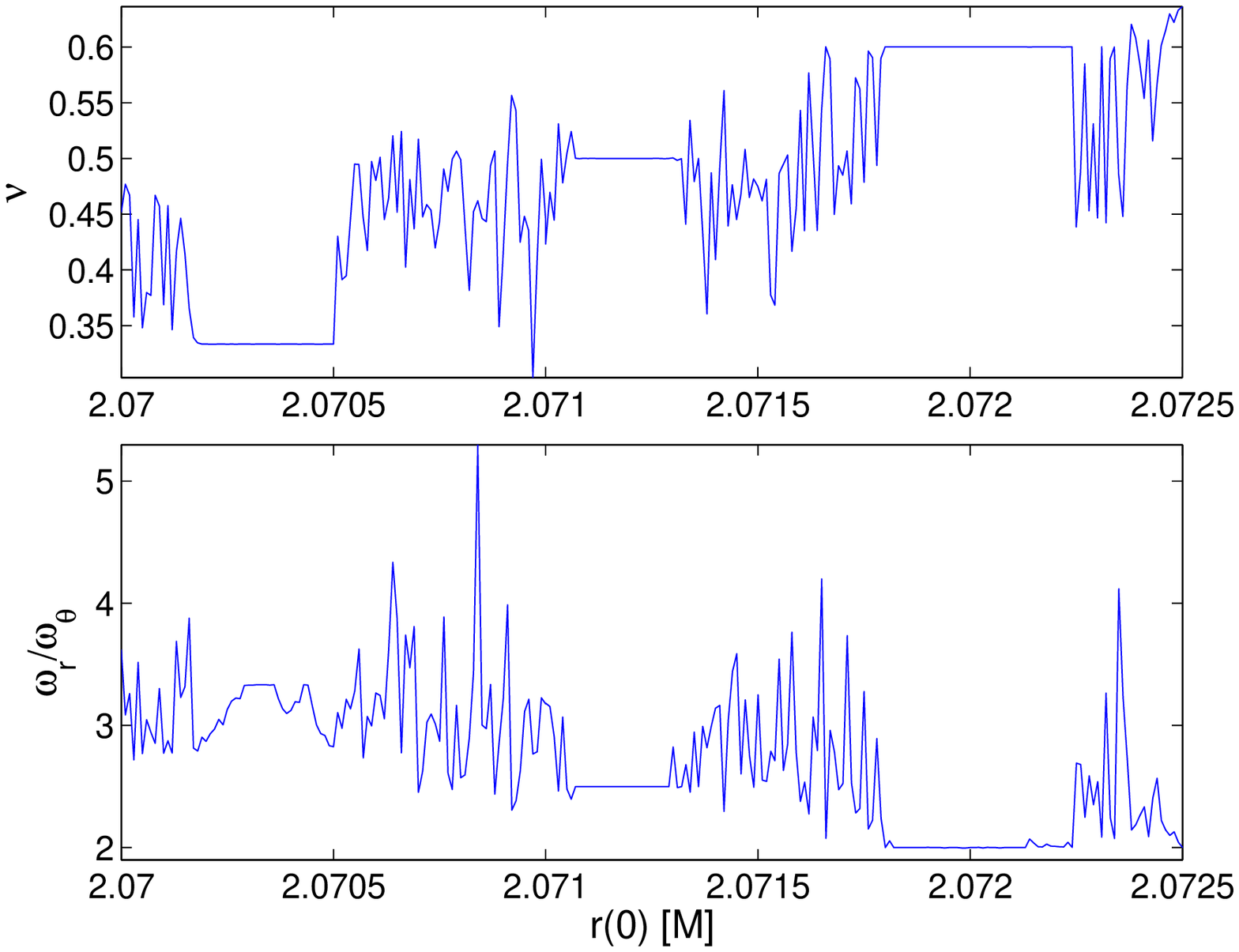}
\caption{Portion of the \rff{rotac_full} is provided in the high resolution revealing the faint resonances producing thin chains of islands whose direct detection in the surface of section is almost impossible (see \rff{rotac_intro}).}
\label{rotac_det}
\end{figure}

However, the critical assumption of the Poincar\'{e}--Birkhoff theorem that the perturbation of the original integrable system is weak is generally not fulfilled in our case. Nevertheless for the trajectories analyzed in figs. \ref{rotac_intro}--\ref{rotac_det} the theorem proved applicable as we managed to identify fragmented tori observed in the surface of section with the intervals of constancy of the rotation number at values given by simple integer ratios $\frac{1}{2}$, $\frac{2}{3}$ etc. Therefore we can identify fragmented tori in the surfaces of section in \rff{rotac_intro} and in analogous closed equatorial lobes with the Birkhoff chains of stability islands anticipated by the theorem. However, we did not manage to do so in the case of potential valley in \rff{corridor} neither for the off-equatorial lobes, e.g. \rff{spin_diskuze}. In both cases we observed fragmented tori in the section but we could not localise the central fixed point $u_0$ which is essential for the evaluation of the rotation number. In the potential valleys the central part of the section is occupied by the escape corridor and in the off-equatorial lobes the structure of the curves in the section is quite different compared to the equatorial lobes. Frequency ratios ${\omega_r}/{\omega_{\theta}}$ corresponding with fragmented tori did not attain any simple integer ratio value here. 

We conclude that the Poincar\'{e}--Birkhoff theorem is only partially applicable in our study. Especially the off-equatorial lobes which are actually supported by the perturbation (there are no such lobes in the pure Kerr without EM field) fail to fulfill the assumption of the perturbation being weak. In the surfaces of section we observe the fragmented curves of a different nature than those of resonant chains detected in the equatorial lobes. We stress that the issue definitely deserves further attention since the analysis of the fundamental frequencies has observational consequences.

\begin{figure}[!ht]
\centering
\includegraphics[scale=0.445,clip=true]{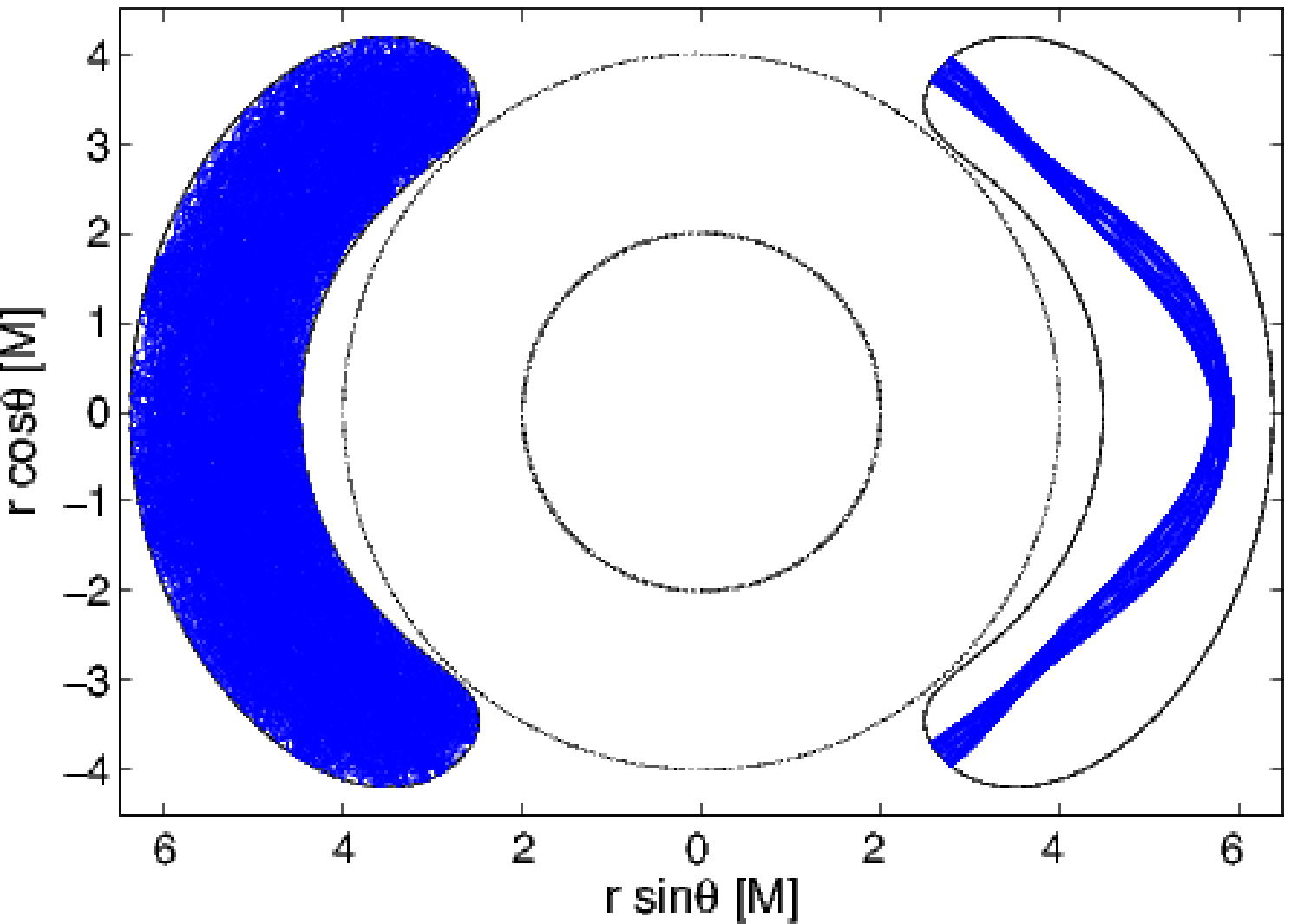}~~\includegraphics[scale=0.3, clip=true]{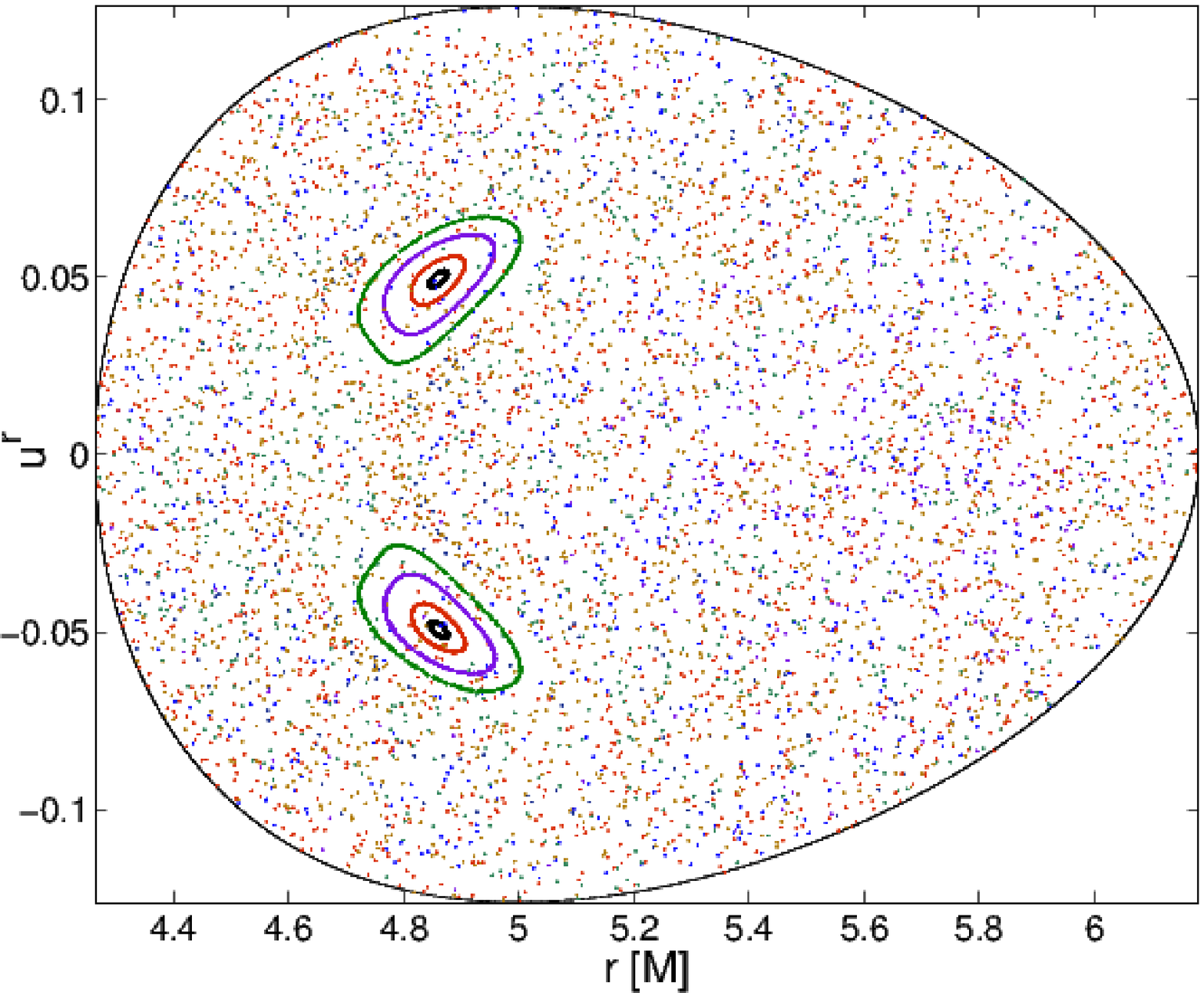}\\[10pt]
\includegraphics[scale=0.36, clip=true]{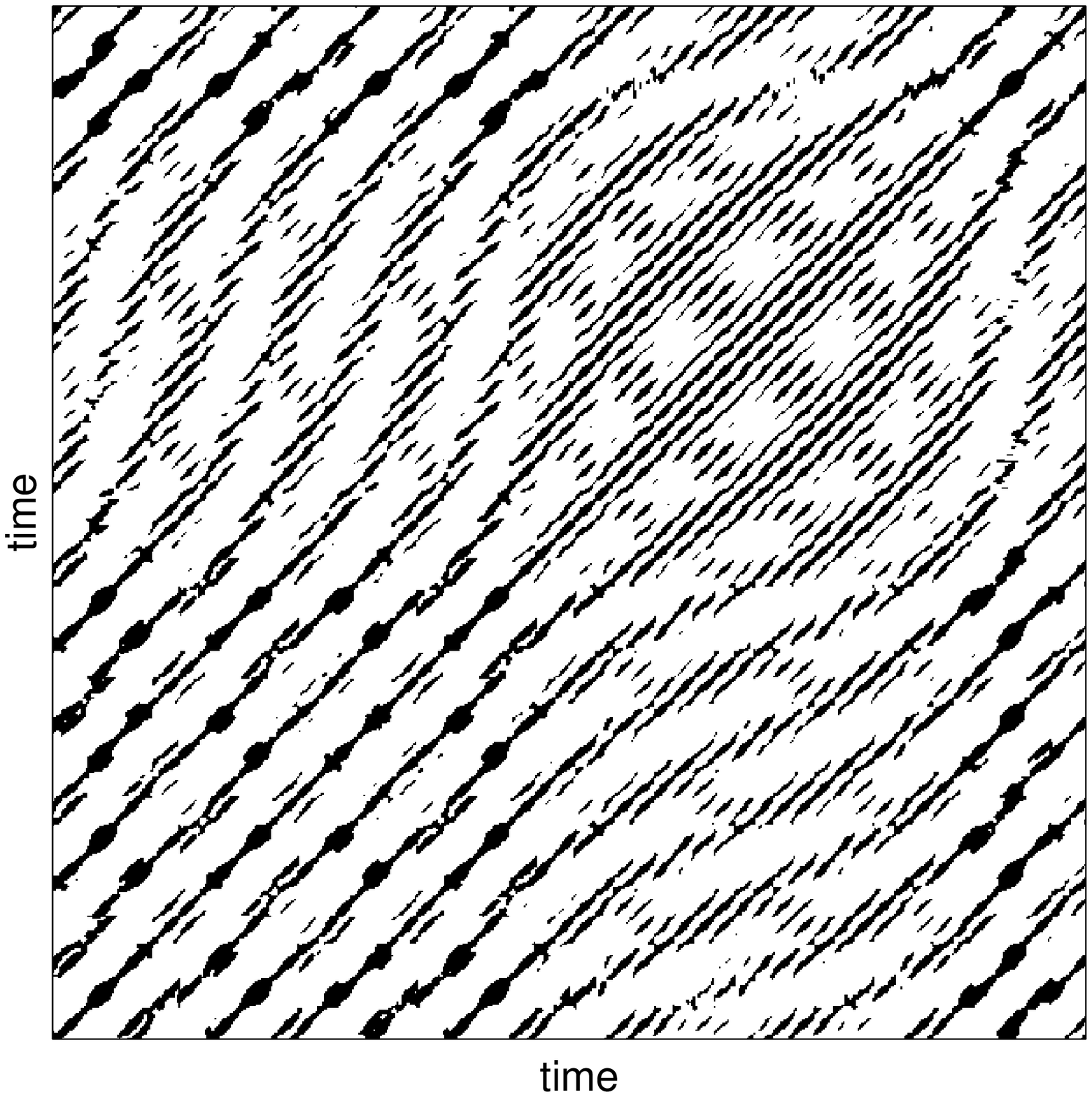}~~\includegraphics[scale=0.34, clip=true]{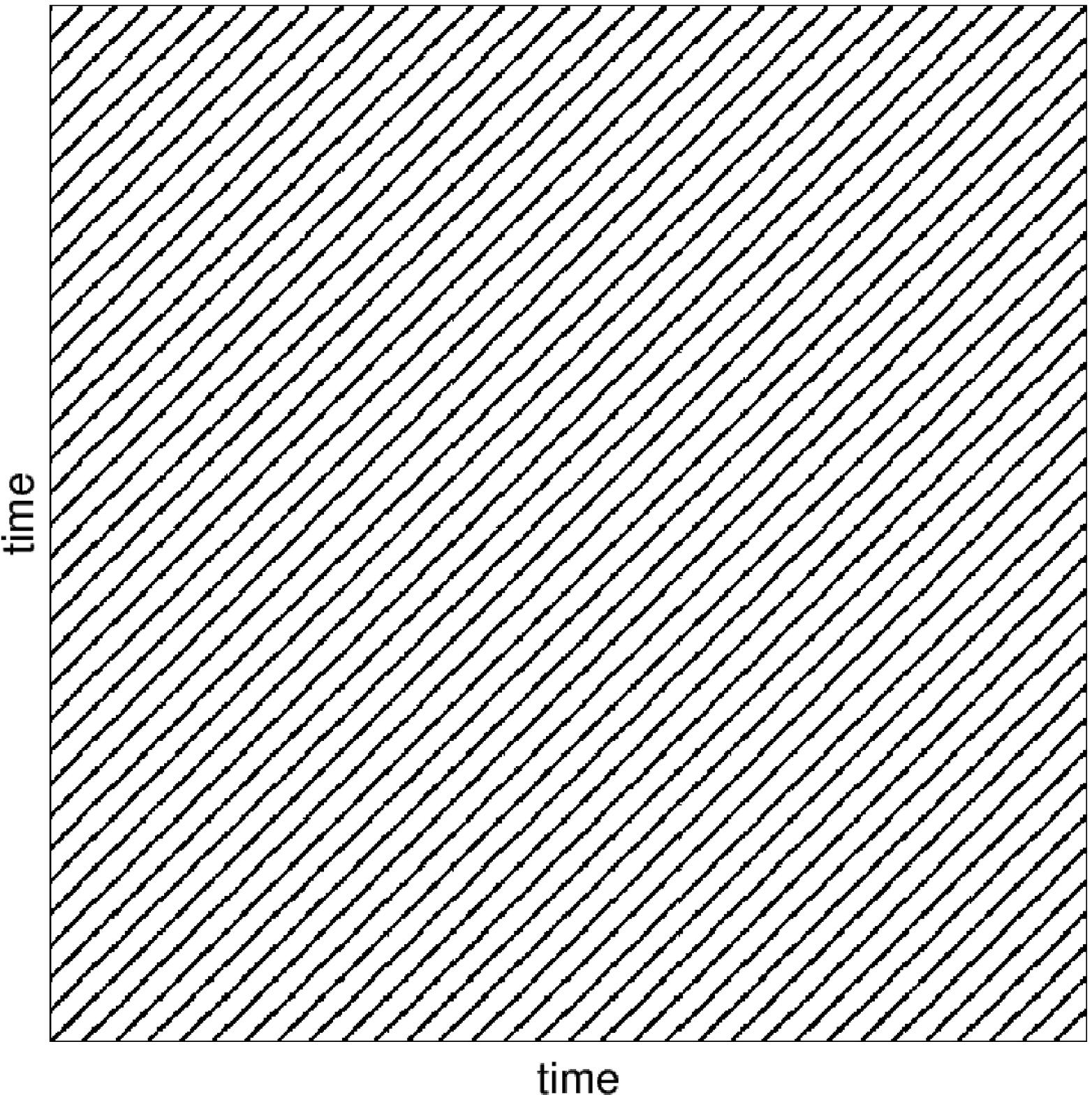}
\caption{For 
the energy level $\tilde{E}=0.857$ we obtain a broad lobe which almost
touches the surface of the star at $r=4\;M$. In the upper-left panel, 
we launch two particles with $\theta(0)=1.0492$, $r(0)=4.75\;M$. The
particle to the left of the star starts with $u^r(0)=0$ and moves
chaotically, while the other one with $u^r(0)=0.03$ follows a perfectly 
regular trajectory. The upper right-panel shows these two types of 
trajectory appear in the surface of section plot. The bottom panels 
demonstrate the difference between the two types in terms of 
Recurrence Plots.}
\label{rotdip1_3}
\end{figure}

\section{A magnetic star}
\label{sectionmagnetized}
Various types of stars exhibit very strong magnetism. Peculiar main sequence stars of Ap and Bp classes may bear large-scale magnetic field of strength $\approx 10^4\:\rm{G}$ \citep{bagnulo06,borra82}. Considerably stronger fields are found in the case of degenerate stars. White dwarfs may reach $\approx10^8\:\rm{G}$ \citep{valyavin03} and magnetars as a magnetically extremal subclass of neutron stars even $\approx10^{15}\:\rm{G}$ \citep{duncan92}. In our context a broad term {\it magnetic star} encompasses all the above objects. In the following, however, we concentrate ourselves mainly
on the case of compact stars.

We describe the gravitational field outside a magnetic star
by the Schwarzschild metric,
\begin{equation}
\label{MetricSchw}
{\rm d}s^2=-\left(1-\frac{2M}{r}\right){\rm d}t^2
+\left(1-\frac{2M}{r}\right)^{-1}{\rm d}r^2+r^2({\rm d}\theta^2+\sin^2{\theta}{\rm d}\varphi^2).
\end{equation}
The associated magnetic field is modeled as a dipole rotating
at angular velocity $\Omega$ \citep{sengupta}:
\begin{eqnarray}
\label{rotdippot}
A_{t}&=&-\Omega{}A_{\varphi}=\frac{3\mathcal{M}\Omega \mathcal{R}\sin^2{\theta}}{8M^3},\\ A_{\varphi}&=&-\frac{3\mathcal{M}\mathcal{R}\sin^2{\theta}}{8M^3},
\end{eqnarray}
where
\begin{eqnarray}
\mathcal{R}=2M^2+2Mr+r^2\log{\left(1-\frac{2M}{r}\right)}.
\end{eqnarray}
The related dipole moment $\mathcal{M}$ is given by \citep{halo2_31}
\begin{equation}
\label{Magnetic}
\mathcal{M}=\frac{4M^3r_{\star}^{3/2}\left(r_{\star}-2M\right)^{1/2}\;B_{0}}{6M(r_{\star}-M)+3r_{\star}\,(r_{\star}-2M)\,\ln{\left(1-2Mr_{\star}^{-1}\right)}},
\end{equation}
where $B_{0}$ is the magnetic field at the neutron star equator,
$r_{\star}$ is the radius of the star surface.\footnote{The existence of
extremely compact stars with $r_{\star}\approx{}3M$ is unlikely, but not
excluded \citep{bah89,stu09}. Most of the realistic equations of state
imply a lower limit $r_{\star}\approx{}3.5M$ \citep{glend97}. On the
other hand, the models of Q-stars do allow a lower limit of
$r_{\star}\approx2.8M$ \citep{mil98}.} 

We assume eqs.\ (\ref{MetricSchw})--(\ref{rotdippot}) to hold
outside the star surface ($r>r_*$) and inside the light cylinder
($u^{\mu}u_{\mu}<0$). We
set $r_*=4M$ as the inner radial boundary of the particle motion.
As for the light cylinder, the mentioned condition results
in a relation $r^2\sin{\theta}^2\Omega^2=1-\frac{2M}{r}$, which
implicitly specifies the outer boundary.
The vector potential (\ref{rotdippot}) is valid inside the rigidly
corotating magnetospheric plasma, which we consider to be an excellent
conductor, so that the force-free condition $F^{\mu}_{\nu}u^{\nu}=0$
holds for the plasma for which $u^{\mu}=(u^t,0,0,u^{\varphi})$ and
$\frac{u^{\varphi}}{u^{t}}=\Omega$.

A general formula for the effective potential \rf{effpot}
simplifies to the form
\begin{eqnarray}
\label{EffectiveNS}
V_{\rm eff}&=&
-\frac{3\tilde{q}\mathcal{M}\mathcal{R}\Omega\sin^2{\theta}}{8M^3}\\
& &+\left(1-\frac{2M}{r}\right)^{\frac{1}{2}}\!\left[1+\left(\frac{\tilde{L}}{r\sin{\theta}}+\frac{3\tilde{q}\mathcal{M}\mathcal{R}\sin{\theta}}{8M^3r}\right)^2\right]^{\frac{1}{2}}.\nonumber
\end{eqnarray}

\subsection{Motion inside the potential lobes}
Our previous analysis \citep{halo2} revealed a number of distinct types
of possible topological structures of the effective potential. However
it appears that the system is not as rich in its dynamical properties.
The test particle trajectories share some similar features across
different classes of the effective potential. Therefore, we only present
surveys of particle dynamics in three exemplary types: Ia, IIa and IIIc 
(\rff{rotdip_abc}; see \citet{halo2} for the complete review).

Class Ia lobes grow with energy increasing. Once the level of the
equatorial saddle point is reached, the lobes merge with each other
across the equatorial plane. The single merged lobe eventually
intersects the surface of the star if the energy level is increased
sufficiently, letting the particle fall onto the surface. Lobes of IIa
type also merge via the equatorial plane but in contrast to the first
type the merged lobe opens toward the light cylinder (beyond which the
model becomes invalid). Lobes of the class IIIc first open via the
off-equatorial saddle points, allowing the particles fall onto the
star, before the lobes merge through the equatorial plane.

Now we study the three selected types of the effective potential
topology in more detail. We are primarily interested whether and how the
dynamic regime changes across the given range of specific energy
$\tilde{E}$. Especially, we shall address what happens with the dynamics
when the particle acquires enough energy to cross the saddle point.

The first survey (type Ia) begins at energy level $\tilde{E}=0.8482$,
corresponding to the closed lobe. In \rff{rotdip1_1p} we observe that
the motion inside the lobe is stable. No chaotic properties are detected
-- neither in Poincar\'e surfaces of section nor in the Recurrence
Plots.

\begin{figure}[htb]
\centering
\includegraphics[scale=0.34, clip=true]{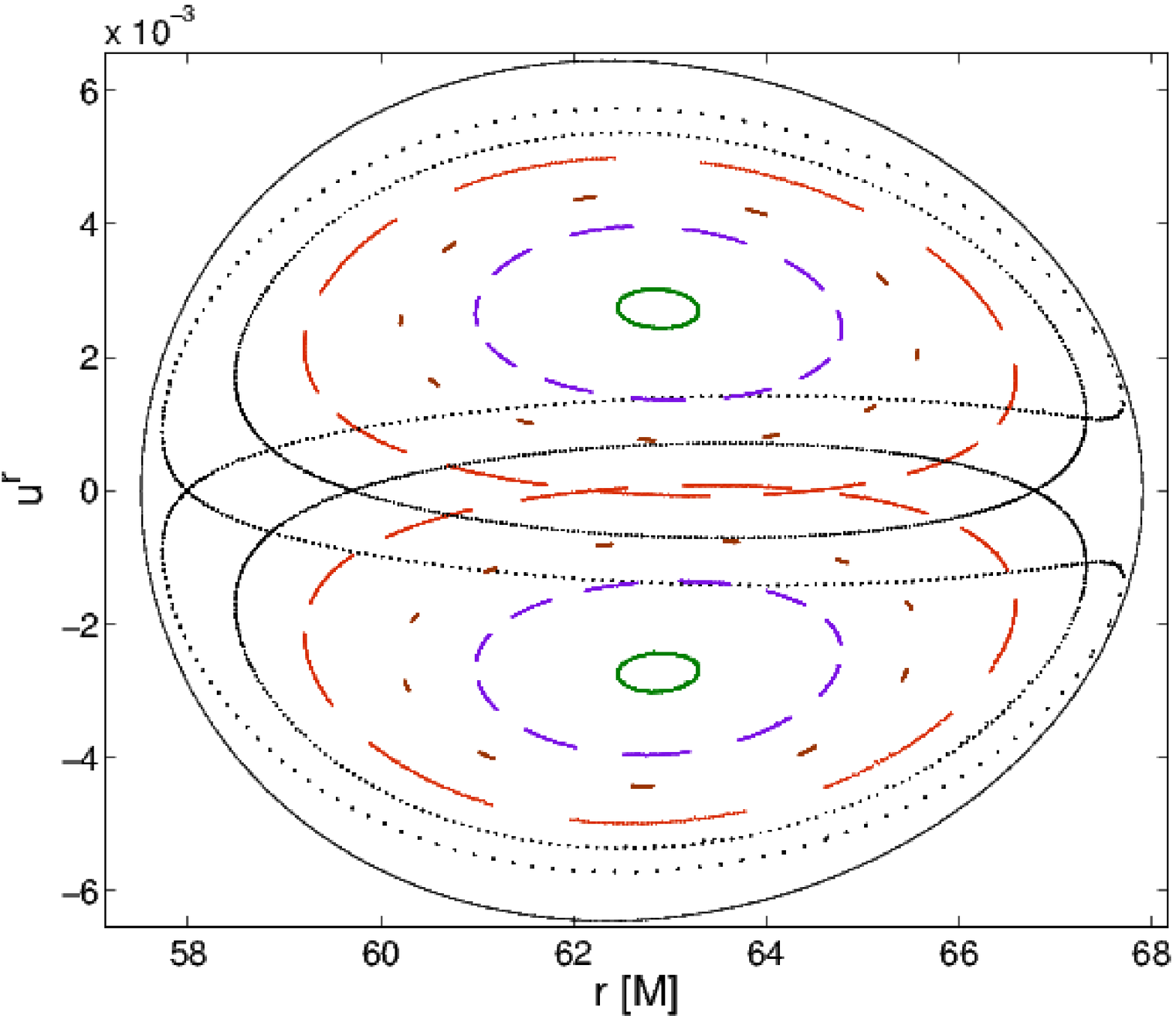}~~ 
\includegraphics[scale=0.51, clip=true]{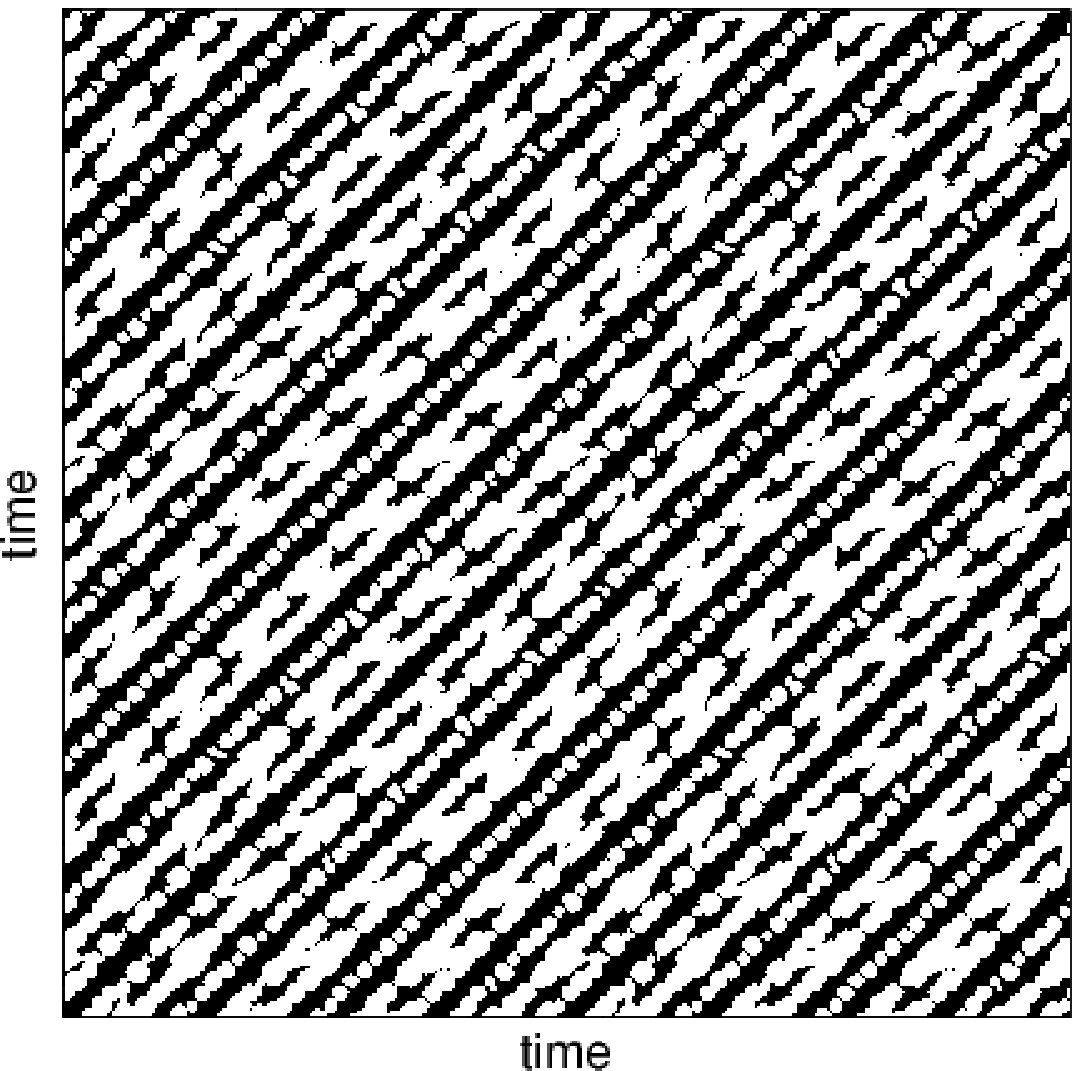} 
\caption{Regular motion in an
off-equatorial lobe of the third type. Parameters used:
$\tilde{E}=0.99579$, $\tilde{L}=6.25382\;M$,
$\tilde{q}\mathcal{M}=45.87368\; M^2$, $\tilde{L}=6.25382\; M$,
$\Omega=0.011485\; M^{-1}$. Particles are launched from latitude
$\theta(0)=\theta_{\rm{section}}=1.0492$.}
\label{rotdip3_1}
\end{figure}

\begin{figure}[hp]
\centering
\includegraphics[scale=0.45, clip=true]{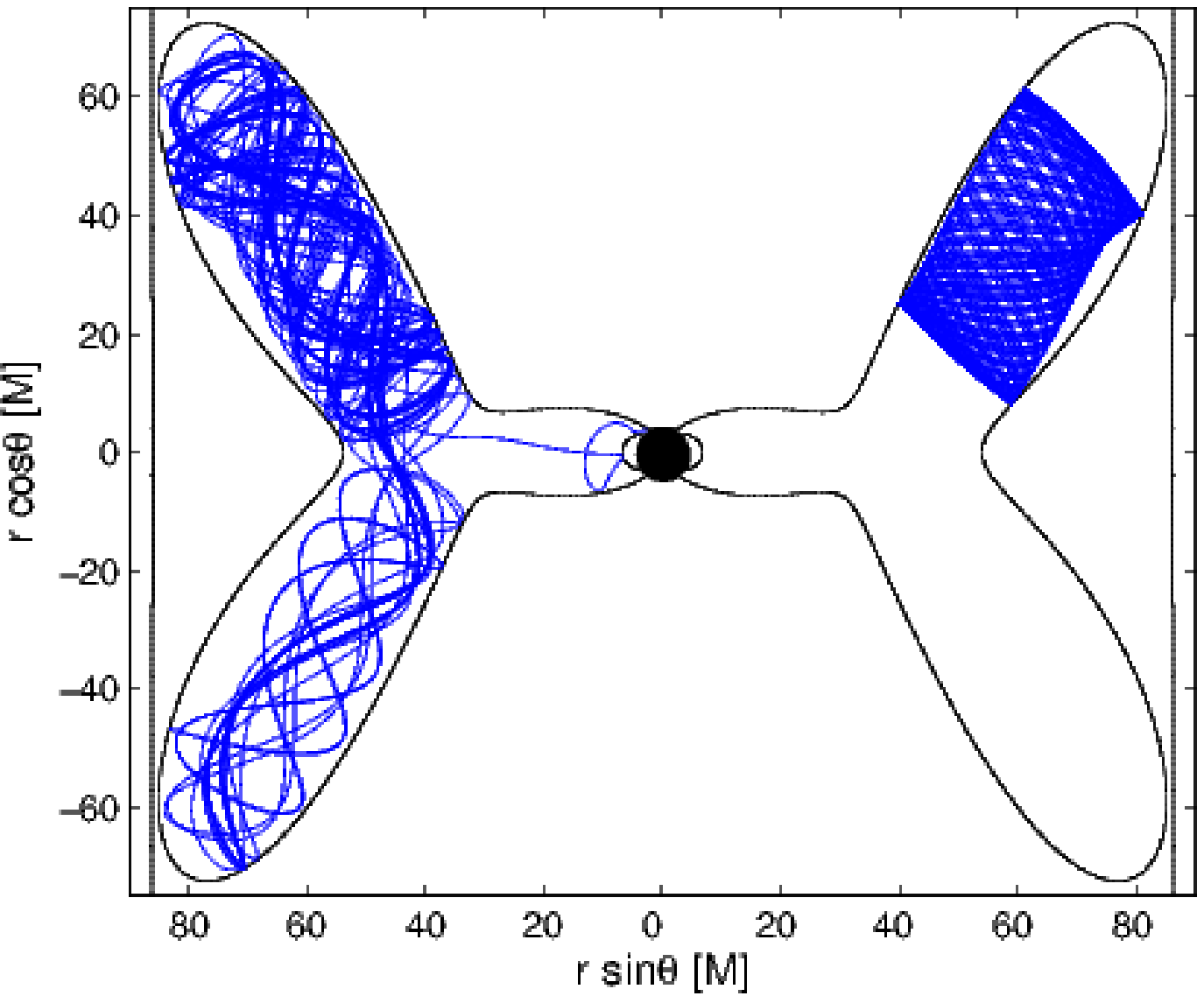}~
\includegraphics[scale=0.295, clip=true]{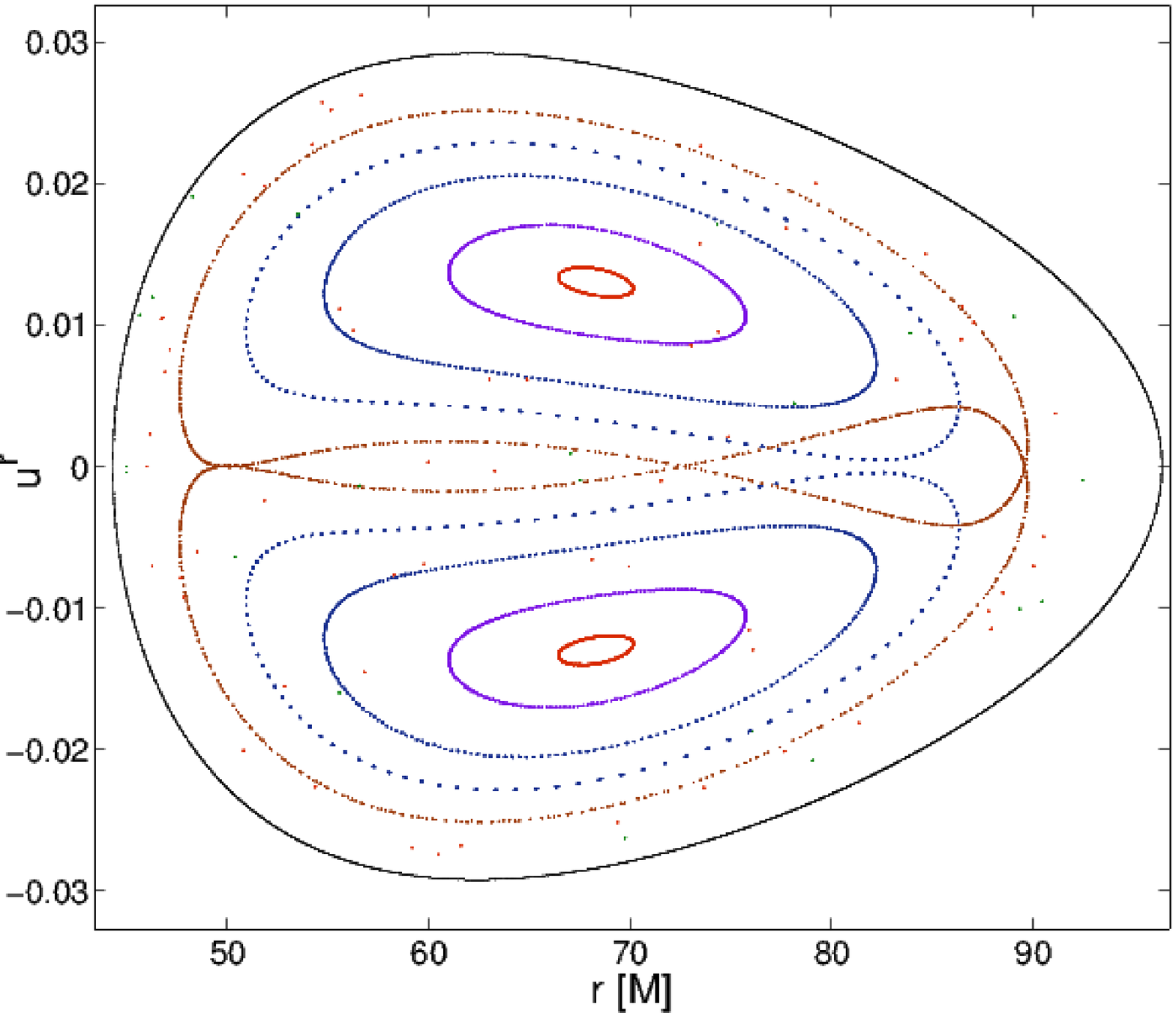}\\[10pt]
\includegraphics[scale=0.51, clip=true]{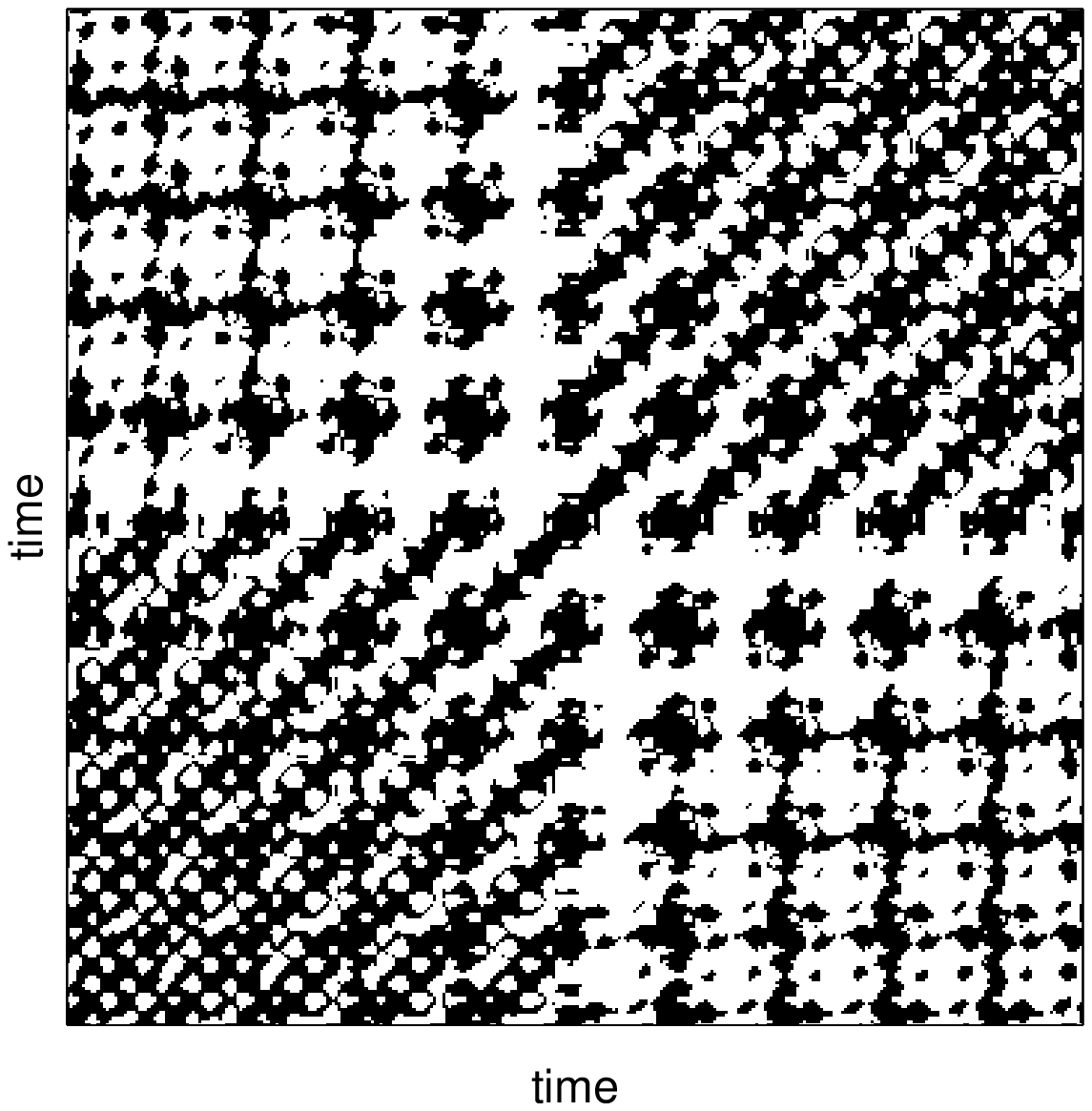}~
\includegraphics[scale=.545, clip=true]{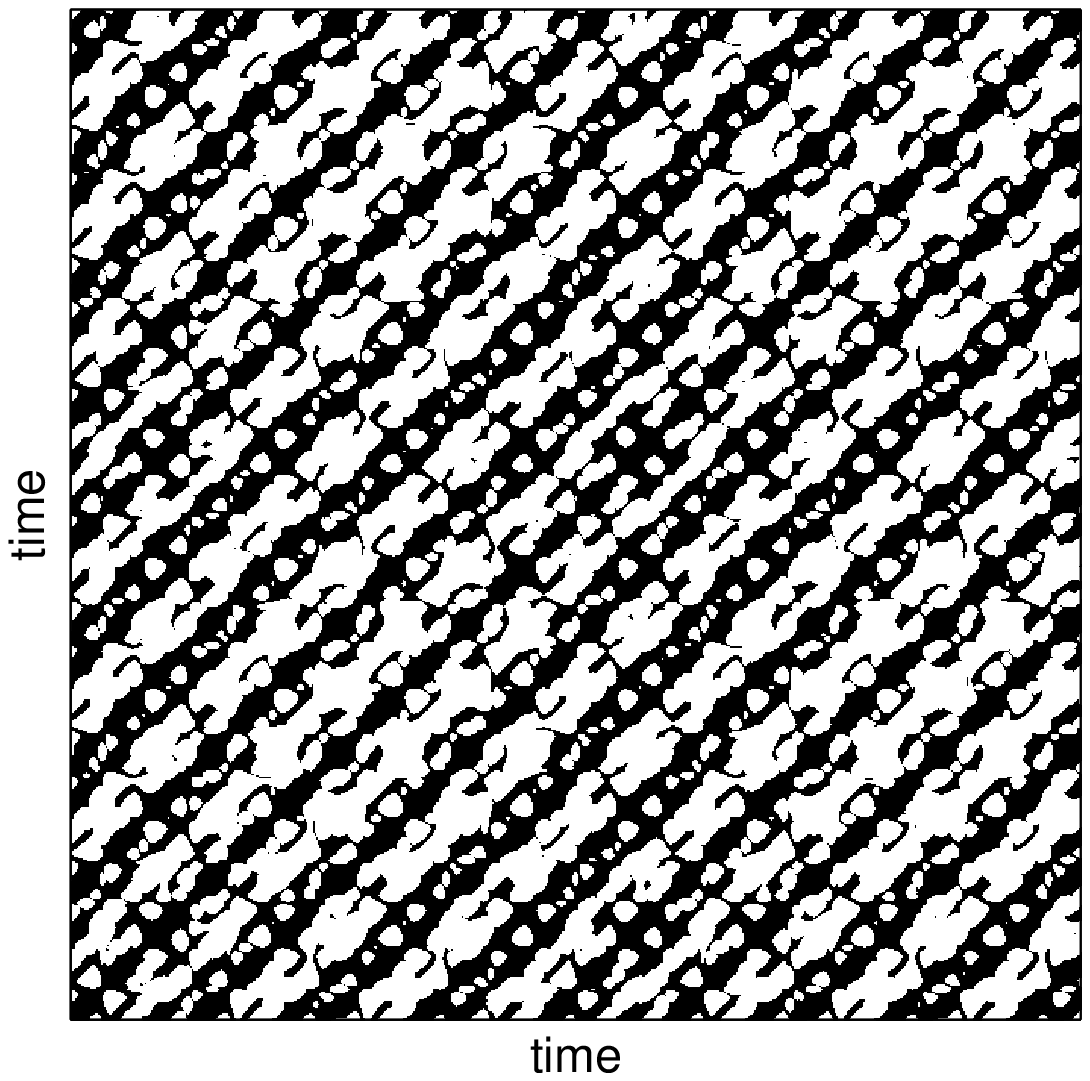} 
\caption{For the energy level $\tilde{E}=0.9962$ (other parameters as in 
\rff{rotdip3_1}) we obtain a large lobe that almost touches the
light cylinder and allows the particle to fall onto the surface of the
star via a narrow passage above and below the equatorial plane. The
upper left panel shows two trajectories launched from
$\theta(0)=1.0497$, $u^r(0)=0$. One of the particles (starting from
$r(0)=61.5\;M$) follows an unstable path and eventually falls on the
star surface. On the contrary, the other particle (starting from
$r(0)=72.5\;M$) moves regularly and never escapes any given part of the
lobe. The upper-right panel shows these two kinds of trajectory depicted
in the Poincar\'e surface of section. Chaotically dispersed points
belong to the escaping trajectory. The bottom-left panel shows the
Recurrence Plot of the escaping particle; this plot does not exhibit
typical chaotic behavior, although the large-scale structures are
present. The bottom-right panel presents the Recurrence Plot of stable
motion.}
\label{rotdip3_3}
\end{figure}

As the energy increases to $\tilde{E}=0.8485$, the symmetrical lobes
merge via the equatorial plane. By inspecting a number of trajectories
in this case we find that chaos starts appearing at this point -- those
particles which notice the gate through the equatorial plane always fill
the entire allowed region and they move chaotically. Nevertheless, there
are still such particles which move regularly in one of the two parts of
merged lobe and never cross the equatorial plane. We can also find
transient trajectories corresponding to regular motion lasting for some
period of time in one part of the lobe, followed by chaotic motion over
the entire lobe once the particle finds and encounters the passage
across the equatorial plane. All of the mentioned cases are illustrated
in \rff{rotdip1_2}.

Increasing the energy further to $\tilde{E}=0.857$, we obtain a broad
potential lobe which almost touches the star surface. The situation
changes from the previous case where the gate connecting the
off-equatorial lobes was narrow. Now we do not find trajectories which
occupy only one part of the lobe, above (or below) the equatorial plane.
Chaotic trajectories densely filling the entire lobe are typical for
this setup. We also encounter perfectly regular trajectories forming
ribbon--like structures spanned between northern and southern borders of
the lobe. In Poincar\'e sections these appear as regular islands
surrounded by a chaotic ocean (upper panels of \rff{rotdip1_3}). We
notice that the RP of this regular trajectory is extraordinarily simple
and consists of almost perfect diagonal lines (bottom panels of
\rff{rotdip1_3}). Thus its dynamic properties are close to those of a
periodic system, which is in contrast with the neighboring fully chaotic
orbits.

The second type (class IIa) of the effective potential topology of
off-equatorial lobes differs from the first one significantly as the
lobes do not open towards the star when the energy is raised
sufficiently. On the contrary, in this case we observe that the lobe's
boundary touches the light cylinder first.

The motion in the off-equatorial lobes proves to be regular while the
merging lobes bring chaos into play. Chaos becomes dominant for broader
lobes, however, stable regular orbits also persist. The results are
similar to those of the first type of potential topology discussed
above.

The last analyzed topology of the lobes (class IIIc) differs profoundly
from the preceding two cases, as can be seen in \rff{rotdip_abc}. We
find that stable motion dominates in this setup. This can be verified by
comparison with \rff{rotdip3_1}.

As we further increase the energy level, we obtain more complicated
shapes of the equipotentials that allow the particle to fall on the star
surface. On the other hand, opening the outflow gate energetically
precedes the merging point of both off-equatorial lobes. In \rff{rotdip3_3}, we discuss the motion governed by the largest possible
lobe which almost touches the light cylinder. We observe that stable
regular orbits are still possible for those particles that do not hit
the passage.

From the above-given discussion we conclude that the motion of charged
test particles in the off-equatorial lobes allowed by the test field of
the rotating magnetic dipole on the Schwarzschild background is largely
regular. Once the off-equatorial lobes merge with each other, chaos
may appear. Increasing the energy, the chaotic motion becomes typical
but, quite surprisingly, very stable orbits also exist under these
circumstances.

%% file: chap4n.tex
\chapter{\sffamily{Conclusions}}
\label{conclusion}
\section{Structure of the electro-magnetic field}
\pagestyle{headings}
In this work we went through various issues concerning the structure of the electromagnetic field which arises from the interplay between the frame-dragging effect and the uniform magnetic field with general orientation with respect to the rotation axis of the Kerr source. We further generalized the model by allowing the black hole to move translationally in a general direction with respect to the magnetic background. Components of the electromagnetic tensor $F_{\mu\nu}$ describing resulting field were given explicitly (in a symbolic way regarding their length) in the terms of the former nondrifting solution. Special attention was paid to the comparison of various definitions of the electric/magnetic field. We also reviewed the construction of three distinct frames attached to the physical observers. 

Before exploring the rich structures arising from the drift and oblique background field we first revisited the issue of the expulsion of the aligned magnetic field out of the horizon of the extremal Kerr black hole (Meissner effect). Since the effect itself has already been discussed thoroughly in the literature we did concentrate on the observer aspect of the problem instead. By comparing alternative definitions of the magnetic vector field given in \rs{lines} combined with the choice of the four-velocity profiles presented in \rs{tetrads} we came to the conclusion that (i) the Meissner effect is observer dependent and (ii) some definitions of the field lines do not fit well into the Boyer-Lindquist coordinate system since they artificially amplify the effect of the coordinate singularity at the horizon. 

Namely we observed that in the ZAMO tetrad the field does not exhibit the Meissner effect while in the frame of freely falling observer (FFOFI) the field is expelled. On the other hand in the renormalized field components the expulsion is observed for both ZAMO as well as for FFOFI test charges. In coordinate components the Meissner effect also appears but we decide not to use them because the coordinate basis is not normalized which causes artificial deformation of the field lines. On the other hand the properly normalized physical components appear problematic since they amplify the effect of the coordinate singularity at the horizon as mentioned above. We found them not convenient for the use in Boyer-Lindquist coordinate system (at least in the region close to the horizon). Asymptotically-motivated (AMO) components which are {\it observer independent} as they reflect the $F_{\mu\nu}$ components directly were also employed and the resulting magnetic lines of force were identified with the section of the surfaces of the constant magnetic flux which represent yet another way to display the field. We note that in the FFOFI frame both magnetic and electric fields are expelled out of the horizon in the case of the extremal spin.

Upon introducing the perpendicular component we observe that generally (i) the magnetic field is not expelled anymore and (ii) both the electric and magnetic fields acquire a tightly layered structure in the narrow zone just above to the horizon. Structure of the field is surprisingly complex in this region, self-similar patterns are observed regardless the choice of the observer proving that the layering is an intrinsic feature of the field rather than a mere observer effect. 

In the case of BH's translational motion through the aligned field we also observe the complex layering of the field which we attribute to the transversal component arising from the Lorentz boost. However, for a sufficiently rapid drift we observe a new effect emerging: as the layers transform they give rise to the formation of the neutral points of both electric and magnetic fields (though not at the same location!). The field structure surrounding such point is characterised by four distinct domains (bundles of the field lines) divided by two separatrices intersecting at the neutral point. Such a topology is known to result from the {\it separator reconnection}, a process which has been studied in the framework of resistive manetohydrodynamics (MHD), see e.g. \citet{priest}. In our electro-vacuum model, however, it arises entirely from the interaction of the strong gravitational field of the rotating BH with the background magnetic field, i.e. it is a mere gravitomagnetic effect. Charged matter injected into the magnetic separator site is prone to the acceleration by the electric field since its motion is not affected by the vanishing magnetic field and thus the acceleration is very effective. 

From the astrophysical viewpoint we regard the topological changes which the drift causes upon the field structure, especially the formation of the neutral points, as our main result demonstrating clearly that the strong gravitation of the rotating Kerr source may itself entangle the uniform magnetic field in a surprisingly complex way. We suggest that the gravity of the rotating black hole could work as a trigger for magnetic reconnection.

\section{Motion of charged matter}
\label{concl_traj}

We studied the regular and chaotic motion of electrically charged
particles near a magnetized rotating black hole or a compact star. We
employed the method of recurrence analysis in the phase space, which
allowed us to characterize the chaoticness of the system in a
quantitative manner. Unlike the method of Poincar\'e surfaces, the
Recurrence Plots have not yet been widely used to study the chaotic
systems in the regime of strong gravity.

The main motivation for these investigations is the question of whether
the matter around magnetized compact objects can exhibit chaotic motion,
or if instead the system is typically regular. One of the main
applications of our considerations concerns the putative envelopes of
charged particles enshrouding the central body in a form of a fall-back
corona, or plasma coronae extending above the accretion disk. While we
concentrated on the specifications of the RP method in circumstances
of a relativistic system, the assumed model cannot be considered as any
kind of a realistic scheme for a genuine corona. We simply imposed a
large-scale ordered magnetic field acting on particles in a combination
with strong gravity.

Various aspects of charged particle motion were addressed throughout
the chapter. First of all, we investigated the motion in off-equatorial
lobes above the horizon of a rotating black hole (modeled
by Kerr metric equipped with the Wald test field), as well as above the
surface of a magnetic star (modeled by the Schwarzschild metric with the
rotating dipolar magnetic field). In both cases we conclude that the
motion of test particles is regular, which was confirmed for a
representative number of orbits across the wide range of parameters over
all topological types of off-equatorial potential wells. This result is
somewhat unexpected because the off-equatorial orbits require a
perturbation to be strong enough (in terms of strength of the
electromagnetic field), so that it can balance the vertical component of
the gravitational force. 

Further, we investigated the response of the particle dynamics when the
energy level $\tilde{E}$ was raised gradually from the potential minimum
to values allowing cross-equatorial motion. We examined various
topological classes of the effective potential and came to the
conclusion that the cross-equatorial orbits are typically chaotic,
although very stable regular orbits may also persist for a certain
intermediate energy range. The classical work of \citet{henhei64} should be recalled in this context since it also identifies the energy as a trigger for chaotic motion in the analysed simple system. More recently the H\'{e}non--Heiles system was revisited in the relativistic context by \citet{vieira96}.

We also addressed the question of spin dependence of the stability of
motion for Kerr black hole in the Wald field. We noticed that this is a
rather subtle problem. The effective potential is by itself sensitive to
the spin value $a$ -- hence, we had to link the potential value roughly
linearly with the energy $\tilde{E}$ to maintain the potential lobe at a
given position. In other words, we did not find any clear and unique
indication of the spin dependence of the motion chaoticness. Most
trajectories exhibited regular behavior, which is also in agreement with
the previous results indicating that motion in off-equatorial lobes is
generally regular. On the other hand, in the case of the
cross-equatorial motion we observed that, for higher spins, more chaotic
features come into play when compared with the case of slow rotation.
This trend might be also attributed to simultaneous adjustments of
$\tilde{E}$. In other words, it appears impossible to give an
unambiguous conclusion about the spin dependence of the particles
dynamics. Instead, one has to deal with a complex, interrelated
dependence.

In the case of a Kerr black hole immersed in a large-scale magnetic
field, we observed the effect of confinement of particles regularly
oscillating around the equatorial plane. Escape of particles from the
plane is allowed for a given range of initial conditions since the
equipotentials do not close; they form an endless axial ``valley''
instead. The escaping trajectories create a narrow, collimated structure
parallel to the axis.

%% file: chap5n.tex
\chapter{\sffamily{Future prospects}}
\label{future}
\pagestyle{headings}
In the following we present our {\it to-do list} comprising of the issues which naturally arose during the previous study of the electromagnetic fields and charged particle dynamics. Most importantly we want to combine ideas of \rs{emfield} and \rs{kaptraj}. In other words we plan to investigate ionized particle motion governed by the generalized {\it oblique and drifting} EM field. Besides that we shall go through several rather technical issues related closely to the topic.
\begin{itemize}
\item {\it More general model of gaseous corona.}\\
We shall enhance our former axisymmetric model by considering oblique (misaligned with the rotation axis) magnetic fields in which the central body may be uniformly drifting in a general direction. Structure of the electromagnetic field is profoundly enriched and we suppose that similarly the dynamics of the particles will become considerably more complex. We plan to discuss the impact of new parameters upon the off-equatorial stable orbits and investigate how do they affect the dynamic regime of motion. We will try to identify a possible trigger of chaotic dynamics among new parameters. Besides standard methods the recurrence analysis will be employed since it proved useful in our previous work.

\item {\it Observational consequences.}\\
We plan to elaborate ideas introduced in \rs{rotacnic} concerning the frequency analysis of the off-equatorial orbits. We have seen that fragmented curves we observed in the Poincar\'{e} surfaces of section corresponding with the trajectories bound in the closed equatorial lobes may be identified with the Birkhoff chains of islands of stability. Such resonant chain is characterized by a single value of a rotation number which is in principle detectable in terms of spectral analysis of the observed signal. Presence of the Birkhoff chains allows us to discriminate between perturbed and regular system. Moreover the position and the width of the chains reflects other properties of the system. Detailed discussion of this approach applied to the different type of system may be found in \citet{vlny10}. However, in our analysis of the off-equatorial trajectories we observed more complicated structures in the surfaces of section which do not allow for the straightforward evaluation of the rotation number, nor the ratio of fundamental frequencies. Therefore we intend to adjust the method for the application to our scenario and infer the possible observational consequences for the system of gaseous corona we studied theoretically in \rfch{kapEM}.


\item {\it Magnetic shift of the ISCO.}\\
Position of the inner edge of the accretion disk is usually identified with the marginally stable geodesic orbit $r_{\rm{ms}}$ (also referred to as innermost stable circular orbit, ISCO) whose position is uniquely determined by the value of the black hole spin $a$ \citep{bardeen72}. Common black hole spin measurement methods are based on this relation as they actually determine $r_{\rm{ms}}$ to evaluate $a$ \citep{mcclintock}. In this context we raise the question whether the presence of the magnetic field may change the position of ISCO noticeably. Recently a similar problem was addressed by \citet{halo2_31} for the case of Schwarzschild source endowed with the dipole magnetic field. An introductory account of the influence of the uniform magnetic field aligned with the symmetry axis of Kerr black hole was brought by \citet{prasanna78}. We shall discuss the effect of the oblique uniform magnetic field around Kerr source upon the marginally stable orbit in detail. 

\item {\it Application of a new method for the computation of Lyapunov spectra.}\\ 
Lyapunov characteristic exponents (LCEs) are the basic indicators of chaos which capture the divergent features of the chaotic orbits straightforwardly. The classical non-covariant definition of LCEs, however, meets serious difficulties in curved spacetimes. Recently \citet{stach10} proposed novel geometrical approach to the computation of the Lyapunov spectra which completely avoids the conventional method of solving the variational equations to obtain the Lyapunov vectors which are periodically Gram-Schmidt orthonormalized along the flow. New algorithm is covariantly formulated and thus seems to be highly convenient for the application in general relativistic systems. We plan to implement this method when inspecting the dynamics of charged particles. This might be beneficial for both the results itselves and also to prove the new method fruitful.
\end{itemize}

%% file: appendix.tex
\chapter{Geometrized units}
\label{gu}

We use geometrized units instead of SI throughout this work. We set the speed of
light $c$, the gravitational constant~$G$, the Boltzmann constant
$k$ and the Coulomb constant $k_c=\frac{1}{4\pi\epsilon_{0}}$ equal
one.

\begin{table}[ht]
\label{gus}
\begin{center}
\begin{tabular}{c|c|c|c}
  constant& SI value & SI dimension & geometrized units\\
  \hline
    $c$ & $2.998\times10^{8}$ & $\rm{m}\,\rm{s}^{-1}$ & 1 \\
  $G$ & $6.67\times10^{-11}$ & $\rm{m}^{3}\,\rm{s}^{-2}\,\rm{kg}^{-1}$ & 1 \\
  $k$ & $1.38\times10^{-23}$ & $\rm{J}\,\rm{K}^{-1}$ & 1 \\
  $k_C$ & $8.988\times10^{9}$ & $\rm{kg}\,\rm{m}^{3}\rm{s}^{-2}\rm{C}^{-2}$ & 1 \\
\end{tabular}
\caption{Redefined constants in SI and in geometrized units.}
\end{center}
\end{table}
We aim to express arbitrary quantity in the terms of meters (thus it
becomes ``geometrized''). To manage that we construct conversion
factor $f$ consisting of constants $c$, $G$, $k$ and $k_{C}$ whose
dimension multiplied by the dimension of the quantity being
converted gives just meters (typically to the power of $1$, $2$ or $-1$). To make it more clear we convert mass
as an example:

\begin{math}
\begin{array}{c}
\left[M\right]_{\rm{SI}}=\rm{kg}\lb{}\rm{while}\lb{}\left[\textit{M}\right]_{\rm{geom}}=\rm{m}\\
\mbox{factor needed:}\lb\left[f_{\rm{M}}\right]=\rm{m}\,\rm{kg}^{-1}\\
\mbox{unambiguously:}\lb{}f_{\rm{M}}=\frac{G}{c^2}=7.43\times{}10^{-28}\;\rm{m}\,\rm{kg}^{-1}\\
M_{\rm{geom}}=f_{\rm{M}}M_{\rm{SI}}=7.43\times10^{-28}\:M_{\rm{SI}}\;\rm{m}\,\rm{kg}^{-1}\\
\mbox{e.g. solar
mass:}\:\:\left(M_{\odot}\right)_{\rm{geom}}=7.43\times{}10^{-28}\;\rm{m}\,\rm{kg}^{-1}\cdot\:1.989\times10^{30}\;\rm{kg}=1472\:\rm{m}
\end{array}
\end{math}

When converting from the geometrized units back to SI
we just need to divide by the same factor. Factors for basic
quantities are given in table A.2.
\begin{table}[ht]
\begin{center}
\begin{tabular}{c|c|c}

  quantity& factor & numerical value\\
  \hline
    time & $f_{\rm{t}}=c$ & $3.00\times10^{8}\;\rm{m}\,{}s^{-1}$\\
    mass &
    $f_{\rm{M}}=\frac{G}{c^2}$&$7.43\times{}10^{-28}\;\rm{m}\,\rm{kg}^{-1}$\\
    charge &
    $f_{\rm{C}}=\frac{\sqrt{G\,k_{\rm{C}}}}{c^2}$ &
    $8.62\times10^{-18}\;\rm{m}\,\rm{C}^{-1}$\\
    momentum & $f_{\rm{m}}=\frac{G}{c^3}$ &
    $2.48\times10^{-36}\;\rm{m}\,\rm{kg}^{-1}\rm{m}^{-1}\,\rm{s}$\\
    angular momentum & $f_{\rm{m}}=\frac{G}{c^3}$ &
    $2.48\times10^{-36}\;\rm{m}^2\,\rm{kg}^{-1}\rm{m}^{-2}\,\rm{s}$\\
    energy & $f_{\rm{E}}=\frac{G}{c^4}$ &
    $8.26\times10^{-45}\;\rm{m}\,\rm{J}^{-1}$\\
    temperature & $f_{\rm{T}}=\frac{G\,k}{c^4}$ &
    $1.14\times10^{-67}\;\rm{m}\,K^{-1}$\\
    magnetic induction &
    $f_{\rm{B}}=\frac{1}{c}\sqrt{\frac{G}{k_{\rm{C}}}}$ &
    $2.87\times10^{-19}\; \rm{m}^{-1}\,\rm{T}^{-1}$
\end{tabular}
\caption{Conversion factors for basic quantities.}
\end{center}
\label{factors}
\end{table}

Finally we show how to convert specific charge $\tilde{Q}$ which we
use in the main text often:
\begin{equation*}
\left(\tilde{Q}\right)_{\rm{geom}}=\frac{\left(Q\right)_{\rm{geom}}}{\left(M\right)_{\rm{geom}}}=\frac{f_{\rm{C}}\left(Q\right)_{\rm{SI}}}{f_{\rm{M}}\left(M\right)_{\rm{SI}}}=\sqrt{\frac{k_{\rm{C}}}{G}}\left(\tilde{Q}\right)_{\rm{SI}}=1.16\times10^{10}\:\rm{kg}\,\rm{C}^{-1}\cdot\left(\tilde{Q}\right)_{\rm{SI}}
\end{equation*}
leaving $\left(\tilde{Q}\right)_{\rm{geom}}$ dimension-less.

For example specific charge of the proton $\tilde{q}_{\rm{p}}$ is in
the  geometrized units expressed as follows:
\begin{equation*}
\left(\tilde{q}_{\rm{p}}\right)_{\rm{geom}}=1.16\times10^{10}\cdot
\left(\tilde{q}_{\rm{p}}\right)_{\rm{SI}}=1.16\times10^{10}\cdot9.58\times10^{7}=1.11\times10^{18}.
\end{equation*}

When dealing with compact objects we often scale all the geometrized quantities by the mass of the central body in order to simplify our equations. The mass of the object only needs to be specified at the very end of calculations when we need to recover actual value of a quantity from it's dimension-less scaled version.

Scaling of an arbitrary quantity $X$ by the mass of the gravitational source $M$ can be formally expressed as follows:
\begin{equation*}
\left[ X\right]_{\rm{geom}}=m^p\;\;\Rightarrow\;\left( X\right)_{\rm{geom,\, scaled}}=\frac{\left( X\right)_{\rm{geom}}}{\left( M\right)_{\rm{geom}}^p}.
\end{equation*}

For instance one reads the particle's orbital proper period from the output of the dimension-less equations of motion to be $T=200$. Setting the mass of the central object as  $M=3\; M_{\odot}$ we obtain following SI value of the proper period
\begin{equation*}
(T)_{\rm{SI}}=\frac{ (T)_{\rm{geom,\, scaled}}(M)_{\rm{geom}} } { f_{\rm{t}} }=\frac{ 200\cdot 3\cdot 1472 } { 3\times 10^8 }=2.94\times10^{-3}\;\rm{s}.
\end{equation*}

Angular momentum of the rotating object $S=a\;M$ is also commonly scaled by the mass of the object. In the case of the Kerr black hole the spin parameter $a$ is restricted to $|(a)_{\rm{geom,\, scaled}}|\le 1$. For example if we consider (highly idealized) Sun as a homogeneous sphere of radius $R_{\odot}=6.96\times 10^{8}\:\rm{m}$ rotating with period of 25 days $\left(T_{\odot}=2.16\times 10^6\:\rm{s}\right)$ we arrive at the following value of the scaled spin parameter $a$
\begin{equation*}
\label{sluncespin} 
(a_{\odot})_{\rm{geom,\, scaled}}=\frac{(S)_{\rm{geom}}}{(M_{\odot})^2_{\rm{geom}}}=\frac{f_{\rm{m}}(S)_{\rm{SI}}}{(M_{\odot})^2_{\rm{geom}}}=\frac{f_{\rm{m}}}{(M_{\odot})^2_{\rm{geom}}}\left( \frac{4\pi}{5}\: \frac{M_{\odot}R^2_{\odot}}{T_{\rm{\odot}}} \right)_{\!\!\rm{SI}}=1.28.
\end{equation*}
Kerr solution for the source with given mass and angular momentum would thus describe a naked singularity rather than black hole.
  
Another example is the product $\tilde{q}B$ of specific charge of test particle (dimension-less in geometrized units) and the asymptotic strength of the magnetic field (of geometrized dimension $m^{-1}$) which acts as one of the parameters determining particle's trajectory in chap. \ref{kaptraj}.  For example setting $(\tilde{q}B)_{\rm{geom,\, scaled}}=10$ in the dimension-less equations would correspond with the following strength of magnetic field $B$ for a given supermassive black hole of mass $M=1\times 10^6 \; M_{\odot}$ if we further specify the test particle to be an electron

\begin{align*}
(B)_{\rm{SI}}&=\frac{(B)_{\rm{geom, \,scaled}}}{f_{\rm{B}}\;(M)_{\rm{geom}}}=\frac{(\tilde{q}B)_{\rm{geom,\, scaled}}}{f_{\rm{B}}\;(M)_{\rm{geom}}(\tilde{q}_{\rm{e}})_{\rm{geom,\, scaled}}}
=\frac{ (\tilde{q}B)_{\rm{geom,\, scaled}} } { (M)_{\rm{geom}}(\tilde{q}_{\rm{e}})_{\rm{SI}} }\frac{ f_{\rm{M}} }{ f_{\rm{B}}f_{\rm{C}} }\\
&=\frac{10\cdot 3 \times 10^8}{1\times 10^6 \cdot 1472 \cdot 1.76\times 10^{11}}=1.16\times 10^{-11}\: \rm{T}=1.16\times 10^{-7}\: \rm{G}.
\end{align*}

%% file: appendix2n.tex
\chapter{Choice of the integrator}
\label{integ}
\pagestyle{headings}
In this section we shall give some details about a rather technical issue concerning the proper choice of the integration scheme which would fit best to our problem. In particular we will compare the performance of the symplectic integrator with several non-symplectic routines and discuss under which circumstances we should choose the symplectic one and when we should switch to some other scheme. We will be basically concerned with two crucial aspects -- accuracy of the integration and CPU time consumption. The latter is generally less critical in our application since we are not facing that computationally intensive problem. 

We are dealing with autonomous Hamiltonian system$^1$ \footnotetext[1]{Equations of motion may be equivalently expressed in terms of Lorentz force \citep[p. 898]{mtw} which leads to the set of four second order ODEs. Numerical experiments, however, led us to the conclusion that this formulation is computationally less effective compared to the Hamiltonian formalism. Generally for a given numerical scheme with the same parameters (resulting in similar accuracy of integration) the integration of Hamilton's equations was roughly two times faster.} whose equations of motion form a specific subclass of first order ordinary differential equations (ODEs). Two fundamental characteristics of the Hamiltonian flow should be highlighted 
 \begin{itemize}
\item conservation of the net energy (Hamiltonian) of the system
\item conservation of the symplectic structure $\bm{\omega}=\mathbf{d}\pi_{\mu}\wedge \mathbf{d}x^{\mu}$.
\end{itemize}
In the classical mechanics the natural choice of the generalized coordinates leads to the Hamiltonian which may be interpreted as a net energy of the system. This is true even for the system of a charged particle in the external EM field where the generalized momenta-dependent potential is introduced \citep[][chap.~ 8]{goldstein}. Time-independance of the Hamiltonian is thus equivalent to the conservation of the net energy of the system. In the general relativistic version of this system, however, we employ super-hamiltonian formalism \citep[][chap. 21]{mtw} in which the energy of the particle $E$, as a negatively taken time component of the canonical momentum $E\equiv -\pi_t$, is conserved by virtue of the Hamilton's equations itselves providing that the super-hamiltonian doesn't depend on the coordinate time $t$. On the other hand the value of the super-hamiltonian $\mathcal{H}=\frac{1}{2}p_{\mu}p^{\mu}$ is by construction equal to $-\frac{1}{2}m^2$ where $m$ is the rest mass of the particle. Conservation of the super-hamiltonian in the system is thus equivalent to the conservation of the rest mass of the particle.

By conservation of the symplectic 2-form $\bm{\omega}$ we mean that its components $\omega_{\alpha\beta}$ in the basis $\left(\mathbf{d}t(\lambda),\mathbf{d}r(\lambda),\mathbf{d}\theta(\lambda),\mathbf{d}\varphi(\lambda)),\mathbf{d}\pi_t (\lambda),\mathbf{d}\pi_r (\lambda), \mathbf{d}\pi_{\theta} (\lambda), \mathbf{d}\pi_{\varphi}(\lambda) \right)$ do not change during the evolution of the system and for arbitrary value of the affine parameter $\lambda$ (i.e. at each point of the phase space trajectory) we obtain
\begin{equation}
\omega_{\alpha\beta}=
\begin{pmatrix}
0&-\mathbb{I}\\
\mathbb{I}&0\\
\end{pmatrix},
 \label{symplmatice}
\end{equation}
where $\mathbb{I}$ stands for the four-dimensional identity submatrix and $0$ is null submatrix of the same dimension. Conservation of the symplectic structure expresses in the abstract geometrical language the fact that the evolution of the system is governed by the Hamilton's canonical equations. See \citet{arnold} for details on the geometric formulation of the Hamiltonian dynamics. 

It would be highly desirable to use such integration scheme which would conserve both quantities which are conserved by the original system. It appears, however, that this is not possible for non-integrable systems and one has to decide whether he employs the scheme which conserves energy or rather the integrator which keeps symplectic structure. The latter are referred to as symplectic integrators and by many accounts provide most reliable results in numerical studies involving Hamiltonian systems. See \citet{yoshida93} for a comprehensive review on symplectic methods.

We list all the schemes we employ in this survey specifying their basic properties. We shall actually compare one symplectic method with several standard integrators. Code names we use for the schemes are those which denote the routines in the MATLAB system.
\begin{itemize}
\item
GLS -- Gauss-Legendre symplectic solver, $s$-stage implicit Runge-Kutta (RK) method, crucial control parameter: stepsize $h$
\item
ODE87 -- Dormand-Prince 8th - 7th order explicit RK scheme, the most precise RK method (local error of order $O(h^8)$), adaptive stepsize -- RelTol is set to control local truncation error  
\item
ODE113 -- multistep Adams-Bashforth-Moulton solver, based on the pre\-dic\-tor-corrector method (PECE), RelTol is set
\item
ODE45 -- Dormand-Prince seven stage 5th-4th order method of explicit RK family, adaptive stepsize, default integration method in MATLAB and GNU OCTAVE, error is controlled by RelTol
\end{itemize}
Apart from ODE113 all other routines are single-step (Runge-Kutta like) methods which means that they express the value of the solution in the next step in terms of a single preceding step. They may be related explicitely or implicitly. Multistep methods in contrast employ more preceding steps to calculate the solution at the succeeding point. RelTol is a parameter which specifies the highest allowed relative error in each step of integration (local truncation error) when the adaptive stepsize methods are used. In the case of exceeding the RelTol the stepsize is reduced automatically to decrease the error. 

\begin{figure}[htb]
\centering
\includegraphics[scale=0.62,clip]{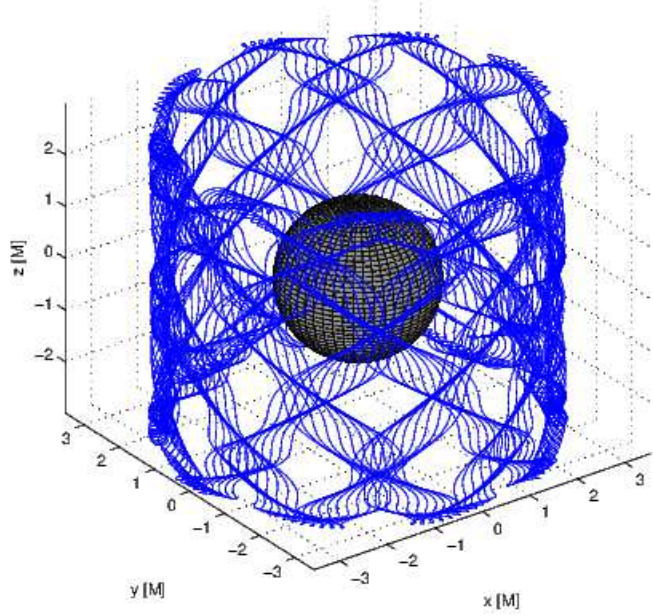}
\caption{Regular trajectory of charged test particle ($\tilde{q}\tilde{Q}=1$, $\tilde{L}=6\;M$ and $\tilde{E}=1.6$) on the Kerr background
($a=0.9\;M$) with Wald magnetic field
($\tilde{q}B_{0}=1M^{-1}$). Particle is
launched at $r(0)=3.68$, $\theta(0)=1.18\:M$ with $u^r(0)=0$.}
\label{traj_reg_3d}
\end{figure}

\begin{figure}[hp]
\centering
\includegraphics[scale=0.61,clip]{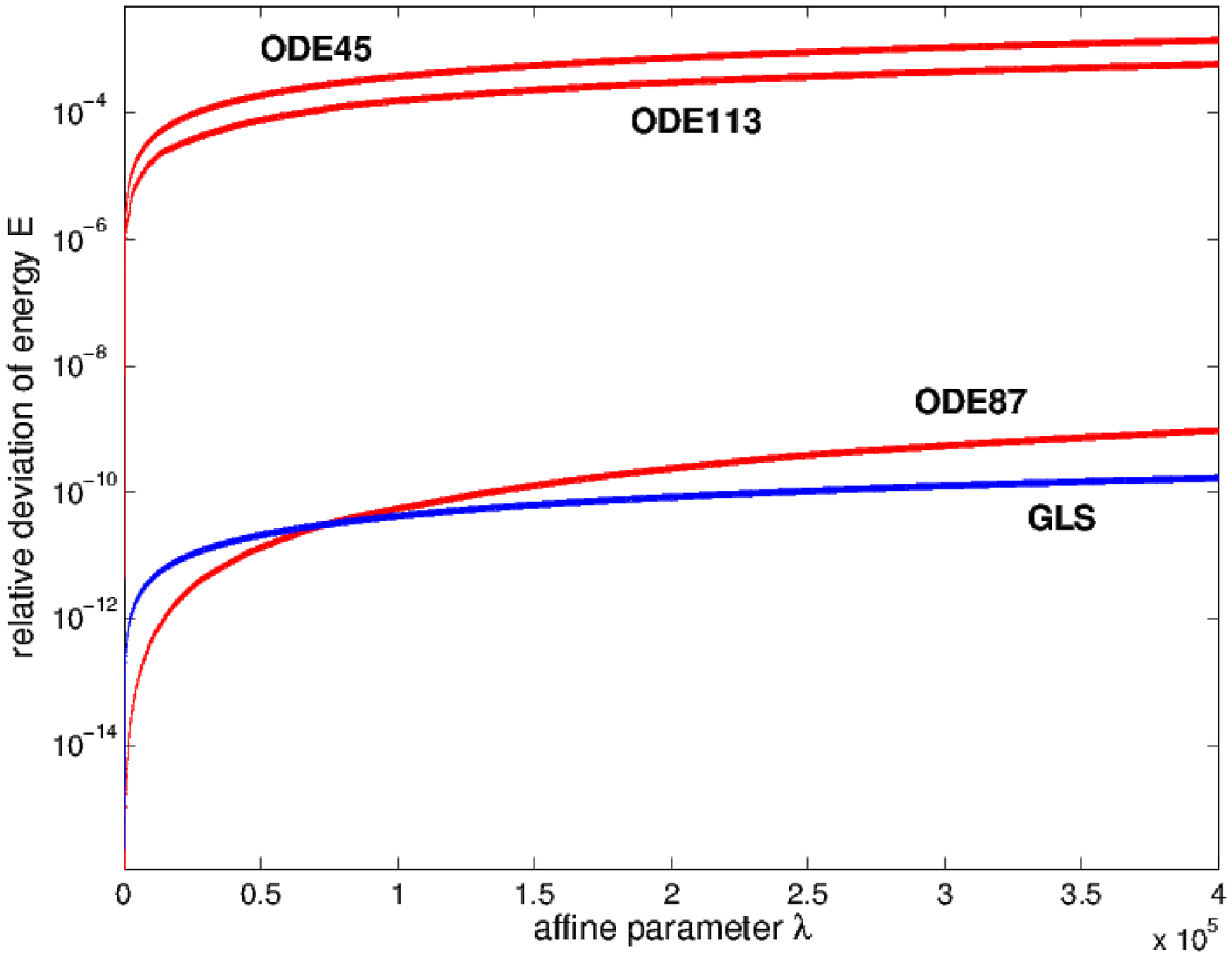}
\includegraphics[scale=0.61,clip]{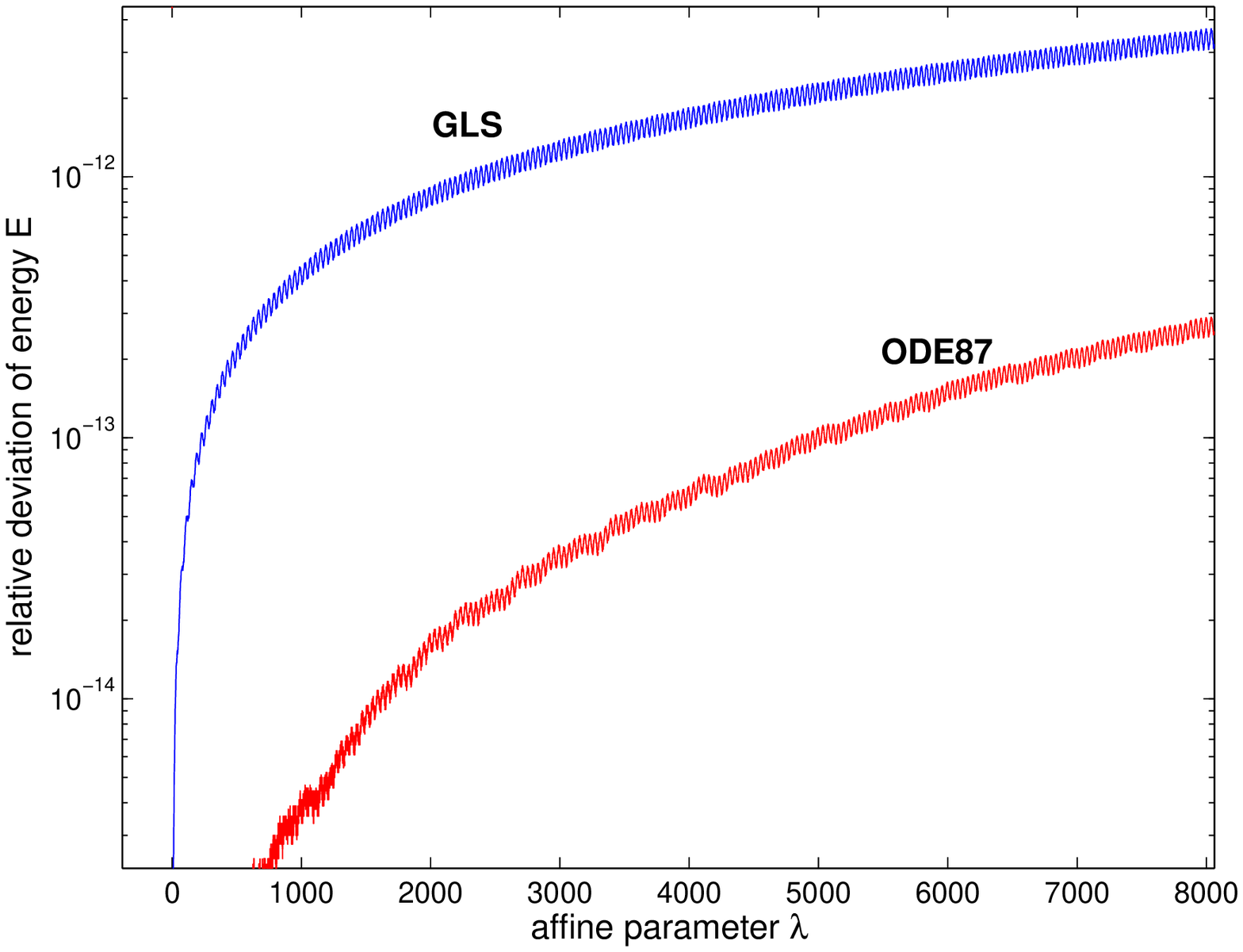}
\caption{Comparison of the integrators in the case of regular trajectory. Symplectic GLS provides the most reliable results for $\lambda\gtrsim10^5$. Bottom panel shows that besides secular drift in energy (artificial excitation or dumping of the system; plot shows absolute values, however) it also oscillates on the short time scale.}
\label{traj_reg}
\end{figure}

We comment that for general non-separable Hamiltonians only implicit symplectic schemes may be found. Explicit methods exist for separable Hamiltonians and for some special forms of non-separable ones \citep{chin09}. Besides other implications of the usage of the implicit methods we note that they necessarily involve some type of iterative scheme which is typically of a Newton's type and thus requires to supply Jacobian of the right hand sides of the equations of motion which is the Hessian matrix of the second derivatives of the super-hamiltonian $\mathcal{H}$ in our case.

Another inconvenience connected with the symplectic methods is their failure to conserve the symplectic structure once the adaptive stepsize method would be used \citep{skeel92}. Therefore the stepsize has to be set rigidly for a given integration segment when using symplectic method. Several workarounds have been suggested to combine benefits of symplectic solvers and variable stepsize algorithms -- e.g. Hairer's symplectic meta-algorithm \citep{hairer97} which is, however, only applicable to the separable Hamiltonians. In our context one would considerably suffer from the fixed timestep only in the case of highly eccentric orbits.

\begin{table}[htb]
\label{stat_reg}
\begin{center}
\begin{tabular}{l|l|l|l|l}
integrator& $\Delta |E|/|E|$ & $t_{\rm{comp}} [h]$& RelTol& stepsize $h$\\
\hline
GLS&$\approx 10^{-10}$ & 14 & N/A & 0.25\\
ODE87&$\approx 10^{-9}$& 14 & $10^{-14}$ & adaptive\\
ODE113&$\approx 10^{-3}$& 1/3  & $10^{-14}$ & adaptive\\
ODE113&$\approx 10^{-3}$& 1/4 & $10^{-6}$ & adaptive\\
ODE45&$\approx 10^{-3}$& 1/4 & $10^{-14}$ & adaptive\\
\end{tabular}
\caption{Comparison of the performance of several integration schemes for the regular trajectory integrated up to $\lambda=4\times 10^{5}$ (see \rff{traj_reg}).}
\end{center}
\end{table}

\begin{figure}[htb]
\includegraphics[scale=0.62, clip]{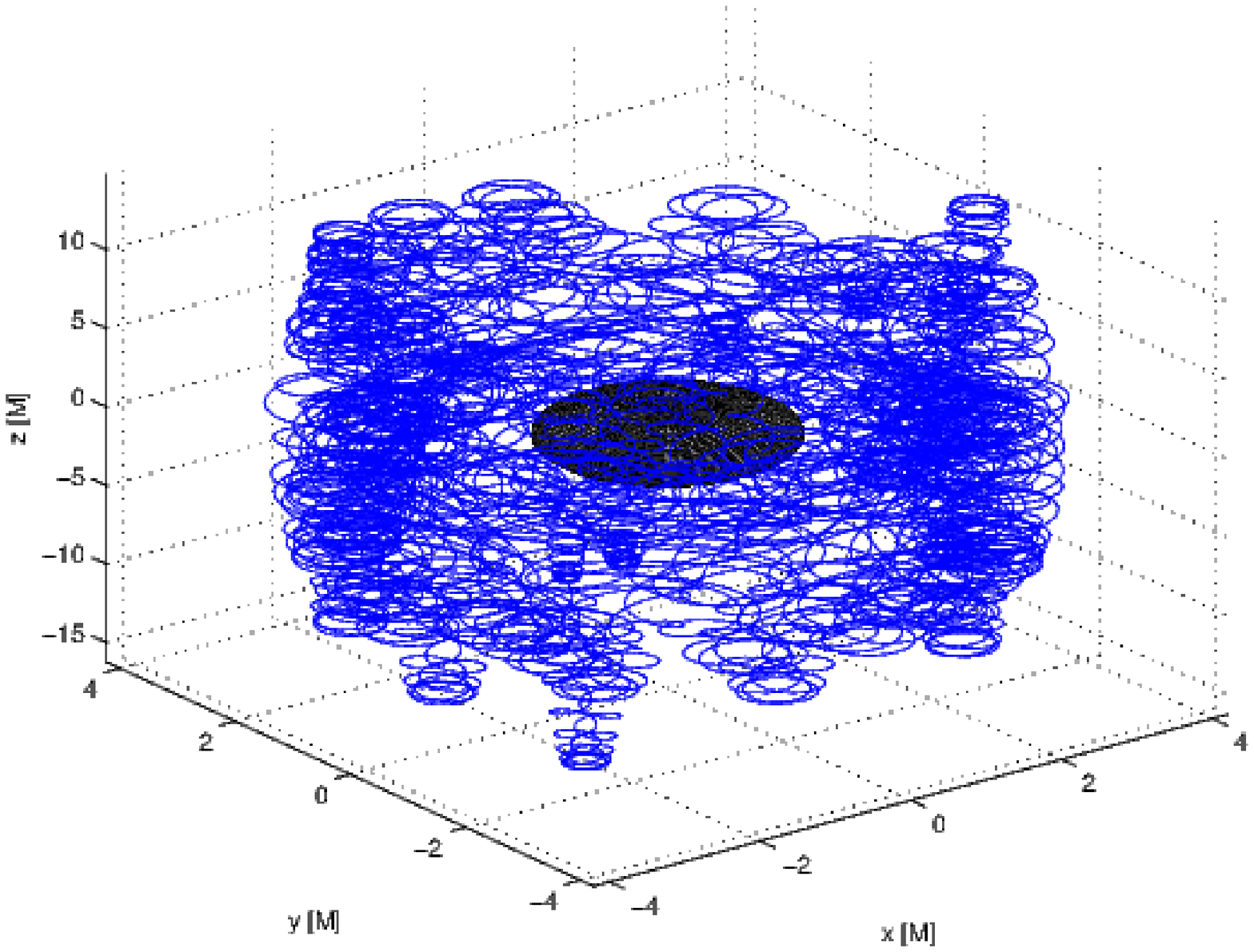}
\caption{Chaotic trajectory of charged test particle ($\tilde{q}\tilde{Q}=1$, $\tilde{L}=6\;M$ and $\tilde{E}=1.8$) on the Kerr background
($a=0.9\;M$) with Wald magnetic field
($\tilde{q}B_{0}=1M^{-1}$). Particle is
launched at $r(0)=3.68\: M$, $\theta(0)=1.18$ with $u^r(0)=0$.}
\label{traj_chaos_3d}
\end{figure}

\begin{figure}[hp]
\centering
\includegraphics[scale=0.6,clip]{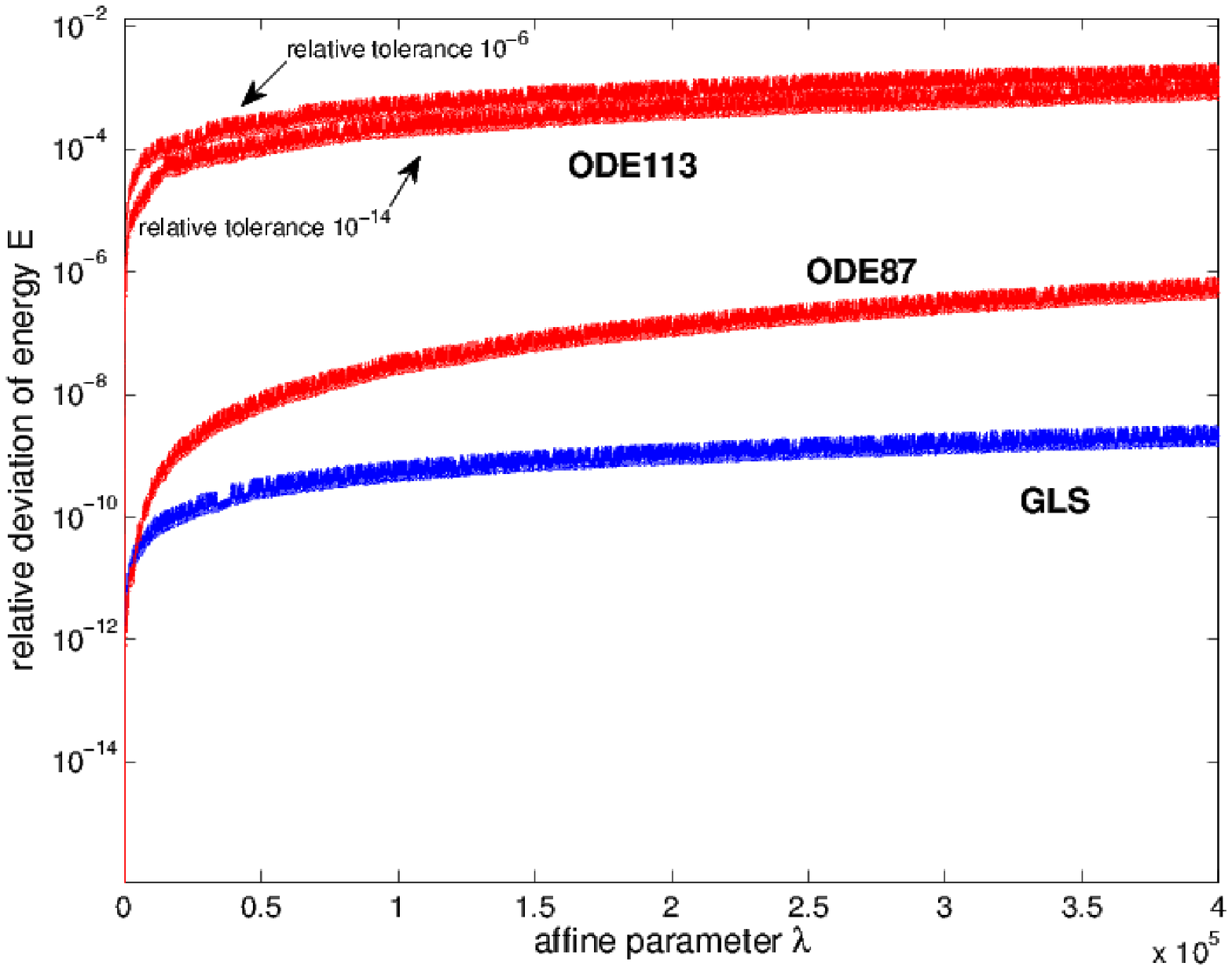}
\includegraphics[scale=0.6,clip]{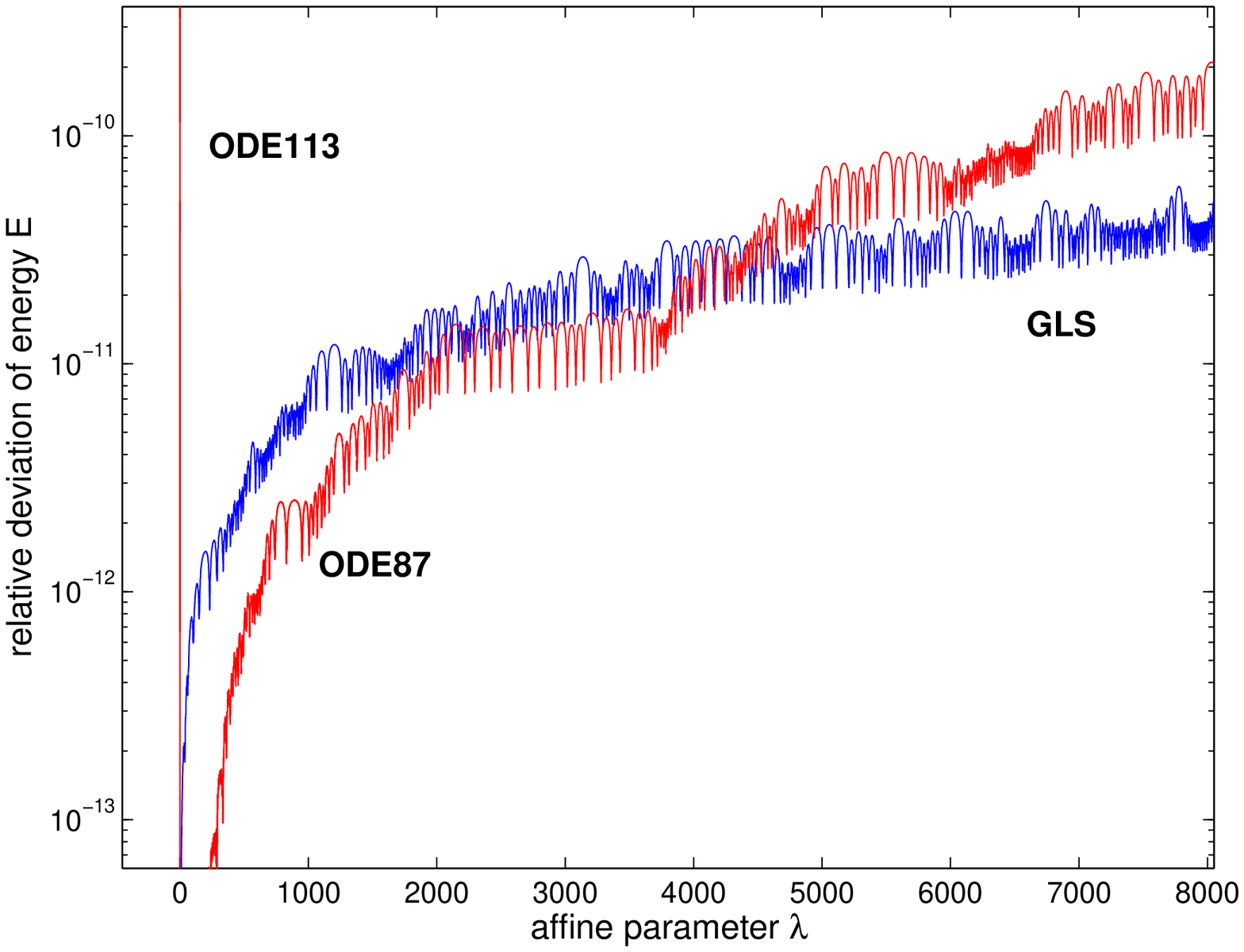}
\caption{Comparison of the integrators in the case of chaotic trajectory. For $\lambda\gtrsim5\times10^3$ the GLS dominates in accuracy over other schemes with the difference rising steadily. In the upper panel we compare ODE113's outcome for two distinct values of RelTol parameter. ODE45 is not shown to avoid overlapping of its plot with ODE133 curves.}
\label{traj_chaos}
\end{figure}

First we integrate the cross-equatorial regular trajectory \rff{traj_reg_3d}. Comparison of the performance of the integrators is plotted in \rff{traj_reg}. We plot relative deviation of the particle's specific energy $\tilde{E}$ from its initial value rather than the error in super-hamiltonian because the discussion of motion in chap. \ref{kaptraj} was mostly held in terms of $\tilde{E}$ whose impact upon the trajectory is thus more familiar to us. We calculate the current value of $\tilde{E}$ from the super-hamiltonian $\mathcal{H}$, while the value of $\pi_t$ remains truly constant regardless the integrator since the Hamilton's equation for its evolution is simply $\mathrm{d}\pi_t / \mathrm{d}\lambda =0$.

Stepsize of GLS is set in such a way that the integration consumes roughly the same amount of the CPU time as it does for ODE87 with $\rm{RelTol}=10^{-14}$ to make the results comparable. The global accuracy of the GLS solver could be further increased by reducing the stepsize while decreasing the RelTol hardly improves the secular accuracy of non-symplectic methods here (we have compared $\rm{RelTol}=10^{-6}$ and $\rm{RelTol}=10^{-14}$ results for ODE113 obtaining global errors of the same orders in both cases). 

We observe that the error of GLS rises steeply at the beginning and ODE87 is considerably better for some amount of time. However then the error of GLS almost saturates while ODE87's error keeps growing significantly. For $\lambda\gtrsim10^5$ which corresponds to $\approx 1000$ revolutions around the center$^2$ \footnotetext[2]{For instance for $M=10^{6}M_{\odot}$ the azimuthal proper period of a given particle reads $T_{\varphi} \approx 10^3 s$ in SI.} the GLS scheme becomes more accurate than ODE87 with the difference further rising steadily. We conclude that in the case of regular trajectory ODE87 is appropriate for short-term accurate integration and GLS for any longer accurate integrations. On the other hand for fast, though inaccurate computations one employs ODE113 on all time scales.

\begin{table}[htb]
\label{stat_chaos}
\begin{center}
\begin{tabular}{l|l|l|l|l}
integrator& $|\Delta{}E|/|E|$ & $t_{\rm{comp}}\; [h]$& RelTol& stepsize $h$\\
\hline
GLS&$\approx 10^{-9}$ & 14 & N/A & 0.25\\
ODE87&$\approx 10^{-6}$& 14 & $10^{-14}$ & adaptive\\
ODE113&$\approx 10^{-3}$& 1/6  & $10^{-14}$ & adaptive\\
ODE113&$\approx 10^{-3}$& 1/6 & $10^{-6}$ & adaptive\\
ODE45&$\approx 10^{-3}$& 1/2 & $10^{-14}$ & adaptive
\end{tabular}
\caption{Comparison of the performance of several integration schemes for the chaotic trajectory integrated up to $\lambda=4\times 10^{5}$ (see \rff{traj_chaos}).}
\end{center}
\end{table} 

In the case of the chaotic trajectory \rff{traj_chaos} the dynamics changes in favor of symplectic solver GLS. In \rff{traj_chaos_3d} we observe that in this case the symplectic scheme is superior to the others in even more convincing manner than it was in the regular case. Although the initial phase when the error induced by GLS rises more steeply than that of ODE87 is also present, it turns over very quickly and for
$\lambda\gtrsim5\times10^3$ ($\approx 50$ azimuthal revolutions) the GLS turns out to be more accurate. 
The difference then rises much faster compared to the regular case. 

Experiments with ODE113 show that here we observe distinct (thought not sharply) errors by changing the RelTol. Difference of eight orders of magnitude in RelTol resulted in roughly one order difference in global error. We also note that chaotic regime induces disorder in short-time oscillations of the global error (see bottom panel of \rff{traj_chaos}). We summerize that the chaotic regime accents the supremacy of the symplectic scheme which is to be applied on all time scale here (except very short integrations where ODE87 dominates) to obtain the most accurate results. For fast though inaccurate calculation one would switch to ODE113 as before.

\begin{figure}[htb]
\centering
\includegraphics[scale=0.365,clip]{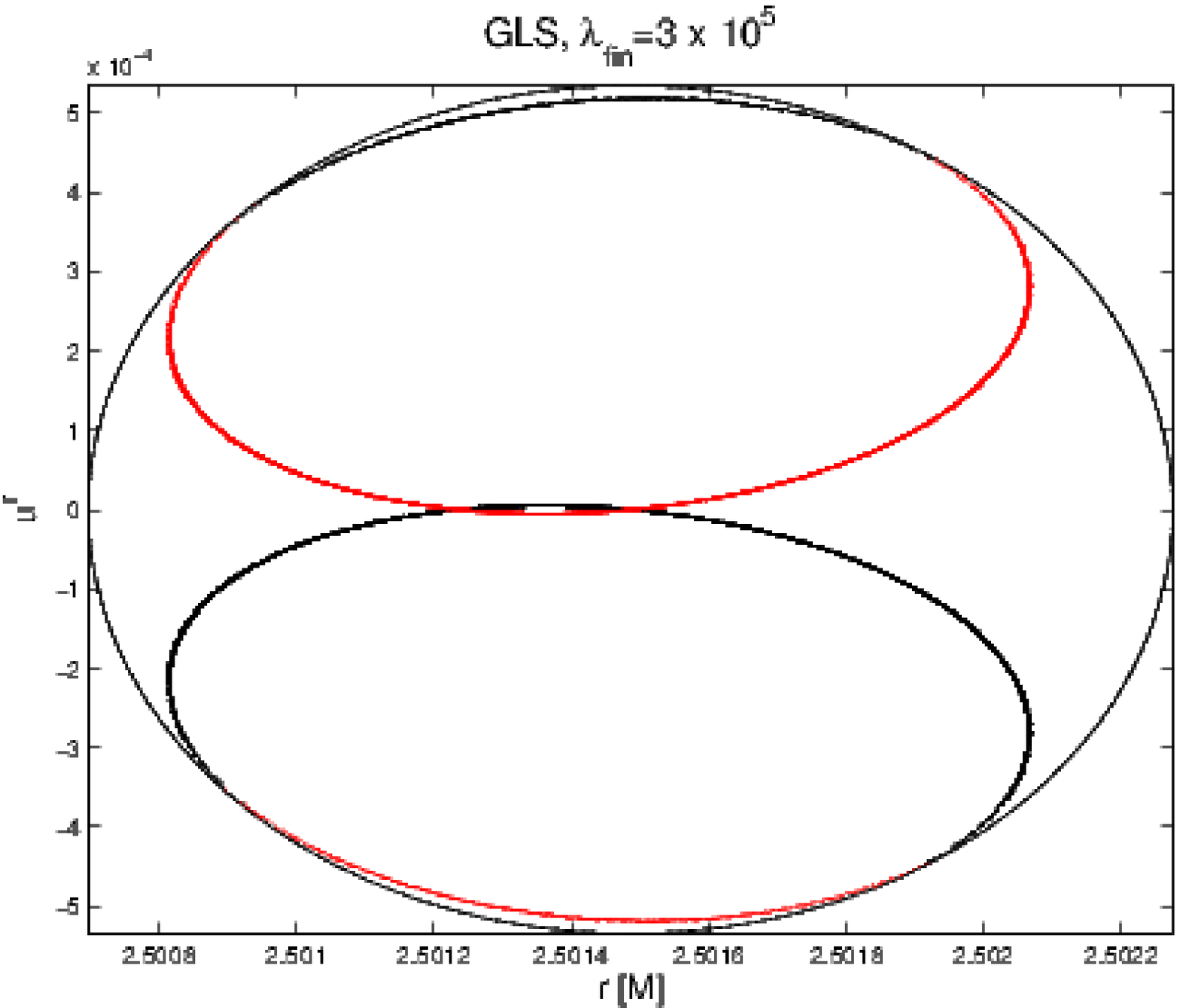}
\includegraphics[scale=0.365,clip]{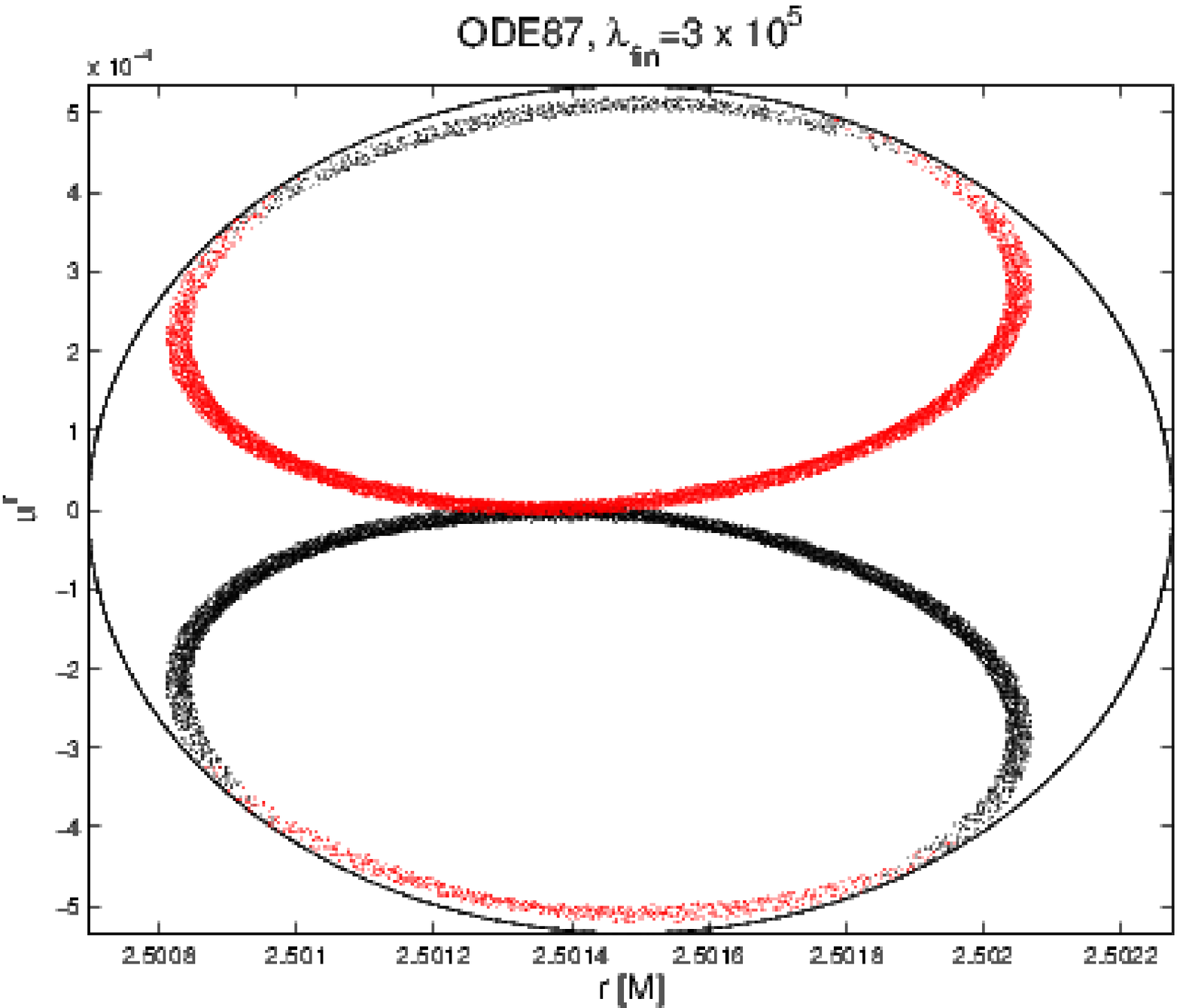}\\
\includegraphics[scale=0.365,clip]{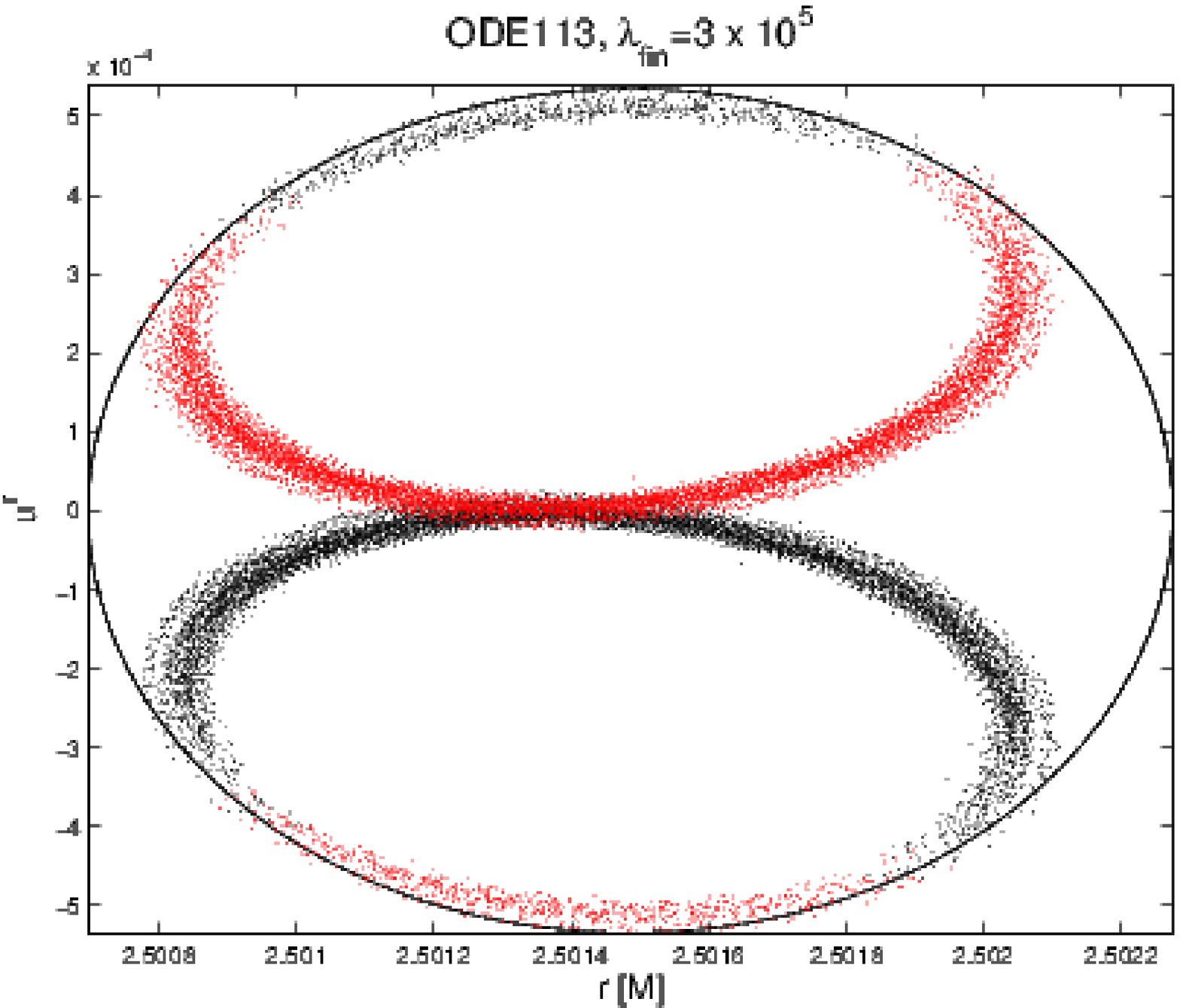}
\includegraphics[scale=0.365,clip]{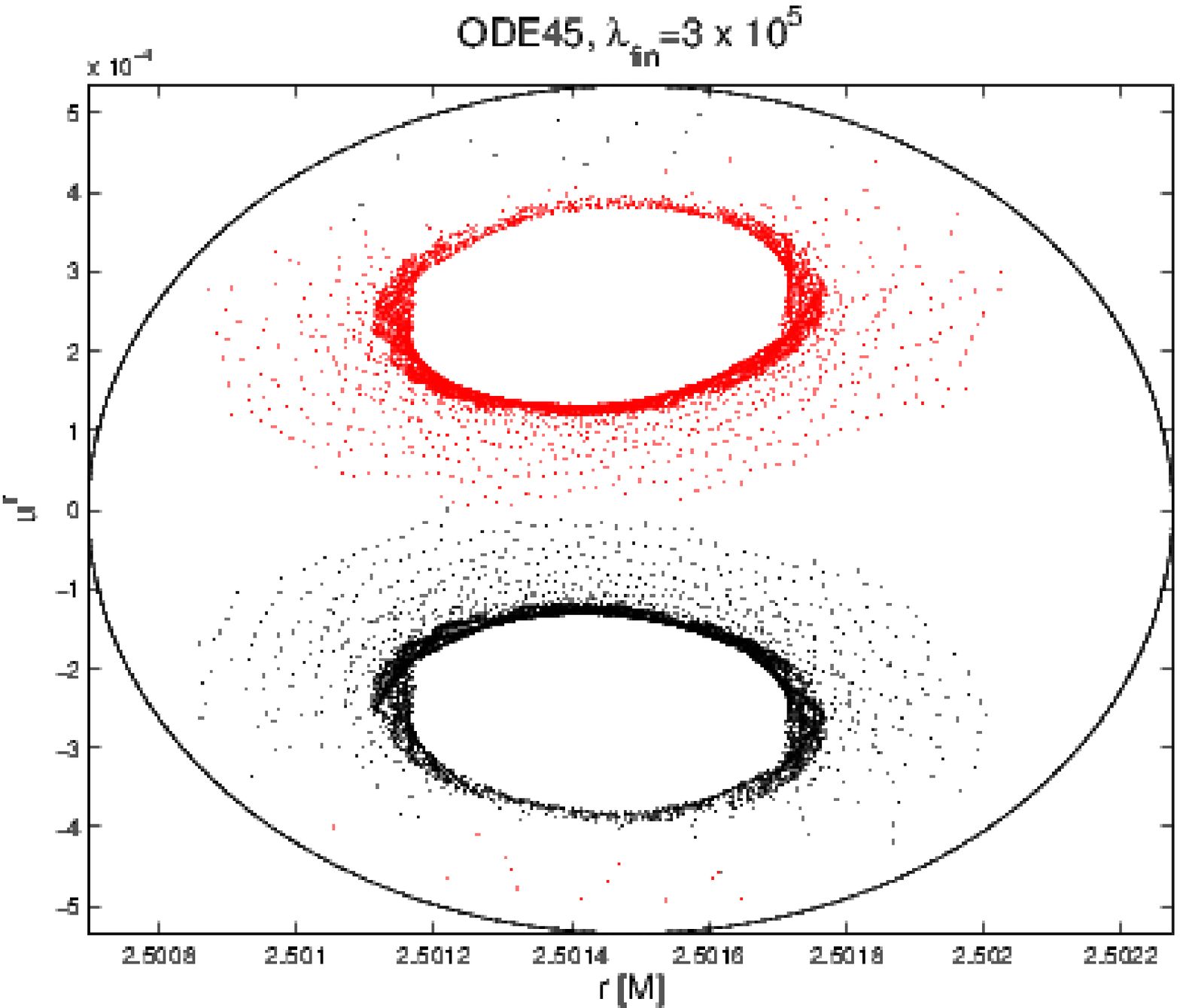}
\caption{We show how the accuracy of the integration crucially affects the appearance of the Poincar\'{e} surfaces of section of a single regular trajectory with $\tilde{q}\tilde{Q}=1.76$, $\tilde{L}=4.02\;M$ and $\tilde{E}=1.619855$ on the Kerr background
$a=0.55\;M$ with Wald magnetic field
$\tilde{q}B_{0}=1.92\:M^{-1}$. Particle is
launched at $r(0)=2.5012$, $\theta(0)=1.0447\:M$ with $u^r(0)=0$. We distinguish $u^{\theta}\geq0$ (black point) from $u^{\theta}<0$ (red point) in the surfaces of section.}
\label{rezy}
\end{figure}

From a practical point of view we demand high accuracy of the long-term integration when constructing Poincar\'{e} surfaces of section. By theory the intersection points with regular trajectory form one-dimensional curve in the section plane. In \rff{rezy} we observe, however, that the points may be dispersed over the considerable area if the global error in energy rises causing artificial excitation/dumping of the system. Symplectic GLS provides the most reliable outcome, with ODE87 the curve is blurred significantly but the interpretation remains unambiguous. With ODE113 the curve is further blurred and using ODE45 solver we obtain completely unreliable outcome which could easily lead to incorrect interpretation of a trajectory as a chaotic one. We note that we intentionally chose such trajectory which is highly sensitive to the relative errors in dynamic quantities since it itself spans small range of coordinate and momenta values.

We confirm that the symplectic integrators are the method of choice in the case of long-term integration of the Hamiltonian system which in our case consists of a charged test particle orbiting around the Kerr black hole with stationary and axisymmetric electromagnetic test field. Its supremacy over non-symplectic methods is even more apparent in the case of chaotic orbits, where the global accuracy of non-symplectic methods decreases rapidly. The accuracy of the symplectic integrator could be further increased by reducing the stepsize (at the cost of the computational time). On the other hand the performance of the non-symplectic solvers is not considerably affected by changing the RelTol parameter across the wide range of the values. Once the integrator doesn't fit the problem (= is not symplectic) there is no effective way to control the global error and even the extremely small local truncation errors do not ensure reliable outcome on a long time scale.

%% file: appendix3.tex
\chapter{Vfexplorer2}
\label{vfexplorer2}
\pagestyle{headings}
Here we briefly introduce the software tool {\it vfexplorer2} we programmed in Matlab system in order to survey electric and magnetic fields studied in \rfch{kapEM}. We set up simple graphic user interface (GUI) for the sake of effective exploration of complex field structures. 

Basic features of {\it vfexplorer2} involve interactive zooming, step-wise navigation and track logging allowing the user to return to any previous step of the exploration. Moreover, he can export the plot to the graphic file directly at any moment. Alternatively he may redirect the matrix output describing given portion of the vector field to the LIC toolbox \citep{shambo05} which encodes the field structure into the texture resembling iron fillings (see e.g. \rff{mag_drift_LIC})  

In \rff{vfexp} we present two screenshots illustrating the survey of magnetic field around a drifting Kerr BH (see \rs{emfield} for details). Interactive zooming feature proves very useful when searching for a particular topological pattern (e.g. separator null points which we studied in \rs{driftingEM}) in the complex field structure.

Currently the {\it vfexplorer2} GUI also handles the problem specific parameters such as spin of the BH, velocity of the drift etc. Nevertheless, it can be adopted to explore any other vector field quite easily. Besides that we plan to include another features, especially we intend to encode the field strength into the color scale of the background of the field lines.

\begin{figure}[hp]
\centering
\includegraphics[scale=.4,trim=2mm 0mm 0mm 0mm,clip]{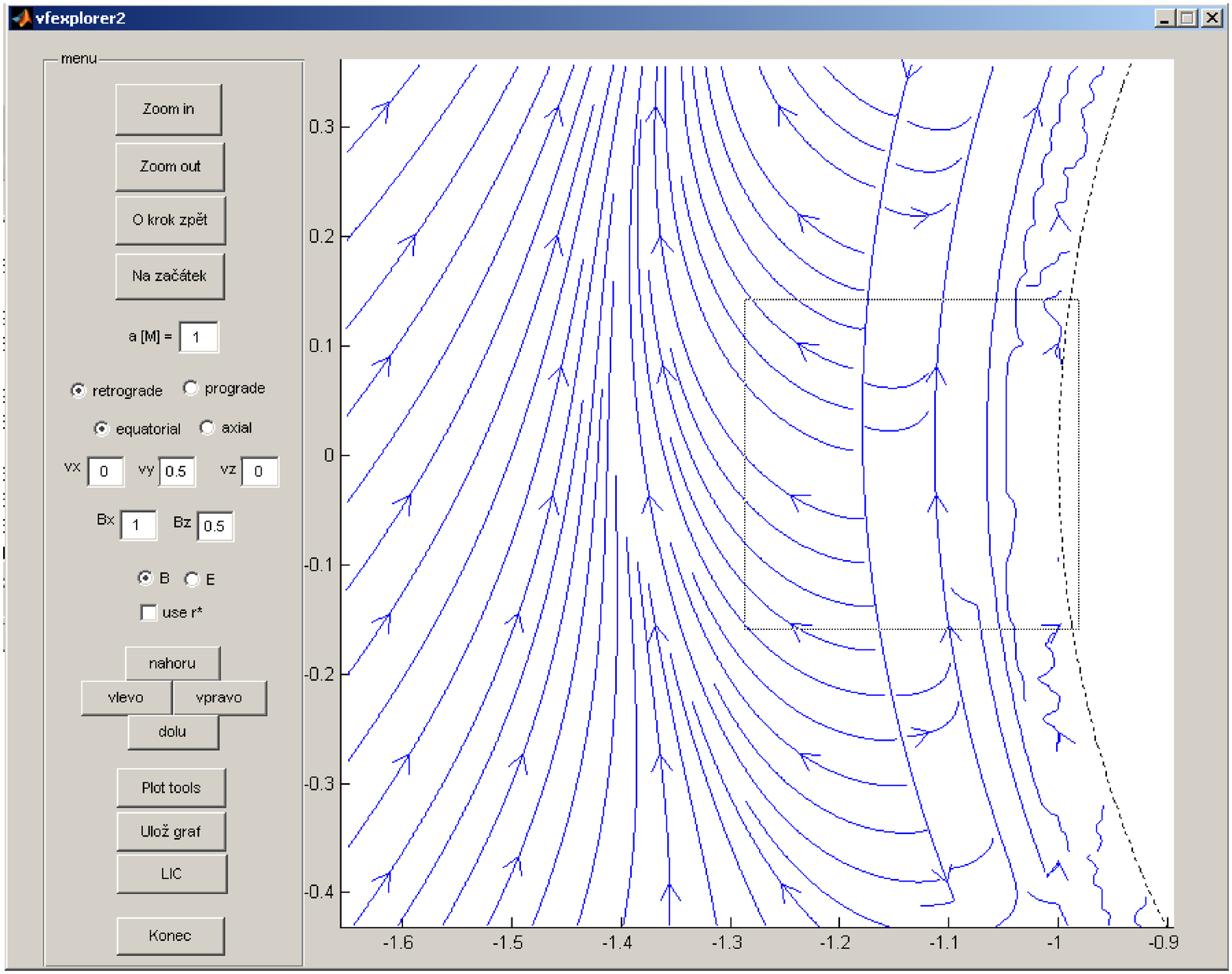}\vspace{3mm}
\includegraphics[scale=.4,clip]{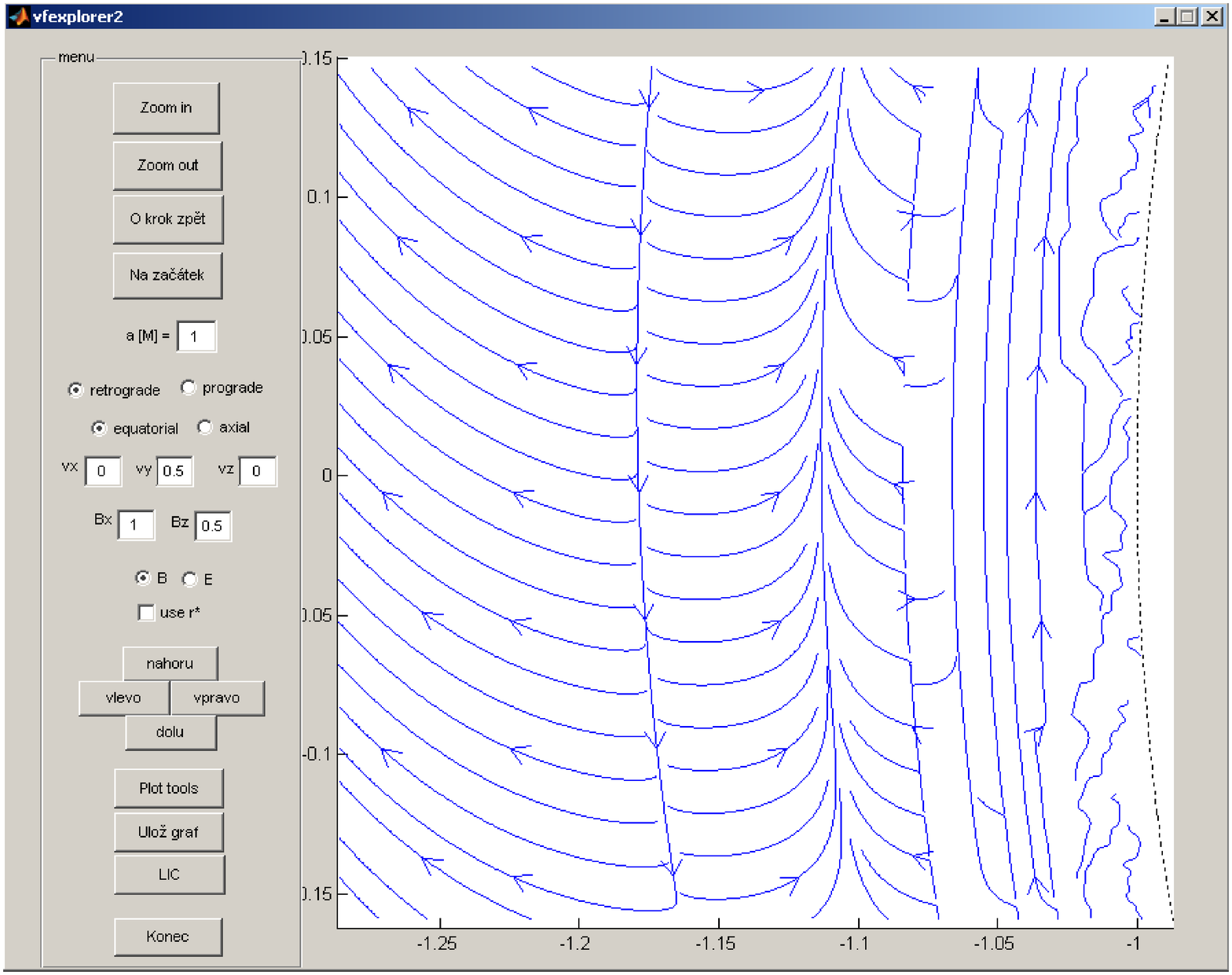}
\caption{Survey of the complex magnetic field above the drifting extremal Kerr BH using the {\it vfexplorer2} GUI. Tightly layered structure near the horizon observed in the upper panel is zoomed in the bottom one.}
\label{vfexp}
\end{figure}